\documentclass[oneside,titlepage]{style/cernyrep}
\usepackage[utf8]{inputenc}
\DeclareUnicodeCharacter{2212}{-}
\counterwithin{table}{section}
\counterwithin{figure}{section}
\counterwithin{equation}{section}

\usepackage[table]{xcolor}

\usepackage{comment} 
\usepackage{amsmath}
\usepackage{color}
\usepackage{parskip} 
\usepackage{array}

\usepackage{graphicx}
\usepackage{caption}
\usepackage{subfig}
\usepackage{subfloat}
\usepackage{subcaption}

\usepackage{multirow}
\usepackage{booktabs}
\usepackage{adjustbox}
\usepackage{rotating}
\usepackage{siunitx}
\DeclareSIUnit\clight{\text{c}}

\usepackage{nth} 

\usepackage[backend=biber,sorting=none, style=numeric-comp]{biblatex}

\usepackage{pdfpages}

\usepackage{epstopdf}
\usepackage{acronym}
\usepackage{balance}
\usepackage{authblk}
\usepackage{fancyhdr}
\usepackage{afterpage}
\usepackage{csvsimple}
\usepackage{pgfsys}

\usepackage{todonotes}

\usepackage{tcolorbox}

\usepackage{lineno}
\usepackage{placeins}

\usepackage{emptypage}
\usepackage{lipsum}

\usepackage{texnames}
\usepackage{ctable}
\usepackage[T1]{fontenc}

\usepackage{blindtext}

\usepackage[bookmarks, colorlinks=true, linktoc=page, pdftex, linkcolor=blue, citecolor=blue, urlcolor=blue]{hyperref}
\usepackage{float}

\usepackage{times,lipsum}
\usepackage[margin=1in]{geometry}
\usepackage[onehalfspacing]{setspace}

\usepackage{arydshln} 
\usepackage{makecell}
\usepackage{listings}
\usepackage{placeins} 

\definecolor{cornflowerblue}{rgb}{0.39, 0.58, 0.93}
\usepackage{moresize}         

\usepackage[hang,flushmargin]{footmisc}
\fancypagestyle{plain}{}
\newlength{\oddmarginwidth}
\newlength{\evenmarginwidth}
\fancyhfoffset[LO,RE]{\oddmarginwidth}
\fancyhfoffset[LE,RO]{\evenmarginwidth}
\bibliography{main} 

\title{\vspace*{0.5cm}\HUGE The Muon Collider\vspace*{0.2cm}

\normalsize Supplementary report to the European Strategy for Particle Physics - 2026 update}
\author{\Large \textit{The International Muon Collider Collaboration}\\
\vspace*{1cm}
\large
\textbf{The most up-to-date version of this document can be found at:}
\url{https://edms.cern.ch/document/3284682/1}    
}
\begin{abstract}
Muons offer a unique opportunity to build a compact high-energy electroweak collider at the 10~TeV scale.  A Muon Collider enables direct access to the underlying simplicity of the Standard Model and unparalleled reach beyond it. 
It will be a paradigm-shifting tool for particle physics representing the first collider to combine the high-energy reach of a proton collider and the high precision of an electron-positron collider, yielding a physics potential significantly greater than the sum of its individual parts. A high-energy muon collider is the natural next step in the exploration of fundamental physics after the HL-LHC and a natural complement to a future low-energy Higgs factory. 
Such a facility would significantly broaden the scope of particle colliders, engaging the many frontiers of the high energy community.

The last European Strategy for Particle Physics Update and later the Particle Physics Project Prioritisation Panel in the US requested a study of the muon collider, which is being carried on by the International Muon Collider Collaboration. In this comprehensive document we present the physics case, the state of the work on accelerator design and technology, and propose an R\&D project that can make the muon collider a reality.
\\ \\ \\ \\ \noindent
\includegraphics*[width=6.6cm]{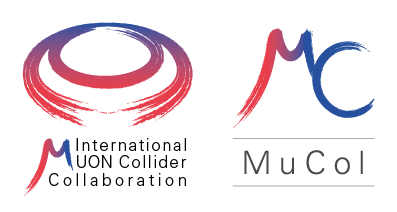}\includegraphics[width=3.3cm]{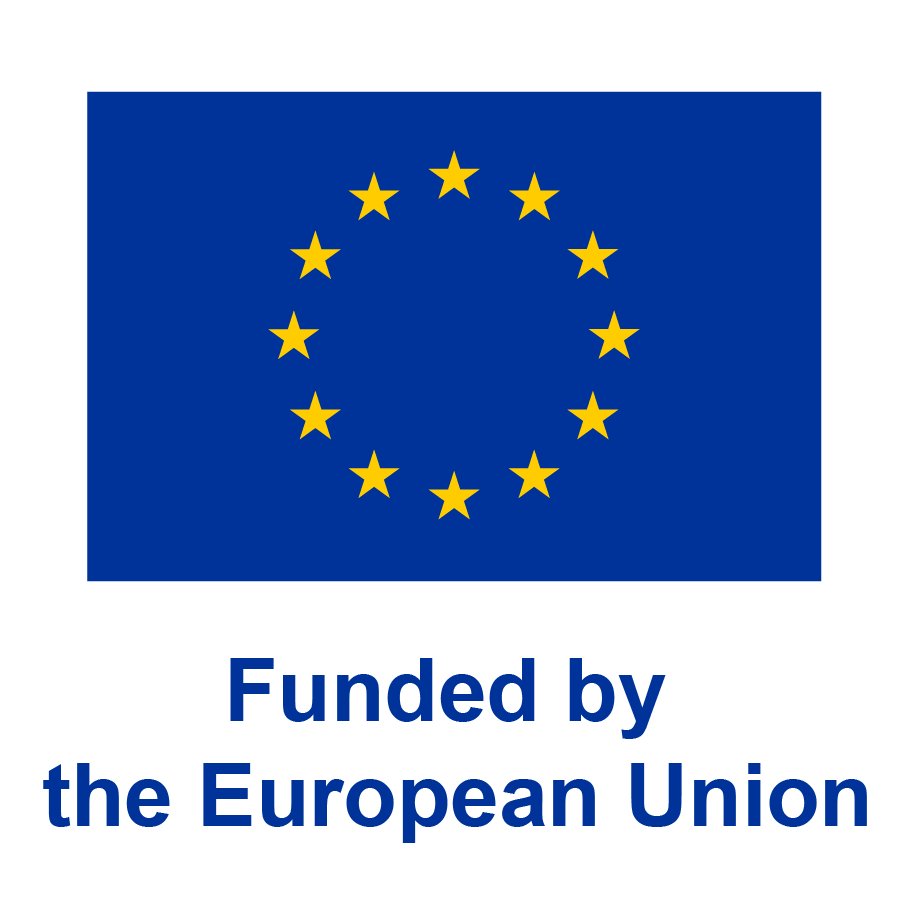}
\end{abstract}

\hypersetup{pdftitle={Muon Collider ESPPU 2026}, pdfauthor={The International Muon Collider Collaboration}}

\begin{document}
\setcounter{tocdepth}{1}

\setlength{\unitlength}{1mm}
\maketitle
\tableofcontents

\clearpage
\pagenumbering{arabic}
\setcounter{page}{1}

\newpage

\vskip 10cm
\begin{center}
{\huge\bfseries
IMCC authors \& supporters }

\end{center}

\vspace{5mm}
{\small
\noindent
Carlotta~Accettura$^{1}$, 
Simon~Adrian$^{2}$, 
Rohit~Agarwal$^{3}$, 
Claudia~Ahdida$^{1}$, 
Chiara~Aime'$^{4, 5}$, 
Avni~Aksoy$^{6, 1}$, 
Gian Luigi~Alberghi$^{7}$, 
Siobhan~Alden$^{8}$, 
Luca~Alfonso$^{9}$, 
Muhammad~Ali$^{10, 11}$
% ...
} 

\vspace{3mm}

\begin{flushleft}

{\em\footnotesize
$^{1}$ CH - CERN, \\ 
$^{2}$ DE - UROS, University of Rostock, \\ 
$^{3}$ US - LBL, Lawrence Berkely National Laboratory, \\ 
$^{4}$ IT - INFN - Pavia,  Istituto Nazionale di Fisica Nucleare Sezione di Pavia, \\ 
$^{5}$ IT - INFN - Pisa, Instituto Nazionale Di Fisica Nucleare - Sezione di Pisa, \\ 
$^{6}$ TR - Ankara University , \\ 
$^{7}$ IT - INFN - Bologna, Instituto Nazionale Di Fisica Nucleare - Sezione di Bologna, \\ 
$^{8}$ UK - RHUL, Royal Holloway and Bedford New College, \\ 
$^{9}$ IT - INFN - Genova, Istituto Nazionale di Fisica Nucleare Sezione di Genova, \\ 
$^{10}$ IT - INFN - Bari, Instituto Nazionale Di Fisica Nucleare - Sezione di Bari, \\ 
$^{11}$ IT - UNIPD, Universit\`a degli Studi di Padova , \\ 
% ...
}
\end{flushleft} 
 
\part{Executive Summary} \label{1:exsum}
\chapter{Introduction and overview} 
\label{0:intro:sec}

The muon collider is a unique, compact, high-energy electroweak collider concept which will produce, cool, accelerate and collide two single-bunch muon beams of opposite charge.
It is a paradigm-shifting tool for particle physics representing the first collider to combine the high-energy reach of a hadron collider and the high precision of a lepton collider, yielding a physics potential significantly greater than the sum of its individual parts.

The muon collider potential motivates a growing community, with substantial international interest and enthusiasm, with many calling for the scientific community to aim high and "\textit{shoot for the muon}".
The innovative nature of the muon collider in accelerator, detector and magnet technologies has attracted a vibrant community of early career researchers, many of whom are key designers of systems throughout the accelerator complex.
The innovative physics exploration methodologies and opportunities enabled by high energy muon collisions similarly revitalised particle collider theory and phenomenology.
These are important assets for the muon collider and for the fields of accelerator and particle physics in general.

This report contains the proposals and steps necessary to make the muon collider a reality, even within the next generation of high-energy particle physics facilities.

The report is submitted as input to the 2026 European Strategy for Particle Physics Update (ESPPU). Beyond this executive summary, it contains four further parts:
\begin{itemize}
    \item \textit{Part~\ref{2:evr}: Evaluation Report} is the most up to date status of the muon collider. This part presents a comprehensive assessment of the physics case and of the experiment and facility design. We have established what is available, and what requires further developments.
    \item \textit{Part~\ref{3:rd}: R\&D Proposal} describes the research \& development roadmap that is needed to bring the muon collider into reality. A ten-years R\&D plan is introduced with emphasis on simulation, design of detectors and accelerators, and technological demonstration.
    \item \textit{Part~\ref{4:impl} Implementation} describes the pathway to build the muon collider. It discusses the overall technically-limited timeline for the project, the cost drivers and scale and initial sustainability assessment. This part includes considerations of site-specific designs.
    \item \textit{Part~\ref{5:ppg} Physics Preparatory Group Benchmarks} further expands the physics case---including very recent results and original work---in answer to the questions posed by the Physics Preparatory Group (PPG) in the context of the 2026 ESPPU and provides detailed technical input to the PPG.
\end{itemize}

\newpage
\section{The muon collider concept}
The baseline muon collider design is a \SI{10}{\tera\electronvolt} centre-of-mass collider providing an integrated luminosity of \SI{10}{\atto\barn^{-1}} \cite{Accettura:2023ked}. An initial stage that can be implemented by around 2050 is also considered.

\begin{figure}[h]
\includegraphics[width=\textwidth]{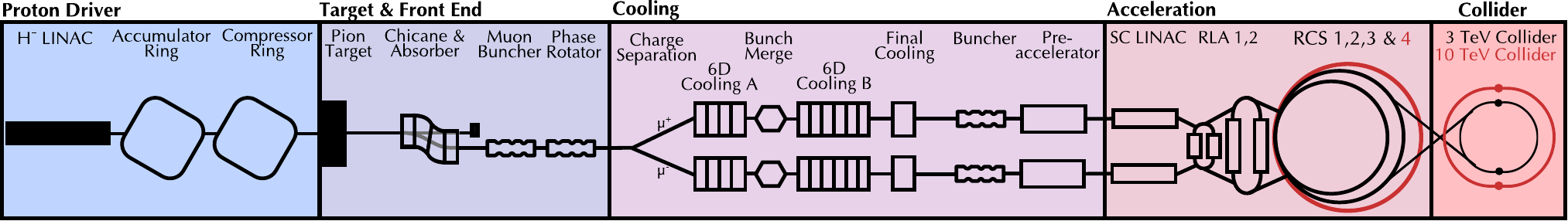}
\caption{Conceptual layout of the muon collider.}
\label{0:intro:fig:layout}
\end{figure}

The design of the muon collider is based on a concept, which was developed by the U.S. Muon Accelerator Programme (MAP) until 2017~\cite{MAP2}.
The design is now being progressed by the International Muon Collider Collaboration (IMCC) \cite{InternationalMuonCollider:2024jyv}. A schematic layout of the collider is shown in Fig.~\ref{0:intro:fig:layout} and
contains the following key areas:
\begin{enumerate}
    \item The \textbf{proton driver} (blue box in the diagram) produces a short, high-intensity proton pulse.
    \item This pulse hits the \textbf{target} (indigo) and produces pions. The decay channel guides the pions and forms a beam with the resulting muons via a buncher and phase rotator system.
    \item Several \textbf{cooling} stages (purple) reduce the longitudinal and transverse emittance of the beam using a sequence of absorbers and RF cavities in a high magnetic field.
    \item A system of a linac and two recirculating linacs \textbf{accelerate} (light red) the beams to 63\,GeV followed by a sequence of high-energy accelerator rings which reach \SI{1.5}{\tera\electronvolt} or \SI{5}{\tera\electronvolt}.
    \item Finally the beams are injected at full energy into the \textbf{collider} ring (red). Here, they will circulate and collide within the detectors until they decay.
\end{enumerate} 

A set of parameters has been defined for \SI{3}{\tera\electronvolt} and \SI{10}{\tera\electronvolt} centre-of-mass collisions. They are reported in Tables~\ref{t:facility_param} and~\ref{t:beam_param} respectively for the collider and beam parameters.
These are target parameters to explore the limits of each technology and design. 
If they can be fully met, the integrated luminosity goal could be reached within five years (or $2.5$~years, with two detectors) of full luminosity operation. 
This provides margin for further design and technology studies and a realistic ramp-up of the luminosity.
A benefit of this design is that it enables initial stages which can be implemented faster, but with stages of reduced luminosity performance.

\begin{table}[!hb]
\caption{Tentative target parameters for a muon collider at different
  energies. Scenario 1 corresponds to Energy Staging, and Scenario 2 corresponds to Luminosity Staging. Both are defined in Section~\ref{s:stagingSCD}. The estimated luminosity refers to the value that can be reached if all target specifications can be reached, including beam-beam effects.}
\label{t:facility_param}
\centerline{
  \begin{tabular}{|*3{c|}|*2{c|}|*2{c|}|}
    \hline
    Parameter & Symbol & Unit & \multicolumn{2}{c||}{Scenario 1} & \multicolumn{2}{c||}{Scenario 2}\\
 & & & Stage 1 &Stage 2 & Stage 1 & Stage 2\\
\hline
    Centre-of-mass energy & $E_{\mathrm{cm}}$ & TeV & 3 & 10 & 10 &10\\
    Target integrated luminosity & $\int{\cal L}_{\mathrm{target}}$ & $\rm ab^{-1}$ & 1 & 10 &10 & 10\\
    Estimated luminosity & ${\cal L}_{\mathrm{estimated}}$ & $10^{34}\rm cm^{-2}s^{-1}$ & 2.1 & 21 & 5 (tbc) & 14\\
    Collider circumference& $C_{\mathrm{coll}}$ & $\rm km$ & 4.5 & 10 & 15 &15\\
    Collider arc peak field& $B_{\mathrm{arc}}$ & $\rm T$ & 11 & 16 & 11 & 11\\
    Luminosity lifetime & $N_{\mathrm{turn}}$ &turns& 1039 & 1558 & 1040 & 1040\\
    \hline
    Muons/bunch & $N$ & $10^{12}$ & 2.2 & 1.8 & 1.8 &1.8\\
    Repetition rate & $f_{\mathrm{r}}$ & $\rm Hz$ & 5 & 5 &5 &5\\
    Beam power  & $P_{\mathrm{coll}}$ & $\rm MW$ &5.3  & 14.4 & 14.4 &14.4\\
    RMS longitudinal emittance& $\varepsilon_\parallel$ & $\rm eVs$ & 0.025 & 0.025 & 0.025 &0.025\\
    Norm.\,RMS transverse emittance& $\varepsilon_\perp$ & \textmu m & 25 & 25 & 25 &25\\
    \hline
    IP bunch length& $\sigma_z $ & $\rm mm$ & 5 & 1.5 & tbc &1.5\\
    IP betafunction& $\beta $ & $\rm mm$ & 5 & 1.5 & tbc & 1.5\\
    IP beam size& $\sigma $ & \textmu m & 3 & 0.9 & tbc & 0.9\\
    \hline
    Protons on target/bunch & $N_{\mathrm{p}}$ & $10^{14}$ & 5 & 5 & 5 & 5\\
    Proton energy on target  & $E_{\mathrm{p}}$ & $\rm GeV$ & 5 & 5 & 5 & 5\\
    \hline
  \end{tabular}
}
\end{table}

\begin{table}
  \caption{Tentative target beam parameters along the acceleration chain.
    A 10~\% emittance growth budget has been foreseen in the transverse and
    longitudinal planes, both for 3 and 10 TeV. This assumes that the
    technology and tuning procedures will have been improved between the two
    stages. The very first acceleration is assumed to be part of the final cooling. This choice allows optimisation of the energy in the last absorber with no strong impact on the acceleration chain.}
  \label{t:beam_param}
\begin{center}
  \begin{tabular}{|*6{c|}}
    \hline
    Parameter & Symbol & Unit &  Final cooling & at 3 TeV & at 10 TeV \\
    \hline
    Beam total energy & $E_{\mathrm{beam}}$ & GeV & 0.255 & 1500 & 5000 \\
    \hline
    Muons/bunch & $N_{\mathrm{b}}$ & $10^{12}$ & 4 & 2.2 & 1.8 \\
    Longitudinal emittance& $\varepsilon_\parallel$ & $\rm eVs$ & 0.0225 & 0.025 & 0.025 \\
    RMS bunch length& $\sigma_z$ & $\rm mm$ & 375 & 5 & 1.5 \\
    RMS rel.\,momentum spread& $\sigma_P/P$ & $\rm \%$ & 9 & 0.1 & 0.1 \\
    Transverse norm.\,emittance& $\varepsilon_\perp$ & \textmu m & 22.5 & 25 & 25 \\
    \hline
    Aver.\,grad.\,$0.2$--$1500\rm\,GeV$&$G_{\mathrm{avg}}$&$\rm MV/m$& --- & 2.4 &\\
    Aver.\,grad.\,$1.5$--$5\rm\,TeV$&$G_{\mathrm{avg}}$&$\rm MV/m$& --- &  & 1.1\\
    \hline
  \end{tabular}
\end{center}
\end{table}

\section{Physics Case}
\label{0:intro:phys}
A muon collider is a high-energy electroweak collider. It can access directly and precisely the underlying simplicity of the Standard Model (SM) in its high-energy regime of “unbroken” electroweak symmetry. This makes it an ideal and unique machine to investigate fundamental questions about our universe – both long-held ones and more recent ones sparked by the LHC.  Simultaneously, it is a paradigm-shifting tool for particle physics representing the first high-energy, high-precision compact collider.  It combines the precision of a lepton collider with the energy reach of a hadron collider, yielding a physics potential significantly greater than the sum of its individual parts. This enables unparalleled exploration of the Electroweak-Higgs Unification era that we have entered since the discovery of the Higgs.   We can now hope to answer the question of why electroweak symmetry breaking (EWSB) occurs by directly probing the transition between the “broken” and ``unbroken” symmetry regimes.  Furthermore, we can address the phase diagram of EWSB and quantitatively investigate the earliest moments in our universe and its ultimate fate. With high-energy EW collisions we can also search for physics beyond the SM directly using a muon collider as an unrivaled EW discovery machine.  These same high-energy high precision collisions allow us also to probe new physics to scales far beyond the collider’s energy and study the Standard Model in new domains where new phenomena emerge. 

For example, a muon collider can make exquisite measurements of TeV-scale vector boson scattering.  This is due to the inherently quantum and relativistic effect of abundant effective vector bosons contained in a high-energy muon, which gives rise to large collision rates with low backgrounds.  Such measurements enable the first precision experimental tests of the ``unitarization” of massive gauge boson scattering, arguably the foremost prediction of Electroweak symmetry breaking.  High-energy, high-precision studies of the Higgs and vector bosons also give the first detailed probes of the ``Electroweak symmetry restoration” realm within the SM.  The high-energy nature of the muon collider also enables new insights into, and tests of, the “quantum compositeness” of particles.  This is ideally studied in Electroweak processes due to their perturbative and thus in principle calculable nature (unlike the strong interactions), and the physical mass gap that makes ``particles'' fully well defined as asymptotic states unlike in Quantum Electrodynamics (QED).  Finally, the intrinsic nature of a high-energy muon collider provides the most abundant and best characterized source of high-energy neutrino-target collisions conceived thus far. The neutrino physics program at a muon collider provides a high-energy complement to current and future long-baseline neutrino experiments.

\begin{figure}[ht]
    \centering
    \includegraphics[width=0.56\linewidth]{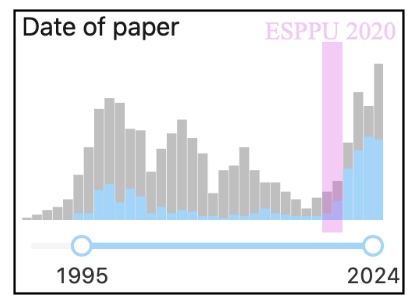}
    \includegraphics[width=0.43\linewidth]{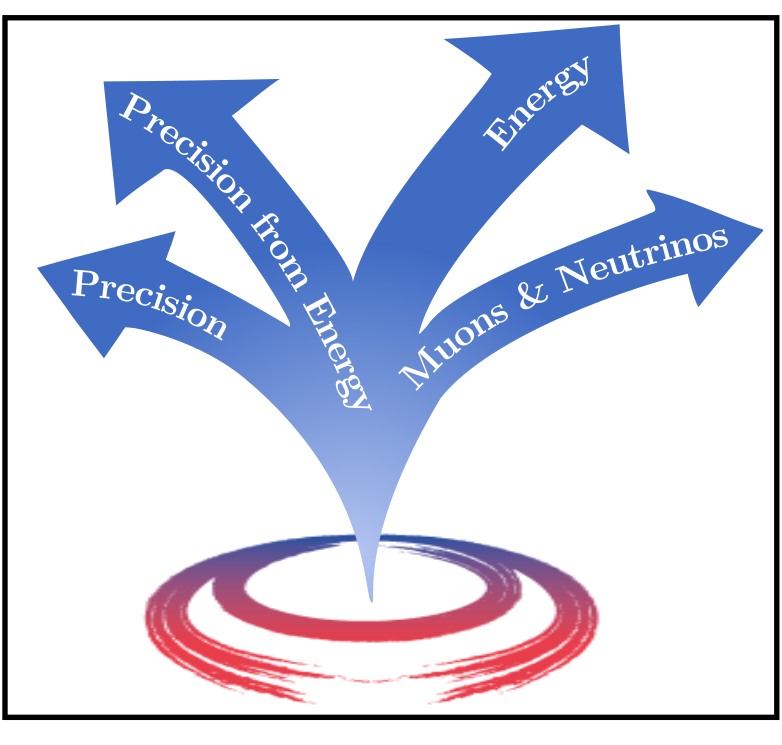}
    \caption{The ground-breaking nature of the muon collider physics perspectives and its contemporary pertinence in the particle physics landscape after the LHC is well represented by the enthusiastic reaction of the theory community to the muon collider study plans initiated by the 2020~ESPPU. The number of papers on muon colliders in the hep-ph category is reported in blue on the left panel. The right panel summarises the core directions of the physics case.
    \label{fig:theory}}
\end{figure}

A muon collider simultaneously offers numerous pathways to searching for Beyond Standard Model (BSM) physics by utilizing the energy reach, precision measurement capabilities, and the combination thereof. 
The same abundant vector boson scattering that enables exploration of new SM phenomena also furnishes a next-generation Higgs factory. In certain channels this allows an order of magnitude or more improvement in the understanding of Higgs properties, including its potential.  This allows one to experimentally probe BSM contributions that could change the nature of the EW phase transition and baryogenesis, the ultimate fate of our vacuum and the ``Higgs portal” to hidden sectors.  Furthermore the energy reach of a muon collider allows one to test the origin of any deviation from SM properties found in a precision measurement at the same collider, and by extension any previous Higgs Factory. The combination of precision and energy also allows one to probe BSM possibilities such as ``Higgs compositeness” up to the $\mathcal{O}(100)$ TeV scale, far beyond the reach of any other proposed collider.  Reaching the 10 TeV scale directly also enables unmatched probes of new EW particles such as those responsible for the simplest dark matter paradigms, or alternatively probing the possibility of new gauge forces in many cases to unrivaled scales.  The combination of energy and precision also revolutionises flavor physics by enabling direct tests of fermions interactions, with a new physics scale reach that is comparable or superior to the one attainable by traditional studies of meson or leptons decays. This includes neutral current flavor tests sensitive to mediators at the 100 TeV scale, while also enabling new windows into SM flavor in the Higgs sector and potential explanations of neutrino masses with Heavy Neutral Lepton searches.

The impressive physics potential of a muon collider is enabled by both its energy reach as well as its high luminosity.  The luminosity goal of 10 TeV with 10~ab$^{-1}$ is naturally consistent with the emittance targets from cooling and the inherent increase of luminosity power efficiency at high-energy.  While most studies have been performed for these parameters, it's important to note that the bulk of the physics potential is only mildly affected by a slight change in energy or an order of magnitude change in luminosity. Furthermore, a muon collider is an innately stageable project in both luminosity and energy given a common front end for muon production and cooling.  With a staged approach, considerable technical and financial risk could be retired and  physics progress can be achieved along the path to the highest energies. Low energy stages could provide direct measurements of the Higgs width or experimentally and theoretically precise measurements of the top mass. Such low-energy muon colliders could be also considered as an independent satellite project at the muon collider complex. A 3 or 3.2 TeV stage can offer the first significant step beyond the HL-LHC in understanding the Higgs potential as well as testing certain Dark Matter (DM) candidates. A 7.6 TeV stage at CERN would extend our understanding of EWSB even further allowing a differential test of EW restoration, especially relevant if luminosity goals are relaxed in the pursuit of the highest energies.  Furthermore, while 10 TeV is the ultimate goal of this study, it is not the final goal of particle colliders nor the highest conceivable energy of a muon collider in the far future.  

Successfully building a muon collider allows us to reset the collider landscape with this paradigm shifting tool.   Muon collider investment now lays the groundwork for coming generations to explore nature {\em directly} at even shorter distances, in likely the only sustainable and technically practical way for the future of High-Energy Physics (HEP).

\section{Detectors}
The design of dedicated experiments to take data at the collider interaction points has sparked a lively environment that allowed the community to make fast progress in just a few years. The detector design is still in its infancy, but it is already possible to make substantive statements about the technological requirements, expected performance, and opportunities for further improvement.

Requirements were spelled out in terms of detector acceptance, particle detection and identification efficiency, as well as to resolutions on the various particle properties inferred by the instrumental measurements. They were outlined in terms of ``baseline'' and ``aspirational'' targets, corresponding, respectively, to the requirements to fully exploit the physics potential of the machine and a more ambitious set of performances comparable to those targeted by Higgs/Top/Electroweak-factories. These targets are summarised in Table~\ref{tab:detector_req_ex}. 

\begin{table}[ht]
\begin{center}
\caption{Preliminary summary of the ``baseline'' and ``aspirational'' targets for selected key metrics for a $10$~TeV muon collider.}
\label{tab:detector_req_ex}
\begin{tabular}{lcc}
\hline\hline
 \textbf{Requirement} & \textbf{Baseline} & \textbf{Aspirational}\\
\hline
Angular acceptance $\eta=-\log(\tan(\theta/2))$ & $|\eta|<2.5$ & $|\eta|<4$ \\
Minimum tracking distance [cm] & $\sim 3$ & $< 3$\\
Forward muons ($\eta> 5$) & tag & $\sigma_{p}/p \sim 10$\%\\
Track $\sigma_{p_T}/p^{2}_{T}$ [GeV$^{-1}$] & $4 \times 10^{-5}$ & $1 \times 10^{-5}$ \\
Photon energy resolution & $0.2/\sqrt{E}$ & $0.1/\sqrt{E}$\\
Neutral hadron energy resolution  & $0.4/\sqrt{E}$ & $0.2/\sqrt{E}$ \\
Timing resolution (tracker) [ps]  & $\sim 30-60$ & $\sim 10-30$ \\
Timing resolution (calorimeters) [ps]  & 100 & 10 \\
Timing resolution (muon system) [ps] & $\sim 50$ for $|\eta|>2.5$ & $< 50$ for $|\eta|>2.5$ \\
Flavour tagging & $b$ vs $c$ & $b$ vs $c$, $s$-tagging \\
Boosted hadronic resonance identification & $h$ vs W/Z & W vs Z\\
\hline\hline
\end{tabular}
\end{center}
\end{table}

\begin{figure}[h]
    \centering
    \includegraphics[trim={0 0 2cm 0},clip,width=0.49\linewidth]{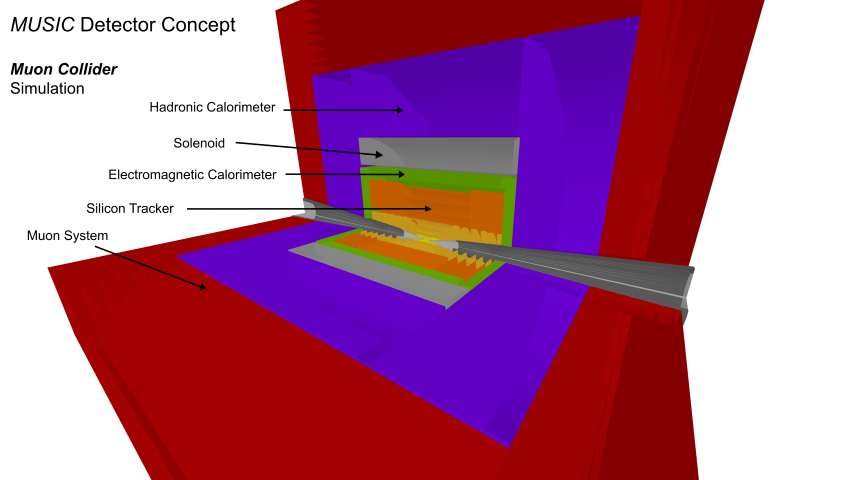} \includegraphics[trim={0 0 3.8cm 0},clip,width=0.49\linewidth]{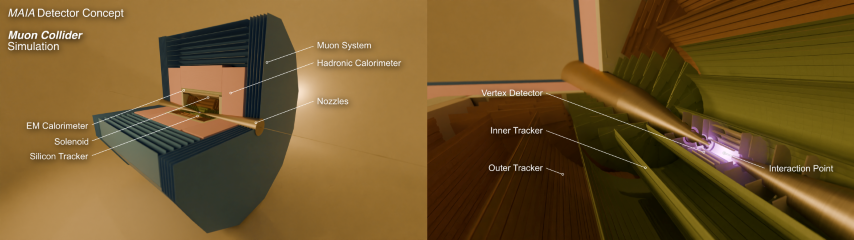}
    \caption{Layout of the MUSIC (left) and MAIA (right) detector concepts
    \label{fig:detector_concepts_ex}}
\end{figure}

An initial detector, optimised for operation at a $\sqrt{s}=3$~TeV muon collider, was extensively studied and demonstrated the feasibility of the physics programme. Recent work focused on developing the first detector designs for a $\sqrt{s}=10$ TeV machine. Two distinct detector concepts have been developed: MUSIC (MUon System for Interesting Collisions) and MAIA (Muon Accelerator Instrumented Apparatus).
Both designs, shown in Figure~\ref{fig:detector_concepts_ex}, envision a multi-purpose detector sharing a similar structure: a cylinder $11.4$~m long with a diameter of $12.8$~m. They comprise typical collider detector instrumentation, such as an all-silicon tracking system, an electromagnetic calorimeter (ECAL), a hadron calorimeter (HCAL), and a muon sub-detector. Superconducting solenoids are used to provide bending power for the measurement of charged particle momenta within the tracking system. The design work follows the concept already developed for $\sqrt{s}=3$ TeV, with modifications to account for the higher energy. The main differences between the two detector concepts lie in the placement of the solenoid (after the tracker in MAIA and between ECAL and the HCAL in MUSIC) and the technologies selected for the ECAL (Si-W for MAIA and semi-homogeneous crystals for MUSIC).

High levels of beam-induced background in the detector pose unprecedented challenges for the reconstruction and identification of particles produced in muon collisions. 
Studies based on detailed simulations of the MAIA and MUSIC detector concepts were carried out. These studies used state-of-the-art FLUKA simulations to account for the beam backgrounds produced by the decays in the latest accelerator lattice, and GUINEA-PIG predictions to study the effects of incoherent pair production at the interaction point. The detector response was studied to develop initial background-mitigation measures. In both cases, the results indicate that the background effects on the detector response can be minimised to a degree that approaches the aspirational goals.

Research and development over the next decade must focus on developing the necessary technologies, tools, and crucial expertise to design and construct state-of-the-art detectors for this future machine. At the muon collider, most detector components must simultaneously optimize position resolution, timing capability, radiation hardness, data transmission, and on-detector background rejection, all while maintaining low-mass and low-power consumption. Many of these technical challenges align with ongoing research in other initiatives, such as the LHC and Higgs/Top/EW-factory R\&D programs. However, the development effort must also address challenges unique to the muon collider, including its harsh radiation environment, the need to suppress significant beam-induced backgrounds, and the requirements for high-precision calorimetry to measure significantly higher energy physics objects.

The R\&D program is organised around three main areas: simulation and performance, technology and computing. In technological R\&D, multiple technologies will be investigated in the first phase of the work. The choices are expected to consolidate after the first four to six years and resources will transfer to the identified technologies needed for the chosen detectors designs. Coverage across areas for items relevant for the same sub-detectors is also ensured. Among all tasks that have been identified, ASICs and detector magnets have been identified as critical components requiring extended development timelines and should therefore be prioritized to ensure readiness for detector construction.

\section{Readiness and status}
The IMCC, the Muon Beam Panel of the Laboratory Directors Group (LDG) and the Snowmass process in the U.S. have all assessed the muon collider challenges with the support of the global community~\cite{Adolphsen:2022ibf, Black:2022cth}.
Key conclusions are that, although the muon collider concept is less mature than several linear collider concepts, no insurmountable obstacles have been identified, and that important design and technical challenges have to be addressed with a coherent international effort. 
Furthermore, past work, in particular within the U.S.\,Muon Accelerator Programme (MAP)~\cite{MAP2}, has demonstrated several key technologies and concepts, and gives confidence that the overall Muon Collider concept is viable. 
Since then further component designs and technologies have been developed that provide increased confidence that one can cool the initially diffuse beam and accelerate it to multi-TeV energy on a time scale compatible with the muon lifetime. 
However, a fully integrated design has yet to be developed and full demonstrations of technology are required. 

Following the last European Strategy, the IMCC has been formed with the goal to establish whether the investment into an important R\&D
programme on the muon collider is justified. A prioritised work programme has been developed by the LDG with
this goal and is being implemented by IMCC. Resources have been made by the 58 member institutions, several other partners and the European
Union. Following discussions with DoE representatives, an addendum to the CERN-DoE agreement is prepared and currently being reviewed by DoE.
At this point, progress in several areas is limited by resources and additional resources are being sought across the collaboration, in parallel
with the processes to include new partners. The R\&D programme thus addresses priority challenges.

The IMCC programme prepares the way towards a conceptual design report (CDR) and a demonstration programme for a muon collider. The IMCC studies physics potential, detector design, accelerator design and performance studies. Studies assess technology maturity and develop critical designs to understand key challenges.

As mentioned above {\bf the demonstrator design at CERN} is progressing and detailed studies of an implementation on CERN site - i.e. in the TT7 tunnel - is ongoing.

The collaboration has developed a baseline parameter set and designed critical components for the muon collider facility. Two detector designs have begun. The current level of R\&D resources has not enabled a fully integrated lattice design. Study of critical technologies has begun and a small set of hardware tests have been performed.

A {\bf second detector design study}, MAIA, has started in addition to an improved and dedicated MUSIC design at 10~TeV. 
Both experiments' design is constrained by beam induced background. Shielding reduces the radiation to levels similar to HL-LHC. Full simulation studies with background, including beam-beam pair production, indicate that the most relevant physics channels can be studied with near-future technology, in many cases thanks to developments for HL-LHC. Further improvement will be possible with more computing power, improved algorithms and better technology. 

{\bf Beam dynamics studies} are progressing.
Two tentative muon cooling system designs now promise a transverse emittance of $<25\;\rm \mu m$, close to the target of $<22.5\;\rm \mu m$. This is a marked improvement
compared to the previous performance of $55\;\rm \mu m$ obtained by the MAP study. For one design, the longitudinal emittance is still slightly above the target (88 vs 64 mm). For the other design it is significantly below (38 mm).
An initial collider ring lattice design achieves the target beta-functions but its energy acceptance is a factor
2--3 below the goal. Further system design to improve the acceptance is ongoing. The muon cooling system with smaller longitudinal emittance would almost half the required acceptance.

A first preliminary estimate of the total {\bf transmission of muons} predicts $1.5\times10^{12}$ at the collision point,
which is 20\% below the target.
The transmission in the muon cooling section is below the initial goal, while it is above in the high-energy part of the complex.
Further studies will address this aiming to improve the sub-system transmission and to increase the initial number of
muons with a higher power target.

In the Rapid Cycling Synchrotrons (RCS), initial assessment shows that the RF cavities could be distributed around the arcs and dispersive effects
could be mitigated by carefully choosing the phase advance between them. Detailed lattice design is ongoing. This gives
more flexibility implementing the {\bf RCS in an existing tunnel} and it would also reduce the small mismatch of the ramping
magnet field strength and the beam energy. This may allow reduction of the injection energy of the first RCS and could reduce the
cost of the initial linacs.

Parameters for the \textbf{proton driver} were developed based on existing facilities and simulations of key systems for a high-intensity 5~GeV or 10~GeV beam. Heat and radiation load in the \textbf{target} area was assessed for 2~MW proton beam power; a target and shielding design was developed that adequately protects the capture solenoid and limits damage to the graphite target rod. Extraction of the spent proton beam at the target has been identified as a potential technical issue. 

The \textbf{muon cooling} system reduces the 6D beam size (emittance) to improve luminosity, and has been optimised from previous designs. The system delivers transverse and longitudinal emittances of \SI{22.5}{\micro\meter} and \SI{7}{\mega\electronvolt\per\second} respectively. The longitudinal emittance is almost a factor 3 better than the target parameters but the transmission is 20 \% lower than target. Optimisation continues by matching between cooling stages, integrating components within the cooling cell and improving the longitudinal beam capture section. Beam-loading and heat-load on the hydrogen absorbers have been identified as technical issues.

Low-energy \textbf{acceleration} solutions have been found for the LINAC and the second recirculating linear accelerator. Simulation of the high-energy acceleration has been performed including single-turn and multi-turn wakefields and with counter-rotating beams.The lattice exceeds the target values for transmission. Suitable emittance control has been demonstrated in the absence of errors. Lattice errors may impact the beam quality and this is a focus of ongoing studies. 

The \textbf{collider} ring at 10~TeV centre-of-mass energy and $\beta^*$=1.5 mm must deliver a short bunch with a large energy spread to avoid luminosity dilution from the hour-glass effect. Challenges arise from controlling chromatic aberration in the ring despite the 0.1\% RMS momentum spread. The current $\beta^*$ lattice has a reduction in energy acceptance compared to the target parameters. Magnet misalignments may reduce performance further, especially as the magnets will be periodically moved $\pm$1~mrad, to dilute the neutrino flux. Designs are ongoing to improve energy acceptance. The potential improvement in longitudinal cooling may alleviate the shortfall.

The \textbf{magnet team} has developed conceptual designs for critical magnet systems, including: a 1.2~m bore and 20~T\textit{ target solenoid};\textit{ cooling solenoids} with achievable values for radial and hoop stress, stored energy, current density and fields up to 40~T; \textit{rapidly pulsed normal-conducting} $\pm$1.8~T magnets for the accelerator system with high-efficiency power converters that require acceptable wall-plug power and; \textit{high-field dipoles}, quadrupoles and combined-function magnets including appropriate shielding from muon decay electrons. A resource-loaded programme for magnet hardware development, which is on the muon collider facility critical path, has been assessed and is discussed below.

The \textbf{RF team} has developed 352 and 704~MHz normal conducting RF cavity designs for the cooling system. The \textit{cooling RF cavities} are challenging owing to tight integration with the solenoids and up to 32~MV/m required electric field. Power couplers have been developed to minimise interference with neighbouring equipment.\textit{ High-energy acceleration cavities} have been developed assuming 1.3 GHz RF structures. The required layout of RF stations and impact on the beam quality has been assessed. It has been integrated within the overall lattice. 

A \textbf{Muon Cooling Demonstration} programme has been proposed based on a typical section of the cooling channel. An integrated engineering design has been developed. Candidate sites at CERN, Fermilab and other laboratories have been identified. Concepts for layout and beam optics design of the beam transport sections have been assessed. The programme, including magnet and RF cavity R\&D, has been developed into a resource-loaded timeline. The Demonstrator is also on the muon collider facility critical path and implementation is discussed below.

{\bf Collective effects studies} are progressing. They show that impedance effects can be mitigated in the RCSs and collider ring. The same is true for beam-beam effects.
The implementation of muon cooling physics into a CERN tracking code is ongoing and will allow to combine single particle
effects in matter with collective effects.
Further efforts to implement all collective effects and develop a start-to-end model of the machine are planned.
The impact of imperfections and their mitigation need to be studied.

The overall facility design has progressed well given the available resources. The transmission and beam emittance has been estimated across the facility. Further development of the lattice design may reveal new issues, but also make possible a global optimisation for cost, power and performance. Some technical issues have been uncovered. None of the issues have impacted the concept feasibility although in some areas the performance relative to target parameters may be degraded. Innovative approaches have enabled the IMCC to exceed target parameters in other areas. {\bf Overall we are convinced that the target luminosity is achievable.} If the target luminosity can be met {\bf IMCC considers that we have a significant contingency in luminosity in hand.}

\section{R\&D plan}
IMCC proposes a comprehensive R\&D programme to reach the maturity required to initiate the approval process. 
The programme requires approximately 300 MCHF material budget and about 1800 FTEy of personnel for the accelerator and about 20 MCHF and 900 FTEy for detectors. 
With timely funding, the programme spans about 10 years. This would enable a first muon collider stage with a start of operation around 2050. Its could thus be the next flagship project in Europe in case no Higgs factory is realised at CERN. 
A slightly longer timescale is envisaged for an implementation in the U.S. taking into account budget constraints. 

The first phase of the R\&D programme contains completion of a start-to-end facility design. 
It also includes hardware development of components that drive the overall timeline for the collider such as the superconducting magnets and the muon cooling technology. After this phase a ramp-up of resources will enable a more detailed facility design to prepare for the start of the decision process. 
This process could start in 2036 and allow investing into industrialisation of the
components before project construction approval. The proposed technically limited timeline is feasible, contingent on the strong commitment from the community and the funding agencies to realise a muon collider at the earliest possible time. 

The proposed R\&D plan will achieve the following:
\begin{itemize}
\item Further development of the {\bf detector} will optimise the performance and minimise the impact of 
  beam induced background.
\item The {\bf muon cooling technology} demonstration programme will develop this novel technology. It will develop the components such as
  the HTS solenoids, the cavities and the absorbers. Test of a full cooling cell with RF power will allow verification of the
  integrated performance. The demonstrator facility will test several of the cooling cells with beam to demonstrate the technology.
\item An intense programme will establish the {\bf superconducting magnet} performance through the construction and test of
  models and protoypes. It will focus in particular on HTS solenoids for the muon production target and the muon cooling.
  These have strong synergy with applications in society such as fusion reactors. A model of the collider ring dipoles will
  also be constructed and establish the field at large aperture.
\item A {\bf start-to-end model} of the collider.
  The completion of the lattice design along the whole complex and its optimisation will guide the component development.
  The development of simulation tools that include the relevant beam physics, such as collective effects and imperfections as well
  as the relevant mitigation techniques, will enable robust luminosity predictions. A study of the machine availability
  will establish the integrated luminosity performance and guide component and accelerator design.
\item Experiments in combination with further design work will verify the {\bf target} robustness.
\item Conceptual designs of the superconducting and normal-conducting {\bf cavities} along the whole complex.
  Experimental verification of the performance limits will allow to optimise the complex design. In
  particular, the construction of an infrastructure to test RF in a high magnet field will enable
  experimental optimisation and verification of the normal-conducting muon cooling RF.
\item The development of high-power, high-efficiency {\bf klystrons} will be instrumental for the muon cooling cell test and enable cost effective design.
\item The performance of the {\bf fast-ramping magnet systems and power converter} for the Rapid-Cycling Synchrotrons (RCS) will be demonstrated.
\item {\bf Site and environmental impact studies}, including civil engineering, will allow optimisation for
  power consumption and material usage as well as minimising the impact of the machine for the local
  environment.
\item An {\bf overall optimisation} of the complex for cost, power consumption and risk will be performed and is particularly
  essential since we cannot base ourselves on experience with previous similar projects. This optimisation will also
  cross the boundaries between the different systems.
\end{itemize}
It will also be important to explore promising alternatives to the current baseline, e.g. the use of a Fixed-Field Accelerator (FFA)  in the proton complex to reduce the linac energy and cost.

\section{Synergies}
\label{1:synergies}
Particle physics and the associated accelerator and detector development have made important contributions to society; both in training of young people and in developing
technologies.

Many young people have developed their scientific and technical
skills in the field; they also learned to work in fully international
collaborations. Because the muon collider is a novel concept it opens opportunities for young
researchers to make original contributions to the development that are much
harder to make in long-established design approaches.

The muon collider needs technologies in several areas that differ
from other colliders. High-field solenoids are a prime example.
In the past low-temperature superconductors such as the very mature NbTi and
still developing Nb$_3$Sn were the technologies of choice for accelerators and
most other applications.
Now HTS are becoming an important technology.
In particular they are of interest for fusion reactors, that have similar requirements to the one for the muon collider target solenoid. Highly-efficient superconducting motors and power generators, e.g., for off-shore windmills, also have strong synergy. Other relevant areas are life and material sciences; in particular, applications exist for nuclear magnetic resonance (NMR) and magnetic resonance imaging (MRI). In addition synergy exists with magnets for neutron spectroscopy, physics detectors and magnets for other particle colliders, such as hadron colliders.

The muon production target is synergetic with neutron spallation sources targets and neutrino targets, in particular the alternative liquid metal concept.

The muon collider RF power sources have synergy with other developments of high-efficiency klystrons and superconducting cavities. Some RF systems need to work in high magnetic fields, an issue that also exists in some fusion reactor designs.
 
The test facility and the collider itself require a high power proton source. This allows sharing technology and potentially even facilities. Neutron spallation sources such as SNS and ESS are major examples; other examples are neutrino facilities, such as NuSTORM, lepton flavour violation experiments, such as mu2e and COMET and low-energy muon beam facilities used for materials science.

\section{Site development}
\label{0:intro:sec:site}

\begin{figure}[h]
    \centering
    \includegraphics[width=0.57\textwidth]{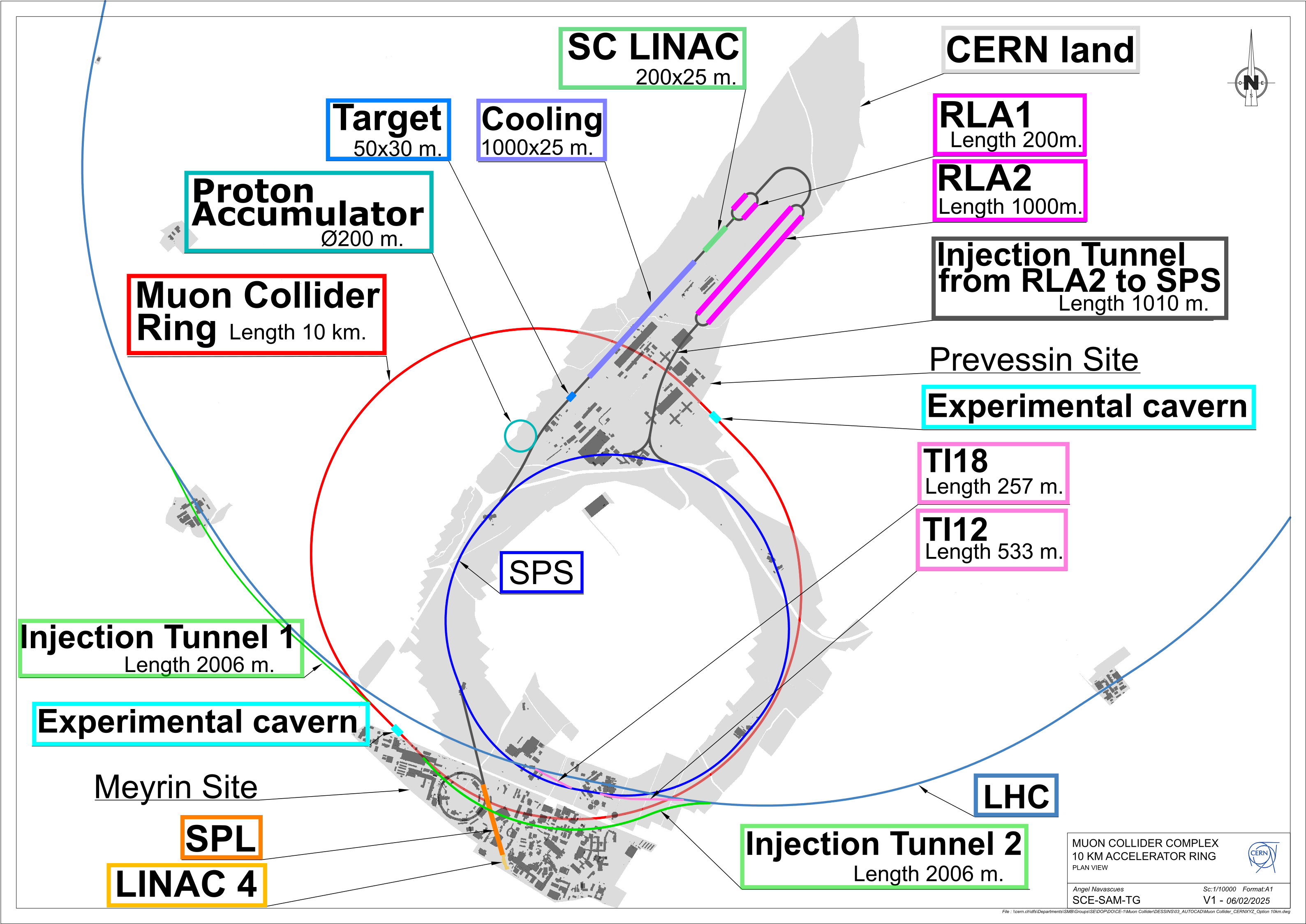}
    \includegraphics[width=0.40\textwidth]{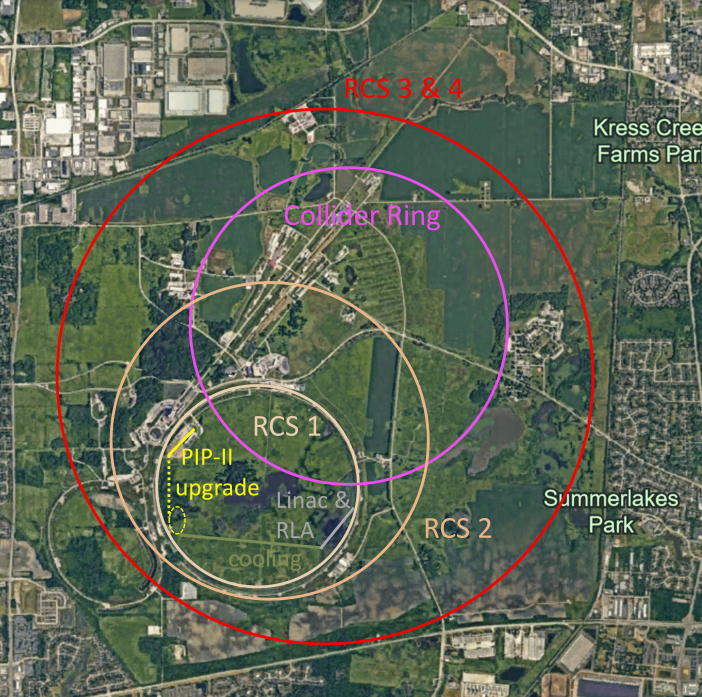}
    \caption{Conceptual layout of the muon collider at CERN (left) and Fermilab (right).}
    \label{f:cern_cv}
\end{figure}

IMCC is developing a site independent design, which could be realised at any location. It is also exploring  two main specific implementations at this moment, one at CERN and one at FNAL. 
The proposed layouts are shown in Figure \ref{f:cern_cv}. At these sites the muon collider can potentially benefit from existing infrastructure, e.g. at CERN from the SPS and LHC tunnels. 
The muon collider can benefit from existing infrastructure but does not depend on it.

Civil engineering studies at CERN indicate that the surface installations of the accelerator facility could be constructed fully on CERN land and that {\bf the SPS and LHC tunnels could be reused} to host the accelerator rings, thus minimising the overall civil engineering. The proton complex would be located on the Meyrin site. The beam would be transported through the SPS tunnel to the Prevessin site where the cooling and initial linacs would be located in cut-and-cover tunnels. The beam is injected into the SPS then the LHC and finally into a new 10 km long collider ring. It may be possible to maintain the current SPS in parallel to the muon collider. 
This scenario would enable collision energy as high as 7.6 TeV with the baseline technologies. It could be envisaged to increase the energy by using higher field HTS fast-ramping magnets in the last RCS in the LHC. 
Alternatively, a second hybrid RCS could be installed in the SPS and the first RCS in the LHC could be be hybrid. 
A detailed study will enable validation of these implementation concepts.

A similar siting study, including investigation of how the existing and planned infrastructure can be used, is underway for Fermilab and the preliminary layout, constrained within the boundaries of the Fermilab site, is shown in Fig.~\ref{f:cern_cv} (right). The exact parameters of the collider sited at Fermilab are to be refined taking into account findings from the study.  

Muon decays produce neutrinos that will exit the earth's surface far from the accelerator facility. These neutrinos will enable a unique experimental programme that goes far beyond current state-of-the-art programmes such as FASER$\nu$.  In this case neutrino detectors would be installed on the surface where the neutrino beam emerges. A suitable implementation has been identified at CERN. The neutrinos arising from the rest of the facility must be diluted so that there is negligible radiation outside the CERN site, comparable to the impact of the LHC. A system of movers are required in collider arcs to displace the magnets vertically to achieve this. Tentative studies indicate that it is possible to have a negligible impact on the environment. Further optimisation and expansion of the studies to the full complex are required.

\section{Timeline and staging}
\label{s:stagingSCD}

A roadmap towards a muon collider must remain flexible, and is an important part of a diverse international research programme in particle physics.
New results, from the HL-LHC or non-accelerator based experiments, and potentially also lower energy experiments, might change the desired parameters for a future muon collider. Societal changes and priorities might also influence resource availability and hence timelines for investments in basic science as needed for future colliders. The potential of a muon collider to reach around 10 TeV parton-parton collisions with high luminosity makes it an exciting opportunity for the near and more distant future.

The potential of a muon collider to reach 10 TeV parton-parton collisions with high luminosity makes it a very exciting future opportunity both on a longer and shorter timescale.

We already know that sustainability and environmental consideration needs to guide the design process much more than projects decades back. Staging scenarios, a clear focus on working with developing technologies (from HTS magnets, to machine learning (ML) and artificial intelligence (AI) methods) with a wide impact and interest outside our field, synergetic studies and projects as described in next subsection, and a broad international collaboration, are all important to develop the muon collider as a scientifically and technically novel but robust future facility when it comes to implementation.

\begin{figure}[h]
\includegraphics[width=\textwidth]{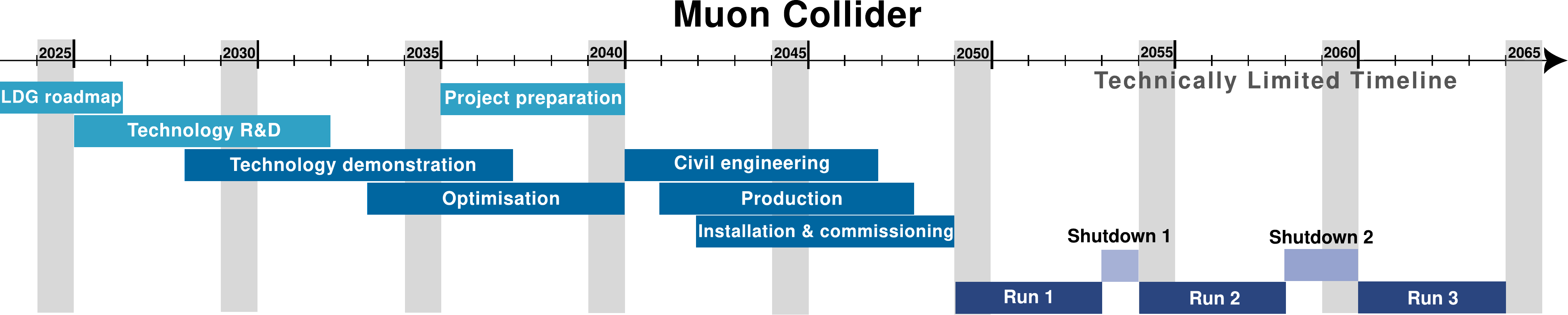}
\caption{Technically limited muon collider timeline.}
\label{f:tl_ex}
\end{figure}

All the possible scenarios cannot be addressed in parallel so IMCC developed a timeline that allows to operate a muon collider by 2050. This demands that all needed technologies be mature in approximately 15 years.  This timeline is essentially driven by technical considerations and assumes that sufficient funding is available and that the R\&D is successful. This technically-limited, success-oriented timeline can serve the community and the decision makers as a basis to define the strategy and the budget for the future of the study. It is shown in Figure~\ref{f:tl_ex}.

The current timeline considerations identified that three technical developments are defining the minimum time to implement a muon collider:
\begin{itemize}
\item The muon production and cooling technology and its demonstration in a test facility. At this moment we expect that this development can be mature enough for a decision in 15 years provided sufficient funding is made available.
\item The magnet technologies, in particular HTS superconductor technology. At this moment, we expect that the different HTS solenoids for the muon production and cooling are available within 15 years; the same is expected for the fast-ramping magnets. For the collider ring, one can expect 11~T Nb$_{3}$Sn magnets with an aperture of 16 cm to be mature. Higher performant HTS or hybrid collider ring magnets may take longer.
\item The detector and its technologies that impact the efficiency of background suppression and the quality of the measurements. At this moment, we expect this technology to be mature in 15 years.
\end{itemize}
Sufficient funding should enable to accelerate the R\&D of the other technologies and designs as to not constrain the timeline.

A staged implementation that anticipates use of Nb$_{3}$Sn collider ring magnets can provide a muon collider by 2050. 
The scenarios are possible, see also Table~\ref{t:facility_param}:
\begin{itemize}
\item
  In the {\bf energy-staging} approach, the initial stage is at lower energy, for example 3\,TeV. There is an important physics case already at this energy. In this strategy, the cost of the initial stage is substantially lower than for the full project. This could accelerate the decision-making processes. The 3\,TeV design is consistent with Nb$_3$Sn magnets at 11\,T. In the second stage the whole complex will be reused with the exception of the collider ring. An RCS to accelerate to full energy and a new collider ring will be added.
\item
In the {\bf luminosity-staging} approach, the initial stage is at the full energy but using less performant collider ring magnets. This leads to an important reduction of the luminosity. If 11\,T Nb$_3$Sn dipoles are used instead of 16\,T HTS dipoles, the resulting increase of collider ring circumference reduces the luminosity by one third. In addition a further reduction arises from the interaction regions. A detailed study is required to quantify the loss, but a factor three reduction might be a good guess.
In the luminosity upgrade, the interaction region magnets will be replaced with more performant ones, similar to the HL-LHC. However one most likely will not replace the other collider ring magnets. In this scenario almost the complete project cost is required in the first stage, which could have important implications on the timeline.
\end{itemize}
The choice between the staging options will depend on physics needs and funding availability. Also the progress in magnet development will play an important role. Faster progress, in particular of HTS for the collider ring magnets, combined with strong funding support will make an earlier start of the 10\,TeV
option more attractive.

\section{Sustainability and environment}
\begin{figure}[h]
    \centering
    \includegraphics[width=0.7\textwidth]{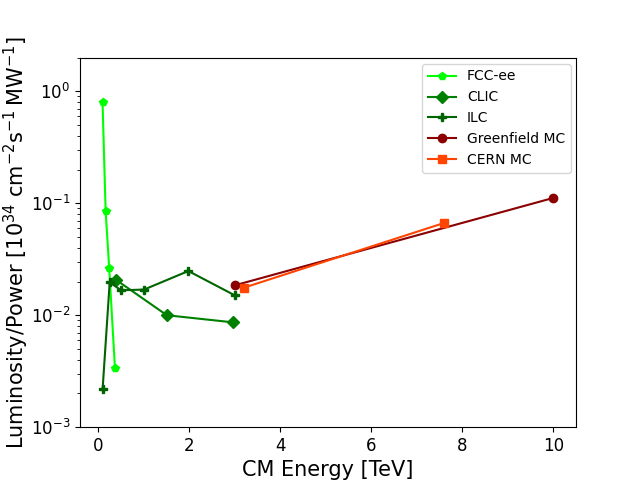}
    \caption{Ratio of luminosity to wall plug power compared to several $e^{+}e^{-}$ machines.}
    \label{f:lumi_vs_power}
\end{figure}

The concept of muon colliders inherently involves less resources (energy, land, materials) than other collider concepts. 
All the energy of the particles is utilised in collisions, therefore it requires less energy to probe physics at the same parton centre of mass as a hadron collider. In addition, the relatively large muon mass strongly suppresses energy losses due to synchrotron radiation. In fact, as shown in Figure~\ref{f:lumi_vs_power}, a muon collider is most efficient known technology to explore multi-TeV energy scales.

A 10~TeV muon collider at 10~TeV requires about half the length of the tunnels (including the injectors) of an 100 TeV $pp$ collider. All the acceleration chain is entirely, or partially pulsed, therefore electrical consumption is reduced, and the collider ring, which is the only CW machine, is only about 10 km long. 
For a green field construction with no re-use of previously existing infrastructure, the CO$_2$ footprint for the construction can therefore be barely estimated as half of that of the FCC-ee, and equivalent to that of a linear electron collider at 3 TeV. 

Furthermore, the IMCC based its design on technologies that should provide a reduced energy consumption with respect to the present state of the art, namely:
\begin{itemize}
    \item \textbf{Magnets based on High Temperature Superconductors (HTS)}. HTS have the potential to greatly lower the energy consumption for the same level of magnetic field (not always achievable with conventional LTS materials). It is fair to say, however, that this technology is not yet demonstrated to work reliably in an accelerator environment, therefore the need for a muon cooling Demonstrator. 
    \item \textbf{High Efficiency Klystrons}. The MC cooling channel requires low duty factor, high power (> 20 MW) RF power sources. No klystron with similar characteristics exist on the market, therefore the IMCC will, if budget is made available, develop a new klystron with a minimum of 80\% efficiency (against a typical 60\% to 65\% efficiency of presently commercially available klystrons. 
    \item \textbf{Superconducting RF cavities}. Wherever possible (for non-pulsed RF) the IMCC has foreseen superconducting RF cavities. This choice allows to reach high electric gradient for a much reduced electrical consumption with respect to room temperature RF structures. 
    \item \textbf{Efficient Cryogenics systems}. By choosing from the start to use high temperature superconductors, the IMCC will invest, as resources will become available into the development of efficient cryogenic systems at temperatures around 20K, and will therefore help pushing the boundaries for such technologies.  
    \item \textbf{Energy recovery linacs}. The IMCC presently considers in its baseline design the use of Energy Recovery Linacs, in the early stage of muon acceleration. While giving advantages in terms of fast acceleration, they also provide energy to the beam in an efficient way.
\end{itemize}

At present the ideal muon collider is being designed considering a green field facility, not yet based in a precise geographical situation, for this reason, no considerations are being done for energy sourcing, energy recovery, water sourcing, waste water treatment, etc. 
Such considerations will be part of the specific implementations in the various labs, in coherence with national legal frameworks and will profit of local efforts already ongoing. 
For instance, an eventual implementation at CERN will build on the experience of the HL-LHC and the FCC-ee feasibility study. For details of the CERN implementation, see Section~\ref{2:implement:sec:cern}. 

\section{Developing the muon collider study} 
The growing collaboration is based on a Memorandum of Cooperation (MoC). The number of institutes signing the MoC has increased to 58 and several more institutes are in the process of joining the collaboration. 
Its structure consists of the International Collaboration Board that represents the partner institutions, the Steering Board that guides the studies and links to the LDG, an International Advisory Committee to review the study and the Coordination Committee that coordinates and executes the study and is chaired by the study leader.

IMCC succeeded in receiving funding for an EU cofunded design study, MuCol, which started 1.3.2023. The total co-funding of MuCol amounts to 3 MEUR, provided by the European Commission, the UK and Switzerland. In the design study, CERN only receives limited contributions for administrative support and travel, but has, in support of the successful Design Study bid, increased its contribution to the muon collider study.
The design study is fully integrated in the overall muon collaboration. The technical meetings and leaders are in common and governance and management are fully synchronised.

The US Particle Physics Project Prioritization Panel recommended in December 2023 to consider hosting a muon collider in the US and joining the IMCC. Discussions with US partners including the Department of Energy's bureau of high-energy physics are on-going. The US community already collaborates strongly in developing the R\&D plan for the next decade and is in the process of setting up their national organisation. An inaugural US muon collider community meeting took place at Fermilab in August 2024 with close to 300 participants.

The overall IMCC goal is to carry out the R\&D together and to develop options to host the collider at CERN or at FNAL and potentially also other sites. On the 2050 timescale, the muon collider is an important option as an alternative to low energy Higgs factories, that can also deliver energy frontier measurements. 

At this point progress in several areas is limited by resources and additional resources are being sought across the collaboration, in parallel with the above-mentioned processes to include new partners.  

\section{Conclusion}
The formation of IMCC has enabled important progress on the muon collider design and technologies, greatly increasing confidence in the concept.
Investment in an experimental programme for the technologies is now essential for further progress. Demonstrating key technologies for muon cooling---such as HTS solenoids, RF cavities, klystrons, absorbers and their integration into cooling cells---will advance the novel aspects of the design. 
Additionally, the development of other critical technologies, including fast-ramping magnet systems, high-field superconducting dipoles, high-power target and superconducting RF technology, will optimise collider performance.
The technology development has strong synergies with other projects, such as FCC-hh, as well as applications with significant societal impact, such as fusion reactors.
Increasing effort on the start-to-end design will enable robust predictions of luminosity performance and availability of the collider, optimise cost, power consumption, and risk, and guide the continued development of technologies.

Global interest in the muon collider as a path to 10 TeV parton-parton collisions is growing, with CERN and Fermilab currently being considered as potential sites.
Initial exploration indicates that a muon collider could be implemented at CERN,
reusing the SPS and LHC tunnels.
The corresponding technically limited timeline indicates that a first collider stage could be available by around 2050. Following HL-LHC or a potential fast Higgs factory, the muon collider could thus become Europe's next flagship project.
\part{Evaluation Report} \label{2:evr}
\chapter{Physics}
\label{1:phys:ch}
The physics opportunities offered by a high-energy muon collider have been investigated extensively in the recent years. The present section summarises the results accomplished. The extraordinary opportunities and perspectives for advancing muon collider theory and phenomenology during the next decade are outlined in Section~\ref{2:physicsr&d}. Technical input and other results in answer to the benchmark questions proposed by the Physics Preparatory Group are reported in Part~\ref{ch:PPGB}.

We start in Section~\ref{1:physics:sec:overview} from a high-level overview of the muon collider physics case, including a quantification of the enthusiastic reaction of the physics community to the IMCC R\&{D} plans. Selected highlights from the muon collider physics case are reviewed in Section~\ref{1:physics:subsec:highlights}. We devote Section~\ref{1:physics:sec:flav} to the opportunities offered by the muon collider to investigate the physics of flavour in the quark and in the lepton sectors, outlining a novel path towards flavour physics by energy-frontier---rather than intensity-frontier---experiments. Opportunities for neutrino physics, including a new forward neutrino experiment parasitical to the 10~TeV collider, synergies with other neutrino experiments at the demonstrator or possibly dedicated muon beams, are described in Section~\ref{1:physics:sec:neutrinos}. 

\section{High-level overview}\label{1:physics:sec:overview}

The Higgs discovery in 2012 marked the dawn of a new era: the {\emph{Electroweak-Higgs Unification era}}. We discovered that electroweak physics is at least approximately described by a unified symmetry---the electroweak symmetry---that is broken by the fundamental Higgs field and the associated Higgs particle. For the first time, we obtained direct experimental confirmation that electromagnetism and the weak nuclear force are manifestations of a spontaneously-broken gauge symmetry, a paradigm at the heart of the Standard Model (SM). The SM accommodates the mass of the electroweak bosons and of the quarks and leptons in a framework that is extremely predictive, potentially providing a microscopic description of electroweak physics up to amazingly short distances and high energies.

It is also the first time that we have encountered the possibility that a spin-zero particle such as the Higgs boson could be point-like and fundamental rather than composite. The Naturalness Paradox famously illustrates the tension between point-like spin-zero particles and the Wilsonian Paradigm of Quantum Field Theories as Effective Field Theories, making the Higgs discovery at once a success and a mystery for theoretical physics.

Current technology enables the construction of low-energy electron-positron colliders to precisely measure the properties of the Higgs and electroweak gauge bosons, but not to access the most direct manifestations of the novel paradigm of broken gauge symmetry, nor to verify directly its validity as the microscopic description of electroweak physics.

High energy is needed to accomplish this goal, as well as leptonic collisions that are not polluted by the overwhelming effect of strong interactions. Muons are the most promising path towards a lepton collider with very high energy, and accessing the high-energy phenomena of Electroweak-Higgs unification is the major (and most unique) physics driver of the muon collider project.

\subsubsection*{Exploring new Standard Model phenomena}

The SM predicts a host of as-yet-unobserved phenomena as the energy rises well above the masses of the electroweak particles. At TeV energies the process of electroweak boson scattering receives a contribution from the virtual exchange of the Higgs that cancels the growth of the vector exchange amplitudes. This is vital to ensure the high-energy viability of the unified Electroweak-Higgs theory, hence this so-called ``unitarization’’ mechanism is arguably its most fundamental prediction. The experimental confirmation of unitarization is a long-sought target of the LHC, which is nonetheless hard to accomplish because of the large backgrounds and the limited event rates in the relevant high-energy regime. A muon collider can instead measure vector boson scattering at TeV energies with precision, thanks to both low backgrounds and the enhanced collinear emission of vector bosons from energetic muons that results in large collision rates.

The effective presence of massive vector bosons inside the massless muon is in itself a striking high-energy manifestation of the broken gauge symmetry paradigm, which the muon collider will establish experimentally. This will intriguingly probe the “quantum compositeness” of particles, effectively made of partons, and its fundamentally different nature from the classical intuition of a composite state. Electroweak physics is ideally suited for these studies because it is perturbative and hence in principle calculable (unlike the strong interactions), and endowed with a physical mass gap that makes ``particles'' fully well defined as asymptotic states unlike in QED. 

By studying electroweak scattering processes at 10~TeV, the muon collider will enable a variety of tests of the most basic properties of the unified electroweak-Higgs sector. This includes establishing with high precision the restoration of the electroweak symmetry, as well as the confirmation of the Goldstone Boson Equivalence Theorem that puts the Higgs boson in the same doublet as the longitudinal polarizations of the electroweak bosons. 

The neutrino partonic content of the muon---an intrinsic feature of electroweak-Higgs unification, as neutrinos and charged leptons combine into unified electroweak multiplets---provides new ways to explore the neutrino sector at muon colliders. Interactions of these partonic neutrinos will enable the first experimental detection of high energy neutrino-neutrino and neutron-muon collisions. Furthermore, neutrinos produced in the decay of muon bunches provide an intense and well-characterized beam of collider neutrinos for a forward detector. The neutrino physics program at a muon collider provides a high-energy complement to current and future long-baseline neutrino experiments.

Studying 10~TeV processes will also offer a unique window on the emergent phenomenon of electroweak radiation. Conversely to the partonic content of the muon particle, this will test the particle content of partons, under the above-mentioned favourable theoretical conditions that are offered by electroweak physics. Among the striking experimental consequences, the effective $W$ boson and electron content of the neutrino parton will give rise to observable ``neutrino jets'', making neutrinos detectable for the first time at a collider experiment.

\subsubsection*{Beyond the Standard Model with Precision, Energy, and Precision from Energy}

The observation of novel Standard Model phenomena is a guaranteed outcome of the muon collider project. An equally interesting and guaranteed outcome is the plenitude of answers to compelling questions about physics beyond the SM (BSM), with a high chance of BSM discovery given the strong improvement in sensitivity relative to current knowledge. Much of this sensitivity emerges in connection with the novel SM electroweak phenomena previously described, and in some cases from the very same measurements that will serve for the observation of the new SM phenomena. 

For instance, the abundance of effective vector bosons in the muons enables a number of BSM tests including the search for new interactions in high-energy vector boson scattering, as well as producing BSM particles coupled through Higgs-portal or gauge-portal type of interactions. Concrete scenarios with Higgs-portal interactions connect with fundamental questions like Dark Matter, Naturalness, and the nature of the electroweak phase transition in the early universe. 

Effective vector bosons are also responsible for the large rate of production for the Higgs, both in the single- and in the pair-production mode. This enables a per-mille level precision measurement of several couplings of the Higgs, and a few-percent precision in the measurement of the Higgs trilinear coupling. The overall prospects for the determination of Higgs boson properties at the muon collider are broadly comparable to those of future low-energy electron-positron Higgs factories, but also widely complementary. The new muon collider observables would improve single-Higgs coupling measurements beyond the capabilities of the Higgs factory and provide a direct and much more accurate determination of the Higgs trilinear coupling. 

Even more importantly, the high energy reach of the muon collider will also allow it to directly probe the new physics that is responsible for coupling modifications that could be observed at a Higgs factory or the muon collider itself. A striking example is provided by minimal BSM scenarios where a first-order electroweak phase transition in the early universe is induced by the presence of new neutral states coupled via the Higgs portal. In a vast region of the relevant parameter space, this scenario will be discovered at the muon collider both by modifications of the single and of the triple Higgs coupling, but also observed directly by the resonant production of the new particles in vector boson fusion. More generally, new physics responsible for Higgs coupling modifications may also contain electroweak-charged particles that can be discovered and precisely characterized at the muon collider up to the kinematic threshold mass of 5~TeV.

The generic 5~TeV direct reach on electroweak-charged particles produced in pairs will strongly advance our knowledge relative to the LHC era. The reach is even higher than that of a 100~TeV proton-proton collider if the new particles do not carry QCD interactions. While the anticipated reach of 5~TeV is for particles that decay promptly into easily detectable final states, comparable sensitivity can also be obtained in more challenging scenarios. Specific careful studies—including the simulation of fully realistic beam-induced background from muon decay—have been conducted to assess the muon collider’s prospects to discover minimal WIMP dark matter candidates such as the Higgsino and Wino. The cosmological abundance of these dark matter candidates is entirely determined by their mass and electroweak charge, so that the observed dark matter abundance in the universe sets sharp mass targets. These targets can be reached (and exceeded) at the muon collider by detecting disappearing or soft tracks. Such search strategies, in combination with many others that leverage clean leptonic collisions, offer a wide coverage of the landscape of minimal WIMP dark matter models at the muon collider.

It bears emphasizing that the muon collider’s unique union of precision measurements and direct searches opens new windows into physics beyond the Standard Model beyond the two in isolation. The very concept of separating precision measurements from direct searches in the pursuit of BSM physics is simply an organizational artifice which does not exist in the underlying quantum field theory framework. The muon collider accesses this underlying coherence: for weakly-coupled BSM physics, the fact that the direct reach of a muon collider is often comparable to precision measurements implies that the full set of experimental observables at high energy probes a complementary space in UV-complete models. In addition, the combined observation of new particles and of new effects in precision observables will be a formidable tool for the characterisation of a discovery. Thus the union of precision and energy shows that the whole of a muon collider physics program is much greater than the sum of its individual parts.

Still, it would be reductive to describe the BSM discovery opportunities of the muon collider as the simple {\emph{union}} of the opportunities offered by precision measurements and by the direct search for new particles, which could also be respectively performed at future leptonic and hadronic colliders with slightly superior or inferior sensitivity depending on the specific BSM scenario at hand. The muon collider uniquely benefits from the {\emph{intersection}} between energy and precision by precisely measuring high-energy observables such as scattering cross sections at 10~TeV. This offers unique opportunities not only for SM electroweak studies as previously emphasised, but also for the search for new physics with very high mass.

The contribution of heavy new physics to low-energy observables is suppressed by some power of the ratio between the characteristic energy scale of the observable and the new physics mass scale. The measurement accuracy that is needed to discover its effects thus depends very strongly on the characteristic scale of the observable under consideration, in a way that very low energy observables are useful only if they can be measured with exquisite accuracy, while a more modest accuracy is sufficient when using observables with higher characteristic scale. Equivalently, the reach on the new physics scale of a measurement with given accuracy increases linearly with the energy of the observable. The muon collider can measure 10 TeV scattering cross sections with order percent accuracy. For typical new physics effects that scale with the second power of the energy over mass ratio, this generically corresponds to an order 100 TeV reach on the scale of new physics.

A historical example of such energy enhancement is the discovery of the finite radius---and hence the composite nature---of the proton. This was accomplished by detecting departures of the electron scattering cross section from the point-like proton prediction. These effects are suppressed at low energy by the square of the scattering energy divided by the proton compositeness scale, which we today associate with the QCD confinement scale of about 300~MeV. The limited accuracy in the cross section measurement prevented these effects from being detected until technological developments enabled scattering energies as ``high'' as 100~MeV, mitigating the suppression. By exploiting this very same strategy, the measurement of 10~TeV cross sections at the muon collider can discover evidence for Higgs compositeness at scales as high as 40~TeV, or in the case of null results exclude compositeness scales as high as 60~TeV. In this respect, the muon collider will probe the compositeness of the Higgs boson up to a scale that is far above not only our present-day sensitivity, but also the projected sensitivity of any other future project or planned measurement.

Another historical example is the observation of neutral current weak interactions in muon production from electron-positron collisions. Here the suppression is the square of the scattering energy divided by the mass of the $Z$ boson, and in this case the observation was enabled by the availability of high-energy electron-positron collisions with around 20~GeV of energy. Correspondingly, the muon collider can discover new weak currents mediated by a heavy $Z^\prime$ particle with a mass possibly in excess of 100~TeV, depending on the coupling. Such discovery would not merely consist of the observation of a tension with the SM prediction for a single observable. Many different cross sections can be measured, as well as differential distributions, enabling the characterisation of the properties of the new discovery. 

Beyond specific models, the competitive advantage of the muon collider can be also illustrated by the sensitivity to new interactions encoded by Effective Field Theory operators. Flavour-diagonal and -universal operators that induce energy-growing effects in the production of two leptons, quarks, vector bosons or Higgs can be probed up to an effective scale that is generically about 50~times higher than the one that is presently tested. The reach is generically higher than the one that can be attained by electron-positron $Z$-boson factories with high luminosity. Unlike $Z$-factories, the muon collider’s sensitivity to the new interactions does not rely on the feasibility of very precise measurements and theoretical predictions with order $10^{-5}$ resolution, but on the detection of percent-level effects on the high energy cross section. This adds a considerable degree of robustness to muon collider projections that is further augmented by the large variety of available measurements---including differential distributions---that can illuminate the discovery of new physics.

By 10~TeV cross-section measurements, vector boson scattering, and Higgs physics, the muon collider can greatly advance the program of precision tests of the electroweak and Higgs physics, commonly referred to as EWPT (electroweak precision tests). Importantly, these advances will be attained by a novel methodology, less prone to systematic uncertainties and more suited for the characterisation of discoveries.

The 10~TeV measurements can also advance flavour physics through their exquisite sensitivity to neutral-current flavour transitions, capable of resolving effects due to mediators at the 100~TeV scale. Unlike for EWPT, the strategy of investigation in this case is not to probe corrections to SM cross sections, but rather to observe phenomena that are very rare in the SM, such as the production of fermion anti-fermion pairs of different flavours. The energy enhancement works as before and boosts the BSM rate of transition to an observable level. The sensitivity to neutral flavour transitions in both the quark and lepton sectors often matches or surpasses the best future prospects for low-energy experiments.

\subsection*{A {\emph{new}} interest in muon collider physics}
The 2020 European Strategy Update process and the creation of the IMCC outlined an R\&{D} path towards the realisation of a muon collider with 10~TeV energy and high luminosity, triggering an enthusiastic reaction from the theory and phenomenology community. Since 2019 (when the muon collider R\&D effort catalyzed by the 2020 European Strategy Update began), INSPIRE-HEP has recorded around 200 entries about muon colliders in the ``Phenomenology-HEP'' subject category~\cite{inspire}. This is more than half of the papers ever written on this topic. The left panel of Figure~\ref{fig:papdist} compares the time distribution of the phenomenology papers (in blue) with the papers about muon colliders in all subjects (in grey). Before 2019, muon collider phenomenology papers comprised a small fraction of the total and the development of the field was almost entirely driven by the advances in accelerator physics. Physics studies are instead a major component and a driver of the activity in the last few years, indicating  an unprecedented enthusiasm for muon colliders physics opportunities. 

Extensive reviews of this growing literature can be found in Refs.~\cite{AlAli:2021let,Black:2022cth} and, most recently, in the IMCC EPJC Review~\cite{Accettura:2023ked} and the IMCC Interim Report~\cite{InternationalMuonCollider:2024jyv}. Major topics of investigation are projections for the direct discovery of new particles~\cite{Buttazzo:2018qqp,Ruhdorfer:2019utl,Liu:2021jyc,Asadi:2021gah,Qian:2021ihf,Bottaro:2022one,Bottaro:2021srh,Mekala:2023diu,Li:2023tbx,Kwok:2023dck,Jueid:2023zxx,Han:2021udl,Kalinowski:2020rmb,Rodejohann:2010jh,Li:2023ksw,Zhao:2024fgc,Braathen:2024lyl,Bandyopadhyay:2024gyg,Bi:2024pkk,Dung:2024ctj,Lu:2024qrf,Barik:2024kwv,Giang:2024gnb,Marcos:2024yfm,Das:2024ekt,Martinez-Martinez:2024lez,Barducci:2024kig,Jiang:2024wwa,De:2024tbo,Parashar:2024obw,Bandyopadhyay:2024plc,Lu:2023jlr,Liu:2023jta,Urquia-Calderon:2023dkf,Ghosh:2023xbj,Dasgupta:2023zrh,Jueid:2023qcf,Vignaroli:2023rxr,Ruhdorfer:2023uea,Belyaev:2023yym,Aboudonia:2024frg,Bhaskar:2024snl,Jia:2024wqi,Li:2024zhw,Gutierrez-Rodriguez:2024nny,He:2024dwh,Samarakoon:2023crt,Chigusa:2023rrz,Martinez-Martinez:2023qjt,Mikulenko:2023ezx,Ouazghour:2023plc,Sun:2023rsb,Li:2023lkl,Belyaev:2023yym,Giang:2023cxb,Guo:2023jkz,Das:2022mmh,Lv:2022pts,Abdullayev:2022uvh,Yang:2022ilt,Inan:2022rcr,Inan:2022rcr,Chakraborty:2022pcc,Li:2022kkc,Haghighat:2022bwj,Bao:2022onq,Chun:2021rtk,Han:2025wdy,Bojorquez-Lopez:2024bsr,Adhikary:2024tvl,Asadi:2024jiy,Han:2022mzp,Chigusa:2025otr,Li:2025ptq} including WIMP dark matter~\cite{Han:2020uak,Bottaro:2021snn,Capdevilla:2021fmj,Han:2020uak,Bottaro:2021snn,DiLuzio:2018jwd,Franceschini:2022sxc,Korshynska:2024suh,Belfkir:2023vpo,Belfkir:2023lot,Inan:2023pva,Liu:2022byu,Sen:2021fha,Haghighat:2021djz,Sahin:2021xzt,Capdevilla:2024bwt,Han:2022ubw,,Black:2022qlg,Fukuda:2023yui}, the measurement of the Higgs couplings including the trilinear and possibly the quadrilinear coupling, and the sensitivity to heavy new physics through precision measurements in both an EFT context and concrete new physics scenarios such as Composite Higgs~\cite{Forslund:2022xjq,Han:2020pif,Bartosik:2019dzq,Bartosik:2020xwr,Costantini:2020stv,Han:2020pif,Buttazzo:2020uzc,Chiesa:2020awd,Costantini:2020stv,Han:2020pif,Buttazzo:2020uzc,Forslund:2022xjq,Gurkanli:2024qaf,Ruhdorfer:2024dgz,Han:2024gan,Ma:2024zjy,Gu:2024wrk,Frigerio:2024pvc,Fortuna:2024rqp,Denizli:2024uwv,Li:2024joa,Celada:2023oji,Cao:2023yeb,Han:2023njx,Cetinkaya:2023wgg,Stylianou:2023xit,Cassidy:2023lwd,Dermisek:2023rvv,Yamatsu:2023bde,Denizli:2023rqe,Jana:2023ogd,Yang:2023gos,Amarkhail:2023xsc,Sun:2023cuf,Chiesa:2021qpr,Alici:2024wjd,Spor:2024nsx,Wang:2024bfc,Homiller:2024uxg,Zhang:2024ebl,Zhang:2023khv,Zhang:2023ykh,Ruzi:2024cbt,Bhattacharya:2023beo,Spor:2023sdk,Sun:2023ylp,Liu:2023yrb,Forslund:2023reu,Dong:2023nir,Zhang:2023yfg,Chen:2022yiu,Ahluwalia:2022qsp,Spor:2022hhn,Senol:2022snc,Yang:2022fhw,deBlas:2022aow,Spor:2022mxl,Chen:2021pqi,Yang:2021zak,Yang:2020rjt,Franceschini:2022veh}. Several ``muon-specific'' opportunities that stem from colliding muons for the first time, in connection with the muon \mbox{g-2} and with lepton flavour physics, have been also investigated~\cite{Capdevilla:2020qel,Buttazzo:2020ibd,Yin:2020afe,Capdevilla:2021rwo,Dermisek:2021ajd,Dermisek:2021mhi,Capdevilla:2021kcf,Bandyopadhyay:2021pld,Dermisek:2022aec,Paradisi:2022vqp,Chakrabarty:2014pja,Huang:2021biu,Huang:2021nkl,Homiller:2022iax,Casarsa:2021rud,Han:2021lnp,Azatov:2022itm,Altmannshofer:2023uci,Ake:2023xcz,Liu:2021akf,Cesarotti:2022ttv,Altmannshofer:2022xri,Bause:2021prv,Allanach:2022blr,Ruzi:2023atl,Qian:2022wxa,Chakrabarty:2014pja,Batell:2024cdl,Cesarotti:2024rbh,Amarkhail:2024kfq,Asadi:2023csb,Cesarotti:2023sje,Lu:2023ryd,Arakawa:2022mkr,Li:2021lnz}. Growing attention is being devoted to theoretical tools and predictions for muon collider physics~\cite{Han:2021kes,Han:2020uid,Ma:2024ayr,Marzocca:2024fqb,Chen:2022msz,Capdevilla:2024ydp,Ciafaloni:2024alq,Garosi:2023bvq,Bredt:2022dmm,Ruiz:2021tdt}.

Several workshops and seminars on muon colliders physics were held in the last few years, including a very successful series of events organised by the ``Muon Collider Forum'' in the context of the Snowmass 2021 Community Planning Exercise. The activities and the work triggered by the Forum~\cite{Black:2022cth} strongly impacted the Snowmass Energy Frontier outcome~\cite{Narain:2022qud}, which recognised the muon collider's potential for the exploration of the energy frontier and advocated R\&{D} investments with the perspective of hosting a muon collider in the US. The P5 panel report confirmed and strengthened this view~\cite{P5}.

\section{Muon collider physics highlights}\label{1:physics:subsec:highlights}

\begin{figure}[t]
   \begin{center}
        \includegraphics[width=0.42\textwidth]{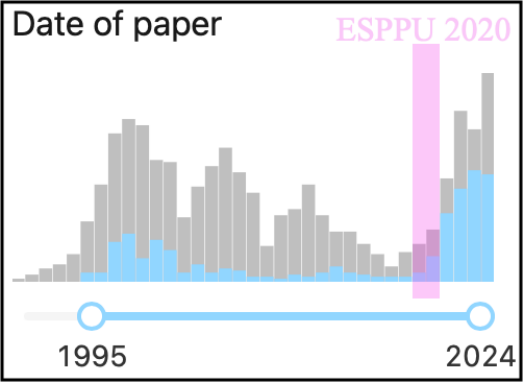}   \hspace{40pt}
        \includegraphics[width=0.38\textwidth]{2_EvaluationReport/A_Physics/Figures/arrows_text_white.png}  
    \end{center}
 \caption{Left: Time distribution of the papers on muon colliders (grey) compared with those (blue) with subject ``Phenomenology-HEP''~\cite{inspire}. Right: Pictorial view of the muon colliders exploration opportunities.}
 \label{fig:papdist}
\end{figure}

The motivations for exploring short-distance physics are strong, but broad. Therefore, a radical advance requires an equally broad and comprehensive program of energy frontier exploration that leverages many diverse strategies of investigation. The many exploration strategies available at a muon collider are depicted as arrows on the right panel of Figure~\ref{fig:papdist} and are described in turn throughout the rest of this section.

\begin{figure}[t]
   \begin{center}
        \includegraphics[width=0.47\textwidth]{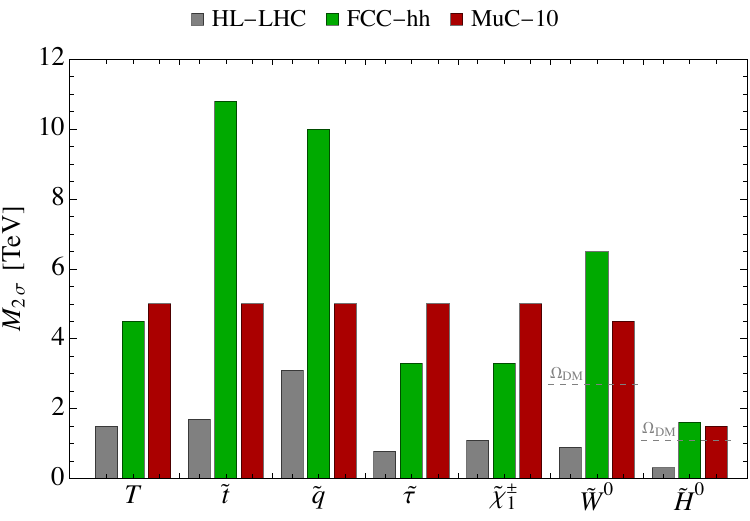} \;
        \raisebox{10pt}
        {\includegraphics[width=0.46\textwidth]{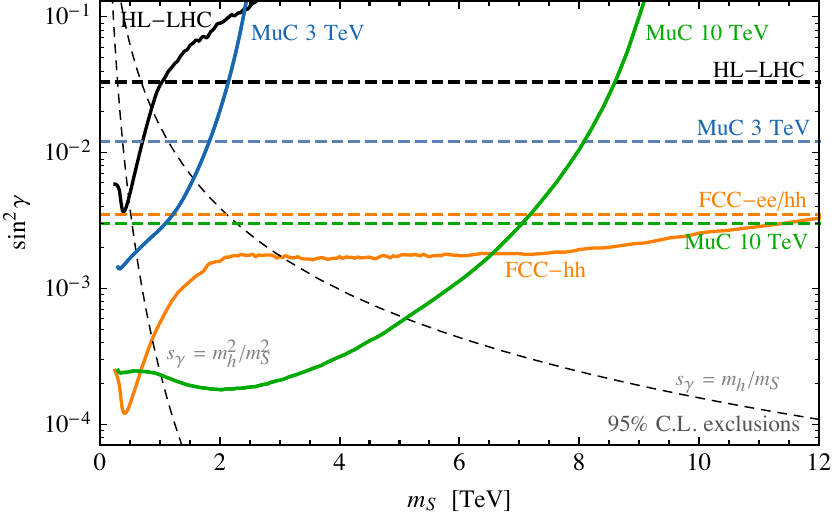}}  
    \end{center}
 \caption{Left: 95\%CL exclusion reach on the mass of several BSM particles at future colliders \cite{CidVidal:2018eel,Saito:2019rtg,EuropeanStrategyforParticlePhysicsPreparatoryGroup:2019qin,Han:2020uak,Capdevilla:2021fmj,Bottaro:2022one}. Only EW pair production is considered to assess the MuC sensitivity. This underestimate the reach in models where single-production is possible (see e.g.~\cite{Belyaev:2023yym}). For the wino and the Higgsino, we label as ``$\Omega_{\rm{DM}}$'' the mass required to reproduce the observed dark matter abundance. Right: exclusion contour~\cite{Accettura:2023ked} for a scalar singlet of mass $m_S$ mixed with the Higgs boson with strength $\sin\gamma$.}
 \label{fig:dir-3h-ch}
\end{figure}

\subsection*{Energy}

The high available energy enables a search for new heavy particles, with a reach in mass that strongly extends that of the LHC. This mass reach owes largely to the fact that the muons are elementary and their collision energy is entirely available to produce new particles. The protons instead are composite and their effective energy reach is limited to a fraction of the collider energy by the steep fall-off of the parton distribution functions. This is the reason why a muon collider with 10~TeV energy can access heavier particles than the 14~TeV LHC, as illustrated on the left panel of Figure~\ref{fig:dir-3h-ch}. 

The figure shows the projected exclusion reach on the mass of a number of hypothetical particles (labelled with a standard BSM notation\footnote{For instance, $T$ is a fermionic top partner, $\tilde{t}$ is the stop and ${\tilde{W}}^0$ and Higgsino ${\tilde{H}}^0$ are the wino and the Higgsino, respectively. The notation is the same as in Ref.~\cite{EuropeanStrategyforParticlePhysicsPreparatoryGroup:2019qin}.}) at the muon collider with 10~TeV energy in the centre of mass, at the HL-LHC, and at the 100~TeV proton-proton future collider FCC-hh~\cite{CidVidal:2018eel,Saito:2019rtg,EuropeanStrategyforParticlePhysicsPreparatoryGroup:2019qin,Capdevilla:2021fmj,Bottaro:2022one}. At a muon collider, these particles are produced in pairs by electroweak (EW) interactions and the corresponding EW production cross sections are determined by the EW and spin quantum numbers of the states. The cross-sections range from $0.1$ to 10~fb at the 10~TeV MuC, for masses almost up to the kinematic threshold of 5~TeV. With the target integrated luminosity of  10~ab$^{-1}$, enough events (more than 1000) will be available for discovery up to the threshold provided the particle decays promptly to an easily-detectable final state. Therefore, all particles considered in the figure with the exception of the wino and the Higgsino (see later) can be discovered up to 5~TeV mass by only exploiting the model-independent process of EW pair-production. An extended mass-reach is possible if BSM interactions mediate the production of the new state. For instance, the 10~TeV muon collider reach on top partners is around $9.5$~TeV from single production~\cite{Belyaev:2023yym}.

The mass reach of the 10~TeV MuC is above the HL-LHC exclusion limit for all of the BSM candidates considered in Figure~\ref{fig:dir-3h-ch}. The 10~TeV muon collider has an even higher reach than a 100~TeV proton-proton collider FCC-hh in QCD-neutral particles such as charginos ${\widetilde{\chi}}_1^\pm$ and tau sleptons $\widetilde\tau$. It surpasses the thermal target (see later) for the Higgsino and the Wino dark matter candidates.

Along these lines, the ``Energy'' arrow in Figure~\ref{fig:papdist} represents the possibility of searching for new heavy particles of very generic nature, or specific well-motivated candidates. Past works have investigated the MuC sensitivity to a number of BSM scenarios ranging from WIMP dark matter, extended Higgs sectors, heavy neutral leptons, composite resonances, solutions to the $\mathrm{g}-2$ anomaly and more~\cite{Buttazzo:2018qqp,Ruhdorfer:2019utl,Liu:2021jyc,Asadi:2021gah,Qian:2021ihf,Bandyopadhyay:2021pld,Bottaro:2022one,Bottaro:2021srh,Mekala:2023diu,Li:2023tbx,Kwok:2023dck,Jueid:2023zxx,Han:2021udl,Kalinowski:2020rmb,Rodejohann:2010jh,Li:2023ksw,Zhao:2024fgc,Braathen:2024lyl,Bandyopadhyay:2024gyg,Bi:2024pkk,Dung:2024ctj,Lu:2024qrf,Barik:2024kwv,Giang:2024gnb,Marcos:2024yfm,Das:2024ekt,Martinez-Martinez:2024lez,Barducci:2024kig,Jiang:2024wwa,De:2024tbo,Parashar:2024obw,Bandyopadhyay:2024plc,Lu:2023jlr,Liu:2023jta,Urquia-Calderon:2023dkf,Ghosh:2023xbj,Dasgupta:2023zrh,Jueid:2023qcf,Vignaroli:2023rxr,Ruhdorfer:2023uea,Aboudonia:2024frg,Bhaskar:2024snl,Jia:2024wqi,Li:2024zhw,Gutierrez-Rodriguez:2024nny,He:2024dwh,Samarakoon:2023crt,Chigusa:2023rrz,Martinez-Martinez:2023qjt,Mikulenko:2023ezx,Ouazghour:2023plc,Sun:2023rsb,Li:2023lkl,Belyaev:2023yym,Giang:2023cxb,Guo:2023jkz,Das:2022mmh,Lv:2022pts,Abdullayev:2022uvh,Yang:2022ilt,Inan:2022rcr,Inan:2022rcr,Chakraborty:2022pcc,Li:2022kkc,Haghighat:2022bwj,Bao:2022onq,Chun:2021rtk,Capdevilla:2024bwt,Han:2020uak,Bottaro:2021snn,Capdevilla:2021fmj,Han:2020uak,Bottaro:2021snn,DiLuzio:2018jwd,Franceschini:2022sxc,Korshynska:2024suh,Belfkir:2023vpo,Belfkir:2023lot,Inan:2023pva,Liu:2022byu,Sen:2021fha,Haghighat:2021djz,Sahin:2021xzt,Capdevilla:2024bwt,Cheung:2025uaz}. A few specific results are outlined below. It should be emphasised that the results described below---as well as in the majority of the muon collider studies in the literature---are based on detailed phenomenological analyses that consider the relevant backgrounds as well as a parametric modelling of the detector effects. The assumed detector performances are those of the IMCC muon collider DELPHES card~\cite{deFavereau:2013fsa,delphes_card_mucol}, which match the performances of the CLIC detector and lie in between the ``Baseline'' and ``Aspirational'' performances described in Section~\ref{1:inter:sec:physics}.

Reference~\cite{Buttazzo:2018qqp} (see also Refs.~\cite{AlAli:2021let,Ruhdorfer:2019utl,Liu:2021jyc}) studied one extra EW-singlet Higgs scalar which is potentially responsible for the generation of a strong first-order EW phase transition in the Early Universe, and is present in other BSM scenarios as well. Such a ``scalar singlet'' is a standard benchmark for future colliders, also in light of its peculiar coupling to the SM, which occurs only through a Higgs-portal interaction. The 10~TeV MuC mass-reach on this BSM scenario is superior to that of the FCC-hh in the most motivated region of the model's parameter space. In fact, the sensitivity is superior in the whole parameter space upon including the indirect MuC reach from Higgs coupling measurements. This is shown on the right panel of Figure~\ref{fig:dir-3h-ch} in the plane formed by the mass of the particle and its coupling to the SM, expressed in terms of the degree of mixing with the Higgs boson. The MuC advantage over FCC-hh stems from the larger MuC cross-section for the production of Higgs portal-coupled new physics in vector boson fusion. Similar findings have been reported in other Higgs portal-coupled BSM scenarios, making the muon collider an ideal option to cover this class of models at the multi-TeV scale.

\begin{figure}[t]
\begin{minipage}{0.55\textwidth}
\renewcommand{\arraystretch}{.89}
\setlength{\arrayrulewidth}{.2mm}
\setlength{\tabcolsep}{0.6 em}
\begin{center}
\begin{tabular}{c|c|c|c}
\hline
& \multicolumn{1}{c|}{\makebox[35pt]{\small{HL-LHC}}} & \makebox[35pt]{\small{HL-LHC}} & \makebox[35pt]{\small{HL-LHC}} \\[0pt]
& \multicolumn{1}{c|}{\ } & \multicolumn{1}{l|}{\makebox[35pt]{+\small{$10\,\textrm{{TeV}}$}}} & \multicolumn{1}{l}{\makebox[35pt]{+\small{$10\,\textrm{{TeV}}$}}}  \\[-2pt]
\ & \ & \multicolumn{1}{c|}{\ } & \multicolumn{1}{l}{\hspace{0.5pt}+
{$e e$}} \\ 
\hline
$\kappa_W$               & 1.7 & 0.1 & 0.1 \\ \hline
$\kappa_Z$               & 1.5 & 0.2 & 0.1 \\ \hline
$\kappa_g$               & 2.3 & 0.5 & 0.5 \\ \hline
$\kappa_{\gamma}$        & 1.9 & 0.7 & 0.7 \\ \hline
$\kappa_{Z\gamma}$       & 10  & 5.2 & 3.9 \\ \hline 
$\kappa_c$               & -   & 1.9 & 0.9 \\ \hline
$\kappa_b$               & 3.6 & 0.4 & 0.4\\ \hline
$\kappa_{\mu}$           & 4.6 & 2.4 & 2.2 \\ \hline
$\kappa_{\tau}$          & 1.9 & 0.5 & 0.3 \\ \hline\hline
$\kappa_t^*$             & 3.3 & 3.0 & 3.0 \\ \hline
\multicolumn{4}{l}{ {\scriptsize $^*$ No input used for the MuC}}\\
\end{tabular}
\end{center}
\end{minipage}
\begin{minipage}{0.45\textwidth}
\vspace{0pt}
\includegraphics[width=1\textwidth]{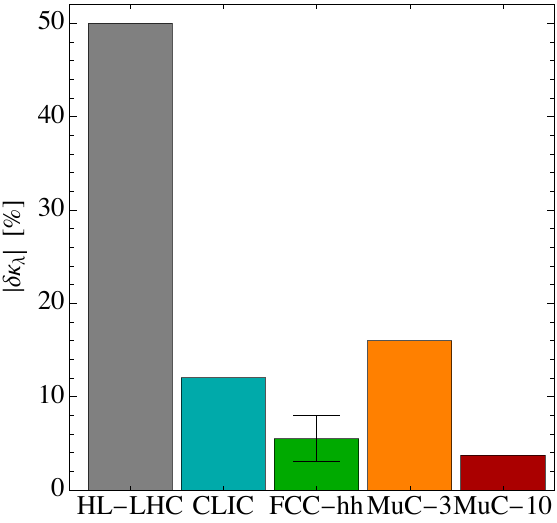}
\end{minipage}
 \caption{Left: $1\sigma$ sensitivities (in \%) from a 10-parameter fit in the $\kappa$-framework at a $10$~TeV MuC with $10$~ab$^{-1}$, compared with HL-LHC. The effect of measurements from a $240$~GeV $e^+e^-$ Higgs factory is also reported. 
 See Section~\ref{sec:EHT} for details.
 Right: Sensitivity to the Higgs trilinear coupling modifier $\delta\kappa_\lambda$ of different future colliders. The sensitivity of the 3~TeV muon collider (MuC-3) and 10~TeV muon collider (MuC-10) is compared with that of the HL-LHC, CLIC, and FCC-hh. Plots adapted from Ref.~\cite{Accettura:2023ked}.}
 \label{fig:3h-ch}
\end{figure}

Several papers~\cite{Han:2020uak,Bottaro:2021snn,Capdevilla:2021fmj,Han:2020uak,Bottaro:2021snn,DiLuzio:2018jwd,Franceschini:2022sxc,Korshynska:2024suh,Belfkir:2023vpo,Belfkir:2023lot,Inan:2023pva,Liu:2022byu,Sen:2021fha,Haghighat:2021djz,Sahin:2021xzt,Capdevilla:2024bwt} studied the observability of a variety of WIMP DM candidates at the muon collider (see Ref.~\cite{Accettura:2023ked} for a summary). Detection strategies include mono-photon (or, more generally mono-$X$) searches, indirect searches from loop effects, and direct searches of the disappearing tracks produced by the charged particle in the dark matter multiplet. The beam-induced-background (BIB) from the decay of the muon beams has a potentially large impact on the disappearing track search, which is difficult to estimate and to parametrize faithfully in a fast detector simulation. For this reason, a complete study of disappearing tracks was performed in Ref.~\cite{Capdevilla:2021fmj}, based on realistic BIB Monte Carlo simulations. The observability of long-sought dark matter candidates such as the Higgsino and the Wino has been established up to the thermal mass. In addition, Ref.~\cite{Capdevilla:2024bwt}, studied the possibility of observing soft tracks and demonstrated that the Higgsino with thermal mass could be also discovered with this second strategy. Using soft tracks, the thermal Higgsino could be discovered even at a first muon collider stage operating with 3~TeV energy in the centre of mass. The mass-reach reported in Figure~\ref{fig:dir-3h-ch} (for the 10~TeV Muc) is the one obtained using mono-photon with one disappearing track~\cite{Han:2020uak,Capdevilla:2021fmj}. 

More energy-frontier opportunities for the muon collider to produce and discover new particles directly are described in Chapters~\ref{sec:BSM} and~\ref{sec:DM}.

\subsection*{Precision}
The ``Precision'' arrow in Figure~\ref{fig:papdist} represents accurate measurements in the EW and Higgs sector. These are possible at the muon collider by exploiting the large rate for EW-scale scattering processes initiated by effective vector bosons, and the small QCD background that is typical of leptonic collisions. The prospects for exploiting the 10 million Higgs boson produced by the collider in the vector boson fusion channel have been investigated in details~\cite{Han:2020pif,Forslund:2022xjq} based on fast simulation but also validated against the available full-simulation studies~\cite{MuonCollider:2022ded}. These sensitivity projections for Higgs signal-strength measurements were used in Ref.~\cite{Accettura:2023ked} for a Higgs couplings fit in the same setup that was employed in Ref.~\cite{deBlas:2019rxi} for a global comparative assessment of future colliders sensitivity in preparation for the 2020 European Strategy Update process. The most updated available input, described in Section~\ref{sec:EHT}, is employed for the couplings fit shown on the left panel of Figure~\ref{fig:3h-ch}.
The result is that a per mille level determination of the single Higgs boson couplings will be possible at the 10~TeV MuC, similarly to future low-energy $\mathrm{e}^+\mathrm{e}^-$ Higgs factories. A detailed inspection of the sensitivity to different couplings reveals the complementarity between MuC and low-energy Higgs factory couplings determinations.

The findings above refer to the ``kappa-0'' Higgs couplings fit~\cite{deBlas:2019rxi}, where no BSM Higgs decay channel is allowed. The closure of the ``kappa-3'' fit, where this assumption is relaxed, requires a measurement that is sensitive to the absolute value of at least one of the couplings of the Higgs, without dependence on the Higgs total width. The determination of the total (inclusive) Higgs cross-section---with excellent $0.5\%$ precision~\cite{Li:2024joa,Ruhdorfer:2024dgz}---in the Z-boson fusion production channel could provide such measurement at the MuC, enabling a very precise coupling determination also in the kappa-3 scheme and an indirect determination of the Higgs width with a precision of $\sim 1\%$
(see Chapter~\ref{sec:EHT}).  However, this measurement relies on the observability of muons in the forward region by a dedicated detector, whose feasibility is still to be assessed. If such a measurement turns out to be impossible, the closure of the kappa-3 fit will have to rely on the inclusive cross-section measurement at a low-energy $\mathrm{e}^+\mathrm{e}^-$ Higgs factory, providing one further element of complementarity with such a machine. The flat direction in the kappa-3 fit could be also lifted by high-energy measurements at the MuC, by relying on extra assumptions~\cite{Forslund:2023reu}.

Unlike low-energy Higgs factories, the MuC can also measure the double-Higgs production cross-section and determine the Higgs trilinear coupling at the percent level, as shown on the right panel of Figure~\ref{fig:dir-3h-ch}. This result~\cite{Han:2020pif,Buttazzo:2020uzc} is based on a parametric modeling of the detector effects---assuming CLIC-like performances---validated against the CLIC full simulation sensitivity projection~\cite{deBlas:2018mhx}. Figure~\ref{fig:dir-3h-ch} also reports the expected sensitivity of FCC-hh~\cite{Mangano:2020sao}, which ranges from $3\%$ to $8\%$ depending on the assumed detector performances. The MuC result has been found less sensitive to detector performances because the background is lower.

\subsection*{Precision from energy}
A unique feature of the muon collider is the simultaneous availability of energy and precision. This allows, very simply, the precise measurement of high-energy scattering cross-sections. The possibility of simultaneously exploiting energy and precision is represented by the ``Energy with Precision'' arrow in Figure~\ref{fig:papdist}. It corresponds to an exploration strategy that is typical of high energy physics: higher energy observables are more sensitive to short-distance physical laws and give access to novel effects that were too tiny to be detected in previous experiments performed at lower energy. This mechanism has produced some of the most striking past discoveries, such as the one of the finite radius---i.e., the composite nature---of the nucleons.

Sufficient statistics will be available at the muon collider to measure 10~TeV cross sections---specifically, the production of quarks, leptons, or vectors and Higgs pairs with 10~TeV invariant mass---with percent-level accuracy. For BSM effects that are proportional to the square of the energy divided by the characteristic new physics mass scale, this translates into sensitivity to new physics at 100~TeV. Notice for comparison that 100~TeV scale new physics produces instead unobservably small one part-per-millon effects on EW-scale (100~GeV) observables. This offers a competitive advantage to the muon collider relative to EW-scale measurements (at or above the $Z$ pole) at future $e^+e^-$ colliders. On the other hand, the 100~TeV scale will not be directly accessible even at a 100~TeV hadron collider.

In spite of the fact that the $\mu^+\mu^-$ initial state is electrically neutral, this high sensitivity is not limited to scattering processes with a neutral hard final state. In a collision at 10~TeV, the emission of soft-collinear charged $W$ bosons is very likely because of the Sudakov enhancement, resulting in large rates for the production of two hard particles with non-vanishing total charge such as $WZ$, $hZ$, $tb$, etc. This gives effective access to charged scattering amplitudes with muon-neutrino initial states, and to scattering processes that are effectively initiated by neutrino collisions. Exploiting the emission of the $W$ and other EW radiation effects allows to define a variety of different observable cross-sections and a rich program of measurements that can probe and disentangle different BSM effects and different EFT interaction operators~\cite{Chen:2022msz}. A recent comprehensive stydy of the different measurements~\cite{EWMuC} is presented in Section~\ref{sec:EHT}.

\begin{figure}[t]
   \begin{center}       
   \includegraphics[width=0.47\textwidth, angle=0]{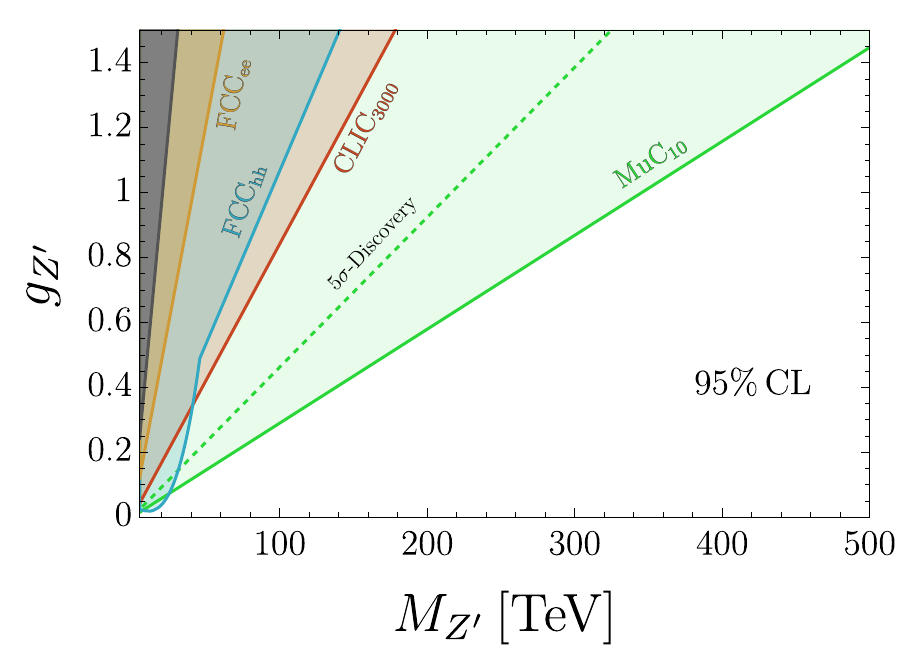} \hfill
   {\includegraphics[width=0.47\textwidth, angle=0]{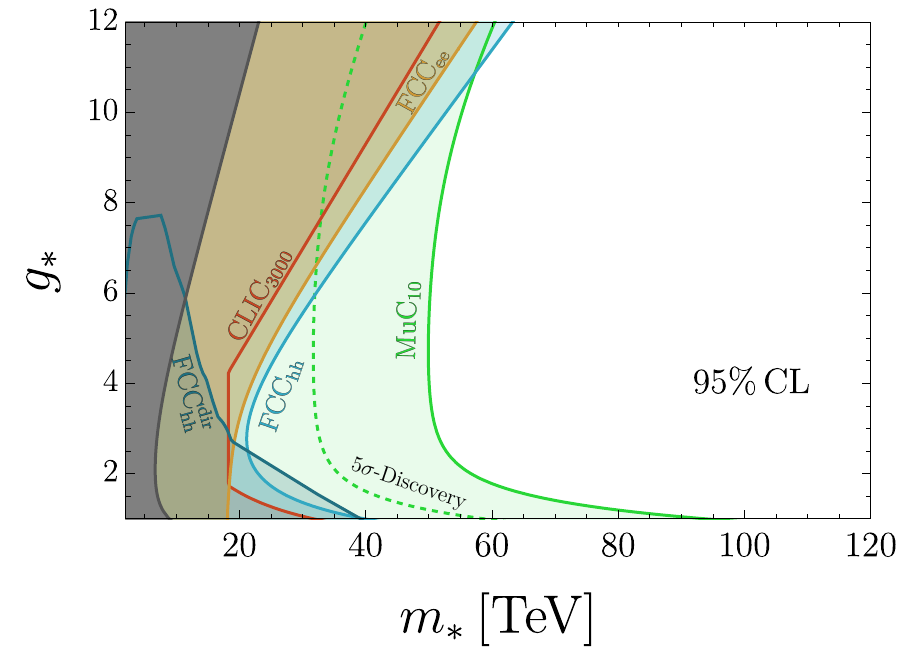}}    
   \end{center}
 \caption{
Left: Future colliders $95\%$~CL exclusion sensitivity to a minimal $Z^\prime$~\cite{Appelquist:2002mw}. In the case of muon colliders, the $5\sigma$ discovery reach is also shown by dashed lines. Right: The sensitivity to Higgs compositeness.}
 \label{fig:fl}
\end{figure}

The sensitivity of high-energy measurements to concrete BSM scenarios (from Ref.~\cite{Chen:2022msz}) are illustrated in Figure~\ref{fig:fl}. The left panel displays the muon collider reach on a new neutral current interaction mediated by a heavy $Z^\prime$ gauge boson coupled to the SM Hypercharge~\cite{Appelquist:2002mw} in the plane formed by the $Z^\prime$ mass and gauge coupling. The 10~TeV MuC mass reach for discovery is around 100~TeV for a coupling of the order of the SM EW gauge couplings. The mass reach for exclusion extends far higher, up to 500~TeV for the maximal value of the $g_{Z^\prime}$ coupling, $g_{Z^\prime}\simeq1.5$, allowed by perturbativity in this specific model. The sensitivity of CLIC, FCC-ee, and FCC-hh (from Ref.~\cite{EuropeanStrategyforParticlePhysicsPreparatoryGroup:2019qin}) is also reported in the figure for comparison. The design and construction of a muon collider appears as the only option to probe this scenario at the 100~TeV scale.

The right panel of Figure~\ref{fig:fl} quantifies the sensitivity of the 10~TeV MuC to Higgs compositeness. The scenario under consideration is that of a pseudo-Nambu-Goldstone Boson (pNGB) composite Higgs (see~\cite{Panico:2015jxa} for a review), which is the only known possibility to explain---at the price of a moderate fine-tuning on a single parameter---the agreement between current measurements of the Higgs couplings and SM predictions. The experimental manifestations of a composite pNGB Higgs can be robustly modelled in terms two parameters $m_*$ and $g_*$~\cite{Giudice:2007fh}, which correspond respectively to the Higgs compositeness scale---i.e., to the inverse of the Higgs particle radius---and to the coupling of the new strong sector that delivers the Higgs as a bound state. This theoretical setup was extensively employed for the comparison of future collider projects in preparation for the 2020 European Strategy Update~\cite{deBlas:2019rxi,EuropeanStrategyforParticlePhysicsPreparatoryGroup:2019qin}.

The muon collider sensitivity to Higgs compositeness emerges from 3 different classes of measurements, whose combined sensitivity is shown in Figure~\ref{fig:fl} in the $(m_*,g_*)$ plane. Higgs coupling modifications are mostly relevant when $g_*$ is large and they dominate the $m_*$ reach for $g_*$ above around $9$. Searches for new effects in the 10~TeV di-fermion production cross section due to the modification of the EW gauge interactions induced by the new strong sector are relevant only when $g_*$ is small, explaining the enhanced sensitivity when $g_*\simeq1$. Measurements in di-boson and boson-plus Higgs final states---again performed at 10~TeV exploiting ``Precision from Energy''---are instead equally relevant for any value of $g_*$, because they probe new interactions of the Higgs doublet with the vector bosons that are directly related to the finite radius of the Higgs. The magnitude of these new interactions thus depends only on the compositeness scale $m_*$ and not on the coupling $g_*$. Such direct manifestations of Higgs compositeness dominate the muon collider sensitivity and they allow the discovery of Higgs compositeness up to around 35~TeV (or exclusion up to around $50$~TeV even for the most unfavourable value of $g_*$). The comparison with other future collider projects (from~\cite{deBlas:2019rxi,EuropeanStrategyforParticlePhysicsPreparatoryGroup:2019qin}) displays the competitive advantage of the muon collider for the study of Higgs compositeness.

\begin{figure}[t]
   \begin{center}       
   \includegraphics[width=1\textwidth, angle=0]{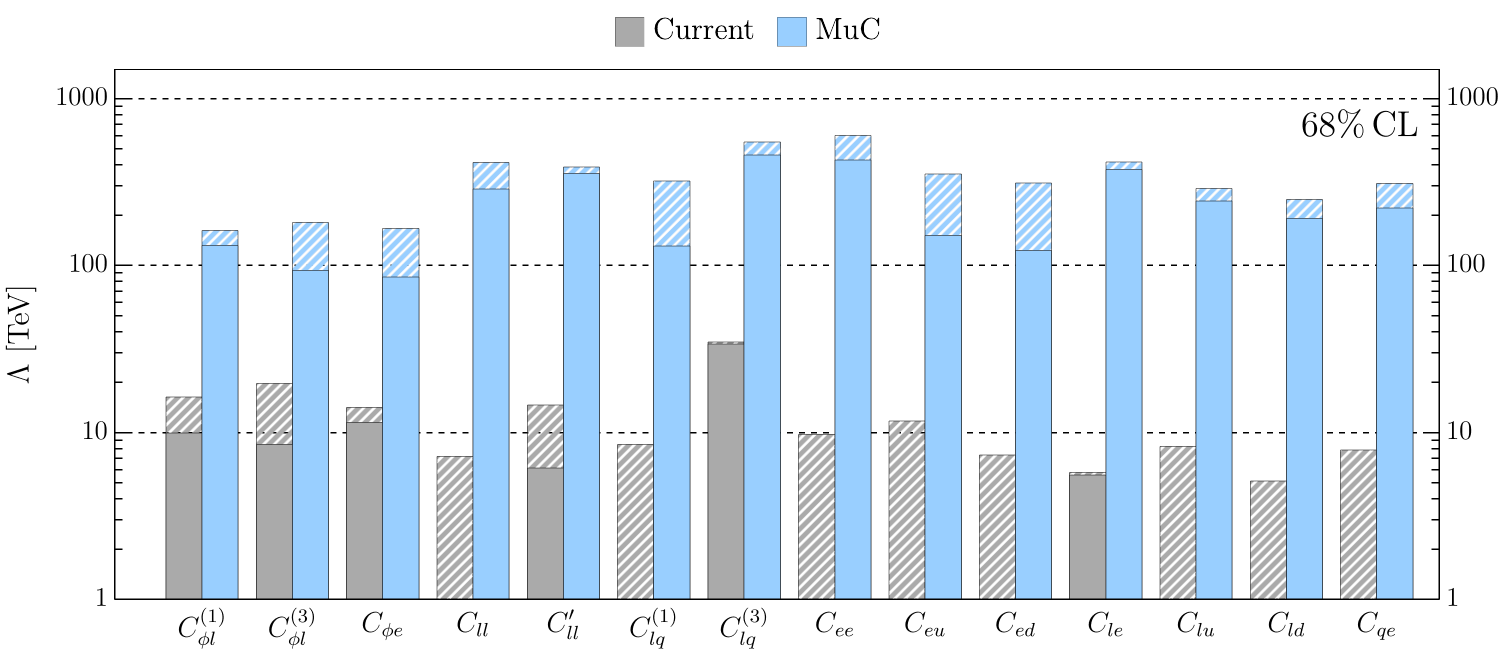}    
   \end{center}
 \caption{Fit to the $CP$-even flavour-blind  Warsaw basis~\cite{Grzadkowski:2010es} operators that grow with the energy and interfere with the SM in di-fermion and di-boson production at the MuC. The MuC projections (blue bars) from the measurements of these observables are compared with current knowledge (gray bars).
 }
 \label{fig:fl1}
\end{figure}

Beyond explicit models, the power of the ``Precision from Energy'' arrow can be also illustrated by the sensitivity to new interaction operators in a SM EFT context. Figure~\ref{fig:fl1} reports the result of a global fit~\cite{EWMuC} on a set of flavour-diagonal, universal and CP-preserving operators of dimension 6 in the SM EFT in the Warsaw basis~\cite{Grzadkowski:2010es}.\footnote{A much larger set of operators is considered in the fit of Section~\ref{sec:EHT}.} The selected operators are the ones whose effects on the di-fermion and di-boson (and boson-plus-Higgs) production amplitudes grow quadratically with the centre of mass energy and that produce a quadratically-growing contribution to the cross section by interfering with the SM amplitude. They form a coherent set of operators for a muon collider fit because they are probed up to much higher scale than the other dimension-six operators, by exploiting the high-energy measurements of di-lepton, di-quark and di-boson production. 

The result for the 10~TeV muon collider (solid blue bars) displays sensitivity to an effective scale of 100~TeV or higher on all the operators. The limited sensitivity degradation of the global fit in comparison with the single-operator reach (dashed blue bars) shows that the muon collider can measure a sufficient number of different observables to probe and disentangle the different operators. The different measurements exploit charged final states and the possibility of distinguishing final states that are ``exclusive'' or ``semi-inclusive'' with respect the the emission of EW radiation~\cite{Chen:2022msz}.

An important technical note is that the global fit results are obtained by marginalising only with respect to the operators considered in the figure, and not with respect to a complete basis of dimension-6 operators. This is equivalent to a complete global fit in the case of the muon collider, because the selected operators are (among the flavour-universal ones) the only ones that can be probed to such high scale. For the ``current''---which includes present knowledge on EW precision observables from LEP/LEP2 and LHC measurements---
instead, this procedure overestimates the global fit sensitivity 
to the first three operators in the figure, as it does not include the marginalisation on other operators that can give comparable contributions to the relevant observables (in this case, the EW precision observables).

It should be also emphasised that the MuC R{\&}D path could eventually enable the construction of a muon collider with more than 10~TeV energy in the centre of mass. If the luminosity grows quadratically with the energy as in the IMCC target scaling, the $\Lambda$ scale sensitivities of Figure~\ref{fig:fl} would grow linearly with the muon collider energy. This offers a path for improving the reach even further, which is instead harder to envisage with $e^+e^-$ colliders because the measurements' precision is limited by systematics and theoretical uncertainties.

\subsection*{Muons and neutrinos}

The ``Muons and neutrinos'' arrow in Figure~\ref{fig:papdist} reminds us that energetic muon beams in the TeV range will be made available for the first time at a muon collider. Studying muon collisions offers self-evident opportunities to discover new physics coupled primarily to muons and not for instance to electrons, whose collisions are instead extensively studied.

In addition to the quest for agnostic exploration, there are strong theoretical motivations for the study of muon-philic new physics. The SM Higgs Yukawa coupling of the muon is larger than that of the electron, hence the muon coupling to new physics in the EW symmetry breaking sector is generically larger than that of the electron. The implications of this larger coupling were studied for instance in Ref.~\cite{Chakrabarty:2014pja} in the context of an extended Higgs sector. The larger Yukawa couplings of the second family leptons and quarks also suggests stronger muon interactions with a new physics sector responsible for the origin of flavour.
For instance, UV models that aim at explaining the hierarchies in Yukawa couplings by introducing flavour non-universal gauge interactions typically predict larger effects in muons than in electrons, see e.g. \cite{Bordone:2017bld,Fuentes-Martin:2020pww,FernandezNavarro:2022gst,Davighi:2022fer,Davighi:2022bqf,Davighi:2023iks,Greljo:2023bix,Barbieri:2023qpf,Greljo:2024zrj}.
Finally, studying high-energy muons interactions is complementary to precise low-energy tests of the SM such as the measurement of the muon anomalous magnetic moment~\cite{Capdevilla:2020qel,Buttazzo:2020ibd,Yin:2020afe,Capdevilla:2021rwo,Dermisek:2021ajd,Dermisek:2021mhi,Capdevilla:2021kcf,Bandyopadhyay:2021pld,Dermisek:2022aec,Paradisi:2022vqp} and  the $\tau \to 3 \mu$ (see Section~\ref{1:physics:sec:flav}) decay. A muon collider will be needed to establish and characterise the BSM origin of possible tensions with the SM of these measurements. 

Beyond collisions, energetic muon beams offer additional unique opportunities. They produce a collimated beam of neutrinos with unprecedented properties to be employed for a neutrino experiment. This and other opportunities for neutrino physics are described in Section~\ref{1:physics:sec:neutrinos}, Section~\ref{sec:CKM} and in Chapter~\ref{sec:SI}.

\section{Flavour physics at the energy frontier}\label{1:physics:sec:flav}

\begin{figure}[t]
   \begin{center}        \includegraphics[width=0.6\textwidth]{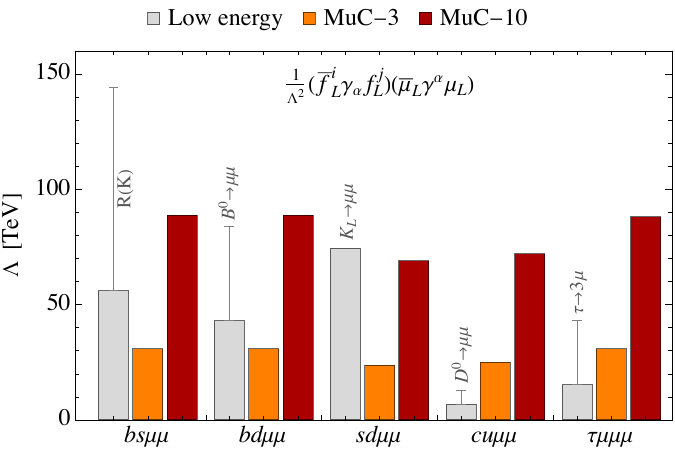} 
   \includegraphics[width=0.36\textwidth]{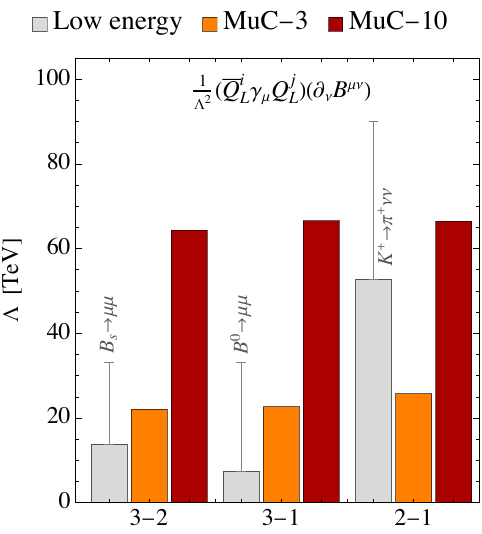} 
   \end{center}
\caption{Sensitivity reach in the effective scale $\Lambda$ [TeV] of effective operators containing a quark or lepton flavour-violating current, coupled to either a muon current (left panel) or a flavour-blind gauge current.
The gray bands show the present  constraints from meson \cite{LHCb:2022vje,Neshatpour:2022pvg,CMS:2024smm,NA62:2024pjp} and tau \cite{Hayasaka:2010np} decays, while the gray lines are the expected future sensitivity at the end of LHCb upgrade II \cite{Cerri:2018ypt}, Belle II \cite{Belle-II:2018jsg}, and NA62 \cite{HIKE:2023ext} runs.}
 \label{fig:muc_flavour}
\end{figure}

High energy also enables novel and powerful tests of flavour physics. On the one hand, the high energy enhances flavour transitions mediated by very heavy new physics, offering the opportunity to discover BSM flavour transitions directly in either the quark or the lepton sector by studying the $\mu^- \mu^+ \to f f^\prime$ production of two fermions with different flavour~\cite{Azatov:2022itm,Altmannshofer:2023uci,Ake:2023xcz}. On the other hand, the high available energy enables to search for the new particles that are potentially responsible for flavour violation. These two strategies for flavour physics investigation at the muon collider are illustrated by examples in the rest of the section in turn.

Figure~\ref{fig:muc_flavour} displays the sensitivity~\cite{2025:MucFlavour} (see Section~\ref{sec:flavor} for more details) to representative flavour-breaking BSM interactions from the search for the high-energy production of flavour-breaking final states such as $bs$, $bd$, $sd$, $cu$ in the quark sector, and $\tau\mu$ in the lepton sector. The analysis takes into account realistic quark-flavour and lepton ID tagging and misidentification efficiencies, which are the main detector parameters that control the sensitivity. On the left panel, we report the single-operator exclusion reach to interactions where the flavour-violating current is only coupled to the muon current and not to the other leptons in a setup that violates Lepton Flavour Universality (LFU) maximally. This assumption does not affect the reach of the muon collider, which is only sensitive to the muon interaction, but boosts the sensitivity of traditional low-energy strategies by exploiting LFU violation tests such as the measurement of R(K)~\cite{LHCb:2022vje}. The 10~TeV MuC reach on the effective operator scale $\Lambda$ is always at least comparable with current bounds (gray bands) and with future prospects for improvement (gray lines), and in some case it is significantly stronger. The right panel of Figure~\ref{fig:muc_flavour} considers instead flavour-violating currents that are coupled universally to all leptons and quarks, through a dimension-6 operator---reported in the figure---involving the EW Hypercharge gauge field $B_\mu$. The 10~TeV MuC sensitivity to this operator is much stronger than both current bounds and future prospects, with the exception of second-first (2-1) families transition, that is probed with the very rare decay $K^+\to\pi^+\nu\bar\nu$ at NA62 \cite{NA62:2024pjp,HIKE:2023ext}.

Building a muon collider will open a new high-energy path towards flavour physics exploration, with a generic EFT interaction scale reach of the order of 100~TeV that is often much higher than what is allowed by traditional methods based on low-energy high-intensity experiments. Furthermore, the muon collider sensitivity is attained by measurements with moderate backgrounds, whose interpretation requires theoretical predictions that do not appear extremely challenging. This has to be contrasted with low-energy measurements that require a precise control over experimental systematics and theoretical uncertainties that are furthermore often polluted by non-perturbative QCD dynamics. In addition, the putative discovery and characterisation of BSM effects at the muon collider will not rely on the measurement of a single quantity but rather on the determination of a variety of cross sections and differential distributions.

A more extensive illustration of the muon collider flavour physics potential through high-energy measurements is presented in Section~\ref{sec:flavor}.

\begin{figure}[t]
   \begin{center}        
   \includegraphics[width=0.5\textwidth]{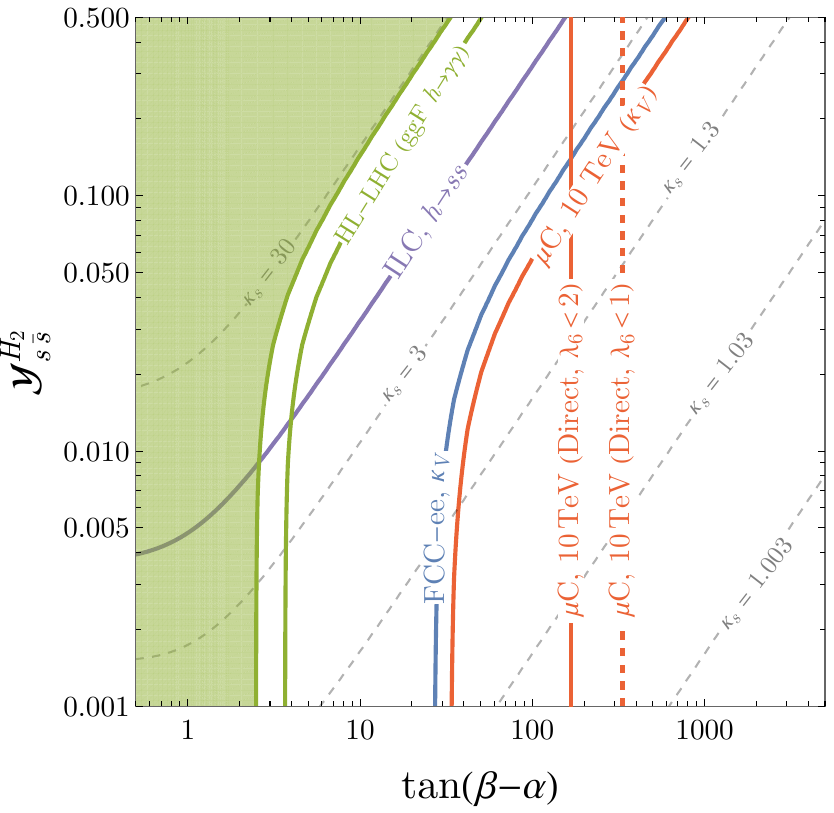} 
   \end{center}
\caption{Sensitivity to a 2HDM~\cite{Egana-Ugrinovic:2018znw,Egana-Ugrinovic:2019dqu,Egana-Ugrinovic:2021uew} which can generate appreciable changes to strange quark Yukawa coupling. The reach is shown in the plane of the mixing between the 2nd Higgs doublet and the SM Higgs, $\tan(\beta - \alpha)$, and the coupling of the heavy Higgs to the strange quark, ${\cal Y}_{s \bar{s}}^{H_2}$. Lines of constant modification to strange Yukawa coupling, $\kappa_s$ are shown as dashed lines. Present LHC Higgs measurements exclude the green shaded region, future prospects for different colliders from Higgs or direct searches (in case of muon collider, $\mu C$) are shown as solid lines.}
 \label{fig:muc_higgsflavour}
\end{figure}

Testing quark flavour in relation to the Higgs can also be a powerful probe of flavour violation.
At low-energy Higgs factories one can test some of the unmeasured quark Yukawa couplings\footnote{Unmeasured here means after HL-LHC, where u,d,s,c quark Yukawa couplings will not be measured if they have their SM values. Flavour tagging of lighter generation quarks could in principle be considered at a muon collider as well like at Higgs factories, but has not been studied experimentally yet.}  with flavour tagging.  This can be done  with precision in the case of the charm Yukawa coupling or approaching observation in the case of the strange Yukawa coupling.  The interpretation of a deviation in a Yukawa coupling at a low energy Higgs factory is through a higher dimensional operator
\begin{equation*}
\frac{1}{\Lambda^2}\bar{Q}Hq(H^\dagger H)\,,
\end{equation*}
and the projected reach on the operator scale $\Lambda$, given the precision of the measurements, ranges from the EW scale to order of the TeV scale at best. Observing departures from the SM thus {\em implies} the existence of new particles in the mass range that the 10~TeV MuC can probe directly.

The maximal scale for the new particles corresponds to scenarios where they contribute to the flavour-dependent process at tree-level. Otherwise, a lower BSM particles' mass would be needed in order to compensate for the loop factor suppression. However, at tree level there are only two classes of UV completions that allow for an uncorrelated Higgs flavour enhancement of the operator under examination: a two Higgs doublet model (2HDM)~\cite{Egana-Ugrinovic:2021uew} or new vector-like quarks (VLQs)~\cite{Alasfar:2019pmn,Bar-Shalom:2018rjs}. In either of these two possible classes of models there is a new EW-charged state. Given that the scale $\Lambda$ of the operator in question is proportional to the mass of the new state and inversely proportional to a new Yukawa coupling, $\Lambda^2 \propto M^2 / {\cal Y}_{s \bar{s}}$, there is a bounded parameter space of interest.  A large mass requires a large coupling to generate the same effect and therefore unitarity and perturbativity set an upper mass in the $\mathcal{O}(1)$~TeV range if $\Lambda$ is small enough to be seen.  A 10~TeV Muon Collider can therefore cover the entire relevant range via pair-production of the EW states and effectively probe Higgs flavour at much higher precision than Higgs couplings alone can conceivably reach. 

To illustrate this we show the results for a strange-flavour-violating (SFV) 2HDM that can generate sizeable strange quark Yukawa deviations that are consistent with flavour physics constraints~\cite{Egana-Ugrinovic:2018znw,Egana-Ugrinovic:2019dqu,Egana-Ugrinovic:2021uew} in Figure~\ref{fig:muc_flavour}. The model has effectively 3 parameters, the new coupling of the strange quark to the 2nd Higgs doublet, ${\cal Y}_{s \bar{s}}^{H_2}$, the mixing between the 2nd Higgs doublet and the SM Higgs one, $\tan(\beta - \alpha)$, and the mass scale of the 2nd Higgs doublet, $M_{H_2}$.  The mixing angle is proportional to an underlying quartic coupling between Higgses in the Higgs basis, $\lambda_6$, and therefore the scale of the operator is $\Lambda^2 \propto M_{H_2}^2 / \left({\cal Y}_{s \bar{s}}^{H_2} \lambda_6\right)$.  In Figure~\ref{fig:muc_flavour}, lines of constant modification to SM strange Yukawa coupling, $\kappa_s = y_s / y_s^{\rm SM}$, are shown as dashed lines. Large modifications to SM Yukawa couplings induce shifts to all branching ratios, and thus the best measured single-Higgs signal strength sets the strongest indirect bound. This is shown for current LHC data as the shaded green region, HL-LHC projection as the solid green line, FCC-ee as the blue line, and the 10 TeV MuC as the curved red line. Direct strange tagging measurements are possible at $e^+e^-$ colliders shown as the purple ILC line~\cite{Albert:2022mpk}. The direct reach for the muon collider comes from searching for the charged Higgs state in the 2HDM which we make the conservative assumption that can probe up to $4.5$~TeV mass. This is very powerful because it only depends on the mass, $M_{H_2}$, and not the other model parameters since it is produced due to its gauge quantum numbers electroweak pair-production.  However, in overlaying the direct search constraints in the parameter space of Figure~\ref{fig:muc_flavour} one has to fix a maximal value of $\lambda_6$ to interpret the direct search as $\tan(\beta - \alpha)\propto M_{H_2}/\lambda_6$.  The direct search constraint is then a vertical line on the plot, and to illustrate different choices in the criteria for perturbativity we show two different values of $\lambda_6$ in Figure~\ref{fig:muc_flavour}.

\section{Neutrino physics opportunities}\label{1:physics:sec:neutrinos}
\label{1:phys:ch:nu}

Understanding the properties of neutrinos and the new physics responsible for nonzero neutrino masses is among the highest priorities of contemporary particle physics (see, for example, \cite{Butler:2023glv}). Associated with these pursuits are long-baseline ``superbeam'' experiments: experiments where the source of neutrinos is the decay in flight of accelerator-produced mesons (mostly pions and kaons). DUNE \cite{DUNE:2020ypp} and Hyper-Kamiokande \cite{Hyper-Kamiokande:2018ofw}, both under construction, are the next-generation, and perhaps the ultimate, superbeam experiments. A muon collider facility contains, unavoidably, a plurality of neutrino ``beams'' of various flavour compositions, energy profiles, and intensities. The source of muons for the muon collider is very similar to that of neutrinos for superbeam experiments and, downstream of the production target, the decay of muons as these are cooled, bunched, accelerated, and stored leads to neutrinos.    

The physics opportunities associated to all possible neutrino sources at different stages of a muon collider facility are still being identified and explored by the community; here we briefly highlight only a few of them. We will concentrate on opportunities allowed by neutrinos produced in the decay of muons with well-characterized energy profiles since these yield neutrino beams with well-characterized energy spectra -- we understand the physics of muon decay very precisely -- and, in the case of negative (positive) muons, both muon-type (anti)neutrinos and electron-type antineutrinos (neutrinos). High-energy, -- energies above a few hundred MeV --  intense, and well-characterized electron-type neutrino and antineutrino beams are not available at any other existing or planned facility. These may play a crucial role in oscillation experiments in the near and the intermediate future. 

By the 2040s, the precision and, in some cases, the sensitivity of both DUNE and Hyper-Kamiokande might be limited by systematics. Among those are uncertainties associated with our understanding of neutrino--nucleus scattering and our ability to reconstruct the incoming neutrino energy \cite{NuSTEC:2017hzk,Balantekin:2022jrq}. Dedicated, high-precision neutrino-scattering experiments may be necessary in order to tame those systematics and fully exploit the potential of DUNE and Hyper-Kamiokande. These, in turn, will need to make use of well-characterized beams of electron-type neutrinos and antineutrinos with energies between hundreds of MeV and a few GeV. NuSTORM (Neutrinos from Stored Muons \cite{nuSTORM:2012jbd}) is a proposal to exploit stored muons with energies of a few GeV for such measurements. There are concrete studies of how NuSTORM can be connected to a muon collider demonstrator~\cite{nuSTORM:2022div}, as discussed in Section~\ref{3:sec:synergy:facilities}. Similar opportunities -- and more ambitious ones -- should also be present at any muon collider facility in the future.  

Whether neutrino oscillation experiments beyond DUNE and Hyper-Kamiokande will be needed in order to understand the properties of neutrinos with the necessary precision is an open question that cannot be answered in any quantitative way at the moment. Neutrinos from a muon storage ring are an excellent source for next-next-generation experiments. These facilities are referred to as ``neutrino factories'' and have been studied at some length in the literature. Recent analyses include \cite{Bogacz:2022xsj,Kitano:2024kdv,Denton:2024glz}. Neutrino factories allow one to circumvent beam-related systematics and backgrounds that plague neutrino superbeams and, most important, allow access to yet-to-be-measure oscillation channels, including $\nu_{e}\to\nu_{\mu}$ and $\nu_{e}\to\nu_{\tau}$. These, in turn, allow for searches for T-invariance and CPT-invariance violation not available at DUNE and Hyper-Kamiokande.  Given constraints associated to the size of the Earth and that of neutrino detectors, neutrino factories require intense stored muon beams with energies below tens of GeV. Previous studies point to $10^{20}$ muon decays per year \cite{IDS-NF:2011swj}. In order to collect these many useful muon decays, the geometry of the storage ring for the neutrino factory is very different from what is being considered for the muon collider and the energies are much smaller energies. Neutrino factories require long straight sections so most muon decays ``point'' towards the neutrino detector. In the past, race-track, bow tie, and triangle designs were considered \cite{ISSAcceleratorWorkingGroup:2008laz}. While neutrino factories and muon colliders have different requirements when it comes to, for example, the energy and the geometry of the storage ring, they share, to leading order, the same front-end: muon production, cooling, and acceleration. The R{\&}D on muon colliders is vital to enable the construction of a neutrino factory in the future.  

In the case of a muon collider of high (3 or 10~TeV) energy, most of the muons decay on the arcs and do not produce usable neutrinos. The neutrino beams are produced only from muons decaying in the relatively short straight sections and the intensity is lower than the one for a neutrino factor. Furthermore, the neutrino energy is too high for Earth-bound oscillation experiments, as the oscillation length scales linearly with neutrino energy. On the other hand, they do allow access to unprecedented, ultra-high energy neutrino fixed-target scattering experiments. In the next subsection we describe the characteristics and the physics potential of this neutrino beam in more detail. 

Finally, $\mu^+\mu^-$ collisions at several to dozens of TeV have the potential to decisively inform the neutrino mass puzzle. The new degrees of freedom may be directly produced and observed at a high-energy muon collider or their indirect impact in other observables may be within reach. Muon colliders also allow, thanks to the neutrino content of the muon, studies of virtually-on-shell neutrino--muon, neutrino--gauge boson, and neutrino--neutrino scattering. For a recent discussion, see, for example, \cite{Capdevilla:2024ydp}. Neutrino--neutrino scattering has never been observed in the laboratory. These parton-level collisions allow access to several different initial states, including those with non-trivial lepton-flavor number or total lepton number (e.g. $\mu^-\nu_e$ collisions have lepton number two, eletron number one and muon number one).

\subsection*{A forward neutrino experiment at the high-energy MuC}

When the muons in the beam decay in the straight section close to the Interaction Point (IP), they produce a collimated beam of very energetic neutrinos that could be used for physics measurements in a dedicated detector. A robust parametric estimate of the properties of this neutrino beam is presented below.

The collider target parameters reported in Table~\ref{t:facility_param} foresee a single bunch of $N_\pm=1.8\times10^{12}$ muons (and one of anti-muons) injected in the 10~TeV muon collider every $1/f_r=0.2~{\textrm{sec}}$. Assuming $10^7\,{\textrm{sec}}$ of operation per year, this corresponds to a current of $9\times10^{19}$ muons per year. The muons circulate until they decay, and those that decay in a straight section of the collider ring give rise to a collimated beam of neutrinos. The fraction of usable muons is the length $L$ of the straight section divided by the collider circumference, of 10~km, and a region of at least $L=10$~m without fields is required for the installation of the detector close to the Interaction Point (IP). This gives $9\times10^{9}$ neutrinos of each species per second and $9\times10^{16}$ per year at the 10~TeV collider. At the 3~TeV collider, a similar estimate gives $2.4\times10^{10(17)}$ neutrinos per second (year). Notice that our estimate only accounts for the muons that decay in the detector region. The total length of the straight section is about 100~m in the current versions of the collider lattice design, and most of the muons decaying in this longer region should produce usable neutrinos. At the real collider we could thus expect few or 10 times more neutrinos than our estimate.

The neutrinos are very collinear to the decaying muons, with an angle that is typically below $0.1~{\textrm{mrad}}$ already at the 3~TeV MuC, and even smaller at 10~TeV because of the larger muon energy.
The muon beam angular divergence is typically much above $0.1~{\textrm{mrad}}$, hence the distribution of the neutrino angle relative to the beam axis is driven by the dynamics of the muon beam and it cannot be estimated without reference to the lattice design, which is still preliminary. Fortunately, the beam angular divergence in the detector region equal to $0.6~{\textrm{mrad}}$, both for the 3 and 10~TeV colliders is a rather robust design parameter because it is directly related to the luminosity at fixed emittance. Our estimate of the neutrino angle (and energy) distribution thus assumes a constant $0.6~{\textrm{mrad}}$ angular spread\footnote{Specifically, we smear $p_{x,y}/p_z$, with $p$ the muon momentum, by independent Gaussians with $\sigma=6\times10^{-4}$.} for the decaying muons, and nominal beam energy of $1.5$ and 5~TeV for the 3 and 10~TeV muon collider, respectively. This simplified simulation of the neutrino angle and energy distribution has been found in qualitative agreement with the complete simulation based on a preliminary design of the lattice.

\begin{figure}[t]
   \begin{center}        \includegraphics[width=\textwidth]{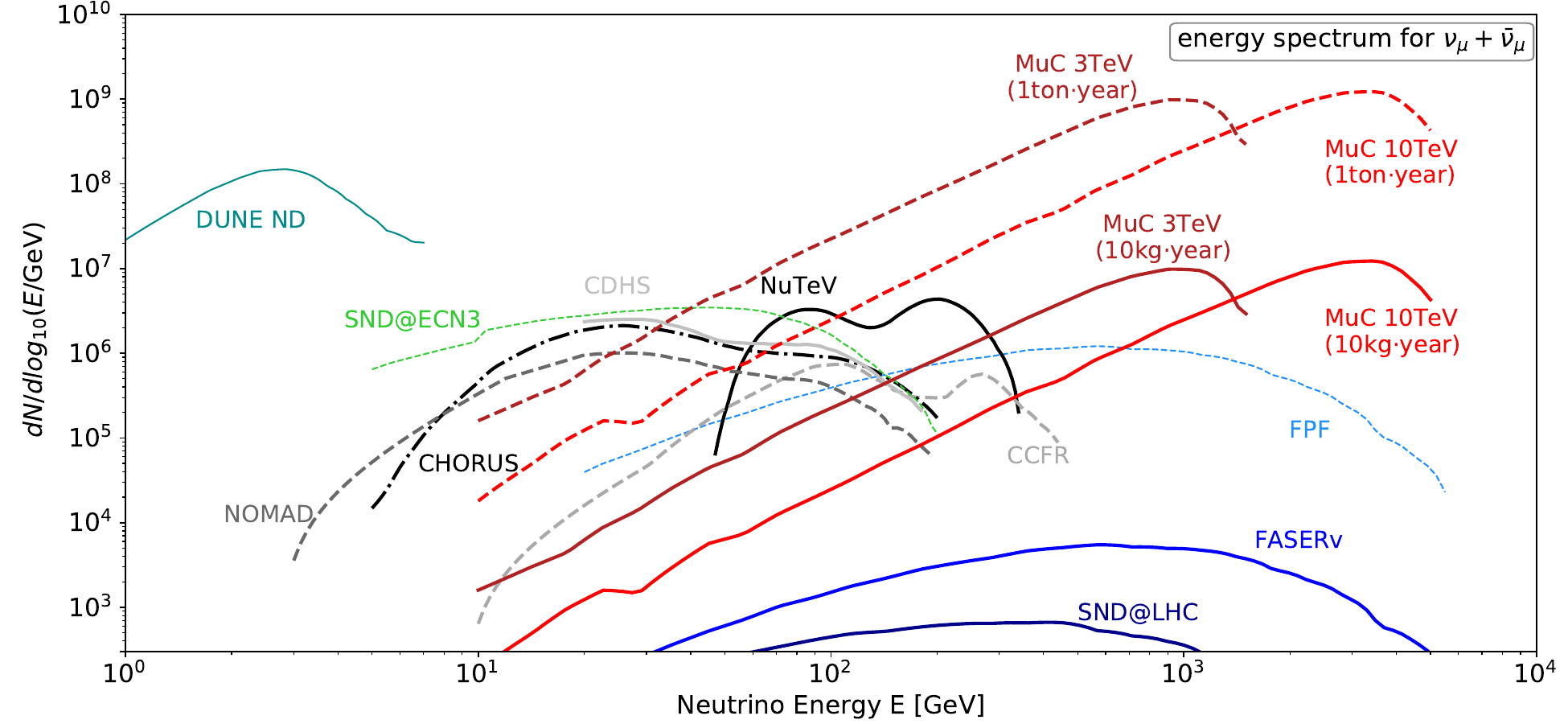}   
   \end{center}
 \caption{The energy spectrum of neutrino interactions produced by the 3~TeV and 10~TeV MuC in one year, overlaid with the summary plot in Ref.~\cite{Ahdida:2023okr} for past and planned neutrino experiments. The solid and dashed lines assume, respectively, a small 10~kg and a realistic 1~ton target mass.}
 \label{fig:nfl}
\end{figure}

The relevant parameter for physics studies is the rate of neutrino interactions, which in turn depends on the geometric acceptance and on the mass per unit area of the neutrino target. There are good perspectives to attain order-one geometric acceptance: around 30\% of the neutrinos would intercept a small 10~cm radius cylindrical target placed at a realistically large (200~m) distance from the IP inside a dedicated forward detector. Each neutrino crossing the target has the chance to produce a detectable interaction, with probability
\begin{equation*}
    p_{\rm{int}}\simeq 6 \times10^{-12}\,
    \frac{\rho\cdot{L}}{{\rm{g\,cm}}^{-2}}\,\frac{E}{\rm{TeV}}\,,
\end{equation*}
where we assumed a typical cross-section on nucleons of $10^{-35} {\rm{cm}}^2\cdot{E}/{\rm{TeV}}$ for all neutrino species, with $E$ the neutrino energy. In the equation, $\rho\cdot L$ is the density of the neutrino target times its longitudinal extension  along the neutrino beam. This parameter---or, equivalently the mass of the target since the radius is fixed at around 10~cm---is not straightforward to estimate, because it is intertwined with the capabilities of the detector to record the products of the neutrino interactions. Two options are considered in what follows, and the resulting interaction rates correspond to the continuous or dashed lines of Figure~\ref{fig:nfl}.

No design is currently available for the neutrino detector. One concept was developed long ago~\cite{King:1997dx}, with the  purpose of studying the neutrinos produced by a dedicated 250~GeV muon ring. The concept foresees a cylindrical target with 1~m length and 10~cm radius composed of 750 silicon CCD tracking planes, which would also act as a vertex detector. This would be followed by a magnetized spectrometer and calorimeter. The design of Ref.~\cite{King:1997dx} ensures excellent tracking and reconstruction performance for all the particles that are possibly produced in the interaction, including electrons, muons (and perhaps $\tau$'s) as well as bottom and charm hadron tagging and discrimination. On the other hand, the target mass per unit area is relatively small: $\rho\cdot{L}=50$~g$\cdot{\textrm{cm}}^{-2}$, for a total target mass of only 10~kg. Contemporary neutrino detectors operating at similar energy and with similar physics needs employ denser and bigger targets and attain masses ranging from one ton (FASERv~\cite{FASER:2022hcn}, SND@LHC~\cite{SNDLHC:2022ihg}, CHORUS~\cite{CHORUS:2005cpn}) to several hundred tons (CHARM~\cite{CHARM:1987pwr}, NuTeV~\cite{NuTeV:2005wsg}). Waiting for dedicated studies, it seems reasonable to assume a target mass of at least one ton for the neutrino detector.

Figure~\ref{fig:nfl} reports the energy distribution and the total number of neutrino interactions that could be recorded by a 10~kg target (corresponding to Ref.~\cite{King:1997dx}) and by a 1~ton target during one year of run at the 10~TeV and 3~TeV muon colliders. The neutrinos available for a number of past and planned experiments (from Ref.~\cite{Ahdida:2023okr}) are also reported for comparison. Notice that only the anti-neutrinos from the decay of the $\mu^+$ are included in the figure. A second detector would be needed to collect the neutrinos from the $\mu^-$ decay, since they fly away from the IP in the opposite direction. Electron (anti-)neutrinos with a similar spectrum are also present.

The figure shows that one single year of muon collider operation would enable to collect orders of magnitude more interactions of TeV-energy neutrinos than current facilities such as FASER$\nu$ and SND@LHC. Even with the very small 10~kg detector, more neutrinos will be collected than at the proposed Forward Physics Facility (FPF)~\cite{Feng:2022inv}. The energy spectrum peaks at 1~TeV for the 3~TeV MuC, and at 4~TeV in the case of the 10~TeV collider, and around thousand million interactions would be recorded at these energies by a 1~ton detector. There are many orders of magnitude more neutrino interactions in this energy range than any other past or planned neutrino experiment.  Furthermore, extremely small uncertainties are expected in the predictions for the energy spectrum of  the muon collider neutrinos,  to be contrasted with the large uncertainties in the spectrum of the neutrinos produced by the LHC arising from forward hadron production. On top of the superior statistics, neutrino physics measurements at a muon collider are thus expected to benefit from reduced systematic uncertainties in comparison with LHC-based experiments such as FASER or FPF.

The physics opportunities offered by the neutrino beams are still to be explored. Ideas discussed long ago~\cite{King:1997dx} include measurements of the CKM quark mixing matrix, nucleon structure, EW precision and charm quark physics. The contemporary relevance of these measurements is being assessed, and the sensitivity projections adapted to the higher-energy neutrino beams that would be available at the 3 and 10~TeV MuC. Preliminary results described in Section~\ref{sec:CKM} and Chapter~\ref{sec:SI} outline unprecedented opportunities for both CKM elements and Parton Distribution Functions (PDFs) determination from neutrino Deep Inelastic Scattering (DIS) measurements.

An extraordinary neutrino DIS experiment could be performed with the high energy muon collider neutrinos. The intense and precisely characterised beam of TeV-energy neutrinos will enable high-statistics and low-systematics DIS measurements, in a region of high transferred momentum $Q^2$ that is well within the realm of perturbative QCD, enabling accurate predictions. Very fine binning in the $x$--$Q^2$ plane with permille-level statistical uncertainties will enable the combined determination of the PDFs and of SM parameters such as the CKM elements. Multi-differential measurements will be possible to access the 3D structure of the proton in terms of non-perturbative quantities such as transverse-momentum dependent PDFs or generalised PDFs. The large statistics can be positively compared with the predicted event yields at the Electron-Ion Collider (EIC)~\cite{AbdulKhalek:2021gbh}. DIS measurements at a dedicated (but fully parasitical) far-forward neutrino detector at the muon collider would provide a charged-current analogue of the EIC for nuclear physics.
\chapter{Interface}
\label{1:inter:ch}

\section{Physics and detector needs}
\label{1:inter:sec:physics}

The vast physics potential described in Chapter~\ref{1:phys:ch} can be translated into requirements for the machine and the experiments that will harvest data from it.

The most notable requirements on the machine are: the mitigation of the beam-induced backgrounds (discussed in Section~\ref{1:inter:sec:mdi}), the amount of integrated luminosity delivered and the ability to measure it with permille-level precision, and a precise determination of the beam energy.

The requirements on the detectors situated at the main interaction points span from detector acceptance, to particle detection and identification efficiency, as well as to resolutions on the various particle properties inferred by the instrumental measurements.
They are outlined in terms of ``baseline'' and ``aspirational'' targets. The latter are specific to a muon collider operating at $\sqrt{s}=10$~TeV. 
The baseline requirements are set by the need to separate collision products from energy depositions from beam-induced backgrounds and the need to identify and measure particles over a wide range of energies with a precision at least comparable to the experiments taking data at the LHC.
The aspirational targets are set with the goal of fully exploiting the physics potential of the machine. They aim at a reconstruction performances comparable to those targeted by Higgs/Top/Electroweak-factories while extending to significantly higher energies. Furthermore, these requirements are set with the goal of maintaining sensitivity to unconventional signatures as well as profiting from the unique potential of muon colliders to use forward muons to study vector boson fusion processes.
The preliminary targets for several key metrics are discussed in more detail in the next Section and summarised in Table~\ref{tab:detector_req}. 

\begin{table}[h]
\begin{center}
\caption{Preliminary summary of the ``baseline'' and ``aspirational'' targets for selected key metrics, reported separately for machines taking data at $\sqrt{s}=3$ and $10$~TeV. The reported performance targets refer to the measurement of the reconstructed objects in physics events after, for example, background subtraction and not to the bare detector performance.}
\label{tab:detector_req}
\begin{tabular}{lccc}
\hline\hline
 \textbf{Requirement} & \multicolumn{2}{c}{\textbf{Baseline}} & \textbf{Aspirational}\\
  & \textbf{$\sqrt{s}=3$~TeV} & \textbf{$\sqrt{s}=10$~TeV} & \\
\hline
Angular acceptance & $|\eta|<2.5$ & $|\eta|<2.5$ & $|\eta|<4$ \\
Minimum tracking distance [cm] & $\sim 3$ & $\sim 3$ & $< 3$\\
Forward muons ($\eta> 5$) & -- & tag & $\sigma_{p}/p \sim 10$\%\\
Track $\sigma_{p_T}/p^{2}_{T}$ [GeV$^{-1}$] & $4 \times 10^{-5}$ & $4 \times 10^{-5}$ & $1 \times 10^{-5}$ \\
Photon energy resolution & $0.2/\sqrt{E}$ & $0.2/\sqrt{E}$ & $0.1/\sqrt{E}$\\
Neutral hadron energy resolution & $0.5/\sqrt{E}$ & $0.4/\sqrt{E}$ & $0.2/\sqrt{E}$ \\
Timing resolution (tracker) [ps] & $\sim 30-60$ & $\sim 30-60$ & $\sim 10-30$ \\
Timing resolution (calorimeters) [ps] & 100 & 100 & 10 \\
Timing resolution (muon system) [ps] & $\sim 50$ for $|\eta|>2.5$ & $\sim 50$ for $|\eta|>2.5$ & $< 50$ for $|\eta|>2.5$ \\
Flavour tagging & $b$ vs $c$ & $b$ vs $c$ & $b$ vs $c$, $s$-tagging \\
Boosted hadronic resonance ID & $h$ vs W/Z & $h$ vs W/Z & W vs Z\\
\hline\hline
\end{tabular}
\end{center}
\end{table}

Initial considerations on the interplay between the accelerator complex and detectors dedicated to the study of neutrino physics are discussed in Section~\ref{1:phys:ch:nu}.

\subsection*{Metrics and benchmarks for detector optimisation}
\label{subsec:detector_challenges}

The requirements on a detector that can successfully extract physics from collision data at a muon collider can be grouped into two categories: the rejection of the reducible beam-induced backgrounds (BIB) and the need to measure physics observables with high precision. 
These two categories can lead to conflicting choices that require an overall optimisation to be performed. 

The optimisation considers a set of key metrics, defined below. In addition, it is required that excellent (>90\%) detection efficiencies must be achieved for energies between $\mathcal{O}(1)$~GeV to $\mathcal{O}(1)$~TeV for all measured particle species. 
\begin{itemize}
    \item Angular acceptance: Expressed here in terms of the pseudorapidity $\eta=-\log(\tan(\theta/2))$. The presence of large absorber nozzles located in the forward region of the detector is key to reducing the high-energy showers from the beam decays to a soft, diffused, energy contribution in the detector volume. At the same time, especially when considering a machine operating at a centre of mass energy of 10~TeV, a sizeable fraction of the collision events will include outgoing particles within the uninstrumented volume of the absorbers. 
    \item Minimum tracking distance: Defined as the minimum distance between the first sensitive tracking detector layer and the IP. The minimum tracking distance affects the resolution with which the track impact parameters can be measured, directly influencing the performance of flavour tagging algorithms and sensitivity to a broad range of non-prompt physics. A shorter distance typically implies a larger number of BIB particles impinging on the detectors.
    \item Forward muons: Expressed here in terms of $\eta$ of dedicated muon detectors placed in the regions upstream and downstream of the interaction point (IP). Because of their highly penetrating nature, outgoing muons produced in the IP could be detected by dedicated instrumentation placed beyond the shielding nozzles and the accelerator components. A realistic study of the placement of the forward muon detectors, their technology and the effects of BIB still needs to be performed.
    \item Track $\sigma_{p_T}/p^{2}_{T}$: The track transverse momentum resolution. This quantity is used to assess the performance of the choices of tracking detector layouts and magnetic field intensity. 
    \item Photon energy resolution: The photon energy resolution, provides a benchmark for electromagnetic calorimetry.
    \item Neutral hadron energy resolution: The neutral hadron energy resolution, provides a benchmark for hadron calorimetry.
    \item Timing resolution (tracker/calorimeters/muon system): The resolution on the timestamps measured by various detector sub-systems. The requirements are mostly driven by BIB suppression considerations. The time of arrival of BIB particles on the sensitive elements of the detector spreads over a few ns after the bunch crossing, providing a powerful handle for rejection. The use of timing and time-of-flight information for particle identification is a potentially promising tool that has yet to be studied.
    \item Flavour tagging: A qualitative metric to track the ability to identify the flavour of the partons initiating the hadronic jets produced at the IP. Several sub-detectors can affect this capability, from tracking to time-of-flight measurements.
    \item Boosted hadronic resonance ID: A qualitative metric to track the ability to identify boosted hadronic resonances. This capability is mostly affected by the granularity of the calorimeter systems and the jet energy resolution.
\end{itemize}

In order to guide the optimisation, a set of physics benchmark channels has been chosen to allow a quantitative assessment of the requirements:
\begin{itemize}
    \item Higgs and multi-Higgs boson: These processes represent high-rate SM benchmarks. They contribute to setting requirements on the angular acceptance of the detectors, particle identification and energy/momentum resolution as well as the particle identification and flavour tagging efficiencies. In particular, the ability to distinguish hadronically-decaying Higgs bosons from Z and W bosons in the whole detector acceptance is crucial.
    \item Vector boson scattering: A comprehensive programme of vector boson scattering measurement and BSM searches (e.g.\,extended Higgs sectors) sets the requirements for measuring not only the presence of the outgoing muons, but also their momenta with good resolution. 
    \item Heavy Vector Triplets (HVT): The identification of boosted hadronic resonances is an important feature. High-mass searches, e.g.\,searches for HVTs, would benefit from the ability to distinguish W and Z bosons in centrally-produced resonances. Furthermore, HVTs provide targets for the performance of all high-energy objects.
    \item Long-lived particles: The search for heavy (meta-)stable massive particles was identified as a benchmark channel that would profit directly from a precise determination of the time of flight by each of the detector systems. A minimum goal of retaining sensitivity to slow-moving particles with a relativistic velocity $\beta \geq 0.5$ was taken as the baseline model-independent benchmark. Unconventional signatures and the search for long-lived states still need to be explored in detail at a muon collider. However, it is reasonable to assume that they would impose requirements on the ability to reconstruct displaced decays using only a subset of the available detector systems. 
\end{itemize}

The requirements on detector acceptance and performance are less stringent when considering an energy stage at $\sqrt{s}=3$~TeV. The differences are driven by the smaller contribution of vector boson fusion processes to the total cross-section and by the lower maximum energy of the collision products.

\subsection*{Recent achievements}

The last few years have marked a steep progress towards understanding and defining the physics requirements of a muon collider detector. This allowed to first establish the feasibility of experimentation at a centre-of-mass energy of 3~TeV~\cite{Black:2022cth,Accettura:2023ked,Casarsa:2023vqx} and recently led to the design of the first detector concepts (described in Chapter~\ref{1:det:ch}) for the 10~TeV collider.

Phenomenological studies (discussed in Section~\ref{1:phys:ch}), have been performed at a variety of levels of sophistication, from background free studies on truth level events, to fast detector simulation in \textsc{DELPHES} with all relevant backgrounds.  These studies are able to cover a large array of phenomena and inform the detector design focusing on the processes produced at the IP.  At the same time, studies based on full simulation were also performed for fewer selected channels to quantify the detector requirements in the presence of beam-induced and beam-related backgrounds, such as the production of incoherent $e^{+}e^{-}$ pairs.  

The results obtained with full- and fast-simulation methods have also been validated against each other in selected scenarios~\cite{Black:2022cth,Accettura:2023ked} finding good agreement between the predictions. This increased the confidence in the use of fast simulations to help better delineate the physics performance goals that have been reported.

Notable highlights were the studies performed on Higgs measurements and dark matter searches with disappearing tracks. As described in the previous Section, Higgs measurements~\cite{Forslund:2022xjq,Forslund:2023reu,Andreetto:2024rra} serve as a general purpose benchmark for low-energy Standard Model processes. These were used to evaluate the effects of the detector acceptance on the final Higgs boson coupling precision, as well as to perform an initial study of the resolution and instrumental effects related to the presence of beam-induced backgrounds. The search for disappearing tracks~\cite{Capdevilla:2021fmj} is a unique probe to definitively test the WIMP dark matter paradigm at a collider experiment. The signature consists of short tracks formed by a few (3-4) tracking detector hits. As such, it is uniquely sensitive to the detector layout, technology and the levels of beam-induced background.
A more recent study~\cite{Capdevilla:2024bwt} further extended the detector considerations by focusing on the reconstruction of charged particle trajectories for particles with transverse momenta below 1~GeV.

Furthermore, fast simulation studies~\cite{Ruhdorfer:2023uea,Forslund:2023reu} have been crucial in demonstrating the need for not just forward muon tagging, but also the aspirational goal of forward muon measurements and how these measurements affect, e.g., the search for exotic Higgs boson decay modes.

\begin{figure}[!h]
   \begin{center}
        \includegraphics[width=0.45\textwidth,angle=270]{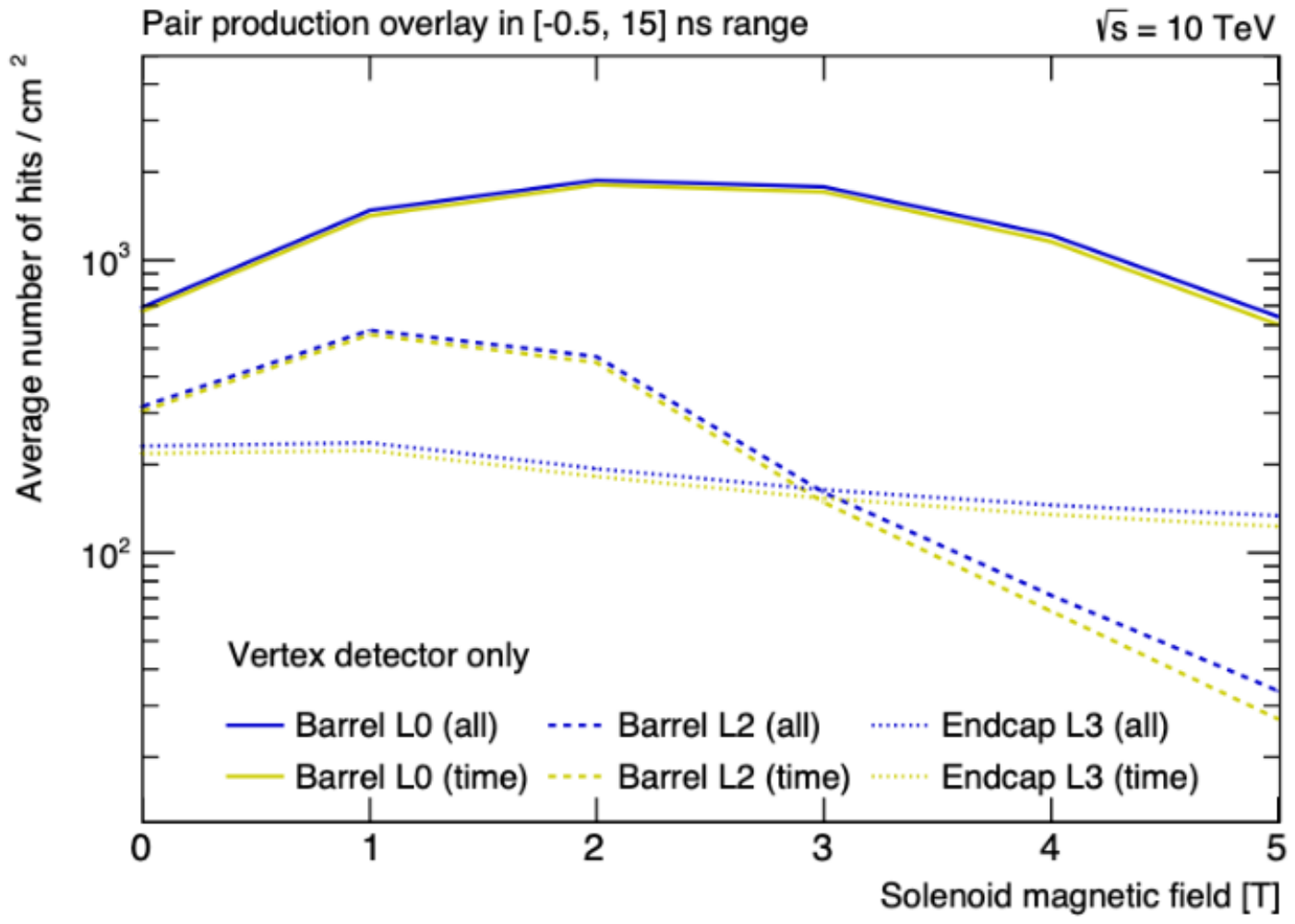}
    \end{center}
 \caption{Average hit density per layer, assuming the 3~TeV detector model (\textsc{MuColl\_v1}), as a function of the solenoid magnetic field.}
 \label{1:inter:fig:Bfield}
\end{figure}

Full-simulation was employed to characterise the tracking detector occupancy as a function of the intensity of the magnetic field present in the central region of the detector. The study propagated in the detector volume the outgoing particle prediction from \textsc{GUINEA-PIG} scanning uniform magnetic field values from 0 to 5~T. The results, shown in Figure~\ref{1:inter:fig:Bfield} illustrate how the highest field intensity is necessary to mitigate the impact of incoherent pairs. The effects of this choice on the tracking of low-momentum particles in the benchmark phenomena (e.g. Higgs boson production) was also tested with fast simulation and found not to lead to sizeable inefficiencies.

\begin{figure}[h]
   \begin{center}
        \includegraphics[width=0.8\textwidth]{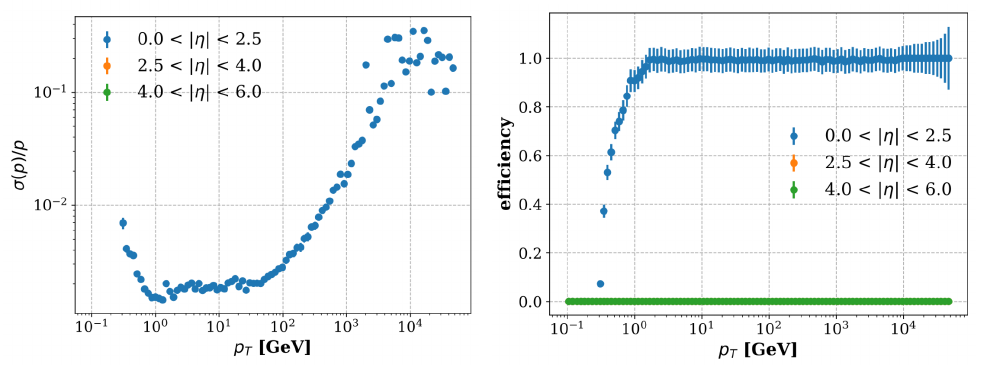}
    \end{center}
 \caption{Track momentum resolution (left) and track reconstruction efficiency (right) as a function of the particle transverse momentum. Taken from~\cite{lucchesi_2024_14001529}.}
 \label{1:inter:fig:DelphesMUSIC}
\end{figure}

Last, a new DELPHES card was recently released~\cite{lucchesi_2024_14001529} to enable fast simulation phenomenology studies of one of the 10~TeV detector designs (MUSIC). Figure~\ref{1:inter:fig:DelphesMUSIC} shows an example parameterisation from the released card.

\section{Machine-Detector Interface}
\label{1:inter:sec:mdi}

The beam-induced background (BIB) poses a significant challenge for the physics performance of a multi-TeV muon collider. The background is dominated by the decay of stored muons, with additional contributions from incoherent electron-positron pair production and possible beam halo losses on the aperture. The incoherent pairs are produced in the collisions of real or virtual photons emitted by muons of the counter-rotating bunches. The particles from the different processes interact with surrounding materials and generate a mixed radiation field composed of secondary electrons, positrons, photons, as well as hadrons (photo-nuclear interactions) and muons (Bethe--Heitler pair production). Without dedicated mitigation measures, the beam-induced background would severely impede the reconstruction of collision events in the detector and would lead to significant radiation damage in detector components. 

An optimized design of the machine-detector interface (MDI) is hence critical for minimizing the impact of machine operation on the physics performance reach, and for reducing the ionizing dose and displacement damage in the detector. The MDI design must include an elaborate absorber configuration, consisting of masks inside the final focus region and a conical shielding, which penetrates deeply into the detector region, up to a few centimeters from the interaction point. The optimization of these masks and absorbers must be done jointly with the detector and interaction region (IR) design. A conceptual MDI layout for a \SI{1.5}{\tera\electronvolt} center-of-mass muon collider has been devised previously within the MAP collaboration using the MARS Monte Carlo code~\cite{Mokhov2011,Mokhov2012,Alexahin2011}. Starting from this configuration, MDI design studies for higher-energy muon colliders (\SI{3}{\tera\electronvolt} and \SI{10}{\tera\electronvolt}) were carried out within the IMCC \cite{Calzolari2022,Calzolari2023,Lucchesi2024} using the FLUKA particle transport code~\cite{Battistoni2013,Ahdida2022,FLUKAwebsite}. The MDI design for the \SI{10}{\tera\electronvolt} machine is based on a new interaction region lattice developed by the IMCC~\cite{Skoufaris2022}, whereas the \SI{3}{\tera\electronvolt} collider studies rely on the MAP optics~\cite{Alexahin2011}. 

\subsection*{Key challenges}

The main challenge for the design of the machine-detector interface is the short lifetime of muons. This intrinsic source of radiation distinguishes a muon collider from other high-energy colliders. One of the key requirements is the design of a sophisticated shielding configuration at the interface between final focus magnets and detector. As shown in previous studies by the MAP collaboration for \SI{1.5}{\tera\electronvolt}~\cite{Mokhov2011,Mokhov2012,Alexahin2011}, the number of secondary particles entering the detector can be suppressed by orders of magnitude by placing massive absorbers in close proximity of the interaction point (IP). The innermost part consists of a nozzle-like shielding, which defines the inner detector envelope and hence the angular acceptance of the detector (10\textdegree \ in MAP). The shape and material budget of the nozzle determines the entry points, directions and energy spectrum of particles reaching the detector. The nozzle must be made of a \mbox{high-Z} material to efficiently shield the electromagnetic showers induced by muon decay products or by beam halo losses on the aperture. In addition, it must embed a layer of borated polyethylene or a similar material to moderate and capture secondary neutrons produced in photo-nuclear interactions. The nozzle must be carefully optimized for different center-of-mass energies. Besides the conceptual design, one also faces engineering challenges including assembly, support and alignment of the nozzle, as well as integration of the nozzle inside the detector. In addition, an adequate cooling system must be devised to evacuate the heat deposited by decay products and beam halo losses. The technical design of the nozzle must also take into account the requirements of other systems and related equipment (e.g.\,beam instrumentation, vacuum system). 

While the characteristics (e.g.\,spectra) of the decay and halo-induced background are mainly determined by the nozzle, the total amount of background particles is to some degree also influenced by the interaction region layout, i.e., by the lattice design and the placement of mask-like absorbers inside IR magnets. For example, the previous MAP studies for a \SI{1.5}{\tera\electronvolt} collider suggested that a dipolar component in the final focus configuration can be beneficial for reducing the background flux into the detector~\cite{Mokhov2012}. Designing and optimizing the IR layout and lattice is one of the key challenges for the collider design, in particular for \SI{10}{\tera\electronvolt} (a \SI{3}{\tera\electronvolt} and \SI{6}{\tera\electronvolt} IR design was previously devised within MAP~\cite{Alexahin2012,Alexahin2016,Alexahin2018}, although no background simulations were published). Besides the beam-induced background, the IR lattice design has to cope with other major challenges (see Section~\ref{1:acc:sec:collider}), like a small $\beta$-function at the interaction point ($\beta^* = 1.5 \mbox{ mm}$ for \SI{10}{\tera\electronvolt}). The $\beta$-function is a measure for the transverse beam size along the accelerator (the $\beta$-function in the collision point is referred to as $\beta^*$). The small $\beta^*$ needed for a muon collider gives rise to significant chromatic aberrations and very large $\beta$-functions in the final focus region. This requires large quadrupole apertures and high field gradients. In addition, a few centimeter-thick shielding is required inside the final focus magnets to reduce the decay-induced heat load and the cumulative radiation damage in the magnets. 

Another important source of background arises from beam-beam interactions, in particular incoherent electron-positron pairs, which are produced in the vicinity of the IP and are not directly intercepted by the nozzle. The flux of the pairs into the detector is influenced by the solenoid field in the detector region.  While the choice of the solenoid field strength is primarily driven by the particle detection efficiency in the detector, it must also take into account the benefits for background suppression.

Even with an optimized MDI and interaction region design, a suitable choice of detector technologies and reconstruction techniques is needed to reduce the effects of the remaining background (see Chapter~\ref{1:det:ch}). The signatures of background particles often exhibit distinct differences with respect to collision products, which can be exploited for background suppression. This concerns, for example, the arrival time of background particles with respect to the bunch crossing, as well as the directions of background particles when they enter the detector. The shielding requirements and MDI design will hence strongly depend on detector R\&D efforts to suppress the effect of background particles. 

\subsection*{Recent achievements}

\begin{figure}[!t]
   \begin{center}
        \includegraphics[width=0.68\textwidth]{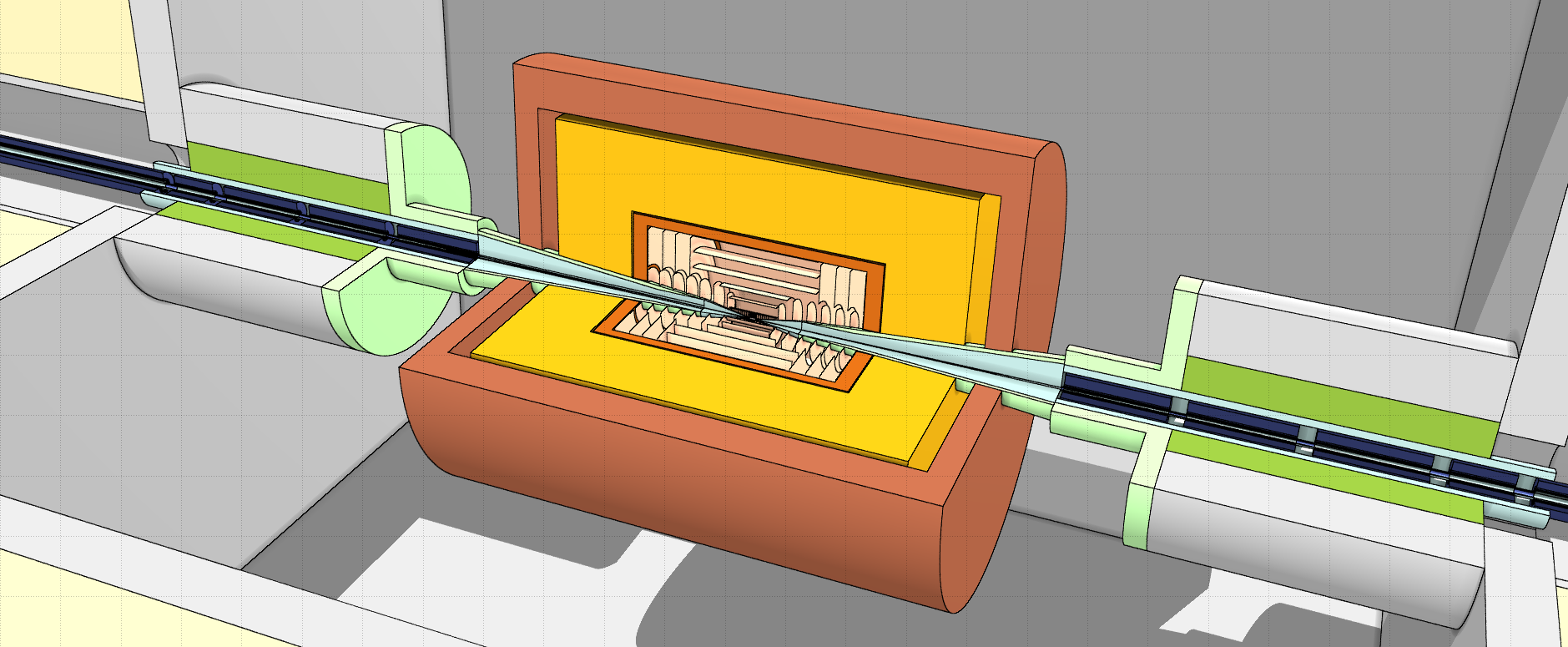}
    \end{center}
 \caption{FLUKA model of the interaction region and detector.}
 \label{1:inter:fig:irflukamodel}
\end{figure}

A full simulation framework based on the FLUKA Monte Carlo code was developed for computing the beam-induced background entering the detector (see geometry model in  Fig.~\ref{1:inter:fig:irflukamodel}). The simulation setup can also be used for assessing the radiation damage in detector components. First muon decay studies for \SI{1.5}{\tera\electronvolt}, replicating the MAP interaction region, exhibited a good agreement with the previous MAP results derived with MARS15 \cite{Collamati2021}. The studies were further extended to \SI{3}{\tera\electronvolt} and, in particular, to a higher collision energy of \SI{10}{\tera\electronvolt}~\cite{Calzolari2022,Calzolari2023,Lucchesi2024}. At the same time various simulation techniques were improved, like the sampling of decays from a fully matched beam phase-space distribution, which provided a more accurate description of the transverse beam tails. Using the FLUKA simulation framework, a comparison of the decay-induced background for \SI{1.5}{\tera\electronvolt}, \SI{3}{\tera\electronvolt} and \SI{10}{\tera\electronvolt} was performed~\cite{Collamati2021,Calzolari2022,Calzolari2023,Lucchesi2024}. As one of the key findings, these studies demonstrated that the spectra and time distributions of secondary particles entering the detector are not substantially different for the considered collider energies. 

\begin{figure}[!t]
   \begin{center}
        \includegraphics[width=0.48\textwidth]{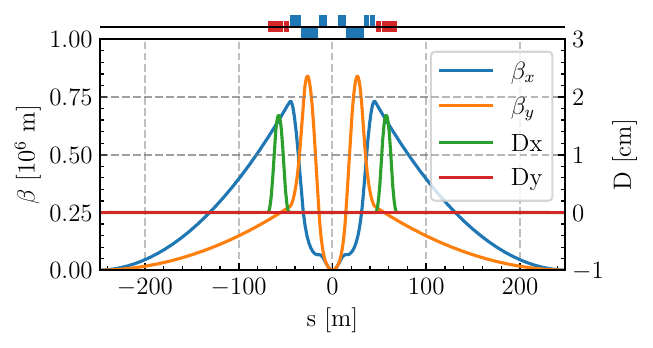}
        \includegraphics[width=0.48\textwidth]{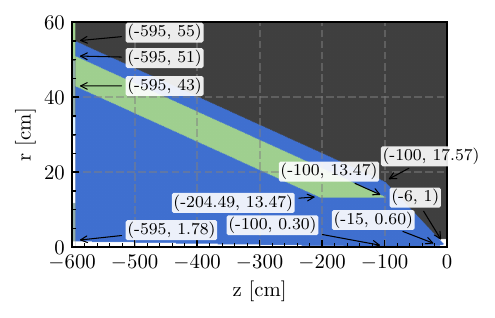}
    \end{center}
 \caption{Optics functions for present \SI{10}{\tera\electronvolt} lattice design (left) and present nozzle adapted from the original MAP design.}
 \label{1:inter:fig:ir}
\end{figure}

While the \SI{1.5}{\tera\electronvolt} and \SI3{\tera\electronvolt} studies were still based on the interaction region lattice from MAP, significant advances have been made in developing an interaction region design for the \SI{10}{\tera\electronvolt} collider. The \SI{10}{\tera\electronvolt} lattice adopts a triplet configuration as final focus scheme \cite{Skoufaris2022}. The $\beta$- and dispersion functions for the present lattice version are shown in Fig.~\ref{1:inter:fig:ir}. As one of the design criteria, the maximum magnetic field at the magnet aperture is limited to \SI{20}{\tesla}. The first quadrupole is divided into shorter segments to maximize achievable gradients while remaining within field limits. The lattice design allows for $\beta^*$ values of a few millimeters and incorporates an adequate chromatic compensation without sacrificing the physical and dynamic aperture. By using the FLUKA simulation setup, the impact of lattice design choices on the decay-induced background was assessed while iterating on the final focus layout. This made it possible to converge on key design choices. 

A key parameter for the interaction region design is the distance between the interaction point and the first magnet, commonly referred to as $L^*$. A longer $L^*$ implies larger beta functions in the final focus magnets, therefore increasing the magnet aperture. On the other hand, the quadrupole gradient decreases if the maximum field at the coil aperture is kept the same, thus implying a longer final focus scheme. Background studies for the \SI{10}{\tera\electronvolt} collider showed that muon decays between the final focus magnets ($s<L^*$) contribute several orders of magnitudes less to the background than those in final focus quadrupoles~\cite{Calzolari2023} (the $s$-coordinate represents the longitudinal coordinate in the curvilinear coordinate system of the beam). The solenoid field in the detector region traps the electrons and positrons, which travel in the beam vacuum until they reach the other side of the interaction region, where they deposit their energy in the machine components. A comparative assessment of two different lattices with $L^*=6$ and $10$\,m showed only a moderate background reduction in the latter case (by a few tens of percent), but at the expense of a more complex lattice design~\cite{Calzolari2023}. Increasing $L^*$ is hence not considered a viable option for reducing the background. As a consequence, $L^*$ is assumed to be \SI{6}{\meter} for the \SI{10}{\tera\electronvolt} collider, as in the \SI{1.5}{\tera\electronvolt} and \SI3{\tera\electronvolt} MAP lattices. 

As shown in Fig.~\ref{1:inter:fig:ir}, a long drift section is needed after the final focus quadrupoles in order to allow for a smoother reduction of the $\beta$-functions towards the chromaticity correction section. This long straight section leads to an accumulation of secondary particles from muon decay, which yield a non-negligible contribution to the overall detector background. In order to suppress this contribution, a dipolar chicane is included upstream of the triplet, which enhances the separation of decay products from the beam. The secondary electrons and positrons are deflected more strongly than the primary beam particles owing to their lower energy and smaller mass. Even high-energy decay electrons and positrons rapidly lose energy in the dipolar field due to synchrotron radiation. As a consequence, these secondaries are swept across the machine aperture before they can reach the detector region. FLUKA studies confirm that the chicane is very effective in reducing the contribution of distant decays. On the other hand, introducing a dipolar component in the triplet itself (through combined-function magnets) suppresses only marginally the contribution of decays in the final focus region~\cite{Calzolari2022}. Such a dipolar component strongly alters the azimuthal distribution of e-/e+ impact positions on the vacuum aperture, yet the massive nozzle dilutes any azimuthal dependence of the particle flux into the detector. Considering the limited benefits, the option of using combined-function dipole-quadrupoles for the final focus has been discarded.

\begin{figure}[t]
   \begin{center}
        \includegraphics[width=0.48\textwidth]{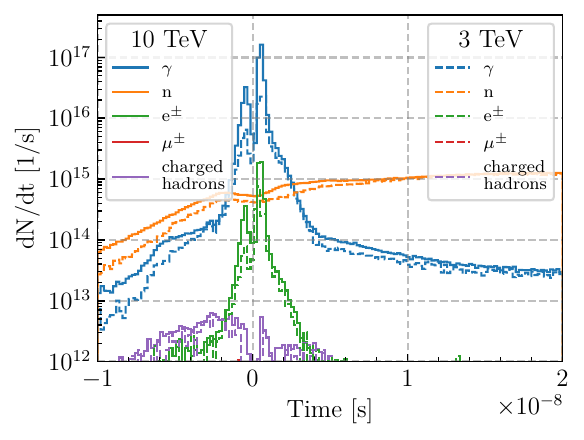}
        \includegraphics[width=0.48\textwidth]{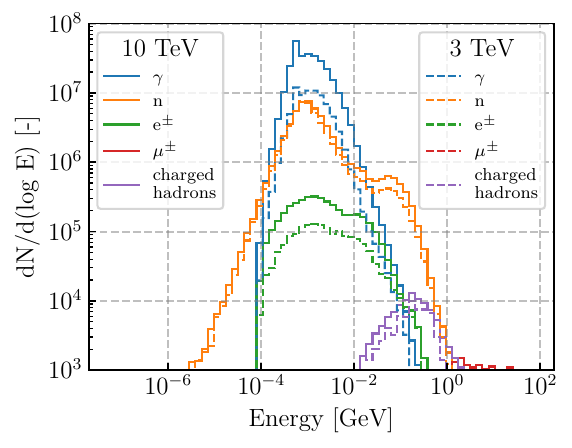}
    \end{center}
 \caption{Distribution of the arrival time of decay-induced background particles entering the detector (left). Energy spectra of particles arriving within the time window $[-1:15]$\,ns with respect to the bunch crossing (right). Photons, electrons and positrons below 100\,keV were discarded. The figures correspond to one bunch crossing.}
 \label{1:inter:fig:mdibibdecay}
\end{figure}

Figure~\ref{1:inter:fig:mdibibdecay} illustrates the distribution of arrival times and energy spectra of decay-induced background particles for \SI{3}{\tera\electronvolt} and \SI{10}{\tera\electronvolt}, respectively. The results for \SI{10}{\tera\electronvolt} were obtained with latest lattice described above, while the \SI{3}{\tera\electronvolt} studies were based on the MAP lattice. The differences between \SI{3}{\tera\electronvolt} and \SI{10}{\tera\electronvolt} are not only because of the different energies, but also because of the different interaction region and nozzle design. 
The \SI{10}{\tera\electronvolt} studies were carried out with a modified nozzle compared to the original MAP design which had been optimized for \SI{1.5}{\tera\electronvolt}. As discussed in Ref.~\cite{Calzolari2023} (for a previous optics version), changing the external shape of the nozzle and adapting the placement of the borated polyethylene layer can reduce the neutron and photon flux into the detector. The presently used nozzle is shown in Fig.~\ref{1:inter:fig:ir}. Compared to the MAP design, the outer radius of the nozzle has been reduced at the non-IP side, where the shielding capabilities are less needed. The borated polyethylene layer has been encapsulated inside the W-alloy in order to reduce the photon background after neutron capture. Furthermore, the nozzle material has been updated using a tungsten heavy alloy with a density of \SI{18}{\gram\per\centi\meter^3} instead of pure tungsten. Optimizing the nozzle further remains one of the tasks for future studies. For example, the simulations showed that particle flux into the inner tracker depends on the internal angle of the nozzle tip~\cite{Calzolari2023}.

In addition to the muon decay studies, a first assessment of background sources other than decay was performed for \SI{10}{\tera\electronvolt}. In particular, a first comparison of incoherent electron-positron pairs against decay-induced spectra was carried out~\cite{Calzolari2023}. Although some electrons and positrons from pair production will be trapped by the solenoid, a non-negligible fraction of particles with energies up to about \SI{1}{\giga\electronvolt} is expected to enter the inner tracker. The studies suggest that contribution of incoherent pairs corresponds to a few ten percent of the total number of electrons and positrons entering the detector in the vicinity of the IP (within $\pm$\SI{50}{\centi\meter}). On average, the incoherent pairs have a higher energy than the decay-induced background component since the latter is strongly diluted by the nozzle.

Through the same simulation framework and by using models for the two different detector options (MAIA and MUSIC, described in Section~\ref{1:det:ch}), the cumulative radiation damage in the different detector sub-systems was calculated. Currently, only the contribution of muon decay was considered in the radiation load studies. Two quantities were evaluated, the total ionizing dose and the 1~MeV neutron-equivalent fluence in Silicon. The former describes the ionization damage in organic materials, while the latter is related to the displacement damage. The highest dose values occur around the vertex detector, reaching about $\sim$\SI{1}{\mega\gray}/year (10\,TeV collider). The 1\,MeV neutron-equivalent fluence is highest in the inner tracker due to increased leakage of neutrons from the nozzle, with a peak value of around $10^{15}$\,n/cm$^{2}$/year (10\,TeV collider). Table~\ref{1:inter:tab:detraddamage_tab} provides a summary of the maximum values in different parts of the detector.

Most of the quantities are very similar in the two detector options, due to the equivalent layout of the tracker structure. The main difference in the radiation damage maps is due the position of the solenoid. In the case of MAIA, the solenoid is placed inside the electromagnetic calorimeter, acting as a shielding for the outermost detector components. As a secondary effect, having such a heavy structure closer to the collision point increases the neutron-equivalent fluences in the outer tracker volume due to the backscattered neutrons.

\begin{figure}[t]
   \begin{center}
        \includegraphics[width=0.48\textwidth]{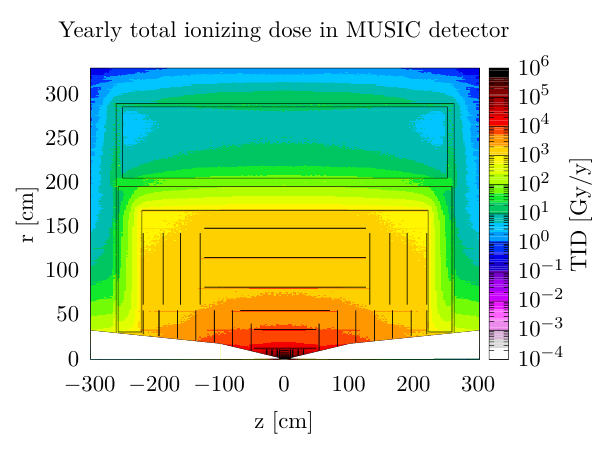}
        \includegraphics[width=0.48\textwidth]{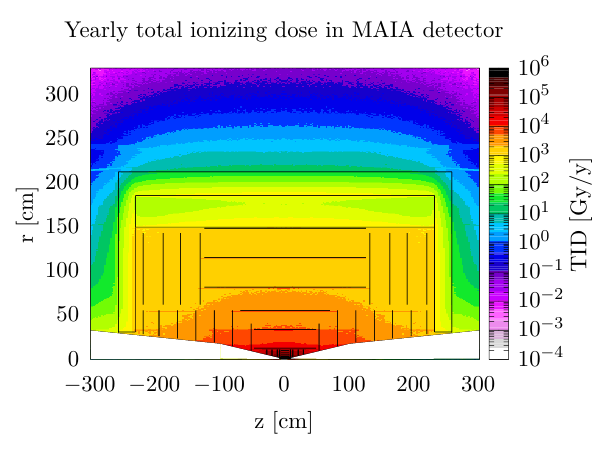} \\
        \includegraphics[width=0.48\textwidth]{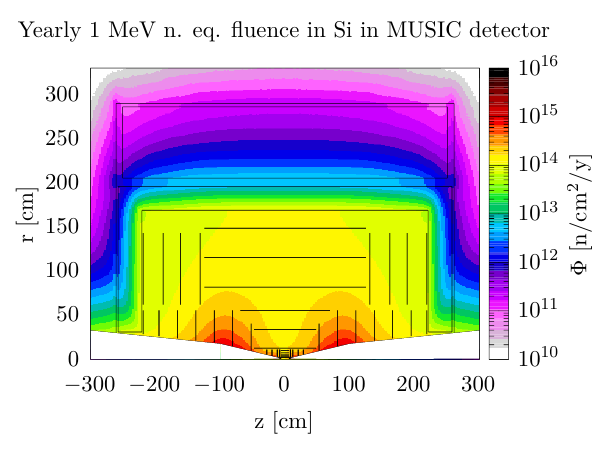}
        \includegraphics[width=0.48\textwidth]{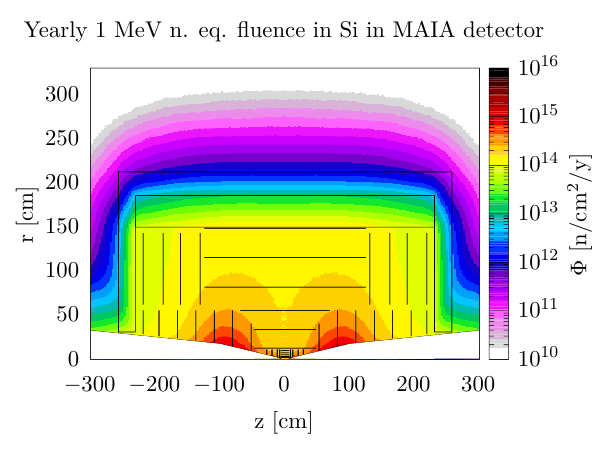} 
    \end{center}
 \caption{Total ionizing dose (top) and $\SI{1}{MeV}$ neutron-equivalent fluence in Silicon (bottom) for the MUSIC detector (left) and the MAIA detector (right). The main difference in the maps is due the solenoid (modelled in aluminium). The maps correspond to one year of operation (\SI{10}{\tera\electronvolt}), assuming 139 operational days.}
 \label{1:inter:fig:detraddamage}
\end{figure}

\begin{table}[t]
\begin{center}
\caption{Maximum values of the ionizing dose and the $\SI{1}{MeV}$ neutron-equivalent fluence (Si) for the two detector options. All values are per year of operation (\SI{10}{\tera\electronvolt}) and include only the contribution of muon decay.}
\label{1:inter:tab:detraddamage_tab}
\centering
\begin{tabular}{c|>{\centering\arraybackslash}m{1.5cm}>{\centering\arraybackslash}m{1.5cm}|>{\centering\arraybackslash}m{2.1cm}>{\centering\arraybackslash}m{1.5cm}}
\hline \hline
\textbf{Component} & \multicolumn{2}{c|}{\textbf{Dose~[kGy]}} & \multicolumn{2}{c}{\textbf{\makecell{1 MeV neutron-equivalent\\ fluence (Si) [\SI[detect-all]{e14}{n/cm^2}]}}} \\ \hline
 & \textbf{MAIA} & \textbf{MUSIC} & \textbf{MAIA} & \textbf{MUSIC} \\
Vertex (barrel) & \multicolumn{2}{c|}{1000} & \multicolumn{2}{c}{\SI{2.3}{}} \\
Vertex (endcaps) & \multicolumn{2}{c|}{2000} & \multicolumn{2}{c}{\SI{8}{}} \\
Inner trackers (barrel) & \multicolumn{2}{c|}{70} & \SI{4.5}{} & \SI{4}{} \\
Inner trackers (endcaps) & \multicolumn{2}{c|}{30} & \SI{11.5}{} & \SI{10}{} \\
ECAL & 0.58 & 1.4 & \SI{0.15}{} & \SI{1}{} \\ \hline \hline
\end{tabular}
\end{center}
\end{table}

\chapter{Detector concepts}
\label{1:det:ch}
\section{Overview}
\label{1:det:sec:overview}

The design of dedicated experiments to collect data at the collider interaction points has sparked a lively environment that allowed the community to make fast progress in just a few years.

The detector design is still in its preliminary phase, but it is already possible to make solid statements about the technological requirements, expected performance, and opportunities for further improvement.

The primary challenge for detector designs at muon colliders is the mitigation of the abundant beam-induced backgrounds (BIB) arising from the in-flight decays of the circulating beams. 
These backgrounds are typically reduced by introducing absorbing elements, which, however, limit detector acceptance, creating a tension that requires careful optimisation.

An initial detector design, referred to as MuColl, based on the CLICdet concept~\cite{CLICdp:2017vju,CLICdp:2018vnx,ILDConceptGroup:2020sfq,ILC:2007vrf} and optimised to be operated at a $\sqrt{s}=3$~TeV muon collider was studied extensively~\cite{Accettura:2023ked} and demonstrated the feasibility of the physics programme.

While further work is needed to optimise this design and enhance the performance of the reconstruction algorithms to achieve the ultimate expected precision, recent efforts have focused on delivering the first detector designs for a $\sqrt{s}=10$ TeV machine.
The design work follows the concept already developed for $\sqrt{s}=3$ TeV with modifications to account for the higher energy. Two distinct detector concepts are presented, MUSIC (MUon System for Interesting Collisions) and MAIA~\cite{MAIADetector} (Muon Accelerator Instrumented Apparatus), to fully exploit the two interaction points of the collider.
Both designs share a similar structure, a cylindrical shape $11.4$ m in length with a diameter of $12.8$ m. The main detector components are: a tracking system, an electromagnetic calorimeter (ECAL), a hadronic calorimeter (HCAL) and a muon system. 
A superconducting solenoid is also envisioned to provide bending power for the measurement of charged particle momenta within the tracking system.

Table~\ref{tab:detector_general} summarises the detector parameters sub-system by sub-system for the two 10~TeV detector concepts, alongside the 3~TeV design. While the tracking system has a similar structure, the MAIA detector has the solenoid just outside the tracking system and before the ECAL, while MUSIC places the solenoid magnet between ECAL and HCAL.

\begin{table}[ht]
\centering
\begin{tabular}{l|ccc} 
Detector Concept & \textbf{MuColl} & \textbf{MUSIC} & \textbf{MAIA} \\
 & $\sqrt{s}=3~\textrm{TeV}$ & $\sqrt{s}=10~\textrm{TeV}$ & $\sqrt{s}=10~\textrm{TeV}$ \\
\hline
\textbf{Inner Trackers} & & & \\
R$_\textrm{min}$ -- R$_\textrm{max}$ [\si{\milli\meter}] & 30 -- 1486 & 29 -- 1486 & 30 -- 1486 \\
z$_\textrm{min}$ -- z$_\textrm{max}$ [\si{\milli\meter}] & 0 -- 2190 & 0 -- 2190 & 0 -- 2190 \\
Angular Acceptance [\si{\degree}] & 10 -- 170 & 10 -- 170 & 10 -- 170 \\
$X / X_0$ & 0.3 & 0.1 & 0.1\\
$L / L_0$ & 0.1 & 0.04 & 0.04\\
\hline
\textbf{EM Calorimeters} & & & \\
R$_\textrm{min}$ -- R$_\textrm{max}$ [\si{\milli\meter}] & 1500 -- 1702 & 1690 -- 1960 & 1857 -- 2125 \\
z$_\textrm{min}$ -- z$_\textrm{max}$ [\si{\milli\meter}] & 2307 -- 2210 & 2307 -- 2577 & 2307 -- 2575 \\
Angular Acceptance [\si{\degree}] & 10 -- 170 & 10 -- 170 & 10 -- 170 \\
$X / X_0$ & 26 -- 32 & 33 -- 38 & 40 -- 42\\
$L / L_0$ & 1.2 -- 1.5 & 1.4 -- 1.7 & 1.8 -- 1.9\\
\hline
\textbf{Hadron Calorimeters} & & & \\
R$_\textrm{min}$ -- R$_\textrm{max}$ [\si{\milli\meter}] & 1740 -- 3330 & 2902 -- 4756 & 2125 -- 4113 \\
z$_\textrm{min}$ -- z$_\textrm{max}$ [\si{\milli\meter}] & 2539 -- 4129 & 2579 -- 4434 & 2575 -- 4562 \\
Angular Acceptance [\si{\degree}] & 10 -- 170 & 10 -- 170 & 10 -- 170 \\
$X / X_0$ & 82 -- 87 & 89 -- 116 & 100 -- 114 \\
$L / L_0$ & 8.8 -- 9.3  & 9.5 -- 12.5 & 10.9 -- 12.3\\
\hline
\textbf{Muon Systems} & & & \\
R$_\textrm{min}$ -- R$_\textrm{max}$ [\si{\milli\meter}] & 4461 -- 6450 & 4806 -- 6800 & 4150 -- 7150 \\
z$_\textrm{min}$ -- z$_\textrm{max}$ [\si{\milli\meter}] & 4179 -- 5638 & 4444 -- 5903 & 4565 -- 6025 \\
Angular Acceptance [\si{\degree}] & 10 -- 170 & 10 -- 170 & 10 -- 170 \\
\hline
\textbf{Solenoid} & & & \\
R$_\textrm{min}$ -- R$_\textrm{max}$ [\si{\milli\meter}] & 3483 -- 4290 & 2055 -- 2862 & 1500 -- 1857 \\
z$_\textrm{min}$ -- z$_\textrm{max}$ [\si{\milli\meter}] & 0 -- 4129 & 0 -- 2509 & 0 -- 2307 \\
$X / X_0$ & -- & 18 & 6\\
$L / L_0$ & -- & 2.7 & 1.4\\
\hline
\textbf{Nozzles} & & & \\
R$_\textrm{min}$ -- R$_\textrm{max}$ [\si{\milli\meter}] & 10 -- 600 & 10 -- 550 & 10 -- 550 \\
z$_\textrm{min}$ -- z$_\textrm{max}$ [\si{\milli\meter}] & 60 -- 6000 & 60 -- 6000 & 60 -- 6000 \\
\end{tabular}
\caption[Detector parameters]{Detector parameters for the MuColl (v1), MUSIC (v2) and MAIA (v0) concepts. Values that are left empty ("--") are not relevant for the specific detector. $X / X_0$ and $L / L_0$ are for a particle travelling from the nominal beam interaction point (IP). The origin of the space coordinates is the IP. The z-axis has direction parallel to the beam pipe, the y-axis is parallel to gravity acceleration and the x-axis is defined as perpendicular to the y and z axes.
}
\label{tab:detector_general}
\end{table}

\section{MUSIC}
\label{1:det:sec:music}
\begin{figure}[t]
    \centering
    \includegraphics[width=0.95\linewidth]{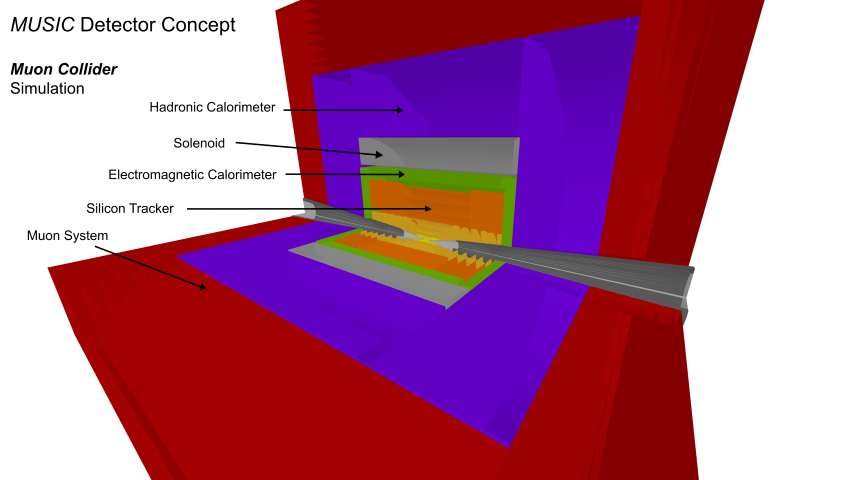}
    \caption{Layout of the MUSIC detector concept: from the center to the outermost region, it includes a Vertex Detector and an Inner Tracker (yellow), an Outer Tracker (orange), an electromagnetic calorimeter (green), a superconducting magnet (gray), a hadronic calorimeter (purple), and muon detectors (red). The shielding nozzles, installed along the axis of the detector, are shown in dark gray.
    \label{fig:MUSIC}}
\end{figure}

The configuration of the MUSIC detector is shown in Figure~\ref{fig:MUSIC}. In the following sections, the main features of the MUSIC sub-systems are briefly described.

The tracking system of the MUSIC detector consists of three sub-detectors, which cover the polar angle range between 10$^\circ$ and 170$^\circ$: a Vertex Detector (VXD), an Inner Tracker (IT), and an Outer Tracker (OT), all structured in concentric cylindrical layers (barrels) in the central region and in a series of disks, perpendicular to the detector axis, in the forward and backward regions (end-caps).

The VXD is composed of silicon-pixel planar modules, arranged in five 26-cm long barrel layers with radii between 2.9 cm and 10.1 cm and four disks on each side at distances between $|z| = 18$ and 36.6 cm from the nominal interaction point (IP).
Both the barrel and end-caps feature $25\times 25$ $\upmu$m$^2$ silicon pixels with hit spatial and time resolutions of 5 $\upmu$m $\times$ 5 $\upmu$m and 30 ps, respectively.
The IT comprises three barrel layers with radii ranging from 16.4 to 55.4 cm and seven disks on each side at $|z|$ positions between 60.4 and 219 cm. The first layer is 96.3 cm long, while the second and third layers are 138.5 cm long.
The OT consists of three barrel layers, each 252.8 cm long, with radii between 81.9 and 148.6 cm, and four disks per side positioned at distances from 141.0 to 219.0 cm.
Both sub-detectors feature 50 $\upmu$m $\times$ 1 mm macropixels, with a hit spatial resolution of 7 $\upmu$m $\times$ 90 $\upmu$m and a hit time resolution of 60 ps.

The MUSIC calorimeter system includes an ECAL, installed inside the magnet bore to provide high energy resolution for electrons and photons, and an HCAL positioned outside the solenoid. The calorimeters hermetically cover the polar angle range from approximately 7$^\circ$ to 173$^\circ$.

The ECAL features a central barrel section that is 4.4 m long with an inner radius of 1.69 m, closed at both ends by two end-caps located at $|z| = 2.3$ m from the IP. Both the barrel and the end-caps are 27 cm thick. It is a semi-homogeneous electromagnetic crystal calorimeter with longitudinal segmentation (CRILIN), consisting of $10\times 10\times 45$-mm$^3$ lead-fluorite crystals arranged in six layers for a total of 26.5 radiation lengths. 

The HCAL consists of a 5-m long barrel with an inner radius of 2.9 meters and two end-caps located at $|z| = 2.58$ meters from the IP, each with a thickness of 1.86 m. It is an iron-scintillator sampling calorimeter comprising 70 layers of 20-mm iron absorber and $30 \times 30$ mm$^2$ scintillator pads, for a total of approximately seven nuclear interaction lengths. Additionally, the iron absorber serves as a return yoke to close the magnetic field flux.

Outside the HCAL, there are seven layers of muon detectors in the barrel and six layers in the end-caps, covering polar angles between about 7$^\circ$ and 173$^\circ$.
A specific technology for the muon detectors has not yet been chosen. Since there is no magnetic field outside the HCAL, the muon detectors are primarily used to identify particles as muons.

The MUSIC detector will feature a superconducting solenoid, capable of a 4-5 T magnetic field at the interaction point. The magnet vacuum tank is 5 m long, with an inner radius of 2 meters and a total thickness of 80.7 cm.

Designs placing the hadronic calorimeter either inside or outside the solenoid have been considered and are currently under investigation.  Based on the experience with previous magnets, especially the CMS solenoid, no technical showstopper is foreseen. The CMS solenoid is longer than the MUSIC magnet in both configurations, and has a comparable diameter with the inner HCAL design. The CMS solenoid cable, if used for the MUSIC solenoid, could relatively easily allow for a 4 T central field, with forces and stresses of the same magnitude.

\section{MAIA}
\label{1:det:sec:maia}
\begin{figure}[h]
\centering
\includegraphics[width=0.9\textwidth]{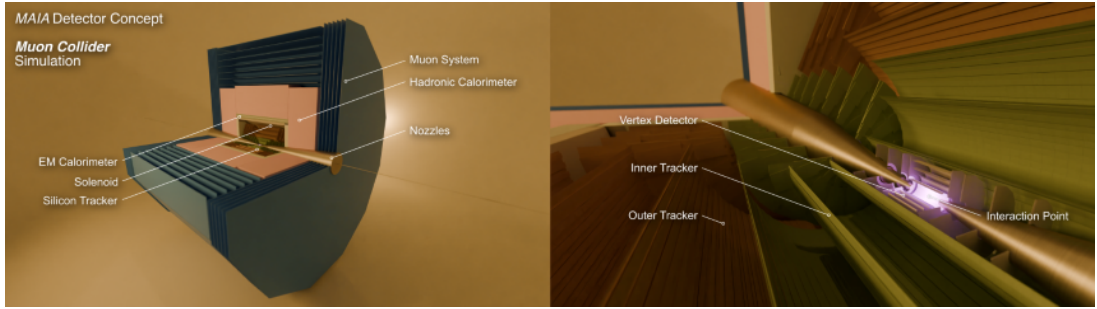}
\vspace{1.2em}
\caption{The layout of the MAIA detector concept is shown with a closeup of the silicon tracker around the interaction point. The detector is shown with a $\pi/2$ cutaway in $\phi$ for illustration \cite{MAIADetector}.}
\label{fig:subdetectors}
\end{figure}

The MAIA detector concept~\cite{MAIADetector} is a new detector concept based on a re-optimisation of the MuColl design to collect data at $\sqrt{s}=10$~TeV. A visualisation of its geometry can be found in Figure~\ref{fig:subdetectors}. 

\begin{figure}[h]
 \centering
  \includegraphics[width=0.49\textwidth]{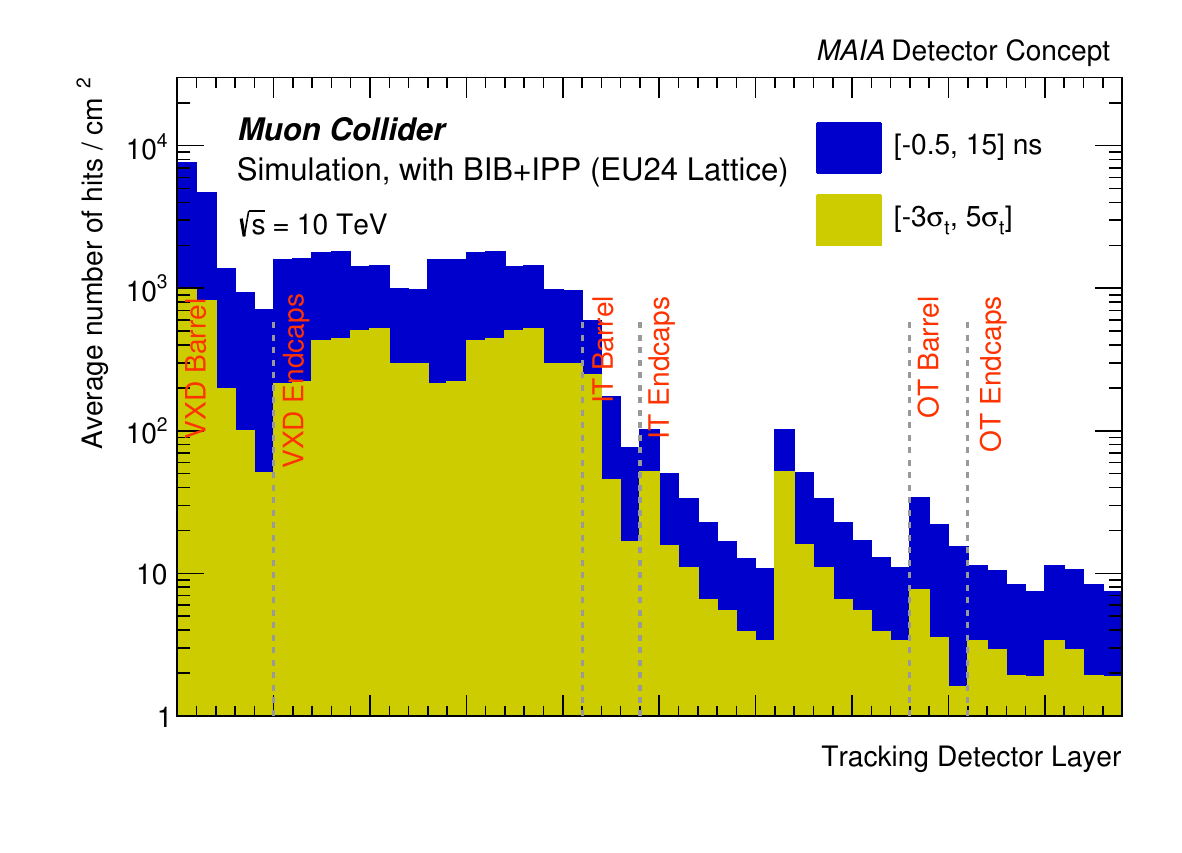}
 \includegraphics[width=0.49\textwidth]{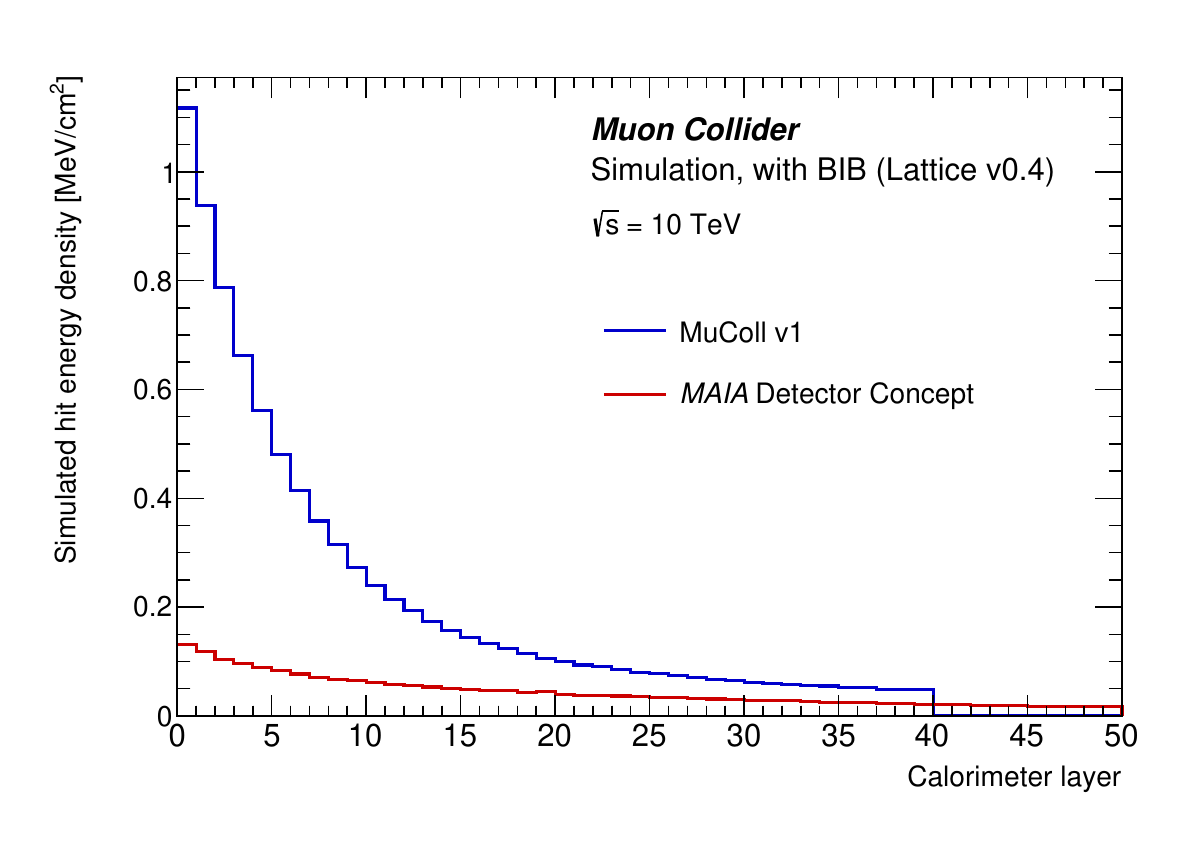}
 \caption{Right: Average hit density (defined here as energy deposits above 3.5~keV) in the MAIA tracking detector, shown separately for each detector layer, for $\sqrt{s}=10$~TeV BIB. Left: Average energy density of simulated BIB hits for each ECAL barrel layer. The layer index increases with radius. The same BIB particles, from a previous lattice design (v0.4), are propagated through the 3~TeV detector (MuColl, v1) and the MAIA concept to illustrate the effects of the solenoid placement \cite{MAIADetector}.}
 \label{fig:ECAL_energy_density}
\end{figure}

The fully-silicon tracking detector comprises a Vertex Detector (VXD), Inner Tracker (IT), and Outer Tracker (OT), whose attributes can be found in Table~\ref{tab:tracker_specs}. The VXD features 4 barrel and 4 endcap layers with pixel sensors with $25\times 25$ $\upmu$m$^2$ pixel size. It has a time resolution of 30 ps and a spatial resolution of 
5 $\upmu$m $\times$ 5 $\upmu$m. The IT, with 3 barrel and 7 endcap layers, uses macro-pixels (50 $\upmu$m $\times$ 1 mm) and has a time resolution of 60 ps and spatial resolution of 7 $\upmu$m $\times$ 90 $\upmu$m. 
The OT shares the same sensor thickness as the Inner Tracker, with 3 barrel layers and 4 endcap layers per side, however, a cell size of 50 $\upmu$m $\times$ 10 mm is chosen.
Advancements in track reconstruction software showed that fewer double layers\footnote{Double layers are closely spaced layers (~2mm) whose signal coincidence can provide directional information of the incoming particle.} than were used in the MuColl detector are sufficient to achieve similar track reconstruction performance, even for the challenging detector occupancies shown in Figure~\ref{fig:ECAL_energy_density} (left).

To account for the larger centre-of-mass energy and maintain the resolution of track $p_\mathrm{T}$ measurements for the highest momentum particles, the magnetic field was increased from 3.6 T to 5 T. This change also prevents incoherent $e^{+}e^{-}$ pairs from saturating the innermost layers of the tracker, and reduces the number of hits produced by low momentum background particles. The size of the calorimeters was also increased to fully contain electromagnetic and hadronic showers. To minimize the cost and challenges associated with the higher-field magnet, the solenoid was reduced in diameter by moving it inside the electromagnetic calorimeter. The placement of the solenoid (along with its shielding) provides a significant reduction of BIB in the calorimeters, shown in Figure~\ref{fig:ECAL_energy_density} (right).

\begin{table}[h]
    \centering
    \begin{tabular}{l|c|c|c}
                        &  \textbf{Vertex Detector} & \textbf{Inner Tracker} & \textbf{Outer Tracker} \\
    \hline\hline
    Sensor type           &  pixels & macro-pixels & macro-pixels \\
    Barrel Layers  &  4  & 3  & 3 \\
    Endcap Layers (per side)  &  4   &  7 & 4 \\
    Cell Size           &  \qty{25}{\um} $\times$ \qty{25}{\um} & \qty{50}{\um} $\times$ \qty{1}{\mm} & \qty{50}{\um} $\times$ \qty{10}{\mm} \\
    Sensor Thickness    &  \qty{50}{\um} & \qty{100}{\um} & \qty{100}{\um} \\
    Time Resolution     &  \qty{30}{\ps} & \qty{60}{\ps} & \qty{60}{\ps} \\
    Spatial Resolution  &  \qty{5}{\um} $\times$ \qty{5}{\um} & \qty{7}{\um} $\times$ \qty{90}{\um} & \qty{7}{\um} $\times$ \qty{90}{\um} \\
    \end{tabular}
    \caption{Assumed spatial and time resolution for MAIA Tracking Detector sub-systems. There is no resolution difference between the barrel and end-cap regions. The first layer of the Vertex barrel and all Vertex endcap layers are implemented as double layers.}
    \label{tab:tracker_specs}
\end{table}

Like its predecessors, the MAIA detector concept makes use of a silicon-tungsten electromagnetic calorimeter (ECAL) and an iron-scintillator hadronic calorimeter (HCAL), both based on the CLIC calorimeter. 
Compared to the calorimeters optimised for lower energy, this detector concept features more layers, each with a slightly thicker absorber. The cells have also been scaled up in size. Specifically, the ECAL has a cell size of 5.1 mm $\times$ 5.1 mm, with a sensor thickness of 0.5 mm and an absorber thickness of 2.2 mm, distributed over 50 layers. The HCAL, on the other hand, has larger cells (30 mm $\times$  30 mm), a sensor thickness of 3.0 mm, and an absorber thickness of 20.0 mm, spread over 75 layers. The details are summarised in Table~\ref{tab:calo_specs}. 

\begin{table}[htb]
    \centering
    \begin{tabular}{l|c|c}
                        &  \textbf{Electromagnetic Calorimeter} & \textbf{Hadron Calorimeter} \\
    \hline\hline
    Cell type           &  Silicon - Tungsten & Iron - Scintillator \\
    Cell Size           &  \qty{5.1}{\mm} $\times$ \qty{5.1}{\mm} & \qty{30.0}{\mm} $\times$ \qty{30.0}{\mm}  \\
    Sensor Thickness    &  \qty{0.5}{\mm} & \qty{3.0}{\mm}  \\
    Absorber Thickness  &  \qty{2.2}{\mm} & \qty{20.0}{\mm}  \\
    Number of layers  &  50 & 75  \\
    \end{tabular}
    \caption{Cell and absorber sizes in the MAIA calorimeter systems, describing both the barrel and end-cap regions.}
    \label{tab:calo_specs}
\end{table}

The Muon System is least affected by BIB, and has thus been a lower priority for optimisation. The current system is identical to the 3 TeV detector concept, with the only difference being an expansion in size to accommodate the changes made to the rest of the detector. In the current MAIA design, there is no dedicated magnetic field for the muon system, so a standalone measurement of $p_\mathrm{T}$ is not possible. Further work is required to re-optimise this detector and integrate its measurements into combined tracks. 

\FloatBarrier
\section{Performance}
\label{1:det:sec:performance}
High levels of beam-induced background in the detector pose unprecedented challenges for the reconstruction and identification of particles produced in muon collisions. 
Studies based on detailed simulations of the MAIA and MUSIC detector concepts were carried out to assess the effects of this background on the detector response and develop appropriate mitigation measures. In both cases, results indicate that the background effects on the detector response can be minimized to a level that does not compromise overall performance of the detector.

For the MAIA concept, tracker, ECAL, and HCAL performance were evaluated in terms of the reconstruction efficiencies and resolution of tracks, photons, and neutral hadrons, respectively. This strategy minimises the dependency on high-level reconstruction algorithms, which remain to be optimised. Performance was evaluated in scenarios with and without BIB overlay.

For the MUSIC concept, track reconstruction performance was evaluated by examining the efficiency and the track parameter resolution. The ECAL performance is determined by evaluating photon and electron reconstruction efficiencies, energy resolution and fake rate. Muons are identified by matching tracks with hits in the outer muon detectors. Jet reconstruction efficiency and resolution reflect the performance of the combined sub-detectors: tracker, ECAL, and HCAL along with the capabilities of the software algorithms.

Beam-induced background causes very high hit multiplicities in the detector’s tracking system, increasing the complexity of the track finding process, since the number of hit combinations to be considered increases exponentially. To mitigate the impact of BIB, tracker hit timestamps, corrected by the time-of-flight, are required to be within a window of [${-}3\sigma_{t}$, $5\sigma_{t}$] from the beam crossing, where $\sigma_{t}$ refers to the detector time resolution. Track reconstruction is performed using a combinatorial Kalman Filter implemented in the ACTS~\cite{ai2022common} library and the \textit{ACTSTracking} processor in the \textsc{MuonColliderSoft} framework. This algorithm was developed for the high-occupancy environment of hadron colliders and re-optimised for the environment expected in a muon collider. Thus, it is better suited to the large BIB present at a muon collider than algorithms designed for $e^+e^-$ colliders. 

\begin{figure}[h]
    \centering
    \includegraphics[width=0.49\textwidth]{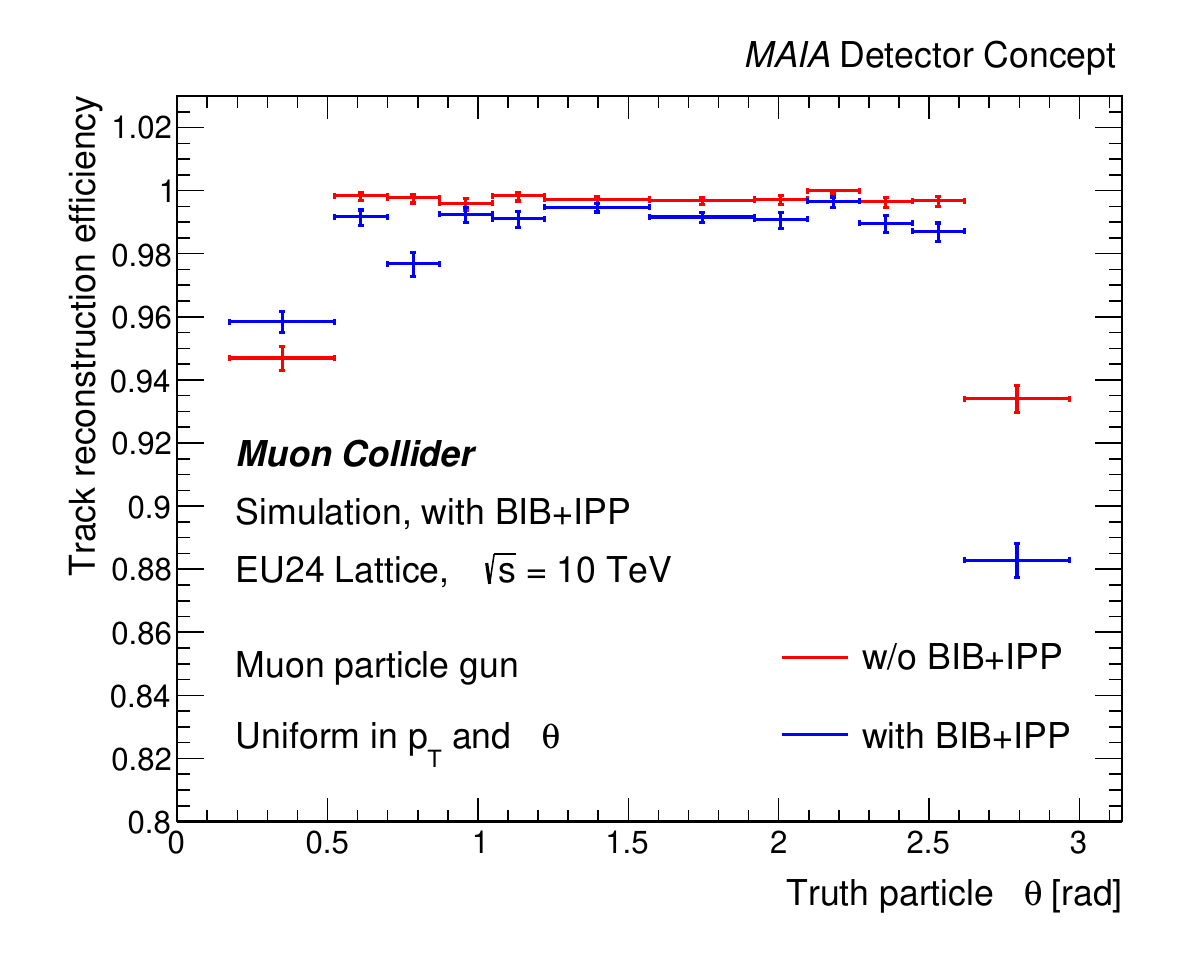}
    \includegraphics[width=0.49\textwidth]{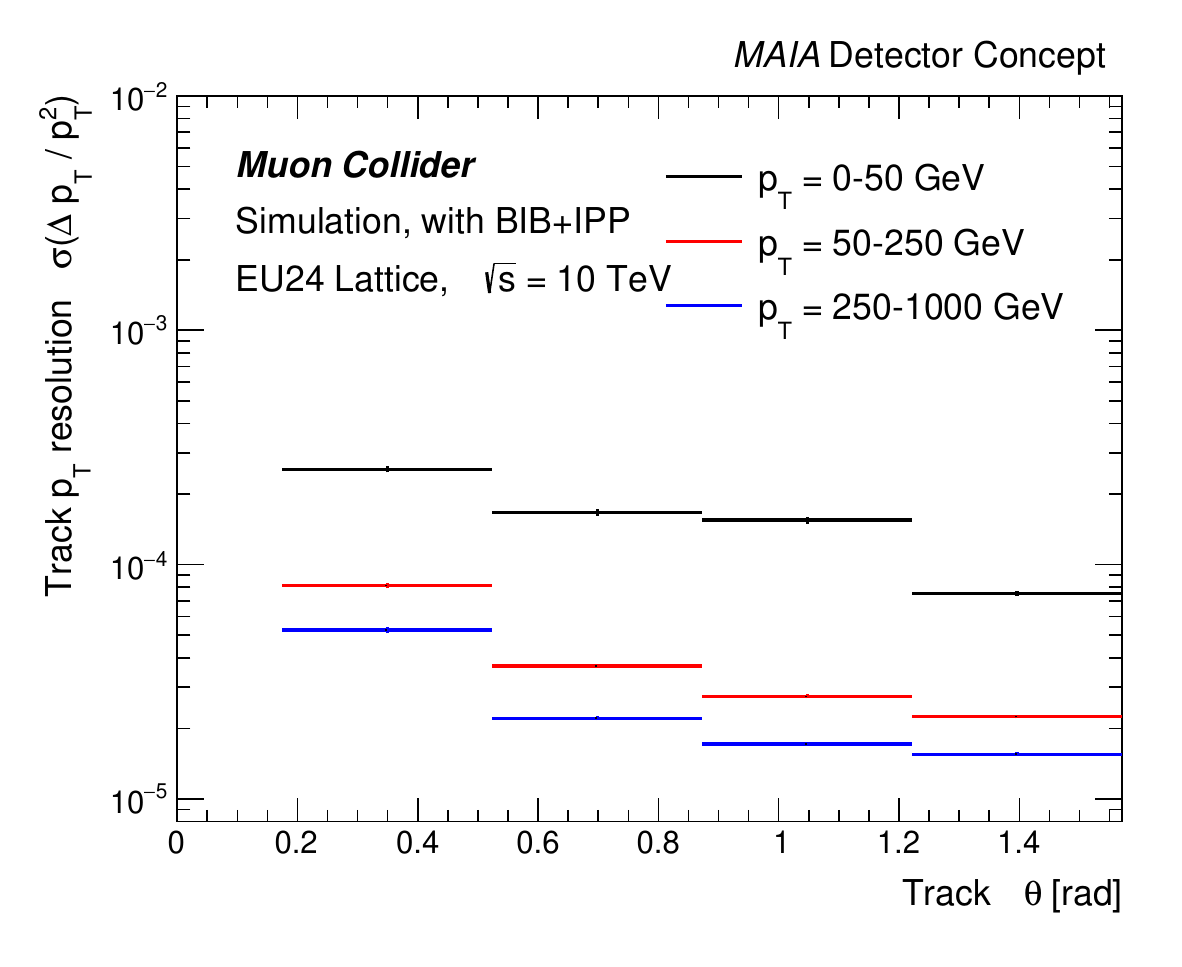}
\caption{Track reconstruction efficiency (left) and transverse momentum resolution $\sigma(\Delta p_{T})/p_{T}^2$ (right) for the MAIA detector concept. To minimize the rate of tracks reconstructed from random combinations of hits, cleaning requirements of $p_T > 0.5$~GeV and $\chi^2/n_{dof} < 3$ are applied~\cite{MAIADetector}.}
    \label{1:det:fig:track_efficiency}
\end{figure}

Figure~\ref{1:det:fig:track_efficiency} shows the track reconstruction performance for the MAIA detector concept. Excellent track reconstruction efficiency and momentum resolution are obtained, even in the presence of BIB. Further optimisation of the detector layout and track reconstruction algorithms in the endcap could improve efficiency, especially for transverse momenta below 1~GeV, as well as reduce the CPU resources required to reconstruct an event. 
 
\begin{figure}[h]
    \centering
    \includegraphics[width=0.49\textwidth]{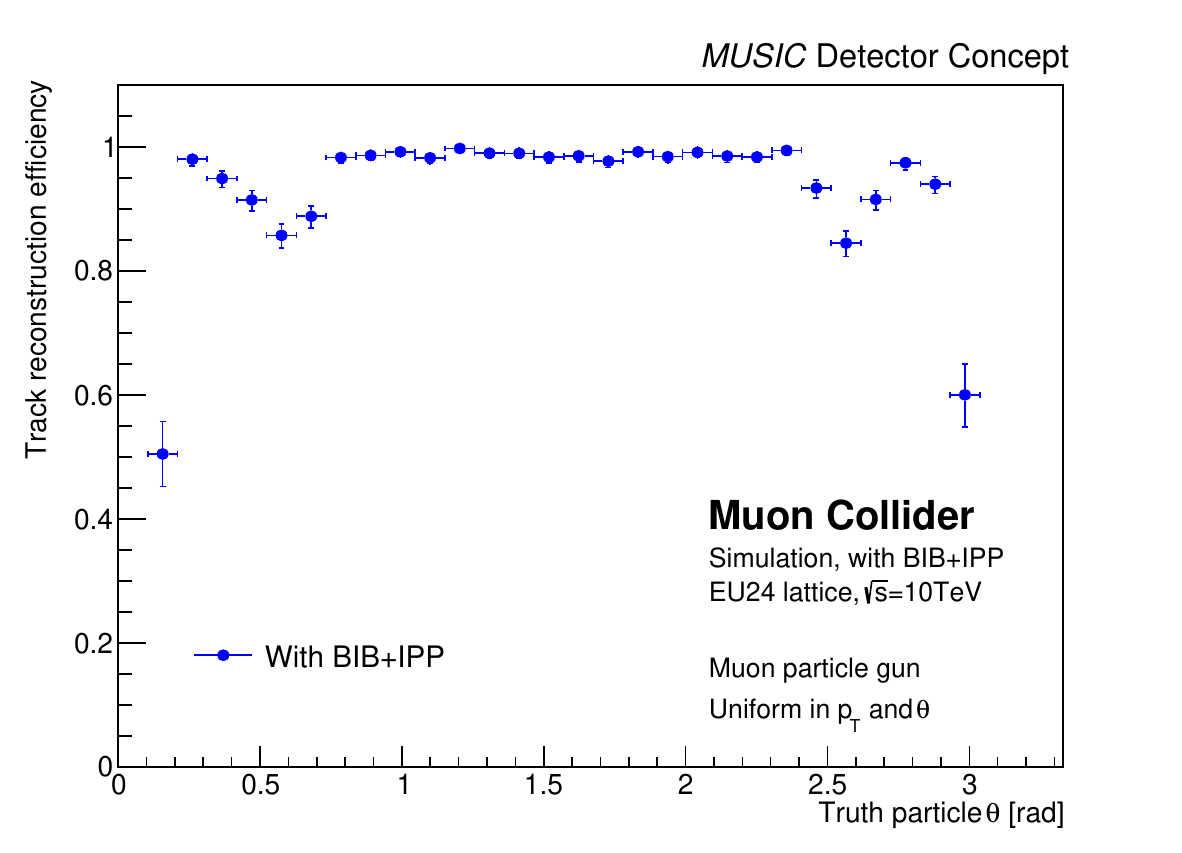}
    \includegraphics[width=0.49\textwidth]{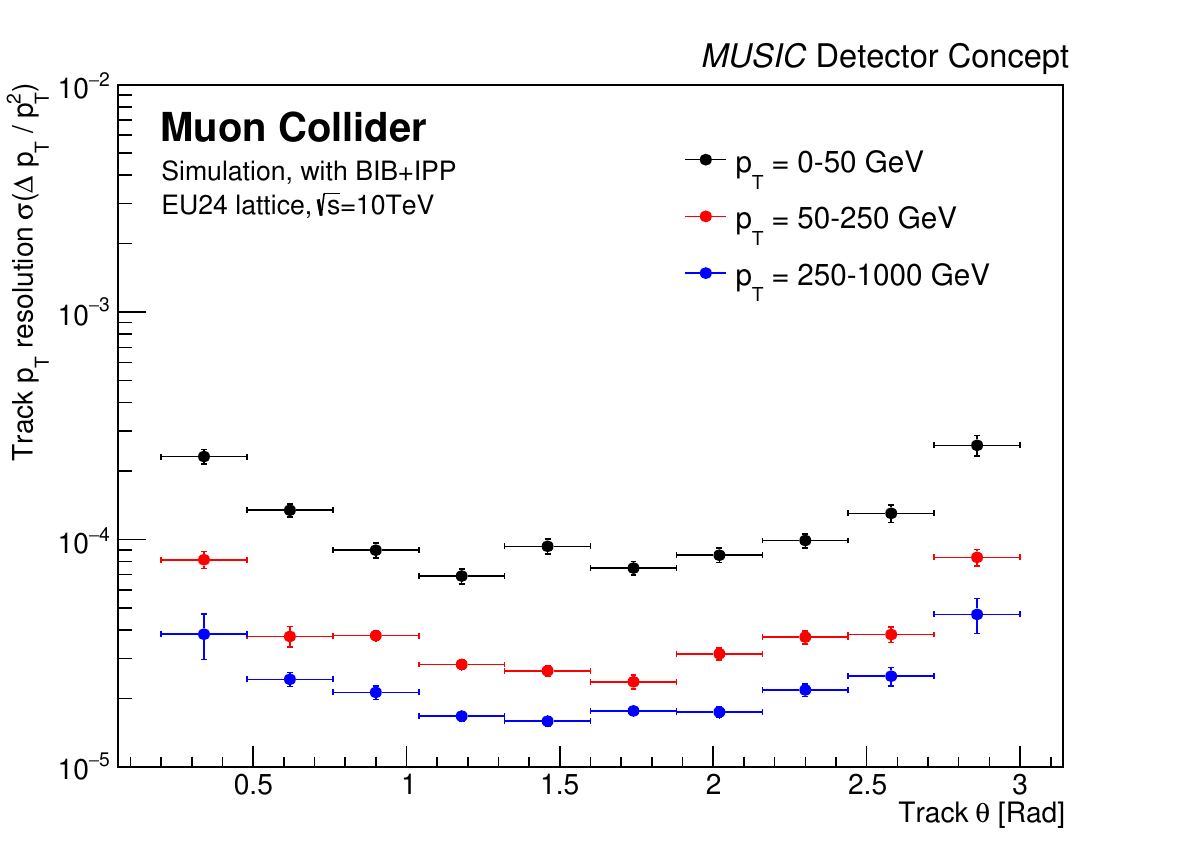}
\caption{Track reconstruction efficiency as a function of particle $\theta$ (left) and transverse momentum resolution $\sigma(\Delta p_{T}/p_{T}^2)$ as a function of track $\theta$ for different $p_T$ ranges (right) for the MUSIC detector concept obtained with prompt muons. To minimize the rate of tracks reconstructed from random combinations of hits, cleaning requirements of $p_T > 1$~GeV and $|d_0| < 0.1$~mm are applied.}
    \label{1:det:fig:MUSICtrack_efficiency}
\end{figure}

Similarly, the track reconstruction performance is also evaluated for the MUSIC detector concept, with the same tools used for the MAIA detector concept. Performance is shown in Figure~\ref{1:det:fig:MUSICtrack_efficiency} 
for both the track reconstruction efficiency and the transverse momentum resolution. The presence of BIB and incoherent pair production (which highly affects the occupancy of the first layers of the VXD) does not compromise the performance of the tracker. In particular, the regions close to the nozzles show very good efficiencies, similar to those in the central region, while two small dips can be seen in the transition region between barrel layers and endcap disks.

Diffused BIB energy depositions in the calorimeters make it difficult to identify and accurately measure the energy of particles from collisions. Standard Pandora algorithms~\cite{Marshall_2012} are used to reconstruct photons and neutrons, which could be further optimised to account for BIB contamination. Currently, no timing information beyond a simple selection (-0.5 to 15 ns for the two calorimeters) is used in the reconstruction process.  

\begin{figure}[h]
\centering
\includegraphics[width=0.49\textwidth]{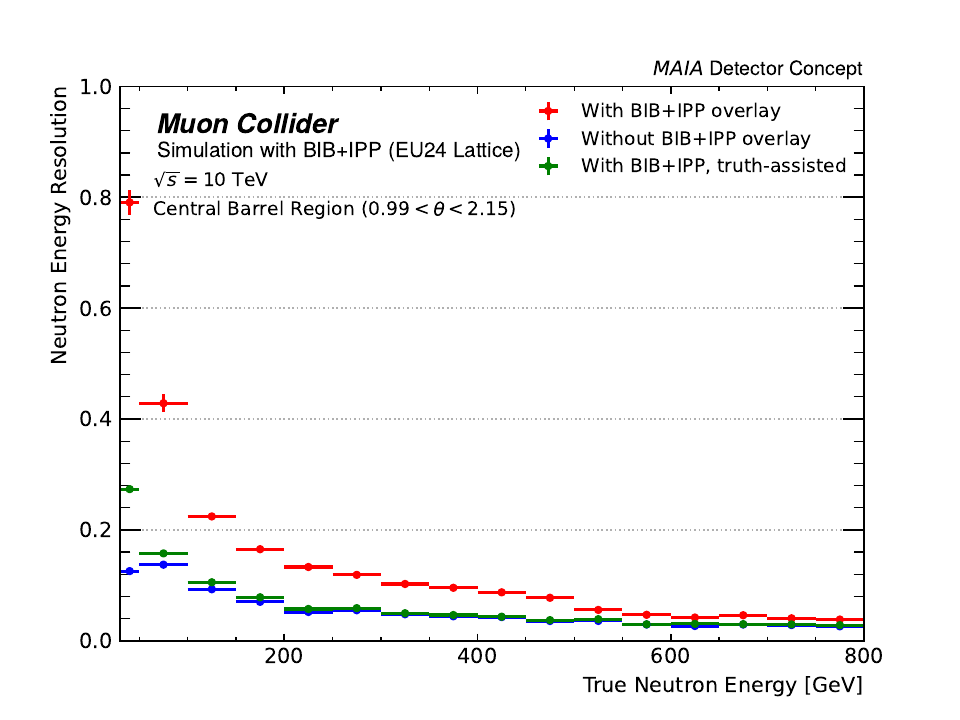}
\includegraphics[width=0.49\textwidth]{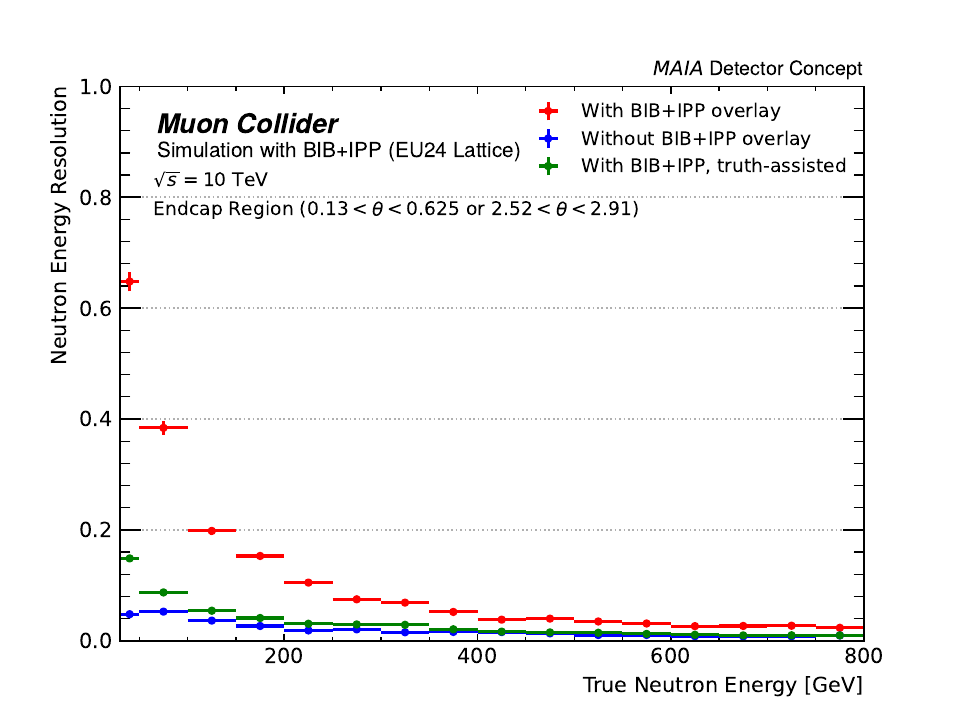}
\includegraphics[width=0.49\textwidth]{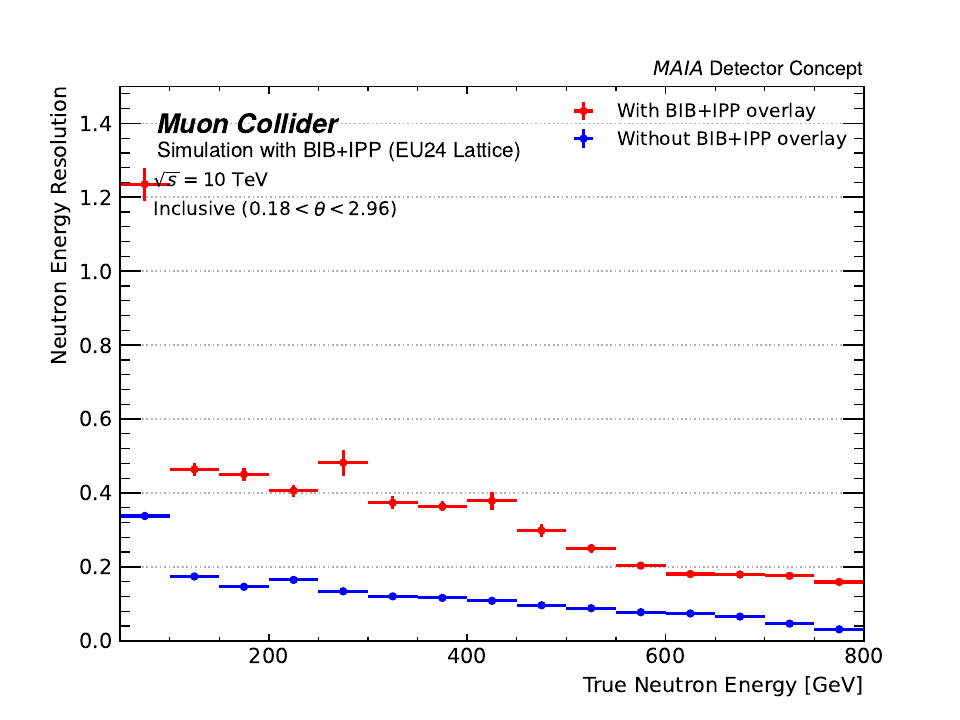}
\caption{Top: energy resolution of reconstructed photons, assessed after 2D energy calibration derived from a no-BIB simulation in the barrel (left) and endcap (right) regions. Bottom: Energy resolution of reconstructed neutrons, assessed after 2D energy calibration derived from a no-BIB simulation~\cite{MAIADetector}.}
\label{fig:maia_calo_res}
\end{figure}

The MAIA concept, with its solenoid before the ECAL, reduces the amount of BIB that reaches the calorimeters. This extra material was found not to significantly affect the resolution for low-energy signal photons. Figure~\ref{fig:maia_calo_res} shows the energy resolution of reconstructed photon and neutron candidates. The intrinsic detector resolution effects at low photon energies are much smaller than those introduced by BIB. The presence of BIB severely limits the performance of the clustering algorithms employed, resulting in a significantly degraded response. Region-based thresholding is applied to individual calorimeter cells to limit the impact of BIB, but dedicated clustering and BIB-substraction algorithms are likely further improve this performance. To visualise the resolution degradation due to the suboptimal clustering algorithms, a set of data points labelled ''truth assisted`` shows the resolution of photon objects reconstructed by summing the energy of all calorimeter cells within a cone of radius 0.05 around the true photon direction. The improved performance suggests that an optimised clustering algorithm should be able to almost fully recover the energy resolution expected in the absence of BIB.

\begin{figure}[h]
\centering
\includegraphics[width=0.49\textwidth]{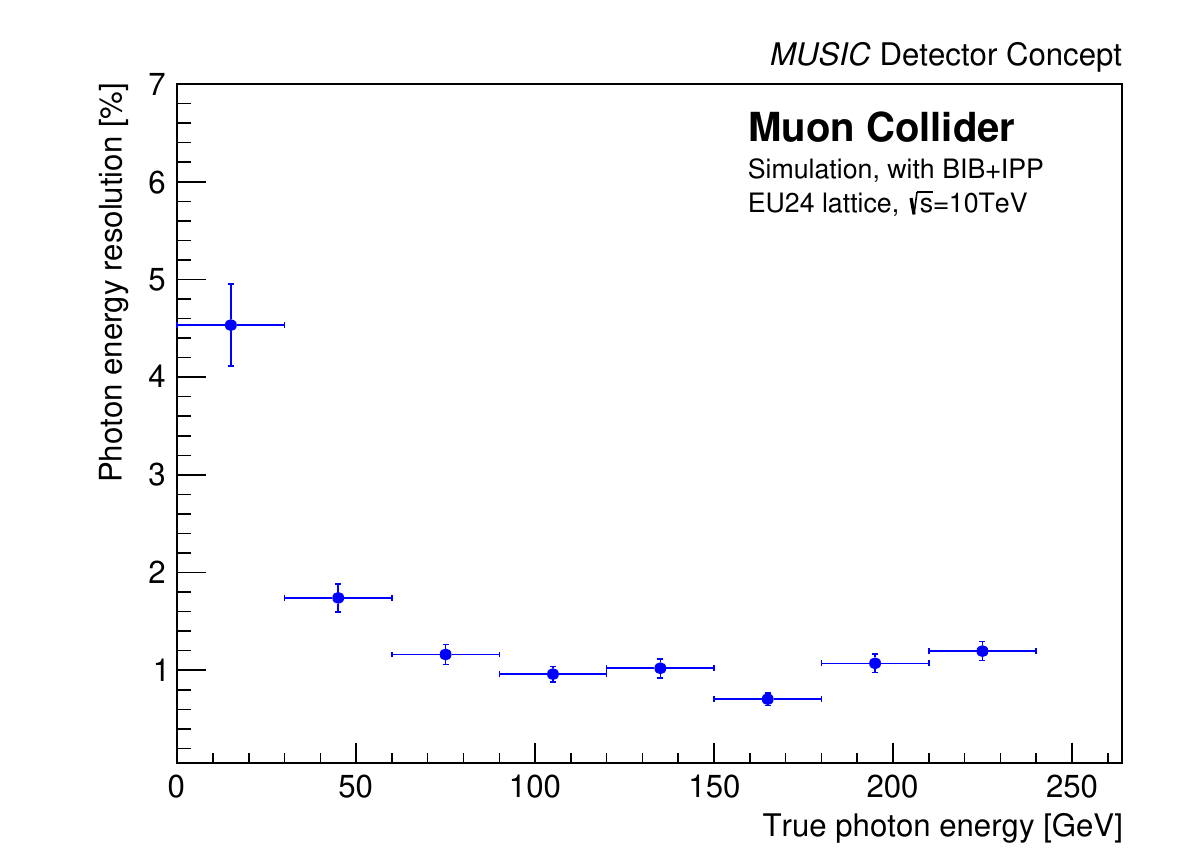}
\includegraphics[width=0.49\textwidth]{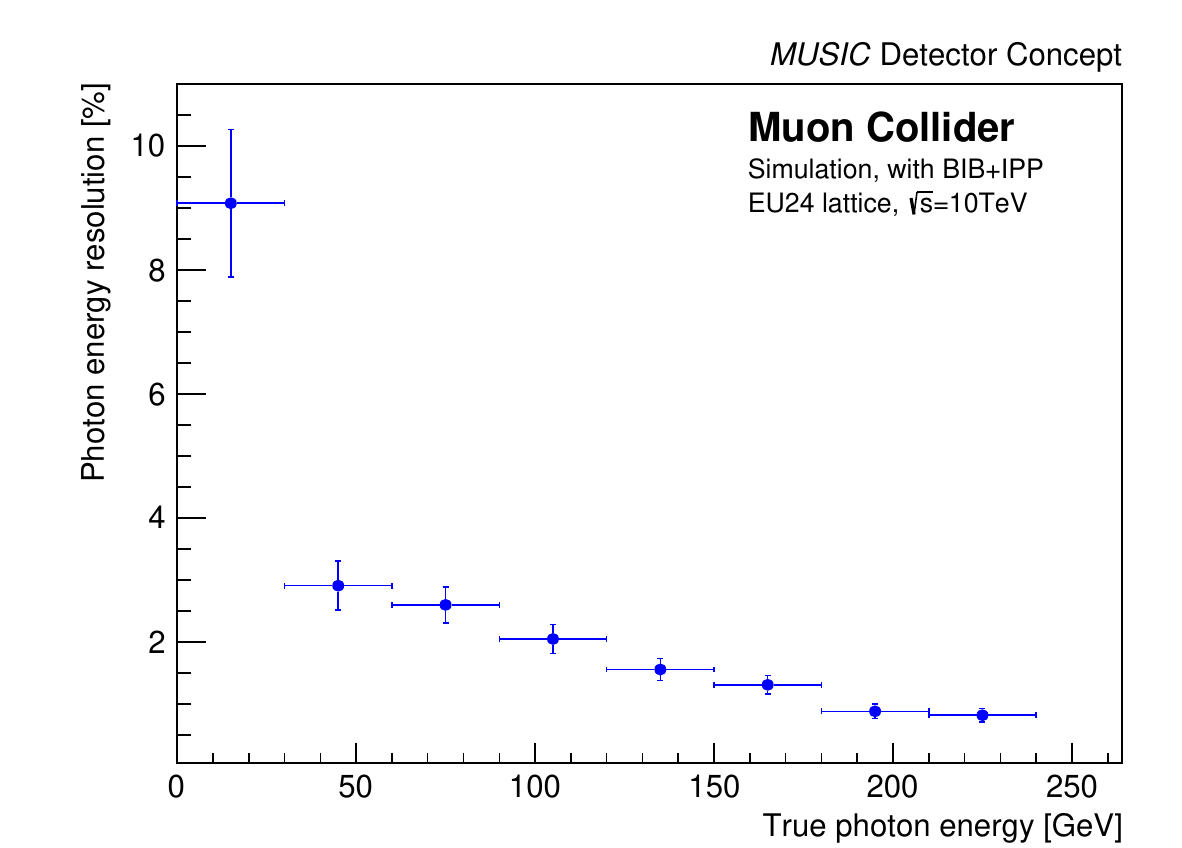}
\caption{Energy resolution for photons in the ECAL barrel (left) and in the ECAL endcap (right). Photons are defined as neutral clusters in the ECAL reconstructed by the PandoraPFA algorithm. }
\label{fig:music_photon_res_ecal}
\end{figure}

\begin{figure}[h]
\centering
\includegraphics[width=0.49\textwidth]{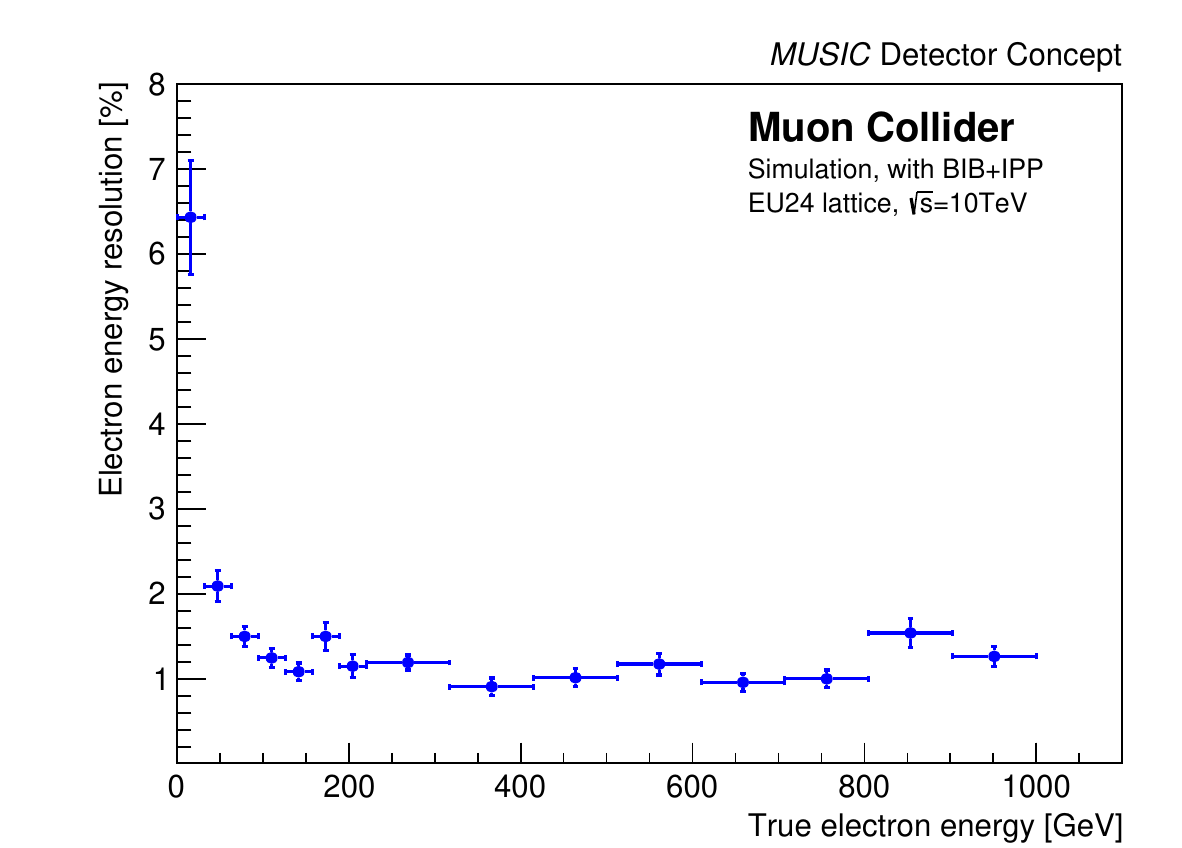}
\caption{Energy resolution for electrons in the ECAL barrel. Electrons are defined as charged clusters in the ECAL reconstructed by the PandoraPFA algorithm.   }
\label{fig:music_electron_res}
\end{figure}

The photon and electron reconstruction performance of the MUSIC detector is determined after an ECAL-dedicated hit digitisation and filter procedure to minimise the effect of the photon background from BIB. The ECAL is divided in $\theta-\phi$ regions and layers to take into account the different BIB spatial distribution. A time window and an energy threshold is defined for each region by studying the BIB and a sample of photons generated with a distribution flat in energy and angle. The figure-of-merit in the optimisation is to keep 95\% of the signal while reducing the BIB contribution as much as possible. Performance plots of the energy resolution for photons and electrons with the MUSIC detector concept are shown in Figures~\ref{fig:music_photon_res_ecal} and ~\ref{fig:music_electron_res}.

\begin{figure}[h]
    \centering
    \includegraphics[width=0.49\textwidth]{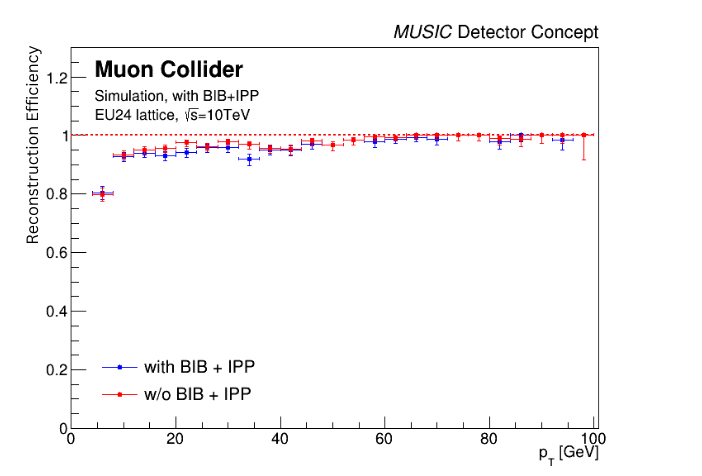}
    \includegraphics[width=0.485\textwidth]{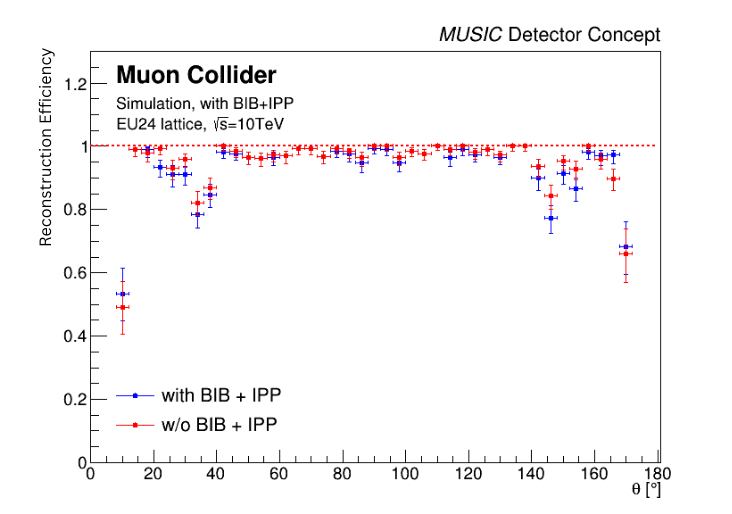}
    \caption{Muon reconstruction efficiency as a function of the transverse momentum (left) and the polar angle (right) for samples of single muons without background (red dots) and with beam-induced background and incoherent $e^+e^-$ pairs overlaid (blue dots). Only tracks with $|d_0|<0.1$ mm and $|z_0|<0.1$ mm are used.
    \label{fig:MUSICmuon_efficiency}}    
\end{figure} 
Muons are identified by extrapolating and matching tracks to hits in the outer muon detectors. Figure~\ref{fig:MUSICmuon_efficiency} shows the MUSIC detector reconstruction efficiency, which takes into account both the tracking efficiency (that dominates the efficiency losses\footnote{Further optimisation of reconstruction algorithms are expected to fully recover this loss of efficiency.}) and the hit-matching efficiency, for prompt muons with $|d_0|<0.1$ mm and $|z_0|<0.1$ mm.

Jet reconstruction performance was assessed with the MUSIC detector concept.
Jets are clustered using the $k_T$ algorithm with a distance parameter $R=0.5$ from the particle-flow objects reconstructed with the PandoraPFA algorithm, which uses calorimeter hits and reconstructed tracks as inputs. 
Jet energy correction functions are also determined with a calibration procedure, and they are applied to the reconstructed jets in order to obtain the measured 4-momentum.
Fake jets originated by BIB and incoherent pair-production (IPP) are removed by applying quality requirements to the jets: jets are required to have at least one track with a $p_{\mathrm{T}}$ larger than 2 GeV. Following this procedure, an average number of 1.96 fake jets per bunch crossing are kept. These fake jets have mostly low $p_{\mathrm{T}}$ (below 80 GeV) and are concentrated in the transition region between barrel and endcap detectors. Further identification, kinematic and analysis requirements can be applied to reduce the contribution from fake jets to a negligible level, depending on the signal channel studied.

\begin{figure}[h]
    \centering
    \includegraphics[width=0.49\textwidth]{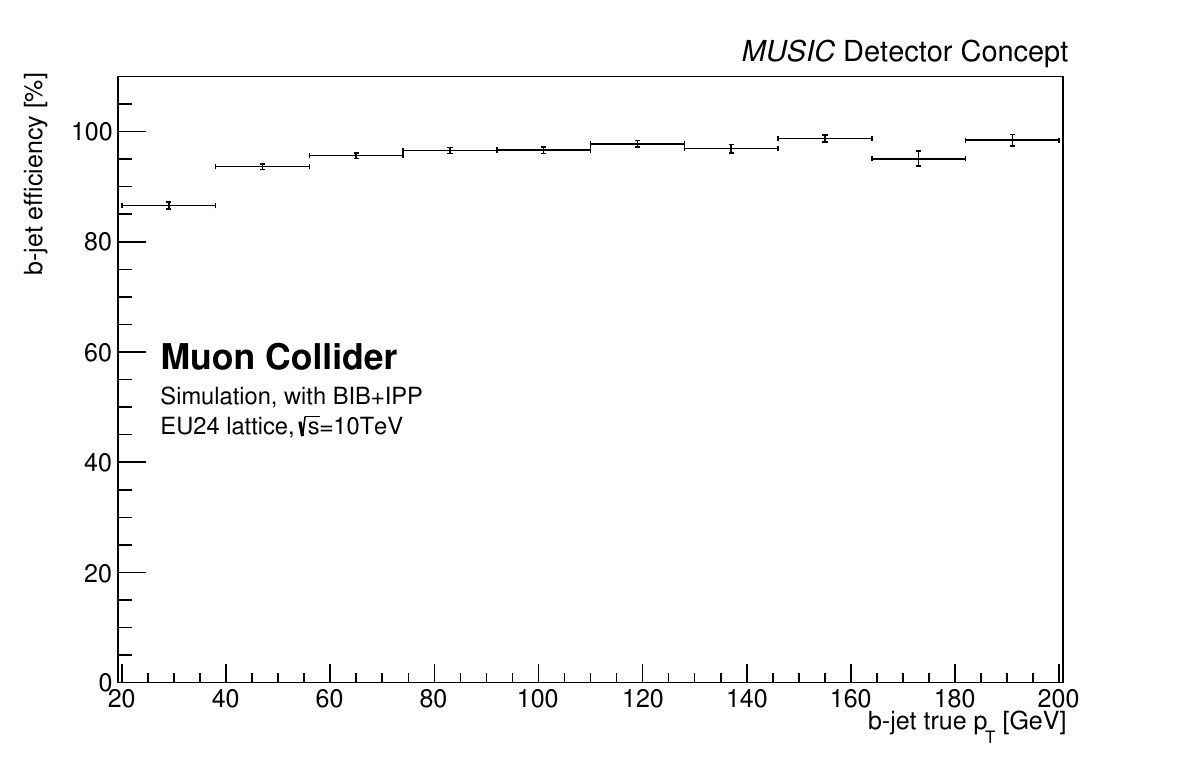}
    \includegraphics[width=0.49\textwidth]{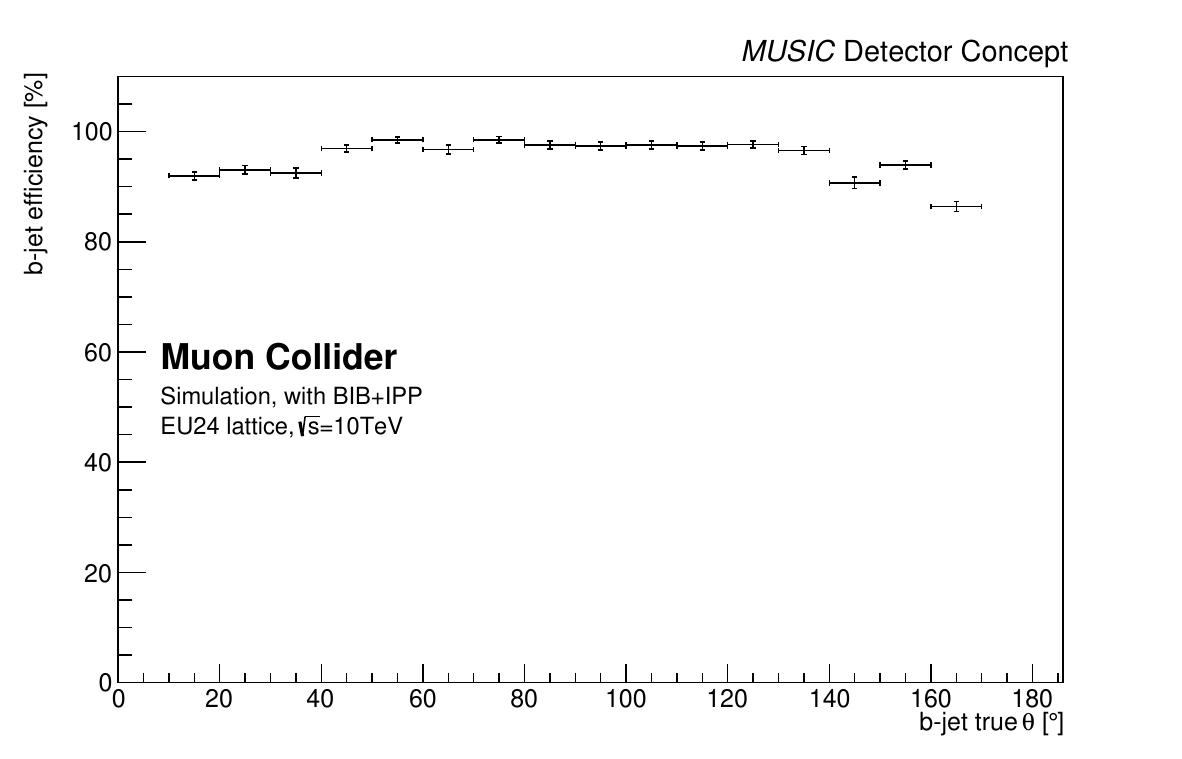}
    \caption{Selection efficiency for $b$-jets as a function of the jet $p_{\mathrm{T}}$ (left) and jet $\theta$ (right). 
        \label{fig:MUSICjet_eff}}    
\end{figure}

\begin{figure}[h]
    \centering
    \includegraphics[width=0.49\textwidth]{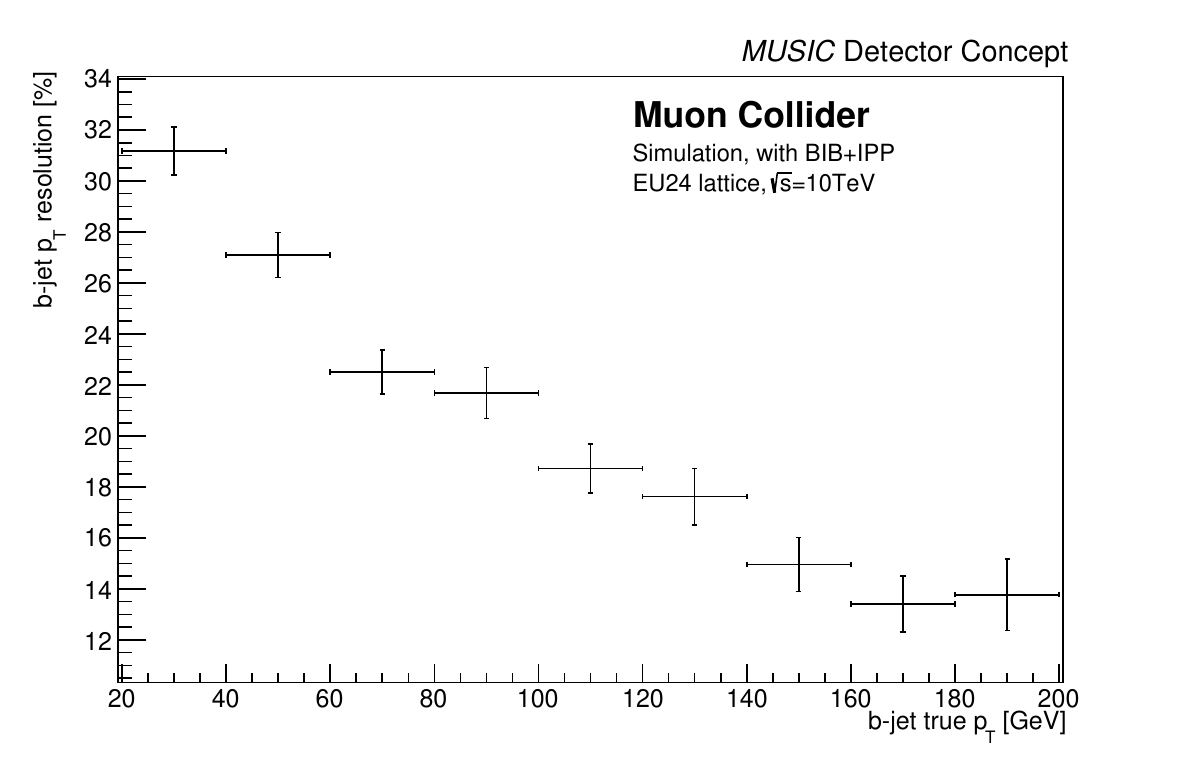}
    \includegraphics[width=0.49\textwidth]{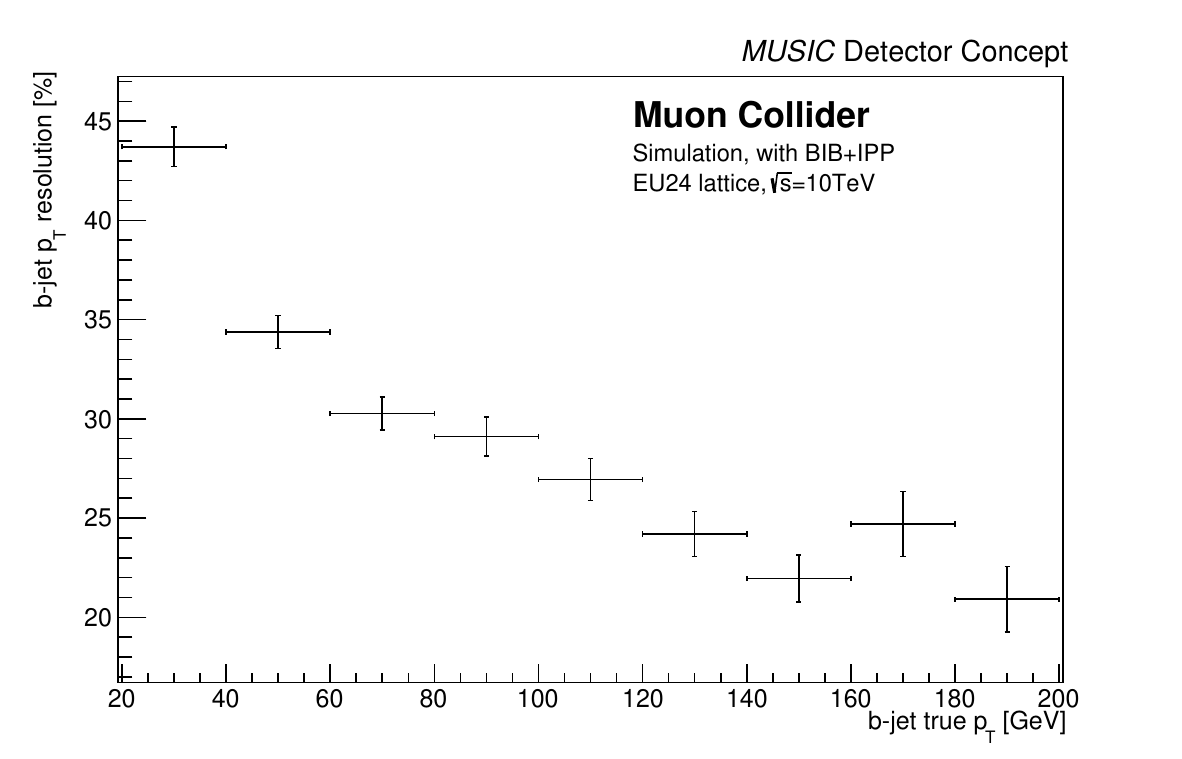}
    \caption{The $b$-jet $p_{\mathrm{T}}$ resolution as a function of the jet $p_{\mathrm{T}}$ for the central region $60^\circ<\theta<120^\circ$ (left) and forward region $10^\circ<\theta<30^\circ$ and $150^\circ<\theta<170^\circ$ (right). 
        \label{fig:MUSICjet_reso}}    
\end{figure}

The jet performance is determined from a sample of simulated $b\bar{b}$ di-jet events for various invariant masses of the jet pairs. The $b$-jet selection efficiency is shown in Figure~\ref{fig:MUSICjet_eff} as a function of the jet $p_{\mathrm{T}}$ and jet $\theta$. 
The selection efficiency reaches about 85\% for a jet $p_{\mathrm{T}}$ between 20 and 40 GeV, and it is above 95\% for jets with $p_{\mathrm{T}}$ higher than 60 GeV. The efficiency in the forward region is similarly high, above 90\%.
The jet efficiency does not appear to be significantly impacted by BIB.
The $b$-jet $p_{\mathrm{T}}$ resolution is shown in Figure~\ref{fig:MUSICjet_reso} as a function of the jet $p_{\mathrm{T}}$, in two different detector regions, central ($60^\circ<\theta<120^\circ$) and forward ($10^\circ<\theta<30^\circ$ and $150^\circ<\theta<170^\circ$). 
In the central region the jet $p_{\mathrm{T}}$ resolution ranges from 31\% at $p_{\mathrm{T}}=20$ GeV to 14\% at $p_{\mathrm{T}}>160$ GeV. In the forward region the $p_{\mathrm{T}}$ resolution is higher, from 44\% at $p_{\mathrm{T}}=20$ GeV to 22\% at $p_{\mathrm{T}}>140$ GeV.
The jet $p_{\mathrm{T}}$ resolution is clearly affected by the presence of the BIB in the tracker and in the calorimeters, a further optimisation is needed. A qualitative comparison with previous studies with the 3 TeV detector concept shows that performance is comparable.

\FloatBarrier
\section{Technologies}
\label{1:det:sec:technologies}
The tracking detector is the most technologically challenging component due to the highest total ionising dose (1 Mrad/year), which drives its radiation hardness requirements, as well as the highest density of BIB particles per collision event, significantly impacting track reconstruction performance.
Extensive full-simulation studies have demonstrated the vital importance of timing resolution down to 30~ps to effectively reject BIB contributions, which is the driving force behind the choice of the ultra-fast all-silicon tracker technology shared by both MAIA and MUSIC detector concepts.
Currently, both concepts only include the generic performance parameters that are relevant for full-simulation studies, without explicit technology assumptions.
These relevant parameters include time resolution of 30~ps (60~ps) in the VXD (IT/OT), in line with the current state-of-the-art technologies based on LGAD sensors~\cite{lgad_cartiglia}, and total silicon thickness of 190~$\mu m$ to account for the material budget.
No dedicated cooling material is introduced at this stage, assuming the feasibility of low-enough thermal power that can be dissipated using air or gaseous-Helium cooling alone.
In terms of granularity we assume pixel pitch ranging from $25 \mathrm{\mu m} \times 25 \mathrm{\mu m}$ pixels in the high-occupancy regions of the VXD to $50 \mu m \times 1 mm$ (or up to 10 mm) macro-pixels in the lower-occupancy IT and OT. Additional rejection power from on- and off-detector shape analysis of the deposited energy has been found to further increase separation of signal from BIB, and reduce to some extent timing requirements.

Instead, the MUSIC concept adopts a novel semi-homogeneous CRILIN design, which is based on high-density Cherenkov-emitting crystals with silicon photo-multiplier (SiPM) readout.
This technology can deliver superior time resolution (down to 45 ps) and similar energy resolution with a much smaller number of channels and about factor 10 lower cost.
Prototypes of this technology are being developed and tested under the dedicated CRILIN R\&D project~\cite{crilin2024}, described in more details in Section~\ref{3:det:sec:technologies}.

Both experiment concepts assume a HCAL made of plastic scintillator 30 $\times$ 30~mm$^{2}$ tiles and iron absorbers, whose thickness is 3~mm and 20~mm respectively. The choice of iron absorbers is mainly due to the need of closing the magnetic field lines, while the use of plastic scintillators is a well-consolidated choice for the active layers as demonstrated by ATLAS and CMS~\cite{ATLASCollaboration_2008,CMSCollaboration_2008}. For example, the ATLAS HCAL, made of plastic scintillators and steel as absorbers, shows an energy resolution for isolated pions of $\sigma/E = 56.4\%/\sqrt{E(\mbox{GeV})} \bigoplus 5.5\%$~\cite{ATLASCollaboration_2008}, which is close to the $55\%/\sqrt{E}$ needed to guarantee a jet energy resolution good enough to separate $Z$ and $W$ bosons in the hadronic channel~\cite{THOMSON200925}.  In addition, such a hadronic calorimeter shows a time resolution of a few ns, for an energy cell deposit below 5~GeV, or even lower for higher energies~\cite{TileATLASRun2}, which helps in distinguishing BIB from hadrons coming from the interaction point. Another possible solution for the HCAL, now under investigation, is the possibility of using Micro-Pattern Gaseous Detectors  (MPGD) as an active layer; preliminary results have shown an energy resolution for impinging pions of $\sim 46\%/\sqrt{E}$, well-below the tile case~\cite{Longo:2024pk}. 

As already mentioned in sections~\ref{1:det:sec:music} and~\ref{1:det:sec:maia}, both MUSIC and MAIA do not have a dedicated magnetic field for the muon system and the muon momentum information will be provided only by the inner tracking detectors. 
Nevertheless, the muon detectors play a crucial role in providing additional information for muon identification. To cope with the high rate in the high-eta region and to match the track from the inner tracker, good spatial ($\approx 100 \mu\mbox{m}$) and time (below 1 ns) resolution are required for the muon system. Dedicated R\&D is therefore needed and preliminary studies on the use of MPGD (GEM and PICOSEC) are ongoing.

\section{Software \& Computing}
\label{1:det:sec:software}

The muon collider software framework~\cite{mccsoft} is based on the Key4hep framework~\cite{Key4hep:2022xly}. Releases are created using the Spack package manager~\cite{Gamblin_The_Spack_Package_2015} and distributed as Docker images. A copy of the image is readily available for Apptainer via the CernVM file system~\cite{cvmfs} unpacked service~\cite{unpacked}.

The releases include as core components the DD4hep toolkit~\cite{Frank:2014zya} to model the detector geometry, the Gaudi framework~\cite{Barrand:2001ny} for event processing, and the LCIO~\cite{Gaede:2003ip} and EDM4hep~\cite{Gaede:2021izq} event data models for data representation. The detector response simulation via Geant4~\cite{GEANT4:2002zbu,Allison:2016lfl} is driven by DD4hep, while event digitisation and reconstruction use the \emph{MarlinWrapper}~\cite{marlinwrapper} to provide a Python interface to configure, execute, and enable EDM4hep output from pre-existing reconstruction algorithms developed outside of Gaudi for earlier studies~\cite{Gaede:2006pj}. 
Third party libraries are leveraged for advanced reconstruction algorithms, including the ACTS~\cite{ai2022common} library for charged particle track reconstruction and the PandoraPFO~\cite{Marshall_2012,THOMSON200925} library for physics object reconstruction using particle flow.
The legacy algorithms in Marlin are gradually being migrated to Gaudi to improve compatibility with the newer key4hep software stack and corresponding support.
The first migration, for the ACTS algorithm~\cite{mcs-acts}, is nearing completion.
Subsequent migrations, beginning in 2025, will initially target digitisation and BIB overlay algorithms, and then other reconstruction algorithms.

Stable particles are generated by external packages and given as input to the aforementioned software. Signal and physics background events are produced by users with common packages like WHIZARD~\cite{Kilian:2007gr} and MADGRAPH~\cite{madgraph}. The beam-induced background bunch-crossings are produced by using the FLUKA~\cite{Battistoni2013,Ahdida2022} and FlukaLineBuilder~\cite{Mereghetti:2012zz} programs, as described in Section~\ref{1:inter:sec:mdi}.

\begin{figure}[htbp]
    \centering
    \includegraphics[width=0.99\linewidth]{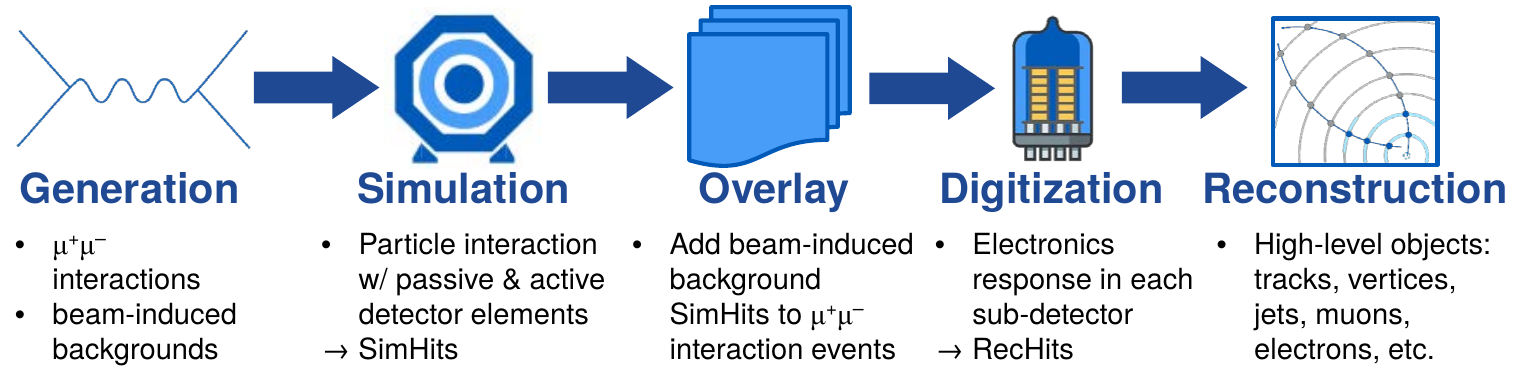}
    \caption{The steps of the muon collider simulation workflow.}
    \label{fig:sim-workflow}
\end{figure}

The simulation workflow proceeds as outlined in Figure~\ref{fig:sim-workflow}.
Intermediary data, with the exception of the generator output, is stored using the SLCIO format.
The reconstructed physics objects are used for analysis measurements. Their performance is summarized in Section~\ref{1:det:sec:performance}. 

The requirements on processing and disk space are driven by the presence of the beam-induced background, which are responsible for largest contribution of SimHits.
A study of the walltime per event of the different Gaudi processors in the recommended workflow with and without a BIB overlay was performed. Consistently with expectations, the walltime of each algorithm is completely dominated by the presence of BIB, with a steep dependency on the collider lattice version taken into account. 

Algorithms that have to deal with combinatorial challenges dominate the walltime: tracking and, to a minor extent, calorimeter clustering. Depending on the collider lattice, reconstructing all the tracks in one event using the standard tracking configuration can take from 5 min/event to multiple hours. Similarly, the clustering of calorimeter hits was observed to range between 1 min/event to approximately an hour/event. 
The overlay of the simulated hits from BIB before digitisation is another notable walltime user with an average of 5 mins/event.
A maximum memory usage of 32 GB was observed throughout the data processing chain. 

The performance strongly depends on the available processing power and the algorithms utilised. Both are expected to evolve significantly in the next decade. In particular, the introduction of multithreaded algorithms and heterogeneous computing resources is expected to drastically reduce the processing time per event.

Similarly to CPU resources, storage is also dominated by the amount of BIB-related information that will need to be stored on disk.
The current average event size for a typical $\mu\mu \rightarrow \nu\nu h$ event at $\sqrt{s}=10$~TeV throughout the simulation, digitisation and reconstruction chain is of about 1 MB/event in the absence of BIB.
When considering the EU24 lattice and the MAIA detector, an average of approximately 20~GB of simulated BIB hits is overlaid before digitisation to each event.
The size of the reconstructed data depends strongly on the amount of information that is needed to carry forward in subsequent processing steps.

\chapter{Accelerator complex concepts}
\label{1:acc:ch}

The baseline final muon collider design is a \SI{10}{\tera\electronvolt} centre-of-mass collider providing a design luminosity of \SI[per-mode=reciprocal]{21E34}{\per\centi\meter\squared\per\second}.
The following chapter produces the concept for a greenfield site, assuming a \SI{10}{\kilo\meter} circumference collider ring with straight sections sufficient for two detector caverns.

The muons will be produced from a proton driver, detailed in Section~\ref{1:acc:sec:proton}, which delivers either a \SI{5}{\giga\electronvolt} beam with \SI{2}{\mega\watt} power, or a \SI{10}{\giga\electronvolt} with \SI{4}{\mega\watt} power, both with \num{1}-\SI{2}{\nano\second} bunch length at \SI{5}{\hertz} repetition rate.
This beam will intersect with a graphite target and produce approximately \num{60E12} $\mu^+$ and \num{45E12} $\mu^-$.
The front-end, discussed in Section~\ref{1:acc:sec:cool}, after the target consists of a capture system of superconducting solenoids, a chicane to separate from the proton beam, and charge separator.
The large transverse emittance of the initially generated beam will be cooled from \SI{17000}{\micro\meter} to \SI{22.5}{\micro\meter} through a combination of the 6D rectilinear cooling and final cooling systems. A pre-accelerator will restore energy to \SI{250}{\mega\electronvolt}.\\
The low-energy acceleration will bring the beams from \SI{250}{\mega\electronvolt} to \SI{63}{\giga\electronvolt} through a combination of LINACS and Recirculating Linear Accelerators (RLAs) and in discussed in Section~\ref{1:acc:sec:acc}.
Each beam will be made of one bunch, and will be counter-rotating after the LINAC.
The high-energy acceleration is performed by a series of four Rapid Cycling Synchrotrons (RCS), described in Section~\ref{1:acc:sec:rcs_chain}, which accelerate the beams to \SI{5}{\tera\electronvolt}.
Then finally the positive and negative beams of $\approx$\num{2E12} muons each, will be focused at the two interaction regions within the collider, as detailed in Section~\ref{1:acc:sec:collider}.
All of these systems feature high-intensity beams, so will have different collective effects throughout the complex. These effects are discussed in Section~\ref{1:acc:sec:collective}. 

The current status and key challenges of each of these stages of the accelerator complex have been evaluated in this chapter.

\section{Proton driver}
\label{1:acc:sec:proton}

\subsection*{Overview}

\begin{figure}[ht]
    \centering
    \includegraphics[width=0.7\textwidth]{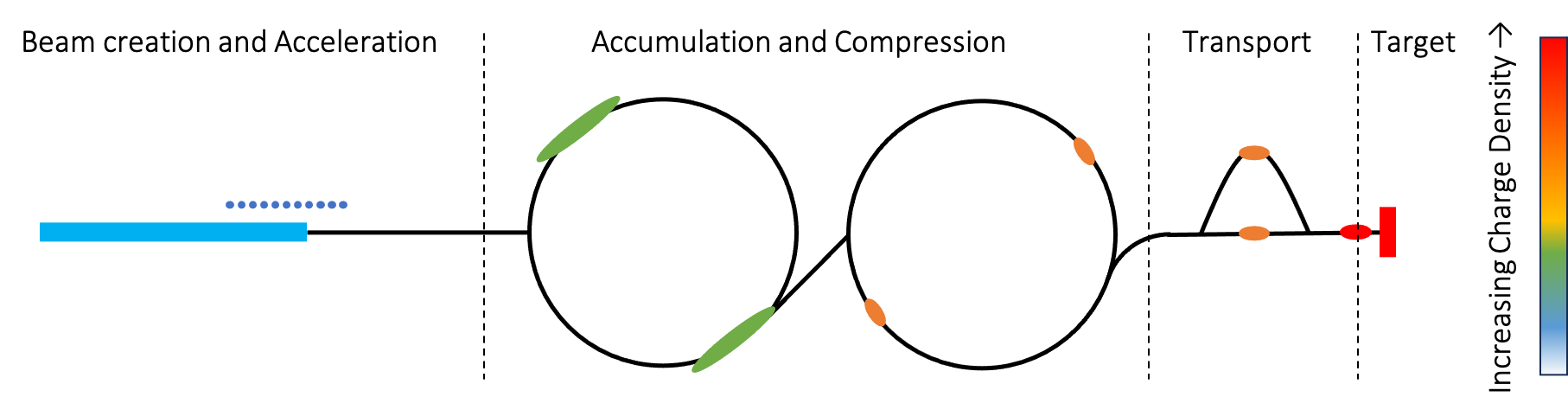}
    \caption{Schematic layout for the proton complex section. The closer to the target the higher the bunch charge density. Reaching these high densities is one the main challenges in the design of the proton complex. 
    \label{2:acc:fig:proton:proton_complex_layout}}
\end{figure}

The proton complex is the first piece in the muon collider complex. It includes a high-power acceleration section, an Accumulator and Compressor, and a target delivery section. A schematic overview can be seen in Figure~\ref{2:acc:fig:proton:proton_complex_layout}.

In our baseline design, we will study a final energy linac of 5~GeV and 10~GeV. This section is responsible for delivering high power, high intensity bunches to the Accumulator and Compressor section. At the end of the high power acceleration section, the pulse total charge is distributed between 1 or 2 bunches with a small energy spread in an Accumulator ring. In order to achieve the high proton intensities in a short pulse, a Compressor ring will be used to compress the protons into very short bunches, with a rms length of the order of $\sim 2$\,ns.  

The first storage ring in the complex, the Accumulator, accumulates the protons via charge stripping of the incoming H$^−$ beam from the linac section, preserving the energy spread. The incoming beam is chopped to allow a clean injection into the ring circumference. The second storage ring, the Compressor, accepts bunches from the Accumulator and performs a 90\textdegree-bunch rotation in longitudinal phase space, shortening the bunches to the limit of the space charge tune shift just before extraction. The Compressor ring must have a large momentum acceptance to allow the storage of the beam with a momentum spread of a few percent, which arises as a consequence of the bunch rotation. 

The Target Delivery Transport Line is used to transport the final intense short bunches from the Compressor ring to the target, ensuring that neither the transverse nor longitudinal properties of the beam are compromised in the process. To achieve optimal luminosity, the final collider will work in single bunch colliding mode therefore, in the multi-bunch solution, short bunches extracted from the Compressor must be recombined and hit the pion production target simultaneously.

The primary requirement of the proton complex is to enable the production of a high number of useful muons at the end of the decay channel after the target. The production rate, to good approximation, is proportional to the primary proton beam power, and (within the 5–15\,GeV range) only weakly dependent on the proton beam energy~\cite{Ding:2011zzf}. Considering a conversion efficiency of about 0.013 muons per proton-GeV~\cite{MAP}, a proton beam in the 1–2\,MW power range at an energy between 5\,GeV and 10\,GeV will provide the sufficient number of muons required.

\subsection*{Key challenges}

The critical challenges for the proton complex include the determination of proton final energy on the target with significant implications for particle creation, the layout of the high power acceleration section, and the choice of compression setup. Two final proton beam energies, 5~GeV and 10~GeV, are now under investigation.

The final energy and repetition rate needed from the linac will drive studies for beam generation and transport (H- sources, RFQ design, H- stripping mitigation issues). Accumulation requirements, such as number of turns and ring circumference will drive the acceleration chain linac repetition rate\footnote{Notice that the linac repetition rate does not need to be the same as the Muon Collider repetition rate. The proton complex, in the end, will deliver proton to the target with a repetition rate that is dictated by the Physics needs and by what the technology on the front-end is able to deliver.} and chopping scheme. Determining the linac chopping scheme is another key parameter, which will directly influence the linac pulse length and current, and subsequently, the Accumulator and further proton complex design.

A comprehensive study is under way to identify key parameters determining the choice of compression scheme and final number of bunches, thus influencing the layout of the Compressor ring lattice and the required number of turns for full compression. The minimum number of bunches needed will affect the layout of the  Target Delivery Transport Line  The beam parameters on the target surface will impact the design of the beam dump, which can be challenging due to target geometry and constraints.

Other important challenges include the design of the Accumulator lattice, investigation of the instability thresholds in the Accumulator and Compressor rings, error analysis across the entire complex with a focus on bunch recombination at the target delivery system, and continuous development of the H$^{-}$ source to meet the needed pulse current and length, including potential upgrades.

\subsection*{Recent achievements}

The most important recent achievements are listed here and discussed in more detail in the rest of this Section:
\begin{itemize}
    \item Definition of the set of preliminary parameters for the study~\cite{PreliminaryParameter_MuCol5} was completed;
    \item A preliminary lattice design for the linac was created and estimations on on H- stripping loss tolerances was done;
    \item A tentative chopping scheme for the linac was defined;
    \item Simulations for Accumulator show no instabilities up to 6000 turns, which is the total number of turns needed for full accumulation of the linac pulse;
    \item Design of a preliminary lattice for the compressor ring and simulations of the compression processes is shows that achieving a 2 ns rms final bunch is possible, and
    \item A preliminary design of a Target Delivery Transport Line was done and simulations of the compressed beam show no loss of beam quality.
\end{itemize}

A preliminary set of parameters for the proton complex was completed and published in~\cite{PreliminaryParameter_MuCol5} and the main parameters are summarized in Table~\ref{2:acc:proton:tab:parameters}. The choice of the 5~GeV baseline energy comes from the SPL design~\cite{baylacConceptualDesignSPL2006}, used as baseline for the linac. Simulations indicate that space charge effects preclude operation at 4 MW and 5 GeV. If this power is required, a proton energy of 10 GeV will be needed and thus was included in the study as a second option.

\begin{table}[ht]
    \centering
    \begin{tabular}{l|ccc}
        Parameters & Unit & Option 1 & Option 2 \\ \hline
        Final Beam Power & MW & 2 & 4\\
        Linac Final energy & GeV & 5 & 10\\
        Repetition rate$^{1}$ & Hz &\multicolumn{2}{c}{5}\\
        Protons on target & \num{E14} & \multicolumn{2}{c}{5.0}\\
        Initial bunch length & ns &  \multicolumn{2}{c}{120}\\
        Initial rms energy spread & \% &  \multicolumn{2}{c}{0.025}\\
        Final rms bunch length & ns & \multicolumn{2}{c}{2.0}\\         
        Final rms energy spread & \% & 1.5 & 0.6\\         
        \end{tabular}
    \caption[Proton Complex parameters]{Summary of main parameters for the proton complex for two power options~\cite{PreliminaryParameter_MuCol5}.}
    \label{2:acc:proton:tab:parameters}
\end{table}

\begin{figure}[htb]
    \centering
    \begin{minipage}[b]{0.48\linewidth}
        \centering
        \includegraphics[width=\linewidth]{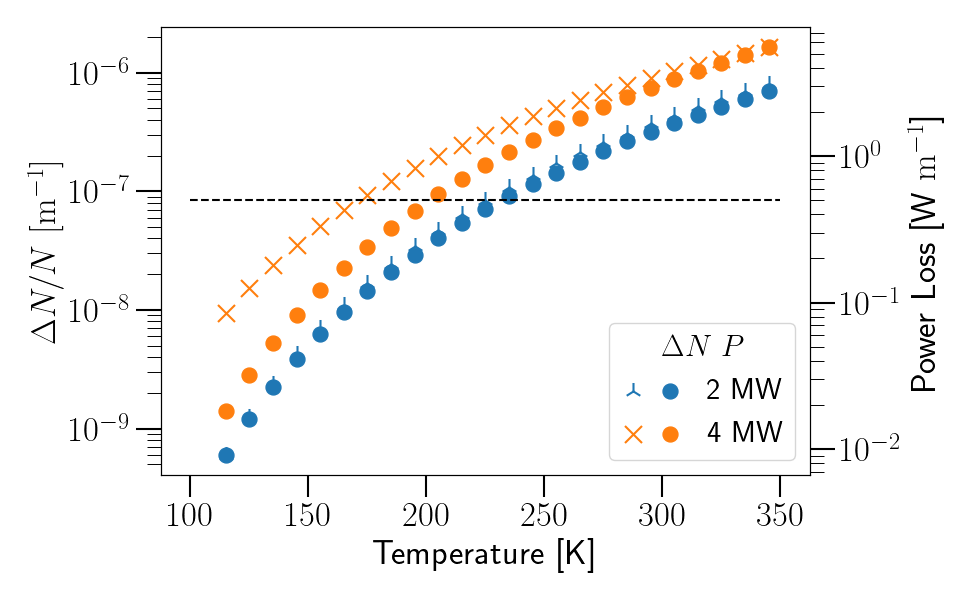}
        \caption{$H^{-}$ stripping loss power components for a full energy linac from blackbody radiation as a function of the temperature of the vacuum pipe. The dashed line represents the 0.5 $W m^{-1}$ threshold.}
        \label{2:acc:proton:fig:blackbody}
    \end{minipage}
    \hspace{0.01\linewidth} 
    \begin{minipage}[b]{0.45\linewidth}
        \centering
        \includegraphics[width=\linewidth]{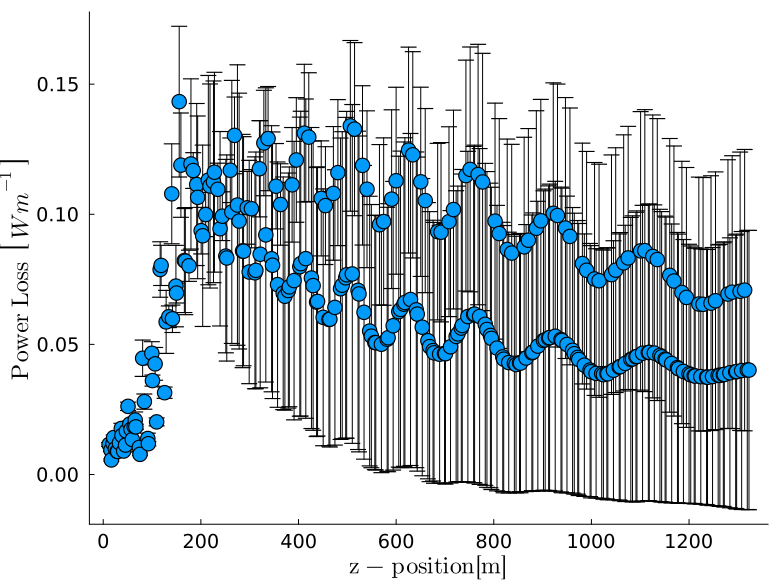}
        \caption{IBS (Intra Beam Scattering) $H^{-}$ stripping for a linac with a final energy of 10 GeV.}
        \label{2:acc:proton:fig:IBS}
    \end{minipage}
\end{figure}

A preliminary design, based on the ESS and SPL lattices, for a full energy linac reaching both energies was presented~\cite{johannesson:ipac2024-wepr24} and a study of the effect of H- stripping on the loss budget of such a machine was investigated~\cite{Hminus_Stripping:Folsom}. For both power options (see~\ref{2:acc:proton:tab:parameters}), the main component for the H- stripping losses comes from blackbody radiation stripping, which could be circumvented by cooling the vacuum chamber in warm areas and transfer lines below 200 K. Alternatively, the use of a coating that reduces secondary emission can be investigated. The stripping $\Delta N/N$ as well as the power loss per length [$W M^{-1}$] as a function of the temperature of the vacuum chamber can be seen in Figure~\ref{2:acc:proton:fig:blackbody}. The intra-beam scattering component is well under control even at higher energies. Lorentz stripping represents a very small effect on the total loss budget as long as the beam sizes throughout the linac are kept below 3 mm rms for both final energies. 

\begin{figure}[htb]
    \centering
    \includegraphics[width=0.8\linewidth]{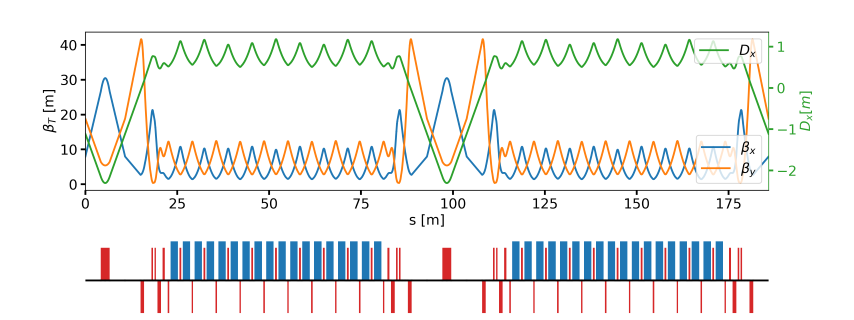}
    \caption{The 5 GeV Accumulator lattice is shown. Dipoles are represented by blue blocks, while quadrupoles are represented by red blocks. Focusing quadrupoles are placed above the reference line, and defocusing quadrupoles are placed below the reference line.}
    \label{2:acc:proton:fig:ACC_5GeV_latt_twiss}
\end{figure}

Baseline lattices for the Accumulator and Compressor for the 5\,GeV  and 10\,GeV cases were revised and ported to XSuite~\cite{bib:xsuite} (Figure~\ref{2:acc:proton:fig:ACC_5GeV_latt_twiss}). For the 5 GeV case the lattices are from the studies for the neutrino factory at CERN~\cite{Aiba:Accumulator,Aiba:Compressor}, while the 10 GeV lattices are scaled lattices for the Accumulator and a brand new lattice for the Compressor. For both energies the Compressor lattice was designed to maximize the slippage factor, thus reducing the time (or number of turns) needed for a full rotation. The ring aperture is mainly determined by the dispersion component in horizontal and since the momentum spread is greatly increased after phase rotation~\ref{2:acc:proton:tab:parameters}, keeping dispersion as small as possible is also a requirement. However, a small dispersion function and a large slippage factor are conflicting given how they are connected through the  momentum compaction:
\begin{equation}
    \eta = \frac{1}{C}\int_{0}^{C}\frac{D(s)}{\rho(s)}ds - \frac{1}{\gamma ^2}
\end{equation}
where $\eta$ is the slippage factor, $\gamma$ the Lorentz factor of the circulating beam, $C$ the ring circumference, $D(s)$ the dispersion function and $\rho(s)$ the dipole bending radius. By introducing negative bending magnets in the lattice, where the dispersion function is negative, the dispersion function could be kept small without reducing the target slippage factor (see Figure~\ref{2:acc:proton:fig:Com5_latt_twiss}). This type of rotation setup requires high voltages from the cavities and most of the ring straight sections will be filled with RF stations in order to be able to reach the 4~MV voltage needed to perform it.

\begin{figure}[htb]
    \centering
    \includegraphics[width=0.4\linewidth]{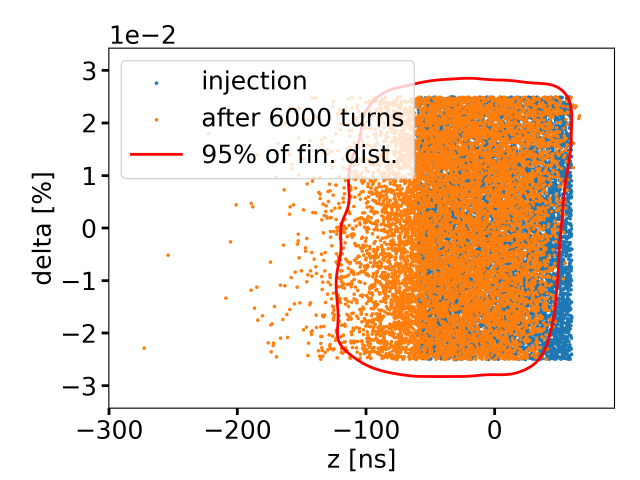}
    \includegraphics[width=0.4\linewidth]{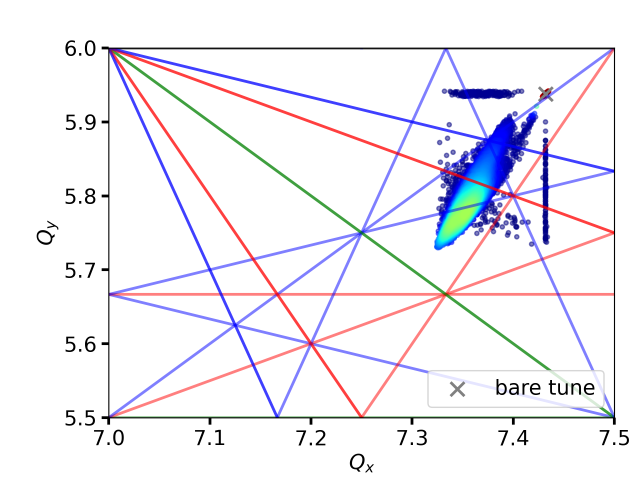}
    \caption{Simulation of the accumulation of 5 GeV. Some numerical features in the tunes calculation are still not understood in XSuite (vertical and horizontal flat lines in the tune plot). No instabilities were observed so far on the trasnverse planes.}
    \label{2:acc:proton:fig:ACC_5GeV}
\end{figure}

\begin{figure}[htb]
    \centering
    \includegraphics[width=0.5\linewidth]{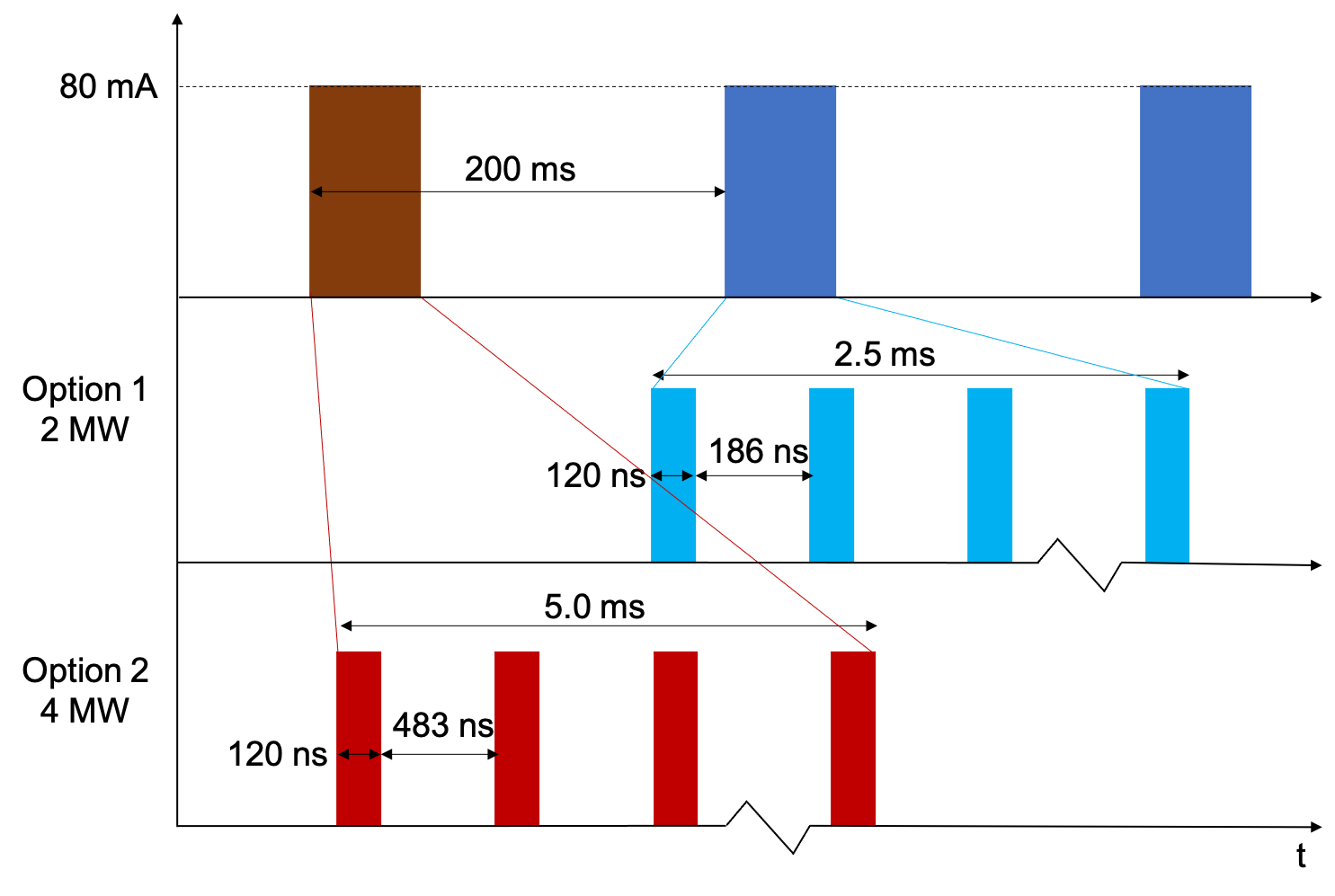}
    \caption{Possible chopping scheme for the two options power options 2 MW and 5 GeV (in blue) and 4 MW and 10 GeV in red assuming that the repetition rate o f the linac is 5 Hz. The accumulator lattice for 5 and 10 GeV have different circumferences of 180 and 300 m respectively. Considering that the maximum length before rotation for the bunches in the accumulator is 120 ns, together with the revolution period for each energy, it it is possible to define the spacing of bunches coming from the linac and show in the picture. }
    \label{2:acc:proton:fig:chopping}
\end{figure}

\begin{figure}[htb]
    \centering
    \includegraphics[width=0.8\linewidth]{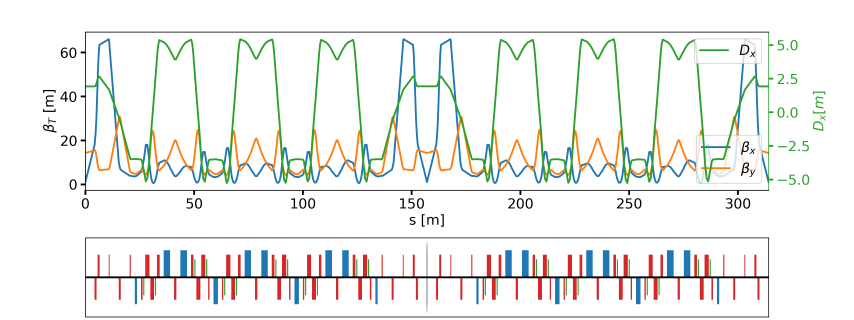}
    \caption{Twiss parameters and lattice of the proposed 5 GeV Compressor lattice design. The color code for accelerator elements is the same as in Figure~\ref{2:acc:proton:fig:ACC_5GeV_latt_twiss}, with the addition of sextupoles represented in green and a cavity represented by a grey line.}
    \label{2:acc:proton:fig:Com5_latt_twiss}
\end{figure}

An example of accumulation for the 5 GeV case can be seen in Figure~\ref{2:acc:proton:fig:ACC_5GeV}. A series of simulations to verify the lattice behaviour over many turns was launched and indicated no major show stoppers for either option when the beam is accumulated for over about 6000 turns, which corresponds to a linac working at a 5 Hz repetition rate. In the Accumulator ring, so far, there is no RF, however, there will be a need for RF barrier buckets to avoid smearing on the bunches over the many turns of accumulation. Currently, the Accumulator lengths and number of bunches envisioned for each energy dictate a possible chopping scheme needed for the linac, which is shown in Figure~\ref{2:acc:proton:fig:chopping}. For the lower energy case, the parameters needed from the source are not too far from what can be achieved today however for the 4 MW case either some R\&D is needed~\ref{4:sec:acc}.

\begin{figure}[htb]
    \centering
    \includegraphics[width=0.32\linewidth]{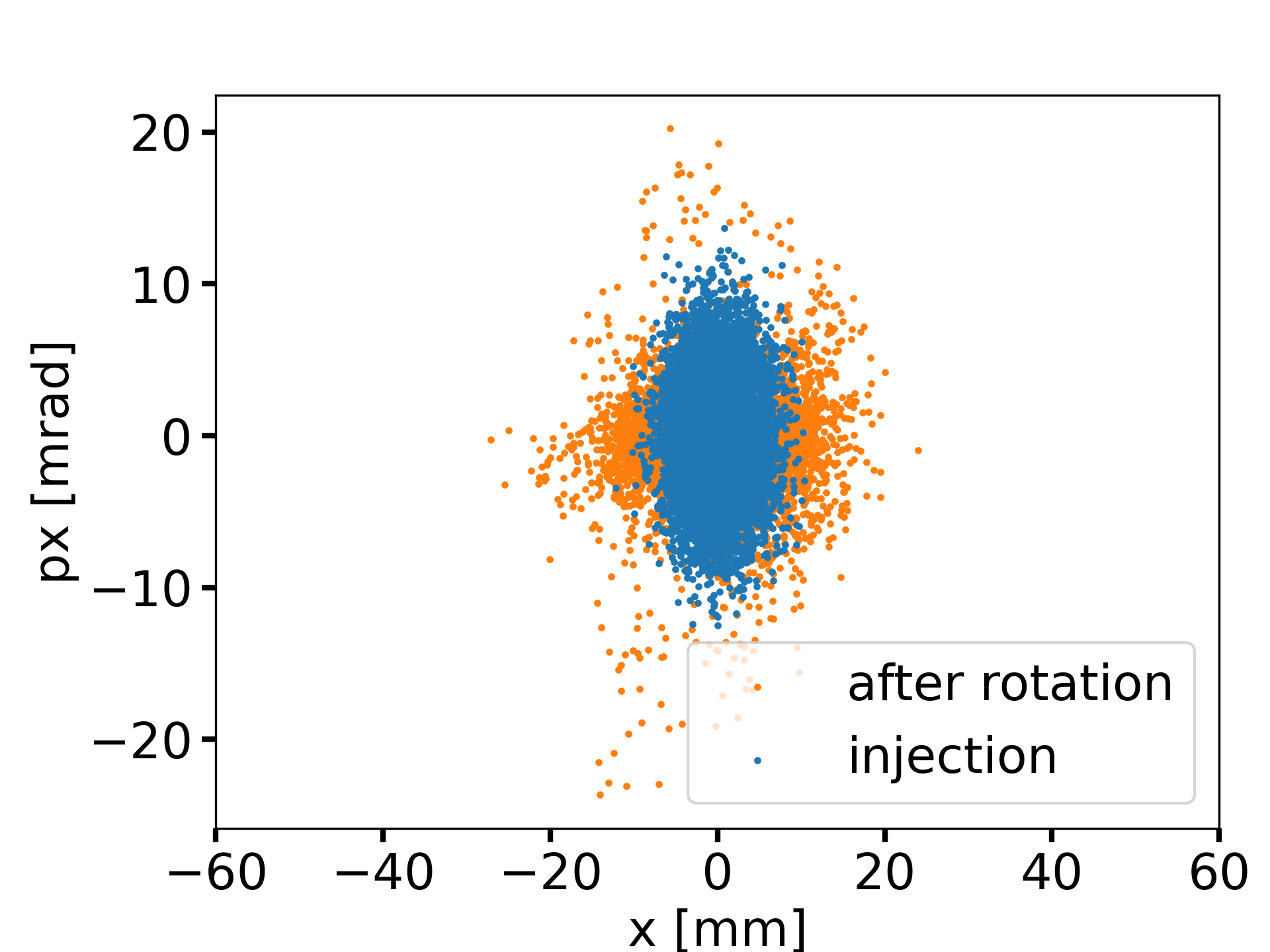}
    \includegraphics[width=0.32\linewidth]{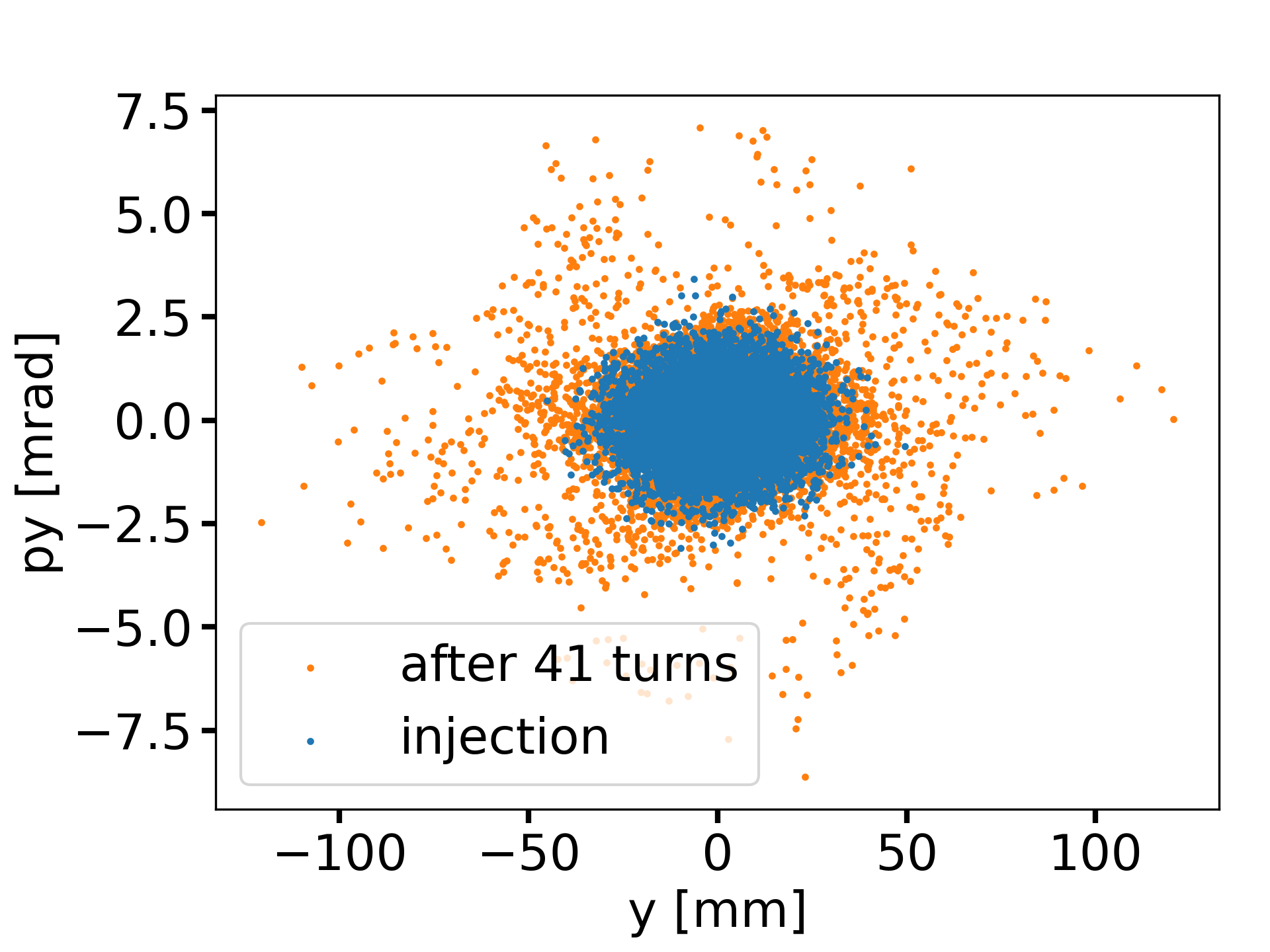}
    \includegraphics[width=0.32\linewidth]{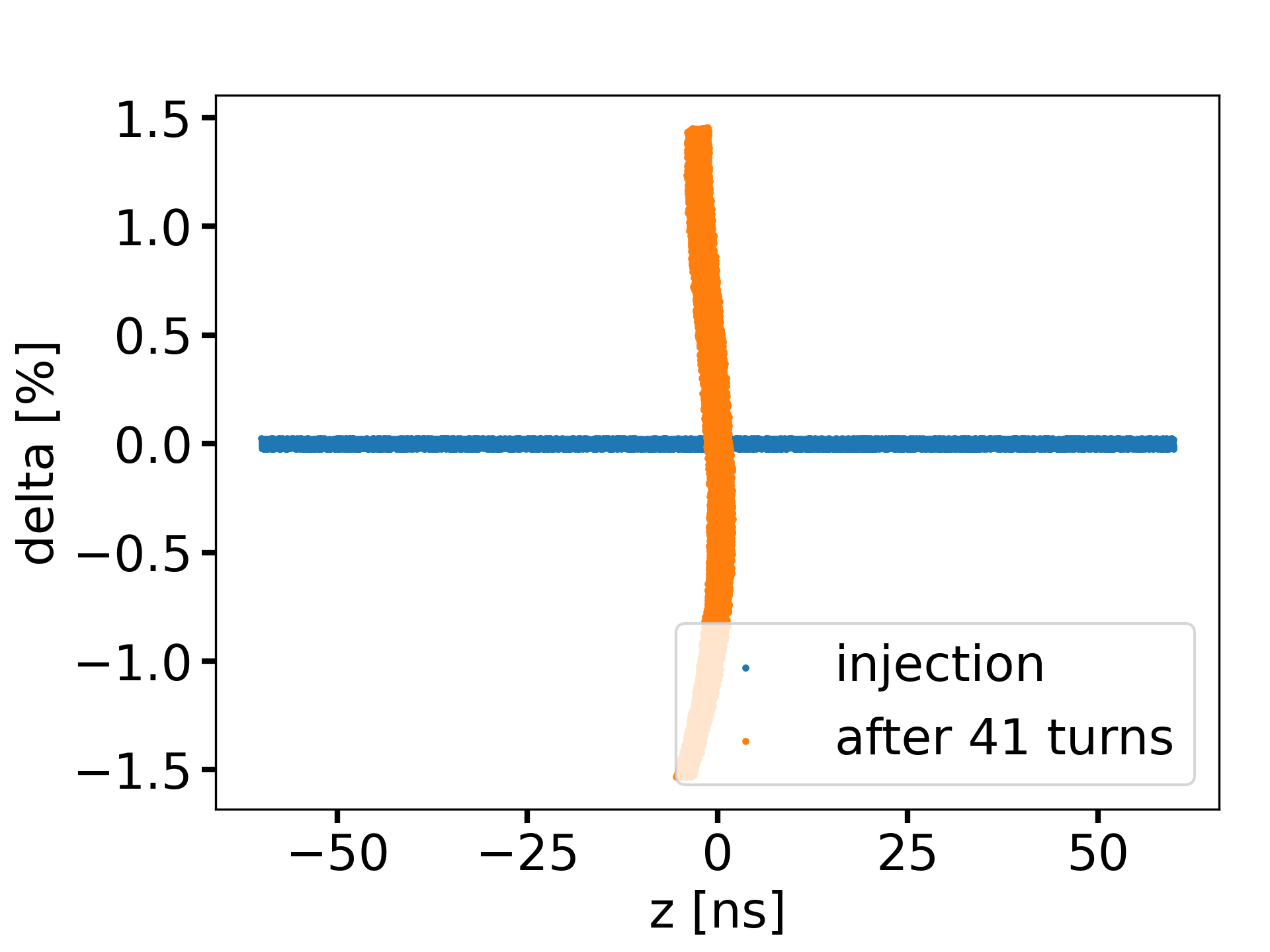}

    \includegraphics[width=0.32\linewidth]{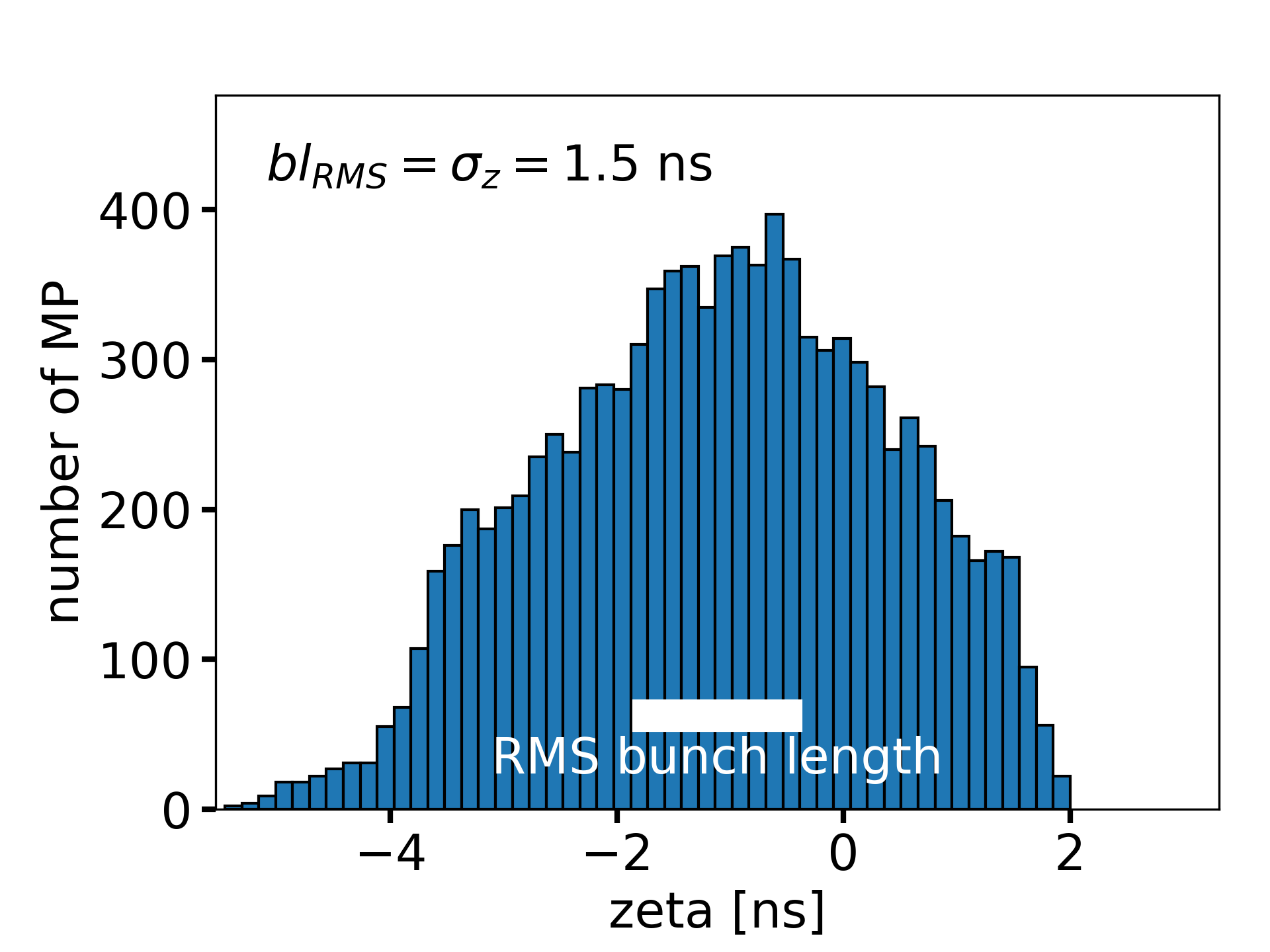}
    \includegraphics[width=0.32\linewidth]{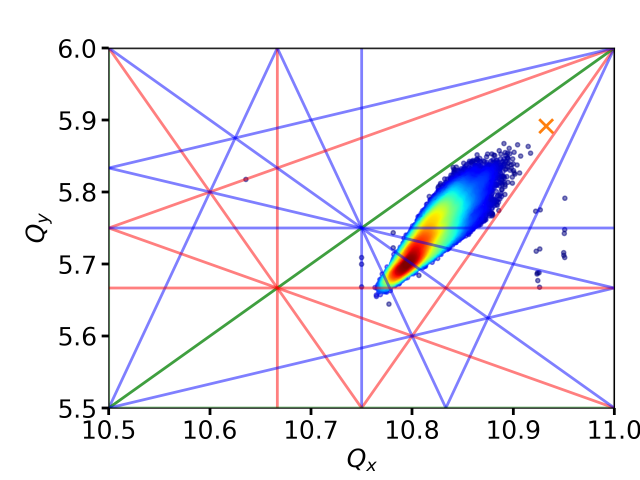}
    \includegraphics[width=0.32\linewidth]{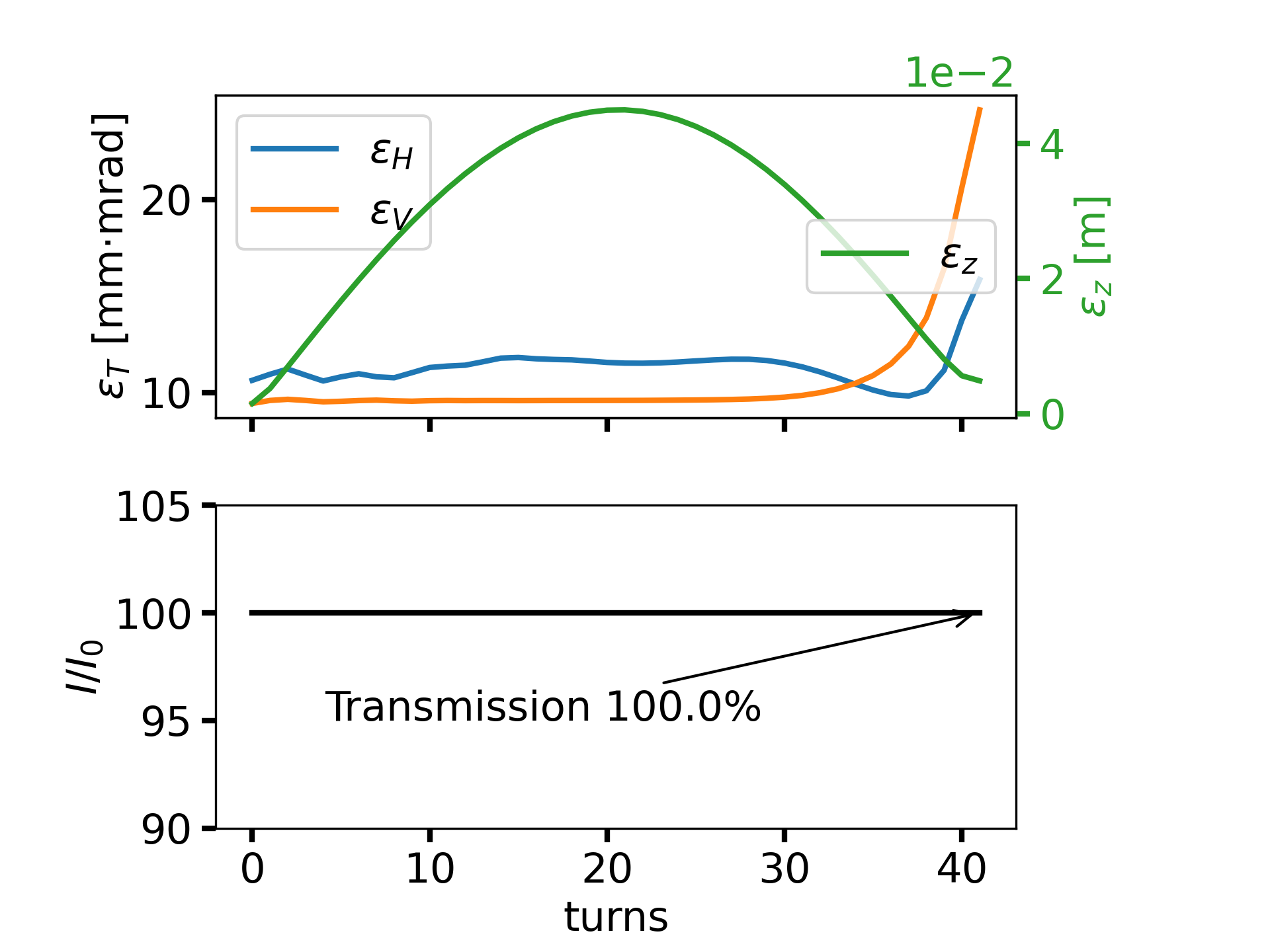}
    \caption{Simulation of the full compression for one bunch at 5\,GeV. Since this requires a 2 bunch solution scheme, this bunch has half of the full intensity shown in Table~\ref{2:acc:proton:tab:parameters}. Notice that at the end of the compression there is still some emittance blow up that need still to be addressed.}
    \label{2:acc:proton:fig:Com_5GeV}
\end{figure}

The Compressor accepts bunches from the Accumulator and performs a 90 \textdegree-bunch rotation in longitudinal phase space. A detailed study of the rotation for both power options was launched and optimization of the bunch length in order to achieve the 2 ns final bunch length, assuming an initial rms momentum spread of $2.5 \times 10^{-4}$, was established.  For the 5 GeV, a lattice (Figure~\ref{2:acc:proton:fig:Com5_latt_twiss}) a  space charge driven lattice resonance still present at the end of rotation, when the tune shift reaches its maximum, causing emittance blow. Figure~\ref{2:acc:proton:fig:Com_5GeV} shows the initial and final 6D phase space of the bunches before and after compression for the 5 GeV case, further study and lattice work is needed in this case still. For the 10 GeV case the rotation of the bunches is clean and no issues were encountered to reach the 2 ns bunch length so far.

A first look at the Target Delivery Transport Line of a 2\,ns high-current bunch has started. The bunches from the Compressor for both energies (and powers) could be transported through a FODO lattice and focused to the specified beam sizes on the target surface without loss of quality. Extra work on the extraction from the Compressor and a design and study of the bunch recombination in the Target Delivery Transport Line is still to be done.

For the remaining years of the study our efforts will be very concentrated on the Compressor lattice design refinement and Target Delivery Transport Line parameters for both power options listed in Table~\ref{2:acc:proton:tab:parameters}. Other planned work regarding the proton complex includes exploring the possibility of using a RCS to bring the linac beam to 5 or 10 GeV, instead a  full energy linac and also a increase the repetition rate in the linac. The latter implies longer accumulation times that will need to be studied. Study on H$^{-}$ injection into the Accumulator ring and alternative chopping schemes to relax the requirements on the source and front-end, should be carried out in more detail. In parallel, an initial investigation of the main instabilities in the Accumulator and Compressor rings will be initiated~\ref{1:acc:sec:collective}. On the experimental side, we are in close contact with SNS and preparing to attempt to perform a 90$^o$ bunch rotation in their accumulator ring. Although their energy is much lower that what is envisioned for the proton driver the beam perveance, given by $Q \propto \lambda/\beta^2\gamma^3$ where $\lambda$ is the line density, is similar the one expected for the 10 GeV option (as presented in~\cite{SNS-experiments}), this means that for similar beam parameters (transverse emittances) the tunes shifts should be of the same order. This results can be used to benchmark our current codes used for simulations. Similar experiments could also be performed at the CERN PS ring in order collect as much data as possible for benchmarking and validation.

\FloatBarrier
\section{Muon production \& cooling}
\label{1:acc:sec:cool}

\begin{figure}[htb]
    \centering
    \includegraphics[trim={0 5cm 0 4cm 0},clip,width=\textwidth]{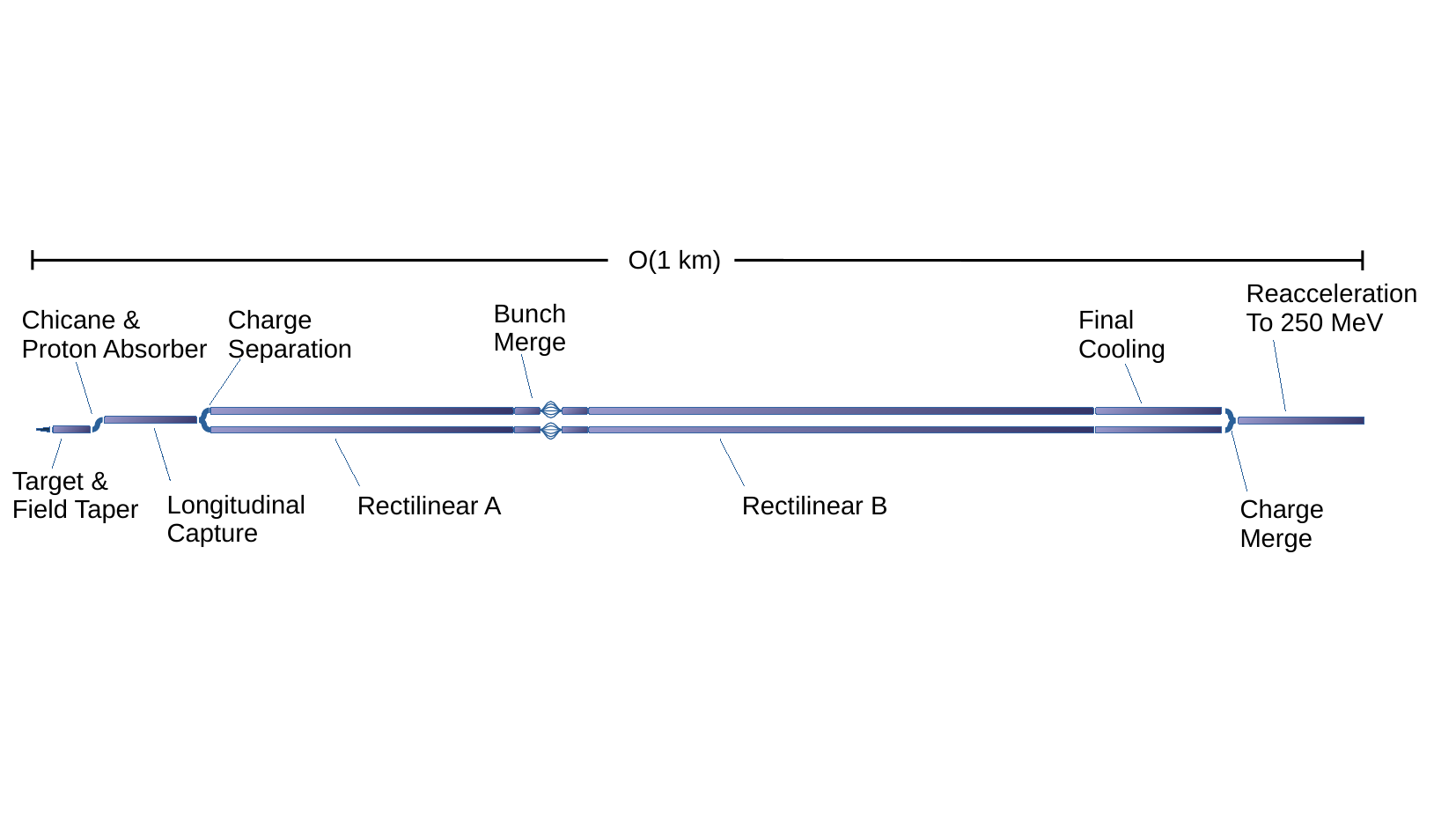}
    \caption{Schematic of the muon production system.}
    \label{1:acc:fig:muon_production}
\end{figure}

The muon production and cooling system is comprised of several subsystems, as shown in Figure~\ref{1:acc:fig:muon_production}. The protons provided by the proton complex (section \ref{1:acc:sec:proton}) intersect a target (Section~\ref{sec:Target}) to produce pions. The target is immersed in a 20\,T solenoid field, yielding a high pion flux. The field is rapidly tapered to 1.5\,T. Pions and muons traverse a solenoid-focused chicane where high momentum beam impurities are removed followed by a Beryllium absorber which ranges out remnant low energy protons. The remaining beam passes through a longitudinal drift where any remaining pions decay. The beam is then captured longitudinally. A multi-frequency RF system that captures the beam into 21 bunches is required owing to the initially large longitudinal beam emittance. Another solenoid chicane system splits the beam by charge species, in preparation for the ionisation cooling system.

The ionisation cooling system is comprised of a series of solenoids, which focus the beam onto energy absorbers that reduce the beam's momentum. The momentum is restored longitudinally using RF cavities, and overall there is a reduction in the beam's emittance. Transverse, longitudinal and 6D emittance is defined as the volume in transverse $(x,p_x,y,p_y)$, longitudinal $(t, E)$ and 6D $(x,p_x,y,p_y$, $t, E)$ phase space. An initial rectilinear cooling system, Rectilinear A, reduces the beam 6D emittance sufficiently that the many initial bunches can be merged into one single bunch. The beam is then cooled further to the final emittance, using a continuation of the rectilinear cooling system, Rectilinear B, and then a final transverse cooling with a sequence of high field solenoids operated with a low momentum (non-relativistic) beam. The beam is finally re-accelerated with a pre-accelerator to semi-relativistic speeds so that it can be delivered to the acceleration system.

\begin{figure}
\includegraphics[width=\textwidth]{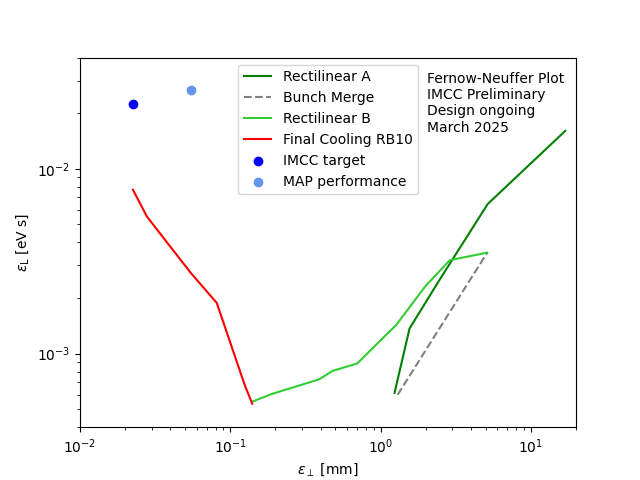}
\caption{Current performance estimate for the muon cooling system. Optimisation is ongoing. Best performance so far yields $\epsilon_\perp$=\SI{22.5}{\micro\meter}, $\epsilon_\parallel$=\SI{7.7}{\electronvolt\milli\second}. \label{fig:fn_plot_dec_2024}}
\end{figure}

The key components in the Muon Production system, as designed by MAP, have been identified and the design has been further optimised by the IMCC team. A significant improvement in performance has been achieved. The overall performance estimate of the ionisation cooling system is summarised in figure \ref{fig:fn_plot_dec_2024}.
This plot shows the beam initial emittance on the right and the reduction in both emittances via the 6D rectilinear cooling (green). Further reduction in transverse emittance comes at the cost of longitudinal emittance via final cooling (red).

\begin{table}[!ht]
    \centering
    \begin{tabular}{l|cccccc}
         Num. bunches&  Sim. $\varepsilon_{\rm{T}}$&  Target $\varepsilon_{\rm{T}}$&  Sim. $\varepsilon_{\rm{L}}$&  Target $\varepsilon_{\rm{L}}$&  Mean $p_z$& N($\mu^-$)\\
 Unit& um& um& mm (eVs)& mm (eVs)& MeV/c& 10$^{12}$\\ \hline
     Front End               &      -& 17000&            -&           46&   288&  45.0 \\
     End of charge separation&      -& 17000&            -&           46&   288&  42.8 \\
            End of 6D Cooling&    140&   140&         1.56&          1.56&  200&  5.3 \\
         End of Final Cooling&   22.5&  22.5&         21.9&            64& 26.6&  3.2 \\
        End of Reacceleration&       -&  22.5&            -&   64 (0.0225)& 339&  2.8 \\ 
    \end{tabular}
    \caption[Cooling system emittance overview]{Target beam parameters through the muon production system. Target performances are based on MAP studies. Where IMCC has performed new simulations, the simulated performance is listed. The number of $\mathrm{\mu}^-$ traversing the system is listed; about 30 \% more $\mathrm{\mu}^+$ are produced at the target than $\mathrm{\mu}^-$}.
    \label{cool:tab:summary}
\end{table}

\subsection*{Key challenges}
The muon production and cooling system has many key challenges, 10 of which have been identified here in bold.\\
Primarily, many subsystems rely on \textbf{novel arrangement} of accelerator components. Solenoid focusing is used throughout most of the system, while particle accelerators in general use quadrupoles and combined function dipoles. The high emittance beams present throughout much of the muon production system require continuous focusing both longitudinally and transversely so equipment is very closely integrated to maintain satisfactory transmission. The cooling system in particular uses beam intersecting devices extensively and, while beam power is relatively low, in some regions energy deposition is of concern.

Regarding the \textbf{target}, the combined handling of target, solenoid and beam is challenging. Pion yield is improved by maintaining the strongest possible solenoid field, so high temperature superconductor solenoids have been chosen as the capture and focusing technology. Protection of this solenoid from the high radiation environment requires significant shielding, which can only be implemented with a suitable thickness of material, and consequently high solenoid bore. The overall design is broadly comparable to those considered for fusion facilities. 

The target itself may be damaged by the \textbf{incident proton beam}. Great care has been taken to ensure that thermal heating, instantaneous shock and integrated dose are within acceptable limits. A significant fraction of the proton beam passes through the target and must be extracted from the solenoid bore; ongoing effort is being invested to develop a solenoid lattice that maintains good beam transport properties while presenting an aperture through which the used proton beam may be extracted. 

Significant \textbf{unwanted radiation} exists within the beam and this may be cleaned by means of a solenoid chicane system followed by a proton-absorbing window. The chicane is planned to employ purely solenoid focusing. This results in a charge-symmetric vertical dispersion, whereby high momentum particles are absorbed on collimators and low momentum particles stay on the beam axis. A relatively clean removal of high momentum particles is possible using this technique while the low momentum portion of the beam ($p$ < 500~MeV/c) is unperturbed. Low momentum protons, arising from spallation of the target, would deliver significant radiation dose in the early parts of the system preventing maintenance. These are removed by means of a Beryllium plug, which stops protons but not muons and pions.

The \textbf{longitudinal capture} system relies on the use of several families of RF cavities, each family having different frequency. The RF cavity frequency is chosen such that RF cavities are synchronous with the several muon bunches passing through the capture system, even when those bunches have different central momenta. As the low momentum bunches fall further behind the high momentum bunches, lower frequency cavities may be employed to maintain synchronicity. Controlling the relative phasing and frequency for such a system may be challenging.

\textbf{Charge separation} is achieved in an additional solenoid chicane similar to the one planned for the beam cleaning system. In the region of highest dispersion, the solenoid will be split so that muons of differing signs travel in different directions. While tracking simulations of this system have been successfully performed, there is no engineering design for such a split solenoid. The charge separation system is planned to be implemented after longitudinal capture, so any system would require RF to be integrated into the system to maintain the bunch structure. A helical dipole has also been considered for this system, but similar issues are unsolved.

The \textbf{cooling system} has received the most attention from IMCC. The rectilinear cooling system employs a compact assembly of absorbers, RF cavities and magnets. The magnets are closely packed and strongly bucking, so that large forces are exerted between adjacent solenoids. Warm RF cavities are placed between the magnets. Implementation of such a system presents a number of challenges, which the IMCC plan to address with the design and construction of a test cooling cell as described in Section \ref{1:tech:sec:cool_cell}. The technological challenges are described in that section. While the integrated optimisation has not yet been performed, preliminary study indicates that the performance optimum may be reached by driving the rectilinear cooling channel to the technological limit and this has been the approach taken by the IMCC. Additional beam physics challenges remain; collective effects have not been fully estimated in these cooling channels although preliminary analyses indicate that attention must be paid to the beam loading and space charge simulation. Even in the rectilinear system heating of liquid hydrogen absorbers is of concern.

The \textbf{final cooling} system employs very high field solenoids enclosing absorbers. Fields as high as 40~T have been envisaged; the IMCC expects that such magnets may be available on the time scale required by IMCC although currently no suitable magnets exist. Delivery of such magnets would be an outstanding achievement for the particle physics community with impact in many other areas of science.
It has been shown that even higher field magnets, up to 60 T, could be built using existing technology and within a reasonable time scale \cite{pugnat:2020}. Fields of this level would improve performance beyond that presented here.
 
The final cooling magnets would require integration with \textbf{absorbers}. The focusing power of the high field magnets is further enhanced by operating the system with a low energy muon beam. Owing to the low energy and emittance of the beam, significant heat load is deposited in a small volume within the absorber. This has the potential to cause large pressure excursions leading to damage on the absorber windows such as cavitation. For this reason hydrogen vapour absorbers are foreseen. Composite windows employing low-Z materials, such as lithium hydride, sandwiched between high-Z materials may provide additional resilience while controlling the amount of scattering suffered by the muons. Studies on the hydrogen absorber are described in Section \ref{1:tech:sec:vac}. 

Significant longitudinal emittance growth occurs at the low energy which must be controlled by requiring a small energy spread in the beam. This results in very long beams and so relatively low frequency RF is required both to reaccelerate the beam and also to perform gymnastics to prevent the momentum spread becoming too large. While the RF is not expected to be challenging, detailed design work is required to understand the likely available RF voltages at the required frequencies.

After the final cooling, the beam must be \textbf{reaccelerated} back to relativistic speeds. For now no design exists for this reacceleration. Acceleration using standard proton accelerator equipment such as radiofrequency quadrupoles and drift tube linacs is possible. Alternately high gradients may be achieved using induction linacs such as those used in high current electron applications. Muons become relativistic more easily than protons owing to their lower mass, so only a relatively short linac is required. Typical available gradients have been used to infer an estimate for the decay losses in such a system.

\subsection*{Recent Achievements - Target \& Front End}

\begin{figure}
    \centering
    \includegraphics[width=1.0\linewidth]{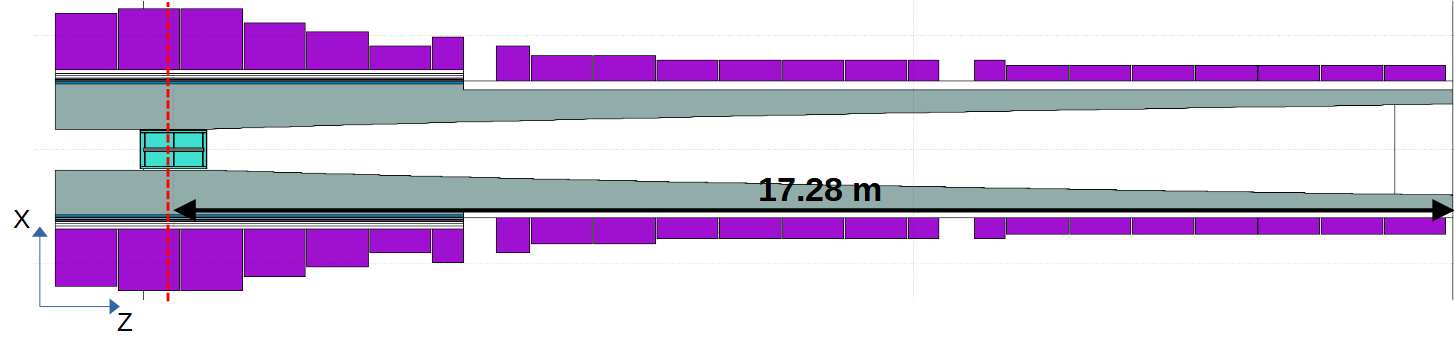}    
    \includegraphics[width=0.5\linewidth]{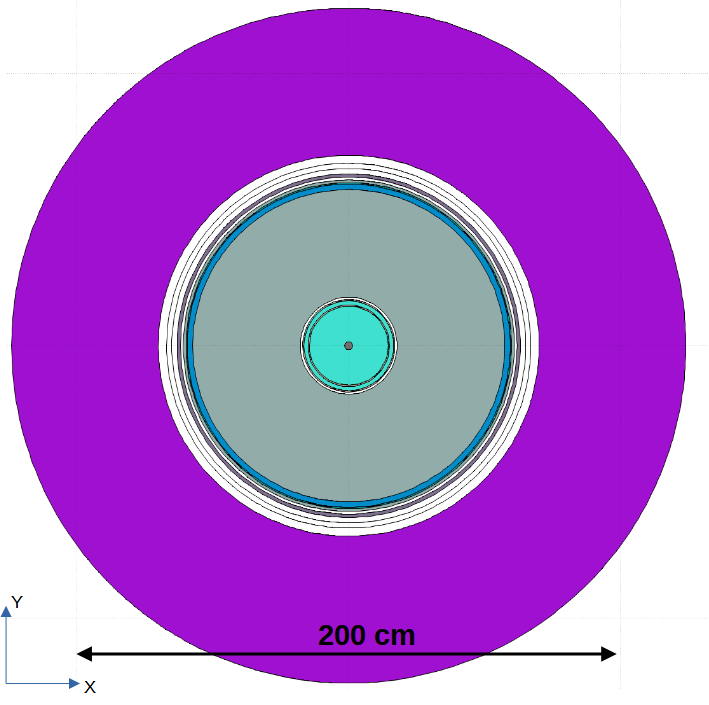}

    \caption{Geometrical model of the target area implemented for FLUKA simulations depicted in XZ plane (top) and XY plane (bottom). The red dashed line indicates the Z position of the XY cross section picture.}
    \label{fig:target_model_fluka}
\end{figure}

The target team has designed a baseline radial build for the pion production target, taking into account radiation flux into the surrounding solenoid, heat load on the graphite target itself and appropriate cooling systems for the target and shielding (detailed in Sections~\ref{1:tech:sec:tar} and \ref{1:tech:sec:rad_shield}). The radial build is explained in Table~\ref{target:tab:radialbuild}.
\begin{table}[!ht]
    \centering
    \begin{tabular}{l|cccc}
     Component& Material& $r_i$ [mm]& $r_e$ [mm]& ${\Delta}r$ [mm]\\ \hline
     Solenoid coils& HTS& 700& -& -\\
     Insulation& Insulation& 690& 700& 10
\\
     Vacuum& Vacuum& 670& 690& 20
\\
     Thermal shield& Copper \& Water& 651& 670& 19
\\
 Vacuum& Vacuum& 631& 651& 20
\\
 Inner supporting tube& Stainless-steel& 619& 631& 12
\\
 Vacuum& Vacuum& 609& 619& 10
\\
 Outer Target shielding& Tungsten& 599& 609& 10
\\
 Neutron absorber& Boron Carbide& 594& 599& 5
\\
 Target shielding and neutron moderator& Stainless-steel& 589& 594& 5
\\
 & Water& 569& 589& 20
\\
 & Stainless-steel& 564& 569& 5
\\
 & Tungsten& 179& 564& 385
\\
 & Stainless-steel& 174& 179& 5
\\
 Vacuum& Vacuum& 173& 174& 1
\\
 Target vessel& Titanium& 168& 173& 5
\\
 & Helium& 155& 168& 13
\\
 & Titanium& 150& 155& 5
\\
 & Helium& 15& 150& 135
\\
 Target& Graphite& 0& 15&15
\\
    \end{tabular}
    \caption{Target System radial build for a graphite target.}
    \label{target:tab:radialbuild}
\end{table}
A 3D FLUKA model for the target area has been constructed and the pion and muon yield through the target system has been assessed. The FLUKA geometry implementation of the target area is shown in Fig~\ref{fig:target_model_fluka}. The model is simplified with respect to Table~\ref{target:tab:radialbuild}. The yield results from simulations are presented in Table~\ref{target:tab:yield}, assuming the standard deviation of the transverse beam position is 5~mm, the target rod has a radius of 15~mm and that all the muon and pions going in the chicane can be captured if their momentum is below \SI[per-mode=symbol]{500}{\mega\electronvolt\per\clight}. The magnet team have identified HTS technology as having greater tolerance to radiation issues (section \ref{3:det:sec:magnet}). The target team have updated the design with reduced shielding and including structural elements (section \ref{sec:Target}).
\begin{table}[!ht]
    \centering
    \begin{tabular}{l|ccccclll}
     \multicolumn{1}{c|}{ }& \multicolumn{8}{c}{Proton beam energy [GeV]} \\ 
     Yield [$10^{-2} $/~\si{\giga\electronvolt}$/p^+$]& 3& 4& 5& 6&  7& 8& 9&10\\ \hline
     $\mu^+$ & 2.8& 2.6& 2.4& 2.3&  2.2& 2.1& 1.9&1.9\\
     $\mu^-$ & 1.8& 1.8& 1.8& 1.8&  1.7& 1.7& 1.7&1.7\\
     $\pi^+$ & 1.3& 1.2& 1.1& 1.1&  1& 0.98& 0.92&0.9\\
     $\pi^-$ & 0.84& 0.81& 0.84& 0.82&  0.83& 0.8& 0.8&0.81\\
    \end{tabular}
    \caption{Yield per unit energy proton beam [$10^{-2}$/~\si{\giga\electronvolt}$/p^+$]}
    \label{target:tab:yield}
\end{table}

Preliminary concepts for the spent proton beam handling have been studied. It is challenging to extract the beam in the high field region and taper owing to the need to maintain a smooth field in this region and so extraction is foreseen in the chicane where fields are relatively lower. The pion and muon beam is deflected here by the solenoid field. The proton beam is much more rigid and so is relatively undeflected. A transition from a high radius solenoid to a low radius solenoid is envisioned to allow space for the protons to leave the beam pipe. It was found that the spent protons are not focused enough to avoid extreme radiation load to the chicane structure. At the same time, increasing the aperture size is challenging in practice.
Studies continue to explore the relative size of the solenoids, magnetic field shape and surrounding shielding to extract the beam at this point. Potential alternative solutions to be investigated are shortening the tapering region and changing the target material.

Following the chicane and proton absorber, the beam is contained transversely by solenoid fields and drifts longitudinally. Slower particles arrive later at the downstream RF cavities. Weakly excited cavities provide a small amount of bunching, followed by gradually more strongly excited cavities, in order to capture the beam adiabatically into RF buckets. Because the beam is not captured, the RF frequency must be lower at downstream cavities to accommodate the late arrival of slower particles. Following capture the lower energy, late particles are accelerated. This is achieved by adjusting cavity frequency so that the late particles undergo an accelerating phase while the early particles undergo a bunching phase. Preliminary work has started to examine the dependence of RF frequency and transverse aperture in this system in order to provide an integrated optimisation with the early stages of the rectilinear cooling system. 

\subsection*{Recent Achievements - Rectilinear Cooling}
\label{1:acc:sec:cool:rec_cooling}
The design of the rectilinear cooling lattice, based on the MAP design \cite{stratakis2015rectilinear}, has recently been updated and optimised by the IMCC team \cite{zhu2024performance}. Composed of two rectilinear cooling channels, one before (Rectilinear A) and one after (Rectilinear B) the bunch merging system, the updated lattice yields an improved performance over the MAP design. While here we briefly describe the lattice design and performance, the detailed study can be found in \cite{zhu2024performance}.

The rectilinear cooling system has a periodic lattice composed of a series of solenoid magnets which have a weak dipole field superimposed to enable emittance exchange. While the previous design used tilted solenoids to generate the dipole field, the updated IMCC design employs separate dipole magnets. This design choice facilitates the independent tuning of the dipole field although more space may be required. Each periodic unit, referred to as a cooling cell, consists of solenoids with opposite polarity for tight focusing at the absorber, dipole magnets for dispersion generation, RF cavities for beam energy loss compensation and liquid hydrogen (LH$_2$) wedge absorbers. 

\begin{figure}[htb]
    \centering
    \includegraphics[width=0.85\textwidth]{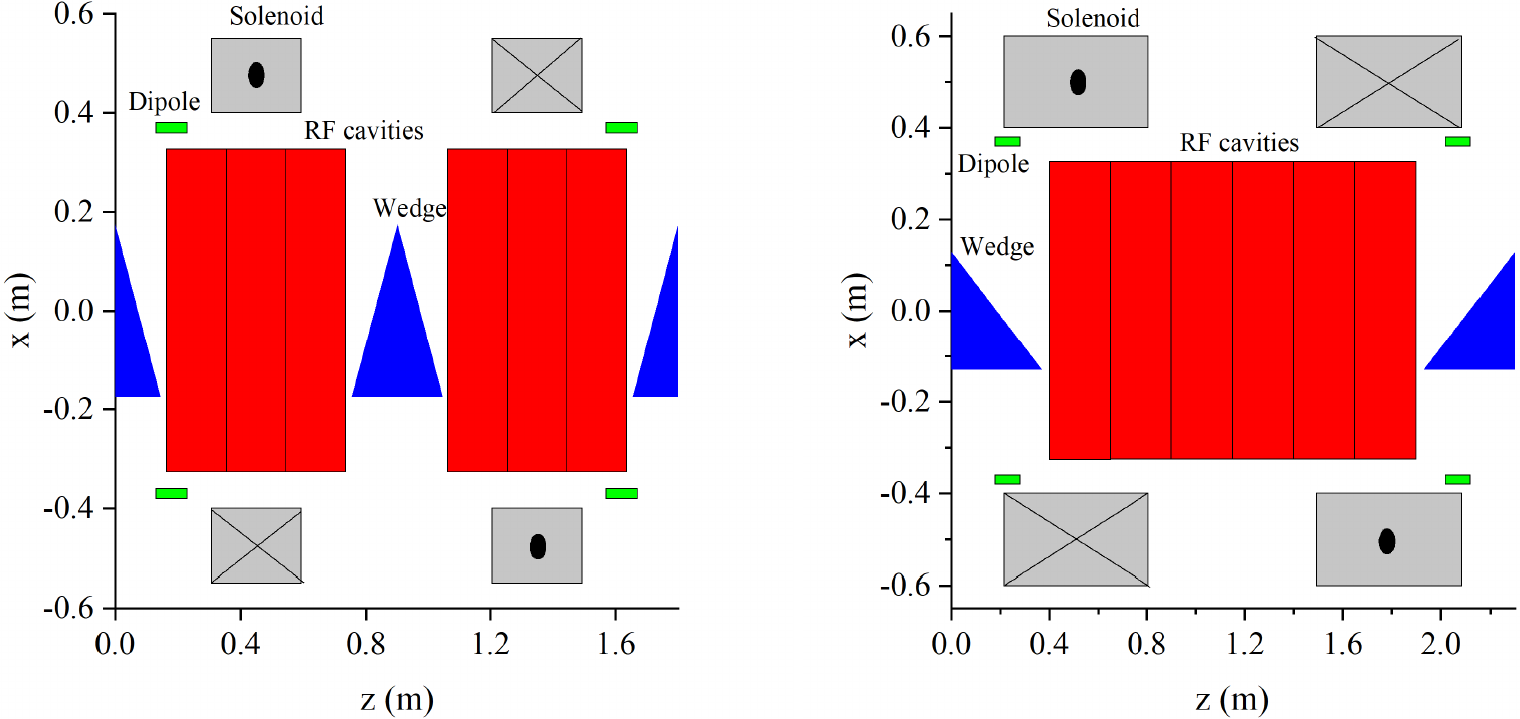}
    \caption{Schematic of the (left) A-type and (right) B-type cooling cell layout.}
    \label{1:acc:fig:rectilinear_cooling_design}
\end{figure}

Two types of cooling cells are used in the present design: A-type and B-type cells, used for the pre-merge and post-merge sections, respectively. The conceptual layout of the two types of cooling cells is shown in Figure~\ref{1:acc:fig:rectilinear_cooling_design}. The primary difference between the two types of cells lies in the cell tune (number of betatron oscillations per cell). A-type cells operate with a cell tune below 1, which allows the lattice to accept a beam with a higher beam emittance, suitable for initial cooling. The A-type cell has a further advantage that the transverse betatron function at the center of the cell is equal to that at the start and end of the cell, which enables the use of an additional wedge absorber in the middle of the cell. B-type cells operate with a cell tune between 1 and 2, in a tighter focusing regime that enables cooling to lower emittances. As a result, these cells do not accept high-emittance beams and have a smaller momentum acceptance.

The RF cavities (section~\ref{1:tech:sec:rf}) are modelled as perfect cylindrical pillbox cavities with thin Beryllium (Be) windows operating in the TM$_{010}$ mode. The Be windows electromagnetically seal the cavities, ensuring that the TM$_{010}$ mode is an appropriate model of the real field, and have been shown to increase the operational RF gradient in multi-Tesla magnetic fields~\cite{Bowring:2018smm}. Multiple independently phased RF cavities are used in each cell. 

Rectilinear A and Rectilinear B lattices are composed of 4 A-type and 10 B-type stages, respectively. Each stage is a series of identical cooling cells, and the design of the cooling cell becomes more challenging (higher fields, more compact assembly, etc.) with increasing stage number. The hardware parameters for the whole rectilinear cooling system are listed in Tables \ref{cool:tab:6d_cell} and \ref{cool:tab:6d_rf}.

\begin{table}[!h]
    \centering
    \begin{tabular}{l|ccccccccc}
         \hline\hline
         &  Cell&  Stage&  Pipe&  Max. $B_z$ & Int.& $\beta_\perp$ & $D_x$ & On-Axis&Wedge\\
         & Length & Length & Radius & On-Axis & $B_y$ & & & Wedge Len.& Angle \\
         &  (m)&  (m)&  (cm)&  (T)& (Tm)& (cm)& (mm)& (cm)&(deg)\\ \hline
         A-Stage 1&  1.8&  104.4&    28&   2.5&  0.102&   70&   -60& 14.5&  45\\
         A-Stage 2&   1.2&  106.8&    16&   3.7&  0.147&   45&   -57& 10.5&  60\\
         A-Stage 3&   0.8&  64.8&     10&   5.7&  0.154&   30&   -40&   15& 100\\
         A-Stage 4&   0.7&  86.8&      8&   7.2&  0.186&   23&   -30&  6.5&  70\\ \hline
         B-Stage 1&   2.3&  50.6&     23&   3.1&  0.106&   35& -51.8&   37& 110\\
         B-Stage 2&   1.8&  66.6&     19&   3.9&  0.138&   30& -52.4&   28& 120\\
         B-Stage 3&   1.4&  84.0&   12.5&   5.1&  0.144&   20& -40.6&   24& 115\\
         B-Stage 4&   1.2&  66.0&    9.5&   6.6&  0.163&   15& -35.1&   20& 110\\
         B-Stage 5&   0.8&  44.0&      6&   9.1&  0.116&   10& -17.7& 12.5& 120\\
         B-Stage 6&   0.7&  38.5&    4.5&  11.5&  0.087&    6& -10.6&   11& 130\\
         B-Stage 7&   0.7&  28.0&   3.75&    13&  0.088&    5&  -9.8&   10& 130\\
         B-Stage 8&  0.65& 46.15&   2.85&  15.8&  0.073&  3.8&    -7&    7& 140\\
         B-Stage 9&  0.65&  33.8&    2.3&  16.6&  0.069&    3&  -6.1&  7.5& 140\\
         B-Stage 10& 0.63& 29.61&    2.0&  17.2&  0.069&  2.7&  -5.7&  6.8& 140\\ \hline\hline
    \end{tabular}
    \caption{Main parameters for the rectilinear cooling cell components, including the cell geometry, solenoid fields, dipole fields, beam optics and wedge absorber geometry. Liquid hydrogen is used as the wedge absorber material for all stages.}
    \label{cool:tab:6d_cell}
\end{table}

\begin{table}[!h]
    \centering
    \begin{tabular}{l|ccccc}
         \hline\hline
         &  RF Frequency&  Num. RF&  RF Length&  Max. RF Gradient& RF phase\\
         &  (MHz)&  &  (cm)&  (MV/m)& (deg)\\ \hline
         A-Stage 1 &  352&  6&  19&   27.4& 18.5\\
         A-Stage 2&   352&  4&  19&   26.4& 23.2\\
         A-Stage 3&   704&  5&  9.5&  31.5& 23.7\\
         A-Stage 4&   704&  4&  9.5&  31.7& 25.7\\ \hline
         B-Stage 1&   352&  6&  25&   21.2& 29.9\\
         B-Stage 2&   352&  5&  22&   21.7& 27.2\\
         B-Stage 3&   352&  4&  19&   24.9& 29.8\\
         B-Stage 4&   352&  3&  22&   24.3& 31.3\\
         B-Stage 5&   704&  5&  9.5&  22.5& 24.3\\
         B-Stage 6&   704&  4&  9.5&  28.2& 22.1\\
         B-Stage 7&   704&  4&  9.5&  28.5& 18.4\\
         B-Stage 8&   704&  4&  9.5&  27.1& 14.5\\
         B-Stage 9&   704&  4&  9.5&  29.7& 11.9\\
         B-Stage 10&  704&  4&  9.5&  24.9& 12.2\\ 
         \hline\hline
    \end{tabular}
    \caption{Rectilinear cooling cell RF parameters.}
    \label{cool:tab:6d_rf}
\end{table}

The performance of the rectilinear cooling system is shown in Table \ref{cool:tab:6d_performance}, in terms of the beam emittance and transmission at the end of each stage. The Rectilinear A section reduces the normalised transverse and longitudinal emittances from 16.96\,mm and 45.53\,mm to 1.24\,mm and 1.74\,mm, respectively, with an overall transmission rate of 49.6\% including the muon decays. The Rectilinear B section reduces the normalised transverse and longitudinal emittances from 5.13\,mm and 9.99\,mm to 0.14\,mm and 1.56\,mm, respectively, with an overall transmission rate of 28.5\% including the muon decays. The final transverse emittance achieved with this design is a factor of two smaller than in the previous design \cite{stratakis2015rectilinear}. This improvement is expected to significantly benefit the design of the final cooling section, which aims to achieve a normalised transverse emittance of 22.5\,$\mu$m. 

The scenario of using coupled `$\pi$-mode' RF cavities has also been investigated. In this case, adjacent cavities operate with a $\pi$ phase difference, and only one power coupler and feed through is required for a group of coupled RF cavities. The impact of using $\pi$-mode RF was studied on a cooling lattice comprised of 'B-Stage 5'-like cells. Comparisons with an identical cooling channel using independently phased RF reveal no significant differences in cooling performance. More details on this work, including a sensitivity analysis can be found in \cite{zhu2024performance}.

\begin{table}[!h]
    \centering
    \begin{tabular}{l|ccccc}
         \hline\hline
         &  $\varepsilon_{\rm{T}}$ &  $\varepsilon_{\rm{L}}$ &  $\varepsilon_{\rm{6D}}$ & Stage & Cumulative\\
         &  (mm)&  (mm)&  (mm$^3$)& Transmission (\%) & Transmission (\%)\\ \hline
         Start&      16.96&  45.53&   13500& &  100\\ \hline
         A-Stage 1 &  5.17&  18.31&  492.60& 75.2 & 75.2\\
         A-Stage 2&   2.47&   7.11&   44.03& 84.4 & 63.5\\
         A-Stage 3&   1.56&   3.88&    9.59& 85.6 & 54.3\\
         A-Stage 4&   1.24&   1.74&    2.86& 91.3 & 49.6\\ \hline
         Bunch merge& 5.13&   9.99&   262.5& 78.0 & 38.7\\ \hline
         B-Stage 1&   2.89&   9.09&   76.07& 85.2 & 33.0\\
         B-Stage 2&   1.99&   6.58&   26.68& 89.4 & 29.4\\
         B-Stage 3&   1.27&   4.05&    6.73& 87.5 & 25.8\\
         B-Stage 4&   0.93&   3.16&    2.83& 89.8 & 23.2\\
         B-Stage 5&   0.70&   2.51&    1.32& 89.4 & 20.7\\
         B-Stage 6&   0.48&   2.29&    0.55& 88.4 & 18.2\\
         B-Stage 7&   0.39&   2.06&    0.31& 92.8 & 17.0 \\ 
         B-Stage 8&   0.26&   1.86&    0.13& 87.9 & 14.9\\ \hline
         B-Stage 9&   0.19&   1.72&    0.06& 85.2 & 12.7\\
         B-Stage 10&  0.14&   1.56&    0.03& 87.1 & 11.1\\ 
         \hline\hline
    \end{tabular}
    \caption{Simulated performance of the rectilinear cooling lattice in terms of emittance reduction (transverse, longitudinal and 6D) and transmission, both per stage and cumulative. The MAP target for the rectilinear cooling was $\varepsilon_{\perp}=$\SI{0.3}{\milli\meter} and $\varepsilon_{\rm{L}}=$\SI{1.5}{\milli\meter}}
    \label{cool:tab:6d_performance}
\end{table}

After undergoing sufficient cooling in the Rectilinear A section, the initial train of bunches are merged into a single bunch by the bunch merging system. The most up-to-date design of the bunch merging scheme has been carried out by MAP and is described in detail in \cite{bao2016conceptual}. In this scheme, 21 bunches are merged in the longitudinal phase space into seven bunches, which then follow seven paths of different lengths and reach a collecting funnel at the same time to be merged transversely.

\subsection*{Recent Achievements - Final Cooling}
\label{1:acc:sec:cool:final}
Final cooling is necessary to provide further reduction of the transverse emittance, while allowing reasonable increases in the longitudinal emittance.
The design and performance of the final cooling is important for the muon collider as it defines the minimum emittance for all downstream systems, including acceleration and collider, which has a direct impact on the available luminosity.
In addition, the final cooling is a hot spot for losses, both from decays - as it has the lowest muon energy across the whole complex - and from losses within the absorber.
This means a lattice which maximises final cooling transmission has a significant impact on the downstream intensity of the beam, and could even relax beam power requirements upstream, at the target.

The final cooling design is similar to that of the rectilinear cooling, but without the use of dipoles and wedge absorbers.
In addition, the lattice designs are shorter, with approximately 10 cells over \SI{100}{\meter} of beamline. This reduces the periodicity condition of the system compared to the rectilinear cooling.
To maximize focusing of the beam throughout the absorbers, large solenoid fields are used. 

\begin{figure}[!h]
    \centering
    \includegraphics[width=0.99\linewidth]{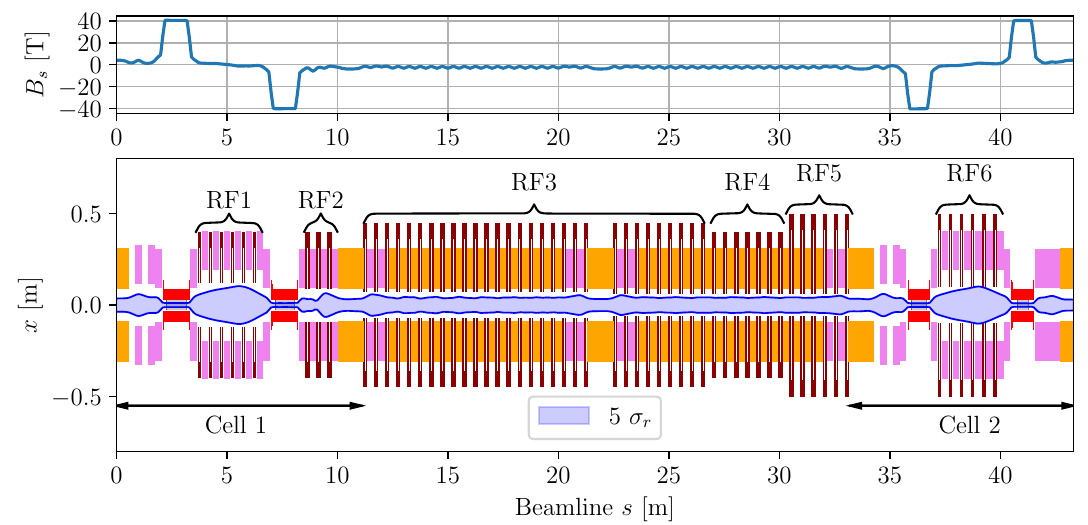}
    \caption{Two-cells of final cooling from Rectilinear Stage B8, including beam envelope, RF cavities and high- and low-field solenoids with a field-flip~\cite{Stechauner:2024ana}}
    \label{1:acc:fig:final_cooling_lattice}
\end{figure}

\begin{figure}[!h]
    \centering
    \includegraphics[width=0.99\linewidth]{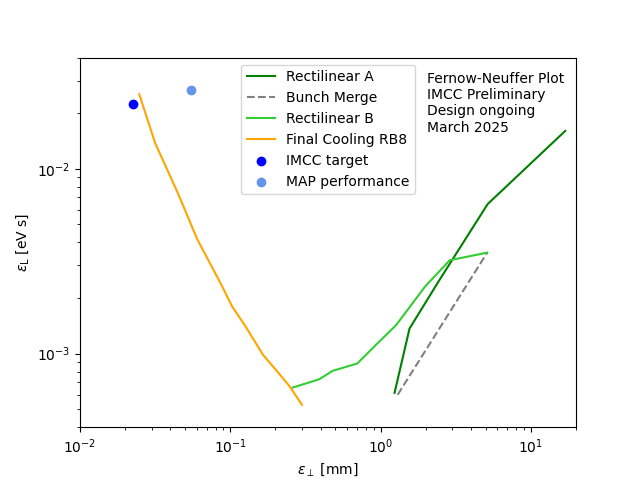}
    \caption{Performance of the muon cooling system assuming that the rectilinear cooling system stops at B-Stage 8.}
    \label{1:acc:fig:fernow_neuffer_rb8}
\end{figure}

The performance of the final cooling lattice is heavily dependent on the initial conditions obtained from the rectilinear cooling channel. Two initial conditions have been considered. In FC-RB10, final cooling starts at the end of rectilinear B-Stage 10 and in FC-RB8 cooling starts at approximately the end of Rectilinear B-Stage 8. The longer rectilinear lattice used by FC-RB10 provides additional 6D cooling, so resulting in a shorter final cooling lattice, and therefore reduced longitudinal emittance increase. B-Stage 10 is technically challenging to implement, so FC-RB8 is studied for risk mitigation and to inform later optimisation. In this study a baseline was taken assuming initially $\varepsilon_\mathrm{T}=0.3 \si{mm}$ and $\varepsilon_\mathrm{L}=1.5 \si{mm}$. The performance of FC-RB8 is shown in Figure~\ref{1:acc:fig:fernow_neuffer_rb8}.
FC-RB8 is simulated in RF-Track~\cite{bib:rftrack}, and FC-RB10 is simulated in G4Beamline.
Thorough benchmarks between the two codes have been performed.
The summary table for both designs is shown in Table~\ref{1:acc:tab:final_cooling_overview}. The decision between the two lattice designs will be made based on the relative cost, performance and technical feasibility inherent in each design.
For further details on the absorber lengths and RF cavity parameters, please refer to the 2024 Preliminary Parameter Report~\cite{PreliminaryParameter_MuCol5}.

\begin{table}
    \centering
    \begin{tabular}{c|cccc|cccc} \hline \hline
 & \multicolumn{4}{c}{FC-RB8}& \multicolumn{4}{c}{FC-RB10}\\ \hline 
         Cell& ($\varepsilon_T$)& ($\varepsilon_L$)& ($N_T$)&  ($E_{kin,F}$)& ($\varepsilon_T$)& ($\varepsilon_L$)& ($N_T$)& ($E_{kin,F}$)\\
         Unit&  \si{\micro\meter}&  \si{\milli\meter}&  \si{\percent}&  \si{\mega\electronvolt} &  \si{\micro\meter}&  \si{\milli\meter}&  \si{\percent}& \si{\mega\electronvolt} \\ \hline
         0&   300&  1.5& 100.0&  82.7&  &  &  & \\
         1&   247&  1.9& 99.1&  80.7&  &  &  & \\
         2&   203&  2.3& 98.3&  85.5&  &  &  & \\
         3&   165&  2.8& 97.4&  43.1&  &  &  & \\
         4&   126&  4.0& 96.6&  43.1&  140&   1.52&  100&  36.4\\
         5&   103&  5.1& 95.7&  12.8&  125&   1.92&  99.5& 26.4\\
         6&  87.2&   6.7& 94.7&  9.4&  81.3&  5.34&  91.9&  9.9\\
         7&  60.3&  11.8& 89.1&  7.5&  54.7&  7.75&  78.9&  6.3\\
         8&  45.0&  20.8& 83.0&  5.9&  39.0&  11.0&  70.5&  4.6\\
         9&  31.6&  39.2& 73.5&  12.4&  27.8&  15.7&  65.5&  3.7\\
         10& 24.8&  71.9& 55.3&   8.8&  22.5&  21.9&  60.4&  3.3\\
         \hline\hline
    \end{tabular}
    \caption{Final cooling lattice performance for two initial starting conditions, from Rectilinear Stage B8 and from Rectilinear Stage B10}
    \label{1:acc:tab:final_cooling_overview}
\end{table}

A lattice schematic of the first two cells of FC-RB8 is shown in Figure \ref{1:acc:fig:final_cooling_lattice}, which shows the magnetic fields (blue line), relative to the high-field solenoids (red), low-field solenoids (orange) and matching solenoids (pink). The RF cavities are represented by the brown boxes. The beam envelope is shown within the elements in blue.

\section{Acceleration}
\label{1:acc:sec:acc}
The acceleration of the muon beam is designed to have maximum acceleration gradient, to minimize decays.
For this reason, the initial acceleration is composed of linacs and recirculating linacs to \SI{63}{\giga\electronvolt}, referred to as the low-energy acceleration.
After that is a series of rapidly cycling synchrotrons until the target energy, referred to as the high-energy acceleration.

\subsection*{Low-energy acceleration}
The low energy section includes three sections: a single-pass linac followed by a pair of multi-pass recirculating linacs (RLA). Previously, a dog-bone accelerator scheme was studied using different lattice configurations such as FODO, Triplet, Theoretical Minimum Emittance (TME) and Double Bend Achromat (DBA). Due to the need for strong focusing in the bending plane to minimize chromatic effects, a symmetrical lattice solution, which is obligatory for dog-bone, could not be found.

\begin{figure}[htb]
    \centering
    \includegraphics[width=0.8\textwidth]{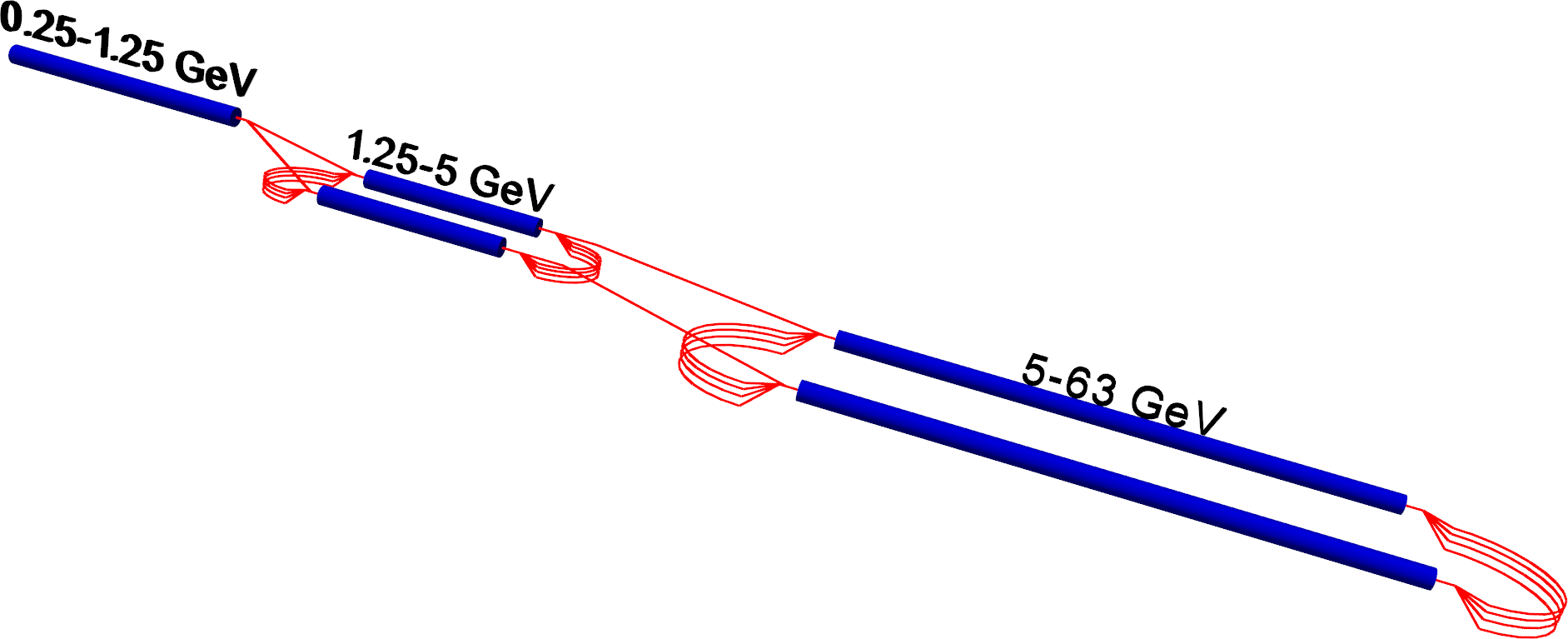}
    \caption{Layout of a two-step-racetrack RLA complex. Pre-accelerator, race-track I and race-track II are stacked vertically; $\mu^{\pm}$ beam is transported between the accelerator sections by the vertical dogleg.}
    \label{1:acc:fig:lowEnSchematic}
\end{figure}

In the presented scenario, acceleration starts after final cooling pre-accelerator at \SI{255}{\MeV} and proceeds to \SI{63}{\GeV}, where the beam is going to be injected into a first rapid cycling synchrotron. A schematic of the low energy section is shown in Figure~\ref{1:acc:fig:lowEnSchematic}.

The linac will use a lower frequency (88 MHz) than the following RLAs to ensure sufficient momentum acceptance for the long beam coming from the final cooling. It will additionally feature a \nth{3} harmonic cavity operating at \SI{264}{\MHz} to minimise uncorrelated energy spread and control the longitudinal emittance.

In subsequent RLAs, superconducting accelerating cavities operating at \SI{352}{\MHz} will be employed. The \SI{1056}{\MHz} cavities are additionally foreseen to linearise the RF waveform to minimise the growth of uncorrelated energy spread in the beam. 
The use of different non-harmonic frequencies in the two acceleration sections requires that the distance between the muon and anti-muon bunches be adjusted with a magnetic array such as a chicane to match the acceleration frequency. 

\paragraph*{Key challenges}

To ensure that the survival rates of muons are sufficient, the acceleration must be done at a high average gradient. 
Since muons are generated as a tertiary beam they occupy a large phase-space volume. In addition to providing a high average gradient, the accelerator must have very large transverse and longitudinal accelerator acceptances; thus, very low-frequency cavities are needed, particularly in the single-pass linac.

In order to accelerate the muon beam within the given transverse and longitudinal emittance tolerances, the beamline must be designed to minimize transverse chromatic effects; thus, tight focusing in the bending plane with weak quadrupoles is required. In addition to the preservation of the longitudinal emittance, the arrival time must be precisely controlled in the arcs to control the acceleration phase. 

Another key challenge is the non-linearity of the RF waveform at the acceleration point. As the particles will be accelerated close to the crest of the waveform, a linearization of the energy gain is necessary to be implemented. 
In the linac and RLAs, this linearization will be provided by 3rd harmonic cavities that have operation frequencies of \SI{264}{\MHz} and  \SI{1056}{\MHz}, respectively. Without it, the energy spread would be too large for the acceptance of the arcs. 

To accelerate $\mu^{\pm}$ bunches using the same linac, the spacing of bunches needs to be adjusted by a path length difference of the injection lines. 

\paragraph*{Recent achievements}
A preliminary study for the pre-accelerator section was carried out to test the evaluation of the longitudinal phase space. It has been shown that an initial beam with an RMS length of \SI{0.4}{\m} and an RMS energy spread of \SI{5}{\percent} can be accelerated by a linac with a length of about \SI{800}{\metre} to \SI{1.5}{\giga\electronvolt} with a final RMS bunch length of \SI{40}{\milli\metre} while maintaining the longitudinal emittance. If one requires higher frequencies than \SI{88}{\MHz} in the linac the bunch length needs to be shortened due to limitation in acceleration of RF waveform. 

A comprehensive study for RLA2 was carried out since the previous report. We used a slightly weak FODO chain in the linac to reduce chromatic effects at low energy, incorporating 2 quadrupoles and 2 sections of 3 superconducting LEP2 cavities. Vertical bending spreaders, which are fully normal conducting, are designed to match the beam to the linac as well as the relevant arcs with increasing energy. For the correction of second-order dispersion in the bending plane as well as to minimise chromatic effects, a second-order achromatic line based on FODO without sextupoles is used for all eight arcs by adjusting the space between the magnets in the cell. 

\begin{figure}[h!]
    \centering
    \includegraphics[width=\textwidth]{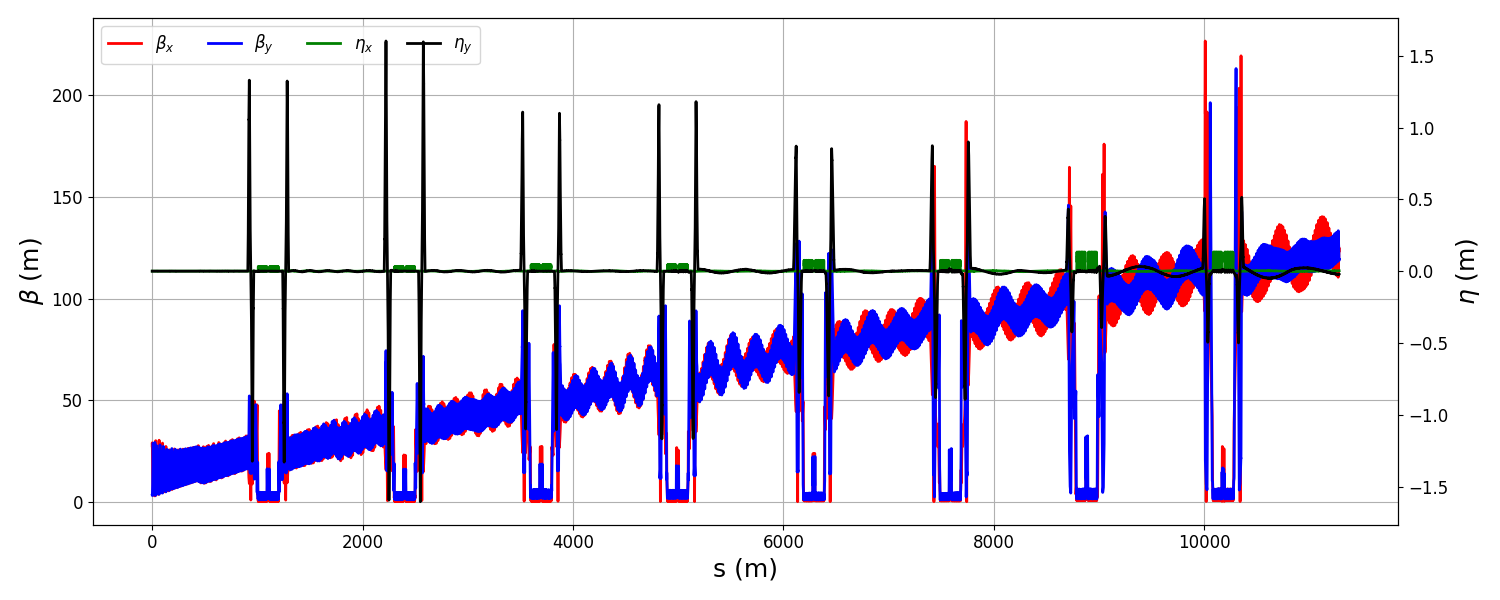}
    \caption{Start-to-end Twiss functions along RLA2}
    \label{1:acc:fig:s2e_twissRLA2}
\end{figure}
An initial bunch with \SI{0.0225}{\eV\s} longitudinal and \SI{20}{\um} transverse emittance could be accelerated from \SI{5}{\GeV} to \SI{63}{\GeV} while preserving longitudinal emittance, with less than  \SI{10}{\percent} increase in transverse plane, and a better than \SI{90}{\percent} muon survival rate. Figure~\ref{1:acc:fig:s2e_twissRLA2} shows the linear lattice functions along the RLA2. Since the quadrupole gradients in the linac are the same for each linac pass, the increasing beam energy in later passes leads to large beta functions in later passes. 

\begin{figure}
    \centering
    \includegraphics[width=0.85\linewidth]{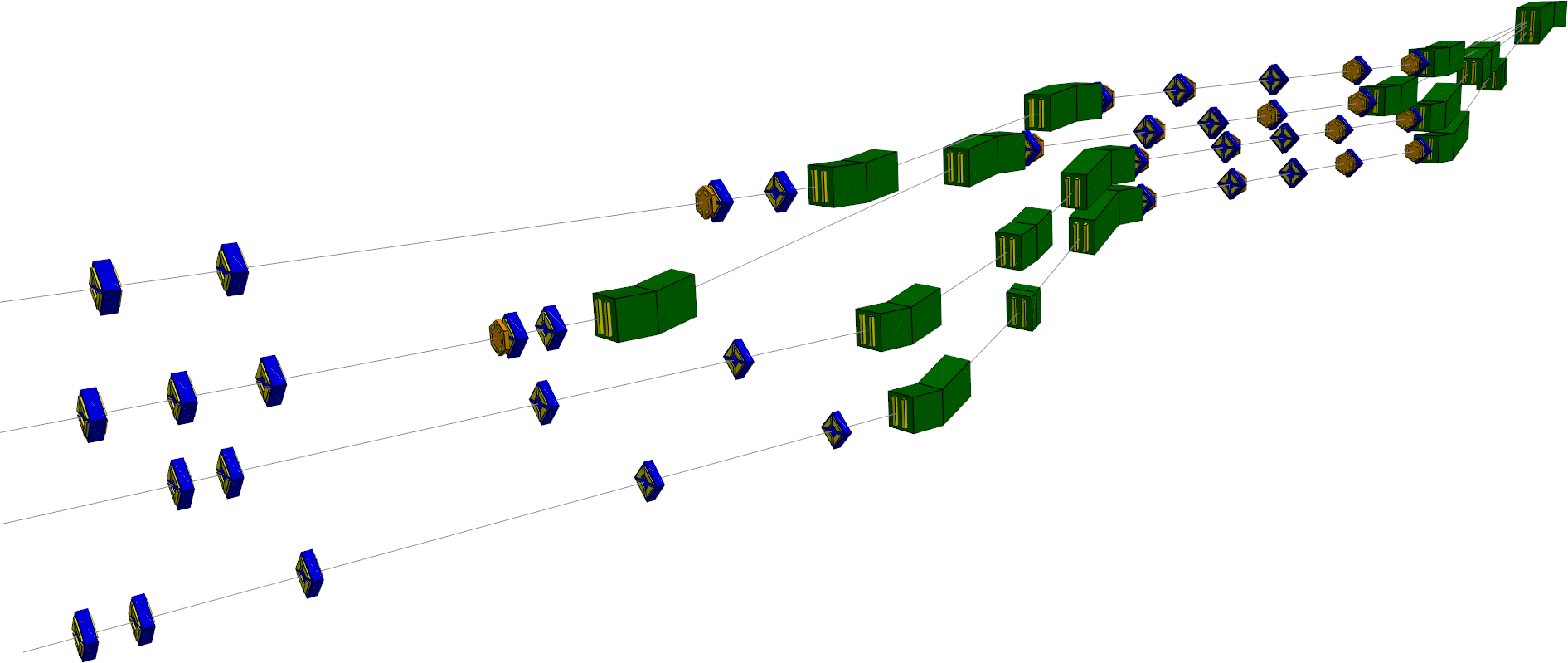}
    \caption{RLA vertical spreader section before the matching section and injection into the arcs, with quadrupoles in blue and the dipoles in green.}
    \label{1:acc:fig:RLA_vert_spreader}
\end{figure}

The arcs feature the same length and can, therefore, be stacked vertically in one tunnel. The arc magnets are superconducting to achieve the required bending strength in higher-energy passes. 
To achieve matching into the arc optics, a vertical spreader with fixed optics is planned to be used, which is presented in Figure~ \ref{1:acc:fig:RLA_vert_spreader}. 
This design achieves a transmission of \SI{93}{\percent} with a final transverse emittance of \SI{21.3}{\um} and a final longitudinal emittance of \SI{0.025}{\eV\s}, meeting the required target parameters. 
The target parameters for RLA2 are a transmission rate of \SI{90}{\percent}, a longitudinal emittance below \SI{0.025}{\eV\s} and a transverse emittance below \SI{22.5}{\um}. 
Table~\ref{1:acc:tab:RLA} contains the main parameters for the simulated RLA2 and the predicted RLA1.

\begin{table}[h]
    \centering
    \caption{Multi-pass recirculating linac parameters. The length of the linac refers to the length of a single linac section. The parameters for the linac are not yet available and the ones of the RLA1 are temporary assumptions, as the detailed designs of both linacs were not performed yet. The arc lengths define the length of an individual return arc, while the linac length refers to the length of one straight section of the RLA.}
    \begin{tabular}{l|c:cl|c:c}
    \toprule
         &  \multicolumn{2}{c}{RLA1} & \ &\multicolumn{2}{c}{RLA2}\\ \midrule
         Initial energy [GeV]&   \multicolumn{2}{c}{1.25} & \ &   \multicolumn{2}{c}{5}\\
         Final energy [GeV]&   \multicolumn{2}{c}{5} &\ &   \multicolumn{2}{c}{63}\\
         Energy gain per pass [GeV]&   \multicolumn{2}{c}{0.85} & \ &   \multicolumn{2}{c}{13.5}\\
         Frequency [MHz]&   352&1056 &\ &   352&1056\\
         No.~SRF cavities&   36&4 &\ &   600&80\\
         RF length [m]&   61.2&3.4 &\ &   1020&68\\
         RF gradient [MV/m]&   15&25 &\ &   15&25\\
         Passes&   \multicolumn{2}{c}{4.5} &\ &   \multicolumn{2}{c}{4.5}\\
         Linac length [m]&  \multicolumn{2}{c}{--} &\ & \multicolumn{2}{c}{915}\\
         Arc lengths [m]&  \multicolumn{2}{c}{--} &\ & \multicolumn{2}{c}{$\approx$ 300}\\
 \bottomrule
    \end{tabular}
    \label{1:acc:tab:RLA}
\end{table}
 
\subsection*{High-energy acceleration}
\label{1:acc:sec:rcs_chain}
\begin{figure}[ht!]
\centering
{\includegraphics[width=0.85\columnwidth]{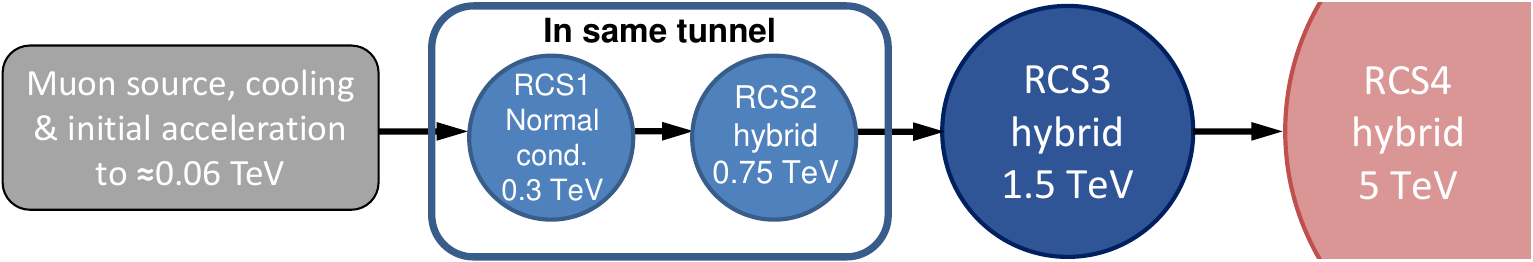}}
\caption{Sketch of the chain of rapid cycling-synchrotrons for the high-energy acceleration complex. From Ref.~\cite{ipac2023Batsch}.}
\label{1:acc:fig:RCSs}
\vspace*{-1\baselineskip}
\end{figure}

Once the muons have been pre-accelerated to an energy in the range of \SI{63}{\GeV}, two muon bunches must be brought simultaneously to an energy of up to \SI{5}{TeV}. 
Several schemes, like fixed-field alternating gradient synchrotrons, have been considered. 
The most promising and straightforward accelerator type in that energy range is the rapid cycling synchrotron (RCS). 
It combines fast acceleration with a cost-efficient implementation up to the highest particle energies. 
It moreover provides the necessary space for the large-scale RF system (section \ref{1:tech:sec:rf:highacc}). 
Nonetheless, compared to existing conventional RCS, the ones for muons will be significantly larger, with much higher accelerating voltages per turn and a very low number of revolutions. 
To achieve extremely fast acceleration with ramp rates of several \unit{kT/s} together with high average bending strength, fixed-field super-conducting and fast-ramping normal conducting magnets will be interleaved. 
The latter can then be driven from negative to positive saturation for a maximum bending field swing, $-B_{\mathrm{nc}}$ to $+B_{\mathrm{nc}}$.
The hybrid RCS concept was inspired by the US Muon Acceleration Program (MAP)~\cite{Berg,MAP2}.
It is foreseen to accelerate two counter-rotating bunches with a repetition of \SI{5}{\Hz}.
The green-field option, also suggested by the US MAP study, includes intermediate stages of \qtylist{0.31; 0.75; 1.5}{\TeV} to finally reach \SI{5}{\TeV} before injecting the muons into the dedicated collider ring. 
This scenario is illustrated in Figure~\ref{1:acc:fig:RCSs}. The acceleration parameters of the RCS chain are given in table~\ref{1:acc:tab:RCS_RFpars}. 
The bending in the first RCS is provided by normal conducting magnets only, while the RCS2 to RCS4 is planned with the hybrid bending scheme. In this configuration, the first two RCSs have the same circumference and layout, which allows their installation in the same ring tunnel. Such a magnet layout allows for a fast acceleration ramp, provided by the normal conducting magnet while simultaneously achieving a high average magnetic field. As a result, high energies can be reached in a relatively compact machine.

\paragraph*{Key challenges}
To keep muon decays at an acceptable level, the RCSs must have high average gradients $G_\text{avg}$, as high as \SI{2.4}{\mega\volt/\metre}. The average gradient is the energy gain in one turn divided by the circumference. Since most of the ring does not contain RF structures, the gradient in the RF sections must be significantly higher than the average gradient. Assuming a survival rate of \SI{90}{\percent} per RCS, ultra-fast acceleration with tens of GeV energy gain per turn is required. 
Hundreds of superconducting cavities are needed to provide high accelerating gradients while withstanding significant transient beam loading from high-intensity muon bunches, with a bunch population of up to \num{2.7e12} muons. For the first simulations, TESLA-like \SI{1.3}{\GHz} cavity structures~\cite{TESLA} have been assumed.

The large RF voltages of the tens of gigavolts per turn result in a very high synchrotron tune. For an RF system at a single location, this even exceeds the conventional stability limit for stable synchrotron oscillations and phase focusing ($1/\pi$~\cite{Pi}). 
As a consequence, the RF system must be distributed over the entire RCS, in order to reduce the synchrotron tune between two consecutive RF stations.

\begin{figure}[htb]
  \begin{center}
    \includegraphics[width=0.5\textwidth]{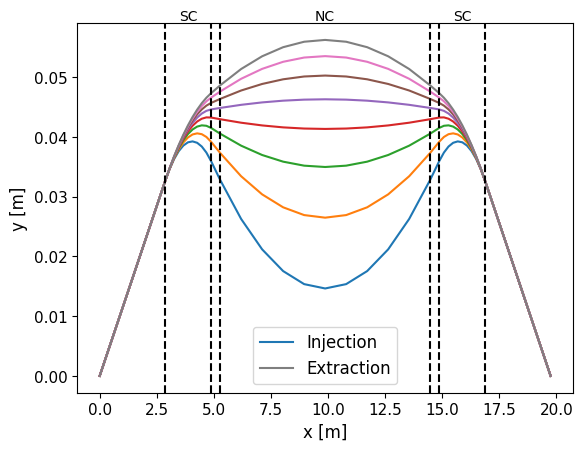}
  \end{center}
\caption{Trajectories in a half-cell of RCS2. The injection/extraction orbits 
are respectively on the inner/outer side (in blue/gray). 
}
\label{1:acc:fig:RCS_orbit_RCS2}
\end{figure}

As a direct result of the decay and required ultra-fast acceleration, the acceleration times are in the millisecond range, and the number of turns in each RCS ranges between 17 and 66. In the RCS chain, the ramp rate of the normal conducting magnets is on the so far unprecedented order of \unit{\kilo\tesla/\s}, posing a significant challenge. Further, due to the hybrid structure of the three downstream RCSs, beam orbits and radii are not constant anymore, but change locally, as depicted in Figure~\ref{1:acc:fig:RCS_orbit_RCS2} \cite{ipacAntoine}.
Special attention is therefore being paid to the lattice design in order to minimize trajectory excursions and orbit length changes during acceleration. 

Because of the orbit length variation, fast cavity detuning on the \unit{\kHz} scale is crucial for matching the RF frequency with the changing revolution frequency. This development addresses the unique challenge posed by the counter-rotating, intense single \textmu$^+$ and \textmu$^-$ bunches. Due to this special beam structure, substantial transient beam loading is anticipated for both fundamental and higher-order modes.

After the acceleration in RLA2, the beam will most likely not be matched to the RF bucket at injection into RCS1. A dedicated simulation with this injection mismatch into RCS1 will be required and likely cause an emittance increase at the start of the acceleration. 

\begin{figure}[htbp]
    \begin{center}
        \includegraphics[width=0.45\textwidth]{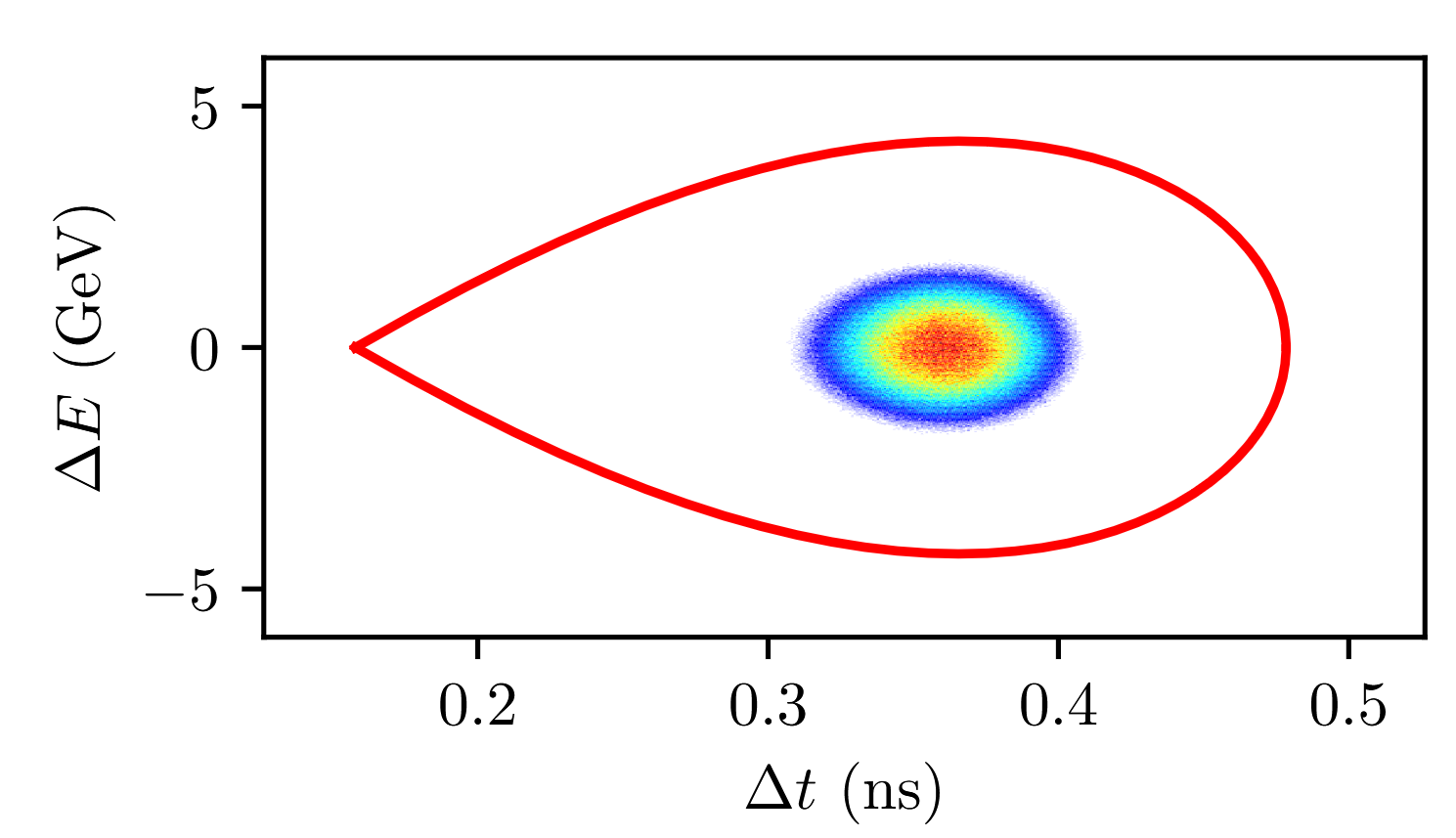}
        \includegraphics[width=0.45\textwidth]{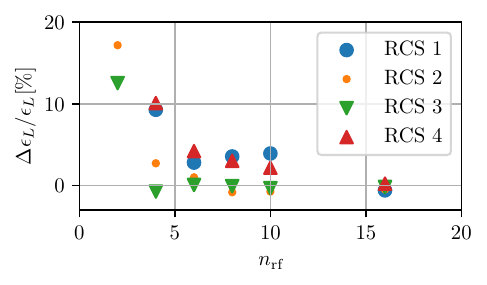}
    \end{center}
    \caption{(a) Longitudinal phase space for a $\mu^+$ bunch in RCS1 for $n_{\mathrm{RF}}=10$ at injection. The red line indicates the separatrix (with continuous energy approximation) for $f_{\mathrm{RF}}=\SI{1.3}{\GHz}$. \\
    (b) Relative emittance growth at the end of the cycle of each RCS with respect to the emittance at injection versus $n_{\mathrm{RF}}$ for RCS1 to RCS4 from simulations including the counter-rotating beams and longitudinal intensity effects.}
    \label{1:acc:fig:RCS_emittance_growth}
\end{figure}

\paragraph*{Single-beam dynamics}
\begin{table}[!htbp]
    \centering
    \begin{tabular}{lc|cccc} 
         Parameter&  Unit&  RCS1&  RCS2&  RCS3&  RCS4\\ \hline
         Hybrid RCS&  -& no & yes & yes & yes \\
         Repetition rate&  Hz&  5&  5&  5&  5\\
         Circumference& m& 5990& 5990& 10700& 35000\\
         Injection energy& GeV& 63& 314& 750& 1500\\
         Extraction energy& GeV& 314& 750& 1500& 5000\\
         Energy ratio& - &  5.0&  2.4&  2.0&  3.3\\
         Assumed survival rate& - & 0.9& 0.9& 0.9&0.9\\
         Cumulative survival rate& - &  0.9&  0.81&  0.729&  0.6561\\     
         Acceleration time&  ms&  0.34&  1.10&  2.37&  6.37\\
         Revolution period&  \textmu s&  20&  20&  36&  117\\
         Number of turns& -& 17& 55& 66& 55\\
         Required energy gain/turn& GeV& 14.8& 7.9& 11.4& 63.6\\
         Average accel.~gradient& MV/m& 2.44& 1.33& 1.06& 1.83\\ \hline
         Number of bunches per species& & 1& 1& 1& 1\\
         Inj.~bunch population& \num{E12}& 2.7& 2.4& 2.2& 2\\
         Ext.~bunch population& \num{E12}& 2.4& 2.2& 2& 1.8\\
         Beam current per bunch& mA& 21.67& 19.5& 9.88& 2.75\\
         Beam power&  MW&  640&  310&  225&  350\\
         Vert.~norm.~emittance& \textmu m& 25& 25& 25& 25\\
         Horiz.~norm.~emittance&  \textmu m&  25&  25&  25&  25\\
         Long.~norm.~emittance&  eVs&  0.025&  0.025&  0.025&  0.025\\
         Bunch length at injection & ps & 31& 30 & 23 & 13 \\
         Bunch length at ejection & ps & 20& 24 & 19 & 9\\ \hline
         Straight section length&  m & 2335 &  2335 &  3977 &  10367\\
         Length with pulsed dipole magnets& m & 3654& 2539 & 4366 & 20376\\
         Length with steady dipole magnets& m & -& 1115 & 2358 & 4257\\
         Max.~pulsed dipole field& T& 1.8& 1.8& 1.8&1.8\\
         Max.~steady dipole field& T& -& 10& 10&16\\
         Ramp rate& T/s& 4200& 3282& 1519&565\\
    \end{tabular}
    \caption{RCS acceleration chain key parameters from \cite{PreliminaryParameter_MuCol5}.}
    \label{1:acc:tab:RCS_RFpars}
\end{table}
To simulate the longitudinal beam dynamics, we use the well-benchmarked longitudinal macro-particle tracking code \mbox{BLonD~\cite{Blond,Blond2}}. The code was successfully extended to model multi-turn wakefields in the multiple RF stations per ring. A plot of the longitudinal phase space at injection for RCS1 is displayed in Figure~\ref{1:acc:fig:RCS_emittance_growth}(a).

Tracking simulations on the influence of the number of RF stations on the longitudinal emittance and beam stability have been performed for all RCSs. 
As shown in Figure~\ref{1:acc:fig:RCS_emittance_growth}, their minimum number is around 6 in all RCS. This allows to keep the longitudinal emittance growth close to the required \SI{5}{\percent} level not taking into account any injection mismatches or powering errors. The minimum number of RF stations depends heavily on the momentum compaction factor, and, therefore, on the choice of the lattice. The simulations in Figure ~\ref{1:acc:fig:RCS_emittance_growth} are based on the last version of the lattices of the four pulsed synchrotrons, shown in \cite{chance_2024_13847233}. Their momentum compactions are included in the preliminary parameter document \cite{PreliminaryParameter_MuCol5}. Hence, the required number of stations will have to be adapted accordingly with the progress on the lattice design.

\begin{figure}[htb]
  \begin{center}
    \includegraphics[width=0.8\textwidth]{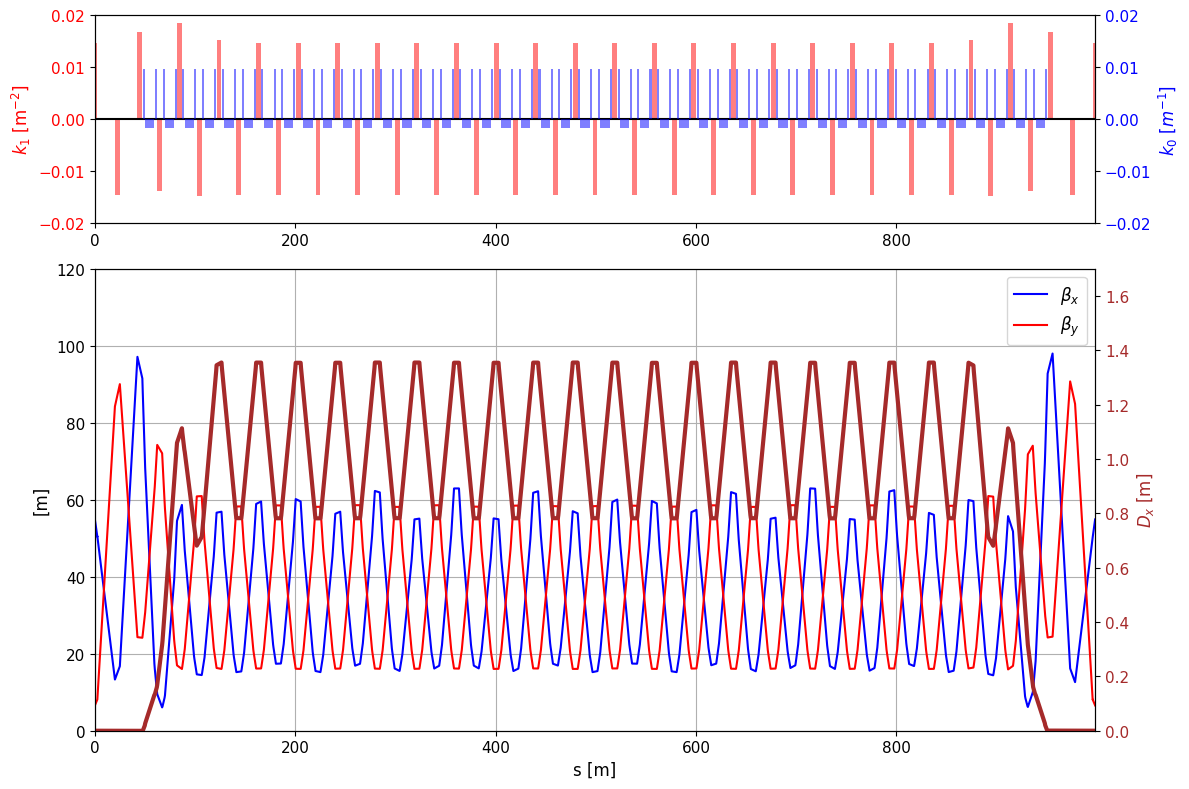}
  \end{center}
\caption{Normalized strength of the dipoles (top, in blue) and quadrupoles (top, in red) and betatron functions (bottom, in blue and red) and dispersion function (in dark red) in one out of 6 arcs of RCS2.}
\label{1:acc:fig:RCS_lattice_RCS2}
\end{figure}

The following in focused on the RCS2 because it is hybrid and is the most sensitive to the trajectory variation due to a reduced curvature radius. However, the 3 other RCS have met the same issues at different levels for the lattice design. In its current configuration (see Figure~\ref{1:acc:fig:RCS_lattice_RCS2}), RCS2 comprises 6 arcs. Of these, the 6 insertions between the arcs house cavities. Each arc half-cell contains three dipoles: two superconducting dipoles (\SI{10}{\tesla}) positioned on the outside and a pulsed normal conductor with a peak field of \SI{1.8}{\tesla} in the center.

Because of the small reachable gradient with normal conducting technology, the length of the quadrupoles becomes non-negligible in the optimization of the lattice. 
For the RCS2, the total length of the quadrupoles corresponds to about one-quarter of the total arc length. 
For these reasons, the cell optimization had to include the quadrupoles to find the best configuration minimizing the orbit variation, the path length variation, and the magnet aperture. Indeed, the path length variation will give the range of tuning required for the RF cavities and their power sources. 
The magnet aperture will bound the magnet design and drive the total magnet cost. 
With the current design, values reach $\Delta \mathcal{C}/\mathcal{C}\approx\num{8.3e-6}$ in RCS2, resulting in a tuning requirement of $\Delta \mathcal{C}/\mathcal{C}\cdot f_\text{RF}\approx\SI{10.7}{\kHz}$, equivalent to a peak tuning speed of approximately \SI{27}{\MHz/\s}. Piezo tuners or fast reactive tuners are under investigation for that purpose. 

At this state, given some tolerances on the magnets, $\Delta \mathcal{C}$ becomes very challenging. 
Due to the fast acceleration, it becomes very difficult to correct any dynamic errors such as a jitter in the powering of the magnets. 
The lattice should enable some passive mitigation of this jitter. 
The task is even more challenging because it should work also for a counter-propagating beam. 
A main concern is also the eddy currents generated in the pulsed normal conducting dipoles, which is a potential source of non-linearities and, thus, of emittance growth. Simulation studies have been launched to investigate this effect and to conclude on mitigation measures. 

Longitudinal tracking simulations, which include the intensity effects due to the single but very intense muon bunches, show the impact of short-range and long-range induced voltages (section \ref{1:acc:sec:collective}). 
The short-range wakefields were calculated using a resonator model for all fundamental and higher-order modes of a TESLA-like cavity and show that the induced voltages are on the order of \SI{2}{\mega\volt/\metre} (and per cavity).

The voltage induced in the HOM (higher-order mode) resonances of the RF structures was studied as a function of their quality factor, to estimate longitudinal stability margins~\cite{HB23Batsch}. Due to the high quality factors involved, it was essential to include multi-turn wakefields in the tracking simulations. However, technically it will be challenging to change the quality factors of all HOMs simultaneously. 

The \lstinline{rcsparameters} simulation tool has been developed to allow for an integrated computation of the base parameters of the RCS chain. This tool calculates all high-level parameters, i.e. the necessary RF and magnet parameters as well as lengths of the different sectors and survival rates. It has been successfully applied to optimize the RCS parameters for the CERN tunnel infrastructure of SPS and LHC, which is an alternative proposal to the green-field approach. More details on this proposal can be found in section \ref{2:site:cern:sec}.

\paragraph*{Counter-rotating beams}

The simulation tools for the longitudinal beam dynamics were extended to include the intensity effects of the two counter-rotating high-intensity muon bunches. This required combining long-range wakefields in both directions with multiple RF stations per turn.

The high quality factor of the fundamental and some HOMs and the associated long decay time of induced voltages, couples the longitudinal beam dynamics of both beams. The bunch in one direction is hence affected by the previous passage of the bunch rotating in the opposite direction through the same RF structure. 
The synchronous phase, $\phi_s$, during acceleration must be identical for both counter-rotating beams, as the effects of the opposite direction and opposite charge cancel out each other. 
Therefore, the induced voltages in the fundamental mode will constructively interfere. 
The longitudinal alignment of the cavity positions in the ring will be essential to maintain an equal voltage vector sum for both beams. Any correction of the phase for one beam will cause a degradation for the other beam.  In this respect, the impact of the HOMs will depend on the exact cavity location in the ring, keeping in mind that the induced voltages in multiple cavities will interfere constructively or destructively.

\paragraph*{FFA as an alternative}
Muon acceleration by an FFA (Fixed-Field Alternating Gradient) accelerator has been investigated for many years since it appeared in the neutrino factory proposal in the 1990s. Recent developments in FFA optics, especially of vertical excursion FFAs, added significant advantages in using an FFA as the muon accelerator. Although there are not many FFA accelerators even for accelerating ordinary particles like protons, an FFA scheme could be an alternative option to an RCS.

An FFA uses DC magnets, as the name indicates. That means both advantages and disadvantages. From the operational point of view, it is ideal for acceleration of short-lived particles because there is no change of magnetic field according to the beam momentum. On the other hand, FFAG magnet apertures are relatively large and that is the main technical challenge, especially with superconducting magnets where the wide aperture means large stored energy.

The optics of a vertical excursion FFA (vFFA) is fully coupled in two transverse planes. The optics design and dynamics analysis are much more complex than the more common horizontal excursion FFA (hFFA), and that in fact discourages consideration of a vFFA as an option. 

Recent work provides an analytical model for vFFA optics, offering insight into parameter dependence and potential optimization for muon colliders\cite{PhD_Topp-mugglestone}. 
 
The analytic model has been used to develop example parameter sets for equivalent vFFA rings to RCS1 and RCS2, showing that vFFA rings with equivalent footprints to the RCS baselines can achieve stable optics whilst maintaining realistic design constraints. In particular, the movement of the orbit from injection to extraction (the excursion) is now less than \SI{120}{\mm} in each case, and the peak dipole fields on orbit are maintained below \SI{8}{\tesla} in the early-stage vFFA and below \SI{16}{\tesla} in the final-stage vFFA.

\section{Collider}
\label{1:acc:sec:collider}
The muon collider ring consists of two straight Interaction Regions (IRs) located at opposite positions, where the two counterrotating muon bunches collide and the detectors are installed and an arc to. The arc consists of several components: a Chromatic Compensation Sections (CCS) adjacent to the IRs to correct chromatic aberrations generated by strong focusing next to the detectors, followed by matching sections connecting to regular arc cells, which are periodic focusing structure over most of the arc.

The primary aim of the muon collider ring design is to maximize the luminosity to the experiments, taking into account the characteristics of the injected beam as determined by the ionization cooling channels and various acceleration stages. Additional constraints are generated by muon decay products. The resulting heat loads for the cryogenic system and the radiation damage to magnets are mitigated by Tungsten absorbers. As described in section \ref{1:tech:sec:rad_shield}, unwanted signals to detectors are suppressed by the Machine Detector Interface (MDI) described in section \ref{1:inter:sec:mdi} and by making use of Tungsten and other shielding materials.

Neutrinos generate radiation at the location where they reach the Earth's surface, which requires careful evaluations described in section \ref{1:tech:sec:rad_protect} and has to be taken into account in the machine design in order to ensure that effective doses to general public remain negligible and below the dose objective of \SI{10}{\micro\sievert}/year 
aimed at by CERN. A large number of neutrinos will be generated in the direction given by the long straight section housing the experiments, developing a significant doses where they reach Earth's surface. This results on the one hand in an opportunity for an experiment profiting from the large neutrino flux, and on the other hand the corresponding flux site should be owned by the organization operating the collider and fenced. Radiation levels from the other parts of the machine will be limited by reducing the length of short straight sections in the arc to the minimum. Thus,  combined functions quadrupoles and higher-order multipoles will be used instead of normal straight quadrupoles and mutipoles and interconnect reagions beween magnets have to be minimized. In addition, peak doses will be reduced by periodic variation of the beam trajectory in the vertical plane and corresponding deformations of the whole machine. These deformations change the vertical beam direction within a range of $\pm$\SI{1}{\milli\radian}, distributing the radiation generated by neutrinos over a larger surface area. This requires the high precision mechanical system described in section \ref{1:tech:sec:movers} to implement the required displacements of the magnets and additional horizontal magnetic fields for vertical deflections in the magnet design described in section \ref{1:tech:sec:mag}. If the collider ring is not located deep underground as with the CERN siting proposed in section \ref{2:implement:sec:cern}, horizontal closed orbit deformations (inside the aperture without deformations of the whole machine in horizontal direction) are required in addition to mitigate neutrino radiation hot spots located in the direction of short straights between arc magnets.

Magnets with as high field strengths as possible are required. Strong chromatic aberrations making the optics design vary challenging are generated with the highest possible gradients and would further increase with lower gradients. Strong dipolar fields in other sections result in a shorter circumference and revolution time, such that less muons decay between bunch encounters and the luminosity increases. As the length of straight sections outside the one housing the IPs must be minimized, combined function quadrupoles and sextupoles are used.

\paragraph*{Recent achievements}
Recent efforts have concentrated on the design of a \SI{10}{\tera\electronvolt} center of mass collider. The basic parameters aim at maximizing the luminosity and, initially, have been determined by extrapolating a procedure from earlier studies on colliders at lower energy and using nominal beam parameters aimed at by optimization of the ionization cooling described in section \ref{1:acc:sec:cool}. Assuming that an rms momentum spread of $\sigma_p/p = 1 \cdot 10^{-3}$ is manageable, the rms bunch length can be as short as $\sigma_z = \SI{1.5}{\milli\meter}$. Furthermore, the Twiss betatron function at the IP has been set to $\beta^* = \sigma_z = \SI{1.5}{\milli\meter}$, such that the luminosity reduction due to the hourglass factor $f_{hg} \approx 0.76$ starts to become relevant. Note that the procedure applied leads to a $\beta^*$ inversely proportional to the energy or beam rigidity. Thus, the optics design becomes significantly more challenging for higher energies. Another challenge is to keep the bunches short for about 1000 turns contributing significantly to luminosity. An RF system helping significantly to keep the bunches short should would require very high frequencies and gradients. Thus, the present baseline foresees only a modest RF system to compensate for energy losses due to synchrotron losses and a stringent control of linear and non-linear momentum compaction factor to keep the bunches short. The latter requires negative momentum compaction arc cells described below. In addition, the path length dependence from the betatron oscillations amplitudes will have to be controlled in the final design.

\begin{figure}[htb]
    \centering
  \includegraphics[width=0.9\textwidth]
    {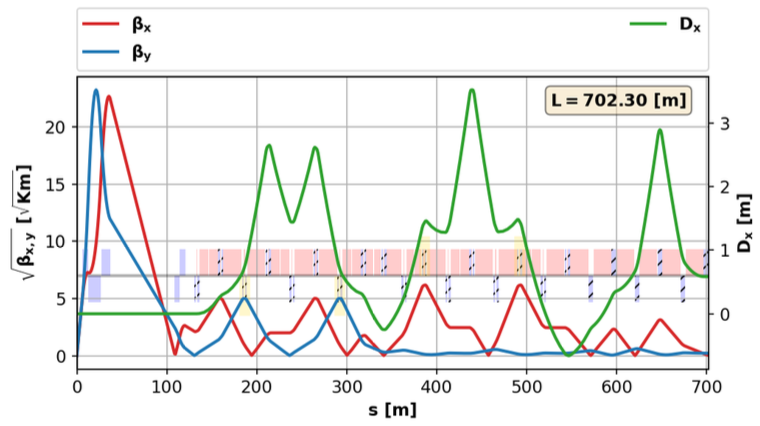}
    \caption{Twiss parameters for part of the collider ring starting at the IP taken from reference \cite{skoufaris:ipac2023-mopl064}. Red and blue traces denote the square root of the horizontal and vertical Twiss betatron functions and the green trace denotes the dispersion.} 
    \label{1:acc:fig:Twiss_collider}
\end{figure}

Twiss parameters for a well advanced version of the optics from the IP through the chromatic compensation section and matching to the arc are shown in Figure~\ref{1:acc:fig:Twiss_collider}. The main components of the muon collider designs are:
\begin{itemize}
  \item Interaction region with the straights housing the detector: an inner triplet is positioned as close as possible to the detector and followed by a chicane (not yet included in the design shown in Figure~ \ref{1:acc:fig:Twiss_collider}) to reduce beam induced background seen by the experiment (see section \ref{1:inter:sec:mdi}). The triplet and chicane is followed by a longer drift with decreasing betatron functions and at least one quadrupole.
  \item A local chromatic compensation section is required to compensate large chromatic aberrations (particles with higher or lower energy than nominal are focused less or more) driven by the small $\beta^*$ and resulting large peak betatron functions at the location of the strong inner triplet quadrupoles. Strong sextupoles are placed pairwise in regions with dispersion. The sextupoles are spaced by betatron phase advances very close to $\Delta \mu = \pi$ and, ideally have identical strength, betatron functions and dispersion to cancel to first order non-linear effects.
  \item Matching section to arc cells.
  \item Negative momentum compaction arc cells (not included in Figure~\ref{1:acc:fig:Twiss_collider}) are required to compensate the unavoidable positive contributions to momentum compaction from the chromatic compensation and matching sections. Dispersion oscillations are provoked and generate negative dispersion over a significantly large portion of the cell, where higher energy muons are displaced towards the inside and take a shortcut.
\end{itemize}
Careful optimizations of the collider design allowed to significantly improve the dynamic aperture over a significant part of the required momentum range for a machine without imperfections as misalignments and unwanted magnetic field components and with periodic machine deformations.

\begin{figure}[htb]
    \centering
    \includegraphics[width=0.6\textwidth]{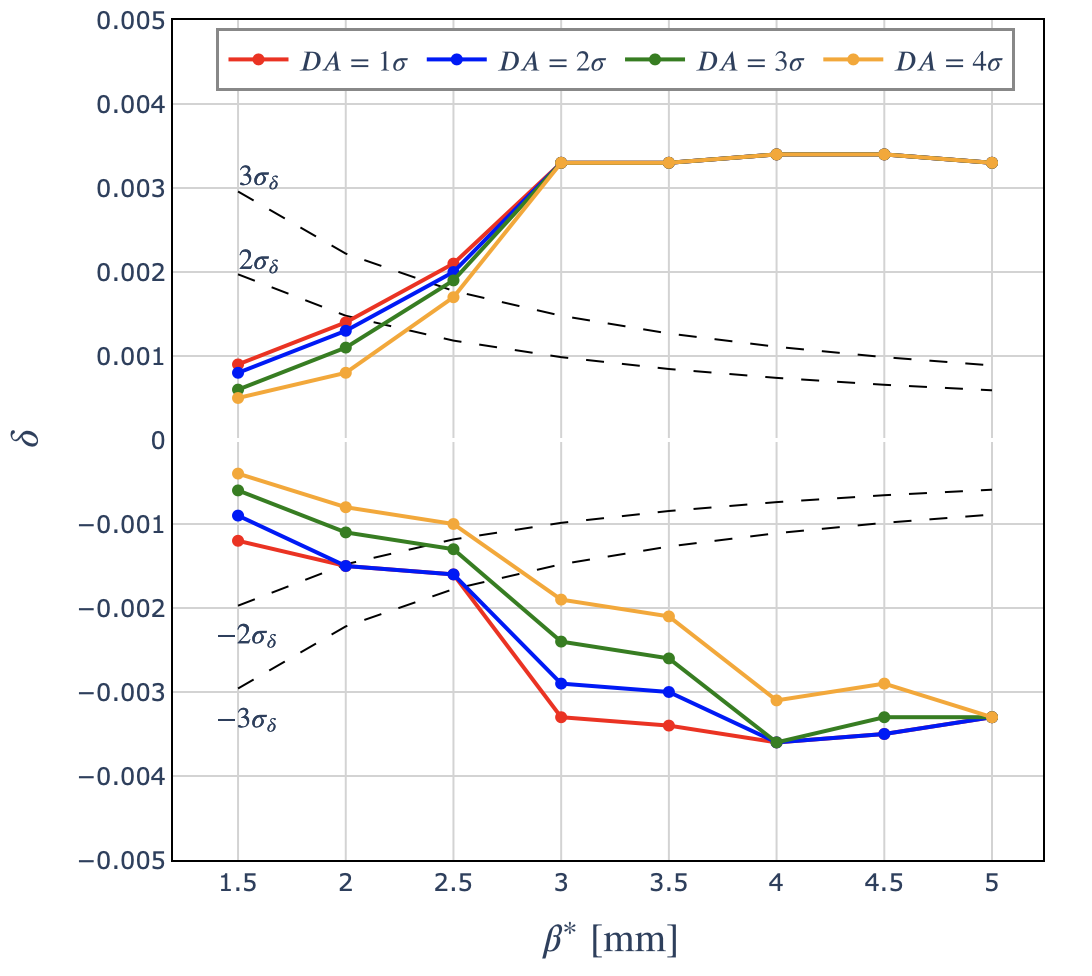}
    \caption{Minimum dynamic aperture of the collider as a function of the relative momentum offset taken from reference \cite{Vanwelde2024a}.} 
    \label{1:acc:fig:MomAcceptance}
\end{figure}

Despite efforts spent and significant progress made, the lattices developed so far for a $\beta^* = \SI{1.5}{mm}$ still suffer from an insufficient momentum acceptance even for a perfect machine. A particular difficulty encountered so far is the large working point excursions due to non-linear effects. The reason is the procedure to extrapolate  beam parameters to higher energies resulting in extremely challenging requirements for a \SI{10}{\tera\electronvolt} centre-of-mass collider. 

In order to allow to come up soon with an initial working lattice (that can be later further optimized based on the design parameters) working at least for a perfect machine, an exploratory study \cite{Vanwelde2024a} on the impact of slightly relaxed $\beta^*$ has been carried out. For each value of $\beta^*$ studied, the lattice had to be redesigned and optimized while keeping the procedure the same for all cases. Figure \ref{1:acc:fig:MomAcceptance} is one of the main outcomes and shows the maximum momentum offset, for which a given dynamic acceptance is achieved, as a function of $\beta^*$. The plot shown is for a version with more realistic reduced inner triplet quadrupole gradients to take into account recent studies on large aperture superconducting magnets. For future iterations of the lattice design, assumptions on maximum fields and gradients of combined function magnets will have to be reduced to more realistic values likely leading to a reduction of the dynamic acceptance for a given moment offset. Another important observation from this study is that the working point variations for a given range in momentum offset are strongly reduced. The strength of local chromatic correction sextupoles is as reduced with larger $\beta^*$. Moreover, stable lattices (first order without study of the dynamic aperture) can be found for larger working point changes due to chromatic effects.

One notes that slightly relaxing $\beta^*$ indeed allows to significantly improve the momentum acceptance (for a given fixed minimum dynamic aperture). Moreover, assuming that the rms bunch length is again set equal to $\beta^*$ and with the given longitudinal acceptance, the momentum spread of the beam is inversely proportional to $\beta^*$. The dashed lines indicate the range $\pm 2 \,\sigma_p/p$ and $\pm 3 \, \sigma_p/p$. Assuming that a dynamic aperture of three times the rms beam size over a $\pm \sigma_p/p$, relaxing $\beta^*$ by a factor two would allow to find a lattice working for the perfect machine. Nevertheless, assumption made so far concerning maximum feasible magnetic fields and gradients for large aperture magnets and aperture requirements for the tungsten absorber have been somewhat optimistic. Updated more realistic assumptions, based on studies on the feasibility of large aperture high field magnets and heat load due to muon decay products, will be made for further iterations of the lattice design and likely render the design even more difficult.

\section{Collective Effects}
\label{1:acc:sec:collective}

Collective effects are an essential topic for accelerators with high-intensity beams.
Several effects must be covered, for example: space-charge effects modelling the interaction of particles with the electromagnetic field generated by the beam distribution itself; wakefield and impedance effects that gather the electromagnetic interaction of the bunch with their surrounding environment such as vacuum chambers or RF cavities; and beam-beam effects modelling the electromagnetic interaction of colliding or separated bunches crossing in interaction points.

After evaluating the magnitude of these effects their impact on the beam properties like emittance evolution or beam losses can be estimated.
If the beam properties are degraded by collective effects, damping mechanisms such as chromaticity, or transverse feedback can be introduced in the accelerators.
Design parameters can also be modified to reduce the collective effects magnitude.
The aim is to ensure there is no showstoppers with regards to intensity limitations.

\subsection*{Proton driver}

In the proton driver, (section \ref{1:acc:sec:proton}) the single bunch intensity will reach \num{5.0e14} protons per bunch, with an emittance \SI{\sim 9}{\micro\metre} and an energy of \SI{5}{\GeV} or \SI{10}{\GeV}.
This intensity and energy regime makes the bunches more prone to coherent instabilities from space charge effects, in particular in the accumulator and compressor rings.
The interplay with beam-coupling impedance effects is also important for the circular machines and it has been a topic investigated for more than two decades.
Since few years quite some progress has been made in the understanding of this intricate mechanism but there are still some open points which need to be clarified.

In this framework, a general comparison of simulation codes for direct space charge simulation has been carried out.
They use different analytical approaches to estimate the coherent mode frequency shifts as a function of bunch intensity.
The codes BimBim, based on the Circulant Matrix Model (CMM); the Effective impedance method for space charge; GALACTIC based on the Vlasov equation; the boxcar model for space charge only; and the ABS model which assumes an Air-Bag bunch distribution in a Square well were compared~\cite{bib:Amorim:ipac2025-space-charge}.
This comparison identified significant differences between models, and will help choose the correct model, understanding its limitations, for the space charge mode frequency shifts estimation in proton driver.

\subsection*{Muon Cooling}
\begin{figure}
\centering
    \includegraphics[width=0.51\textwidth, trim={0, 0, 0, 2cm}, clip]{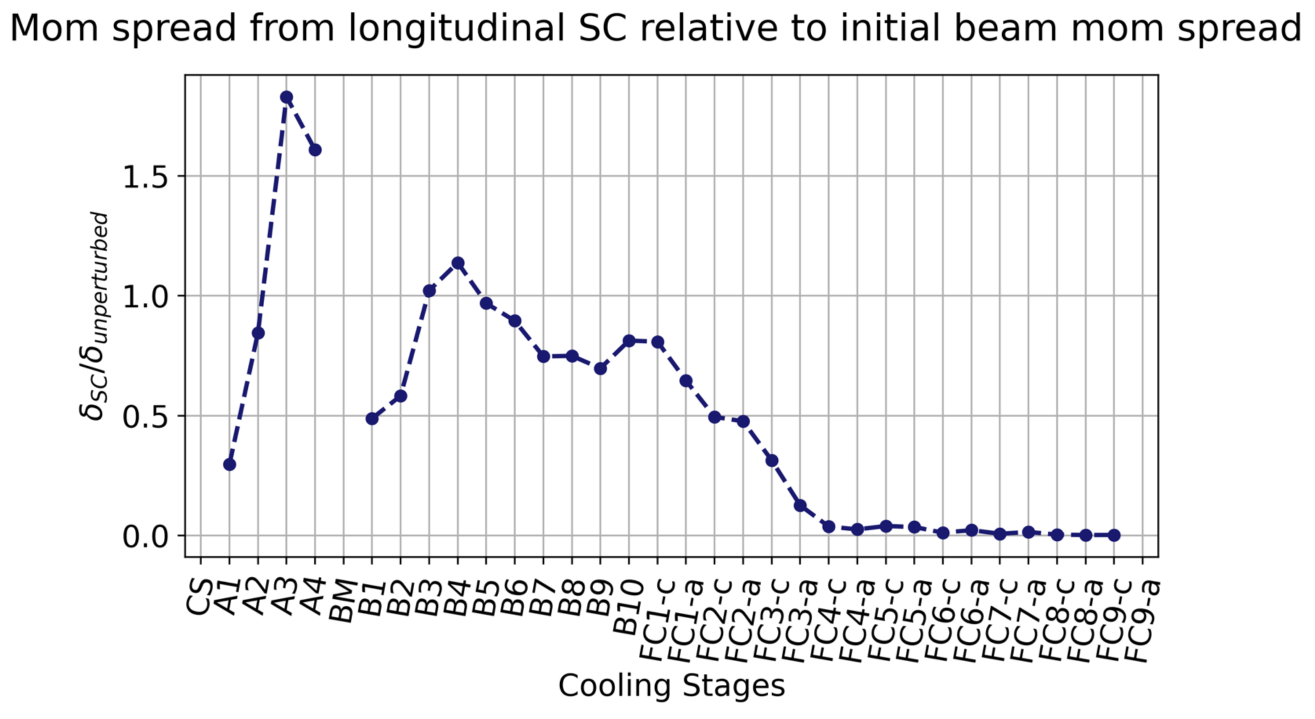}
    \hfill
    \includegraphics[width=0.45\textwidth,  trim={0, 0, 0, 2cm}, clip]{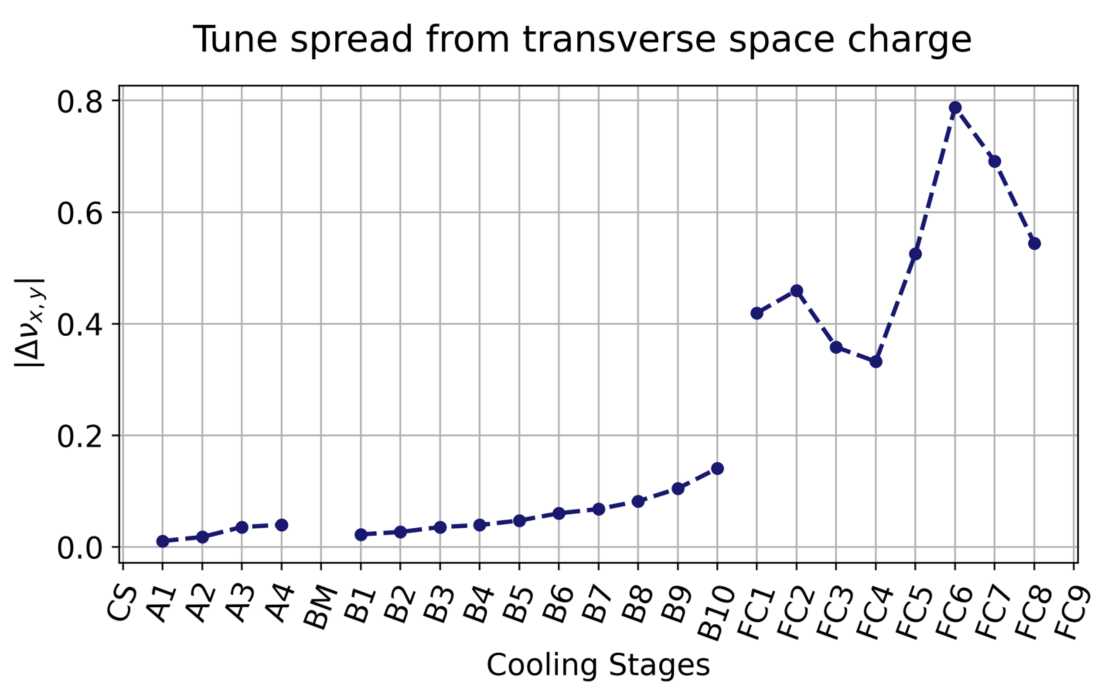}
\caption{Analytical estimate of the momentum spread from longitudinal space charge, relative to the initial beam momentum spread (left plot) and the tune spread from transverse space charge (right plot) along the rectilinear and final cooling.}
\label{1:acc:fig:collective_cooling_momentum_and_tune_spread}
\end{figure}

A first theoretical assessment of the magnitude of collective effects in the muon cooling (section \ref{1:acc:sec:cool}) has been performed~\cite{bib:potdevin_imcc_annual_meeting_2024}.
Three main effects have already been identified. The first two are linked to variations of the longitudinal focusing fields and are caused by space charge and beam loading in the rectilinear cooling. Indeed, the high bunch intensity (starting with \num{1e14} muons per bunch), combined with the high RF frequencies (in the \SIrange{325}{1056}{\mega\hertz} range) generate a high-amplitude beam-induced voltage. The rectilinear cooling lattice and the cavities design might require specific adaptation to mitigate these effects, in particular to maintain the required linear part of the fields. Limitations may then arise from the non-linearity of the collective fields, potentially increasing the tune spread. Pessimistic estimates are shown in Figure~\ref{1:acc:fig:collective_cooling_momentum_and_tune_spread}. \\
The third element is the strong transverse space charge fields in the final cooling caused by the reduced transverse beam sizes $\sigma_{x, y}$. The non-linearity of these forces may lead to reductions of the dynamic aperture. The expected tune spread along the rectilinear and final cooling cells has been estimated analytically, based on the current cooling lattice properties.

The beam break-up instability generated by the resistive wall wake of the beam chambers was also estimated.
Thanks to the large chamber radius the growth parameter for this instability is small.
Moreover BNS (Balakin, Novokhatski and Smirnov) damping was not included in the estimation and should further mitigate this type of instability.

\subsection*{High-energy acceleration}

The two Recirculating Linacs (RLA) (section \ref{1:acc:sec:acc}) will bring the muon bunches energy from \SI{1.25}{\GeV} to \SI{63}{\GeV}.
Beam-beam interactions will occur in each of the four droplet arcs and could lead to emittance growth and beam losses.
Tracking simulations with Xsuite~\cite{bib:xsuite} were performed for the RLA~1, assuming a linear transfer map between each beam-beam interaction point (IP).
The beam-beam interaction is a non-linear effect and is modelled with Bassetti-Erskine formula, which assumes Gaussian profile for the transverse beam distribution.
Simulations show a large emittance growth can occur if the phase advance between IPs is not carefully chosen.
This emittance growth can be mitigated by introducing \SI{\sim 1}{\metre} of dispersion at the IPs.
Combined with the proper choice of longitudinal and transverse phase advance between IPs, the emittance growth from beam-beam interaction can be kept below the percent level~\cite{bib:buffat_rla_beam_beam}.

\begin{figure}
\centering
    \includegraphics[width=0.65\textwidth,  trim={0, 0, 0, 0cm}, clip]{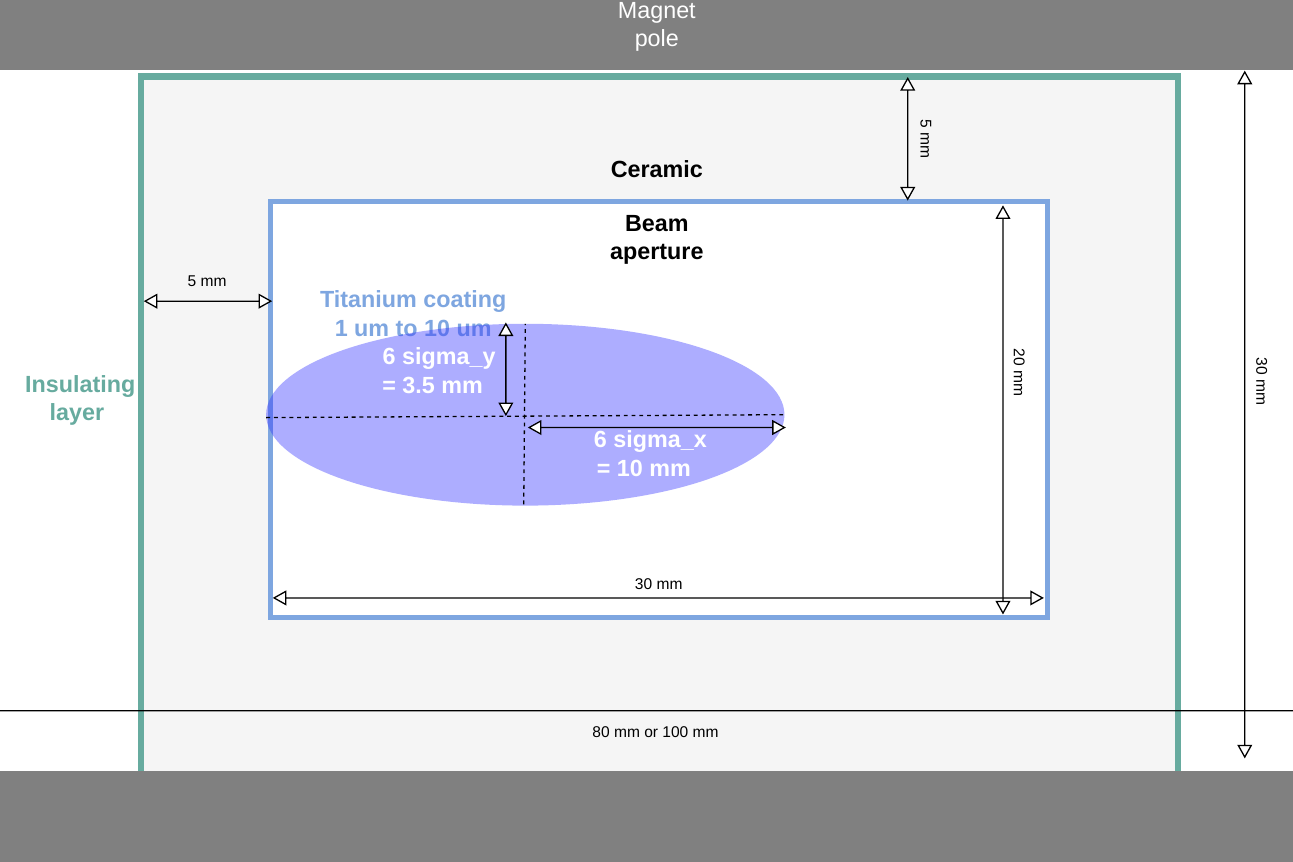}
\caption{Proposed radial build of the RCS 1 normal conducting magnets vacuum chamber.}
\label{1:acc:fig:collective_rcs_nc_magnet_radial_build}
\end{figure}

To meet the muon survival rate target in the high-energy acceleration chain, large acceleration gradients are required in the Rapid Cycling Synchrotrons.
This requires a large number of RF cavities able to reach a high accelerating gradient.
For example in RCS 1 the energy gain per turn must be \SI{14.7}{\GeV} to lose only \SI{10}{\percent} of the intensity to muon decay.
To reach this acceleration, $O(700)$ TESLA superconducting RF cavities operating at \SI{1.3}{\GHz}~\cite{TESLA}, with an individual accelerating gradient of \SI{30}{\mega\volt\per\m} will be needed, as described in Sections~\ref{1:acc:sec:acc} and~\ref{1:tech:sec:rf}.
High-order modes (HOMs) in the cavities will generate short and long-range wakefields which combined to the high muon bunch intensities could disrupt the beam motion and lead to emittance blow-up and particle losses.
Additionally the fast ramping of the normal conducting magnet in the RCS constrains the vacuum chamber design in order to minimize eddy current induced power losses.
Ceramic must be used as the main bulk material of the chamber, together with an inner metallic RF shielding to reduce impedance effects.
An initial radial build of the vacuum chamber is proposed and pictured in Figure~\ref{1:acc:fig:collective_rcs_nc_magnet_radial_build}, meeting constraints set from magnets, powering, vacuum and impedance.

\begin{figure}
\centering
    \includegraphics[width=0.45\textwidth,  trim={0, 0, 0, 1cm}, clip]{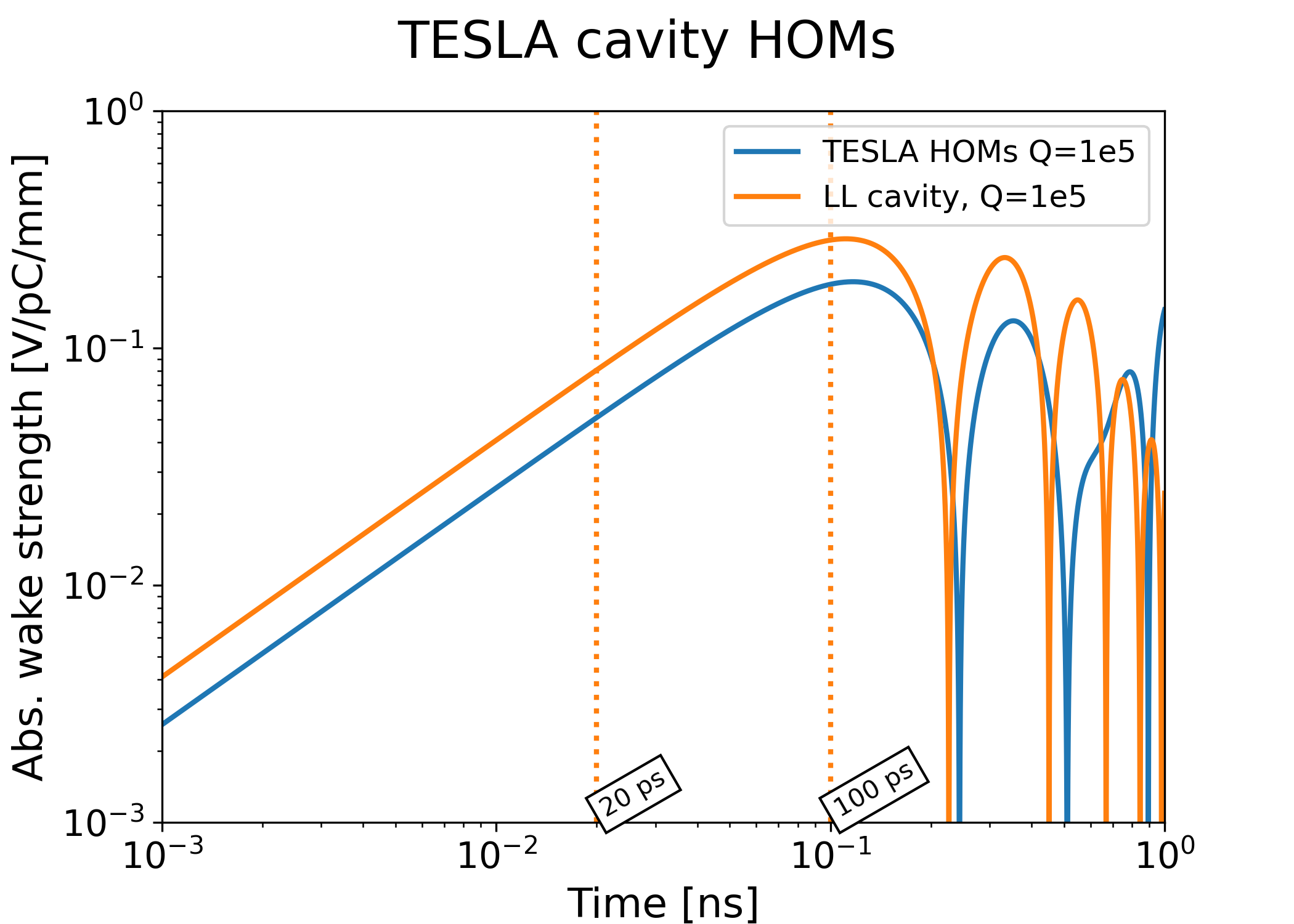}
    \includegraphics[width=0.44\textwidth,  trim={0, 0, 0, 0cm}, clip]{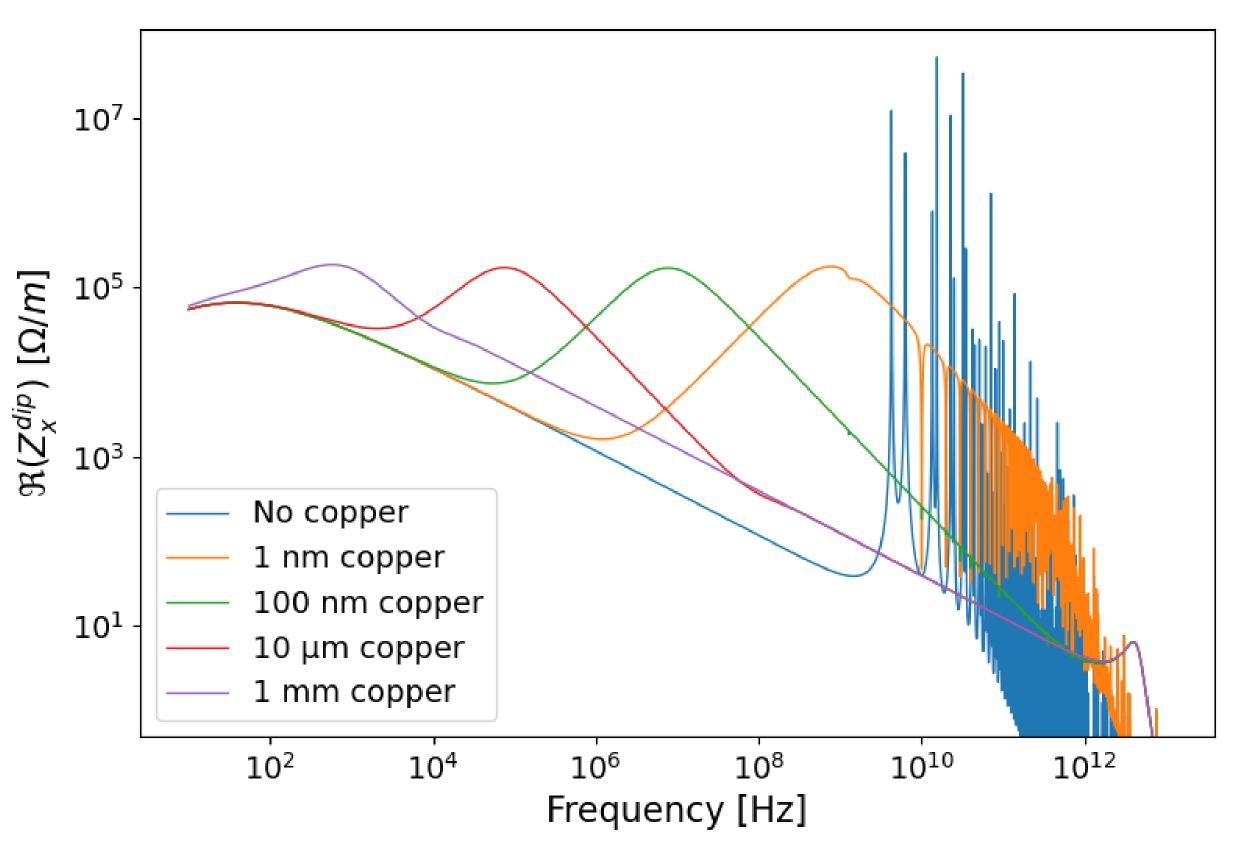}
\caption{Left: wakefield strength generated by the HOMs of a TESLA (blue) and a Low Loss (orange) \SI{1.3}{\GHz} superconducting RF cavity. Right: real part of the horizontal dipolar impedance of the RCS normal conducting magnet vacuum chamber, for different RF shielding thickness, from no shielding (blue) to \SI{1}{\mm} thick copper shielding (purple).}
\label{1:acc:fig:collective_rcs_wakefields}
\end{figure}

Impedance models have been developed for the four RCS of the high energy acceleration chain.
They include the RF cavity HOMs and the normal conducting magnets vacuum chamber design.
Two types of TESLA cavity were considered: the classic TESLA cavity~\cite{TESLA} and the Low-Loss type cavity~\cite{bib:LL_cavities}.
Figure~\ref{1:acc:fig:collective_rcs_wakefields} shows on the left hand side the wakefield generated by the HOMs of a single TESLA cavity or a Low Loss cavity, highlighting the reduction in wake strength obtained by using the classic TESLA cavity.
On the right hand side the figure shows the simulated transverse dipolar impedance of a \SI{1}{\metre}-long section of a normal conducting magnet vacuum chamber for different RF shielding thicknesses.
The ceramic chamber introduces large resonances in the impedance that are reduced by the RF shield.
Macroparticle tracking simulations using the Xsuite and PyHEADTAIL~\cite{bib:pyheadtail} codes were performed to evaluate the impact of impedance on single bunch transverse coherent stability, from injection in the RCS 1 through ejection of the RCS 4.
The effect of an initial transverse offset of the beam at the injection in each ring was also included.
Instability mitigation measures such as a transverse damper and chromaticity were also investigated.
Simulations found out that a positive chromaticity of $Q^{\prime} = +20$ is required in the four RCS to preserve the transverse beam emittance.
An initial transverse offset resulting from injection jitter can also be tolerated, up to \SI{10}{\micro\metre} in each RCS~\cite{bib:amorim_rcs_imcc_annual_meeting_2024, bib:amorim_rcs_update_2024}.

\subsection*{Collider}
\label{1:acc:sec:collective:collider}
In the collider ring (section \ref{1:acc:sec:collider}) the beam chamber will generate short-range resistive-wall wakefields.
To reach the highest possible magnetic field in the dipole, the magnet aperture should be as small as possible and must host a \SI{4}{\centi\metre} thick tungsten shielding to intercept muon decay products, as described in Sections~\ref{1:acc:sec:collider} and~\ref{1:tech:sec:rad_shield}.
The inner radius of the tungsten shield and its material properties, such as the operating temperature or the use of a low electrical resistivity coating, will influence the resistive-wall wakefield and determine the coherent stability threshold.
At the interaction points, strong beam-beam effects will arise from the high bunch intensity of \num{1.8e12} muons per bunch at injection and the small $\beta^*$ target of \SI{1.5}{\milli\metre}.
The short bunch length of $\sigma_z = \SI{1.5}{\milli\metre}$ could also lead to beam-induced heating in some devices.
However the strength of collective effects will reduce over the beam storage time in the collider thanks to the muon decay.
With a \SI{5}{\tera\eV} beam, the Lorentz factor $\gamma=47323$ and the muon lifetime in the laboratory frame $\tau$ is $\tau = \gamma \tau_0 = 47323 \cdot \SI{2.2}{\micro\second}=\SI{104}{\milli\second}$, equivalent to $3121$ turns in the \SI{10}{\kilo\metre} long ring.

\begin{figure}
\centering
    \includegraphics[width=0.48\textwidth,  trim={0, 0, 7cm, 2.45cm}, clip]{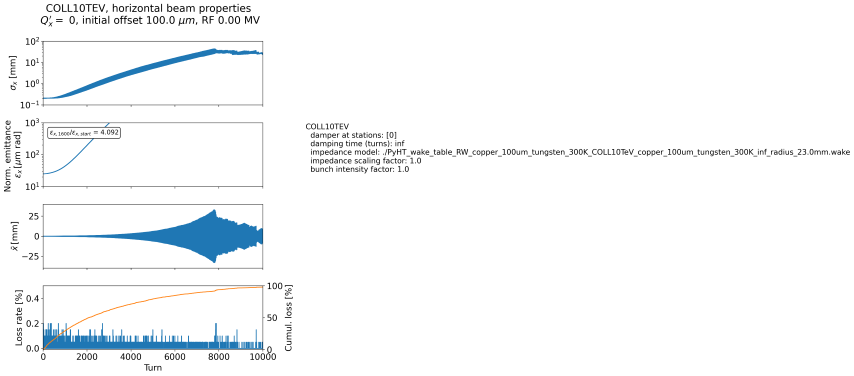}
    \includegraphics[width=0.48\textwidth,  trim={0, 0, 7cm, 2.45cm}, clip]{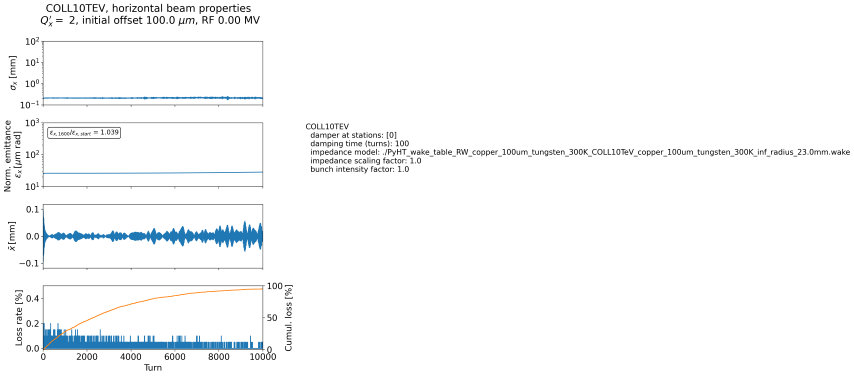}
\caption{Bunch centroid position over time (top plot) and muon loss rate (bottom plot, blue curve) and cumulative losses (bottom plot, orange curve) for the \SI{10}{\TeV} collider including the \SI{23}{\milli\metre} radius beam chamber.
The left plot shows the situation with a chromaticity corrected to $Q^{\prime} = 0$ and the right plot with a chromaticity of $Q^{\prime} = +2$.
Additionally an initial horizontal offset of \SI{100}{\micro\metre} is introduced at bunch injection which is mitigated by 100-turn transverse damper.}
\label{1:acc:fig:collective_coll10tev_instability}
\end{figure}

The impedance model for the collider assumes a circular tungsten chamber with a \SI{23}{\milli\metre} radius (slightly more conservative than the \SI{23.5}{\milli\metre} radius foreseen), with a \SI{10}{\micro\metre} copper coating on the inner side.
The chamber is assumed to be at \SI{300}{\kelvin} with a total length of \SI{10}{\kilo\metre}.
The collider ring parameters assume that no RF is present, and that the slippage factor $\eta=0$.
Single bunch simulations were performed with Xsuite and PyHEADTAIL to assess the impact of the impedance on transverse beam stability.
They found that with a chromaticity $Q^{\prime} = 0$ a strong instability develops for any transverse damper setting.
This instability can be mitigated by introducing a small amount of positive or negative chromaticity, such as $Q^{\prime} = \pm 2$~\cite{bib:amorim_collider_imcc_annual_meeting_2024}, as shown in Figure~\ref{1:acc:fig:collective_coll10tev_instability}.
This value of chromaticity is rather small for such machine and should not impact luminosity significantly: for example in the LHC, the precision on chromaticity setting is in the order of $\pm 2$ units and the operational chromaticity is set around $Q^{\prime} \sim 15$.
Beam-beam simulations were also performed for the collider ring using Xsuite.
Given the transverse emittance of \SI{25}{\micro\metre} and the intensity of \num{1.8e12} muons per bunch, the linear beam-beam parameter is $\xi = 0.078$ at injection.
When the two bunches are colliding head-on, a transverse instability is developing, and leads to large transverse emittance growth.
This beam-beam instability can be mitigated with a 20-turn transverse damper~\cite{bib:amorim_beambeam_workshop_2024}.

\chapter{Accelerator technologies}
\label{1:tech:ch}

\section{Introduction}

\label{1:tech:sec:intro}

This chapter introduces the different  technologies that have been studied within the IMCC and MuCol frameworks. It is important to underline that the feasibility and cost of the muon collider depend to a large extent on the choices made on magnets and RF components, therefore a large part of the chapter is  dedicated to those two technologies. Different options are studied for the target, going from the "traditional"  graphite rods, considered feasible if the beam power stays below 2 MW, to more complex options such as fluidised tungsten or liquid metal curtains to achive a beam power of 4~MW if necessary. 
The efficiency and feasibility of the cooling channels proposed, depend on the other hand not only by the single components of the cooling cell (namely absorbers, RF and magnets), but also on the possibility to integrate them in the shortest and tighter possible space, to limit as much as possible the probability of decay of the muons in the cooling channel itself or in the subsequent section. A paragraph describes the progress made so far in the understanding of the constraints, and the iterative process ongoing to clarify what is reasonable to expect in this sense. 
Other technologies which are very important for the implementation of an ambitious programme on the muon collider are  Beam instrumentation, vacuum and the absorbers. Cryogenics will have an impact once we will be able to demonstrate that magnets working at 20K are a viable option. In that case the possibility to use liquid Hydrogen as a coolant may bring to significant simplification of the system and savings, but will raise well founded safety constraints, therefore it will be important to study this option in the future. Meanwhile, we describe what is our present understanding of the possible cryogenic system.   
A paragraph is dedicated to the mover system for the collider magnet, used to gently move the magnets axis several times during the year in order to paint the spot of the neutrino radiation on surface to reduce the density of integrated dose at the surface. At the moment only some considerations are presented and a programme of R\&D will be necessary to show that  such a system and all the cryogenic and vacuum  connections from magnet to magnet can sustain such movements.  

\section{Magnets}
\label{1:tech:sec:mag}

The Muon Collider poses extraordinary challenges to magnet technology, and meeting them will benefit not only the most efficient accelerator at the energy frontier, but also several other fields of science and societal applications. Through the integrated study and conceptual design activities of the last three years (2022-2024) we have identified the following grand challenges that have driven magnet R\&D activities:
\begin{itemize}
    \item Steady state superconducting solenoids for
    \begin{itemize}
    \item Target, decay and capture channel
    \item 6D cooling channel
    \item Final cooling channel
    \end{itemize}
    \item Fast pulsed normal conducting magnet systems, including the power converter and management, for the rapid cycled synchrotrons
    \item Steady state superconducting accelerator magnets, dipoles, quadrupoles and combined functions, for the rapid cycled synchrotrons and collider arc and interaction region.
\end{itemize}

The sections below describe the main achievements of the work performed in the period since the last Strategy Upgrade, in 2021. We provide in particular a description of the concepts selected, the details of the engineering design and supporting analysis, an evaluation of the challenges to magnet technology, and a reasoned summary of target performance for on-going and future developments.

\subsection*{Target, decay and capture channel solenoid}
\label{sec:Target}
\paragraph*{Magnet design and engineering}

The solenoids that host the target and capture channel, where the muon beam is produced, pose the first grand challenge. The magnetic field profile along the axis of the channel has a shape derived from studies of optimal generation and capture, with peak field of 20 T on the target, and a decay to approximately 1.5 T at the exit of the channel, over a total length of approximately 18 m. The characteristic length of the field change is about 2.5 m, i.e. much larger than the gyration radius of the muons in the field so that the beam expands adiabatically in the channel. Such field profile can be generated

The interaction of the proton beam with the target produces a considerable amount of radiation, which needs heavy shielding to avoid heating and damaging the materials of the superconducting coils of the target solenoid. A free bore of at least 1.4 m is necessary to host the nuclear shield around the target. Such large bore dimension result in high stored magnetic energy, which in turn affects magnet protection, and electromagnetic forces.

We have developed a fully superconducting solution for the 2 MW target variant, based on a HTS cable inspired by recent developments in the field of magnetically confined thermonuclear fusion. Field levels of 20 T are at the upper limit of performance for small bore Nb3Sn, and arguably out of reach for LTS with the bore dimension required. More important, the choice of HTS gives the possibility to set an operating point at a temperature higher than liquid helium. This brings the benefit of increased cryogenic efficiency, reduced wall-plug power consumption, and reduced helium inventory. We have set a reference an operating temperature in the range of 20 K, which has an efficiency advantage of a factor 5 with respect to cooling at liquid helium, 4.5 K.

\begin{figure}
\centering

\includegraphics[width=\textwidth]{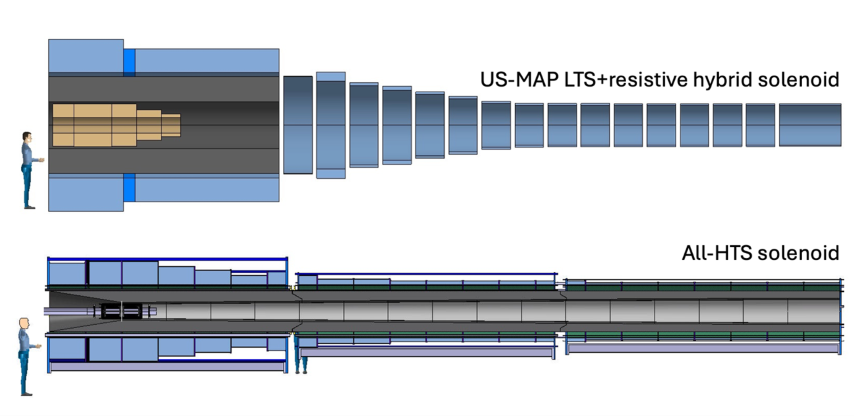 }
\caption{Comparison (to scale) of the solenoid coils of the target, decay and capture channel of a Muon Collider, as produced by the MAP study (top) \cite{MAP} and resulting from the optimization of an all-HTS solution (bottom) \cite{accettura2024conceptual}}
\label{fig:Target_fig1}
\end{figure}

The solution reached is shown in the schematic view of Figure~\ref{fig:Target_fig1}, contrasted to the design originally proposed by US-MAP \cite{MAP}. Thanks to the choice of HTS, operated at high cryogenic temperature, the stored energy is reduced by a factor three from the US-MAP value of 3 GJ to 1.4 GJ of the present design, and the cold mass is similarly reduced by a factor two from the US-MAP value of 200 tons to about 100 tons of the present design. This has significant impact on system cost. In addition the elimination of the resistive insert in the US-MAP proposal, and operation at 20 K, yield to an estimated wall-plug power consumption below 1 MW, to be compared to the estimated 12 MW of the US-MAP proposal.

The reference configuration (December 2024) is reported in Table~\ref{tab:Target_tab1}, where we give the details of the coil geometry, the number of turns and pancakes, and the operating current of each solenoid. To be noted that this configuration was obtained with equal current in all conductors (61.15 kA), the file profile being the result of the geometry optimization. This greatly simplifies powering and protection, allowing to have several of the low-field modules in series, thus reducing the number of circuits and the number of leads.

\begin{table}[h!]
\centering
\begin{tabular}{lccccccc}
\toprule
Solenoid & $R_c$ & $Z_c$  & DR  & $D_z$ & Turns & Pancakes & $I_\text{coil}$ \\ 
Module & & & & & & & \\
 &  (m) & (m) & (m) &  (m) & (-) & (-) & (MA-turn) \\
\hline
SC1  & 0.970 & -1.185 & 0.540 & 0.830 & 13 & 20 & 15.899 \\
SC2  & 0.970 & -0.335 & 0.540 & 0.830 & 13 & 20 & 15.899 \\
SC3  & 0.970 & 0.515  & 0.540 & 0.830 & 13 & 20 & 15.899 \\
SC4  & 0.887 & 1.365  & 0.374 & 0.830 & 9  & 20 & 11.007 \\
SC5  & 0.825 & 2.215  & 0.249 & 0.830 & 6  & 20 & 7.338  \\
SC6  & 0.783 & 3.065  & 0.166 & 0.830 & 4  & 20 & 4.892  \\
SC7  & 0.825 & 3.708  & 0.249 & 0.415 & 6  & 10 & 3.669  \\
SC8  & 0.704 & 4.603  & 0.208 & 0.415 & 5  & 10 & 3.058  \\
SC9  & 0.642 & 5.245  & 0.083 & 0.830 & 2  & 20 & 2.446  \\
SC10 & 0.642 & 6.095  & 0.083 & 0.830 & 2  & 20 & 2.446  \\
SC11 & 0.642 & 6.945  & 0.083 & 0.830 & 2  & 20 & 2.446  \\
SC12 & 0.621 & 7.795  & 0.042 & 0.830 & 1  & 20 & 1.223  \\
SC13 & 0.621 & 8.645  & 0.042 & 0.830 & 1  & 20 & 1.223  \\
SC14 & 0.621 & 9.495  & 0.042 & 0.830 & 1  & 20 & 1.223  \\
SC15 & 0.642 & 10.138 & 0.083 & 0.415 & 2  & 10 & 1.223  \\
SC16 & 0.621 & 11.033 & 0.042 & 0.415 & 1  & 10 & 0.612  \\
SC17 & 0.621 & 11.675 & 0.042 & 0.830 & 1  & 20 & 1.223  \\
SC18 & 0.621 & 12.525 & 0.042 & 0.830 & 1  & 20 & 1.223  \\
SC19 & 0.621 & 13.375 & 0.042 & 0.830 & 1  & 20 & 1.223  \\
SC20 & 0.621 & 14.225 & 0.042 & 0.830 & 1  & 20 & 1.223  \\
SC21 & 0.621 & 15.075 & 0.042 & 0.830 & 1  & 20 & 1.223  \\
SC22 & 0.621 & 15.925 & 0.042 & 0.830 & 1  & 20 & 1.223  \\
SC23 & 0.621 & 16.775 & 0.042 & 0.830 & 1  & 20 & 1.223  \\
\bottomrule
\end{tabular}
\caption{Reference geometry and winding configuration for the solenoids of the target, decay and capture channel, also reporting the operating current for each solenoid module.}
\label{tab:Target_tab1}
\end{table}

The design developed has progressed significantly in terms of magnet engineering, and we have reached the stage of initial engineering details on:
\begin{itemize}
\item	conductor design and performance, including cooling, operating margin, quench detection and protection analysis~\cite{BOTTURA2024VIPER};
\item	mechanical analysis, down to the level of the HTS tapes in the conductor \cite{accettura2024conceptual};
\item	coil manufacturing, including winding technology, joints and terminations, and impregnation;
\item	mechanical structures, supports and screens, cryostat and integration with thermal screen and target.
\end{itemize}

A double pancake winding using a force-flow cooled HTS superconductor seems to be a good solution, meeting most design criteria. The force-flow conductor proposed for the study is largely inspired by the VIPER developed for magnetically confined fusion, and has already an experimental basis of proven performance~\cite{MITcable}. The conductor is made by a hollow copper core hosting soldered stacks of REBCO tape. This cable is then enclosed in a steel jacket that only has structural functions. Most interesting, it seems indeed possible to achieve high field, 20 T peak field on axis, at high operating temperature, 20 K, which has benefits of lower capital and operation expenditure (CAPEX and OPEX) compared to previous solutions.

Cooling at high temperature, 20 K, with gaseous helium is not a trivial extrapolation of force-flow supercritical helium near liquid conditions, 4.2 K. High operating pressure, e.g. 20 bar, and larger temperature increase than usual, e.g. 3 K, will be mandatory to avoid excessive distribution losses, and achieving the gain in cryogenic efficiency associated with the higher operating temperature. More studies, integrating the refrigeration cycle, will be necessary to produce an optimal system.

Some additional features have been identified, that could make construction and operation simple. One such example is the reinforcement jacket which has no leak tightness requirement. The studies reported here also show that thermal stability will not be an issue. At the same time quench detection and protection can likely rely on well-established precise voltage measurement, reasonable detection threshold, in the range of 100 mV, and dump voltages within state-of-the-art technology, 5 kV. The hot spot temperature remains well below 200 K in all cases analyzed. It will be very interesting at this point to realize and test samples of the conductor designed here, to confirm manufacturing features, validate the performance reach and margins, and characterize the behaviour during quench.

On the side of mechanical design, the overall criteria at the coil level can be satisfied within the allowable limits of common material grades. However, looking at the details of the stress and strain distribution within the cable we may have identified locations and conditions where loads could exceed allowable limits. Tensile and shear stresses at the level of the single tapes could reach values in the range of 60 MPa, whereby it is well known that the internal structure of REBCO tapes is not very resilient to this type of loading, with a wide spread of maximum allowable in the range of a few MPa and up to few tens of MPa. Note that while the analysis was performed for the specific geometry considered here, this may be a result of general applicability to soldered and twisted stacks of tapes. The analysis on this topic is only at the beginning, some avenues have been suggested to resolve this issue, and more optimization work is required, also considering the on-going work in the R\&D program being pursued for fusion reactors \cite{bykovsky2017cyclic}. Also in this case, some strong experimental evidence will be necessary to advance understanding and validate the solutions found.

\begin{figure}
\centering
\includegraphics[width=\textwidth]{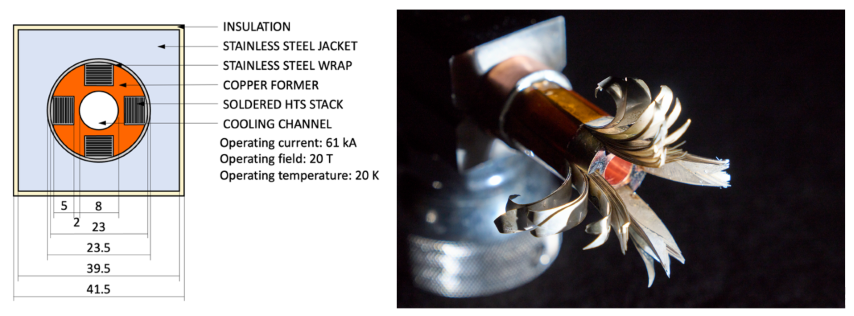 }
\caption{Schematic view of the conductor configurations selected for the solenoid coils of the target, decay and capture channel (left) \cite{MITcable}, with an image of a mock-up produced on real size (right) showing the HTS tapes, central copper former and jacket.}
\label{fig:Target_fig2}
\end{figure}

\begin{figure}
\centering
\includegraphics[width=\textwidth]{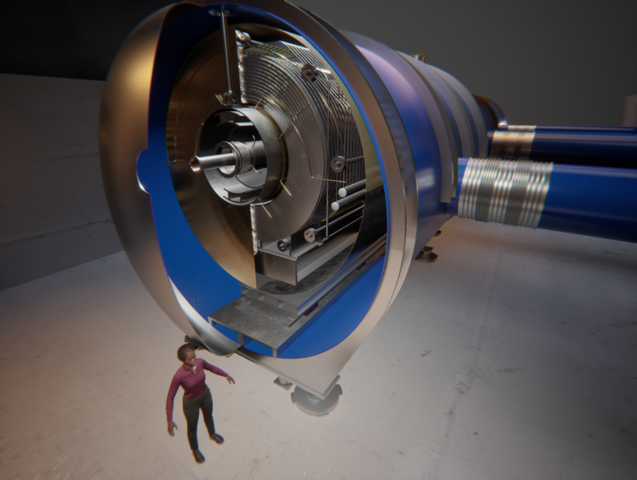 }
\caption{Rendering of the magnet system for the target, decay and capture channel integrated in the cryostat, showing details of the winding, joints, cooling channels, thermal screen and supports.}
\label{fig:Target_fig3}
\end{figure}

The level of detail reached can be appreciated by sample views of the conductor, winding and 3D coil model shown in Figures~\ref{fig:Target_fig2} and~\ref{fig:Target_fig3}. While management of weights and forces remains challenging, we could find valid engineering and integration solutions for all above aspects, and we can be reasonably confident to proceed further with this baseline.

Work is in progress at present to advance in the engineering design of the chicane magnets, whose configuration and geometry has only been sketched. Achieving the required magnetic field profile in the large bore required will be a particular challenge because the radiation load from the spent proton beam is very high, possibly limiting the technology to resistive electromagnets. 

In parallel, we are evaluating the implications of an increase of the beam power on the target, up to 4 MW which would allow to double the number of muons generated. While this will require additional shielding, and thus an increase of the bore dimension, first evaluations seem to indicate that it would still be possible to accommodate such an increase in performance with the present concept.

\paragraph*{Challenges Identified}
As a result of the studies performed, we could identify a number challenges, to be addressed in priority:
\begin{itemize}
\item	High-current HTS conductors qualified for operation in high field and helium gas. Although this class of conductors is being developed for fusion applications, the geometry selected needs experimental validation, especially to address the concerns of internal strain and stresses;
\item	Winding technology. The specific solution envisaged is well established, based on double pancakes wound from an insulated conductor, wrapped with fiber glass, stacked and vacuum impregnated with resin to form a coil module. Still, this was never applied to a HTS coil of this size, and this field level. Furthermore, novel solutions need to be developed for the soldering of the tape stacks in the cable, performed after winding, joints and terminations, as well as diagnostics for operation and protection;
\item	Radiation hardness of magnet materials, foremost polymers (insulation) and superconductor (REBCO). According to the expected radiation loads in a muon collider, both material classes will be at the expected limit of degradation. In addition, for HTS we lack a well-established material database and physical understanding of degradation mechanism.
\end{itemize}

\begin{figure}
\centering
\includegraphics[width=\textwidth]{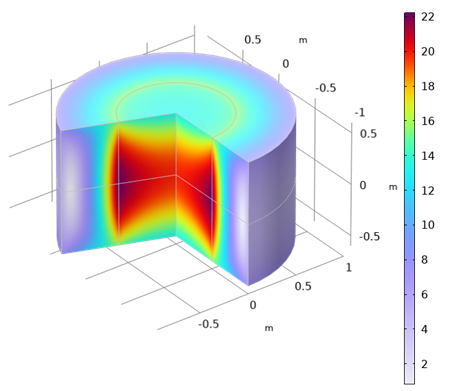 }
\caption{A first result of configuration optimization for the design of a Target Solenoid Model Coil and corresponding field map in nominal operating conditions.}
\label{fig:Target_fig4}
\end{figure}

\paragraph*{Target Solenoid Model Coil}
A magnet system of this field and dimension is a very challenging realization, depending on the success of a new technology, HTS, which is not yet deployed on large scale. This is why, as part of the next study phase, we propose to design, build and test a Target Solenoid Model Coil that shall demonstrate HTS force-flow cooled magnet technology at relevant scale, addressing two of the challenges listed above. The optimal model coil configuration is presently under study, balancing performance in relevant conditions vs. affordable cost. A suitable configuration for the model coil is shown in Figure~\ref{fig:Target_fig4}, a solenoid with a 1 m inner bore diameter, 2.3 m outer diameter and 1.4~m height. Preliminary targets chosen to map closely the operation of the coils of the target, decay and capture channel, are:
\begin{itemize}
\item	Bore field of 20 T at 20 K operating temperature;
\item	Electromagnetic pressure J B R in excess of 500 MPa;
\item	Stored energy in excess of 100 MJ;
\item	Operating voltage of 2.5 kV;
\end{itemize}

Achieving above on a magnet of this size will give sufficient confidence in the realization of the full magnet system. The details of the proposal, including timeline, milestones, deliverables and resources, are detailed later, as well as in the companion paper.

\subsection*{6D cooling channel solenoids}
\label{sec:6DCooling}

\begin{figure}
\centering
\includegraphics[width=\textwidth]{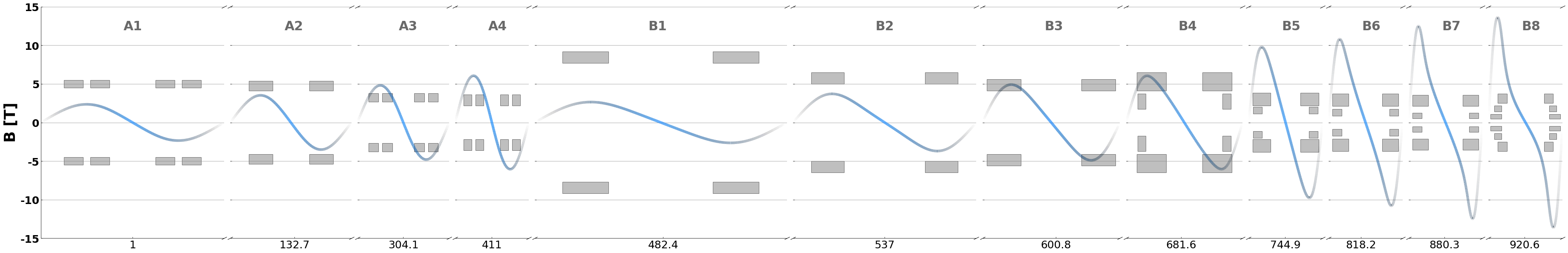}
\caption{Condensed schematic of the 12 types of cooling cells (A1 to B8) of a muon collider from the MAP configuration \cite{stratakis2015rectilinear}, with solenoid cross sections and the on-axis $B_z$ field assuming each cell is in a lattice of neighboring cells of the same type. z-axis values shown correspond to the the middle of the first cell of each type.}
\label{fig:6DCooling_fig1}
\end{figure}

The 6D “beam cooling” process occurs over a ~ 1 km long sequence of tightly integrated absorbers, alternating polarity solenoids, and RF cavities. The US-MAP design study provided a baseline configuration of a 6D cooling channel, consisting of 826 cooling cells over an approximate 970 m distance, with a total of almost 3000 solenoids \cite{stratakis2015rectilinear}. These cells can be divided into 12 unique types termed A1-A4 and B1-B8 and ranging in length of 0.8 m to 2.5 m. Each cell type contains between 2 to 6 solenoids, with a total of 18 unique solenoid types. For example, cell A1 has 4 solenoids, all of the same type, labeled A1-1, while cell B8 has 6 solenoids, of three different types, labeled B8-1, B8-2, and B8-3. The solenoids exhibit a diverse range of parameters, from small-bore to large-bore (90\,mm to 1.5\,m) and modest field to high field on-axis (2.6\,T to 13.6\,T). Each cell repeats a certain number of times (Ex. cell A1 repeats 66 times), before progressing to the next cell type. Fig. \ref{fig:6DCooling_fig1} displays the on-axis field of each cell type (assuming it is nested in a lattice of same-type neighboring cells), and the solenoid cross-sections. 

We performed an analysis on these solenoids, considering them operating individually and within their respective lattice. The results are reported in Table \ref{tab:6DCooling_tab1}. We found substantial stresses (large hoop and tensile radial stress), forces (37 MN axial force), and quench management challenges (energy densities up to 91 MJ/m$^3$ in a single coil).

\begin{table}[h!]
\centering
\begin{tabular}{ccc|ccccc}
\hline
Cell & $E_\text{Mag}$ & $e_\text{Mag}$ & Coil & \textbf{$J_E$} & $B_\text{peak}$   & $\sigma_\text{Hoop}$  & $\sigma_\text{Radial}$  \\

& (MJ) & (MJ/m$^3$) & & (A/mm$^2$) & (T)  & (MPa) & (MPa) \\
\hline
A1 & 5.4 & 20.5 & A1-1 & 63.25 & 4.1   & 34 & -5/0 \\
A2 & 15.3& 75.8 & A2-1 & 126.6 & 9.5   & 137 & -28/0 \\
A3 & 7.2 & 72.8 & A3-1 & 165  & 9.4    & 138 & -29/0 \\
A4 & 8.4 & 91.5 & A4-1 & 195  & 11.6   & 196 & -49/0 \\
B1 & 44.5& 55.9 & B1-1 & 69.8 & 6.9    & 95 & -14/0 \\
B2 & 24.1& 61.8 & B2-1 & 90   & 8.4    & 114 & -20/0 \\
B3 & 29.8& 88.1  & B3-1 & 123  & 11.2   & 173 & -37/0    \\
B4 & 24.1& 42.4 & B4-1 & 94   & 9.2    & 231 & 0/20 \\
   &     &      & B4-2 & 70.3 & 7.8    & 66 & -24/0 \\
B5 & 12  & 86.3 & B5-1 & 157  & 13.9   & 336 & 0/21 \\
   &     &      & B5-2 & 168  & 12.3   & 159 & -55/0 \\
B6 & 8.2 & 68.3 & B6-1 & 185  & 14.2   & 314 & -1/22 \\
   &     &      & B6-2 & 155.1 & 10.3  & 118 & -43/0 \\
B7 & 5.6 & 58.6 & B7-1 & 198  & 14.2   & 244 & -1/21 \\
   &     &      & B7-2 & 155  & 10.1   & 118 & -37/0 \\
B8 & 1.4 & 20.3 & B8-1 & 220  & 15.1   & 255 & -3/22 \\
   &     &      & B8-2 & 135  &  6.2   & 110 & -2/5 \\
   &     &      & B8-3 & 153  & 6.2    & 41  & -23/0 \\
\hline
\end{tabular}
\caption{Table of various parameters for 12 cell types and 18 unique solenoid types in the MAP configuration. Values correspond to solenoids operating in their respective cells within a lattice. Note that if the solenoid is operating stand-alone or in a single cell, some parameters take on higher or lower values.}
\label{tab:6DCooling_tab1}
\end{table}

Such values suggest that the solenoid configuration may need further optimization to reach engineering level. This is a non-trivial task, because the beam optics in the solenoids of the muon cooling channel is far from being formalized as well as that of a collider, and any change in magnet engineering may have dramatic effects on beam transmission. Recognizing this challenge, we have developed solenoid design rules that implement simple engineering limits on operating margin, stress and stored energy density. These rules are integrated already at the stage of the beam optics design, thus anticipating magnet performance limits. Having the design rules as part of the beam optics optimization has reduced the iteration time and improved effectiveness of each design optimization. In parallel, we have developed numerical optimization tools which can scan design variants and improve the solenoid configuration given a desired field profile. Such tools allow to converge to optimal engineering solutions once the initial solenoid configuration is close enough to being feasible. These two advances, the solenoid design rules and optimization tools, are described in Ref.~\cite{Fabbri6D} and in the next sections.

\paragraph*{Solenoid design rules and limits}
Solenoid engineering design rules have been defined and are presently used by the IMCC to improve upon the configuration originated by US-MAP and constrain new evolving optics studies and further magnet optimization studies. The design rules are analytical or semi-analytical expressions (see \cite{Fabbri6D}) that provide solenoid performance limits, based on material and engineering parameters such as the superconductor critical current density ($J_c$) and the required operating margin, the known superconductor behavior under mechanical stresses $\sigma_{r}$, $\sigma_{\theta}$ and $\sigma_{z}$ and strain, the associated mechanical limits on structural materials, and magnet protection for given stored magnetic energy density ($e_m$). To assess solenoid performance limits we generally assume that the superconductor is HTS (ReBCO) using critical current and stress limits that can be obtained from state-of-the-art industrial production \cite{Fujikura}, \cite{fujita2019flux}. The $J_c$ dependence on the operating temperature $T_{op}$ and the field $B_{op}$ was based on values obtained from measurement in field perpendicular to the tape plane (in a solenoid this is $B_r$) \cite{Bordini2024}, which is a conservative assumption. 

For the analyses reported here, we have set an operating temperature at $T_{op} = 20$ K and request an operating margin of 2.5 K. The maximum average hoop stress ($\sigma_{\theta}$) is limited to 300 MPa \cite{Fujikura} \cite{weijers2010high}. Although HTS has showed resilience at higher values, we take a considerable margin to account for 3D stress distribution and the potential of induced stresses during quench, from magnetization currents, and other engineering uncertainties. For the radial stress ($\sigma_{r}$), we consider a maximum compressive stress of 300 MPa and a maximum tensile stress of 20 MPa \cite{Fujikura}. The maximum tolerated tensile $\sigma_{r}$ before degradation of the superconductor is approximately 10-100 MPa \cite{maeda2013recent}. To avoid any tensile $\sigma_{r}$, a coil can be wound in tension generating a compressive pre-stress, such that there is no tensile $\sigma_{r}$ when energized \cite{song2017engineering}, making our initial tensile $\sigma_{r}$ limit possibly conservative. Although we have no analytic description of the stress parallel to the axis of the solenoid ($\sigma_z$), we note that a compressive $\sigma_z$ can be tolerated up to 100 MPa \cite{Fujikura}.

Lastly, the energy stored in these superconducting solenoids will be very large, and in the event of a quench (loss of superconductivity in the conductor), this energy will dissipate into heat. To prevent damage to the magnet during a quench from excessive temperature rise and induced stresses from non-uniform material expansion, managing this stored magnetic energy is crucial. At this preliminary stage, a simplified estimation of the temperature rise during a quench can be computed assuming the magnetic energy is deposited homogeneously in the solkenoid. For a magnetic energy density in the range of 150 MJ/m$^3$ (corresponding to about 17 kJ/kg), the temperature rise of a HTS tape would be about 130 K. Although we are aware that much more effort is required for quench modeling and that the uniform temperature distribution is an overoptimistic simplification of the system, such temperature rise is modest, and we take this range of energy density as an initial acceptable upper limit. Future detailed quench analysis studies will be absolutely necessary.

The engineering limits are summarized in Table \ref{tab:6DCooling_tab2}. They provide an initial framework for iterating the design study of all 6D cooling solenoids, while in parallel more detailed engineering analysis is being carried out for a proof-of-principle 6D cooling cell demonstrator with HTS solenoids. 

\begin{table}
\begin{center}
\begin{tabular}{lcccccc}
\hline\hline
\textbf{Parameter} & Unit & Lower bound & Upper bound \\
\midrule
\textbf{$\sigma_\theta$} & MPa & 0 & 300\\
\textbf{$\sigma_{r}$} & MPa & -300 & 20 \\
\textbf{$e_m$} & MJ/m$^3$ & - & 150 \\ 
\hline\hline
\end{tabular}
\caption{Limits on select single solenoid parameters.}

\label{tab:6DCooling_tab2}
\end{center}
\end{table}

\begin{figure}
\centering
\includegraphics[width=\textwidth]{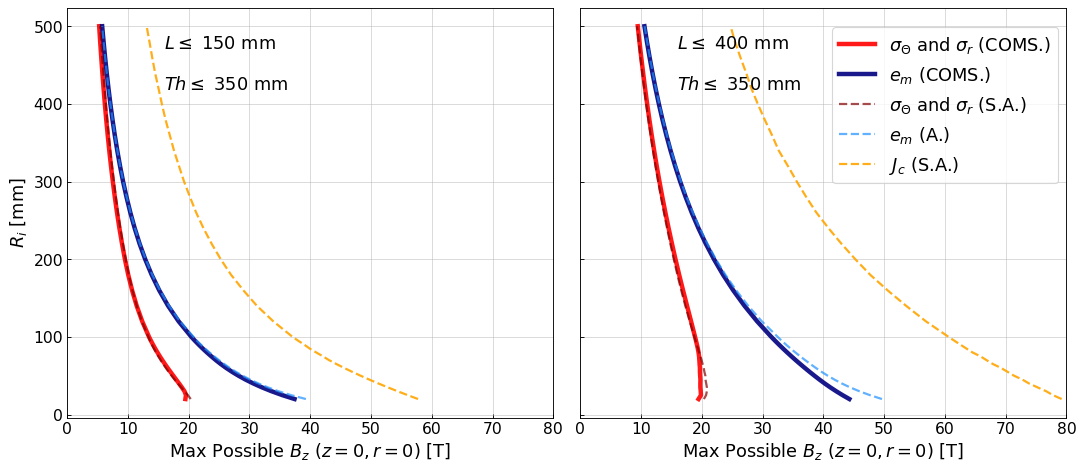}
\caption{The approximate maximum possible $B_z$ on-axis versus bore radius ($R_i$) of a single solenoid with a maximum thickness of 340 mm and a length up to 150 mm (left) or 400 mm (right), for different limiting parameters: red curves correspond to stress limits, blue to the magnetic energy density limit, and orange to the critical current density. Solid lines correspond to results found numerically with COMSOL (COMS.), and dashed lines to analytic or semi-analytic Eqs.}
\label{fig:6Cooling_fig2}
\end{figure}

A useful visualization tool of the key design parameters and their corresponding imposed limits described above is a plot of magnet aperture ($A$) vs. magnetic field ($B$), which we dubbed ($A$-$B$)-plot, where $B$ is the maximum possible field on-axis. Such plots have been part of the conceptual design process for the dipoles of the collider \cite{Novelli2024}, and here we extend this concept to the solenoids of the 6D cooling. However, when considering solenoids there is added complexity because both the length and thickness of a solenoid can vary at a specified aperture. Figure \ref{fig:6Cooling_fig2} shows two example A-B plots for different limits of $L$, with limit curves of $\sigma_r$ $\sigma_{\theta}$ obtained semi-analytically.

\paragraph*{Solenoid Optimization Tools}

While initial configurations resulting from the beam optics studies may satisfy the design guidelines described above, they are not necessarily optimized engineering solutions. To improve the magnet configuration, we have created a numerical tool, partly written in-house and partly based on proprietary software (COMSOL), termed the Solenoid in-Cell Optimization program (SiCO). This program is built to optimize solenoids that produce a desired field profile and tolerance. It can be broken down into three steps, characterized by set-up, computation, and filtering based on the desired field profile and design rules. With this tool millions of solutions can be computed very quickly, allowing the choice of the best solenoids depending on weighted design criteria such as stress, stored energy, or coil volume (and associated cost). In addition, the code can cope with considerations of standardization (choosing solenoids with identical geometry across cells), or powering and quench protection.

\begin{figure}
\centering
\includegraphics[width=0.49\textwidth]{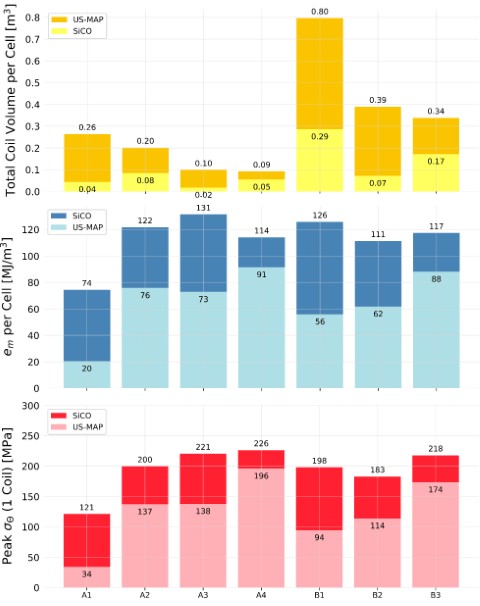}
\caption{Results of the minimum possible coil volume per cell (top) for cells A1 to B3, using the numerical optimization procedure (SiCO) compared to MAP values. The corresponding cell magnetic energy densities (middle) and peak average hoop stresses in each coil type (bottom) are shown, where values were computed with COMSOL for cells nested in lattices.}
\label{fig:results}
\end{figure}

We used this code to analyze cells A1 to B3 of MAP, considering 2 coils per cell. Analysis of cells with coils at multiple radii (B4-B8) is ongoing and will be presented in the future. Figure \ref{fig:results} presents the minimum achievable volume of conductor per cell for A1-B3, and the corresponding cell $e_m$ and single coil stress $\sigma_{\theta}$ (values computed with COMSOL for solenoids in a lattice).

To achieve the minimum volume while maintaining the field profile, the current densities increase to values ranging from 160 A/mm$^2$ (B1-1) to 402 A/mm$^2$ (A3-1). As expected, the hoop stresses and stored magnetic energies also increase (see Tab. \ref{tab:6DCooling_tab1} for comparison). However, other parameters stay similar or improve. The compressive radial stresses are low (maximum of 35.4 MPa in coil A4-1, compared to 49 MPa in MAP), with no tensile radial stress. The peak fields in the conductor are similar, with a maximum fraction of $J_c$ reached in A3-1, with $J = 0.57 J_c$ ($B_{\text{peak}} = 8.2$ T). The net longitudinal forces are significantly smaller across all the solenoids ($F_z = 36.8$ MN in MAP compared to $F_z = 12.4$ MN here for B3-1).

This solution set demonstrates the power of this numerical optimization tool to search for solenoid configurations depending on weighted design criteria and technology options. It provides an excellent starting point for more detailed mechanical analysis, where parameter limits can be easily changed to generate new solution spaces depending on evolving understandings.

\paragraph*{First evaluation of present baseline cooling optics from IMCC}
We have applied the above procedure to the new optics developed recently for the 6D cooling channel. The new otpics achieves an output transverse emittance that is half of what was achieved in previous studies \cite{zhu2024performance}. This will aid in reaching a lower overall final emittance before acceleration and collision, a substantial gain of performance for the collider. During these beam dynamics studies, solenoid geometries and their corresponding field maps (among other parameters) are iterated on. To constrain the allowable magnet geometries and current densities, the `design rules' described above were directly integrated into the beam optics optimization routine. This yielded a final optics with assumed solenoid geometries within or near allowed design limits, summarized in Table \ref{tab:6DCooling_tab3}. 

\begin{table}[!h]
\begin{center}
\begin{tabular}{lccc|lcccc}
\hline\hline
Cell & No. of Cells & $E_\text{Mag}$ & $e_\text{Mag}$ & Coil & \textbf{$J_E$} & $B_\text{peak}$   & $\sigma_\text{Hoop}$ &$\sigma_\text{Radial}$ \\

& & (MJ) & (MJ/m$^3$) & & (A/mm$^2$) & (T)  & (MPa) & (MPa)  \\
\hline\hline
A1 &  58& 5.4   & 21     & A1-1   & 57.6  & 5.2   & 42   & -8/0 \\
A2 &  89& 22.1  & 106.1  & A2-1   & 149.5 & 11.6  & 194 & -48/0 \\
A3 &  81& 5.0   & 49.5   & A3-1   & 131.5 & 10.1  & 121 & -25/0 \\
A4 & 124& 8.0   & 92.3   & A4-1   & 193.2 & 13.8  & 225 & -51/1 \\
B1 &  24& 9.1   & 49.8   & B1-1   & 96.9  & 7.7   & 104 & -24/0 \\
B2 &  34& 15.6  & 64.2   & B2-1   & 102.1 & 9.2   & 131 & -32/0 \\
B3 &  54& 36.9  & 105.9  & B3-1   & 127.9 & 12.9  & 208 & -57/0 \\
B4 &  55& 75.6  & 149.9  & B4-1   & 88.5  & 16.1  & 260 & -1/29 \\
B5 &  55& 17.3  & 88.9   & B5-1   & 179.6 & 14.7  & 295 & -2/17 \\
B5 &    &       &        & B5-2   & 154.0 & 14.7  & 212 & -57/1 \\
B6 &  55& 8.3   & 96.6   & B6-1   & 214.4 & 15.3  & 339 & -5/18 \\
B6 &    &       &        & B6-2   & 211.5 & 12.0  & 214 & -6/6 \\
B6 &    &       &        & B6-3   & 212.7 & 12.4  & 162 & -46/0 \\
B7 &  51& 8.2   & 87.7   & B7-1   & 183.3 & 14.7  & 264 &  0/25 \\
B7 &    &       &        & B7-2   & 153.9 & 11.1  & 175 & -4/10 \\
B7 &    &       &        & B7-3   & 210.3 & 13.2  & 180 & -45/1 \\
B8 &  69& 8.8   & 92.1   & B8-1   & 193.7 & 16.5  & 270 & -6/38 \\
B8 &    &       &        & B8-2   & 202.1 & 15.4  & 270 & -6/29 \\
B8 &    &       &        & B8-3   & 212.8 & 13.2  & 187 & -50/0 \\
B9 &  53& 7.5   & 76.5   & B9-1   & 256.4 & 17.2  & 281 &  0/37 \\
B9 &    &       &        & B9-2   & 88.4  & 10.0  & 95  & -2/12 \\
B9 &    &       &        & B9-3   & 204.9 & 13.2  & 184 & -46/0 \\
B10 & 49& 5.0   & 68.6   & B10-1  & 326.8 & 19.2  & 378 &  0/49 \\
B10 &   &       &        & B10-2  & 146.1 & 11.1  & 105 & -4/13 \\
B10 &   &       &        & B10-3  & 207.8 & 12.5  & 158 & -43/1 \\
\end{tabular}
\end{center}
\caption[Solenoid types in the latest 6D cooling optics]{Table of various parameters for the 14 cell types and 26 unique solenoid types in the latest 6D cooling optics \cite{zhu2024performance}. Number of cells refers to 1 channel. Hoop stress values reported correspond to the maximum, while radial stress values reported are the minimum/maximum. All values correspond to solenoids operating in their respective cells within a lattice. Note that if the solenoid is operating stand-alone or in a single cell, some parameters take on higher or lower values.}
\label{tab:6DCooling_tab3}
\end{table}

This latest initial optics configuration has a total of 3030 solenoids in one 6D cooling chain, with a peak field on-axis broadly increasing from 2.6 T to 17.9 T. The total number of solenoids (6060) represents the largest percentage cost contribution considering magnet materials, consumables, and labor, of a 3 TeV muon colliders, driving significant optimization efforts to minimize the cost. There are 26 unique solenoid types, with bore radii ranging from 25 mm to 400 mm, lengths from 75 mm to 287 mm, and current densities from 58 A/mm2 to 327 A/mm2. As seen in Table \ref{tab:6DCooling_tab3}, the solenoids experience substantial peak fields at the conductor (up to 19 T), large stresses and forces. However these values are not far off target limits, and can be optimized further. The initial solenoid configurations also exhibit tight spacing, and need to better factor in room for RF structure and waveguides. These additional parameters (magnet spacing, RF required spacing), will be factored into the next iteration of the beam optics optimization. 

This is presently work in progress, but demonstrates already that the preparatory work is paying back with beam optics solutions closer to engineering feasibility. Our plan is to proceed towards an updated configuration, taking into account spacing requirements, and apply the SiCO numerical optimization to search for more ideal solenoids given desired field profiles.

\paragraph*{Solenoid design applied to the demonstrator}
Identified as crucial technology demonstrators are a 6D cooling cell demonstrator and an RF test stand (a first simplified HTS split coil for demonstration of HTS technology and RF for a cooling cell). The 6D cooling cell demonstrator, described in Chapter \ref{1:tech:sec:cool_cell}, will address a slew of magnet engineering difficulties seen across the various types of cooling cells, including large forces, stresses, stored energies and fields at the conductor. The magnet design guidelines and optimization procedure described above has been implemented, together with other optimization tools from INFN, in the magnet design of the 6D cooling cell demonstrator. This demonstrator will provide valuable feedback on the design process for all 6D cooling channel solenoids, including further simulation tools such as detailed quench analysis.

\paragraph*{Challenges identified}

The studies have highlighted two major challenges to be addressed in priority:
\begin{itemize}
\item	Compact solenoid windings, achieving performance at minimum cost. The field reach of the single solenoids of the 6D cooling channel is not extraordinary. Indeed solenoids of this class have already been built. But the number of solenoids required is large. High current density, hence minimal use of superconducting material, is a key to making the 6D cooling practical and affordable. This implies mastering large forces and quench in compact windings, which needs demonstration. Note that high current density is also a key to reaching the required field gradients, see also below;
\item	Integration. A unique feature of the solenoids in the cooling cell is the need to generate an alternating field profile with high gradient, and host RF cavities and absorbers. This imposes opposing constraints on distancing and spacing, including the management of large electromagnetic forces among solenoids of opposite polarity, access requirements, and effective thermal management in tight space. Also this challenge requires practical demonstration;
\end{itemize}

\subsection*{Final cooling solenoids}
\label{sec:FinalCooling}
Among the solenoids in the cooling channel, the final cooling solenoids are most challenging in terms of field performance, with a target of 40 T or higher. The bore dimension is relatively small, 50 mm, which makes them an ideal development vehicle to implement new technology such as non-insulated windings, and probe performance limits. Indeed, this solenoid is an ideal vehicle for R\&D, allowing fast turn-around models and tests that are relevant to magnets of larger dimension, such as the solenoids for the 6D cooling. This is why we have advanced very swiftly in the conceptual and engineering design of the 40 T final cooling solenoid. 

We have proposed a conceptual design at the early stage of the study \cite{Bordini2024}. The solenoid concept is based on soldered single pancakes, stacked with stress-management plates, and joined electrically. The coil is pre-compressed radially by a solid mechanical structure, supporting the electro-magnetic loads, and necessary to avoid tensile stress in the coil at nominal operating conditions.

\paragraph*{Concept and Engineering Design}

\begin{figure}
\centering
\includegraphics[width=\textwidth]{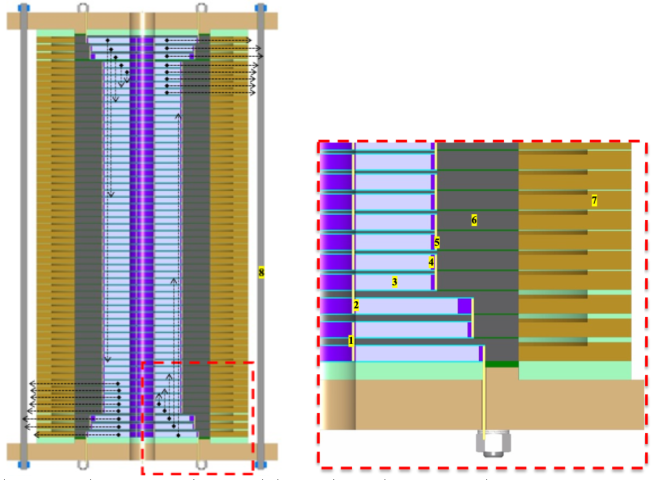}
\caption{Cross-section of the 40 T solenoid; the arrows indicate the axial and radial Lorentz forces acting on each pancake. The lengths of the radial arrows have been scaled down by a factor of 3.}
\label{fig:FinalCooling_fig1}
\end{figure}

The concept for the final cooling solenoid is shown schematically in Figure~\ref{fig:FinalCooling_fig1}, and it was developed at the outset of the study profiting from experience in other fields of science and societal applications. In the past twelve months we have focused on the development of engineering solutions for the realization \cite{bordini2025development}. Figure \ref{fig:FinalCooling_fig1} shows 46 identical modular pancakes and three pairs of thicker single pancakes at both ends of the solenoid. In Figure~\ref{fig:FinalCooling_fig1}, the arrows indicate the axial and radial Lorentz forces acting on each pancake, with their lengths proportional to the magnitude of the respective forces. To enhance readability, the lengths of the radial arrows have been scaled down by a factor of 3 relative to the axial arrows. Despite this difference in arrow lengths, the radial forces are nearly equal to the axial force on the outermost pancake, which is around 300 tons.

Moving radially outward from the solenoid axis and ignoring the two axial extremities, the components are as follows:
\begin{enumerate}
\item	Internal Electrical Connection: This is a superconductor carrying an axial current, represented by yellow lines in Figure~\ref{fig:FinalCooling_fig1}, which connects two adjacent pancakes in series.
\item	Internal Joint Ring: A ~0.5 mm thick normal conducting ring that is electrically connected to the Pancake Coil and the Internal Electrical Connection.
\item	Pancake Coil: This coil consists of ReBCO tape wound around the Internal Joint Ring, with adjacent turns soldered together to form a continuous solid block. For the Modular Pancakes, the coil features an inner radius of 30 mm and an outer radius of 90 mm, while the End Pancakes have a larger outer radius.
\item	External Joint Ring: A ~ 5-mm-thick normal conducting ring electrically connected to the Pancake Coil and the External Electrical Connection.
\item	External Electrical Connection: A superconductor carrying an axial current, represented by yellow lines in Figure~\ref{fig:FinalCooling_fig1}, which connects two adjacent pancakes in series or links the solenoid extremities to the current leads.
\item	Support Cup: A single stainless steel (SS) piece, shown in dark grey, consisting of a 12 mm high disk and a ~2 mm thick radial plate that separates two adjacent Pancake Coils. The Support Cups serve a dual purpose: (1) providing radial stiffness and pre-compression to counteract the radial expansion of the coils caused by Lorentz forces, and (2) intercepting axial Lorentz forces \cite{Bordini2024}. The pre-compression on each pancake coil is achieved through shrink fitting at room temperature after heating the Support Cup to approximately 100-200°C.
\item	Pre-Compression Disk : A SS (or another structural material) disk, depicted in light brown, with a height of 12 mm in the central part and approximately 14 mm at the periphery. The Pre-Compression Disk delivers most of the radial pre-compression through shrink fitting at room temperature. In this scenario, the Pre-Compression Disk, located relatively far from the Pancake Coil, would be heated to temperatures above 200°C.
\item	Axial Rods: Stainless steel bars that ensure good contact between adjacent Support Cups through the two flanges at the magnet’s axial extremities.
\end{enumerate}

Mechanical calculations \cite{accettura2024mechanical} indicate that, when neglecting the contribution of magnetization and assuming sufficiently stiff radial plates, this design can maintain stress and strain within safe limits for the superconductor during regular operating conditions (excluding quenches). Table \ref{tab:FinalCooling_tab1} provides an example of the calculated mechanical loads on the conductor at various stages: Room Temperature (RT), Step 1; at 4.2 K with no current in the conductor, Step 2; and at 4.2 K with the magnet energized to 40 T, Step 3. The table specifically refers to a case with a RT pre-compression of 200 MPa and an Internal Joint Ring thickness of 0.5 mm. These findings, along with other case studies, are discussed in detail in \cite{accettura2024mechanical}.

\begin{table}[h!]
\centering
\begin{tabular}{l|ccc}

 & Step 1 & Step 2 & Step 3 \\ \hline
Radial Stress [MPa]: Min/Max & -205/-8 & -190/-5 & -290/10 \\ 
Hoop Strain [\%]: Min/Max & -0.25/-0.10 & -0.20/-0.12 & -0.04/0.28* \\ 
\end{tabular}
\caption{Radial Stress and hoop Strain values across steps.}
\label{tab:FinalCooling_tab1}
\end{table}

\paragraph*{Quench Protection}
The proposed design achieves a relatively low energy per unit length of approximately 5.4 MJ/m, thanks to the extreme compactness of the coil \cite{Bordini2024}. However, the magnetic energy per unit volume, or magnetic energy density, is relatively high at 300 J/cm³. If this energy were uniformly dissipated within the winding, the temperature would rise from 4.2 K to around 200 K. While 200 K is not considered a threat to the coil, the localized dissipation of this energy during a quench could irreversibly damage the magnet.

In traditional insulated, unprotected, low-temperature superconducting (LTS) coils, a localized quench typically results in conductor melting. To prevent this, detection systems are used to monitor resistive voltages and initiate a fast discharge, redirecting the energy to an external resistor via high voltage. An alternative strategy involves triggering a uniform quench across the entire coil, typically using resistive heating. In LTS accelerator magnets, this is achieved with internal heaters or current/field oscillations.

However, high-temperature superconducting (HTS) coils present unique challenges. Their substantial enthalpy margin and low quench propagation speed make quench detection using voltage measurements difficult. Furthermore, using quench heaters to uniformly quench the entire coil is impractical.

These challenges make it well known that insulated HTS coils with high magnetic energy density are very difficult to protect effectively against localized quenches. Consequently, the choice of non/metal-insulated (N/M-I) coils for the proposed conceptual design was strongly influenced by these considerations.

These coils have low thermal and electrical resistance between turns, enabling rapid quench propagation and reducing localized energy dissipation. N/M-I coils could also allow quench detection based on fast magnetic flux variations resulting from a quench due to the sudden increase/decrease of radial and azimuthal currents in the quenched region. Upon quench detection, the entire solenoid could be quenched by injecting a pulsed current between its terminals, with the current mainly flowing radially, leading to uniform rapid heating of the entire solenoid. The primary advantage of this quench strategy is achieving a symmetric and controlled quench, where the generated mechanical forces are known and reproducible \cite{Bordini2024}. 

Various protection strategies are currently under investigation \cite{mulder2025thermo, mulder2024quench}. We recently investigated the potential of a capacitor discharge (CD) method for magnet protection, which, upon quench detection, injects a large current pulse into the full stack or individual pancakes, generating heat through the coil’s turn-to-turn resistance. Most of this current flows through the low-inductance, radial turn-to-turn paths between the terminals of each pancake coil. The energy from the current pulse is dissipated as heat within the pancakes, using the turn-to-turn resistances as internal quench heaters. This approach rapidly raises the conductor temperatures above the critical temperature within milliseconds and requires no additional internal components, as it can use the magnet’s existing current leads \cite{mulder2025thermo, mulder2024quench} 

\paragraph*{Magnetization and current distribution}

To assess the impact of persistent currents on the mechanical loads acting on the Pancake Coils, a 2D thermal-electromechanical model was developed using COMSOL Multiphysics. The electromagnetic analysis was performed using a T-A formulation \cite{mulder2025thermo}, and the resulting Lorentz forces were then applied as input to a thermo-mechanical model that simulates all 750 turns of the Modular Pancake, with a level of detail similar to the 2D model presented in \cite{accettura2024mechanical}. For the electromagnetic model, it was assumed that adjacent turns of the winding are electrically insulated, which, although not entirely accurate, is considered conservative as it tends to yield higher mechanical load values.

The model shows that, due to persistent currents, the Modular Pancake closest to the End Pancakes experiences a 30\% increase in the maximum strain on the conductor. This increase significantly diminishes in the subsequent pancakes, with the sixth Modular Pancake from the End Pancakes showing an increase  of  no  more  than  1\%.  The  End  Pancakes,  however, exhibit an increase in maximum strain on the conductor well above 30\%. Regarding the axial Lorentz forces, magnetization tends to reduce the axial Lorentz forces by straightening the magnetic field lines (Magnetization currents generate a radial field that opposes the radial field produced by the transport current.). For example, the aforementioned calculations show that the axial Lorentz forces acting on the outermost pancake are reduced from approximately 300 tons to around 260 tons. To mitigate conductor magnetization and the associated hoop strain we are exploring the use of striated tapes or the End Pancakes and for a few adjacent Modular Pancakes. 

Progress has also been made in assessing the impact of quenches on the mechanical loads of the conductor. For this purpose, a 2D axisymmetric electrical and thermal network model developed in Python was coupled with a simplified 2D mechanical model in Comsol. Preliminary results indicate that in some quench scenarios, the maximum hoop strain could increase by up to 30\%  \cite{mulder2025thermo}.

Currently, as noted in \cite{HahnEuCAS}, there is no 3D model capable of accurately describing the transient behavior of large non-insulated (NI) ReBCO superconducting coils during quench events. However, a thorough understanding of these transient phenomena is essential to fully exploit the potential of this technology. While 3D Finite Element Method (FEM) models are the most promising approach for accurately capturing the magneto-thermal dynamics of these magnets, the widely used H formulation of Maxwell’s equations for superconductors \cite{HoellInternal} remains computationally too intensive for simulating large-scale systems like accelerator magnets. To the authors' knowledge, 3D models based on the H-formulation are currently limited to very short lengths of superconducting tapes (a few meters) and even shorter lengths in the case of multifilamentary wires.

To address this issue, a novel mathematical formulation has recently been developed at CERN and integrated into the commercial FEM software COMSOL. Initial results suggest that this new approach could significantly reduce the computational time required for large-scale 3D FEM transient simulations of superconductors, demonstrating its potential to streamline these complex analyses (2). The model has already produced significant results for the project, enabling the quantification of the time required to energize the magnet to its target field as a function of the contact resistance between adjacent turns.

In addition to the COMSOL model, we have developed a 3D model that uses the H formulation for the electromagnetic analysis coupled with a thermal model, utilizing the open-source FEM software GetDP. This model has already been successful in studying quench evolution in a 20-layer pancake \cite{wozniak2024influence}, and ongoing studies aim to reduce the computational time required by this approach, potentially enabling the simulation of larger systems.

\paragraph*{Experimental Studies}

The engineering studies outlined above are complemented by an intense campaign of electrical, mechanical and thermal measurements, necessary to establish the thermo-physical and mechanical properties of single tapes and stacks of tapes. 

\begin{figure}
\centering
\includegraphics[width=0.7\textwidth]{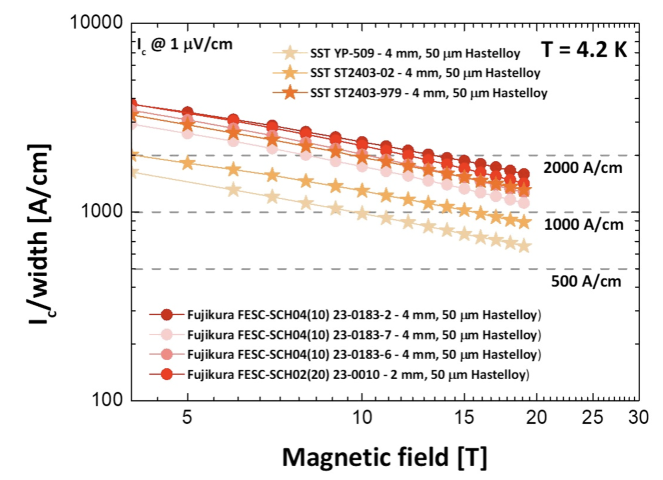}
\caption{Critical current measurements (triangles) at 4.2 K of procured ReBCO tape (B // c-axis).}
\label{fig:FinalCooling_fig2}
\end{figure}

We have procured over 10 km of 4-mm-wide tape from three different companies: Faraday Factory Japan, Fujikura, and Shanghai Superconductor Technology. The goal is to begin producing smaller coils in the initial phase of technological development to manage costs effectively. We have initiated the characterization of the superconducting properties of the procured tape through critical current measurements at 4.2 K in a background magnetic field, oriented perpendicular to the transport current direction and the wide face of the tape, with fields up to 19 T. These measurements were conducted at the University of Geneva. Figure~\ref{fig:FinalCooling_fig2} shows the results of the first sample measured which are outstanding, with an engineering current density exceeding 2 kA/mm2 at 16 T. This exceeds the performance requirements for the 40 T final cooling solenoid, proving that from this point of view the technology is accessible.

\begin{figure}
\centering
\includegraphics[width=\textwidth]{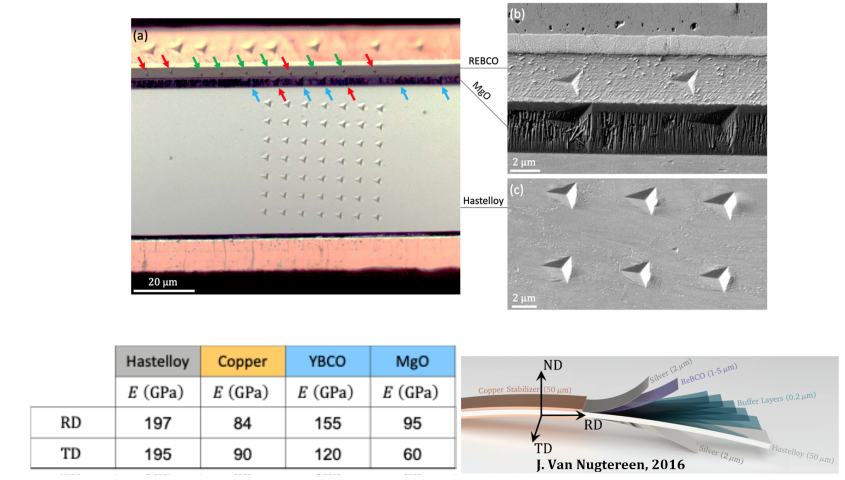}
\caption{(a) Optical image of the residual indentation imprints in the Hastelloy, copper, MgO (blue arrows) and REBCO (green arrows) layers. Red arrows indicate tests rejected due to being too close to other layers because of poor targeting. (b) SEM images of the residual indentation imprints in the REBCO and MgO layers. (c) SEM images of the residual indentation imprints in the Hastelloy layer. The measured values are summarized in the table.}
\label{fig:FinalCooling_fig3}
\end{figure}

\begin{figure}
\centering
\includegraphics[width=\textwidth]{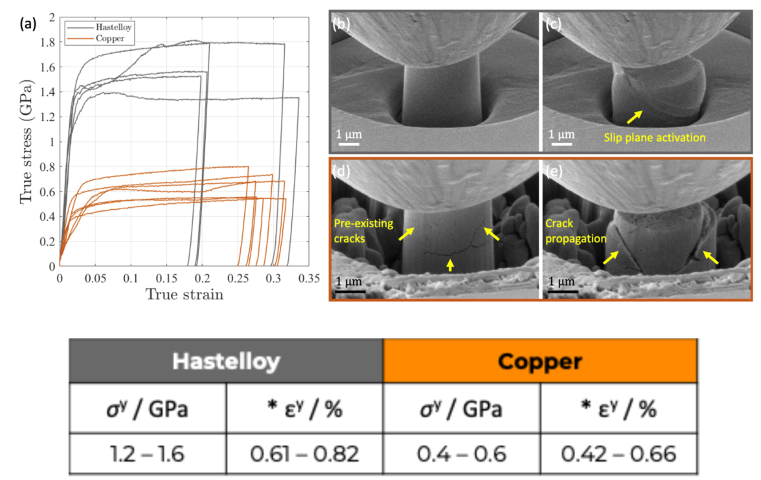}
\caption{(a) True stress-strain curves of Hastelloy and copper layers obtained by micropillar compression. (b)-(c) SEM images of a representative Hastelloy pillar before and after compression, respectively. (d)-(e) SEM images of a representative copper pillar before and after compression, respectively. The measured values are summarized in the table.}
\label{fig:FinalCooling_fig4}
\end{figure}

Accurate knowledge of the elastic-plastic properties, fracture toughness, and thermal expansion of ReBCO materials is crucial for precise stress analysis in superconducting magnet systems. To address this, we initiated a comprehensive measurement campaign to determine the thermomechanical properties of ReBCO tapes and stacks. The campaign includes both macro and micro mechanical characterizations. The macro-scale samples consist of individual tapes or stacks a few centimeters long \cite{vernassa2025elasticity}, while the micro-scale samples are micrometric pillars (micropillars) obtained through focused ion beam (FIB) milling of the individual layers of REBCO tapes \ref{fig:FinalCooling_fig4}. Additionally, nanoindentation measurements were performed \ref{fig:FinalCooling_fig3}. Although we are just at the beginning, this campaign has already yielded valuable results, including data on the thermal expansion properties of various REBCO tapes \ref{fig:FinalCooling_fig6}, as well as insights into the elastic modulus, yield stress, plastic flow behavior, and fracture toughness of the different layers within the REBCO tapes.

\begin{figure}
\centering
\includegraphics[width=\textwidth]{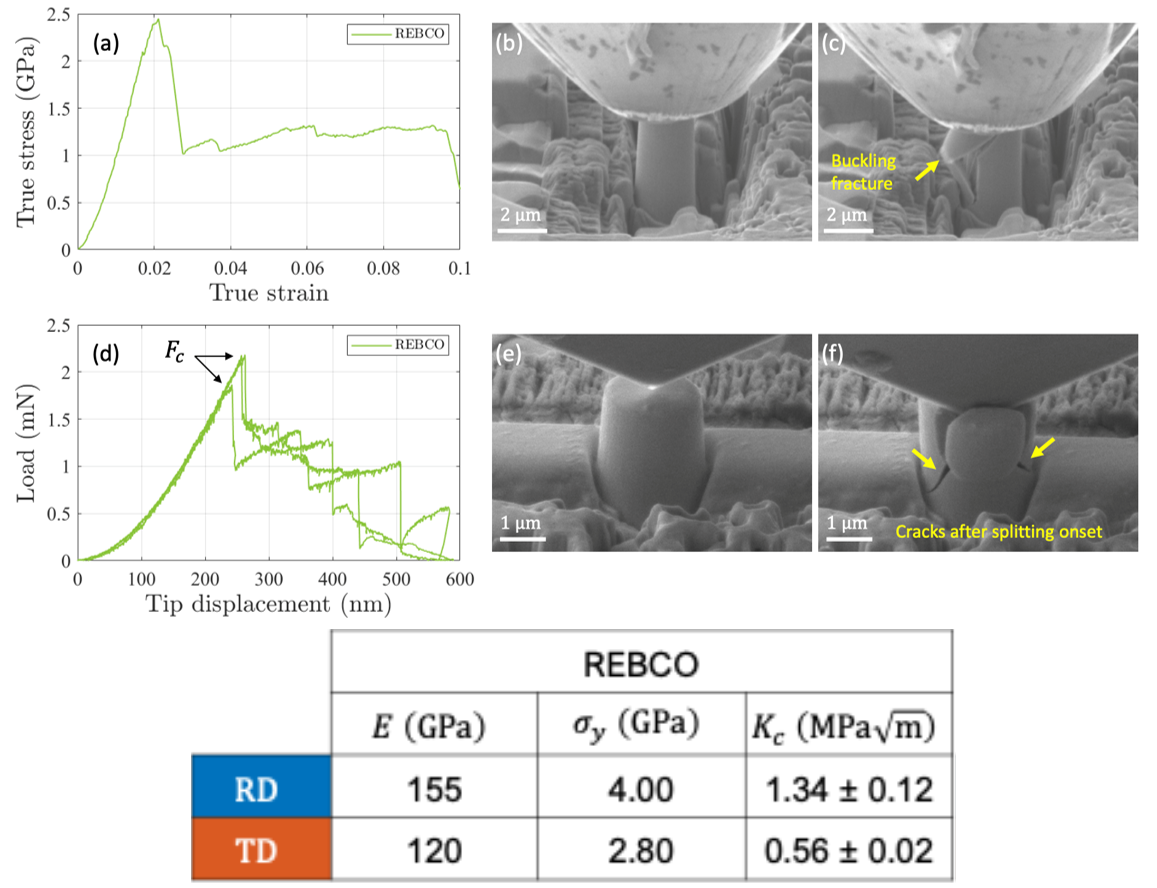}
\caption{(a) True stress-strain curve obtained by compression of a REBCO micropillar. (b)-(c) SEM images of the REBCO pillar before compression and after the first fracture, respectively. (d) Load-displacement curves obtained from splitting REBCO micropillars. (e)-(f) SEM images of a representative REBCO pillar before splitting and after the first fracture, respectively. The measured values are summarized in the table.}
\label{fig:FinalCooling_fig5}
\end{figure}

\begin{figure}
\centering
\includegraphics[width=\textwidth]{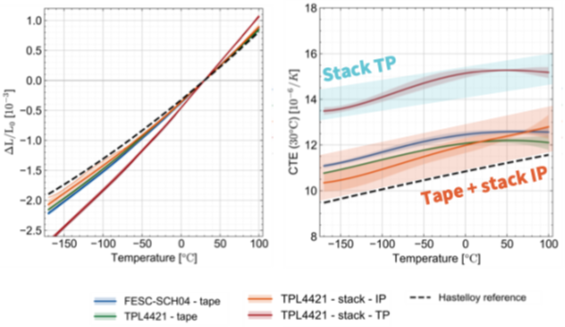}
\caption{Thermal contraction measurements of 2 ReBCO tapes (Theva TPL4421, and Fujikura FESC-SCH04) and of a soldered tape stack made with TPL4421. The measurements were carried out only along the In Plane (IP) direction for the tapes, while for the stack also the Through Plane (TP) measurements were performed }
\label{fig:FinalCooling_fig6}
\end{figure}

In parallel, we are starting the winding of single pancakes that match well the dimensions and properties of the final cooling solenoid design. These pancakes, tested singularly or stacked in small coils, will serve as the main R\&D vehicle to develop and validate engineering solutions. We are exploring two approaches for soldering the coils: either after the winding is completed or during the winding process. For the latter, we have designed and manufactured a custom winding machine, which is currently being commissioned. Meanwhile, we have initiated a testing campaign to evaluate the quality of different soldering techniques on tape stacks and small pancakes. After soldering, the samples were analyzed using X-ray Computed Micro Tomography and micrographic techniques to assess the level of residual porosity. For the pancakes, critical current measurements were also performed. The results obtained so far are encouraging, demonstrating that it is possible to completely fill the gaps between the tapes when precompression is applied during soldering. Moreover, the critical current measurements indicated that the tapes can be soldered without degrading their superconducting properties.

As a final remark on this aspect of the work performed, the methods developed and the data collected constitute a unique knowledge database which is useful also for other HTS magnets, relevant both to HEP as well as other fields of science and societal applications. This additional result, driven by the IMCC activities, deserves special recognition.

\begin{figure}
\centering
\includegraphics[width=\textwidth]{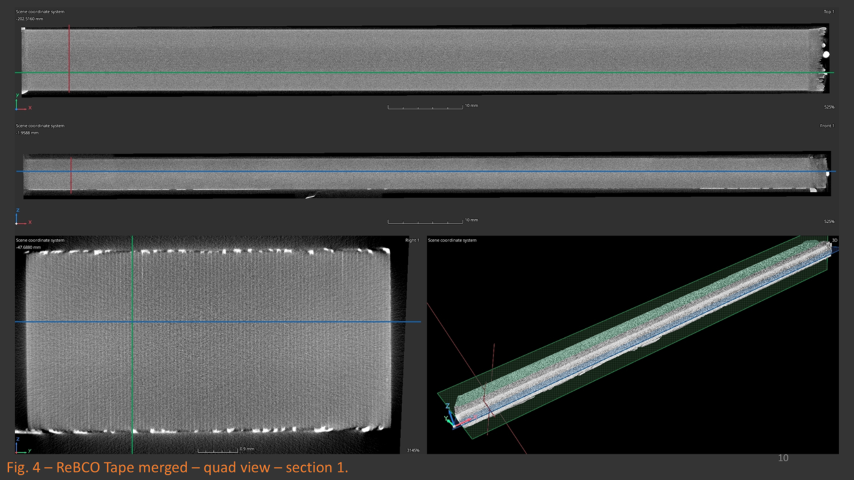}
\caption{Tomography of a stack of ReBCO tapes soldered under vacuum – negligible porosity achieved.}
\label{fig:FinalCooling_fig7}
\end{figure}

The plan in the coming months is to further increase experimental activities, especially the manufacturing and testing of pancakes and conducting critical current measurements at 4.2 K. Once the new winding machine is commissioned, we expect to produce a large number of pancakes in 2025. The inner radius of the superconducting winding is fixed at 30 mm, while for the outer winding diameter (OWD) different values will be tried. 

\begin{enumerate}
\item	We will start producing coils with OWD equal to 35 mm in order to minimize the use of conductor; the goal of these coils will be 
\begin{enumerate}
\item	Verify that we can solder together adjacent turns with a negligible porosity and without degrading the superconducting properties (tomography, microscopies, critical current measurements at 77 T);
\item	Asses the transverse contact resistance associated with fully soldered coils when using the 3 conductors procured.
\item	Study possible solution to modify the transversal contact resistance;
\item	Study the electrical properties of different joints solutions
\end{enumerate}

\item	We will then increase the winding thickness to 50 mm OWD, so that piling up a few of these coils will allow to reach significant field in the centre of the solenoid ( about 20 T at 4.2 K); main goals
\begin{enumerate}

\item	Verify the feasibility of applying a radial precompression via shrink fitting without damaging the superconductor and the joints (tomography , critical current measurements at 77 K) of a single pancake
\item	Verify that the coils do not degrade because of the Lorentz forces once energized at 4.2 K and with a current value that would allow to generate a field of 20 T when piling up several coils. 
\end{enumerate}

\end{enumerate}

Further increase of winding thickness, and higher field levels, will depend on the result of the above phases.

\paragraph*{Challenges identified}
Achieving a bore field of 40 T in the final cooling solenoid is a grand challenge, no such magnet exists today. Success will depend on mastering the following aspects:
\begin{itemize}
\item	Mechanics. The design stress is at the expected material limit, and we are aware that ReBCO tapes can stand minimal shear stresses along the superconductor plane and minimal tensile stress in the direction perpendicular to the superconductor plane. The design shall demonstrate excellent control of the large variation of the strain/stress state on the conductor before and after energization, with no stress concentration;
\item	Quench protection. This will depend on the ability to control the transverse resistance, as well as early quench detection to trigger a power supply trip, so to discharge the energy in a controlled fashion, preventing excessive temperatures and mechanical stresses/strains. A controlled and reproducible electrical and thermal transverse ensuring is crucial for magnet protection, while allowing for reasonable magnet energization times;
\item	Novel magnet technology. Several features need to be demonstrated, like the ability to control winding geometry and a soldering between winding with a minimal amount of porosities, to control pre-compression via shrink fitting and/or alternative methods making sure to avoid coil buckling or degradation, and joints that can properly distribute the current in the coil while avoiding excessive heating and mechanical stresses.
\end{itemize}

\subsection*{Accelerator magnets}
\label{sec:Accelerator}

The largest number and most challenging magnets of the acceleration chain are those in the Rapid Cycled Synchrotrons (RCS) and Hybrid Ramped Synchrotrons (HCS). Among the several configurations studied, we have settled on common specifications for the dipole magnets of all stages, namely:
\begin{itemize}
\item	Resistive dipoles with 1.8 T peak field and 30 mm (H) x 100 mm (W) rectangular aperture. These dipoles are pulsed with ms time scale at a frequency of 5 Hz. In the RCS they are ramped from injection to peak field (two quadrant), while in HCS they swing from negative to positive peak field (four quadrant);
\item	Superconducting dipoles with 10 T peak field and the same 30 mm (H) x 100 mm (W) rectangular aperture. These dipoles operate in steady state and provide the field offset for the HCS.
\end{itemize}

Both magnet types require field homogeneity in the range of few $10^{-4}$ in the good field region. Quadrupole magnets are still a subject of study, whereby we are scanning designs producing gradients in the range of 30 T/m in an aperture in the range of 40 mm to 80 mm, as discussed later.

The above magnet specifications require care and optimal design, and possibly better knowledge of magnetic and resistive properties of materials in the range of ramp-rates and frequencies required, but they appear well within the capabilities of present magnet technology. In fact, the main design driver for RCS and HCS is the management of the magnetic energy and reactive power, which should be highly efficient to minimize losses and very precise to meet beam performance specifications. To set orders of magnitude, the stored energy of a resistive magnet with the above characteristics is of the order of 6 kJ/m. The RCS have lengths in the range of 7 km to 27 km, and the total stored magnetic energy will hence be in the range of 30 MJ to 120 MJ (considering a dipole filling factor of 0.75). Pulsing these circuits in the range of fractions of ms to ten ms will hence involve managing reactive powers in the range of tens of GW. 

Below we describe schematically the powering solution taken as baseline, which includes the crucial component of energy storage, and follow-up with the description of resistive and superconducting magnet optimization and initial engineering design. Further details on power converters optimization and engineering can be found in Section~\ref{1:tech:sec:power}. 

\paragraph*{Powering concept and waveforms for RCS and HCS and magnet configuration}

To power the resistive magnets of RCS and HCS we have chosen a solution relying on resonant power converters and capacitor-based energy storage. The powering system is described in detail in Section~\ref{1:tech:sec:power}. The magnet design work was part of a fully integrated optimization work, where we have scanned extensively design parameters to find optimal configurations that minimize energy storage, reactive and active power, and the need for active filters, which represent one of the most costly systems in the powering scheme. 

\begin{figure}
\centering
\includegraphics{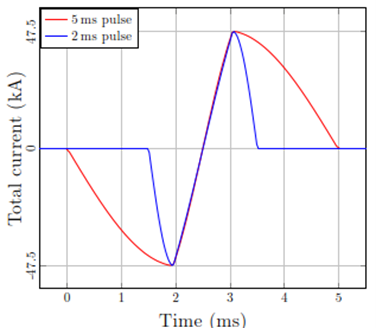}
\caption{Typical resonant current pulses for dipole magnets.}
\label{fig:Accelerator_Fig1}
\end{figure}

An example of a range of optimal ramps is reported in Figure~\ref{fig:Accelerator_Fig1}, where we show the requested linear ramp duration (1 ms, from peak negative current to peak positive current), preceded by a “preparation” phase and followed by a “recovery” phase of different duration (0.5 ms to 2 ms). Tuning of this time mainly depend upon the optimization of the power converter design and components, as well as loss (see also later).

A notable variant with respect to typical architectures is the fact that the power converter, consisting of energy storage and power electronics, is distributed along the accelerator. Few hundreds of simple units called “PEcells” are interleaved in series with the coils of the dipole magnets. The approach is like what done in the SPS but with much higher number of circuits.There are two advantages with this connection style:

\begin{itemize}
\item	There is only one circuit in the full accelerator. The currents among independent sectors is balanced by design. Indeed, tracking sectors as done in the LHC may be a serious problem with the desired small ramp times;
\item	The coils are interleaved with the PEcells and consecutive magnets. This guarantees a relatively small and balanced voltage to ground of all coils and power converters. 
\end{itemize}

The above choice has consequences on magnet design. In order to improve the magnetic efficiency and simplify the design, the dipole magnets will have coils built as single turns (or few turns) to reduce the total inductance, and they will have no heads. The coils are formed by bars that exit the magnet and traverse the magnet-to-magnet transition till the next PEcell or magnet. 

The baseline circuit configuration described above was used to provide a preliminary evaluation of size and cost evaluation of the magnets and power converters. More details on the powering scheme, optimization and technical solutions are reported in Section~\ref{1:tech:sec:power}.

\paragraph*{Resistive dipole magnet}
As for other areas of magnet design, we have started from the results of the US-MAP study. The US-MAP resistive dipole was designed for a bore field of 1.5 T and an aperture of 25 mm (H) x 156 mm (W). The stored energy of this dipole has been calculated at 4.2 kJ/m, and the total loss per cycle, assuming a 1 ms cycle is 112 J/m (see later for details). If we modify the design to meet the IMCC RCS and HCS specifications of 1.8 T in a 30 mm (H) x 100 mm (W) aperture, as shown in Figure 5, the stored magnetic energy rises substantially to 7.08 kJ/m, and the total loss per cycle is also much increased to 277 J/m. Assuming a 5 Hz repetition rate, this loss per cycle corresponds to 1.385 kW/m.

\begin{figure}
\centering
\includegraphics[width=0.5\textwidth]{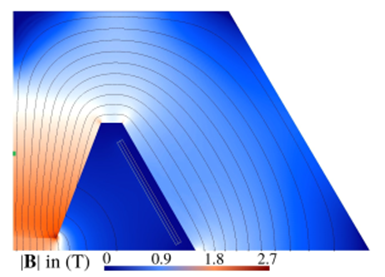}
\caption{MAP design with increased coil length and gap dimensions 30x100mm}
\label{fig:Accelerator_Fig4}
\end{figure}

We have then explored alternative designs. The analysis performed mainly aimed at limiting the magnet stored energy, as this limits the reactive power required from the power converter during fast ramps and reduces the size of the energy storage. To investigate optimal magnet configurations for the RCS resistive dipole magnets, three geometric designs were considered, namely the hourglass, from the US-MAP studies, the window-frame, and the H-type \cite{Breschi2024}. The best configurations found are shown schematically in Figure~\ref{fig:Accelerator_Fig4}. These configurations were optimized to minimize stored energy, subject to the constraint of achieving a specified magnetic flux density in the magnet air gap. The optimization procedure involved varying the peak engineering current density in the coils from 10 to 20 A/mm². The optimization uses an interactive routine based on Matlab for the optimizer part and FEMM for the 2D magnetic field and loss analysis. The routine scans geometric and electric variables and computes the electromagnetic field, searching for the configuration with minimum cost function which is a weighted sum of stored energy, difference from the specified 1.8 T bore field, and field errors \cite{Breschi2024}.

Two commercial ferromagnetic materials were selected for the resistive dipole’s magnetic circuit: Supermendur for the pole pieces and M22 steel for the remainder of the yoke. Supermendur exhibits a high magnetic permeability up to 2.0 T, which is advantageous for minimizing the total ampere-turns and the Joule losses in the conductor. Its linear magnetic characteristics also determines a reduction of the iron losses during rapid field transients. However, Supermendur contains Cobalt, which may get activated in a strong irradiation environment. M22 steel, while less permeable, is more cost-effective, radiation-resistant, and suitable for lateral yoke branches where lower magnetic flux densities are present.

\begin{figure}
\centering
\includegraphics[width=\textwidth]{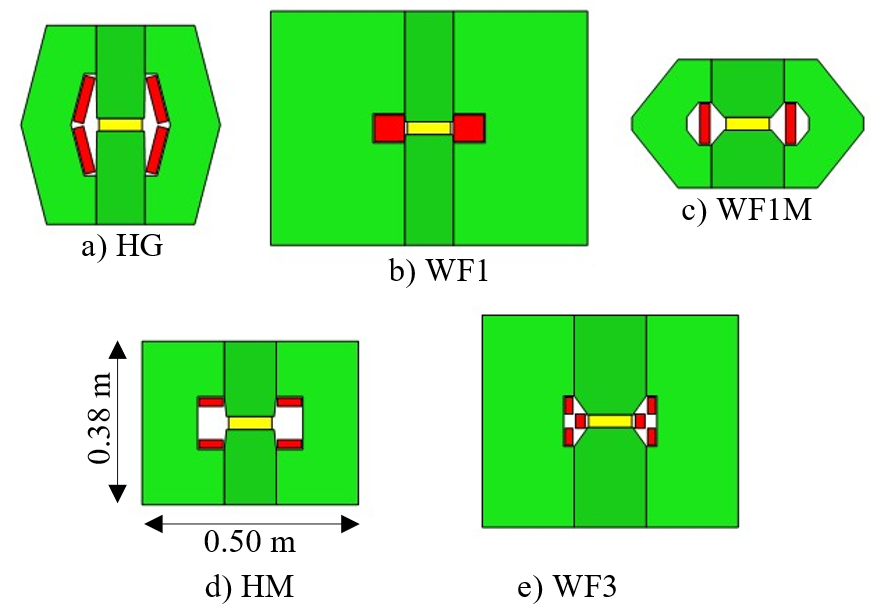}
\caption{Summary of the optimized geometries (the cross sections are to scale): a) Hourglass, J = 10 A/mm2, b) Window-frame WF\#1, J = 10 A/mm2, c) Window-frame WF\#1M, J = 20 A/mm2, d) H-type HM, J = 20 A/mm2; e) Window-frame WF\#3, J = 20 A/mm2.}
\label{fig:Accelerator_Fig5}
\end{figure}

Among the different analyzed configurations, the Hourglass (HG) and H type magnet (HM) exhibit the best results both in terms of stored magnetic energy and losses. Specifically, the HG dipole has stored magnetic energy of 5.77 kJ/m, and total loss per cycle of 406 J/m. This corresponds to an average power loss of 2.03 kW/m. The HM dipole has stored magnetic energy of 5.74 kJ/m, and total loss per cycle of 423 J/m, which corresponds to an average power loss of 2.12 kW/m. In both cases, as done earlier, we have computed the average loss assuming a pulse repetition frequency of 5 Hz. The values for HG and HM dipoles are very close, but, while the stored energy is much smaller than the hourglass US-MAP dipole adapted to the IMCC specifications, the loss is much larger. This is because the US-MAP design allows for a larger conductor window, reducing resistive loss at the expense of a larger magnetic circuit, and correspondingly larger stored magnetic energy.

For a final comparison, it is also interesting to consider the SPS dipole, which has similar gap dimensions and maximum bore field. The SPS dipole has a stored magnetic energy of 19 kJ/m, over three times that of the optimized muon collider accelerator dipoles. The reactive power is 6.8 kW/m, also three times higher than projected for the muon collider accelerator dipoles, though in this case we need to recall that the SPS is operated continuously at low frequency while the RCS have a duty cycle of less than 5 \% but with current frequency of the order of 500 Hz. Still, this comparison demonstrates that the optimization was very effective in reducing both active and reactive power, as well as energy storage needs.

In this initial magnet optimization step the resistive losses were not part of the cost function, and the optimizer always tried to reduce the magnetic circuit area as much as possible. While this is surely advantageous for the power converter, it may not be the best solution and losses in the conductors may be too high. We will come back later on this, showing how this is addressed. 

To speed up the calculation of the magnetic field produced by the resistive dipoles, an alternative method to the FEMM simulation was developed and applied to the analysis of the H-magnet configuration. This method is based on an equivalent lumped parameters non-linear magnetic circuit of the resistive dipole. The topology of the magnetic circuit is obtained from the geometry of the magnetic configuration. The introduction of magnetic reluctances at given locations of the magnetic circuit is based on the analysis of the flux lines obtained through 2D simulations performed with FEMM. An example of magnetic circuit obtained is shown in Fig. 2.10. The reluctances of the ferromagnetic structure are computed accounting for the non-linear characteristics of the magnetic materials used for the difference parts of the structure itself, namely Vacoflux 48 for the poles and the M235-35A for the yoke. The magnetic circuit obtained is then solved by means of the mesh analysis.

The magnetic circuit method has a very good accuracy for a first assessment of the resistive dipole design, with errors on field, stored energy and losses within few \%, with a substantial reduction of the computation time. To give orders of magnitude, the computation time of a magnet configuration drops by more than two orders of magnitudes compared to 2D FEM, from minutes to seconds.

\begin{figure}
\centering
\includegraphics[width=\textwidth]{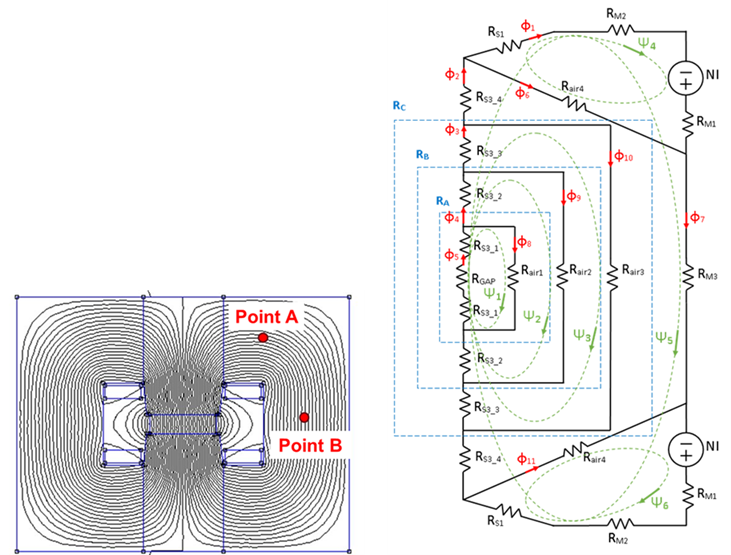}
\caption{a) Magnetic flux density map of the H-type dipole and b) corresponding equivalent magnetic circuit. The magnetic flux density calculated in Point A and Point B is used as a reference for the validation of the results of the circuit model.}
\label{fig:Accelerator_Fig6}
\end{figure}

\paragraph*{Magnetization, resistive and eddy currents losses }
Losses in the resistive magnets are one of the main concerns in the design of the dipoles, as can be inferred from the evaluation of losses reported in the previous section. In fact, evaluating the loss in the specific conditions of the RCS and HCS is not trivial. Losses originate from magnetization hysteresis and eddy currents in the iron laminations, and resistance in the copper coils. Iron losses are well understood within classical electrical engineering. In our case the challenges are the specific geometry and end effects, the fact that the iron is saturated in a large portion of the yoke, and the fact that the database of loss in the frequency regime of interest, about 1 kHz, is not as established as would be necessary. For the copper coils, the frequency is such that the skin depth is in the range of mm. A single conductor bar would not be fully penetrated during a pulse, and the skin current would result hence in much higher resistive loss than would be the case for uniform current distribution. This requires appropriate design devices to drive current distribution in the copper conductors.

Given the challenge, we have performed some benchmarking of loss evaluation using different numerical and semi-analytical transient simulation methods, taking the design values from US-MAP as a starting point. The US-MAP resistive dipole was designed for a bore field of 1.5 T and an aperture of 25 mm (H) x 156 mm (W). This has been reproduced using Maxwell 2D and proprietary software at the Technical University of Darmstadt. The magnet geometry in the latter case was slightly modified to allow for higher field, 1.8 T as specified for the IMCC RCS and HCS. The result of the loss calculation are reported in Tab. \ref{tab:Accelerator_tab1}, where the various components and total loss are given. A cycle of 1 ms was assumed for the calculations. We see from the figures reported there that there is consistency of values, but the spread is significant, typically ± 20\% around the average of all results. Experimental data would be necessary at this stage to progress further.

\begin{table}[h!]
\centering
\begin{tabular}{l|cccc}

ANALYSIS & US-MAP & Maxwell 2D & TUDa & TUDa   
\\ \hline
Bore field (T)  & 1.5& 1.5& 1.5& 1.8\\ 
Aperture (mm x mm)  & 25 x 156& 25 x 156& 25 x 154& 25 x 154  \\
Stored energy (kJ/m)& 4200& 4900& 4551& 6644\\ 
Static loss per cycle&&&&\\ 
Iron yoke (J/m)& 33& 59& 21.8& 32\\ 
Iron pole (J/m)& 63& 61& 40.8& 58.5\\ 
Coil (J/m)& 16& 33& 17.8& 31.5\\ 
Dynamic loss per cycle&&&&\\ 
Coil (J/m)&&& 25& 37\\ 
Total loss per cycle (J/m) & 112& 153& 105& 158\\ 
\end{tabular}
\caption{Benchmark of loss calculations for the geometry of the resistive dipole defined by US-MAP. Calculations for 1.5 T \cite{Berg2016} and 1.8 T (reference IMCC design). A cycle time of 1 ms is assumed.}
\label{tab:Accelerator_tab1}
\end{table}

An example of loss evaluation is shown in Figure~\ref{fig:Accelerator_Fig5}, where we show the influence of the preparation and recovery phases of different length outlined in Figure~\ref{fig:Accelerator_Fig1}, applied to the magnet design in Figure~\ref{fig:Accelerator_Fig4} (US-MAP) The total loss, iron and coil, depends on the total duration of the cycle, and we see from Figure~\ref{fig:Accelerator_Fig5} that there is a clear advantage in maintaining the total cycle time, including preparation and recovery, as short as possible. This result appears rather trivial, powering for longer time increases Joule heating. In reality, what cannot be seen by such an analysis is the effect on the power converter, which has added complexity and cost when demanding faster ramps. We expand on this in the next section.

\begin{figure}[h!]
\centering
\includegraphics[width=0.5\textwidth]{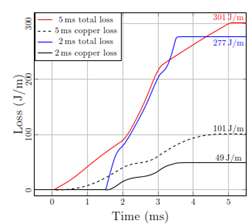}
\caption{Resistive and total loss evaluation for two cycle alternatives with the same ramp time of 1 ms, but either 2 ms or 5 ms total time, see Figure~\ref{fig:Accelerator_Fig1}. The geometry of the modified US-MAP dipole was used for this calculation.}
\label{fig:Accelerator_Fig7}
\end{figure}

\paragraph*{Global optimization and specifications for power converter}
It should be clear from the discussion so far that it is not possible to design an optimal resistive dipole circuit by separately optimizing for the various components and issues. This is why we have created a combined optimizer model of the magnet and the power converter, and used this model to study several optimization directions towards minimal the capital and operational expenditures (CAPEX and OPEX).

The model considers a dipole magnet with a generic geometry inspired by the US-MAP Hourglass concept. This geometry was identified in the optimization exercise presented earlier as one of the most cost-effective configurations and is thus considered representative for system-level optimization.

The optimization process is based on the magnetostatic and harmonic solvers in FEMM, which are used for magnetic sizing and for evaluating copper conductor losses. In parallel, an Artificial Neural Network (ANN) model is being developed to estimate conductor losses without relying on finite element simulations, similarly to the approach previously implemented with the reluctance-based magnetic model.

Additional features have been introduced to support power converter design, such as the possibility to model coils composed of multiple parallel conductors and to include cooling holes within the conductors.

\begin{figure}[h]
\centering
\includegraphics[width=\textwidth]{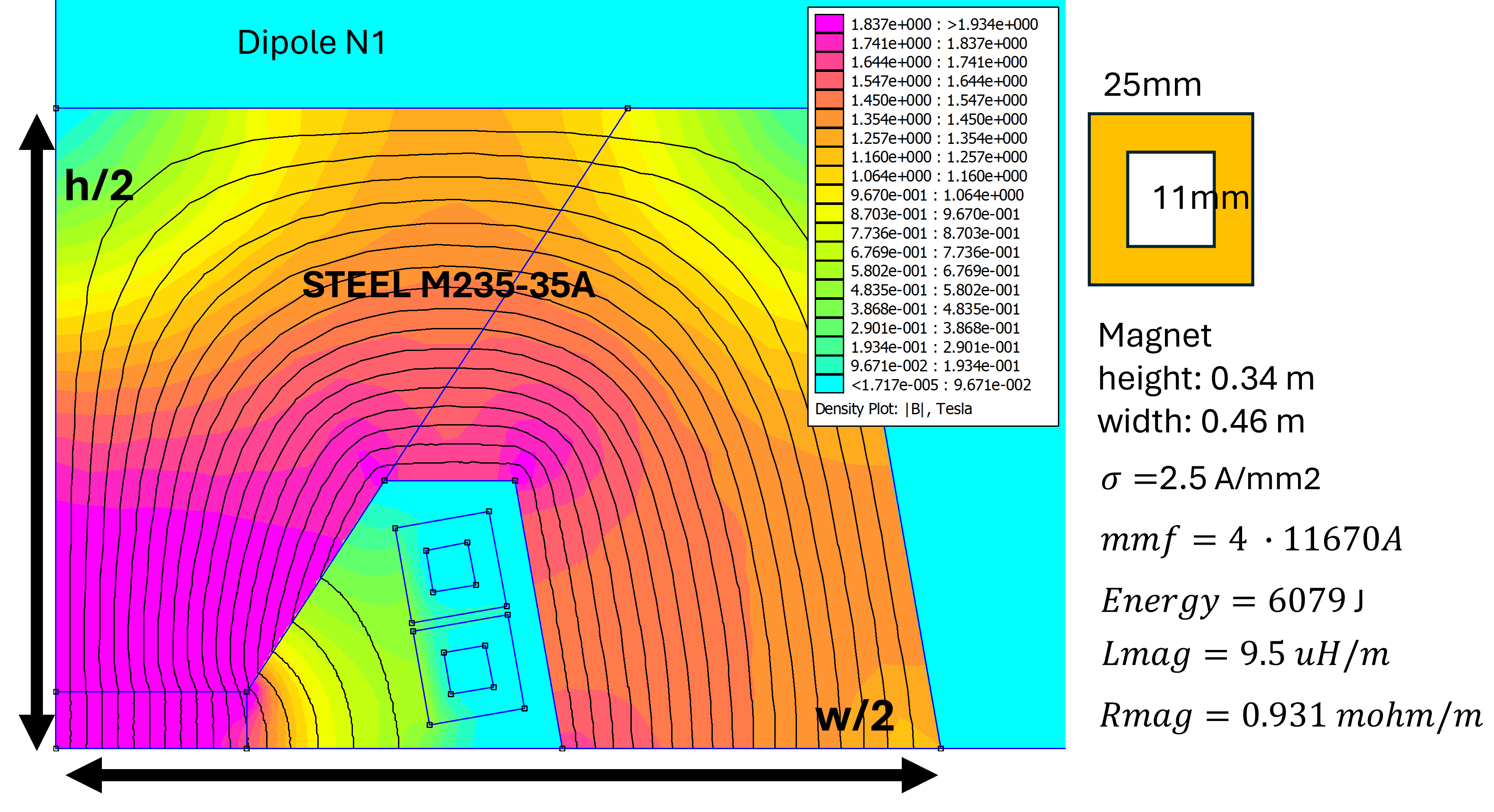}
\caption{CAPEX+OPEX optimization on the dipole for short accelerators}
\label{fig:Accelerator_Fig8b}
\end{figure}

\begin{figure}[h]
\centering
\includegraphics[width=\textwidth]{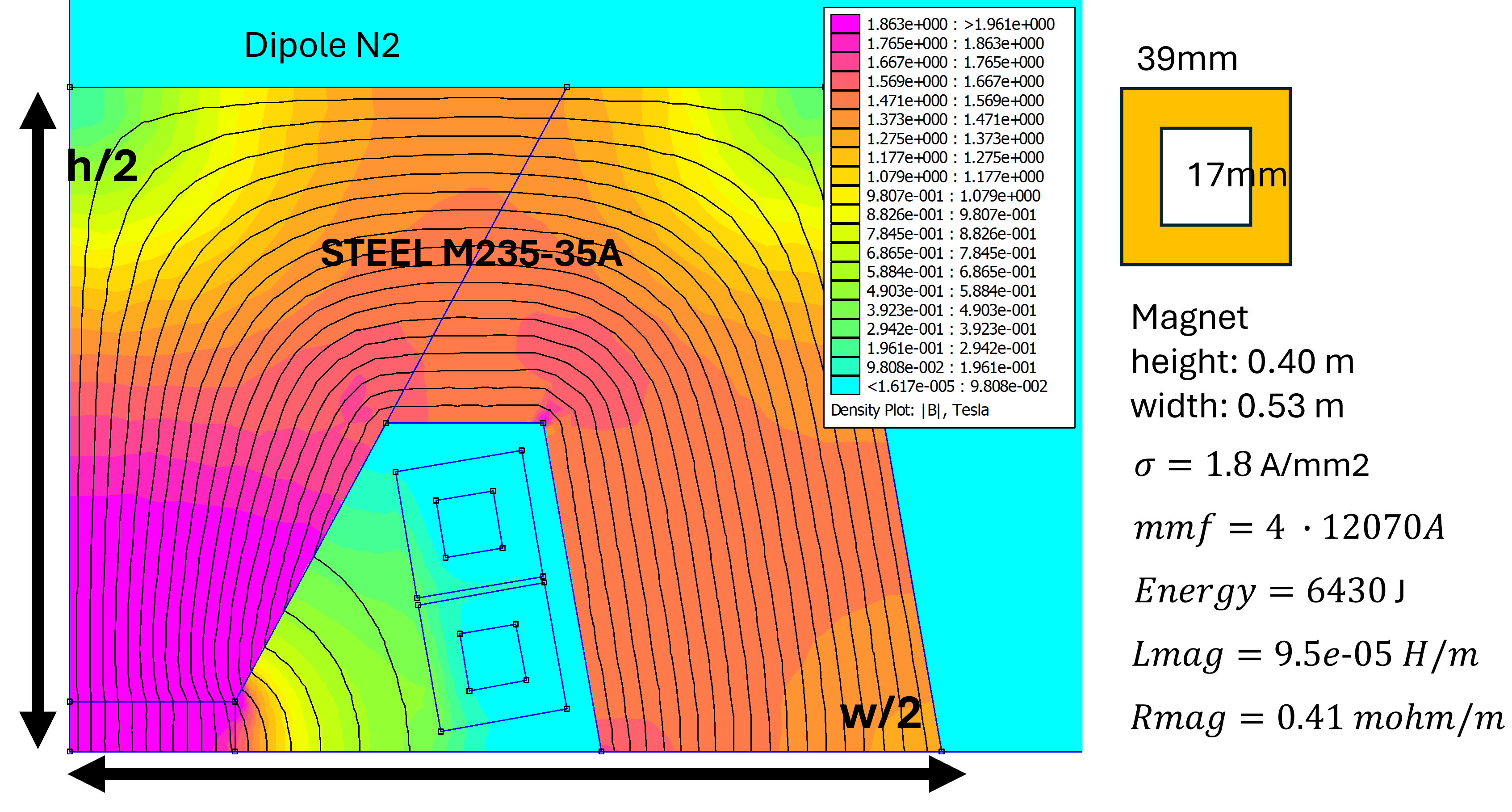}
\caption{CAPEX+OPEX optimization on the dipole for long accelerators}
\label{fig:Accelerator_Fig8a}
\end{figure}

The optimization is highly sensitive to the current duty cycle. Ideally, each magnet–power converter pair should be optimized specifically for the given accelerator scenario. In this report, two designs are proposed based on CAPEX+OPEX optimization for short and long accelerator configurations. These are shown in Figure~\ref{fig:Accelerator_Fig8b} and Figure~\ref{fig:Accelerator_Fig8a}, respectively. The designs, referred to as Dipole1 and Dipole2, differ primarily in current density.

Notably, when loss-related costs become significant—such as in long accelerators with several kilometers of resistive magnets—the optimizer reduces the current density and reshapes the conductor and magnetic circuit to meet the performance targets within acceptable cost limits.

These two designs have been used to size the power converters across all accelerator variants in both the CERN and Green Field scenarios. The optimization framework automatically computes the equivalent inductance and resistance per meter using the following formulas:

\begin{equation} L=\frac{2 E_{\text{stored}}}{I^2} \quad [\text{H/m}] \end{equation}

\begin{equation} R=\frac{E_{\text{loss}}}{I^2 T} \quad [\Omega/\text{m}] \end{equation}

Here, $I^2T$ is the integral of the current squared over the pulse duration and is dependent on the current waveform.

While minimizing the total magnetic energy stored in the air gap and conductor window generally reduces costs, additional constraints—such as the limit on the number of series conductors to keep total voltage within bounds—lead to relatively bulky, liquid-cooled conductor assemblies. Reducing energy by shrinking the conductor window comes at the expense of higher losses as conductors are placed closer to the gap. This tradeoff is handled automatically by the optimizer.

The large return yoke cross-sections are another result of the integrated optimization, which seeks to reduce magnetic saturation and minimize the required magnetomotive force (mmf). The mmf directly influences the number and rating of power semiconductor devices (IGBTs and diodes), which are among the most expensive converter components. Higher mmf leads to a higher count of IGBTs and, consequently, increased cost.

These findings highlight the importance of system-level optimization. Designing magnets in isolation does not necessarily yield the lowest overall cost.

The two designs yield the following magnet parameters:
\begin{itemize}
\item \textbf{Dipole 1:} \quad $L_{\text{mag}} = 95~\mu\text{H/m}$,\quad $R_{\text{mag}} = 0.93~\text{m}\Omega/\text{m}$ 
\item \textbf{Dipole 2:} \quad $L_{\text{mag}} = 95~\mu\text{H/m}$,\quad $R_{\text{mag}} = 0.41~\text{m}\Omega/\text{m}$
\end{itemize}

These values are used to size the corresponding power converters and to estimate the total accelerator losses for the different design scenarios (see the power converter section).

It is worth noting that these values are conservative. Due to iron saturation, the energy stored and switched by the power converter is overestimated. 
\begin{figure} 
\centering 
\includegraphics[width=\textwidth]{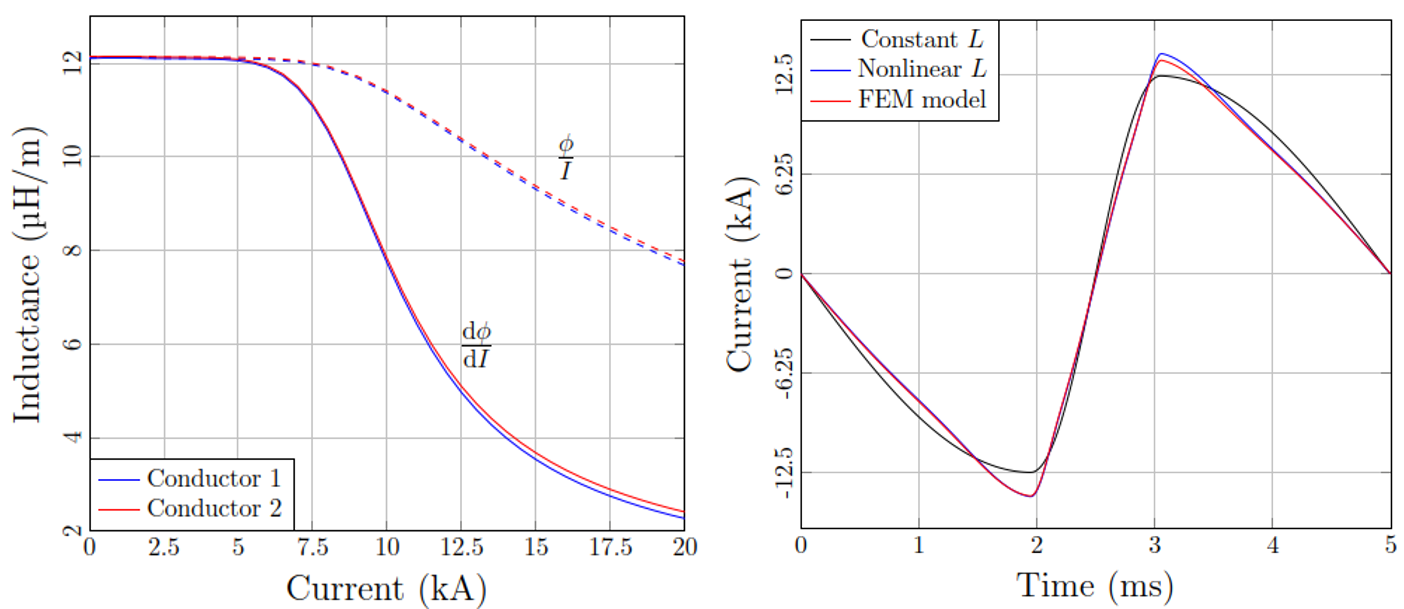} 
\caption{Circuit–magnet transient simulation comparing constant and non-linear inductance} 
\label{fig:Accelerator_Fig8_Alt} 
\end{figure}\\
\begin{figure}
\centering
\includegraphics[width=\textwidth]{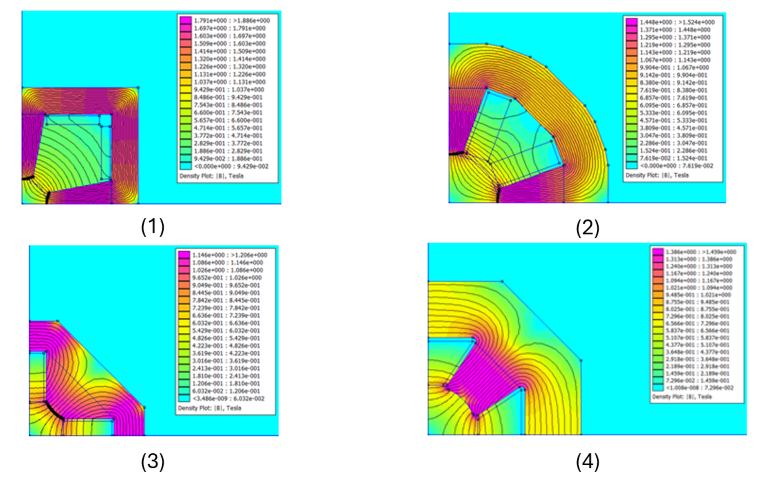}
\caption{Optimized quadrupole configurations for a bore aperture of 60 mm.}
\label{fig:Accelerator_Fig9}
\end{figure}

This effect is illustrated in Figure~\ref{fig:Accelerator_Fig8_Alt}, where the actual non-linear inductance and its differential (right), and the resulting current ramp (left), are shown. The transient response with constant inductance (black curve) is significantly faster than those computed using non-linear inductance (red and blue curves). Taking into account the non-linear differential inductance helps avoid excessive design margins.

\FloatBarrier
\paragraph*{Resistive quadrupole magnet}
A similar study to the one performed to identify the most suited dipole magnet configuration was conducted for the quadrupole magnet design. Four configurations were analysed to understand which could better reduce the power losses, the magnetic energy stored, or both. The quadrupole magnets were designed to have a field gradient of 30 T/m and a specified field quality of $10^{−4}$.

The four configurations were optimized for three values of the air gap diameter, 40 mm, 60 mm and 80 mm, and were compared in terms of total losses, real and reactive power absorbed and stored magnetic energy. A synoptic view of optimal configurations for a bore aperture of 60 mm is shown in Figure~\ref{fig:Accelerator_Fig9}. The most suitable configuration to reduce the losses is the configuration proposed by US-MAP, configuration 1 in Figure~\ref{fig:Accelerator_Fig9}. For a bore aperture of 60 mm, this configuration has a stored energy of 2.4 kJ/m and an average power loss of 212 W/m. The best alternative is the configuration with trapezoidal coils, and smallest magnetic circuit, configuration 4 in Figure~\ref{fig:Accelerator_Fig9}. For a bore aperture of 60 mm, this configuration has a stored energy of 0.83 kJ/m and an average power loss of 352 W/m.

All the optimized configurations achieve the specified field gradient of 30~T/m. However, none of them is yet able to satisfy the field quality requirement. Therefore, further investigations are required to improve this aspect of the quadrupole design.

\paragraph*{Superconducting dipole magnets}

In parallel to the work on the normal-conducting pulsed magnets, we are progressing with the design of the superconducting magnets of the hybrid cycled synchrotrons (HCS). These magnets provide a field offset, and allow using the full field swing of the normal conducting magnets, from negative to positive field values, effectively making the synchrotrons shorter. The work has focused on HTS dipoles, operated in gaseous helium (10 K to 20 K) generating a 10~T steady state field in a rectangular aperture identical to that of the resistive dipole magnets, i.e. 30~mm $\times$ 100~mm. The choice of HTS was driven by the intent to have a large operating margin and reduce consumption, especially in light of the effect of the bursts of muon decay that will cause periodic heating of the magnets. Also, operation at temperature significantly higher than liquid helium will ease the engineering of the transitions from resistive to superconducting magnets that takes place in each cell.

\begin{figure}
\centering
\includegraphics[width=0.5\textwidth]{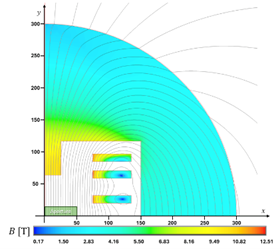}
\caption{Conceptual design of a 10 T HTS, 30 mm x 100 mm aperture dipole for the hybrid cycled synchrotrons.}
\label{fig:Accelerator_Fig10}
\end{figure}

Following initial conceptual studies, the configuration presently studied in detail is shown in Figure~\ref{fig:Accelerator_Fig10} \cite{Levi2025}. The geometric requirements for this dipole have led to a design featuring six planar racetracks, three above and three below the midplane. To simplify both the mechanical structure design and the assembly process, each racetrack consists of 205 turns, and all are perfectly aligned. The magnetic design includes a circular iron yoke that surrounds the coils. The yoke has an outer radius of 300 mm and an inner square window, optimized to maximize magnetic performance and field quality while ensuring at least 30 mm of clearance between the racetracks and the yoke to accommodate the mechanical structure. Additionally, the yoke features a pole with dimensions optimized to further enhance the field quality. The field quality requirements specify that all harmonic components must remain below 10 units within a good field region of 50~mm $\times$ 20~mm. Field quality has been assessed using two methods: the first involves four harmonic expansions, each with a 10~mm radius, positioned at x = 0~mm, 5~mm, 15~mm and 25~mm; the second method uses four paths to calculate the field magnitude from x = 0~mm to 25~mm, evenly distributed in y between 0 mm and 10 mm. This design largely meets all field quality requirements: all harmonic components range between -4.5 and 0 units, achieving a field homogeneity of 0.03\%. The Table below reports all the main parameters of the optimized design of the dipole.

\begin{table}[h!]
\centering
\begin{tabular}{l|c}
Parameters                              & Value      \\ 
\hline
Central field                           & 10 T       \\ 
Peak field                              & 12.51 T    \\ 
Current                                 & 2314 A     \\ 
Engineering current density             & 714 A/mm\(^2\) \\ 
Margin on loadline                      & 25.7\%     \\ 
Operating temperature                   & 20 K       \\ 
Temperature margin                      & 2.5 K      \\ 
Magnetic length                         & 1.3 m      \\ 
Mechanical length                       & 1.6 m      \\ 
Iron yoke radius                        & 300 mm     \\ 
Number of racetracks per quadrant       & 3          \\ 
Number of turns per racetrack           & 205        \\
Number of HTS tapes per turn            & 2          \\
Inductance                              & 1.3 H      \\ 
Stored energy                           & 2.24 MJ    \\ 
\end{tabular}
\caption{Main parameters of the optimized design of the dipole.}
\label{tab:Accelerator_tab2}
\end{table}

The mechanical analysis was conducted using an ANSYS APDL macro, where Lorentz forces were imported node by node from the electromagnetic model. The simulation includes an infinitely rigid structure surrounding the racetracks, with stainless steel cases having an inner over thickness of 0.1 mm. This additional thickness allows the racetracks greater freedom to deform under the influence of Lorentz forces, preventing over constraining of the conductors and yielding more realistic results. The peak Von Mises stress, with a value of 172 MPa, occurs at the lower left corner of the first block. It is notable that the net force in the $y$ direction on the first block is positive (upward). This outcome, which was a secondary goal of the magnetic optimization, facilitates structural designs that keep the midplane region clear, reducing heat deposition from radiation in components likely to be in direct contact with the superconducting material. Given that most of the mechanical load arises from the x component of the Lorentz forces, a stress management strategy has been implemented and optimized to address these forces effectively. The baseline configuration was improved by introducing a 5 mm thick septum, modelled as infinitely rigid, consistent with the rest of the structure. The position of each septum was individually optimized, resulting in slight variations in the number of conductors on either side of the septum across different blocks. This modification reduced the peak Von Mises stress by half, down to 85 MPa. The adjustment to the cross-section has minimal impact on magnetic performance: compared to the baseline configuration, the required current is 2\% higher, the peak field is 1.4\% lower, the new load line margin is 22.8\%, and the stored energy increases by 5\%. Furthermore, all harmonic components remain within the range of -2 to 2 units, with a field homogeneity of 0.04\%.

In addition to the efforts reported here on the steady state dipoles, we have performed a conceptual study on the possibility to use pulsed HTS dipoles as an option for the last and highest energy synchrotron in the accelerator chain. The interest is that the last synchrotron has field ramp-rates of the order of few hundreds T/s, which have been shown to be achievable with  superferric magnets. Although there seems to be scope for further study, it is not yet clear whether a fully superconducting and rapic cycled synchrotron in the range of energy and speed of RCS4 is feasible with HTS magnets.

\paragraph*{Challenges identified}

As we anticipated, the main perceived challenge for the RCS and HCS magnets is the aspect of system optimization. It is the mainly interplay of magnet design, energy storage and power conversion that needs to be understood and mastered to yield finally to an optimal global design. We have directed our main efforts in this direction, towards a baseline design that will result in minimum CAPEX and OPEX. Still, there is a definite need to demonstrate pulsed dipole circuit performance in conjunction with energy storage and power conversion, validating the tracking of ramp current and field reference, the evaluation of losses, the energy recovery efficiency, as well as transient field quality. Such demonstration should be also relevant to the large number of pulses planned throughout the lifetime of the accelerator.

At the same time, there are technical aspects that deserve special attention, namely:
\begin{itemize}
\item	Hysteresis, eddy current and coil resistive losses in the regime of frequency of interest, specific to the resistive magnets of the RCS and HCS. We believe that for this the magnetic material database is not sufficient for accurate prediction;
\item	Field quality in pulsed magnets, especially in presence of other magnetic and/or conducting components (e.g. beam pipe) which will affect field lags and time constants;
\item	Demonstration of accelerator-level performance, specific to the superconducting HTS dipoles of the HCS. This challenge is shared with that of the collider magnets.
\end{itemize}

\subsection*{Collider magnets}
\label{sec:Collider}
The magnets in the collider are the final big challenge that we have identified. Besides the difficulty in the magnet technology, muon beams require optics solutions that are far from standard practice, and integrating the specifications from beam optics poses an additional challenge. In order to provide quick feedbacks to the beam dynamics, cryogenics and energy deposition study requirements, an analytic evaluation of the maximum magnet performances as a function of the magnet aperture was performed (see \cite{Novelli2023} and \cite{Novelli2024}). This preparatory work has then led to the choice of specific design points for the main dipoles and IR quadrupoles of the collider, which we have used to initiate conceptual and engineering design of the collider magnets. We describe below the results of this work.

\paragraph*{A-B plots}

\begin{figure}
\centering
\includegraphics[width=\textwidth]{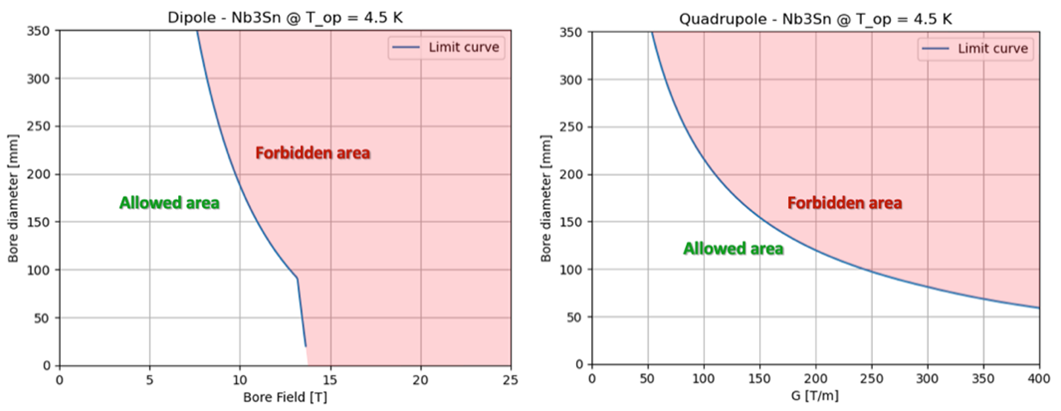}
\caption{A-B plot for dipoles and A-G plot for quadrupoles built with Nb$_3$Sn and operated at 4.5 K.}
\label{fig:Collider_fig1}
\end{figure}

\begin{figure}
\centering
\includegraphics[width=\textwidth]{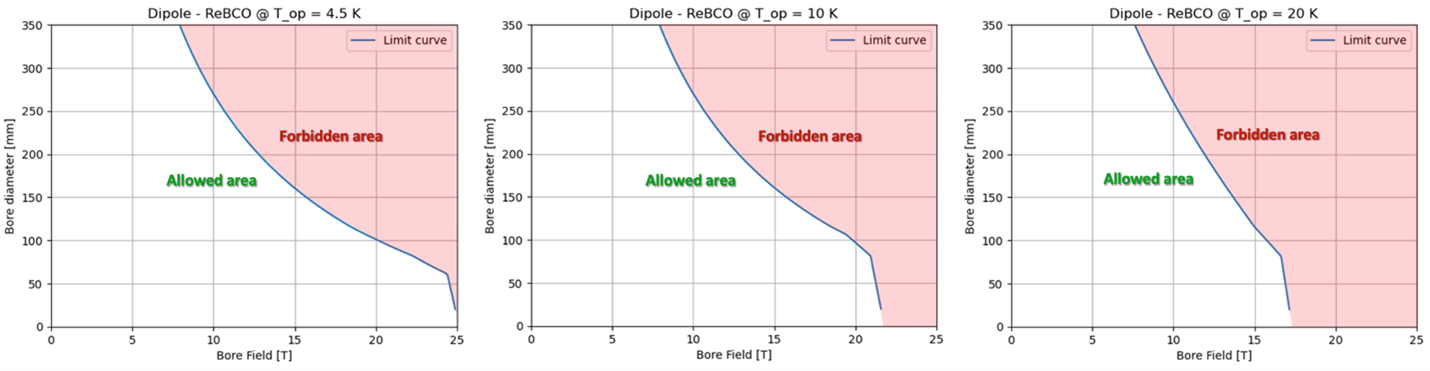}
\caption{A-B plot for dipoles built with REBCO and operated at 4.5 K, 10 K and 20 K.}
\label{fig:Collider_fig2}
\end{figure}

\begin{figure}
\centering
\includegraphics[width=\textwidth]{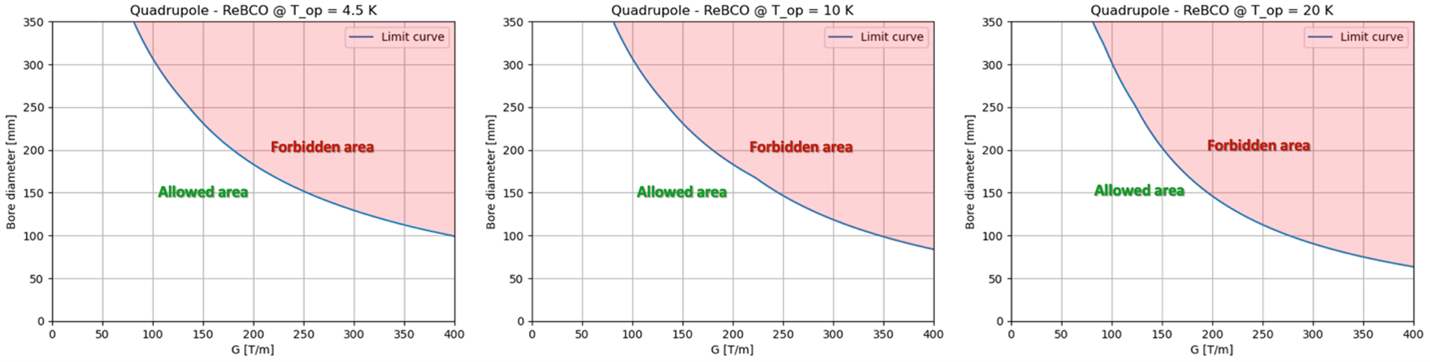}
\caption{ A-G plot for quadrupoles built with REBCO and operated at 4.5 K, 10 K and 20 K.}
\label{fig:Collider_fig3}
\end{figure}

To evaluate the maximum dipolar field or quadrupolar field gradient obtainable, a sector coil approximation was assumed and all the most important constraints were included in the calculation, namely:
\begin{itemize}
\item	Margin on the load line. A temperature margin of 2 K was assumed for Nb-Ti, while 2.5 K was considered for Nb$_3$Sn and ReBCO. While in the case of Nb$_3$Sn the margin is required to ensure stable operation and limit training, in the case of ReBCO we would expect that such margin would be largely more than what is needed. However, considering that we plan to design for operation in gas, we have kept the same margin to accommodate for temperature fluctuations that may come from the cryogenic system.
\item	Feasibility of the protection system. A 40 ms time delay between the quench and the firing of the protection system were assumed and a maximum hot-spot temperature at the end of the discharge of 350 K for Nb-Ti and Nb$_3$Sn and 200 K for ReBCO were set, as explained in details in \cite{Novelli2023}.
\item	Mechanics: the average stress on the midplane was estimated analytically considering only the E.M. forces as explained in \cite{Novelli2023} and the limit was set to 100 MPa, 150 MPa and 400 MPa respectively for Nb-Ti, Nb$_3$Sn and ReBCO.
\item	Cost: the target budget was set to 400 kEur/m for the arc magnets, more than twice the limit set for the FCC-hh project, and 800 kEUR/m for the IR magnets considering the total dimension of the collider ring and available budget for the entire accelerator complex. These are costs comparable to present HL-LHC magnet manuifacturing.
\end{itemize}
The result of this analysis are “A-B” plots of maximum aperture for a given field, and “A-G” plots of maximum aperture for a given gradient, satisfying all requirements above. We report in Figures~\ref{fig:Collider_fig1} (A-B and A-G plots for Nb$_3$Sn operated at 4.5 K, 400 kEUR/m cost limit), Figure~\ref{fig:Collider_fig2} (A-B plots for REBCO dipoles operated at 4.5 K, 10 K and 20 K, 400 kEUR/m cost limit) and Figure~\ref{fig:Collider_fig3} (A-G plots for REBCO quadrupoles operated at 4.5 K, 10 K and 20 K, 800 kEUR/m cost limit).

\begin{figure}
\centering
\includegraphics[width=\textwidth]{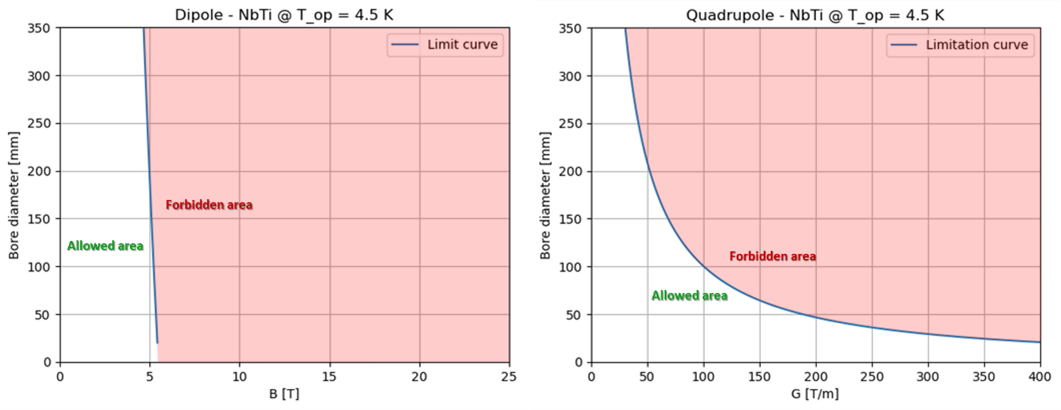}
\caption{A-B plot for dipoles and A-G plot for quadrupoles built with Nb-Ti and operated at 4.5 K.}
\label{fig:Collider_fig4}
\end{figure}

For Nb-Ti, the same analysis performed for other superconducting materials has been carried out. While this material does not achieve the required performance for a 10 TeV collider, it remains a viable alternative for a 3 TeV collider and provides an excellent reference for the work performed. Figure~\ref{fig:Collider_fig4} shows the A-B and A-G plots for Nb-Ti magnets, assuming an operating temperature of 4.5 K.

\begin{figure}
\centering
\includegraphics[width=\textwidth]{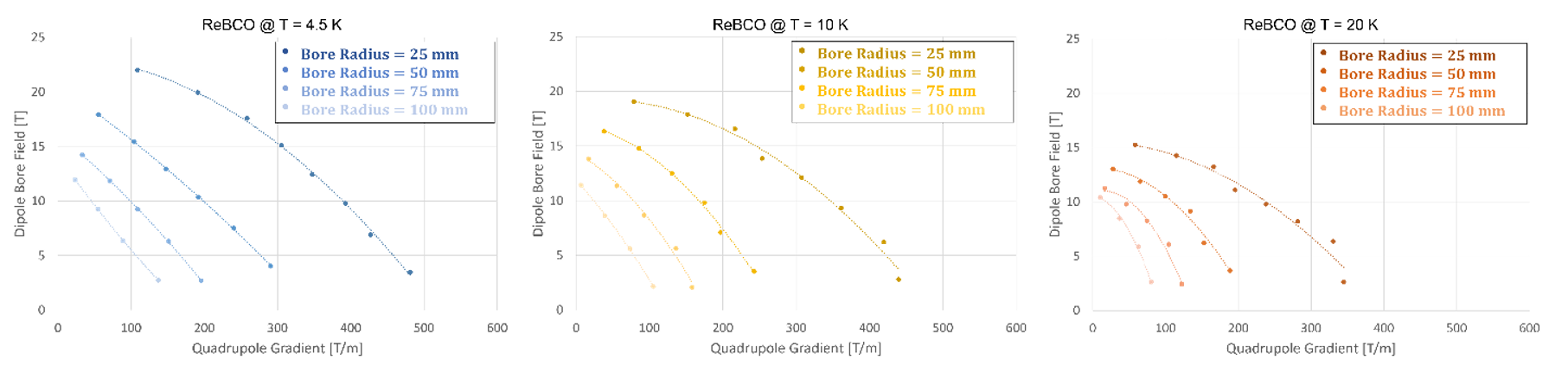}
\caption{B-G plots for nested combined dipole and quadrupoles with REBCO conductor operated at 4.5 K, 10 K and 20 K.}
\label{fig:Collider_fig5}
\end{figure}

Using the data presented in the plots, additional performance limits for combined-function arc magnets have been derived, illustrating the maximum achievable combination of dipolar field and quadrupole gradient as a function of the magnet bore aperture (see Figure~\ref{fig:Collider_fig5}). 
The results are based on a nested magnet configuration, with the quadrupolar coil positioned within the dipole bore. Also in this case the reported performance limits satisfy the constraints of maximum cost minimum temperature margin, and the maximum mechanical stress on the conductor.

The minimum aperture of the arc dipoles have been obtained from considerations of beam optics, impedance, radiation shielding, cryogenics and vacuum integration. In particular, the magnet aperture in the collider arc needs to be at least 158 mm for a cold mass at 4.5 K and 138 mm for a cold mass at 20 K. Using the above plots, we see that a dipole built with Nb$_3$Sn, with an aperture of 158 mm, operated at 4.5 K, can reach fields of the order of 11 T. The same evaluation of a dipole built with REBCO, with an aperture of 138 mm, operated at 20 K, can reach fields of 14 T. We have hence set these as magnet performance targets representative of the challenges and pushing the limit of present technology. We note at this point that these magnet performance targets imply dipole coil dimensions that are significantly larger than what has been done in HL-LHC, and what is planned for FCC. Typically, the stored energy is a factor three to four larger. For the 3 TeV stage, using Nb-Ti at 4.5 K, a field of 5 T appears within reach. 

The quadrupole performance limits are especially crucial for the IR, which is a major challenge of a muon collider. We see from the plots above that the performance limits follow an approximate A = Bpeak/G dependence, where the parameter Bpeak is approximately 5 T for Nb-Ti, 10~T for Nb$_3$Sn and 15 T for REBCO. The quadrupole magnets used in the present IR 10 TeV optics scale with a similar dependence, and we argue that developing a quadrupole of this class would be relevant to the whole IR. In this case we can focus on the largest gradient magnet, 300 T/m, which requires an aperture in the range of 140 mm.

Finally, we see that when considering combined function dipole-quadrupole magnets the magnet performance is limited by the interaction of the two fields, as expected. Using the results reported above, we see that targeting a field of 14 T, and operating at 10 K to have some additional margin, we can only reach an additional field gradient of 100 T/m. Further iterations with beam optics are necessary at this stage, to integrate the results of this study.

To the best of our knowledge these results represent a truly novel approach to magnet design, which was not formalized earlier. The above plots define performance limits of all main accelerator magnets with unique clarity. Still, we recall that the above performance limits should be taken only as guidelines for the choice of parameters combination. Indeed, being purely analytical, they are powerful scaling and scoping tools, but cannot substitute for actual engineering design which may require additional margins to cope with actual geometric and material constraints, or may offer optimization windows that allow exceeding the analytical scaling, see below.

\paragraph*{Conceptual design of dipole options}

Starting from the analytical evaluation described above, we have initiated a detailed FEM-based design work on REBCO dipole magnets. The aim is to address critical aspects of implementing this technology in accelerator-grade superconducting magnets. So far we have not considered Nb$_3$Sn, which is the main focus of the High Field Magnet R\&D programme, nor Nb-Ti, which is industrially available magnet technology.

\begin{figure}
\centering
\includegraphics[width=\textwidth]{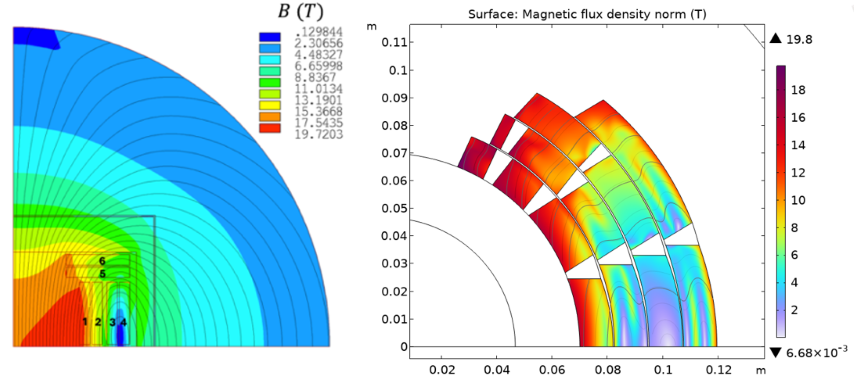}
\caption{Design of the cross section of HTS dipoles.}
\label{fig:Collider_fig6}
\end{figure}

Two dipole geometries are considered for this work, blocks and cos-theta. Figure \ref{fig:Collider_fig6} shows the preliminary cross sections design of the most promising candidate configurations for HTS dipole to be implemented in the ARC cell of the collider \cite{Alfonso2025, Mariani2025}. For both configurations, a non-twisted stacked tapes cable, co-wound with a stainless steel strip is assumed. This choice is primarily due to its ease of scaling to high currents and its flexibility in cable design. Both designs satisfy the objectives of margin and stress, though many issues such as coil winding technology, ends, joints, magnetization and loss, field quality still need to be addressed in detail.

The block coil configuration consists of three double pancakes with a hybrid cable arrangement: in the four blocks on the magnet mid-plane (1-4 in Figure~\ref{fig:Collider_fig6}), the broad side of the cable is parallel to the horizontal-axis, while in the racetracks above the bore aperture (5-6 in Figure~\ref{fig:Collider_fig6}), it is oriented vertically. This configuration minimize the need for hard-way bending of the two first double pancakes while simplifying the upper racetracks winding procedure as they are positioned above the bore. The 2D magnet cross-section electromagnetic design is entirely developed using ANSYS finite element software, with the primary objective of conducting a sensitivity analysis of the geometrical parameters to meet the requirements in accordance with the A-B plots presented above. All simulations include a circular iron yoke with a mid-plane thickness of 200 mm (the outer radius is 326.7 mm), a vertical pad with a thickness of 20 mm and a horizontal pad with a thickness of 40 mm. The cos-theta coil configuration is composed by 4 layers with a 4-4-3-2 blocks arrangement. To enhance magnetic efficiency and achieve more compact designs, the block coil is graded using a two stacked tapes cable for the blocks 3 and 4 and 4 stacked tapes cable for the blocks 1, 2, 5 and 6.

Using the same cable configuration of the block coil design without any keystoning (common in LTS cable design), each block is aligned in the cross-section to minimize the geometrical field error produced in the magnet bore area while maximizing the cable temperature margin by aligning the broad face of the tape to the magnetic flux lines. The preliminary cross-section optimization is performed in Roxie without the use of iron yoke to evaluate the maximum achievable performances of the HTS winding in agreement with the analytical evaluations previously mentioned. A grading approach is used also for the cos-theta coil configuration, using instead a 50 micron stainless steel strip for the two inner coil layers and 25 micron stainless steel strip for the two outer layers. The numerical results of the electromagnetic optimization are presented in the table below.

\begin{table}[h!]
\centering
\begin{tabular}{l|ccc}
Parameter & Unit & Block Coil Design & Cos-theta Coil Design \\
\hline
Operating temperature & K & 20 & 20 \\
Temperature margin & K & 2.5 & 9 \\
Bore field & T & 17.5 & 16\\
Max. peak field & T & 19.7 & 19.3 \\
Current & A & 2481 & 1702 \\
Current density & A/mm$^2$  & 383 (blocks 1-2-5-6) & 546 (layer 1)-612 (layer2) \\
 & & 766 (blocks 3-4) & 780 (layer 3)-746 (layer 4) \\
No. of tapes & -&  524 (blocks 1-2-3-4) & 546 (layer1)-612 (layer 2) \\
& & 540 (blocks 5-6) & 780 (layer 3)-746 (layer 4) \\
Lorentz force along $x$ & MN/m & 16.03 & 11.0 \\
Lorentz force along $y$ & MN/m &-12.13 &-9.52\\
Stored energy & MJ/m & 7.58 & 4.9\\
Inductance  & H/m & 2.46 & 3.4\\
\end{tabular}
\caption{Numerical results of the electromagnetic optimization of two considered configurations.}
\label{tab:Collider_tab1}
\end{table}

\begin{figure}
\centering
\includegraphics[width=\textwidth]{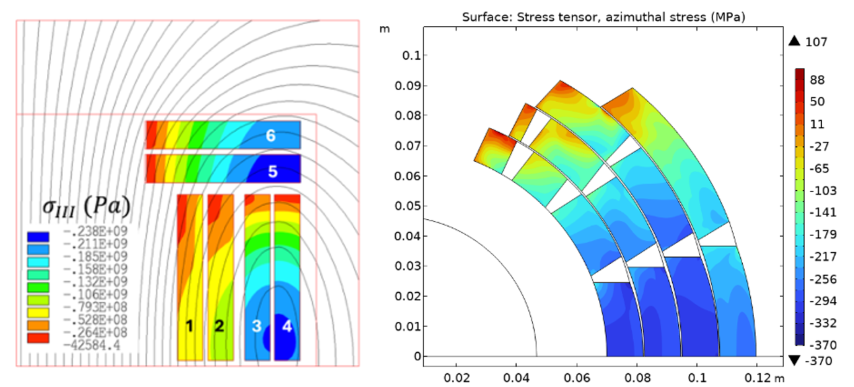}
\caption{Mechanical stress on conductor under Lorentz Forces at nominal current for both block coil and cos-theta magnet configurations.}
\label{fig:Collider_fig7}
\end{figure}

A preliminary mechanical study has been performed using ANSYS for the block coil design and COMSOL Multiphysics for the cos-theta coil configuration, both at nominal operating current. The mechanical structure has been designed to intercept the high electromagnetic forces. So far, in the evaluation of the maximum stresses in the conductors we did not consider contributions from assembly (e.g. pre-load), cool-down effects and energization phase. For this first estimation, an assumption of an ideal, infinitely rigid structure surrounding the coils is made, meaning zero nodal displacements are imposed on this element. To avoid over-constraining the conductors, and to obtain more realistic results, a gap of 0.3 mm is maintained between the infinitely rigid structure and each pancake of the block coil design, with standard frictionless contact type applied between different elements. Since the blocks structure is considered infinitely rigid, the thickness of the infinitely rigid ribs and cases is selected arbitrarily and the definition of material and mechanical properties of the structure is not relevant. For the cos-theta coil configuration a similar infinitely rigid domain boundary condition is applied to the collars surrounding the winding to evaluate the maximum stresses on the conductor under Lorentz Forces. Each layer is considered as independent, with a frictionless contact applied to the inter-layer boundary. For both coil configurations a Young’s modulus of E=174 GPa and Poisson’s ratio of $\nu=0.3$ has been considered for the conductors. The results of the preliminary mechanical simulations are shown in Figure~\ref{fig:Collider_fig7}. 

For the block coil design, we have reported the third principal stress component. This closely matches the vertical stress distribution along the for the lower four blocks and the horizontal stress distribution for the upper racetracks. These are the main components and directions of compressive stress on the tapes, which is one of the design limits. For the cos-theta magnet design, the azimuthal stress on the conductor is reported for comparison. 

All maximum stress values evaluated are within the maximum allowable compressive stress of 400 MPa for ReBCO tape along the direction perpendicular to the broad face. In the block coil design, the compressive stress perpendicular to the narrow face of the tape remains below 100 MPa. For the cos-theta coil configuration, a dedicated stress management strategy will be developed and implemented to mitigate radial stress accumulation on the coil, to avoid degradation of the conductor performance.

One of the issues of REBCO tapes to be understood and addressed is the effect of screening currents on field and AC loss. This is a complex matter, involving multi-physics phenomena. We are participating to the wider efforts to advance understanding and test solutions, but it will take time before accepted design methods and relevant experience are accumulated. This is why in parallel we have initiated evaluation of hysteresis losses in REBCO coils using different analytical and numerical approaches to obtain bounds. We report here first results of this work in progress.

For the block coil design we have performed an estimate of hysteresis loss using the Bean’s critical state model. The evaluation assumes complete penetration of the magnetic field (corresponding to the full saturation of each tape in the model) and the field oriented perpendicular to the broad face of the tape. This approach overestimates the losses for the ramping phase by neglecting field penetration and shielding, and represents therefore a worst case-scenario. The losses per unit length due to the magnet ramping without the contribution of the transport current are (for the entire cross section) Q/L=528 KJ/m per ramp. Considering also the contribution of the transport current they increase to Q/L=652 KJ/m per ramp. These values are high, indicating that we need to dwell in finer level of detail.

\begin{figure}
\centering
\includegraphics[width=\textwidth]{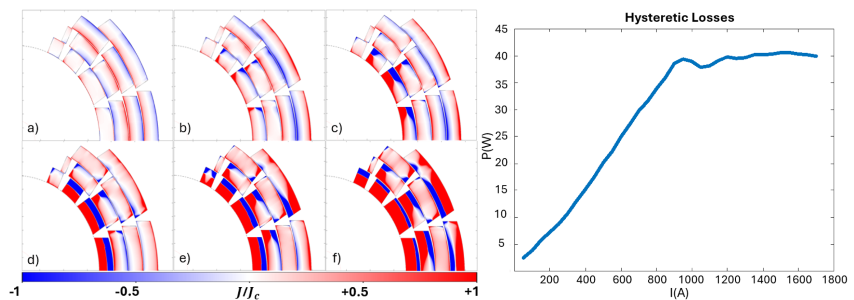}
\caption{Distribution of the current density in the coil cross-section at different step of the powering cycle and corresponding hysteretic losses profile as function of the magnet operating current.}
\label{fig:Collider_fig8}
\end{figure}

To demonstrate how to improve on the above estimate, we report below the results of a calculation performed for the cos-theta coil configuration using a MATLAB optimization routine based on the Brandt Hysteresis Model \cite{brandt1993type}. The routines compute the current density distribution within HTS tapes, and the resulting hysteretic losses in the coil. The calculated current density profile for each HTS tape is driven by the external field change, but also incorporates the effect of the superconductor layer magnetization and of the transport current at each step of the magnet powering cycle. Figure \ref{fig:Collider_fig8} illustrates the redistribution of current density within the cross-section’s superconducting tapes. The saturation of the outer layers, caused by the transport current, becomes evident at high operating current levels, while the inner layers remain unsaturated, continuing to contribute to the coil’s hysteretic losses. The hysteresis losses during magnet operation up to the nominal current were found to reach a value of Q/L=34.8 kJ/m. This is more than an order of magnitude lower than the analytic estimate described above, and is not far from values that could be acceptable for magnet ramp to nominal. There is clearly more work to be done, but as demonstrated here we have methods and indications on how to improve the design.

\paragraph*{Challenges identified}
The challenges of the muon collider magnets, based on any of the possible technology discussed above, are driven by the combination of field (or gradient) and large apertures. Nb-Ti is an exception, as solutions for large aperture magnets of field in the range of 5 T are readily available (e.g. the HL-LHC D1 magnet). Indeed, comparing the magnet specifications set earlier to the state-of-the-art and on-going developments for both Nb$_3$Sn and REBCO, it is evident that the magnet performance targeted is well above what has been achieved and planned so far. Still, we can profit from the fact that the lower center-of-mass collision energy required for the Muon Collider translates directly into a shorter collider ring. We can thus allow a higher cost per unit length, compared to other collider options, and design for larger coils, with more superconducting material, enabling higher dipole magnet performances, pushing the limits of magnet design.

Besides the field reach and aperture demands, both Nb$_3$Sn and REBCO share the difficulty of:
\begin{itemize}
\item	Managing large forces and stress, whereby REBCO has an advantage because of the higher resilience to compressive transverse stress and longitudinal tensile stress;
\item	Quench protection of magnets with large stored energy, where again REBCO may have the advantage of winding magnets using the non-insulated technology which, although to be demonstrated, may open the way for a next step for accelerator magnets;
\item	Cost. Any engineering solution needs to be affordable to scale up to the required series production of accelerator magnets, which calls for compact coils making the best use of the minimum amount of material.
\end{itemize}
Demonstrators will be necessary in any of the selected technologies. In addition, REBCO accelerator magnets will need to address:
\begin{itemize}
\item	The issues originating from the large shielding currents: AC loss during ramps, field distortions, and internal stresses developed as the current distribution changes inside a tape;
\item	Coil ends and terminations, which still remain a delicate region, and for which winding shape optimization studies, development and tests are required.
\end{itemize}

\section{Power converters}
\label{1:tech:sec:power}
\subsection*{Power converters for the muon accelerator}
\begin{figure}[h]
    \centering
    \includegraphics[width=0.55\linewidth]{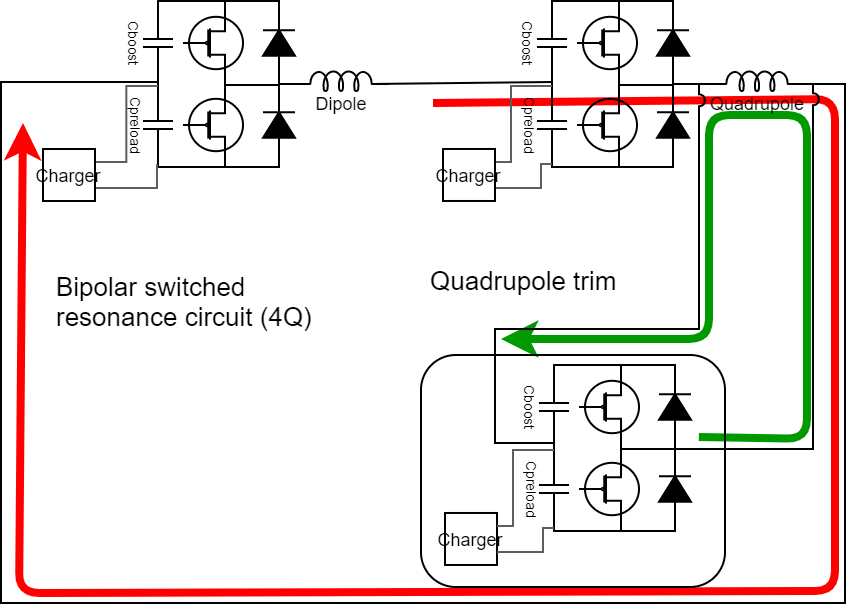}
    \caption{Possible quadrupole powering concept}
    \label{fig:PC_Fig1_QuadPower}
\end{figure}

Quick acceleration is made possible by the significant electrical power supplied by the power converters. The high $\frac{dB}{dt}$ values are associated with large voltage swings according to the total impedance of the magnets that are to be powered.
Concerning dipoles and quadrupole pulsed magnets, they could be connected together in series and powered by the same power converters with the Quadrupoles having an additional fast trim converter to be able controlling the tune as illustrated in Figure \ref{fig:PC_Fig1_QuadPower}.

Quadrupoles will not be considered further in this section, as their contribution to the total impedance of the power converters is expected to be minimal.

The remainder of this chapter focuses exclusively on the resistive pulsed dipole magnets. Based on the preliminary magnet designs presented in the magnet section, two representative configurations are used for sizing the power converters across all accelerator scenarios:

\begin{itemize}
\item \textbf{Dipole 1:} \quad $L_{\text{mag}} = 95~\mu\text{H/m}$,\quad $R_{\text{mag}} = 0.93~\text{m}\Omega/\text{m}$ 
\item \textbf{Dipole 2:} \quad $L_{\text{mag}} = 95~\mu\text{H/m}$,\quad $R_{\text{mag}} = 0.41~\text{m}\Omega/\text{m}$
\end{itemize}

Using these parameters, the corresponding peak voltages and powers required from the power converters are computed. The results are reported in Table~\ref{rf:tab:PC_table1_CERN_Scenario} for the CERN scenario and Table~\ref{rf:tab:PC_table2_GF_Scenario} for the Green Field scenario.
\begin{table}[h]
    \centering
    \begin{minipage}{0.43\textwidth}
        \centering
        \resizebox{\textwidth}{!}{%
        \begin{tabular}{l|ccc}
             & RCS & RCS & RCS\\
              & SPS &  LHC1 & LHC2 \\ \hline
             Length NC magnets [m]  & 4103 & 18650 & 12940\\
             Gap dimensions [mm] & 100 x 30 & 100 x 30 & 100 x 30\\
             Acceleration time [ms] & 0.45 & 2.60 & 4.42\\
             Inductive pk Voltage [MV] & 10.0 & 7.9 & 6.4\\
             Resistive pk Voltage [MV] & 0.029 & 0.133 & 0.092\\
             Resistive / Inductive ratio [\%] & 0.29 & 1.7 & 1.4\\
             pk Power [GW] & 110 & 87 & 70\\
             duty cycle [\%] & $\approx 0.45$ & $\approx 2.8$ & $\approx 4.42$\\
        \end{tabular}
        }
        \caption{CERN scenario}
        \label{rf:tab:PC_table1_CERN_Scenario}
    \end{minipage}%
    \hfill
    \begin{minipage}{0.49\textwidth}
        \centering
        \resizebox{\textwidth}{!}{%
        \begin{tabular}{l|cccc}
             & RCS1 & RCS2 & RCS3 & RCS4\\ 
              &  &  & & \\ \hline
             Length NC magnets [m]  & 3654 & 2539 & 4366 & 20376\\
             Gap dimensions [mm] & 100 x 30 & 100 x 30 & 100 x 30 & 100 x 30\\
             Acceleration time [ms] & 0.34 & 1.10 & 2.37 & 6.37\\
             Inductive pk Voltage [MV] & 11.8 & 5.0 & 4.0 & 7.0\\
             Resistive pk Voltage [MV] & 0.026 & 0.018 & 0.031 & 0.146\\
             Resistive / Inductive ratio [\%] & 0.22 & 0.36 & 0.77 & 2.08\\
             pk Power [GW] & 130 & 55 & 44 & 77\\
             duty cycle [\%] & $\approx 0.34$ & $\approx 1.1$ & $\approx 2.37$ & $\approx 6.37$\\
        \end{tabular}
        }
        \caption{Green Field scenario}
        \label{rf:tab:PC_table2_GF_Scenario}
    \end{minipage}
\end{table}

\begin{figure}[h] \centering \includegraphics[width=0.7\linewidth]{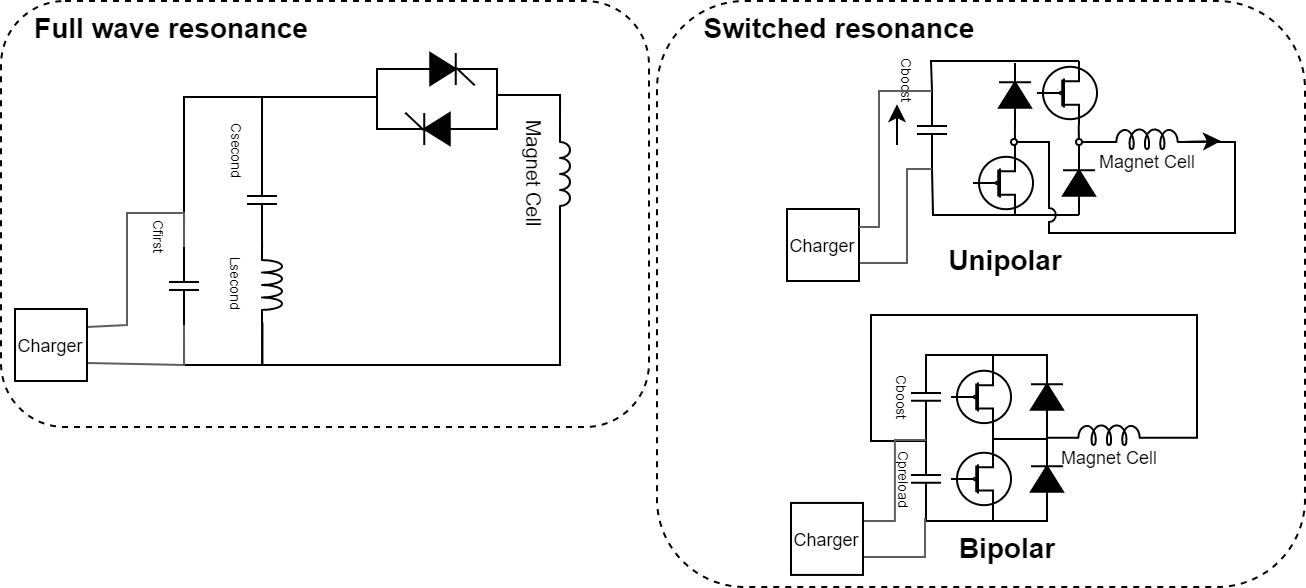} \caption{Full-wave resonance (left) and switched resonance (right) circuits} \label{fig:PC_Fig2_ResCircuits} \end{figure}

To meet the high power and voltage demands of the RCS, the power converters operate using pulsed resonant circuits. Two main types are considered: the full-wave resonance circuit and the switched resonance circuit. These configurations are illustrated in Figure~\ref{fig:PC_Fig2_ResCircuits}.

Both topologies rely on pre-charging one or more capacitors to an initial voltage, followed by activating a switch to discharge the energy into the load. As the load is almost purely inductive, the capacitors are nearly fully recharged at the end of the pulse. The switch is then opened and remains off until the next pulse cycle begins. The pulse typically lasts a few percent of the total repetition period, which is approximately 200~ms.

Given the extremely high voltage and power levels required across the full accelerator, it becomes necessary to divide the system into many sub-converters, referred to as Power Electronics cells (PE cells) in the remainder of this document.

In addition to the scale of the electrical power, an important challenge lies in achieving accurate current control across all PE cells—particularly in systems where the cells operate independently, as in the LHC sector model. For example, in the full-wave resonant topology, implementing current control would require each PE cell to be equipped with a fast, high-power active filter. This adds substantial complexity and cost, especially when regulation must occur within less than 1~ms.

An alternative approach, inspired by the CERN SPS configuration, is to connect all PE cells in series within a single circuit. This guarantees the same current through each cell, simplifying control to only ensuring repeatability from one pulse to the next. However, this architecture is not compatible with the full-wave resonant circuit, as it would require simultaneous switching of all cells—potentially hundreds—which would be impractical and unreliable.

\begin{figure}[h]
    \centering
    \includegraphics[width=0.85\linewidth]{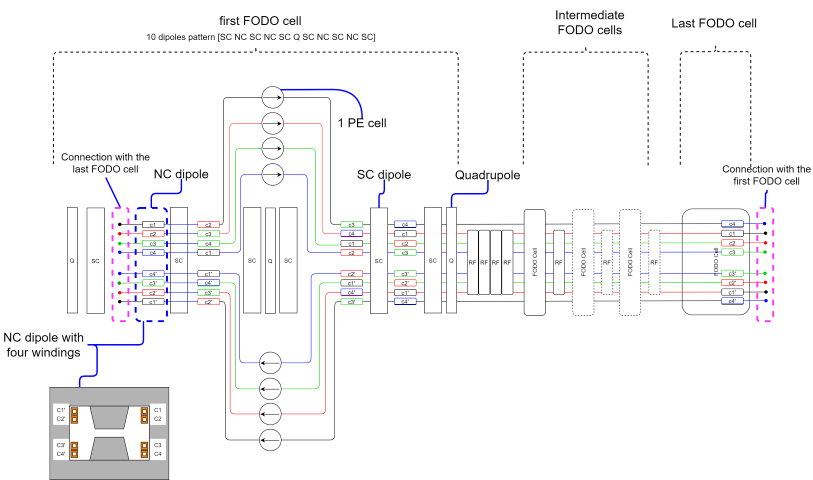}
    \caption{Series connection of all PE cells interleaved with the magnets' windings}
    \label{fig:PC_Fig3_SeriesConnection}
\end{figure}

Instead, the switched resonance circuit supports this series configuration more naturally. A conceptual diagram of such a multi-series arrangement is shown in Figure~\ref{fig:PC_Fig3_SeriesConnection}.
The series connection illustrated in Figure~\ref{fig:PC_Fig3_SeriesConnection} is based on treating each individual winding of the magnet as a separate load. These windings are connected in series with those of adjacent magnets, with power electronics (PE) cells interleaved throughout the chain. This arrangement continues around the full circumference of the accelerator.

While this configuration may require routing cables through sections where they are not strictly needed from a magnetic standpoint, it offers several advantages. First, it simplifies the design of the magnet coils. More importantly, it ensures that the same current flows through all magnets, eliminating any discrepancies between units. Additionally, since the PE cells are embedded in the load chain, their insulation and that of the magnets can be designed for similar voltage stress levels—namely, the differential voltage of the PE cells—simplifying the overall electrical insulation design

\subsection*{Operation of the pulsed switched resonance power converters}

As shown in Figure~\ref{fig:PC_Fig2_ResCircuits} (right), two main variants of the switched resonance circuit are used at the PE cell level: the Unipolar and the Bipolar configurations.

The Unipolar topology is employed when the RCS uses only pulsed magnets. In this case, the dipole current increases from a low injection value to a higher extraction value, following a unidirectional current profile.

The Bipolar topology is used in hybrid RCS configurations, which include both fixed-field and pulsed dipole magnets. Here, the pulsed magnets must produce negative currents at injection energy and swing up to positive values at extraction. This requires the power converter to generate a bidirectional current waveform, hence the need for the Bipolar configuration.

\begin{figure}[h]
    \centering
    \includegraphics[width=1\linewidth]{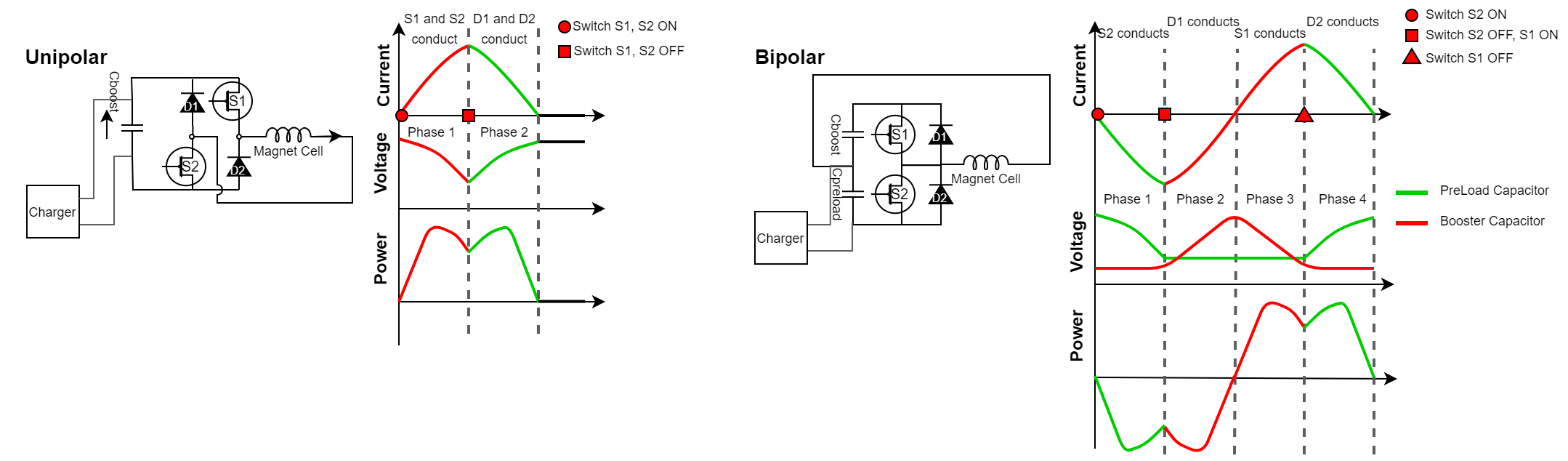}
    \caption{Operation principle of the Unipolar (left) and Bipolar (right) circuits}
    \label{fig:PC_Fig4_ResCircuitsConduction}
\end{figure}

The operating principles of the two circuit types are illustrated in Figure~\ref{fig:PC_Fig4_ResCircuitsConduction}.
In the Unipolar configuration, a single capacitor—referred to as Cboost—is used. This capacitor discharges into the load during Phase\,1 and is recharged by the returning current during Phase\,2.

In the Bipolar configuration, two capacitors are involved: Cpreload and Cboost. The Cpreload capacitor is discharged into the load during Phase\,1 to generate the initial negative current. It is recharged during Phase\,4. The Cboost capacitor is charged during Phase\,2 and discharged during Phase\,3 to provide the positive current swing.

\vspace{1em}

\begin{figure}[htbp] \centering 
\subfloat[\centering The unipolar resonance circuit (RCS1 CERN).]{\includegraphics[width=0.48\textwidth]{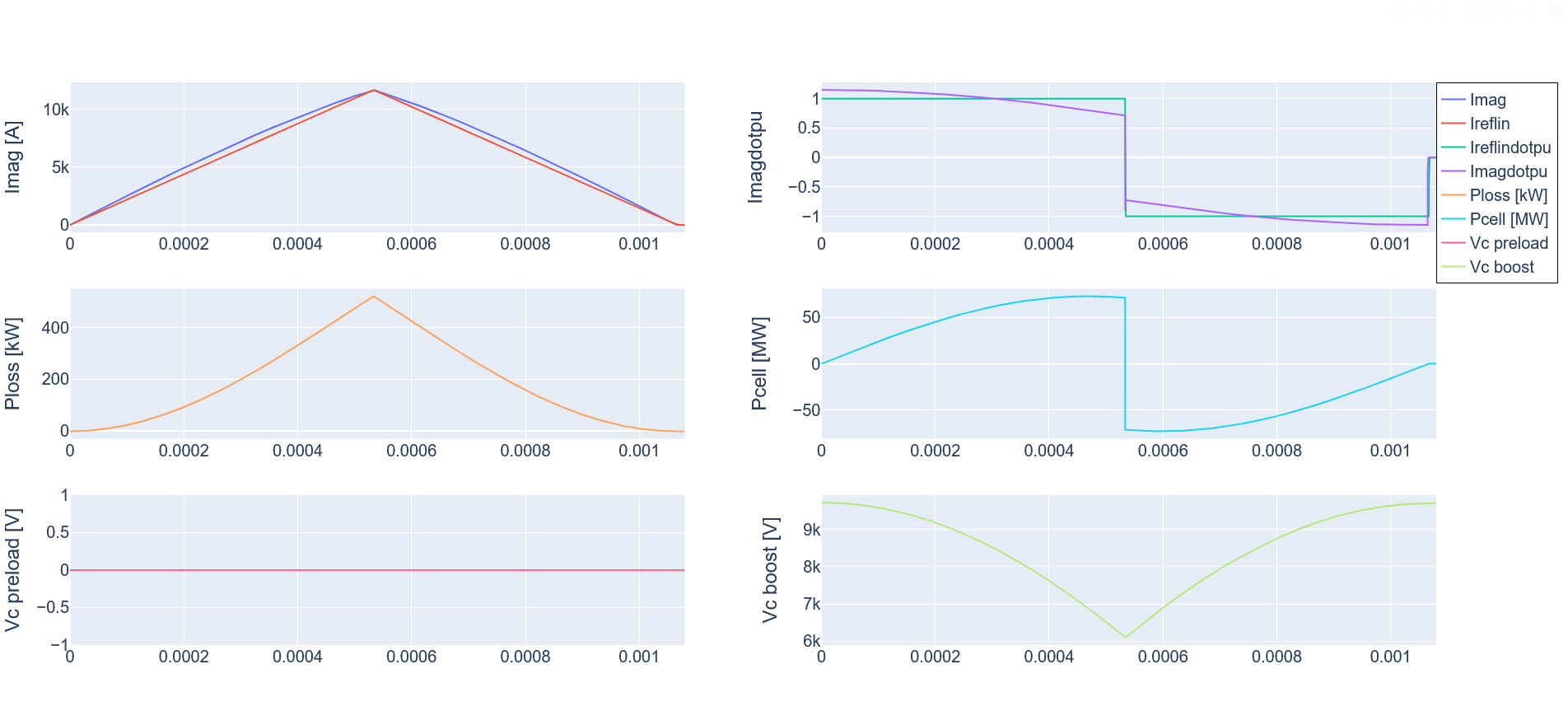} }
\subfloat[\centering The bipolar resonance circuit (HCS3 CERN).]{\includegraphics[width=0.48\textwidth]{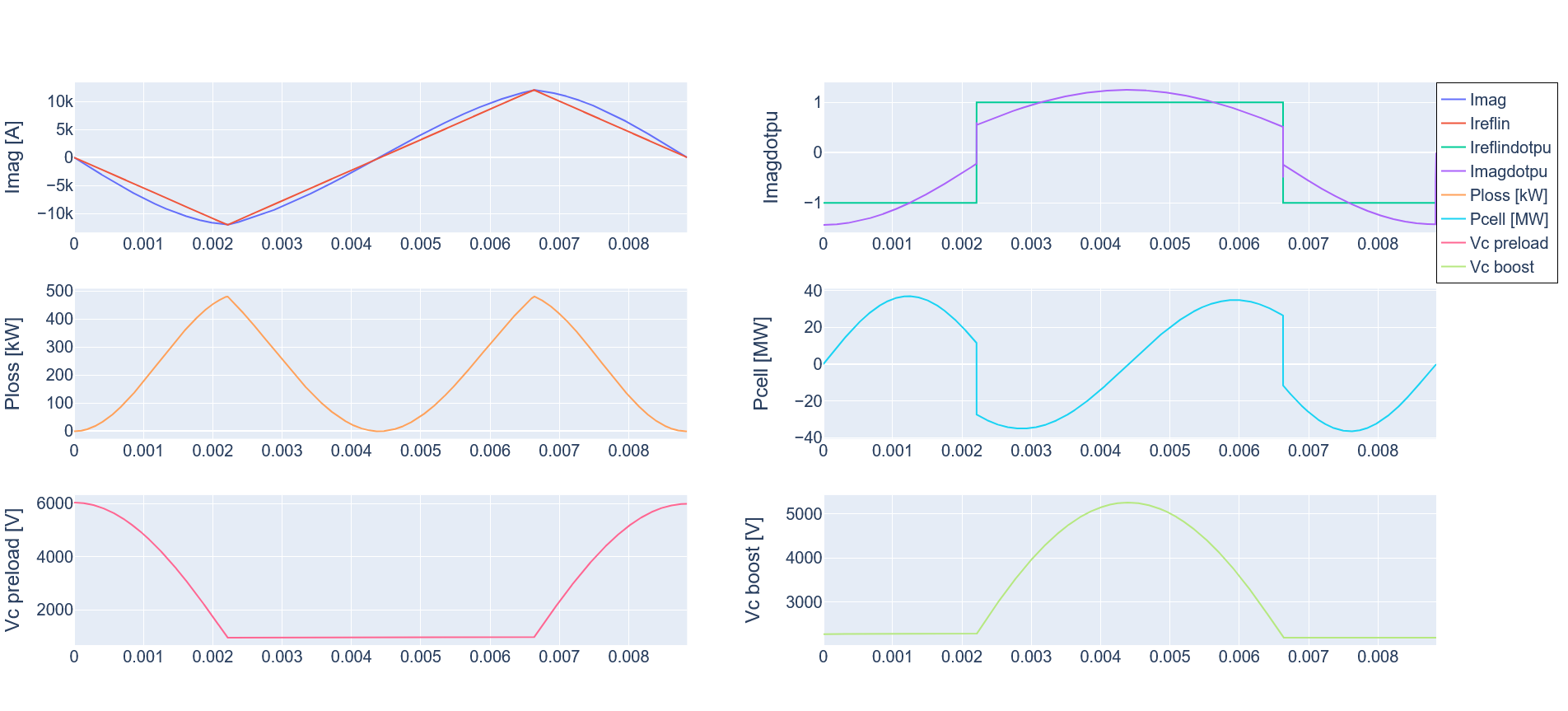} }
\caption{\label{pc:fig:Simulations} Simulation of the resonance circuits.}
\end{figure}

Figure~\ref{pc:fig:Simulations} shows simulations of the two resonance types for the RCS1 and HCS3 configurations in the CERN scenario. The results demonstrate that the presence of a switching device allows control over the linearity of the current ramp by adjusting the energy stored in the resonant capacitors.

Furthermore, nearly all of the energy delivered to the magnets during the pulse is recovered into the capacitors. The small voltage difference between the beginning and end of the pulse corresponds to system losses—primarily within the resistive magnets.

\subsection*{Preliminary dimensioning results}

The dimensioning of the power converters is closely linked to the characteristics of the resistive dipole magnets. The results presented here are based on Dipole designs N1 and N2, as described earlier, and applied across all accelerator scenarios. 

Final converter specifications will evolve alongside the ongoing design of the RCS magnets and their integration into the accelerator layout. This process will ultimately allow for system-level CAPEX+OPEX optimization, including the design of cabling between the power electronics (PE) cells and the magnets.

Tables~\ref{pc:tab:PC_table4_EnergyandLosses_summary} and~\ref{pc:tab:PC_table3_PEcells_Dimensioning} present a possible dimensioning scenario for the CERN and Green Field cases. The reported losses are primarily associated with the resistive magnets. While values may vary depending on alternative design criteria, the general order of magnitude remains valid—particularly the high number of units required to manage peak power demands.

As an example, RCS1 would require approximately 1000 PE cell cabinets, each with dimensions around 3\,m $\times$ 1.2\,m $\times$ 2.5\,m.

\begin{table}[h]
    \centering
      \resizebox{0.95\textwidth}{!}{%
        \begin{tabular}{l|ccccccc}
        & RCS1 CERN & RCS2 CERN & HCS3 CERN
              & RCS1 GF &  HCS2 GF & HCS3 GF & HCS4 GF \\ \hline
        Magnet type  & Dipole N1 & Dipole N2 & Dipole N2 & Dipole N1
                     & Dipole N1 & Dipole N1 & Dipole N2   \\
        Ramping time [ms] & 0.45 & 2.6 & 4.42 & 0.34 & 1.1 & 2.37 & 6.37 \\
        Initial capacitors energy [J] & 43536 & 164308 & 70812 & 50799 & 17068 & 25390 & 112197 \\
        Final capacitors energy [J] & 43327 & 162310 & 68971 & 50608 & 16811 & 24582 & 108032 \\
        Magnetic energy [J] & 26346 & 96443 & 56037 & 30873 & 13588 & 20028 & 88239 \\
        Loss energy x PEcell (integral of RI2) [J/cycle] & 207 & 1971 & 1827 & 187 & 253 & 804 & 4143 \\
        Total average power losses [MW] & 1.03 & 13.20 & 14.62 & 0.71 & 1.52 & 5.63 & 33.15 \\
        \end{tabular}
        }
        \caption{Storage Energy and losses computation for the CERN and Green Field Scenarios}
        \label{pc:tab:PC_table4_EnergyandLosses_summary}
\end{table}

\begin{table}[h]
    \centering
      \resizebox{0.95\textwidth}{!}{%
        \begin{tabular}{l|ccccccc}
        & RCS1 CERN & RCS2 CERN & HCS3 CERN
              & RCS1 GF &  HCS2 GF & HCS3 GF & HCS4 GF \\ \hline
        Magnet type  & Dipole N1 & Dipole N2 & Dipole N2 & Dipole N1
                     & Dipole N1 & Dipole N1 & Dipole N2   \\
        Ramping time [ms] & 0.45 & 2.6 & 4.42 & 0.34 & 1.1 & 2.37 & 6.37 \\
        Pulse time [s] & 0.9 & 5.2 & 8.84 & 0.68 & 2.2 & 4.74 & 12.74 \\
        Number of PE cells & 1000 & 1340 & 1600 & 760 & 1200 & 1400 & 1600 \\
        Number of Dipoles & 500 & 2010 & 1600 & 380 & 600 & 700 & 1600 \\
        Number of coils x Dipoles & 4 & 4 & 4 & 4 & 4 & 4 & 4 \\
        Number of coils x PEcell & 2 & 6 & 4 & 2 & 2 & 2 & 4 \\
        Total cabinet width [m] & 3.0 & 2.5 & 3.0 & 4.1 & 2.5 & 1.9 & 3.5 \\
        Caps cabinet width [m] & 0.2 & 0.9 & 0.7 & 0.3 & 0.1 & 0.2 & 1.1 \\
        Stacks cabinet width [m] & 2.2 & 1.1 & 1.8 & 3.3 & 1.8 & 1.1 & 1.8 \\
        Peak power of one cell [kW] & 113565 & 69046 & 100366 & 172381 & 97448 & 67024 & 110058 \\
        Capacitor max energy [J] & 43536 & 164308 & 126416 & 50799 & 30594 & 45229 & 199466 \\
        Capacitor max voltage [V] & 9728 & 5725 & 8322 & 14766 & 8347 & 5741 & 9126 \\
        Magnet resistance x PEcell [ohm] & 0.003819 & 0.005704 & 0.003314 & 0.004475 & 0.001970 & 0.002903 & 0.005219 \\
        magnet inductance x PEcell [H] & 0.000387 & 0.001326 & 0.000770 & 0.000453 & 0.000199 & 0.000294 & 0.001213 \\
        Peak magnet current [A] & 11674 & 12060 & 12060 & 11674 & 11674 & 11674 & 12060 \\
        RMS magnet current [A] & 520 & 1314 & 1660 & 458 & 802 & 1177 & 1992 \\
        max dI/dt in pu & 1.15 & 1.15 & 1.25 & 1.15 & 1.25 & 1.25 & 1.25 \\
        Cpreload [uF] & 0 & 0 & 3172 & 0 & 759 & 2391 & 4184 \\
        Cboost [uF] & 920 & 10026 & 4955 & 466 & 1186 & 3735 & 6536 \\
        Uc0boost [V] & 9727 & 5725 & 2275 & 14765 & 2273 & 1573 & 2505 \\
        Uc0preload [V] & 0 & 0 & 6046 & 0 & 6073 & 4167 & 6620 \\
        Number of stacks & 6 & 3 & 6 & 15 & 6 & 3 & 6 \\
        Presspacks per stack & 8 & 8 & 6 & 4 & 6 & 8 & 6 \\
        IGBTs/Diodes in series & 4 & 2 & 3 & 5 & 3 & 2 & 3 \\
        IGBTs/Diodes in parallel & 3 & 3 & 3 & 3 & 3 & 3 & 3 \\
        Total presspack components & 48 & 24 & 36 & 60 & 36 & 24 & 36 \\
        I RBSOA [A] & 4800 & 4800 & 4800 & 4800 & 4800 & 4800 & 4800 \\
        U RBSOA [V] & 3000 & 3000 & 3000 & 3000 & 3000 & 3000 & 3000 \\
  
        \end{tabular}
        }
        \caption{Dimensioning of the power converter components for the CERN and Green Field scenario}
        \label{pc:tab:PC_table3_PEcells_Dimensioning}
\end{table}

Routing and connecting all these units into a coherent, series-connected system poses a significant challenge. This aspect will require a dedicated optimization study once the accelerator layout becomes more clearly defined.

\begin{figure}[h]
    \centering
    \includegraphics[width=1\linewidth]{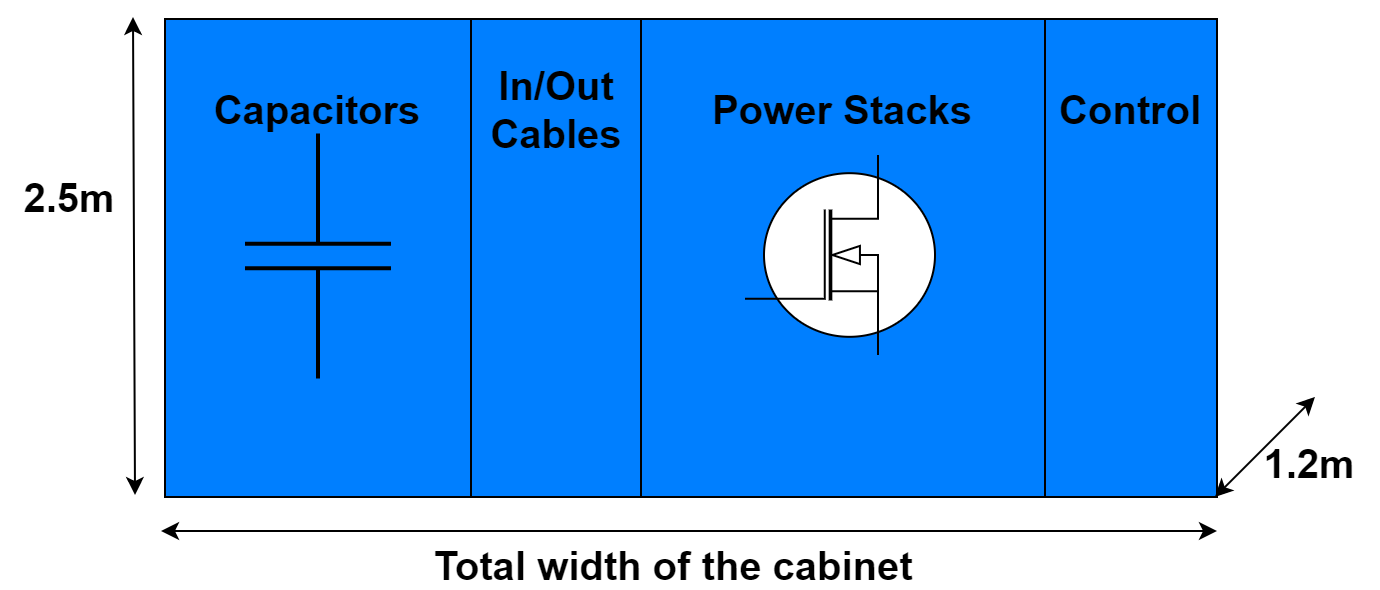}
    \caption{PE cell cabinet structure}
    \label{pc:fig:PC_cabinet}
\end{figure}

The tables also highlight the need to divide the accelerator powering system into a large number of PE cells—often exceeding 1000. The cost evaluation for each PE cell assumes the cabinet structure shown in Figure~\ref{pc:fig:PC_cabinet}. Dimensional data and unit counts for the different accelerator variants are reported in Table~\ref{pc:tab:PC_table3_PEcells_Dimensioning}.

Given the high number of power cabinets involved, a global optimization strategy encompassing power converters, magnets, and civil engineering will be essential for an efficient and cost-effective installation.

\FloatBarrier
\section{Radio-frequency systems}
\label{1:tech:sec:rf}
\label{sec:Section7_3}
\subsection*{RF system for muon cooling}

The RF system for the muon cooling complex consists of several 1000s RF cavities operating in the frequency range starting from a few hundred MHz. The majority of the cavities operate at 352 and 704~MHz in the initial and 6D-cooling channels. The cooling channel design foreseen a scheme where the acceleration induced by the RF counterbalances the absorbers' effect, and a strong solenoidal magnetic field is applied to focus the muons. The requirement on the maximum possible value of the RF electric field, which is necessary to re-accelerate muons as fast as possible due to their short lifetime, is limited by the maximum allowable fields that may be sustained without breakdown effects in the particular electromagnetic structure foreseen.
The fact that the RF cavities will operate in extremely high magnetic fields (that is a peculiarity of this scheme) poses further severe limitations on the performances, due to the well-known effects on maximum high electric fields obtainable under the action of magnetic fields.
Due to the requirement to operate in the presence of magnetic fields, the RF cavities must be of the normal conductive type. Beam dynamics design dictates high gradient values of a few tens of MV/m, which results in high peak RF power on the order of a few MWs per single cell cavity. This leads to a very high peak power requirement of a few tens of GW. On the other hand, the single (or few) bunch operation mode results in relatively short pulse lengths of $\approx 10 - 40$ \textmu s and relatively low duty factor of $\approx10^{-4}$.

Based on the input from the beam dynamics design, a consistent set of parameters for all of the RF cavities and their associated RF systems is elaborated below for the 6D rectilinear muon cooling channel, creating the backbone of the RF system for the muon cooling complex. 
A preliminary design of the RF cavities for each stage of the rectilinear cooling channel was developed, taking as a reference the performances outlined in the design of the cooling channel and trying to minimize the side effect of these requirements. The study of the cavities evolved starting from a design based on a single cell, pill-box type, up to a multicell magnetic coupled structure. The single-cell model has been studied considering the optimization of the shape to enhance the shunt impedance ($R/Q \cdot Q_{0}$ ) and minimize surface power dissipation ($P_{\mathrm{diss}}$) on the cavity walls.  The cavity design is very particular with the iris apertures electrically closed using thin conductive foil, and a beam window. The foil limits the number of muons that travel through it along the acceleration path. This results in a strict requirement on their thickness (with a maximum considered value of the order of tens of microns) and on the atomic number of the material proposed. Beryllium has been suggested as an initial option, but due to its peculiarity (mainly health hazards), Aluminium is under analysis too. The peak surface electric field on the copper wall ($E_{\mathrm{peak,Cu}}$) as well as on the Be window ($E_{\mathrm{peak,Be}}$) was minimized to mitigate the risk of RF breakdown.

\begin{figure}[!htb] 
\centering 
\includegraphics[width=0.7\columnwidth]{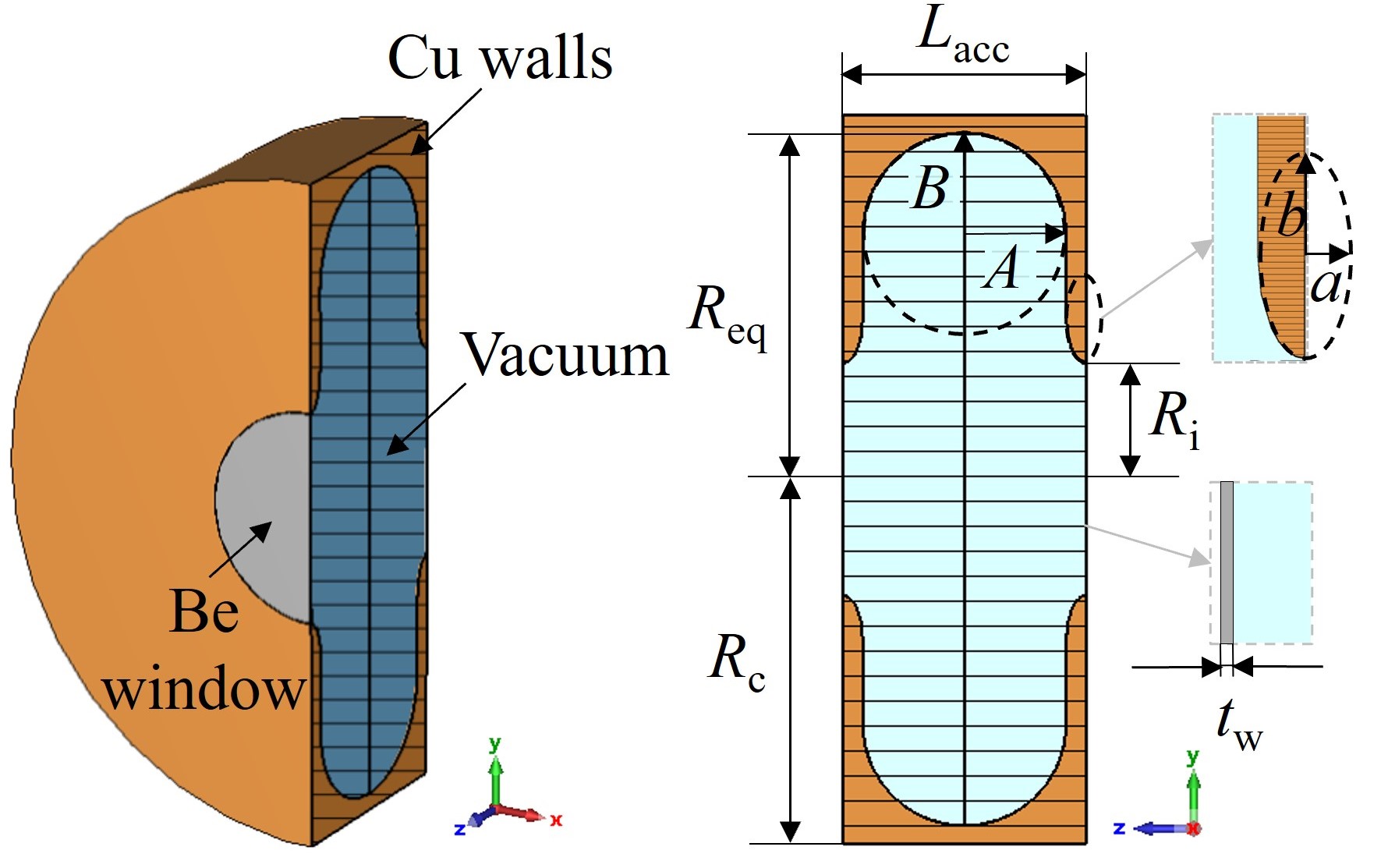} 
\vspace{0.1\baselineskip} 
\caption{Design of the 704 MHz cavity for muon cooling.} 
\label{fig:cavity_parametrization}
\end{figure}

Two distinct RF cavity models have been developed. The first model is based on the geometry described in \cite{Barbagallo:2024zak}, as illustrated in Figure~\ref{fig:cavity_parametrization}. The cavity design was carried out following the beam dynamics specifications outlined in Table~\ref{cool:tab:6d_rf}.

\noindent
This model has been used to carry out a complete analysis of the different cavity parameters foreseen in the beam dynamics studies of the cooling channel and reported below. Table~\ref{cool:tab:6d_rf} lists the RF cavity frequency ($f_{0}$), cavity length ($L_{\mathrm{cav}}$), and nominal RF gradient along the cavity axis ($E_{\mathrm{nom}}$) for the studied cavities. Additionally, Table \ref{rf:tab:cool_dyn} provides the RF cavity window radius, window thickness, and the relativistic beta of the muon beam at each stage.

\begin{table}[!h]
    \centering
    \begin{tabular}{l|ccc}
         & Window&  Window&Relativistic\\
          & radius &  thickness &$\beta$ \\
      & mm&  $\mu$m&- \\ \hline
         Stage A1 & 240&   120& 0.92\\
         Stage A2 & 160&   70&  0.89\\
         Stage A3 & 100&   45&  0.89\\
         Stage A4 &  80&   40&  0.90\\
         Stage B1 & 210&   100& 0.88\\
         Stage B2 & 190&   80&  0.88\\
         Stage B3 & 125&   50&  0.88\\
         Stage B4 &  95&   45&  0.89\\
         Stage B5 &  60&   30&  0.89\\
         Stage B6 &  45&   20&  0.89\\
         Stage B7 &  37&   20&  0.89\\
         Stage B8 &  27&   20&  0.88\\
         Stage B9 &  23&   10&  0.88\\
         Stage B10&  21&   10&  0.88\\
    \end{tabular}
    \caption{Beam dynamics specifications for the RF cavities in the rectilinear cooling channel}
    \label{rf:tab:cool_dyn}
\end{table}

Table~\ref{rf:tab:cool_cavity} summarises the key RF figures of merit calculated for the operational frequencies of these cavities. Most of the power dissipation occurs in the cavity walls.
The filling time, \( t_{\mathrm{f}} \), refers to the time needed to charge the cavity to the nominal voltage \( V_{\mathrm{nom}} = E_{\mathrm{nom}} L_{\mathrm{cav}} \).
\( Q_0 \) is the intrinsic quality factor of the cavity operating mode. The cavity Duty Factor (\( DF \)) is defined as the ratio of the average power to the energy dissipated on the cavity surface in one RF pulse times the repetition rate $f_\text{b}$ and \( V_{\mathrm{acc}} =~TTF \cdot V_{\mathrm{nom}} \) is the accelerating cavity voltage, with the \(  TTF  \) being the Transit-Time Factor (TTF).

Table~\ref{rf:tab:cool_power} presents the power requirements for each stage of the cooling channel. The peak input RF power is expressed as: \( P_{\mathrm{g}} = P_{\mathrm{diss}} \beta_{\mathrm{c}}.\)
The duty factor of the RF power source (\( DF_{\mathrm{g}} \)) is the ratio of the generator's average power to the peak input RF power.
The total plug power for the RF systems is calculated by considering the efficiencies of the generator (\( \eta_{\mathrm{G}} \)) and modulator (\( \eta_{\mathrm{M}} \)), as reported in Table~\ref{rf:tab:cool_const}, using the following expression: \(
P_{\mathrm{plug}} = {N_{\mathrm{cav}} P_{\mathrm{ave,g}}}/{\eta_{\mathrm{G}} \eta_{\mathrm{M}}}\) where \( N_{\mathrm{cav}} \) is the total number of cavities in each stage.

\begin{table}[!h]
    \centering
    \begin{tabular}{c|cccccccc}
          & $Q_0$& $t_f$& $DF$& R/Q&  ${P_{\mathrm{diss}}}$ & $E_{\mathrm{peak,Cu}}$&$E_{\mathrm{peak,Be}}$\\
         Stage & $10^4$& \textmu s& $10^{-4}$& $\Omega$ & MW/cavity & MV/m & MV/m\\ \hline
         A1& 3.06& 31.2& 1.17& 172& 4.25& 12&   27\\
         A2& 3.14& 32.1& 1.21& 150& 4.34& 23&   27\\
         A3& 2.20& 11.2& 0.43& 160& 2.06& 21&   32\\
         A4& 2.22& 11.3& 0.43& 150& 2.21& 28&   32\\
         B1& 3.91& 40.0& 1.51& 184& 2.78& 12&   22\\
         B2& 3.56& 36.3& 1.37& 170& 3.24& 16&   23\\
         B3& 3.15& 32.1& 1.21& 141& 4.07& 26&   24\\
         B4& 3.59& 36.7& 1.38& 154& 3.92& 28&   23\\
         B5& 2.23& 11.4& 0.43& 141& 1.18& 24&   22\\
         B6& 2.22& 11.4& 0.43& 137& 2.34& 37&   30\\
         B7& 2.22& 11.4& 0.43& 137& 2.11& 36&  27\\
         B8& 2.22& 11.3& 0.43& 137& 1.92& 36&  24\\
         B9& 2.22& 11.3& 0.43& 138& 2.14& 39&  23\\
         B10& 2.22& 11.3& 0.43& 139& 1.51& 32& 18\\
    \end{tabular}
    \caption{RF figures of merit for the RF cavities in the rectilinear cooling channel}
    \label{rf:tab:cool_cavity}
\end{table}

\begin{table}[!h]
    \centering
    \begin{tabular}{ccc|c}
         Parameters&  Symbol&  Unit&  Value\\ \hline
         Coupling factor&  $\beta_{\text{c}}$&  -&  1.2\\
         Bunch repetition frequency&  $f_\text{b}$&  Hz&  5\\
         Generator efficiency&  $\eta_\text{G}$&  -&  0.7\\
         Modulator efficiency&  $\eta_\text{M}$&  -&  0.9\\
    \end{tabular}
    \caption{RF system parameters for the rectilinear cooling channel.}
    \label{rf:tab:cool_const}
\end{table}
\noindent

\begin{table}[!h]
    \centering
    \begin{tabular}{c|cccccc}
          & $P_g$& $DF_g$& $N_{cav}$& $P_{g,tot}$& $P_{g,av}$&$P_{plug}$\\
         Stage & MW/cavity& $10^{-4}$& -& MW& kW&kW\\ \hline
         A1&  5.1&  1.6&  348&  1773& 277&   440\\
         A2&  5.2&  1.6&  356&  1855& 298&   473\\
         A3&  2.5&  0.6&  405&  999&  57&    90\\
         A4&  2.7&  0.6&  496&  1317& 75&    120\\
         B1&  3.3&  2.0&  144&  480&  96&    153\\
         B2&  3.9&  1.8&  170&  660&  120&   191\\
         B3&  4.9&  1.6&  216&  1055& 170&   270\\
         B4&  4.7&  1.8&  183&  860&  159&   252\\
         B5&  1.4&  0.6&  275&  390&  22&    36\\
         B6&  2.8&  0.6&  220&  618&  35&    56\\
         B7&  2.5&  0.6&  204&  516&  30&    47\\
         B8&  2.3&  0.6&  276&  636&  36&    58\\
         B9&  2.6&  0.6&  212&  545&  31&    50\\
         B10& 1.8&  0.6&  196&  354&  20&    32\\
         \\
         All    & -  &  -  & 3701&12058& 1426& 2268\\
    \end{tabular}
    \caption{RF power requirements in the rectilinear cooling channel.}
    \label{rf:tab:cool_power}
\end{table}

\begin{figure}[!htb] 
\centering 
\includegraphics[width=0.48\columnwidth]{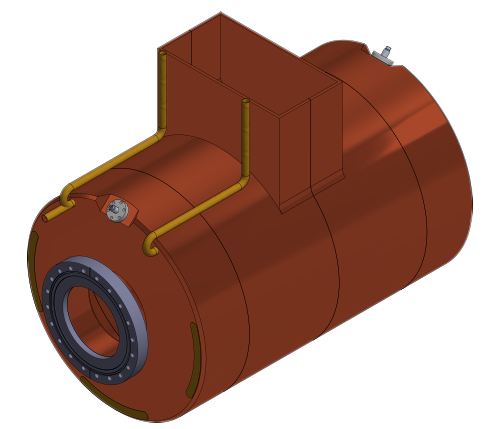} 
\vspace{0.1\baselineskip} 
\caption{Flat-top cavity design for the muon cooling demonstrator.} 
\label{fig:cavity_parametrization_flat_top}
\end{figure}

The second model foresees a flat-top profile, as described Figure~\ref{fig:cavity_parametrization_flat_top}. 
It was designed based on the beam dynamics specifications for the muon cooling demonstrator, which is close to Stage B5 but uses only 3 RF cells with an RF phase advance of 180 degrees. This allows the use of a multicell standing wave RF structure. The shape has been optimized to accommodate a multicell structure with magnetic coupling and provide space for the insertion of a magnetic power coupler in the middle RF cell. 
The coupler foresees the use of a standard WR1150 waveguide brazed to the structure. The coupling of the cells is obtained through 4 coupling slots machined on the lateral walls and optimized using a numerical simulation model.
Figure~\ref{fig:cavity_parametrization_flat_top} reports a sketch of the simulated model along with the fields in the structure.
A thermo-mechanical analysis has been carried out to study the efficiency of the designed cooling system, based on 8 longitudinal cooling channels with a 12 mm diameter, and to verify that the overall increase of temperature in the slab walls would be limited to a few degrees. The water channels will be collected as a bundle in the space around the power coupler.
A study of the machining of the RF cells and their brazing to result in a single 3-cell structure has been completed. The cavity weight has been computed as a total of 180 kg of copper.
A pair of RF pick-ups has been foreseen to monitor the fields in the structure.

\begin{figure}[!htb] 
\centering 
\includegraphics[width=0.48\columnwidth]{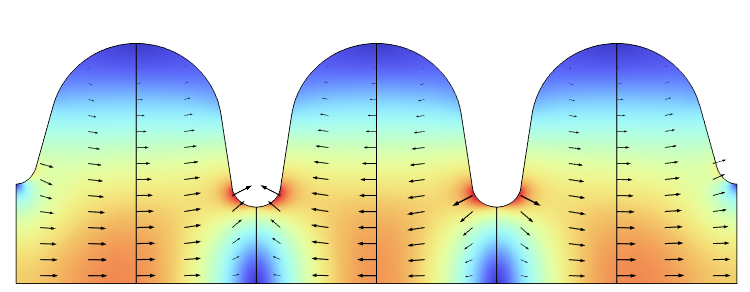} 
\vspace{0.1\baselineskip} 
\caption{Electric field pattern in a 3 cell elliptical cavity with electric coupling.} 
\label{fig:ESS_type_cavity}
\end{figure}

In parallel, a cavity with electric coupling has been designed by CEA, to allow comparison with the magnetic coupling structure. A main advantage is to avoid the presence of metallic sheet (beryllium or any other metal) between the cells.
This 3-cell cavity, based on a $\pi$-mode, has been adapted from an ESS, 704 MHz and $\beta$ = 0.86 cavity \cite{cenni2019}, and adapted to the MuCol stage B5 requirement: $\beta$ = 0.89 and iris diameter of 60~mm. The elliptical shape allows minimization of the surface electric field, responsible for breakdown issues. The cavity length is $L_{acc}=3 \beta \lambda/2$~=~0.56~m, the transit time factor is 0.72, the quality factor is $Q_0~=~3.83\times10^4$ and the shunt impedance is 12.3~M$\mathrm{\Omega}$. For a cavity gradient of 22.5~MV/m, the maximal field on cavity surface is 29.5~MV/m, slightly higher than for a pill-box type cavity.
The electric field in the cavity is shown in Figure~\ref{fig:ESS_type_cavity}.

\paragraph*{Key challenges}

Several experiments on breakdown rates in cavities under a high magnetic field have been carried out in the framework of Muon Accelerator Program at Fermilab (MAP)~\cite{jinst_muons}). Higher breakdown rates leading to lower achievable accelerating gradients were observed in copper cavities operating at 805~MHz. A model based on the "beamlet" principle obtained a good correlation with experiments \cite{Bowring:2018smm}. That model predicts the appearance of breakdown when the local temperature rise exceeds a threshold beyond which plastic deformation and surface damage may affect cavity behavior. Additionally, it also relates the local temperature rise to the magnetic field. Although the study predicted an improvement for cavities made of beryllium, which is corroborated by measurements: a maximal gradient of nearly 50\,MV/m is kept in the presence of 3\,T, the high gradient operation in a strong magnetic field remains the key challenge for the muon cooling RF system which must be addressed experimentally in the most urgent manner.

Integration of the RF cavities together with high-field superconducting solenoids and absorbers in the common vacuum vessel is another engineering challenge. The superconducting solenoids operate at cryogenic temperature whereas RF cavities and the associated RF network operate at room temperature. Combining all the subsystems in a compact way is a key challenge in the engineering design of the muon cooling channel.

 Extremely high muon bunch charge results in very strong beam loading effect in the RF cavities. The impact of the beam loading on the muon energy spread and in general on the muon cooling process must be investigated and appropriate mitigation measures to be found. This remains one of the key challenges in the design of the RF system for the muon cooling complex.

\subsection*{RF system for acceleration} 
\label{1:tech:sec:rf:highacc}
After the cooling channel, a linear accelerator and two Recirculating Linear Accelerators (RLAs) provide an initial acceleration up to \SI{63}{\GeV}. 
The main RF frequency of the RLAs is \SI{352}{\mega \hertz} with an accelerating gradient of \SI{15}{\mega\volt\per\meter}. In addition to the accelerating cavities, 3rd harmonic linearizing cavities with a frequency of \SI{1056}{\mega \hertz} and a gradient of \SI{25}{\mega\volt\per\meter} will be used. 
Downstream from the RLAs, four Rapid Cycling Synchrotrons (RCSs) will gradually increase the energy of the two muon bunches within a few tens of turns each up to the collision energy of \SI{5}{\tera \electronvolt}. 
During the transition from the pre-accelerator into the first RLA, the bunches are split and continue as counter-rotating bunches. 
The design choices for the RF system will be guided by the requirements resulting from the beam dynamics simulations and the short muon lifetime. 
As a result, a high RF voltage per turn is required, supplied by hundreds of superconducting \SI{1.3}{\giga \hertz} cavities per RCS, which operate at an accelerating gradient of \SI{30}{\mega\volt\per\meter}. 
A more detailed description of the acceleration chain can be found in Section~\ref{1:acc:sec:acc}. 

\paragraph*{Key challenges}
All accelerators following the first RCS will be implemented as hybrid RCSs with both normal- and superconducting magnets.  
Due to the nature of this hybrid magnet design, the orbit length and, thus, the revolution period changes during the acceleration, leading to the necessity of fast frequency tuning capabilities for the cavities. 

The large bunch charge of up to $2.7\times10^{12}$ muons per bunch in the RCS chain will lead to significant transient beam loading and higher order mode (HOM)-induced power within the cavities. 
While the instantaneous requirements for cavity powering and HOM power extraction are high, the machine's duty cycle is low. 
The cavities will be operated in a pulsed mode. 

The requirements for cavity powering are further exaggerated in the RLAs, with a higher peak and lower average power. 
The number of passes is $4.5$, while the beam current is higher than in the first RCS due to the higher bunch charge and lower travel time between cavity passages. The induced voltage in both the fundamental mode (FM) and HOMs will be significantly lower as the aperture of the \SI{352}{\mega \hertz} cavity is significantly larger. 

The cavities in the RCS chain will need to be distributed around the ring in several stations to mitigate the influence of the high synchrotron tune. 
The impact of the beam loading will, therefore, not be consistent in all cavities due to the differing time between the passages of the counter-rotating $\mu^+$ and $\mu^-$ bunches.
The same challenge applies to the extraction of the HOM power and cavity powering. 
Due to the distribution of the RF system, there will additionally be a need to either distribute the powering and cryogenic infrastructure around the ring or to distribute the generated RF power over long distances. 
With long delay times, the operation of cavity feedback will be significantly more challenging. 

Currently, the baseline includes the usage of one klystron for multiple cavities at the same time. 
Within this scheme, also cavity feedback will only be able to act on multiple cavities at the same time. 
The effect of the counter-rotating beam and the distributed RF system and control infrastructure will need to be studied in detail. 

The design process of the cavity shape for the high energy acceleration has to carefully balance the cavity HOM performance against the fundamental mode performance, which might lead to different designs for each ring due to the different beam currents and bunch lengths. 
In the cavity design for the low-energy acceleration, one also has to take the particle speed into account, as some of the accelerators operate in an energy regime where the particles are not ultra-relativistic. 

\paragraph*{Recent achievements}
Table~\ref{2:tech:rf:highacc:tab:RCS_RFpars_low_E} lists a first power estimation of the RF system in RLA2. 
For the distribution and klystron losses, the same number as for the high-energy acceleration was applied. 
The beam current is averaged over the passage through the complete accelerator, assuming the injection bunch charge. 
No estimates are presented for RLA1 or the PA, as no design is currently available. 
\begin{table}[!h]
    \centering
    \begin{tabular}{lc|cc}
         Parameter &  Unit &  RLA2 acc& RLA2 lin\\ \hline
         Synchronous phase &  \textdegree &  95& 275\\
         Frequency&  MHz&  352& 1056\\
         Number of cavity cells&  -&  4& 6\\
         Active cavity length&  mm&  1686& 845\\
         Total cavity length&  mm&  1851& 1010\\
         Number of bunches/species &  - &  \multicolumn{2}{c}{1} \\
         Combined beam current ($\mu^+$, $\mu^-$) &  mA&  \multicolumn{2}{c}{134} \\ \hline
         Total RF voltage &  GV&  15.2& 1.69\\
         Total number of cavities &  - &  600& 80\\
         Total number of cryomodules & -  & 200&16\\
         Total RF section length & m& 1110.6&80.8\\
         External Q-factor & \num{E6}& 0.38&0.21\\
         Cavity detuning for beam loading comp. & kHz& 0.04&0.21\\ \hline
         Beam acceleration time & \textmu s& \multicolumn{2}{c}{35.5}\\
         Cavity filling time & \textmu s& 344&65\\
         RF pulse length & ms& 0.38&0.1\\
         RF duty factor & \%& 0.19&0.05\\
         Peak cavity power & kW& 3425&2965\\
         Average RF power & MW& 5.16&0.16\\
    \end{tabular}
    \caption[RF parameters for the RLA2.]{RF parameters for the RLA2 RF system, both for acceleration (RLA2 acc) and linearization (RLA2 lin). The synchronous phase, \SI{90}{\degree} is defined as being on-crest.}
    \label{2:tech:rf:highacc:tab:RCS_RFpars_low_E}
\end{table}

A first approximation of the power requirements for the RCS chain has been performed using the ILC cavities, cryomodules, and powering infrastructures \cite{ch12:ILC-tdr} as a baseline. 
The results of which can be found in Table~\ref{2:tech:rf:highacc:tab:RCS_RFpars}. 
While the ILC features the same repetition rate (5\,Hz) as the muon collider, the beam current and bunch structure differ significantly. 
The requirements do not consider HOM power contributions, cryogenic loss, impact of orbit change detuning, and counter-rotating beams. 
To reduce the impact of the transient beam loading, a stronger cavity detuning according to \cite{2:technologies:rf:highacc:Karpov_tb} is proposed. 
This will, in turn, lead to parts of the power being reflected from the cavity and the need for a larger terminating load. 
\begin{table}[ht]
\begin{center}
\caption{RF parameters for the RCS chain. The average and peak RF power includes losses from the cavity to the klystron, while the wall plug power also includes the klystron efficiency. The parameters for the cryogenic system are calculated from the estimated parameters of the European XFEL \cite{2:technologies:rf:highacc:xfel-tdr} and include a scaling with the duty factor of the different machines, the beam current and the power in the fundamental power coupler. A scaling of the HOM-power contribution is not included. The values refer to the equivalent cooling power required at room temperature and not the actual values at the respective temperatures. For the conversion, the same assumptions are taken as in the XFEL TDR. The value of the installed cryogenic plant size includes a factor of $1.5$ for the cool-down of the machine in comparison to the operational power requirement. }
\label{2:tech:rf:highacc:tab:RCS_RFpars}
\begin{tabular}{lcccccc}
\hline\hline
& Unit & {\textbf{RCS1}} & {\textbf{RCS2}} & {\textbf{RCS3}} & {\textbf{RCS4}} & {\textbf{All}}\\
\midrule
Synchronous phase & \textdegree& 135& 135& 135& 135& - \\
Number of bunches/species & - & 1& 1& 1& 1& - \\
Combined beam current  ($\mu^+$ and $\mu^-$)  & mA& 43.3& 39& 19.8& 5.49& - \\
Total RF voltage & GV& 20.9& 11.2& 16.1& 90& 138.2\\
Total number of cavities & - & 683& 366& 524& 2933& 4506\\
Total number of cryomodules & -  & 76& 41& 59& 326& 502\\
Total RF section length & m& 962& 519& 746& 4125& 6351\\
\midrule
Combined peak beam power ($\mu^+$ and $\mu^-$)  & MW& 640& 310& 225& 350& - \\
External Q-factor & $10^6$ & 0.696& 0.775& 1.533& 5.522& - \\
Cavity detuning for beam loading comp. & kHz& -1.32& -1.186& -0.6& -0.166& - \\
Max. detuning due to orbit length change & kHz& 0& 10.8& 2.6& 2.2& - \\
Beam acceleration time & ms& 0.34& 1.1& 2.37& 6.37& - \\
Cavity filling time & ms& 0.171& 0.19& 0.375& 1.352& - \\
RF pulse length & ms& 0.51& 1.29& 2.73& 7.77& - \\
RF duty factor & \% & 0.19& 0.57& 1.22& 3.36& - \\
\midrule
Peak cavity power & kW& 1128& 1017& 516& 144& - \\
Total peak RF power & MW& 1020& 496& 365& 561& - \\
Total number of klystrons & - & 114& 53& 38& 57& 262\\
Cavities per klystron & - & 6& 7& 14& 52& - \\ 
Average RF power & MW& 1.919& 2.84& 4.43& 18.92& 28.1\\
Average wall plug power for RF System & MW& 2.95& 4.38& 6.811& 29.1& 43.25\\
    \midrule
    Required power for static cooling per cavity &   W  & 750   & 750 & 750 & 750   &  - \\
    Required power for dynamic cooling per cavity  & kW     & 0.7 & 1.3 & 2.0 & 3.9 & - \\
    Installed RF cryogenic plant capacity & MW    &1.4 & 1.1 & 2.1 & 20.1 & 24.7 \\
    Required total power for static cooling & MW & 0.5 & 0.2 & 0.3 & 1.9 & 3.0 \\
\midrule\midrule 
\end{tabular}
\end{center}
\end{table}
As pointed out previously, the bunch charge in the RCS chain will cause a significant induced voltage in the fundamental mode. Additionally, due to the presence of only two bunches, the ring is sparsely populated with beams. As a result, the cavity voltage in the beam gaps will vary significantly, with no equilibrium steady-state being reached. The transient beam loading in this regime was recently investigated with a semi-analytical approach \cite{2:technologies:rf:highacc:lthiele}. The preliminary results indicate that the cavity voltage can be kept stable without a surrounding cavity loop by adjusting the initial conditions and by using the stronger detuning mentioned above. The impact of the changing cavity voltage on the beam dynamics will have to be studied in detail.  

\section{Target}
\label{1:tech:sec:tar}

The production of muons in the muon collider front-end involves the collision of a proton beam with a target material. This interaction triggers deep inelastic reactions that generate kaons and pions, subsequently decaying into muons. To effectively capture these particles and maintain control over their emittance, a robust solenoidal magnetic field is crucial. This magnetic field confines the charged particles along helical trajectories within the production target and subsequent beamline. To sustain the megawatt-class production target, active cooling is essential, requiring separation from the primary vacuum through beam windows. Moving downstream through the tapering sector before reaching the muon-cooling section, provision for an extraction channel for the unspent proton beam is required to accommodate a high-power dump absorber.

Initially, a graphite target is being considered as the primary option due to its suitability. Graphite allows operation at high temperatures and exhibits remarkable resistance to thermal shock. Its extensive usage in various facilities attests to its reliability for \SI{2}{MW} operation. However, it was identified that achieving the required muon bunch intensity may necessitate increasing the primary beam power up to 4 MW. Two other distinct technological alternatives are currently under investigation. The first alternative involves a heavy-liquid metal (HLM) target made of pure lead. This option addresses concerns regarding radiation damage and could potentially eliminate the necessity for integrated active target cooling within the cryostat of the superconducting (SC) solenoid. The second alternative explores a fluidized tungsten target that circulates micro-scale spheres in a closed loop. Notable features include high thermal-shock resistance and reduced susceptibility to cavitation, corrosion, and radiation damage. Both options present a possible pathway for even higher beam power delivery.

\paragraph*{Key challenges} 
The MAP study has laid the groundwork for a muon collider but has left certain crucial aspects unexplored. One significant gap is the lack of an integrated system design and optimization for the target systems for production of a high-charge, high-quality muon beam essential for achieving the desired luminosity. Similarly, detailed specifications for other front-end beam-absorbers, such as beam windows and a proton dump, were not addressed until now within the IMCC. 

\begin{figure}[h]
    \centering
    \includegraphics[width=0.95\linewidth]{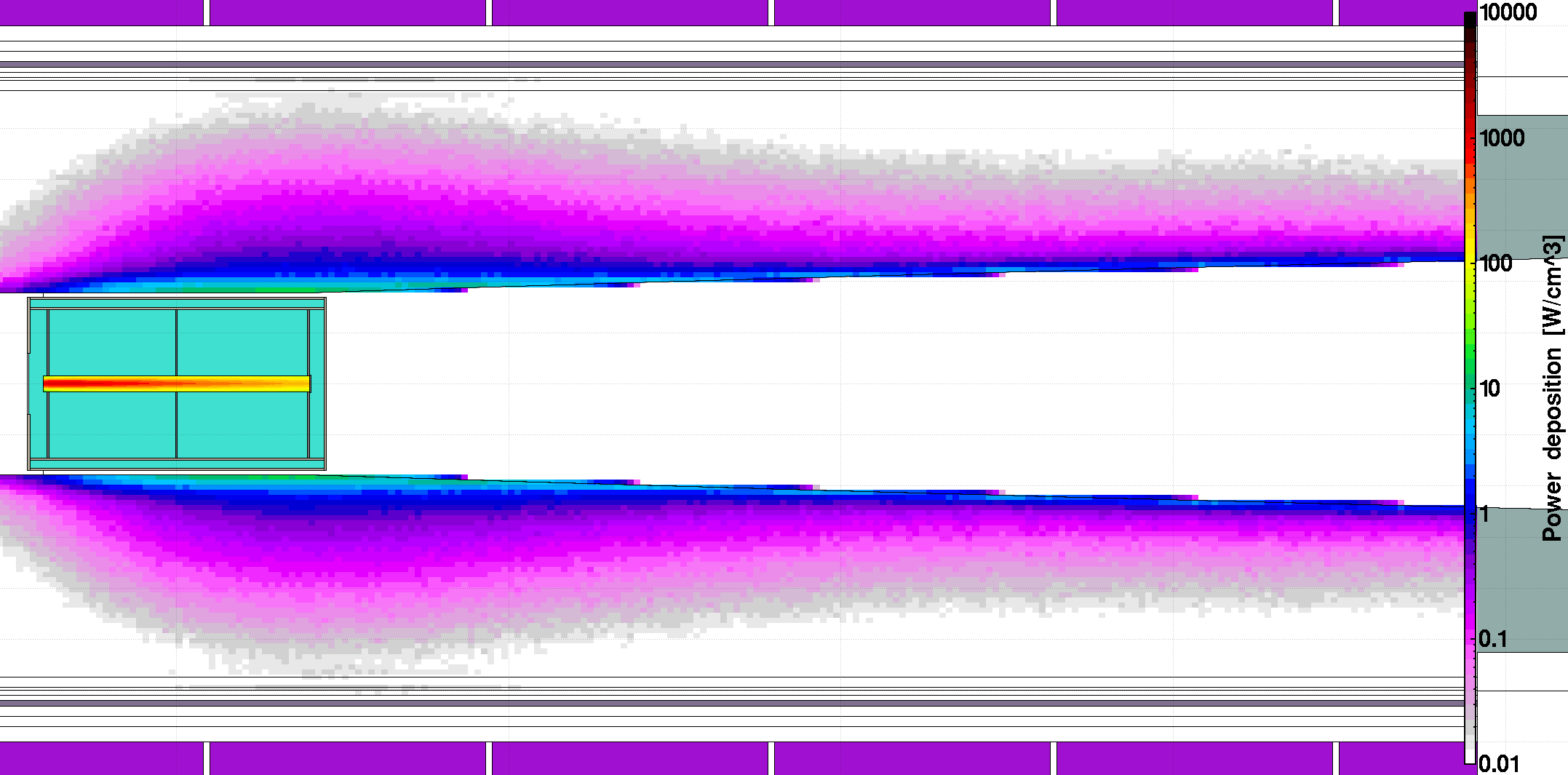}
    \caption{Power deposition in the graphite target and in the tungsten shielding around the target from a 2~MW beam of 5~GeV protons.}
    \label{fig:TargetEdep}
\end{figure}

The integration of a MW-class production target, radiation shielding, and their respective cooling systems within the cryostat of the SC Solenoid poses significant challenges. For instance, in a \SI{2}{MW} facility utilizing a graphite muon production target, around \SI{110}{kW} of thermal power are deposited on the target. Additionally, up to 34\% of the stored beam energy is deposited onto tungsten shielding near the target, necessitating efficient heat extraction and isolation from the surrounding solenoid. Addressing these issues requires dedicated vessels and the routing of multiple services. The ability to align and perform target exchanges independently, without removing the surrounding shielding of over 20 tonnes, needs consideration~(Figure~\ref{fig:TargetEdep}). Furthermore, residual radiation doses on the target systems necessitate designs compatible with remote handling.

Whether employing a carbon, HLM, or fluidized tungsten target, the high-intensity \SI{2}{ns} pulse on target results in a very high adiabatic temperature rise of the bulk material. This temperature rise can lead to dynamic stress-strain responses in solid targets, potentially surpassing material limits~\cite{InternationalMuonCollider:2024wxm}. HLM targets, though free of structural concerns, may induce extreme loads on the containment vessel due to pressure waves in the liquid, necessitating in-depth design and modelling of this complex multiphase problem. Understanding erosion management and powder handling for a fluidized tungsten target is imperative, particularly due to limited operational experience in present facilities.

The substantial levels of radiation exposure, encompassing hadronic and electromagnetic showers, as outlined in Section~\ref{1:tech:sec:rad_shield}, pose a significant challenge for the target system and its adjacent components. A comprehensive study of target and beam window materials against radiation damage is required. High integrated fluence of protons can induce atomic changes in the crystalline structure of the target materials, resulting in conductivity loss and increased brittleness, potentially reducing the lifespan of the device. For graphite, radiation annealing via operation of the target at high-temperature, coupled with larger beam sizes on target, is considered a mitigation strategy. Proton beam windows, separating the primary vacuum from the target confinement atmosphere, benefit from low Z materials such as beryllium, in order to reduce their integrated dpa (displacement per atom).

Effective cooling is essential to maintain reasonable temperature conditions for the entire system and minimize heat dissipation towards the superconducting magnet. Simultaneously, designing a sustainable facility requires optimal sizing of auxiliary services. 

Lastly, a major challenge involves designing a proton absorber capable of fully absorbing the entire beam if necessary at a regular dumping power, alongside the intricate design of a beam extraction channel.

\begin{figure}
    \centering
    \includegraphics[width=0.8\linewidth]{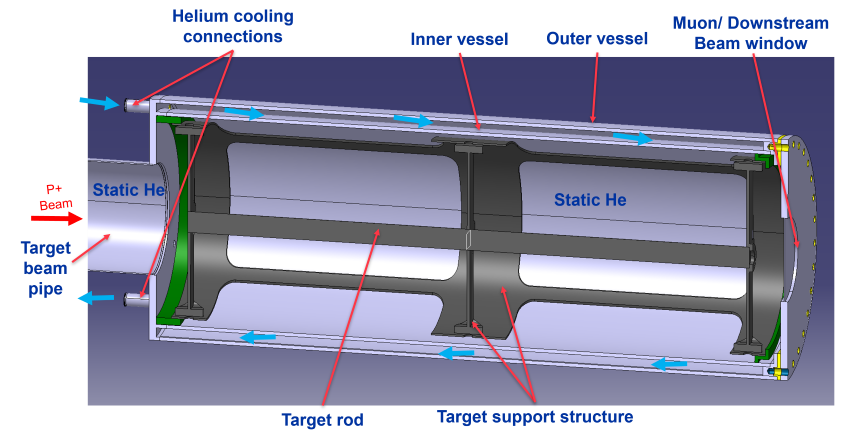}
    \caption{Conceptual design of the 2MW graphite target.}
    \label{fig:target_cad}
\end{figure}

\paragraph*{Recent achievements} 
The culmination of current studies for a 2~MW facility has resulted in a baseline target design featuring an \SI{80}{cm}-long isostatic graphite rod of \SI{30}{mm} in diameter, enclosed within a titanium vessel for hermetic confinement (see Figure~\ref{fig:target_cad}). A carbon composite frame holds the rod in the central position of the vessel. Internally, static helium gas at \SI{1}{bar} prevents sublimation of the graphite during expected high-temperature operation (around 2000$^\circ$C). Additionally, it facilitates heat dissipation through natural convection, as opposed to a radiation-only cooled target. While direct cooling (in contact with the rod) could lower the graphite temperature, the baseline solution was preferred due to its capacity for enhancing radiation defect annealing and reducing erosion concerns.

Active cooling of the inner vessel utilizes a helicoidal helium cooling circuit operating at room temperature and \SI{20}{bar}. Although this setup offers a lower heat transfer coefficient compared to a liquid coolant, it mitigates the formation of pressure waves due to energy deposition on the coolant and concerns regarding water radiolysis. Positioned within the bore of the shielding assembly, the target vessel is planned to be supported on the cryostat's upstream side, allowing easy replacement of the target without interfering with the shielding assembly.

The assembly's protection from radiation has been developed in detail (see also Section~\ref{1:tech:sec:rad_shield}). A tungsten shielding surrounding the pion capture solenoid is required. To address the dissipation of heat (\SI{680}{kW} for a \SI{2}{MW} facility), the approach shifted from contemplating water cooling, which posed complexities and handling limitations, towards utilizing helium gas cooling. This transition aims to resolve technological challenges associated with corrosion, erosion, and hydrogen embrittlement on the shielding material. The tungsten shielding design has evolved to adopt pie-shaped segments, perforated for improved cooling efficiency, together with the engineering design of the shielding vessel. 

The IMCC studies have been focused on refining the target design and optimizing the pion/muon yield towards the cooling section. After careful consideration, a beam size of \si{5}{mm} ($1\sigma$) was determined as the optimal compromise to ensure that the graphite's dynamic response remains within acceptable limits. This solution aligns with the requirements of a \SI{2}{MW} production target facility and is compatible with a proton injection of \SI{5}{GeV} and \SI{2}{ns} ($1\sigma$) pulses at a frequency of \SI{5}{Hz}.

The proton beam window, currently envisioned to be made of beryllium due to its low density and reduced interaction with the primary beam, will be located outside the cryostat. This component will undergo extensive radiation damage and necessitates dedicated cooling. Initially, materials such as titanium were considered for windows but were dismissed due to an expected accumulation of 10s of displacement per atom (DPA) per year. After several iterations, the decision was made to position it outside the solenoid assembly, ensuring improved accessibility for inspections, potential replacements, and increased availability of service space.

\begin{figure}
    \centering
    \includegraphics[width=0.8\linewidth]{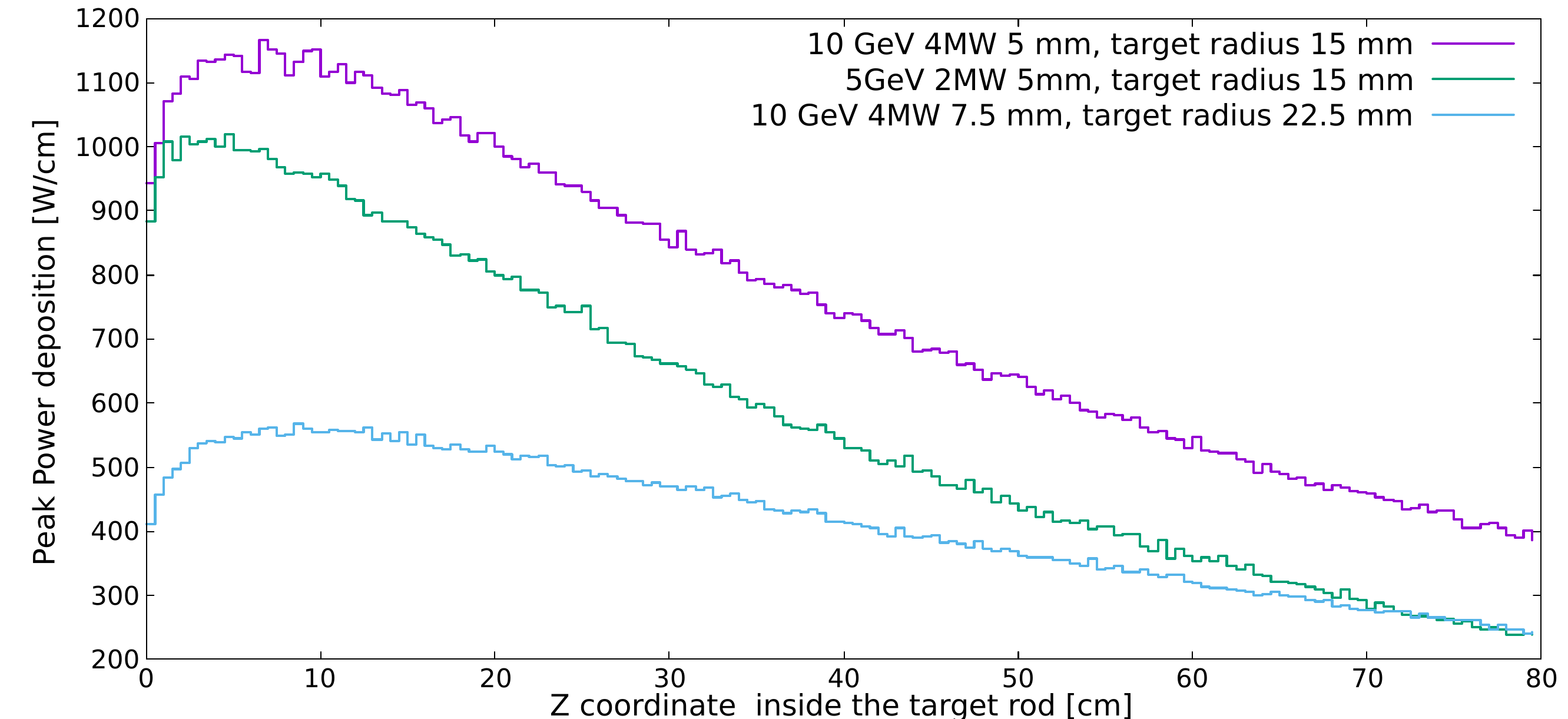}
    \caption{Peak power deposition along the graphite target rod for several proton beam configurations. The target radius is always set to 3 times the width of the Gaussian beam profile (dimensions indicated in the legend).}
    \label{fig:target_edep_peak_power}
\end{figure}

As previously discussed, increasing the primary beam power to 4 MW could improve the collider's performance. Beyond 2 MW, employing a graphite target technology requires active cooling with He (as opposed to indirect cooling). Additionally, the increased thermal power imposes a revision of the cooling parameters and the target vessel design. Preliminary studies suggest a feasibility path by further increasing the beam size on target and having a 10 GeV/c proton injection~(Figure~\ref{fig:target_edep_peak_power}).

\begin{figure}
    \centering
    \includegraphics[width=0.8\linewidth]{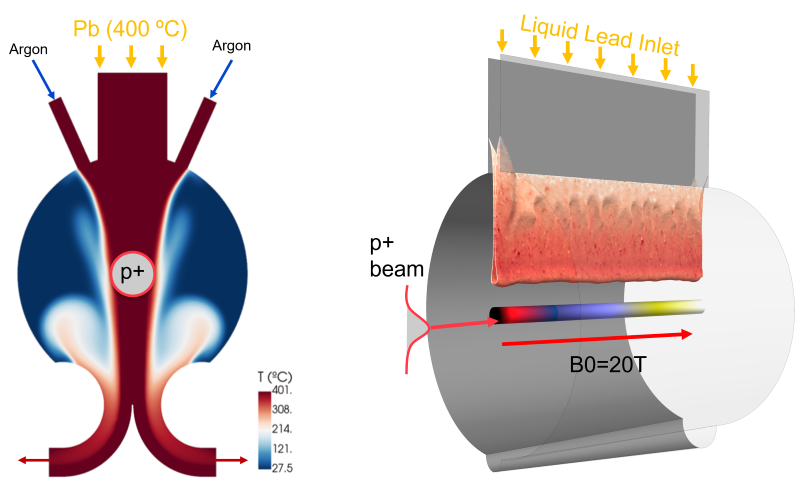}
    \caption{Liquid lead target curtain concept.}
    \label{fig:target_lead}
\end{figure}

Focused research on an HLM lead target emphasized defining a high-level concept while addressing critical obstacles. An initial concept involving flow in a pipe was discarded due to high magnetohydrodynamic (MHD) losses and concerns regarding shock waves on the retaining target pipe. The current model under consideration involves a free-flow of lead in the vertical direction (similar to a curtain), aiming to eliminate the concerns posed by the initial concept~(Figure~\ref{fig:target_lead}). Ongoing work involves further modelling and development of this concept which, together with a fluidised tungsten target, represent the two other muon collider target alternatives for a 4~MW proton-driver.

\section{Radiation shielding}
\label{1:tech:sec:rad_shield}

This section describes the different radiation shielding requirements and challenges for the front-end and the high-energy complex, in particular the collider ring. The shielding design activities are closely related to other technologies described in this chapter, in particular the magnet development (see Section~\ref{1:tech:sec:mag}), the target design (see Section~\ref{1:tech:sec:tar}), cryogenics (see Section~\ref{1:tech:sec:cryo}), and the vacuum system design (see Section~\ref{1:tech:sec:vac}). 

Radiation to equipment poses a significant challenge for all stages of the muon collider complex, from the front-end up to the collider ring. Dedicated shielding configurations must be designed for the different machines in order to dissipate the radiation-induced heat and protect equipment against long-term radiation damage. This concerns in particular the protection of superconducting magnets in different parts of the complex, like the target and capture solenoids of the muon source as well as the superconducting magnets in the RCS and collider rings. The radiation shielding must prevent beam-induced magnet quenches, reduce the thermal load to the cryogenic system, and prevent magnet failures due to the ionizing dose in organic materials (e.g.~insulation, spacers) and due to atomic displacements in the superconductors. In most cases, dedicated absorbers need to be embedded inside the magnet aperture in order to attenuate secondary radiation showers before they can reach the coils. A careful optimization of the absorber design and thickness is necessary since the absorbers can significantly impact the aperture requirements for magnets. In addition, the magnet systems must be designed such that they can support the weight of these inserts.

\textbf{Radiation shielding in the front-end.} The proton-driven MW-class muon source is one of the areas in the muon complex where massive shielding elements are needed for equipment protection. The solenoids near the production target are exposed to the secondary radiation showers generated by inelastic collision products in the target. The front-end design elaborated within MAP considered a combination of resistive copper magnet inserts and larger-aperture solenoids made of low-temperature superconductors. The latter are more sensitive to radiation than resistive magnets and were foreseen to be shielded by a thick layer of helium gas-cooled tungsten beads. A new solenoid configuration, based entirely on high-temperature superconductors (HTS), has been developed within the IMCC, combined with a new arrangement of tungsten shielding inserts for absorbing electromagnetic and hadronic showers. 

\begin{table}[t]
\begin{center}
\caption{Muon decays in the collider ring (one bunch per beam), assuming an injection frequency of \SI{5}{\hertz} and \SI{1.2E7}{s} (=139 days) of operation per year. }
\label{tab:parameters_radcollider}
\begin{tabular}{lcc}
\hline\hline
& \textbf{3~TeV} & \textbf{10~TeV} \\
\hline
Muons per bunch & 2.2$\times$10$^{12}$ & 1.8$\times$10$^{12}$ \\
Circumference & 4.5\,km & 10\,km\\
Muon decay rate per unit length & 4.9$\times$10$^{9}$~m$^{-1}$s$^{-1}$ &  1.8$\times$10$^{9}$~m$^{-1}$s$^{-1}$\\
Power per unit length carried by decay $e^\pm$ & 0.411\,kW/m & 0.505\,kW/m\\
Total decays per unit length and per year & 5.87$\times$10$^{16}$~m$^{-1}$ & 2.16$\times$10$^{16}$~m$^{-1}$\\
\hline\hline
\end{tabular}
\end{center}
\end{table}

\textbf{Radiation shielding in the accelerators and collider.} Dedicated shielding configurations are also needed in the accelerators and the collider, protecting equipment against the radiation load due to muon decay. The power carried by decay electrons and positrons is on average about 35\% of the energy of decaying muons (the rest is carried away by neutrinos, which are irrelevant for the radiation load to the machine). With the presently assumed beam parameters, this amounts to about 500\,W/m dissipated in the collider ring; the collider parameters are chosen such that the power load is about the same at 3\,TeV and 10\,TeV (see Table~\ref{tab:parameters_radcollider}).
The decay electrons and positrons can have TeV energies and emit synchrotron radiation while travelling inside the magnets. The energy is then dissipated through electromagnetic showers in surrounding materials. In addition, secondary hadrons can be produced in photo-nuclear interactions, in particular neutrons, which dominate the displacement damage in magnet coils. Shielding studies for muon colliders have been previously carried out within MAP, indicating that a continuous liner (few centimeters of tungsten) is needed inside magnets and magnet interconnects~\cite{Mokhov2011PAC,Kashikhin2012IPAC,Mokhov2014IPAC}. The shielding requirements for collider energies up to 10\,TeV have been studied more recently within the IMCC~\cite{Calzolari2022IPAC,Lechner2024IPAC}.

Absorbers and shielding configurations might also be needed for the protection of other accelerator systems like RF cavities. In addition, massive absorbers are also required for the Machine-Detector Interface (MDI), in order to reduce the beam-induced background and radiation damage in the detector. The MDI shielding is discussed in more detail in Section~\ref{1:inter:sec:mdi} and will not be covered here. 

\paragraph*{Key challenges}
Designing shielding configurations for a harsh radiation environment like in the muon collider target complex requires a good understanding of radiation damage limits for equipment components like superconductors, insulation materials, etc. The shielding requirements will strongly depend on R\&D efforts towards more radiation-hard solenoids and accelerator magnets. Radiation damage effects in superconductors, in particular in HTS, are also of critical interest for other applications like fusion reactors. The irradiation tests and theoretical studies carried out by the fusion community are highly beneficial for the muon collider design study. One of the key challenges is to derive a relation between radiation damage quantities, e.g., displacement per atom (DPA), and relevant macroscopic material properties like the critical temperature. This is crucial for establishing a relationship between irradiation experiments and radiation environments in the muon collider complex. Despite the still existing uncertainties, it is expected that the understanding of radiation effects in HTS magnets, including the possible benefits of annealing, will significantly improve in the coming decade. 

\textbf{Radiation shielding in the front-end.} The shielding blocks inside the superconducting solenoids around the pion/muon production target need to be tens of centimeters thick (radially), in order to sufficiently reduce the heat load and radiation damage in the magnets. The shielding has to be efficiently cooled since it has to dissipate a significant fraction (about 35\%) of proton drive-beam power (2\,MW). Other important aspects are the machining, handling, assembly and support of the shielding blocks. Optimizing the shielding configuration around the target and in the downstream capture section is crucial, since the shielding will determine the final aperture requirements for the solenoids. Another challenge is the overall integration of the heavy shielding inside the cryostats. Furthermore, the shielding needs to be thermally insulated from the cryostats, considering the significant heat dissipation in the shielding.  

\textbf{Radiation shielding in the collider.} One of the key challenges for the collider shielding design is the overall optimization of geometrical aspects (beam aperture, shielding thickness, coil aperture) and thermal aspects (shielding and coil temperature, thermal insulation, cooling scheme). This optimization critically depends on technology choices, e.g., low-temperature versus high-temperature superconductors, and requires a multi-disciplinary design approach including radiation and beam physics, magnet engineering, cryogenics and vacuum.
A careful optimization of the absorber thickness is important since the absorbers significantly impact the aperture requirements for magnets.
An essential design parameter for the shielding is the maximum allowed heat deposition in the cold mass of collider magnets and the resulting cooling requirements, which will be an important cost factor for facility operation due to the associated power consumption (see Section~\ref{1:tech:sec:cryo} for details). Besides the conceptual shielding design, one also faces different engineering challenges. The shielding needs to be equipped with a cooling circuit in order to extract the heat deposited by the particle showers. In addition, the vacuum chamber needs to support the weight of the shielding absorbers ($>$ \SI{100}{\kilo\gram\per\meter}). Another important aspect is the raw material and manufacturing cost of the shielding. Considering the length of the collider (\SI{10}{\kilo\meter} for the \SI{10}{\tera\electronvolt} machine), the shielding cross section hence needs to be carefully optimized.  

\paragraph*{Recent achievements}

\begin{figure}
    \centering
    \includegraphics[width=0.46\linewidth]{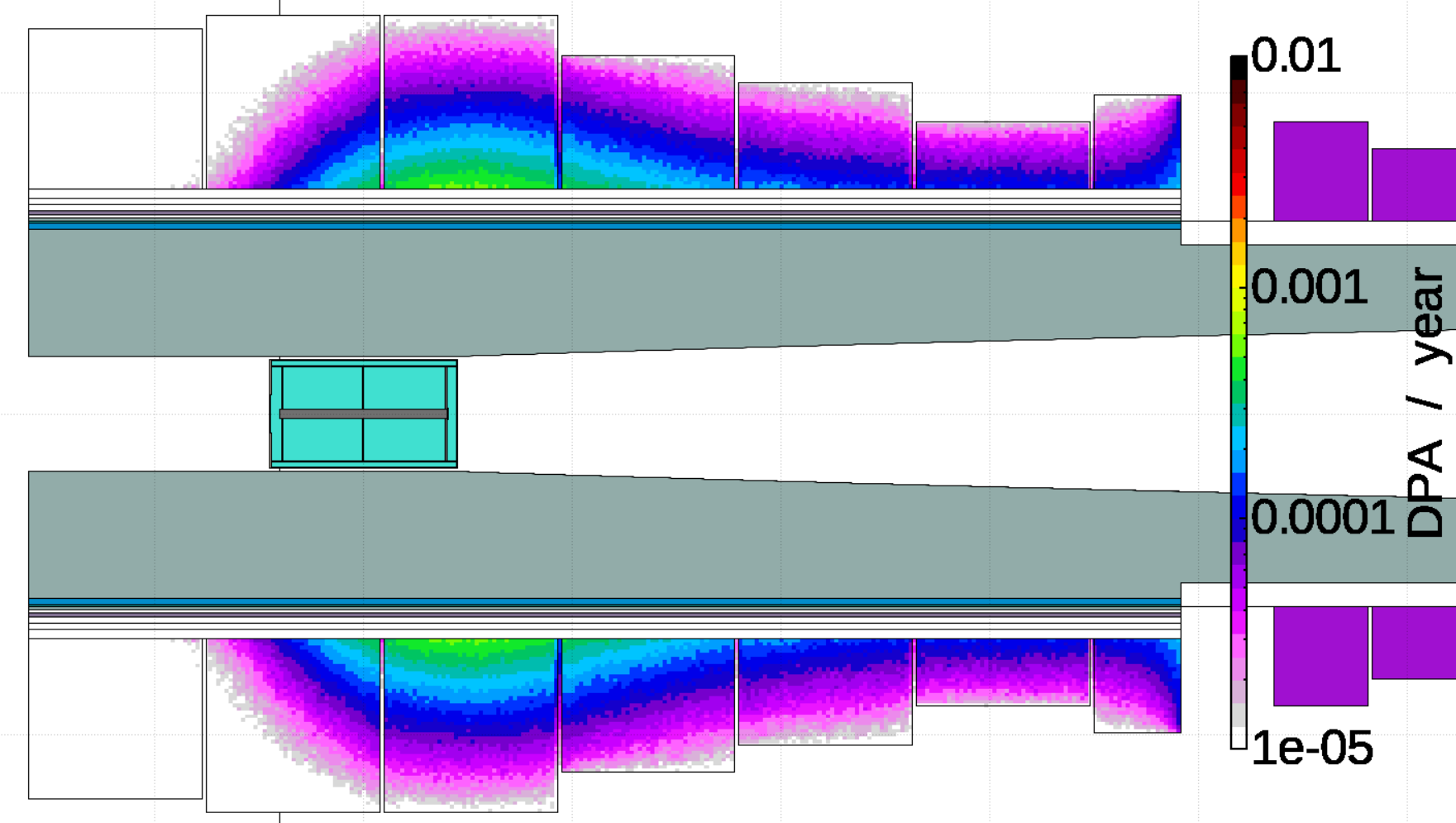}
    \hfill
    \includegraphics[width=0.52\linewidth]{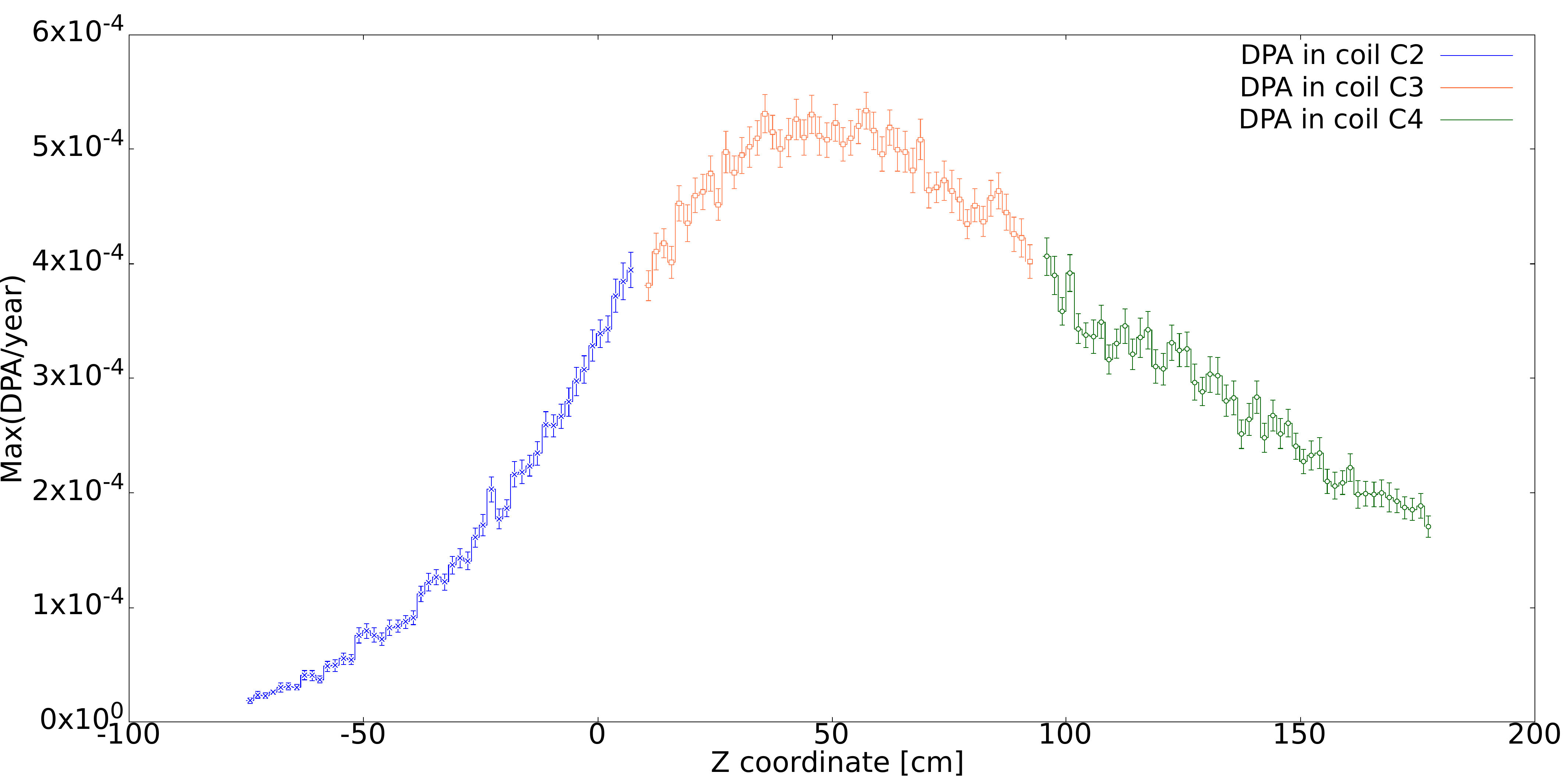}
    \caption{2D map of the displacement per atom (DPA) in the superconducting magnets of the target area (left) and the peak DPA in the coils most exposed to radiation (right).}
    \label{fig:target_DPA_2D_and_peak}
\end{figure}

\textbf{Radiation shielding in the front-end.} A conceptual radiation shielding design was conceived for the target area and the downstream capture section and chicane. Dedicated shielding studies were performed with the FLUKA Monte Carlo code, in order to quantify the heat load and radiation damage in the solenoids. The shielding was assumed to be composed of helium gas-cooled tungsten segments. In order to thermalize and capture neutrons, the shielding around the target was assumed to embed a layer of water and boron carbide. This can reduce the cumulative displacement damage in the coils by about a factor of two. Table~\ref{tab:dpadosetargetsolenoid} shows the annual DPA and dose in the target solenoid for different tungsten shielding thicknesses. The studies were carried out for a graphite target rod, assuming a proton drive beam energy of \SI{5}{\giga\electronvolt} and beam power of \SI{2}{\mega\watt}. The results indicate that a thickness of about \SI{40}{\centi\meter} is needed for reducing the DPA in the coils to \SI{1E-3}{DPA\per year} and the dose to about \SI{5}{\mega\gray\per year}. Considering the beam aperture requirements, the target vessel size and the space required for supports and thermal insulation, the minimum coil aperture (radius) for the target solenoid is hence expected to be about \SI{60}{}--\SI{70}{\centi\meter}. The maximum allowed DPA in the coils per year of operation needs to be further assessed, and will depend on the expected effect of annealing cycles. A dose of \SI{5}{\mega\gray\per year} could possibly be acceptable for insulation materials. The study concluded the target radial build design detailed in Table~\ref{target:tab:radialbuild}. The displacement per atom in the HTS coils in the vicinity of the target for this design is depicted in Figure~\ref{fig:target_DPA_2D_and_peak}.

\begin{table}[t!]
\caption{Radiation load on the target solenoid in terms of the maximum displacement per atom (DPA) and the maximum absorbed dose per year of operation for various shielding configurations. The studies considered a graphite target. The inner aperture of the shielding around the target vessel was \SI{17.8}{\centi\meter}. A gap of \SI{7.5}{\centi\meter} was assumed between the shielding and the coils (supports, thermal insulation).}
\label{tab:dpadosetargetsolenoid}
\centering\resizebox{1.0\textwidth}{!}{%
\begin{tabular}{cccc}
\hline\hline
Inner radius of the magnet coils & Shielding thickness around the target & DPA/year [$10^{-3}$] & Dose [MGy/year] \\
\hline
\SI{60}{cm} & {W \SI{31.2}{cm} + H\textsubscript{2}O  \SI{2}{cm} + B\textsubscript{4}C \SI{0.5}{cm} + W \SI{1}{cm}} & $1.70  \pm 0.02$ &  $10.0 \pm 0.3$\\
\SI{65}{cm} & W \SI{36.2}{cm} + H\textsubscript{2}O  \SI{2}{cm} + B\textsubscript{4}C \SI{0.5}{cm} + W \SI{1}{cm} &  $0.90 \pm 0.02$ &  $5.6 \pm 0.2$\\
\SI{70}{cm} & W \SI{41.2}{cm} + H\textsubscript{2}O  \SI{2}{cm} + B\textsubscript{4}C \SI{0.5}{cm} + W \SI{1}{cm} & $0.49 \pm 0.01$ &  $3.1\pm 0.1$\\
\SI{75}{cm} & W \SI{46.2}{cm} + H\textsubscript{2}O  \SI{2}{cm} + B\textsubscript{4}C \SI{0.5}{cm} + W \SI{1}{cm} & $0.29 \pm 0.01$ & $1.9 \pm 0.1$ \\
\SI{80}{cm} & W \SI{51.2}{cm} + H\textsubscript{2}O  \SI{2}{cm} + B\textsubscript{4}C \SI{0.5}{cm} + W \SI{1}{cm} &  $0.16 \pm 0.01$ &  $1.0 \pm 0.1$ \\
\SI{85}{cm} & W \SI{56.2}{cm} + H\textsubscript{2}O  \SI{2}{cm} + B\textsubscript{4}C \SI{0.5}{cm} + W \SI{1}{cm} & $0.09 \pm 0.01$ &  $0.6 \pm  0.1$ \\
\hline\hline
\end{tabular}
}
\end{table}

\textbf{Radiation shielding in the collider.} In order to estimate the general shielding requirements for superconducting arc magnets in the \SI{3}{\tera\electronvolt} and \SI{10}{\tera\electronvolt} colliders, generic shielding studies were performed with FLUKA. The main focus of these studies was on muon decay, whereas other source terms like beam halo losses still have to be addressed in the future. As in the MAP studies, tungsten was assumed as absorber material due to its high atomic number and density. For engineering reasons, pure tungsten may be substituted by tungsten-based alloys without significantly affecting the shielding efficiency if the alloy has a similar material density. The simulations indicated that the total power penetrating through the shielding is similar for all collider energies, despite the harder decay spectrum at 10\,TeV. On the other hand, the maximum power density and dose in the coils was found to be a factor of 1.5--2 higher at \SI{10}{\tera\electronvolt} than at \SI{3}{\tera\electronvolt}. 

Table~\ref{tab:parameters_radloadcollider} summarizes the calculated power load and radiation damage in collider arc magnets as a function of the radial tungsten absorber thickness (for the \SI{10}{\tera\electronvolt} collider option)~\cite{Lechner2024IPAC}. The power penetrating the absorber, mostly in the form of electromagnetic showers, amounts to almost \SI{20}{\watt\per\meter} in the case of a \SI{2}{\centi\meter} shielding, and decreases to \SI{4}{\watt\per\meter} in the case of a \SI{4}{\centi\meter} shielding. Most of this power is deposited in the cold bore and cold mass of the superconducting magnets. 
As a design criterion for the 10\,TeV collider, the beam-induced heat load in cold elements due to muon decay shall not exceed O(\SI{5}{\watt\per\meter}) for magnet operation in the vicinity of liquid helium (\SI{4.2}{\kelvin}), but could be up to O(\SI{10}{\watt\per\meter}) for operation in the range of \SI{20}{\kelvin}. This is to limit the cooling requirements and power consumption to acceptable levels, also considering additional static heat loads (see Section~\ref{1:tech:sec:cryo}). 
This suggests that a shielding thickness of 2\,cm is not sufficient for \SI{10}{\tera\electronvolt}; the tungsten shielding layer likely needs to be in the range of \SI{3}{}--\SI{4}{\centi\meter}, for which the escaping power decreases to \SI{8}{}--\SI{4}{\watt\per\meter}.
With such a shielding thickness, the cumulative radiation damage in magnets (dose and DPA) is expected to be acceptable. For 10~years of operation, the dose reaches the typical dose limit of \SI{30}{\mega\gray} for Kapton tapes; developing more radiation-resistant insulation materials could be beneficial to increase the operational margin. This shows that the heat load in the cold mass remains the driving factor for the shielding design. Together with the required beam aperture of \SI{2.35}{\centi\meter} (see Section~\ref{1:acc:sec:collider}) and other space requirements (supports etc.), the performed shielding studies indicate that the inner coil aperture (radius) of the collider arc magnets needs to be about \SI{7}{}-\SI{8}{\centi\meter}, which is a key figure for the magnet design.  

\begin{table}[t]
\begin{center}
\caption{Power load and radiation damage in collider ring arc magnets (10\,TeV) as a function of the radial tungsten absorber thickness. The power penetrating the shielding does not include neutrinos, since they are not relevant for the radiation load to the machine; the percentage values are given with respect to the power carried by decay electrons and positrons. The results include the contribution of both counter-rotating beams.}
\label{tab:parameters_radloadcollider}
\begin{tabular}{lcc c}
\hline\hline
        &  \textbf{2~cm} & \textbf{3~cm} & \textbf{4~cm} \\
\hline
Beam aperture (radius) & \SI{23.5}{\milli\meter} & \SI{23.5}{\milli\meter} & \SI{23.5}{\milli\meter}\\
Outer shielding radius & \SI{43.5}{\milli\meter} & \SI{53.5}{\milli\meter} & \SI{63.5}{\milli\meter}\\
Inner coil aperture (radius) & \SI{59}{\milli\meter} & \SI{69}{\milli\meter} & \SI{79}{\milli\meter} \\
\hline
Power penetrating tungsten absorber & \SI{18.5}{\watt\per\meter} (3.7\%) & \SI{8}{\watt\per\meter} (1.6\%) & \SI{4}{\watt\per\meter} (0.8\%)\\
Peak power density in coils & \SI{6.3}{\milli\watt\per\centi\meter^3} & \SI{2.1}{\milli\watt\per\centi\meter^3} & \SI{0.7}{\milli\watt\per\centi\meter^3} \\
Peak dose in Kapton (1 year) & \SI{10.6}{\mega\gray} & \SI{3.3}{\mega\gray} & \SI{1.3}{\mega\gray}\\
Peak dose in coils (1 year) & \SI{8.5}{\mega\gray} & \SI{2.8}{\mega\gray} & \SI{1}{\mega\gray} \\
Peak DPA in coils (1 year) & 1.5$\times$10$^{-5}$ DPA& 1.2$\times$10$^{-5}$ DPA& 1$\times$10$^{-5}$ DPA\\
\hline\hline
\end{tabular}
\end{center}
\end{table}

Similar studies were conducted for the final focusing magnets in the collider ring. In this case, the long drift section following the local chromaticity correction results in the buildup of secondary decay $e^\pm$, which create hotspots in the subsequent elements. The results are summarized in Table~\ref{2:tech:tab:TID_in_FF_magnets}. Under the current assumptions, a thicker tungsten shielding is required in the dipole preceding the final focusing magnets. In the subsequent quadrupoles, the shielding requirements are more challenging to implement, as the beam aperture must accommodate the large betatron functions necessary for the final focus. For these studies, the minimum beam clearance was assumed to be $5\max{\sigma_x,\sigma_y}$, where $\sigma_x$ and $\sigma_y$ are the beam sizes in the horizontal and vertical directions, respectively.

\begin{table}[h!]
\centering
\caption{Final focusing elements, starting from the ones furthest away from the IP. The limiting quantity for the long term survivability of the elements is the TID. For the shielding design, the assumed time of operation per year is $\SI{1.2E7}{s/y}$.}
\label{2:tech:tab:TID_in_FF_magnets}
\small
\begin{tabular}{l c c c c}
\hline\hline
\textbf{Name} & \textbf{L [m]} & \textbf{Beam aperture (radius) [cm]} & \textbf{Coil aperture (radius) [cm]} & \textbf{Peak TID [MGy/y]} \\ \hline
IB2    & 6  & 10   & 16   & 1.3  \\ 
IB1    & 10 & 10   & 16   & 3.1  \\ 
IB3    & 6  & 10   & 16   & 4.9  \\ 
IQF2   & 6  & 10   & 14   & 7.7  \\ 
IQF2\_1 & 6  & 9.3  & 13.3 & 4.6  \\ 
IQD1   & 9  & 10.5 & 14.5 & 1.1  \\ 
IQD1\_1 & 9  & 10.5 & 14.5 & 3.7  \\ 
IQF1B  & 2  & 6.2  & 10.2 & 6.4  \\ 
IQF1A  & 3  & 4.6  & 8.6  & 3.6  \\ 
IQF1   & 3  & 3    & 7    & 3.5  \\ \hline\hline
\end{tabular}
\end{table}

\FloatBarrier
\section{Muon cooling cell}
\label{1:tech:sec:cool_cell}

\begin{figure}[h]
    \centering
    \includegraphics[width=1\linewidth]{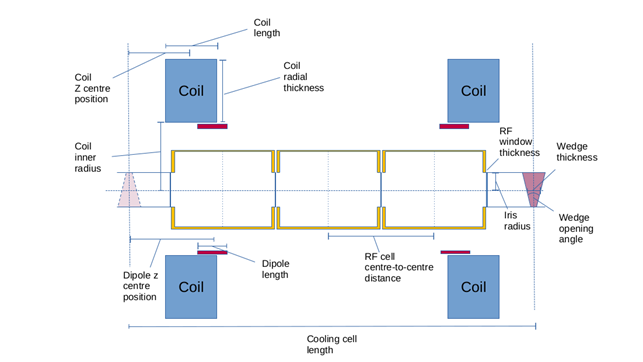}
    \caption{Schematics of the MuCol Cooling cell}
    \label{fig:B5 Cooling cell}
\end{figure}

The purpose of a cooling cell is to reduce the transverse and longitudinal emittance of the muon beam before acceleration. The muon beam is generated by the collision of a proton beam with a production target, which results in a shower of pions that will then decay into muons. These muons are generated with a large angular spread and a large momentum spread. The capture and cooling section will maximize the number of particles captured, and reduce their transverse momentum through interactions with an array of cooling cells. A cooling cell is composed of a low-Z material absorber,  which will reduce both the longitudinal and the transverse momentum with minimal particle loss, a Radio-Frequency (RF) cavity will re-accelerate the particles to restore their longitudinal momentum, and solenoids, that help focusing the beam and maintain the right (small) value of the beta function at the absorber position. Going through the sequence of cells as described in  Table \ref{cool:tab:6d_cell} , the beam is cooled, meaning the particles decrease their transverse energy and their energy spread.  The cooling performance is critical for the luminosity of the collider.
A pre-conceptual design of cooling cells for rectilinear and final cooling had been developed before 2016 by the Muon Accelerator Programme (MAP) in the US \cite{stratakis2015rectilinear}. The IMCC is pursuing, through the MuCol EU funded project, a technical design of a cooling cell to understand the issues of integration of the three main components and prove that the assumptions of the design are correct (see figure \ref{fig:B5 Cooling cell}). 

\paragraph*{Key challenges}
In order to have a good performance for the entire cooling section, peak RF fields as high as 32 MV/m at 704 MHz, magnetic fields on axis as high as 17 T and possibly liquid or gaseous hydrogen should be used as absorber and interact with a very intense beam of muons ($10^{12}$ particles per pulse), that will release a level of energy in that material that could create sudden pressure increase and fast evaporation of liquid if not carefully managed.     

The main challenges for the design are:
\begin{itemize}
    \item \textbf{Tight space:} cooling has to take place in the shortest possible time to avoid excessive decay of muons in the acceleration sections, therefore the drift space between adjacent RF cavities and absorbers has to be minimised;  
    \item \textbf{High RF field:} the required RF fields are at the edge of what has been demonstrated in real structures. A dedicated test programme must be performed to ensure the feasibility of such fields with a reasonable breakdown rate;
    \item \textbf{High Magnetic field:} the magnetic field required can only be achieved using High Temperature Superconductors, and we identified REBCO as the best material for this kind of magnet. The large bore required to accommodate the RF cavities inside them, brings the peak field to even higher values, and the forces that are generated on the coils and their support structure become very high, requiring a careful optimisation of the mechanical structure, and continuous iteration with the layout design;
    \item \textbf{High radiation levels:} although not challenging as in the target area, radiation levels will require careful selection of materials to be used, in order to avoid additional shielding to the magnets and to the sensitive part of the other devices (cables, connectors, sensors);
    \item \textbf{High level of energy deposited in the absorber:}  due to the high density of the beam, the energy deposited in the absorber might be incompatible with the use of liquid hydrogen. An R\&D programme has to determine the limits of application of the different absorbing materials, and feedback such information for a re-optimisation of the cooling section layout.
\item \textbf{Fringe fields of the solenoids:} fringe fields from solenoids exert large forces on adjacent cells and need to be balanced, especially in case of quench of one of the solenoids. 
\end{itemize}  

Other aspects that need to be clarified are how to efficiently cool the solenoids at 20K (for instance how feasible it is to consider Liquid Hydrogen as coolant), and what mitigation strategies have to be implemented to reduce the breakdown rate in RF cavities in high magnetic fields. Some of the possible solutions have been tested in the past, in particular by the MAP programme (see for instance  \cite{Freemire:2017htp}), involving the choice of specific materials for vacuum windows, using high pressure gases inside the RF volume etc... New promising ideas also are coming up (for instance it would be interesting to explore the  $C^3$ technology  \cite{Bai:2021rdg} to verify whether it brings a sufficiently high  advantage in terms of  RF gradients  and  breakdown rate). 

\begin{figure}[h]
    \centering
    \includegraphics[width=0.5\linewidth]{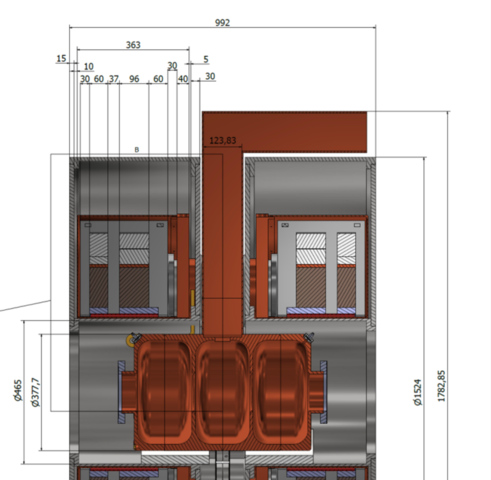}
    \caption{Preliminary design of the MuCol Cooling cell}
    \label{fig:MuCol Cooling cell}
\end{figure}

\paragraph*{Recent Achievements}
A preliminary design of the solenoid, the cavities and of the integrated model  for a cell of the type B-Stage 5 has been presented at the Demonstrator workshop held in Fermilab in October 2024 (see figure \ref{fig:MuCol Cooling cell}), \cite{RossiFermiDem2024}. The design highlighted some critical challenges linked  to the strength of the forces excited by the magnetic fields on the support structure, and by beam loading in the RF cavities which is exacerbated by installation of beryllium windows in the irises of each RF cell.. 

\section{Cryogenics}
\label{1:tech:sec:cryo}

A cryogenic infrastructure that enables the operation of superconducting magnets and radio-frequency (RF) cavities is an essential part of the Muon Collider complex. Systems requiring cryogenics will be present at most stages of the complex, including but not limited to:
\begin{itemize}
    \item Proton Driver: cooling of high-Q RF cavities;
    \item Front End: cooling of the target solenoid;
    \item Cooling Channel: cooling of the magnet array, operation of the liquid and/or gaseous H$_2$ absorbers;
    \item Accelerator and Collider rings: cooling of superconducting magnets.
\end{itemize}

Among these systems, the collider ring will consist almost exclusively of superconducting magnets operating in an especially harsh environment regarding radiation dose and unprecedented beam-induced heat-loads due to muon decay. This makes the collider ring the most demanding part of the complex from a cryogenic infrastructure standpoint. 
With sustainability in mind, one of the main drivers for the design of the cryogenic system will be the overall electrical power consumption. The various challenges of each part of the complex shall be tackled while keeping the highest possible thermodynamic efficiency (\textit{i.e.}, temperature levels as high as feasible) and the complexity to a minimum.

\paragraph*{Key challenges}
Several key challenges have been identified at the level of the Cooling Channel, and the Accelerator and Collider rings. For the moment, the Proton Driver and the Front End systems have not been assessed with respect to cryogenic requirements.

\par
\textbf{Cooling channel:} the aim is to economize on the overall cryogenic system investment and operation by harmonizing the demands of the various magnets that make up the channel. The most demanding magnets of the chain are the final cooling solenoids, where a substantial heat load needs to be extracted from each structure. Here, the challenge is rather on carefully optimizing local heat extraction rather than coping with the overall generated heat load. Solutions are strongly dependent on chosen time constant and ramp scheme for the magnets. The heat load profile and peak heat load generated during ramping varies significantly with these two parameters; a lower time constant reduces the heat load during ramp, but also reduces thermal and electrical stability during operation.

\par
\textbf{Accelerator ring:} the focus is on the investigation of sustainable cryogenic system options in a heterogeneous magnet environment such as the accelerator ring, which will have interleaved normal conducting (warm) and superconducting (cold) magnets, and to design a cooling layout to cope with such arrangement. The sheer number of cold-to-warm transitions calls for careful design to keep heat in-leaks manageable. Another challenge is that of magnet cooling over long distances: with the aim of minimizing cost and complexity, one should strive to maximize the distance required between the accelerator's continuous-cryostat's cryogenic-service feed points and re-cooling stations. Considering the rather short field-free regions between magnets (\textit{i.e.}, interconnects), the internal routing of cryogens in the cryostat while satisfying minimum magnet interconnect distances needs careful consideration.

\par
\textbf{Collider ring:} Convergence on a radial build that can be achieved from both the magnet point of view (aperture \textit{vs.} magnetic field) and cryogenic/thermal design is crucial. It has to support appropriate mitigation strategies for the heat load deposited on the coil, such as a heat intercept between the coil and absorber and the shielding thickness of the absorber so that the total heat load stays below \qty{10}{\W \per \m}. As with the accelerator ring, magnet cooling over long distances needs to be optimized for the same reasons. Optimization of very short magnet interconnects compliant with magnet movement for neutrino flux mitigation is necessary: the internal routing of cryogens in the continuous collider cryostat and the integration of cryogenic-service feed-points has to be solved while satisfying minimum magnet interconnect distances and allowing for magnet movement.

\begin{figure}[htbp]
    \centering
    \includegraphics[width=0.7\textwidth]{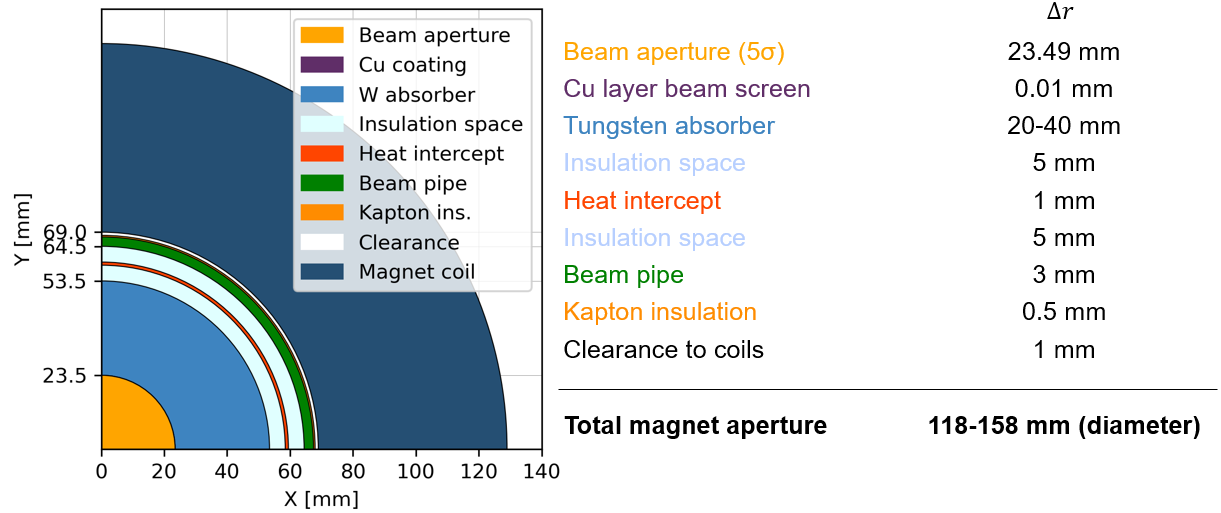}
    \caption{\label{fig:radialbuild} Tentative radial build for the collider ring magnets (from~\cite{BorgesdeSousa2024}), showing the various space allocations inside the coil aperture, shown for an absorber thickness of \qty{30}{\mm}. Note that the coil thickness (shown in dark blue) is arbitrary and for illustration purposes only.}
\end{figure}

\paragraph*{Recent achievements}
Since the definition of the road map, the principal achievement has been the identification of the main drivers for the cryogenic design and operation of the Muon Collider complex. The collider ring has been the subject of a more focused study, due to the stringent conditions regarding power deposition on the superconducting magnets.
A tentative radial build for the collider ring magnets has been outlined by the different stakeholders (vacuum, cryogenics, collective effects, radiation shielding), see Table~\ref{tab:parameters_radloadcollider} highlighting the need for space allocation for an intermediate thermal screen or heat intercept between the tungsten absorber and the magnet cold bore, see Figure~\ref{fig:radialbuild}.

\begin{figure}[htbp]
    \centering
    \includegraphics[width=\textwidth]{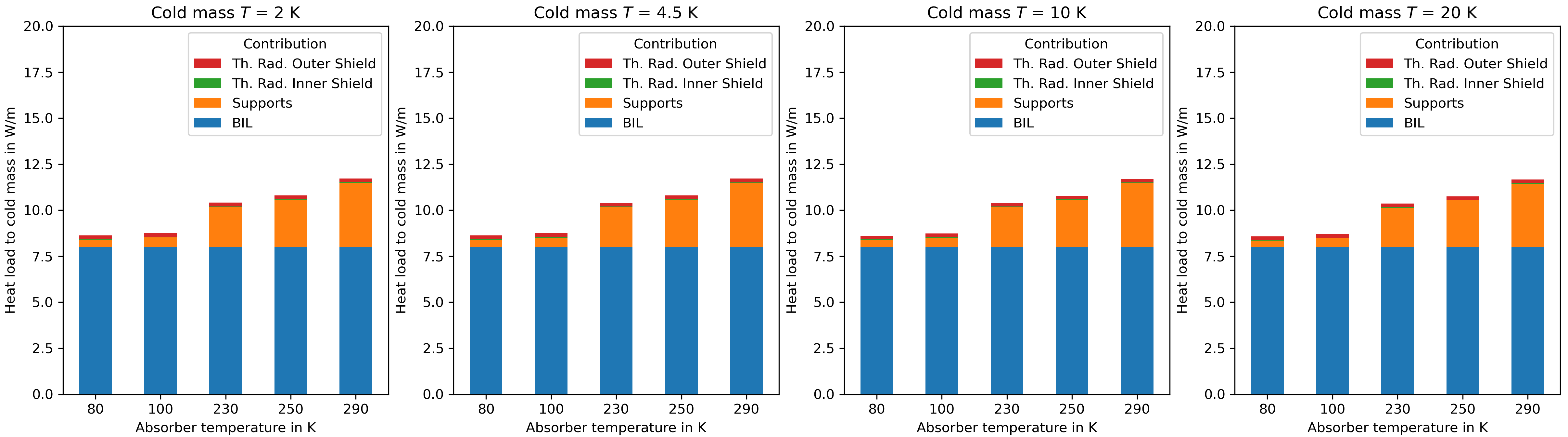}
    \caption{\label{fig:heatload30mm} Estimated heat load deposited on the cold mass in \unit{\W\per\m}, as a function of absorber temperature, for an absorber thickness of \qty{30}{\mm}, for nominal magnet operating temperatures of \qtylist{2; 4.5; 10; 20}{\K} (from~\cite{BorgesdeSousa2024}). The thermal shields (inner shield between absorber and coil, and outer shield) are assumed to be at \qty{80}{\K}. In the legend, ``BIL'' stands for ``beam-induced losses''.}
\end{figure}

The main sources of steady-state (static and beam-induced) heat loads in the different parts of the magnet system for the different temperature levels of the cryostat, as well as their dependence on temperature and absorber dimensions, have been identified~\cite{BorgesdeSousa2024} and are shown in Figure~\ref{fig:heatload30mm}. The main source of heat load to the cold mass is the power penetrating the absorber from muon decay, which depends on the absorber thickness (see Table~\ref{tab:parameters_radloadcollider}). As such, a reduction in absorber thickness, while beneficial in reducing the overall magnet aperture, should be carefully considered as it has a visible impact on the heat load on the magnet, making local heat extraction a challenge. 

An estimation of the heat loads has been carried out for collider ring magnets as function of operating temperature and absorber thickness, including static heat loads from radiation and conduction via mechanical supports. Under the current assumptions, the heat load to an \qty{80}{\K} thermal shield varies between \qtylist{4;11}{\W\per\m} depending on absorber temperature. The calculated heat loads deposited on the cold mass, considering an absorber thickness of \qty{30}{\mm}, vary between \qtylist{8;12}{\W\per\m} depending on absorber temperature.

This assessment will be completed with the presently missing contribution from resistive splices, current leads and ramping losses. Magnet ramping losses can be at least of the same order of magnitude of currently estimated heat loads and might have a determining impact on the collider ring's cooling strategies. The heat load estimation evidenced the need for a heat intercept (thermal shield) in the radial build between the absorber and the coils. This intercept is a necessary component to reduce the heat in-leak from thermal radiation to manageable levels. Also from the heat load estimation, the electrical power-consumption was estimated for the collider ring at the refrigerator interface as a function of the temperature level of the magnets and absorber circuits. To comply with the electrical consumption budget tentatively allocated for the collider ring cryogenics, a \qty{10}{\tera \eV}, \qty{10}{\km} ring, the electrical power consumption of the cryogenic infrastructure should remain below \qty{2.5}{\kW \per \m}.

Some points considering the definition of a cooling scheme have been outlined. As few cryogenic plants as possible should be considered while keeping availability to a maximum and cost to a minimum; however, longer cooling sectors have implications on distribution losses and overall cooling capacity of each cryogenic plant. In turn, the size and complexity of the distribution line will depend on the acceptable temperature gradient along a magnet sector, a parameter that depends heavily on the choice of conductor.

Operation of a muon collider ring at \qty{2}{\K} can be excluded on the basis of power consumption, cryogenic system availability, He inventory, and recovery time following a quench. The chosen cryogenic layout shall be based on cooling schemes that use reduced amounts of cryogenic fluid, namely by switching from a bath-type configuration to forced flow along cooling channels embedded into the cold mass.

For a cold mass using LTS such as Nb$_3$Sn, operating at around \qty{4.5}{\K}, cooling schemes relying on forced flow inside cooling channels can be envisaged using either two-phase helium at \qty{4.5}{\K} or using supercritical helium between \qtylist{4.5;5}{\K} at \qty{0.3}{\mega \Pa}. These schemes need to be carefully designed to cope with transient heat loads such as hysteresis losses from ramping the magnets up and down.
Options using HTS (operation at or above \qty{10}{\K}) involve using helium gas at high pressure or, for temperatures at or above \qty{21}{\K}, two-phase hydrogen, again in a configuration using forced flow in cooling channels. Using helium gas at high pressure is a well-established technique, heat transfer and available enthalpy difference are rather poor above \qty{10}{\K}, requiring large temperature differences along a magnet sector. Hydrogen has a high available enthalpy (at \qty{21}{\K}, 12 times higher by volume than helium at \qty{4.5}{\K}), which would mean comparably smaller mass flow rates circulating in the cryogenic system including the cold masses. The use of Hydrogen for accelerator magnet cooling is, however, far from being trivial and an established technology, and significant effort must be put into it before it can be considered a competitive alternative.

\section{Vacuum}
\label{1:tech:sec:vac}
The vacuum technology of a muon collider will encompass the vacuum system of the different elements of the accelerator complex: proton driver, cooling cells, beam acceleration and collider. Each element of the accelerator complex will have different vacuum requirements and specific constrains. Due to the short storage lifetime of the muons, the pressure requirement of a muon collider will be more relaxed than in existing electron and proton storage rings \cite{gallardoMuMuCollider1996}.

The accelerator complex will require a vacuum system for beam transmission, including superconducting cavities with specific dust-free requirements, thermal insulation for the superconducting magnets, and several barriers between vacuum and absorbers in the cooling cells.

Starting at the proton driver, the pressure requirement would be $<10^{-9}$\,mbar, assuming a similar concept as the SPL machine \cite{baylacConceptualDesignSPL2006}. To achieve this pressure level, only conventional vacuum equipment is required and no specific challenges are expected in this area. The presence of superconducting cavities would require dust management during the assembly and clean pumpdown and venting procedures.

After the muon production at the target, the cooling section will be dominated by the presence of drift pipes, absorbers and RF cavities inside strong solenoidal fields. Despite the pressure requirement may not be challenging, the configuration and integration of the cooling cells will be very demanding. The presence of strong magnetic fields will limit the use of mechanical pumps and difficult the integration of ion pumps and vacuum instrumentation. This area can be pumped using chemical pumping, i.e., with NEG coatings or cartridges, cryo-pumping (taking advantage of the availability of cryogenics for the superconducting magnet cooling), or ion pumps in special configurations like shown in \cite{knasterOptimizedAnnularTriode2004}, taking advantage of the magnetic fields available in the area.

The cooling section will need to integrate either lithium hydride or high density hydrogen (liquid, vapour or gas) as absorber to reach the target emittance (see Section~\ref{1:tech:sec:cool_cell}). Multiple windows will be required to separate the absorbers from the vacuum of the beam line along the cooling channel.

Finally the machine(s) required for the beam acceleration and the collider shall also integrate conventional vacuum equipment to reach a relatively modest vacuum, as showed in reference~\cite{gallardoMuMuCollider1996}. The vacuum chamber of the collider can be integrated with the thick tungsten shielding (see Section~\ref{1:tech:sec:rad_shield}) that have to dissipate a significant amount of the muon decay power. Assuming an aperture of 47\,mm, as shown in Table~\ref{tab:parameters_radloadcollider}, the optimal pump distribution requires 10\,l/s pumps spaced 90\,m to reach an average pressure  <10$^{-6}$\,mbar, not considering beam induced desorption. The stimulated gas desorption produced by the muon decay products impinging the surface of the vacuum chambers and any other dynamic effect (electron cloud, photon stimulated desorption, etc.) shall be considered in the outgassing budget of the future facilities, but no showstoppers have been identified.

\subsection*{Key challenges}
\label{1:tech:sec:vac:2}
The main challenges that have been found are the production of very thin windows for the final stages of the final cooling, the integration of liquid hydrogen as absorber in contact with these thin windows, and the integration of the collider beam screen to intercept the power produced by the muon decay, having low resistive wall impedance (see Section~\ref{1:acc:sec:collective:collider}). The requirements of the window will be conditioned by the pressure excursion inside the absorber after the beam has deposited its energy (see Section~\ref{1:tech:sec:abs}).

A very challenging proposal that has to be studied in detail is the development of thin windows for liquid hydrogen wedge absorbers as proposed in Section~\ref{1:acc:sec:cool:rec_cooling} for the rectilinear cooling.The development of such thin windows with a sharp angle and a controlled deformation compatible with the beam requirements, has to be addressed in order to consider liquid hydrogen a credible absorber at the rectilinear cooling.

The normal conducting magnets of the Rapid Cycling Synchrotrons (RCS) will pulse very fast, with cycles of less than 10 ms (see Table \ref{1:acc:tab:RCS_RFpars}). This rapid pulsing requires a design that avoids the generation of eddy currents on the vacuum chamber and, at the same time, ensure an acceptable impedance. This will involve, either the use of ceramic vacuum chambers with coatings, or copper liners to reduce impedance, or  or the integration of the magnet under vacuum. Despite the integration of the vacuum system can be challenging, no showstoppers have been identified.

Finally, the superconducting dipole magnets of the muon collider will require a thick tungsten shielding to protect the coils from the muon decay products. The vacuum chamber could integrate this shielding and it shall be able to evacuate a significant heat load from the muon decay (in the order of 500\,W/m), and have an acceptable resistive-wall impedance. The high heat load excludes running this beam screen at cryogenic temperature. On the other hand, the modest vacuum required is achievable with distributed pumping with lumped pumps separated by tenths of meters, as mentioned in the previous section.

\subsection*{Recent achievements}
\label{1:tech:sec:vac:3}
Different materials with thickness less than 15 \textmu m, used for the production of thin windows for x-ray transmission, like Be, C, and Si$_3$N$_4$ are available. The potential of these materials have been evaluated \cite{ferreirasomozaIMCC2022,ferreirasomozaMuCol2024}. Among them, silicon nitride windows are commercially available and quite cheap, making them the fist candidate to be evaluated. During the last two years several windows were mechanically characterized and irradiated with high brightness proton beams.

\begin{figure}[h]
    \centering
    \includegraphics[width=1\textwidth]{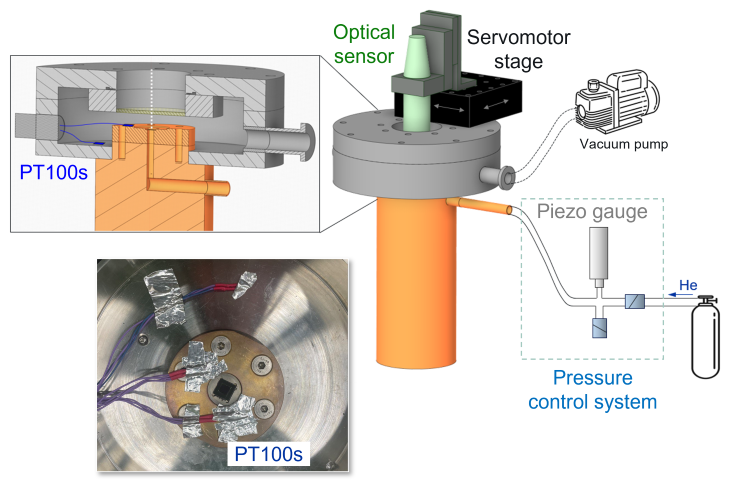}
    \hfill
    \caption{Experimental setup for mechanical characterization of Si$_3$N$_4$ by bulge testing}
    \label{fig:experimental_setup}
\end{figure}

A setup for the mechanical characterization of silicon nitride windows at different temperatures, from 77\,K up to 500\,K, by bulge testing has been built. The setup (see Figure~\ref{fig:experimental_setup}) consists on a window holder made of copper that can be heated or immerse in liquid nitrogen. A cover flange allows to create a vacuum for thermal insulation and a confocal optical sensor measures the deflection of the membrane when it is pressurised injecting helium. With this setup it was possible to measure the mechanical performance of silicon nitride membranes at different temperatures.The mechanical properties were inferred applying analytical formulas and ANSYS simulations that showed an excellent agreement with the observations \cite{giovincoipac2024}. The tested windows, 1\textmu m thick and 6$\times$6\,mm aperture, were able to sustain differential pressures of more than 4\,bar at any temperature. The main drawback of these windows is that are produced on top of a fragile silicon 200 \textmu m frame that has to be glued on the flange holder. The gluing method has an important impact on the observed performance of the windows. Any future application would require to develop a radiation hard joining and/or  a way to remove the membrane from the silicon frame. A new technique has been identified for transferring the Si$_3$N$_4$ to a metallic substrates. During 2025, a feasibility study will be carried out at CERN with the possible application of these windows to separate the Rb plasma cells at AWAKE phase 2.

Several silicon nitride windows have been irradiated at HiRadMat with SPS beams in 2023 and 2024. In 2023 a parasitic experiment was carried out to irradiate one window and test if it could survive proton beams with brightness equivalent to those of the muon final cooling, with a mechanical load of 1\,bar of differential pressure. Assuming all the power remains in the window (Bethe-Bloch equation) $4\times10^{12}$ muons at 5\,MeV and $\sigma_{\mathrm{RMS}}$ 0.6\,mm is equivalent to 440\,GeV/c protons with 0.25\,mm beam size and $3\times10^{13}$ protons. The membrane survived intensities almost 10~times higher than this threshold, with a significant deformation, but still leak tight.

As a continuation of the previous experiment, an experimental proposal was submitted to the HiRadMat Scientific Board and approved. This experiment (SMAUG2) was carried out in 2024. Another set of twelve silicon nitride windows were irradiated at different intensities and with different repetition rates. Two of them failed at maximum brightness at 5 and 50 shots. The results are still being processing, but they confirm Si$_3$N$_4$ windows were able to sustain high brightness beams beyond the muon collider requirements under 1\,bar of differential pressure.

\section{Absorbers}
\label{1:tech:sec:abs}

The ionization cooling of muon beams requires the energy deposition on different absorbers alternated with accelerating cavities. These two areas have to be separated by windows. As the emittance of the beam decreases, the brightness increases and also the power density deposited in the material. At the final cooling the equilibrium emittance is proportional to the beam energy, as shown in Equation~\ref{eq:emittance}, where $E$ is the beam energy, $B$ the magnetic field, $L_{R}$ the radiation length of the absorber and $dE/dS$ the beam stopping power). Beam kinetic energies in the order of few MeV are required at the last cooling stages \cite{Sayed:2015zta, palmerMuonColliderFinal2011}. At this energy the effect of a conventional window on the beam transmission will be significant, so very thin windows are required (see Section~\ref{1:tech:sec:vac}). On the other hand a large radiation length and stopping power favours low Z materials like hydrogen or lithium hydride as absorber material.

\begin{equation}
    \label{eq:emittance}
    \epsilon_{eq}\propto\frac{E}{B\cdot L_{R}(dE/dS)} 
\end{equation}

\subsection*{Key challenges}
\label{1:tech:sec:abs:2}
The kinetic energy of the muon beam at the final cooling will be as low as 20\,MeV at the entrance of final stages \cite{kamalsayedHighFieldLow2015, palmerMuonColliderFinal2011}, reaching a few MeV at the exit. At this energy, the typical thickness used for accelerator windows (tenths of mm) will significantly perturb the beam. 1.3\,mm of beryllium are enough to stop the muon beam completely at 5\,MeV. The presence of liquid hydrogen in the last stages of the muon cooling, as described in \cite{kamalsayedHighFieldLow2015, palmerMuonColliderFinal2011}, poses a big challenge. A cooling cell where the beam kinetic energy inside the absorber is reduced from 20 to 5\,MeV involves the deposition of 9.6\,J. For a small beam of $\sigma_\textrm{RMS}$ equal to 0.6\,mm, it will produce a pressure excursion inside the hydrogen of larger than 800\,bars generating a shock wave. Under these conditions the integrity of a thin window is questionable. If the muon beam has a repetition rate of 5\,Hz, the power deposited in the liquid hydrogen is approximately 50\,W. The evacuation of this power will be also challenging, considering the very limited bore radius inside the solenoid. Forced liquid hydrogen circulation could dissipate tenths of watts and a prototypes able to evacuate up to 20\,W have been built \cite{cummingsMucoolHydrogenAbsorber2006}.
Despite the amount of power deposited in the absorber is considerable lower in previous cooling phases, like in the rectilinear cooling, due to the larger beam size and higher energy, the use of liquid hydrogen wedges (see Section~\ref{1:acc:sec:cool:rec_cooling}) are challenging because the difficult geometry that has to be implemented using thin windows. These absorbers are still in a conceptual phase. A technical proposal of how to build a wedge absorber respecting the  geometrical tolerances, still to be defined, does not exist yet.

\subsection*{Recent Achievements}
\label{1:tech:sec:abs:3}
\begin{figure}[h]
    \centering
    \includegraphics[width=10cm]{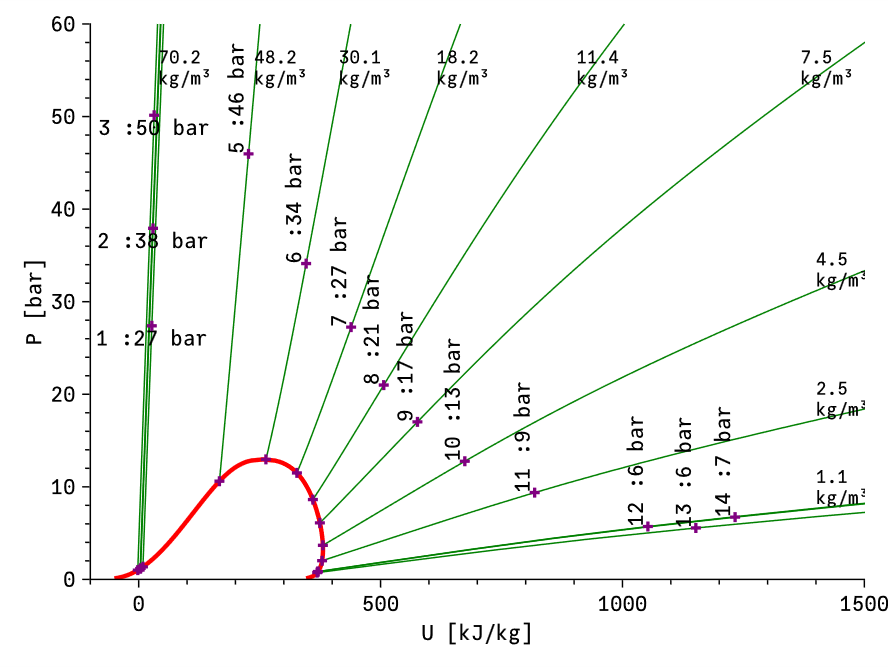}
    \hfill
    \caption{H$_2$ Thermodynamic Diagram. The saturation line is shown in red. The green lines represent isochoric lines and the + signs represent initial and final conditions at each stage \cite{ferreirasomozaIMCC2024,palmerMuonColliderFinal2011, CoolProp}}
    \label{fig:phase_diagram}
\end{figure}

In most of the proposals for the final cooling stages, the absorber material is defined a priori. The length of the absorber is calculated at each stage, assuming a maximum absorber length defined by the solenoid. Using the scheme proposed by \cite{palmerMuonColliderFinal2011} as a baseline, the expected pressure excursions on the liquid hydrogen absorber at the last stages of the final cooling will be higher than 800\,bar. As an alternative, it is possible to keep the maximum absorber length constant at the different stages, and change the density to obtain the same absorbed energy per stage (see Reference \cite{ferreirasomozaIMCC2024}). The absorber density will vary from saturated liquid hydrogen at 1\,bar at the first longer stages (approx. 75\,cm long) down to saturated vapour at 1\,bar at the last stages as shown in Figure~\ref{fig:phase_diagram} \cite{CoolProp}.

\begin{figure}[h]
    \centering
    \includegraphics[width=10cm]{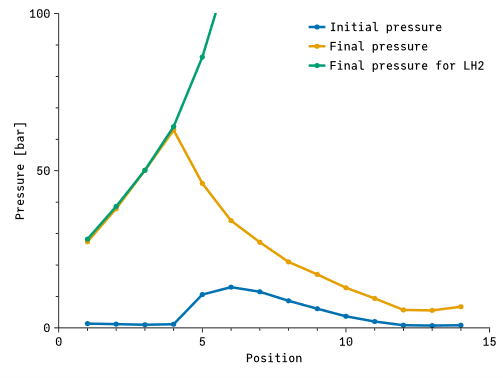}
    \hfill
    \caption{Absorber pressure for each stage of the final cooling lattice described in \cite{palmerMuonColliderFinal2011}. (Blue) initial pressure at each stage; (Green) final pressure at the original scheme; (Yellow) final pressure with new proposal.}
    \label{fig:pressure_evolution}
\end{figure}

It is not possible to tailor the hydrogen density keeping the initial pressure at 1\,bar. To keep a single phase absorber it is necessary to follow the saturation curve. Figure~\ref{fig:pressure_evolution} shows the pressure evolution at the different stages. The first 4 stages are identical to the original proposal, but after the 5\textsuperscript{th} stage the initial pressure has to be higher moving towards the critical pressure at 13\,bar. This change will considerably reduce the final pressure after energy deposition. In the revised scheme the maximum expected pressure is 64\,bar at the 4\textsuperscript{th} stage. Previously the maximum expected pressure was 874\,bar at the last stage, showing a significant improvement. The maximum pressure is located at the intermediate stages of the cooling channel, where the beam energy is higher, rather than at the final stages. The final stages show pressure excursions of less than 10\,bar that allow the use of thin (<15\,\textmu m) windows. On the other hand, the higher energy of the intermediate stages open the option to combine the liquid hydrogen absorber with thick lithium hydride windows. Although lithium hydride has poor mechanical properties, a thick window (>5\,mm) has enough resistance to sustain 64\,bar excursions. Such a window would likely require to be embedded into a thin membrane to be vacuum tight.

Tailoring the absorber density it is possible to mitigate the expected pressure excursions after the beam energy deposition. This gives a roadmap for the optimization of the final cooling channel combining liquid hydrogen, lithium hydride and thin windows.

\section{Instrumentation}
\label{1:tech:sec:inst}

The beam instrumentation for the Muon Collider study has not yet been examined in detail.  The facility different sections — the proton driver, target delivery system, muon cooling channels, accelerator and collider — each have distinct beam parameters and specific constraints, pose a wide range of challenges for beam instrumentation.

\subsection*{Proton Driver}

For the proton complex — consisting of a high-intensity H- linac, accumulator ring, and compressor ring — beam instrumentation will be fundamental to ensure the beam reaches the target with the desired parameters, as outlined in Table~\ref{2:acc:proton:tab:parameters} for the 2 and 4 MW options. This will require comprehensive diagnostics to monitor and control beam transmission (i.e., with minimal losses) while maintaining the required longitudinal and transverse beam parameters.

\paragraph*{Key Challenges} 
The linac's diagnostic challenges will include specific issues related to H- ion acceleration. Experience from CERN LINAC4~\cite{roncarolo2014overview}  and other H- facilities worldwide, such as SNS and JPARC, will provide valuable insights. The development of laser stripping techniques could be particularly important for probing both transverse~\cite{hoffman2017laser, goldblatt2016linac4} and longitudinal~\cite{PhysRevAccelBeams.26.042801} distributions. 

Monitoring H- stripping efficiency at the linac's end is vital for ensuring the proton driver's reliability and minimizing injection losses. The monitoring technique currently used at LINAC4 is detailed in ~\cite{NavarroFernandez:2816210}. Additionally, at LINAC4 and other modern hadron linacs (ESS, SNS, JPARC), bunch shape monitors~\cite{tan2013bunch} proved essential during initial RF cavity tuning and remain vital for regular monitoring of longitudinal beam properties and energy spread.

Secondary Emission Monitors (SEM) represent a well-known technique for beam intensity, position, and size diagnostics. These are widely used in CERN facilities, from LINAC4 (45 keV to 160 MeV H- beams) to the fixed target beam lines in the SPS North Area complex (400 GeV protons), passing through many intermediate proton energy stages in synchrotrons, transfer lines, and experimental facilities. 

SEM monitors are based on thin wires, strips, or foils that are either movable through or fixed on the beam trajectory. They produce a net electric charge when the beam interacts with the material. Due to potential damage from high-power beams, their use is often limited to a maximum beam power density. 

As an example of SEM use, since 2022 a specific set of SEMs, in the form of horizontal and vertical wire grids, has been permanently installed a few meters upstream of the CERN nToF facility target. This system successfully monitors beam power density in combination with a beam current transformer, and inhibits beam transfer to the target if the power exceeds the predefined safe limit. 

Recently (2024), a master's student from the University of Malta conducted a project at CERN studying how secondary emission monitors could be used in the proton driver up to the target delivery system. The study particularly examined different materials and monitor layout options compatible with the various proton driver beam options. This work was presented at the 2024 CERN Beam Instrumenation Day~\cite{biday24-camilleri} and the thesis report is expected to be published before mid-2025.

The monitoring of high-intensity bunches up to 5-10 GeV in the accumulator and compressor rings could be done with similar techniques as used in the CERN proton synchrotrons, ranging from fast flying wires for transverse profile measurements to standard beam current, beam position, and beam loss monitors. While no show-stoppers have been identified yet, detailed attention must be given to specific challenges: the 5Hz repetition rate and the compression of up to \num{5e14}
protons in a single short (\SI{120}{ns} or less) bunch. These parameters could damage intercepting devices like wires and could require data acquisition systems with very high bandwidth.

Measuring beam parameters at the target delivery system will be challenging due to the multi-MW proton power expected at the end of the proton driver. The diagnostics development can build upon the continuously evolving experience at CERN fixed target facilities, as reviewed in ~\cite{Roncarolo:2886629}, and other high-power proton facilities like SNS and ESS. Recent diagnostic implementations at CERN~\cite{camilleri2024target} include a beam imaging system in front of the Antiproton Decelerator (AD) target that continuously monitors the beam position and uses feedforward control of upstream magnets to maintain beam centering on the target.

Recent developments at ESS include thin scintillating coatings for beam imaging at the target entrance~\cite{fackelman2023performance} and a multi-purpose target diagnostics plug featuring 2D imaging of the rotating target through monitoring gamma decay of the tungsten target bricks~\cite{Borghi_2018}. 

In all cases (i.e., linac, rings, all transfer lines and target diagnostics), minimally invasive techniques should be considered and further studied. Apart from the laser stripping suitable for H- in the linac, this includes considering beam-gas ionization \cite{BGI} or fluorescence monitors and low-density materials (i.e., for SEM or imaging systems).

\subsection*{Cooling Channels}
In the ionisation cooling channels, as the Muon beam is getting bunched with bunch length of 100\,ps and bunch intensity of \num{e12} particle per bunch, standard beam instrumentation can be used. The beam position and intensity could be measured using electromagnetic monitors such as button pick-ups and fast beam current transformers. Those pick-ups could also provide a timing reference that will be used to adjust the phasing of the RF cavities. Measuring the beam transverse size could be done using standard imaging systems relying on Optical Transition Radiation or Scintillating screens. And the longitudinal beam profile can be measured using Cherenkov radiator and streak camera \cite{StreakCamera}. 
Installing such instruments upstream and downstream of the cooling channel will be relatively straightforward. The main challenge will then be to design monitors that can be integrated in between cooling cells due to space restriction or inside the cooling cells as the environment is quite challenging, i.e. strong magnetic field, limited space and accessibility. 
As the beam position, size and timing must be known precisely at the location of the absorbers, specific solution must be studied in great details, this could involve optical method such as electro-optical techniques \cite{DEOS}for beam position monitoring, scintillation and Cherenkov radiation for transverse and longitudinal profile respectively.

\subsection*{RCS, other rings and collider}
The beam instrumentation in the rest of the accelerator complex, i.e. RCS and main synchrotron rings, could potential used standard diagnostic solutions as mentioned previously. As the Muon beam energy increases, Synchrotron radiation monitors would also provide another option for transverse and longitudinal beam profile as typically used in high energy synchrotrons (electrons or hadrons).

\section{Radiation protection}
\label{1:tech:sec:rad_protect}

The Radiation Protection (RP) aspects of the Muon Collider can be separated into two main, intrinsically different topics. One is related to the conventional RP aspects connected to prompt and residual radiation, air, water and soil activation, as well as radioactive waste produced in the whole facility starting from the proton driver up to the collider ring. These aspects are principally well understood and can be mitigated to acceptable levels as long as they are addressed in the design phase. The second topic is related to the high flux of high energy neutrinos emitted from the Muon Collider that have a very small probability to interact far away in material near the Earth’s surface producing secondary particle showers. This aspect is most relevant for the collider ring. 

\paragraph*{Key challenges}
    
Muons circulating in the muon collider decay and generate neutrinos within a small solid angle that may lead to a large local flux of neutrinos in the plane of the collider ring. Due to their particularly low interaction cross section, the attenuation of TeV-scale neutrinos traversing the Earth’s matter is very low (e.g.~0.01\% for \SI{100}{km}). A high flux of neutrinos may therefore reach the Earth’s surface even far away from the collider, where rare single events of these high-energetic neutrinos create secondary particle showers concentrated in small areas. 
One of the challenges of a high energy muon collider is therefore to reduce the local radiation flux created by these showers so that it's radiological impact would be negligible.

A refined neutrino dose model has been developed.  
It comprises FLUKA Monte Carlo simulations that allow study of the expected neutrino and secondary-particle fluxes and evaluation of the main parameters for the effective dose predictions. The results are then combined with a realistic neutrino source term taking into account the given real collider lattice. The findings are further used to develop a map in which, together with a given collider placement (i.e. position, depth and inclination) the dose is projected on the Earth’s surface and subsequently used for the dose assessment. Next to the dose assessment also the uncertainties are to be studied and methods allowing to demonstrate compliance. This is an iterative process where optimization is addressed at the various working steps.

The goal is to ensure that the neutrino-induced impact would not increase the natural radiation noticeably as it is the case of the LHC, for example. In terms of the radiation protection, the projected doses to members of the public due to the operation of the facility should not exceed about 10\,\textmu Sv per year. Below this constraint, further dose reduction would not be justified under the ALARA principle~\cite{IAEAGSG10}.

\paragraph*{Achievements: neutrino dose model}

\begin{figure}[!h]
    \centering
    \includegraphics[width=0.49\textwidth]{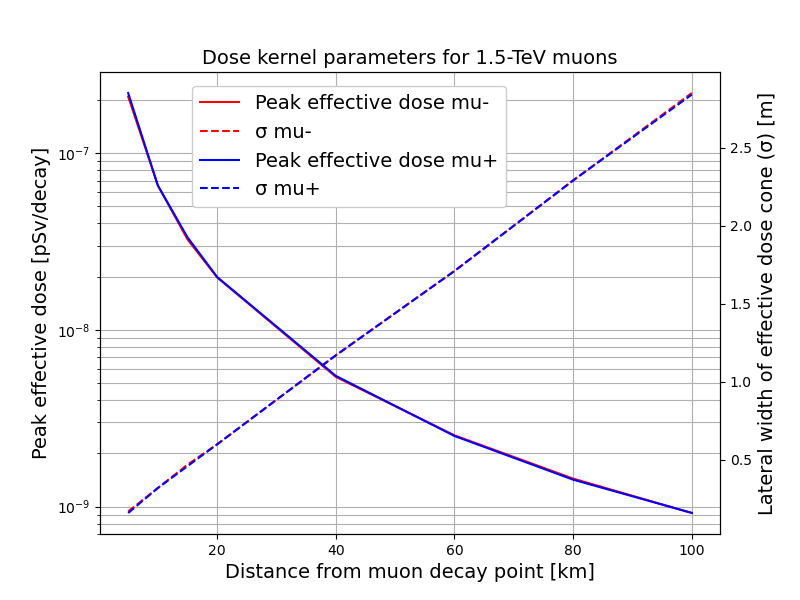}
    \includegraphics[width=0.49\textwidth]{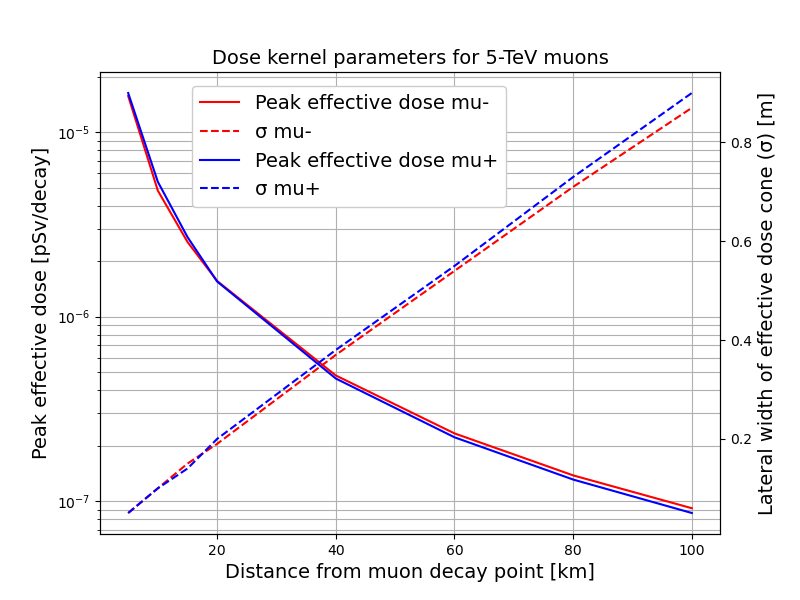}
    \caption{Effective dose kernel parameters within soil as a function of the baseline distance from the muon decay point, for muon energies of $1.5$~TeV (left) and $5$~TeV (right). }
    \label{fig:dosekernel_summary}
\end{figure}

Exhaustive FLUKA simulations were performed to evaluate the effective dose in soil resulting from the interaction of the neutrinos emerging from the decay of 1.5\,TeV and 5\,TeV mono-directional muons. The angular and energy distributions of the neutrinos were sampled according to the respective distributions expected from the muon decays. 
It was found that after only a few meters of path in soil the sampled neutrino-induced showers reach a plateau condition, where the effective dose saturates and remains constant farther ahead. This condition is considered as the conservative, worst-case scenario since the neutrino-induced showers rapidly decrease in the transition between soil and air, as further demonstrated in additional simulations. 
The effective dose kernel has been evaluated on the basis of a Gaussian fit to the radial projection of the saturated worst-case scenario. 
Peak doses and lateral widths of these kernels were calculated for a wide range of neutrino interaction distances from the decay point ($5, 10, 15, 20, 40, 60, 80, 100~\mathrm{km}$) and further interpolated to obtain a kernel shape for any distance between $5$ and \SI{100}{km}. The results of the FLUKA simulations are shown in Figure~\ref{fig:dosekernel_summary}. It was found that the widening of the neutrino radiation cone due to the lateral extension of the secondary particle shower is very small compared to the effect of the neutrino flux angular divergence.  The results additionally allowed for a general verification of simplified analytical expressions showing good agreement.
The obtained values shall be used as the basis for the calculation of the effective dose taking into account realistic lattice parameters, i.e. factoring the associated distribution of muon trajectories in the sections of interest of the accelerator in, as well as the reduction due to realistic geometries and exposure scenarios. Additionally, mitigation methods, such as optimization of the source term, location and orientation of the collider, should be considered.

In order to evaluate the effective dose due to neutrinos reaching the Earth's surface, the source term given by the effective dose kernel needs to be further combined with the relevant properties of the accelerator lattice taking the divergence of the muon beam into account. 
This has been performed based on numerical evaluations for half an arc cell of the latest version of the 10\,TeV center-of-mass energy collider. The results show that sharp dose peaks are expected to be caused by short straight sections between magnets.
Focusing and chromaticity correction based on combined function magnets helps avoiding even higher peaks. Even though the length of all straight sections is identical, the height of the peaks are lower when they originate from the beginning of a cell rather than from its end. The dipolar magnetic field of combined function magnets is lower than the one of pure dipoles leading to slightly increased radiation levels. The variations of dose from different positions along the quadrupoles is caused by variations of the beam divergence.
Obvious mitigation measures are to minimize the length of straight sections in regions outside the long straight section housing the experiments and to install the device deep underground leading to large neutrino distance from the Earth's surface so that the divergence of the neutrino flux would dilute the resulting showers. Careful lattice design avoiding straight sections without horizontal beam divergence due to dispersion derivative may allow the radiation dose peaks from the arcs to be reduced. 

\begin{figure}
    \centering
    \includegraphics[width=0.95\linewidth]{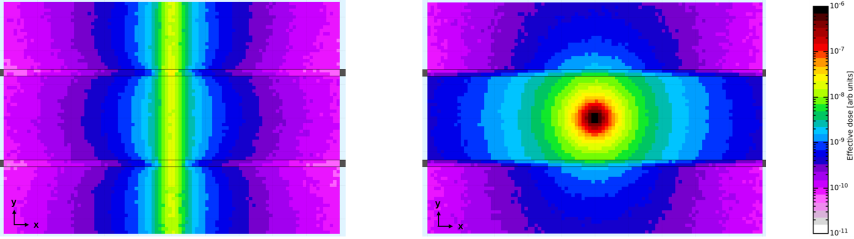}
    \caption{Cross-sectional view of the effective dose (in arb. units) for an above-ground building structure exposed to the electron neutrino flux from the decay of negative muons in a straight section with (left) and without (right) a deformation by the movers within ±1~mrad.}
    \label{fig:kernesVSmovers}
\end{figure}

The density of the neutrino flux arising from the collider ring arcs is expected to be further reduced by deforming the muon beam trajectory achieving a wide-enough angular spread of the neutrino flux. 
Periodic orbit variation of the muon beam within the beam pipe would be sufficient for 1.5 TeV muon beam energy.
For 5~TeV muon beam energy it has been proposed to mount the beamline components in the arcs on movers to periodically deform the ring in small steps. This would cause the muon beam direction to change over time and consequently reduce the integrated neutrino flux at any surface exit point. The foreseen machine movers shall enable vertical beam deformation within $\pm$1~mrad, reducing the saturated dose kernels in the soil by a factor 80-90. The number of discrete deformation steps over a period of a year of operation required to reach the dose reduction plateau, depends on 
the distance from the collider at which the dose is evaluated. As an example, approximately 160~steps would be needed to achieve a reduction factor of 80 at a distance of 60~km. The concept of such movers is described in more detail in Section~\ref{1:tech:sec:movers}. Figure \ref{fig:kernesVSmovers} shows an example effective dose distribution in an above-ground building structure induced by the neutrino flux emerging from the collider arcs with and without the movers.

\begin{figure}
    \centering
    \includegraphics[width=1.0\linewidth]{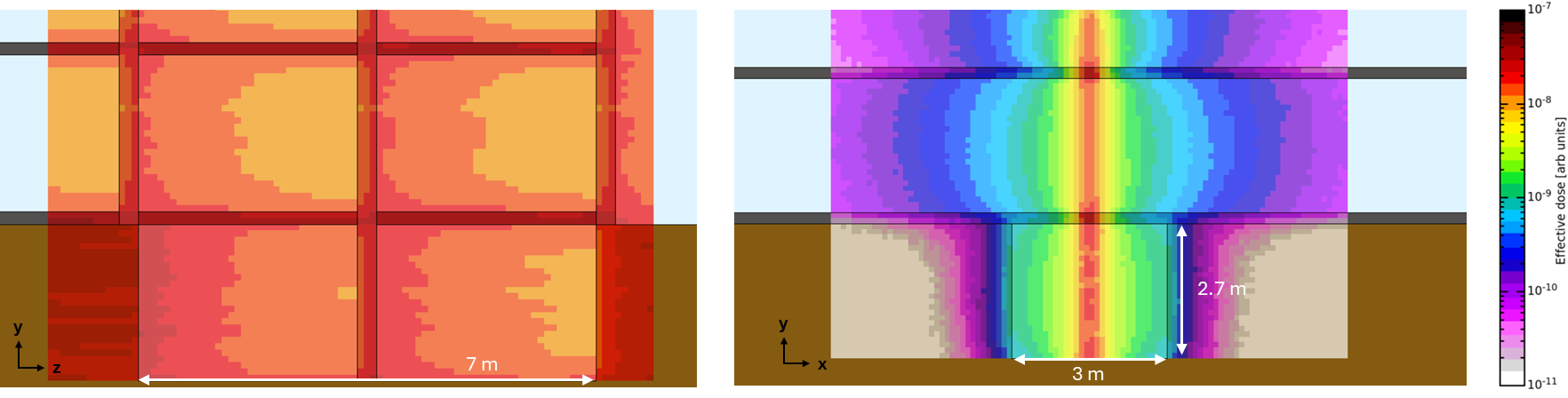}
    \caption{Side (left) and cross-sectional view (right) of the effective dose (in arb. units) for an underground building structure exposed to the neutrino flux from the decay of negative muons in a given straight section, after the vertical deformation by the movers.}
    \label{fig:doble_cellar_neutrinos}
\end{figure}

An additional effect that needs to be considered in the final effective dose predictions arises from the exposure scenarios.
Instead of considering the saturated effective dose within soil, which is unrealistic for an annual exposure of a person, various more realistic, yet conservative scenarios were investigated.
The scenarios include exposure in building structures both below and above ground. The representative scenario for a specific dose assessment depends on the location of the neutrino flux exit point e.g., whether it is situated in a inhabited area, and on the neutrino flux emission region within the collider ring, such as the arcs equipped with movers or the long straight sections housing the experiments.
Figure \ref{fig:doble_cellar_neutrinos} shows a geometry implemented in FLUKA together with the spatial distributions of the effective dose obtained from simulations for a distance of 15\,km from the collider ring. It shows a building structure with underground rooms, illustrating a worst-case exposure scenario in a possibly inhabited area for neutrinos originating from the collider arcs with movers. The neutrino flux is dispersed vertically due to the effect of the movers, and the dose inside the cellar rooms results from the combined contributions of secondary particles generated by neutrino interactions in the soil and structural elements of the building. 

To estimate more realistic yet still conservative annual doses, specific assumptions regarding occupancy within the considered geometries are necessary. Each geometry was divided into two regions: a smaller area (1×1×0.9~m³) corresponding to the region with the highest dose, and a larger area (3×1×3.6~m³) for an average exposure. For the general dose estimates, a 100\% occupancy factor within the defined geometry was assumed with the worst-case exposure occurring for one-third of the total time and average exposure for the remaining two-thirds.
Table \ref{tab:cellar_wobbling_dose_5TeV} shows the annual effective doses for the worst-case exposure scenario in an underground building structure (see Figure~\ref{fig:doble_cellar_neutrinos}), provided as effective doses for annual exposure scenarios. 

\begin{table}[htbp]
    \centering
        \caption[Effective dose due to neutrino-induced radiation for an underground building structure]{Effective dose of neutrino-induced radiation for an underground building structure at different distances from the muon decay when the vertical deformation by the movers is applied. The muon beam energy is $5$~TeV.        }
    \label{tab:cellar_wobbling_dose_5TeV}
    \begin{tabular}{rcc}
         \hline\hline
         \multicolumn{1}{c}{}& \textbf{$\mu^-$} & \textbf{$\mu^+$} \\ 
         \textbf{Distance} & \textbf{Mitigated dose [pSv/decay]} &  \textbf{Mitigated dose[pSv/decay]} \\ \hline 
        $15$~km & $6.6 \cdot 10^{-9}$ & $6.7 \cdot 10^{-9}$ \\
        $20$~km & $4.8 \cdot 10^{-9}$ & $4.9\cdot 10^{-9}$ \\
        $30$~km & $3.0 \cdot 10^{-9}$ & $3.1 \cdot 10^{-9}$ \\
        $60$~km & $1.3 \cdot 10^{-9}$ & $1.4 \cdot 10^{-9}$ \\
       \hline\hline
    \end{tabular}
\end{table}

Higher neutrino fluxes arising from the long straight sections are unavoidable. A tool to determine the location on the surface corresponding to the neutrino flux position and to easily adjust the positioning of the machine was developed. This tool can be used to optimize the position of the collider ring such that higher radiation levels are localized in regions such as rock face, which may be owned by the laboratory and fenced, or at the sea. Additionally the tool shall to identify the most convenient exit points for the entire collider and their respective distances.

\begin{figure}[!h]
    \centering
    \includegraphics[trim={0cm, 0cm, 0cm, 0.5cm}, width=0.75\textwidth, clip]{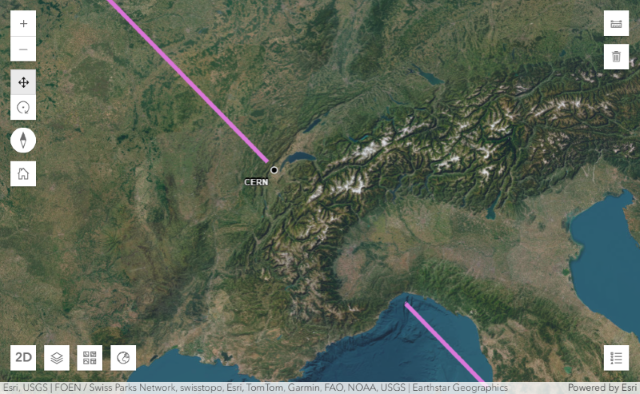}
    \caption{3D visualization of the exit points in the mountainous non-built area (left) and the sea (right) of the potential collider placement option in the local CERN area. }
    \label{fig:geo_exit_points_2D}
\end{figure}

\begin{figure}[!h]
    \centering
    \includegraphics[width=0.49\textwidth]{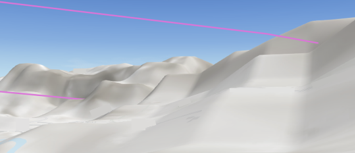}
    \includegraphics[trim={0cm, 0cm, 0cm, 0.5cm}, width=0.49\textwidth, clip]{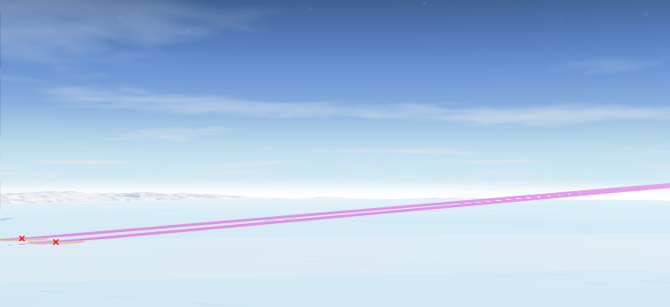}
    \caption{3D visualization of the exit points in the mountainous non-built area (left) and the sea (right) of the potential collider placement option in the local CERN area. }
    \label{fig:geo_exit}
\end{figure}

A first Proof of Concept (POC) to identify the intersection points with the Earth's surface with such a tool has been achieved by simulating straight line exit points in the case of the SPS and LHC.
It was a static simulation to validate the methodology and the potential results, but without allowing dynamic change of the accelerator position. Following this POC, it was decided to add this functionality to ’Geoprofiler’, a decision-support web application for future accelerator localisation study. This functionality allows users to move dynamically the position of the collider via 2D placement with specification of the altitude and to tilt the collider around two axes. It also allows to analyse radiation lines corresponding to the straight sections of the collider for a given collider position. The objective was to show the intersection between these lines and the surface of the Earth (Digital Elevation Model). Due to the extensive range of the radiation lines, the European Space Agency Copernicus DEM was chosen, as it provides complete coverage of the European territory.
To enhance performance, it was decided as a first step, to restrict the number of radiation lines to the two main ones. 
The tool allowed identifying at least one potential option in the local CERN area that would direct the neutrino flux of the IPs to a limited non-built-up area in the mountains on one side and a far away area in the Mediterranean Sea on the other side as shown in Figure~\ref{fig:geo_exit_points_2D}. In addition, this configuration has the advantage of having a single exit point in the mountains from each of the two IPs located with a steep exit angle thanks to the collider inclination and ground profile (see Figure~\ref{fig:geo_exit}).

Another aspect of collider placement optimization involves the assessment of neutrino paths emerging from the collider arcs. The starting points of these neutrinos are associated with each individual magnet within the arcs. For the preliminary location demonstrated in Figure~\ref{fig:geo_exit_points_2D}, a simulation was conducted to determine suitable exit points across Europe. The distances between these exit points and the emission point from the magnet range from 16~km to over 300~km. Instead of modelling the exit points as a single spot, the impacted area is assumed to have the shape of a vertical stripe projected on the landform, as expected from the effect of the movers (see again Figure~\ref{fig:kernesVSmovers}). The length of each stripe is calculated as 2 times the distance from the collider multiplied by 0.001, which corresponds to $\pm$1~mrad beam direction deformation. This procedure allows for an accurate estimation and classification of the areas affected by the neutrinos originating from the collider arcs for any given collider placement option.

 
\begin{figure}
\centering
    \includegraphics[width=0.49\textwidth]{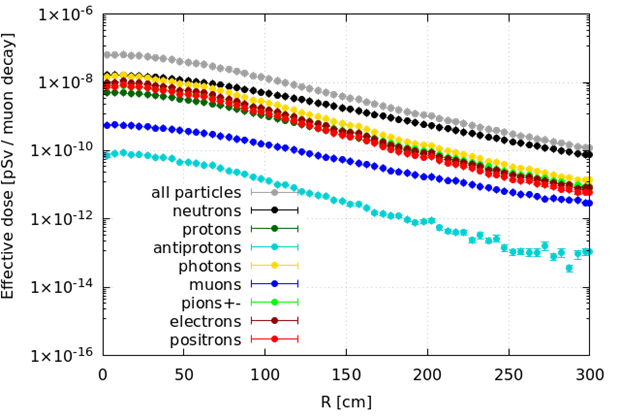}
    \includegraphics[width=0.49\textwidth]{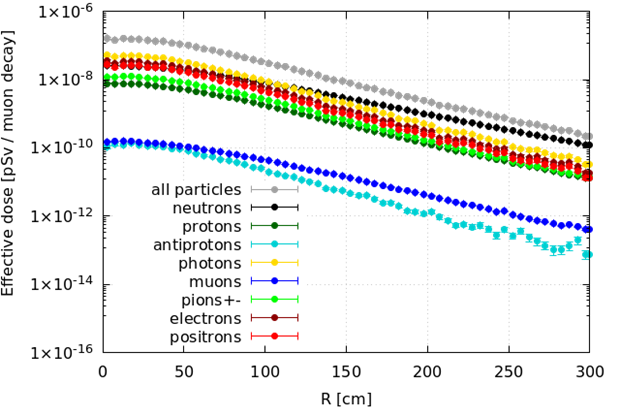}
        \caption{Example of the effective dose distributions as a function of the dose kernel radius within soil for $\overline{\nu}_\mu$ (left) and $\nu_e$ (right) from 5~TeV positive muon decays at 60~km distance. }
    \label{fig:dose_breakdown}
\end{figure}

Generally, for the dose assessment all possible exposure pathways have to be taken into account. In the case of the neutrino induced radiation far away from the collider, it has been found that the only pathway to consider is the external exposure directly from the secondary particle showers induced by the neutrinos. It has been demonstrated with FLUKA simulations that the neutrino-induced soil activation and transfer of activation products to groundwater are negligible. To evaluate the latter, very conservative design limits were applied. These limits are based on the assumption that the pore water should not exceed the Swiss immission limits~\cite{SWISSRPO} for the longer-lived radionuclides, $^3$H and $^{22}$Na, which are both soluble radionuclides likely to be transported by groundwater and therefore of interest for the protection of groundwater resources. Even when applying the limits, the $^3$H and $^{22}$Na activity concentrations were estimated to be of several orders of magnitude below the conservative design limits. The exposure due to activated or contaminated soil and air can therefore be considered as negligible.  
FLUKA simulations were furthermore used to investigate the composition of the secondary particle showers created by the neutrino interactions, which is relevant for identifying methods to demonstrate compliance. As shown in Figure~\ref{fig:dose_breakdown}, it was found that the effective dose stemming from the neutrino interactions is dominated by neutrons and the electromagnetic component. Moreover, various dosimetric quantities were compared to each other allowing to evaluate the field-specific conversions factors. The effective dose was found to be a few percent higher than the ambient dose equivalent, while the effective dose was a factor 1.6-1.9 higher than the absorbed dose in the peak area of the saturated dose kernels for 1.5 and 5~TeV.
Finally, it was concluded that the doses to the reference animals and plants would be orders of magnitude below the reference levels~\cite{ICRP108}.

\paragraph*{Achievements: conventional RP aspects}

A first investigation of the conventional RP aspects of the collider ring arcs was performed based on FLUKA simulations. 
The studies comprised the evaluation of the residual ambient dose equivalent levels in the arc dipoles region for different cool-down times. The radiation levels after 5~years of beam operation were found to be better than in the inner triplet region of the HL-LHC and are thus considered acceptable assuming optimization of interventions.  
The residual dose rates after the commissioning phase assumed to be of three months with 20\% beam intensity were found to be more than one order of magnitude lower and complying with a Supervised Radiation Area ($< 15$\,\textmu Sv/h) after one week of cool-down, thereby enabling easier access for interventions. Furthermore, the soil activation around the tunnel was investigated and found to be similar to that of the LHC. With suitable collider placement deep underground the level of soil activation is expected to be acceptable.

\section{Movers}
\label{1:tech:sec:movers}
\begin{figure}[htb]
    \centering
    \includegraphics[width=10cm]{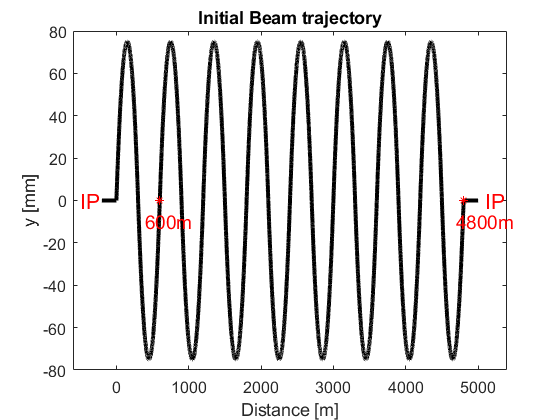}
    \caption{Typical muon collider deformation pattern to mitigate radiation from neutrinos reaching Earth's surface.}
    \label{fig:Mover-Patt1}
\end{figure}

Muon beam circulating in the collider ring creates neutrino radiation that would pass the allowed radiation limit on surface. To mitigate this, the beam trajectory should be replaced periodically, roughly every 12\,hours.
In the mitigation concept the beam trajectory follows a set of opposing parabolas between the two interaction points of the collider, as depicted in Figure~\ref{fig:Mover-Patt1}. 

To change the beam trajectory, the parabola is stepped forward between the interaction points. During operation, the vertical position and the pitch of all collider magnets (pure bendings and combined function magnets) must be changed to vary the orientation of the tangent to the beam trajectory, that represent the cone of radiation induced by the neutrino flux.This leads to requirements never seen for an accelerator or collider and thus the solutions must be proofed in practice.
 
\begin{figure}[htb]
    \centering
    \includegraphics[width=10cm]{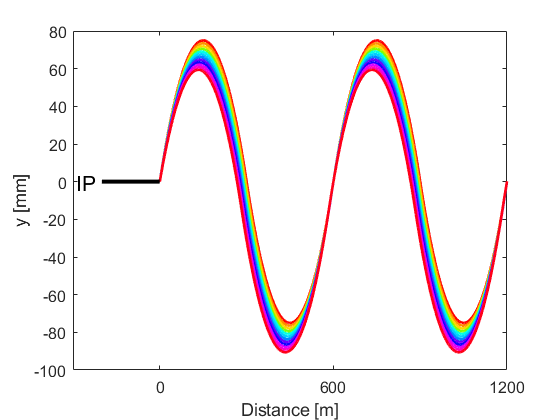}
    \caption{Several collider ring vertical deformation patterns.}
    \label{fig:Mover-Combs1}
\end{figure}

 Several collider ring vertical deformation patterns, each to be used during typically 12\,hours, are depicted in Figure~\ref{fig:Mover-Combs1}.

\begin{figure}[htb]
    \centering
    \includegraphics[width=10cm]{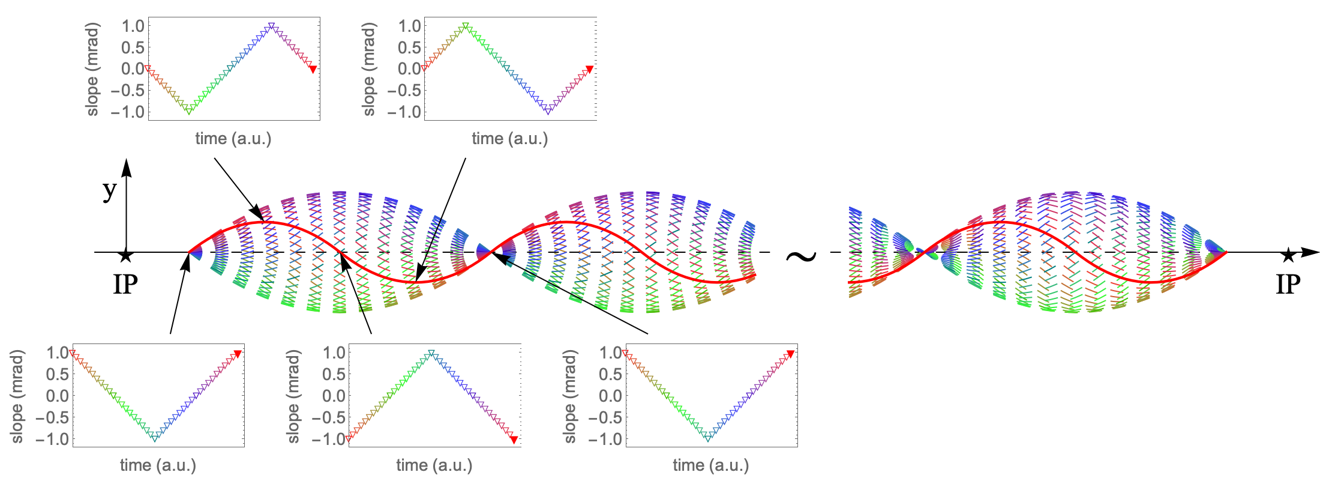}
    \caption{Deformation patterns to cover the whole period of the wave between -1\,mrad and +1\,mrad.}
    \label{fig:wobbling}
\end{figure}

To keep the radiation level at an acceptable level, the tangent to the beam trajectory should change between -1\,mrad and +1\,mrad. 
The deformation patterns have to cover a whole period the wave. At certain locations, the slope will transition from its maximum value to the minimum and then back to the maximum. To fully cover the range of slopes within ±1 mrad, with a step size of 0.02 mrad, approximately 200 distinct deformation patterns are required, as shown in Figure~\ref{fig:wobbling}.

\begin{figure}[htb]
    \centering
    \includegraphics[width=10cm]{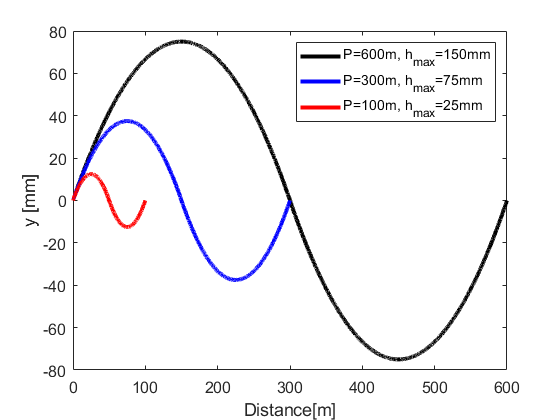}
    \caption{Proposed deformation patterns with different amplitudes and periods.}
    \label{fig:DeformationsVars1}
\end{figure}

Several machine deformation patterns resulting in $\pm 1$\,mrad variations of the slope of the tangent are plotted in Figure~\ref{fig:DeformationsVars1}.

If we assume a period of 100\,m, thus correspond to vertical movement required of the magnets is $\pm25$\,mm with respect to the horizontal beam trajectory.One of the advantages of this option is the limitation of magnet displacement, allowing the use of more conventional support and actuation systems. Additionally, increasing the required magnet movement would lead to a larger tunnel diameter and a greater distance between cryogenic supply points, whose feasibility still needs to be assessed.

However, if the deformation period is reduced, the magnetic field for vertical deflection increases from 0.111\,T (600\,m period) to 0.667\,T (100\,m period). 

In any case, it is important to note that two adjacent magnets must not be set to opposite maximum deviations, as this could overstress the interconnection bellows and compromise their integrity.

\section{Safety}
\label{1:tech:sec:safety}
The safety of people and environmental protection are priorities in the concept of a muon collider complex. Collaboration with CERN's Occupational Health 
\& Safety and Environmental Protection (HSE) Unit is established to assure the satisfaction of the regulations and applications of best practice in these domains.

Many standard industrial hazards will need to be considered for working in such a facility, such as noise, lighting, air quality, working in confined spaces, which can be satisfied by applying existing CERN safety practices and the Host State's regulations for workplaces. The ambient temperature in the facility will be of particular interest for access and work.

 The creation of a muon collider complex will involve civil works, including tunnels, concrete structures and buildings. At CERN, all infrastructure shall be designed in accordance with the applicable Eurocodes to withstand the expected loads during construction and operation, but shall also consider accidental actions, such as seismic activity, fire, release of cryogens and the effect of radiation on the concrete matrix and other tunnel construction fabric. 

All buildings, experimental facilities, equipment and experiments installed at CERN shall comply with CERN Safety Code E and other relevant fire safety related instructions. In view of the special nature of the use of certain areas, in particular underground, and the associated fire risks, CERN's HSE Unit is considered to be the authority for approving and stipulating special provisions. As the muon collider complex study progresses, detailed fire risk assessments will have to be made, in line with the evolution of the study of the technical infrastructures. The most efficient protection strategy is one that uses complementary 'safety barriers', with a bottom-up structure, to limit fires at the earliest stages with the lowest consequences, thus considerably limiting the probability and impact of the largest events. In order to ensure that large adverse events are only possible in very unlikely cases of failure of many barriers, measures at every possible level of functional design need to be implemented:
\begin{itemize}
    \item In the conception of every pieces of equipment (e.g. materials used in electrical components, circuit breakers, etc.);
    \item In the grouping of equipment in racks or boxes (e.g. generous cooling of racks, use of fire-retardant cables, and fire detection with power cut-off within each racks, etc.);
    \item  In the creation and organisation of internal rooms (e.g. fire detection, power cut-off and fire suppression inside a room with equipment);
    \item In the definition of fire compartments;
    \item In the definition of firefighting measures, including smoke extraction and fire suppression systems;
    \item Access and egress sufficiently sized in an acceptable range for evacuation and fire service intervention.
\end{itemize}

With the proposed technology solutions for equipment items, personal and process safety aspects, as well as environmental concerns, must be taken into account across the full life cycle from design, fabrication, testing, commissioning, use and dismantling.

For example, extensive use of cryogens will require considerations of cryogenic and mechanical safety, as well as risk evaluations with regard to potential oxygen deficiency hazards and potential mitigation means compatible with the complex. 

The potential use of gaseous and/or liquid hydrogen will also require particular study, in terms of explosion and flammable gas risk assessments, potential ATEX zoning and particular mitigation measures. 

The electrical hazards present in the complex will need to be assessed and mitigated through sound design practice and execution. The CERN Electrical Safety Rules, alongside NF C18-510, shall be followed throughout the design process; where exceptions are required, this shall be subject to an appropriate level of risk assessment to evaluate the residual risk, and determine the mitigation strategies required. For all aspects concerning non-ionising radiation (magnetic fields, electomagnetic compatibility, etc). the relevant CERN Safety Rules shall be followed.

For environmental aspects, efforts will be made to economise of the use of energy and water:
\begin{itemize}
    \item Atmospheric emissions shall be limited at the source and comply with the regulations in force.
    \item There will be a rational use of water with the discharge of effluent water into the relevant networks in accordance with the regulations in force.
    \item Use of energy will be made as efficiently as possible, with thermal efficiency for all new buildings
    \item The natural physical and chemical properties of the soil will be preserved. All the relevant technical provisions related to the usage and/or storage of hazardous substances to the environment shall be fulfilled to avoid any chemical damage to the soil. Furthermore, the excavated material shall be handled adequately and prevent further site contamination. All excavated material must be disposed of appropriately in accordance with the associated waste regulations. 
    \item The selection of construction materials, design and fabrication methods shall be such that the generation of waste is both minimised and limited at the source, with the appropriate waste handling and traceability of the waste in place.
\end{itemize}

Noise generated at CERN shall comply with safety requirements of the CERN Safety Rules. Emissions of environmental noise related to neighbourhoods at CERN shall comply with CERN's Noise footprint reduction policy and implementation strategy.
\\

\part{R\&D Proposal} \label{3:rd}
\chapter{Overview}
\label{2:obj:ch}

\section{Overall objectives and timeline}

\begin{figure}[h]
\centering
\includegraphics[width=0.6\textwidth]{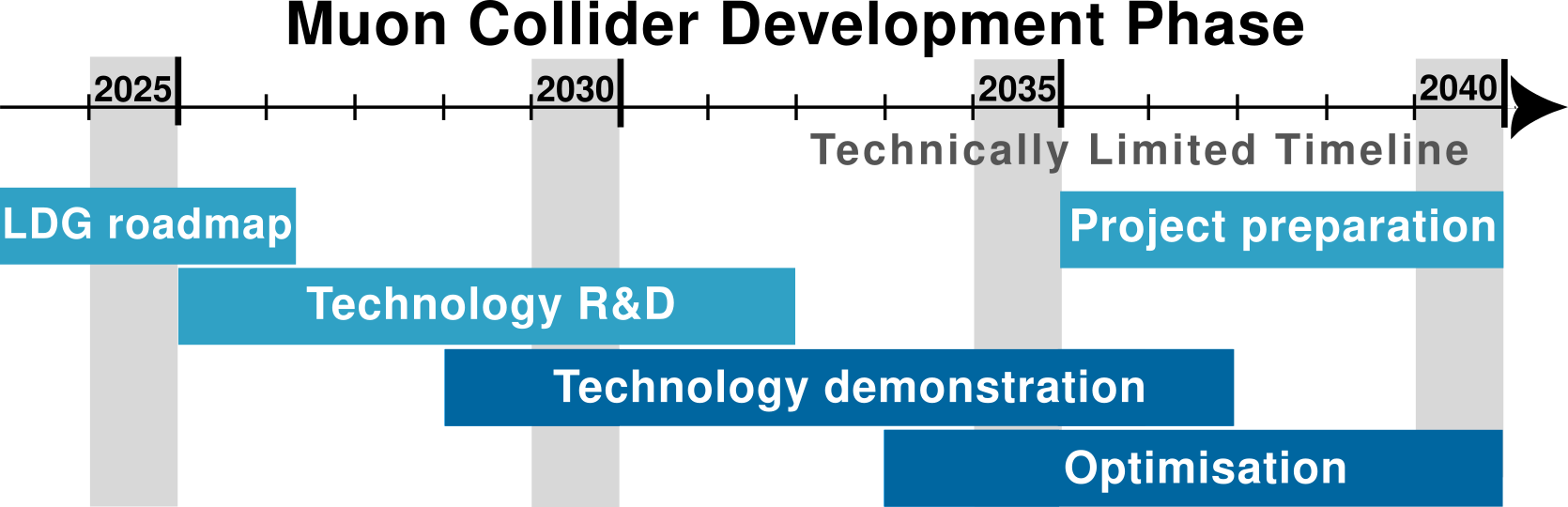}
\caption{Technically limited muon collider R\&D timeline.}
\label{fig:rd_tl}
\end{figure}

The goal of the muon collider R\&D programme for the period 2026-2036 is to reach technical decision readiness by 2036-37 for a staged implementation of a muon collider. The programme requires about 320 MCHF material budget and around 2700 FTEy of personnel for the collider and detector R\&D part over a ten-year time period. The technically limited timeline for the project is shown in Figure~\ref{fig:rd_tl}, assuming a prompt start of the R\&D funding in 2026.
This would enable a first muon collider stage with a start of operation around 2050. 
The timescale is driven by the construction and test of the most essential magnets needed for a staged collider implementation, and it is consistent with having results of a cooling demonstrator. The magnet and cooling demonstrator programmes are dominant parts of the R\&D programme and are considered schedule drivers.    

Technical decision readiness requires mature baseline design(s) documentation, technologies reaching TRL values 6 or above, comprehensive site and environmental studies (e.g. for CERN, FNAL), and cost/power/LCA/schedule studies at mature levels.  

A project decision also requires physics and detector studies to advance significantly, hosting and funding models to be agreed, governance discussions, construction project organisation models and formation of detector collaborations, international discussions and more, in most cases falling outside the scope of the following chapters. 

\section{Overview of the R\&D programme objectives}

The R\&D programme is described in the following six chapters. 

\textbf{Chapter 8} describes the physics R\&D goals. The activities will pursue the consolidation of the broad physics programme and the development of theoretical tools for predictions and simulations. The physics projections will be updated with full detector simulation, the neutrino-related physics case will be studied further, and novel tools, for example for resummation, will be developed.

\textbf{Chapter 9} describes the detector R\&D priorities. The development programme revolves around three pillars: simulation studies and performance, technological R\&D and software and computing. These developments will maximise the physics output, while minimising the impact of beam induced backgrounds. In technological R\&D, multiple technologies will be investigated in the first phase of the work. The choices are expected to consolidate after the first four to six years and resources will transfer to the identified technologies needed for the chosen detectors designs.

\textbf{Chapter 10} describes the magnet development programme, which aims to establish the performance of superconducting magnets through the construction and testing of models and prototypes. The focus will be on HTS solenoids for both the muon production target and muon cooling, leveraging strong synergies with societal applications such as fusion reactors. Additionally, a model of the collider ring dipoles will be built to characterize the magnetic field at large apertures. The performance of the fast-ramping magnet systems and power converter for the RCS will be demonstrated.

\textbf{Chapter 11} addresses key design challenges and critical technical developments essential for the overall performance of the complex. A high-priority task is the establishment of a start-to-end collider model within the initial years of the R\&D phase. The completion and optimization of the lattice design across the entire complex will guide component development. Advanced simulation tools incorporating key beam physics aspects—such as collective effects, imperfections, and mitigation techniques—will enable robust luminosity predictions. Additionally, a study of machine availability will assess integrated luminosity performance and inform both component and accelerator design.\\
Site and environmental impact studies, including civil engineering assessments, will focus on optimizing power consumption, material use, and minimizing the machine’s impact on the local environment.\\
A comprehensive optimization of the complex—covering cost, power efficiency, and risk—is particularly crucial, as there is no direct precedent for this type of project. This optimization will integrate performance considerations across different systems.\\
Experiments, combined with further design efforts, will cover all technological aspects related to the facility.

\textbf{Chapter 12} details the Muon Cooling Demonstrator Programme, which aims to validate the integration and operation of cooling equipment with the beam—both critical for achieving the required beam physics performance. This programme will advance the development of novel muon cooling technology by refining key components, including HTS solenoids, cavities, and absorbers. Testing a fully operational cooling cell with RF power will enable verification of its integrated performance. Additionally, the demonstrator facility will evaluate multiple cooling cells with beam to confirm the technology’s performance. Initial and preparatory phases include the construction of an infrastructure to test RF in high magnet field that will enable experimental optimisation and verification of the normal-conducting muon cooling RF. The development of a high-power, high-efficiency klystron will be instrumental for the muon cooling cell test
and enable cost effective design.

\textbf{Chapter 13} describes the synergies between the IMCC R\&D programme and other projects, technologies and studies within and beyond particle physics. The muon collider R\&D effort will drive the development of key technologies with applications not only in particle physics and related facilities but also in broader scientific domains and society at large. These synergies extends to very relevant and applicable training opportunities for future young scientists and engineers.

\section{Resources summary}

The resources needed for the programme outlined below in Table~\ref{tab:RD_overview} 
has been evaluated and scrutinised by the collaboration, and will be reviewed by the International Advisory Committee in the near future.

The resources are listed in terms of personnel and material costs and itemized according to the major areas of development of the project. The personnel estimate includes senior scientists, postdoctoral researchers as well as PhD students.
The resources are split in the following areas:
\begin{itemize}
    \item Cooling and Demonstrator - including: construction of first cooling cell, five-cell module, and beam demonstrator construction at the TT7 beamline. This category includes the construction of the magnets needed for the cooling modules and beam demonstrator facility.
    \item Detector - including: simulation and performance studies, detector technology R\&D, software and computing resources.
    \item Magnets - including: the pursuit of eight technology milestones (described in detail in Chapter~\ref{3:rd:sec:mag}) and the related material and methods.
    \item Accelerator design and technologies - including: accelerator design, machine-detector interface, RF systems, targets, absorbers, cryogenics, beam instrumentation, vacuum and radiation protection.
\end{itemize}
The resources necessary for phenomenology studies and theoretical work are not included in the estimate. A more detailed resource breakdown is provided in chapters~\ref{3:det:ch} for detectors, and in chapters~\ref{3:rd:sec:mag} and~\ref{2:demo:ch} for the projects schedule drivers (magnets, and cooling technology).

\begin{table}[h!]
\centering
\begin{tabular}{|l|c|c|c|c|c|c|c|c|c|c|}
\hline
\rowcolor{cornflowerblue!80}
\textbf{Year}&\textbf{I}&\textbf{II}&\textbf{III}&\textbf{IV}&\textbf{V}&\textbf{VI}&\textbf{VII}&\textbf{VIII}&\textbf{IX}&\textbf{X} \\
\hline
\rowcolor{cornflowerblue!30}
\multicolumn{11}{|l|}{\textbf{Accelerator Design and Technologies}} \\
Material (MCHF) & 1.6 & 3.2 & 4.8 & 6.4 & 9.6 & 10.8 & 12.0 & 12.0 & 12.0 & 12.0 \\ 
FTE & 47.1 & 60.6 & 75.0 & 85.0 & 100.0 & 120.0 & 150.0 & 174.6 & 177.2 & 185.1 \\ 
\hline
\rowcolor{cornflowerblue!30}
\multicolumn{11}{|l|}{\textbf{Demonstrator}} \\
Material (MCHF) & 0.6 & 2.2 & 3.9 & 5.4 & 7.8 & 15.1 & 25.9 & 32.4 & 31.8 & 12.6 \\ 
FTE & 9.5 & 11.0 & 12.5 & 29.2 & 29.7 & 30.5 & 25.5 & 27.7 & 26.7 & 25.5 \\ 
\hline
\rowcolor{cornflowerblue!30}
\multicolumn{11}{|l|}{\textbf{Detector}} \\
Material (MCHF) & 0.5 & 1.1 & 1.6 & 2.1 & 2.1 & 2.1 & 2.1 & 2.6 & 3.1 & 3.1 \\ 
FTE & 23.4 & 46.5 & 70.0 & 93.0 & 93.0 & 93.0 & 93.0 & 116.4 & 139.5 & 139.5 \\ 
\hline
\rowcolor{cornflowerblue!30}
\multicolumn{11}{|l|}{\textbf{Magnets}} \\
Material (MCHF) & 3.0 & 4.9 & 10.1 & 10.0 & 11.0 & 13.4 & 11.7 & 7.2 & 6.6 & 4.7 \\ 
FTE & 23.3 & 28.4 & 36.4 & 40.9 & 44.3 & 47.1 & 46.2 & 37.7 & 36.1 & 29.4 \\ 
\hline
\hline
\rowcolor{cornflowerblue!30}
\multicolumn{11}{|l|}{\textbf{TOTALS}} \\
\hline
Material (MCHF) & 5.7 & 11.4 & 20.3 & 23.9 & 30.6 & 41.4 & 51.7 & 54.2 & 53.5 & 32.4 \\FTE & 103.3 & 146.5 & 194.0 & 248.1 & 267.0 & 290.6 & 314.8 & 356.3 & 379.4 & 379.6 \\\hline
\end{tabular}
\caption{Summary of R\&D resources, divided by area, over a ten year period. Personnel is given as average Full Time Equivalent (FTE), where students count as 50\% FTE. Material is given in millions of CHF (MCHF). The total is the sum of the four areas.}
\label{tab:RD_overview}
\end{table}

\section{Summary of objectives and deliverables}

The most important objectives and deliverables of the R\&D programme, which are discussed in detail in this Part of the document, are summarised in the following tables and listed according to the area the belong to.
For those area that plan to build and/or operate demonstrator devices (magnets, cooling and radiofrequency), a brief description is also given.

\begin{table}
\centerline{
\renewcommand{\arraystretch}{1.25}
\begin{tabular}{|c|p{4cm}|p{7cm}p{5cm}|}

\hline
& {\textbf{Technologies}} & {\textbf{Objectives and Deliverables} } & {\textbf{Demonstrators}} \\
\hline
\multirow{20}{*}{\rotatebox[origin=c]{90}{\textbf{Magnets}}} & {Solenoid for target, decay and capture channel}&
{Develop conductor, winding and magnet technology suitable for a target solenoid, generating a bore field of 20 T, and operating at a temperature of 20 K. } & {Target solenoid model coil (20@20)} \\
& \multirow[t]{2}{*}{Solenoids for cooling} & Demonstrator of HTS split solenoid performance, including integration in its support structure submitted to mechanical and thermal loads  representative of a 6D cooling cell. & Split Solenoid integration demonstrator for 6D cooling cell (SOLID) \\
& & Build and test a demonstrator HTS final cooling solenoid, producing 40 T in a 50 mm bore, and total length of 150 mm  & Final cooling UHF solenoid demonstrator (UHF-Demo) \\
& RCS fast pulsed field system & Build and test a string of resistive pulsed dipoles, including powering system and capacitor-based energy storage. & RCS fast pulsed magnet string and power system (RCS-String) \\
& LTS accelerator magnets & Demonstrate LTS dipole performance for collider arc & Wide-aperture, steady state Nb3Sn dipole for the collider (MBHY) \\
& \multirow[t]{3}{*}{HTS accelerator magnets} & Demonstrate performance of rectangular aperture HTS dipole for the accelerator & Rectangular aperture HTS dipole (MBHTS) \\
& & Demonstrate wide aperture HTS dipole for the collider arc & Wide aperture HTS dipole (MBHTSY) \\
& & Demonstrate wide aperture HTS quadrupole for the collider IR & Wide aperture HTS IR quadrupole (MQHTSY) \\
\hline
\multirow{16}{*}{\rotatebox[origin=c]{90}{\textbf{Muon Cooling}}} & Muon cooling RF cavities and sources &
Determine achievable gradient of RF cavities immersed in strong magnetic fields and demonstrate suitable high-power and high-efficiency sources for powering them. & One or more RF Test Stands capable of studying parameter dependence of RF breakdown for a variety of frequencies. \\
& 6D muon cooling one-cell module & Demonstrate integration of key components for a muon cooling cell, including appropriate handling of magnet stresses and implementation of RF in the operational magnetic environment. & One-cell module of a rectilinear cooling system \\
& 6D muon cooling multi-cell module & Demonstrate construction of a multicell module, to show integration of absorber, RF and magnets including beam instrumentation and handling of dynamic forces in a quench.  & Multi-cell module of a rectilinear cooling system \\
& 6D muon cooling demonstrator & Show correct operation of the muon cooling system in beam. & Muon cooling Demonstrator with beam passing through the multi-cell module \\
& Final cooling absorber & Demonstrate hydrogen absorber with thin windows. & Demonstrate hydrogen absorber with thin windows. \\
\hline
\end{tabular}
\renewcommand{\arraystretch}{1}
}
\end{table}

\begin{table}
\centerline{
\renewcommand{\arraystretch}{1.25}
\begin{tabular}{|c|p{4cm}|p{7cm}p{5cm}|}
\hline
& {\textbf{Technologies}} & {\textbf{Objectives and Deliverables} } & {\textbf{Demonstrators}} \\
\hline
\multirow{6}{*}{\rotatebox[origin=c]{90}{\textbf{Radiofrequency}}} & Muon cooling RF cavities & Design, build and test RF cavities &  \\
& \multirow[t]{2}{*}{Klystron prototypes} & Design/build with industry 704 MHz klystron & 20 MW peak power, 704 MHz \\
&  & Design/build with industry 352 MHz klystron & 20 MW peak power, 352 MHz \\
& RF test stands & Assess cavity breakdown rate in magnetic field & 20-30 MV/m, 704 MHz–3 GHz cavities in 7–10 T \\
& SCRF & Design SRF cavities, FPC and HOM couplers, fast tuners, cryomodules &  \\
\hline
\end{tabular}
\renewcommand{\arraystretch}{1}
}
\end{table}

\begin{table}
\centerline{
\renewcommand{\arraystretch}{1.25}
\begin{tabular}{|c|p{4cm}|p{12cm}|}
\hline
& {\textbf{Technologies}} & {\textbf{Objectives and Deliverables} } \\
\hline
\multirow{34}{*}{\rotatebox[origin=c]{90}{\textbf{Design and Technology}}} & \multirow[t]{4}{*}{Target} & Simulation of entire frontend \\
& & Technical design of graphite target \\
& & Prototyping and beam test setups \\
& & Study alternative target materials and designs \\
& \multirow[t]{2}{*}{Beam instrumentation} & Instrumentation for proton beams and non-invasive diagnostics \\
& & Instrumentation for cooling channel \\
& \multirow[t]{2}{*}{Machine-detector interface} & Devise a conceptual MDI design and mitigate the beam-induced background \\
& & Define the technical specifications for MDI-related equipment, coordinate the relevant engineering design studies, and guide the overall MDI integration \\
& Neutrino flux mitigation & Develop a comprehensive, site-independent dose assessment model to optimize collider ring designs \\
& \multirow[t]{2}{*}{Radiation shielding} & Study the radiation impact on complex (collider, accelerators and muon production source) \\ 
& & Elaborate a technical shielding design, perform integration studies for the shielding, suggest prototyping for shielding elements. \\ 
& \multirow[t]{2}{*}{Cryogenics} & Investigate cooling strategies focusing on superconducting magnet structures at 20K \\ 
& & Provide functional requirements for local heat extraction on magnets, SRF cavities, and radiation shielding. \\ 
& \multirow[t]{4}{*}{Vacuum} & Development of thin beam windows for final cooling \\
& & Design and prototyping of wedge-shaped absorbers\\
& & Design of integration of vacuum equiment in cooling channels \\
& & Design and prototyping of vacuum chambers for RCS and collider \\
& \multirow[t]{4}{*}{Radiation protection} & Develop a mature concept of the neutrino flux mitigation technology. \\
& & Assess the neutrino flux mitigation for an implementation of the collider in the Geneva area or elsewhere. \\
& & Optimization of the demonstrator to ensure that the exposure of personnel to radiation and the radiological impact are as low as reasonably achievable \\
& & Optimization of the whole facility to ensure that the exposure of personnel to radiation and the radiological impact are as low as reasonably achievable \\
& \multirow[t]{3}{*}{Facility Design} & Study of all beam dynamics phenomena relevant to ensure that a useful luminosity can be reached \\
& & Development of a scenario to commission and operate the machine from first beam to maximum performance (luminosity) \\
& & A start-to-end model of the machine consistent with realistic performance specifications \\
\hline
\end{tabular}
\renewcommand{\arraystretch}{1}
}
\end{table}

\begin{table}
\centerline{
\renewcommand{\arraystretch}{1.25}
\begin{tabular}{|c|p{4.5cm}|p{11.5cm}|}
\hline
& {\textbf{Technologies}} & {\textbf{Objectives and Deliverables} } \\
\hline
\multirow[c]{12}{*}{\rotatebox[origin=c]{90}{\textbf{Detector and Physics}}} & \multirow[t]{3}{*}{Simulation and performance} & Develop conceptual detector design suitable for the needs of the physics program \\
& & Define set of requirements and later technical specifications \\
& & Implement a set of algorithms for event reconstruction and performance evaluation to support specification, requirements and physics projections studies \\
& \multirow[t]{3}{*}{Detector technology} & Identify key technologies for each subsystem, implement low-level simulation (digitization) \\
& & Prototyping and beam test setups that prove the needed technology \\
& & Identify and propose engineering solutions for the technology, the placement and the integration of various sub-systems\\
& \multirow[t]{3}{*}{Software and computing} & Builds and support the tools needed to carry out simulation studies \\
& & Assess and manage computing resources \\
& & Identify and propose engineering solutions for the technology, the placement and the integration of various sub-systems\\
\hline
\end{tabular}
\renewcommand{\arraystretch}{1}
}
\end{table}

\chapter{Physics R\&D}
\label{2:physicsr&d}

\section{Introduction}
\label{2:physicsr&d:sec:introduction}

The muon collider will first reach the multi-TeV energy scale, and will probe it with precision. It will be the first one colliding leptons at this very high energy, and the first one with unstable beam particles. The physics opportunities, the experimental methodologies and detector concepts and technologies, and the theoretical tools for predictions, do not emerge from an extrapolation of past colliders such as LEP and LHC but rather require new ideas and dedicated work. Based on experience with past colliders, and on the higher degree of novelty of the muon collider project, it will probably take around two decades to ensure the optimal exploitability for physics of the future muon collider data. It is thus important that the work in this direction has started already and that it strongly advances after the next Strategy Update.

Muon collider physics studies are not only functional to the muon collider project and thus an integral part of the muon collider R{\&}D plan. The novelty of the problems and the lack of established solutions requires and enables innovative work within the well-established field of particle colliders physics. Muon collider physics is the ideal playground for the incorporation of new talents advancing the field at large during the next decade, on top of paving the way towards a future muon collider.

The work needed on muon collider physics follows two main directions, which are described in Sections~\ref{2:physicsr&d:sec:consolidatingthephysicscase} and~\ref{2:physicsr&d:sec:theoreticaltools} in turn. 

Not only the already broad physics program will be further expanded, but also consolidated by sensitivity projections that systematically include detector effects and theory mis-modeling uncertainties.  This will enable a complete assessment of the target detector performances and theoretical precision, as well as devising mitigation strategies if the targets could not be met. A similar path should be followed for those opportunities of the muon collider project that do not resort directly to the study of muon collisions. This includes opportunities for neutrinos physics at a dedicated forward detector that collects the neutrino beam from the decay of muons close to the interaction point. Further opportunities that can emerge during the process of muon collider demonstration or construction should be identified and investigated. 

The second item is the development of the theoretical tools for predictions and simulations that will be needed to exploit the muon collider data at best. The novel challenges in comparison with state-of-the-art tools stem from electroweak interactions, which will be probed in a previously unexplored regime featuring phenomena akin to those governed by QCD interactions at high-energy hadron colliders. Since the electroweak interactions are weakly-coupled, there are good perspectives to attain a very high level of theoretical accuracy. On the other hand, the prominent role of radiative corrections at the high muon collider energy will require resummations methodologies that are still to be developed and demonstrated at the required level of accuracy. The problem is at the right level of challenge and development to be fruitfully addressed in the coming decade. 

\section{Expanding and consolidating the physics case}
\label{2:physicsr&d:sec:consolidatingthephysicscase}

The new enthusiasm on muon collider physics has triggered---see Section~\ref{1:phys:ch}---the rapid development of a broad physics case for the muon collider project. This work must continue in the future, with the twofold aim of defining new analysis targets and of improving the level of accuracy and realism of the sensitivity projections beyond the one of currently available results.

Among the several missing studies, one can mention for instance a complete assessment of the muon collider perspectives to probe new EFT interactions by the combination of Higgs couplings determination, high-energy measurements of direct muon annihilation cross sections, and Vector Boson Fusion (VBF) or Scattering (VBS) processes. Partial available results---described in Chapter~\ref{sec:EHT}---already include a large variety of interactions but still lack an optimized analysis of the very many relevant experimental measurements. The rich phenomenology of VBS is almost unexplored in the perspective of an EFT fit, but also as a probe of SM Higgs Unitarization and electroweak symmetry restoration. More work is also needed on the sensitivity to concrete BSM models or scenarios, including the perspectives to discover and characterize the new states directly and in combination with indirect probes. Enriching the muon collider physics case along these lines will unveil novel observables and experimental challenges offering new targets to the experiment and detector design.

Consolidating the physics case requires progress in two separate directions, which are discussed below in turn. One is to assess the quality of the relevant experimental measurements or analyses beyond purely-statistical sensitivity projections and including instead fully realistic experimental effects and uncertainties. The other is to assess the quality of the theoretical predictions needed for their interpretation. 

Several muon collider sensitivity projections---see Sections~\ref{1:phys:ch} and~\ref{1:inter:ch}---already include at least a parametric modeling of the detector effects. More such studies should be carried out in the future in order to identify the key detector performance parameters and the adequacy of the current parameter targets defined in Section~\ref{1:inter:ch}. In addition, detailed studied based on the full detector simulation should be performed, and compared with the parametric simulations. This will enable to assess the state-of-the-art detector performances in comparison with the target and their impact on physics. It will also provide robust sensitivity projections for those analyses where detector and BIB effects cannot be modeled parametrically as is typically the case for ``unconventional'' type of signatures. A future work plan for full simulation physics studies is planned for the benchmark channels detailed in Section~\ref{1:inter:ch}.

Physics studies that exploit very forward muons with rapidity above~5 are particularly delicate. Unlike the main detector, the forward muon detector has not yet been designed, and not even a detector concept is currently available. On the other hand, measuring very energetic muons that cross a thick layer of dense material is a new challenge so that the expected performances of this forward detector cannot be guessed based on extrapolation of previously built systems. Available parametric studies have outlined critical performance thresholds that would enable important important measurements such as the one of the inclusive production cross-section, the invisible branching ratio and other properties of the Higgs boson, as well as enabling BSM searches for new invisible particles produced in VBF. Work on the design of the detector is mandatory in order to put these results on firmer ground.

We outlined in Sections~\ref{1:physics:sec:neutrinos} the opportunities for neutrino physics that stem from the muon collider and from the muon collider R\&{D} program. One of them is the exploitation of the neutrino beam that is produced by the high-energy (3~or 10~TeV) collider for a parasitical experiment. A systematic investigation of the physics potential of such an experiment has started (see Section~\ref{sec:CKM} and Chapter~\ref{sec:SI}), and must be accompanied by the design of the corresponding detector. Neutrino factories that exploit instead lower energy muons should be also considered as a complement to superbeam experiments. They could be built in synergy with the muon collider demonstration program. Additionally the technology for muon colliders may enable other physics opportunities, for example muons could be collided with other particles for example the $\mu$Tristan proposal~\cite{Hamada:2022mua} ($\mu^+e^-$) or muon-ion collisions~\cite{Acosta:2022ejc}.  The physics opportunities of these other possibilities still needs further investigation, but they generally range from physics cases that are similar to a muon collider to better understanding the underlying structure of nucleons.

Not much work has been performed so far to assess the quality of the theory predictions that are needed to accomplish the physics goals in comparison with state-of-the-art predictions accuracy. This is well-justified by the fact that the relevant muon collider measurements are statistically limited to the permille level accuracy. This is the case for Higgs physics measurements, given the total of 10 million Higgs boson produced, and more generally for 100~GeV scale processes in VBS or VBF. The data statistics is even lower at higher scattering energy and the expected statistical accuracy for 10~TeV $2\to2$ cross section measurements is at the level of the percent. A much lower theoretical accuracy below permille is generically expected to be possible at a lepton collider and was achieved already long ago for the LEP collider. 

However, while it is indeed expected that the required accuracy targets will be met and theory errors will not be a limiting factor for physics, the muon collider probes a much higher energy than past lepton colliders and this enhances electroweak radiative corrections, entailing new challenges for theory. For $2\to2$  processes at 10~TeV, the enhancement is so pronounced that predictions performed at fixed order in the loop expansion become inadequate, requiring the resummation of at least the leading (Sudakov) infrared logarithms. Promising methodologies have been developed---see Section~\ref{2:physicsr&d:sec:theoreticaltools}---for the resummation of electroweak infrared logarithms. However, their deployment for muon collider physics and their perspectives to attain the required accuracy are far from obvious. Radiative corrections to scattering processes with a lower characteristic scale such as the production of the Higgs boson in VBF are completely unexplored, and it is not guaranteed that these corrections will be small and calculable at fixed order because the collision energy of the initial muons is still much larger than the electroweak scale. The design and implementation of methods to perform these calculations is among the new tools that need to be developed for muon collider physics, described in Section~\ref{2:physicsr&d:sec:theoreticaltools}.

\section{Theoretical tools}
\label{2:physicsr&d:sec:theoreticaltools}

Electroweak (EW) interactions at energies much above the EW mass of around 100~GeV are a topic of self-sending theoretical interest. Investigations in this directions started around four decades ago, producing textbook results such as the Goldstone boson equivalence theorem~\cite{Chanowitz:1985hj,Yao:1988aj,Bagger:1989fc,Kilgore:1992re,He:1992nga,He:1993yd,Pozzorini:2001rs,Bohm:2001yx} and the effective vector boson approximation~\cite{Kane:1984bb,Dawson:1984gx,Kunszt:1987tk}, which later evolved into the notion of EW Parton Distribution Functions~\cite{Bauer:2017isx,Fornal:2018znf,Bauer:2018xag,Chen:2016wkt}. During the following decades, the attention moved to the study of (virtual or real) EW radiation effects that are enhanced by the large separation between the scattering energy $E$ and the EW mass $m=100$~GeV. Famous results in this area are the discovery~\cite{Ciafaloni:2000df,Ciafaloni:2000rp} that real and virtual EW radiation effects do not cancel out---unlike in QCD and QED---in inclusive observables due the violation of the KLN theorem hypotheses in massive gauge theories such as the SM, and the development of strategies to include EW radiation effects by all-orders resummation~\cite{Ciafaloni:2000df,Ciafaloni:2000rp,Fadin:1999bq,Melles:2000gw,Chiu:2007yn,Chiu:2007dg,Chiu:2009ft,Manohar:2014vxa,Manohar:2018kfx} with a variety of techniques including Soft-Collinear Effective Field Theory, or at fixed order up to two loops~\cite{Denner:2000jv,Denner:2001gw}. EW Monte Carlo showering has been also investigated and implemented in Monte Carlo generators~\cite{Christiansen:2014kba,Christiansen:2015jpa,Brooks:2021kji}.

Past work was motivated by pure theoretical curiosity. The perspective of a muon collider offers a concrete target and phenomenological motivation to this already established line of investigation, boosting its pertinence in the present and in the coming years.~\footnote{The annihilation of heavy Dark Matter particles is another motivation for EW radiation studies.} The growing recent literature on high-energy EW physics specifically targeted for muon colliders is reviewed in Section~\ref{1:phys:ch}. The accelerated development of the field will rapidly address the remaining challenges for predictions at a muon collider.

The quest for precise predictions in a regime where higher-order EW effects are amplified by large logarithms will have to be addressed under multiple perspectives. First, by developing fixed-order predictions reaching NNLO EW, which in turn requires multi-loop calculations in the presence of multiple mass scales. Second, by further developing all-orders resummation that is mandatory at 10~TeV energy at least for the leading Sudakov double logarithms. A scheme for matching fixed-order and resummed predictions will most likely be needed in order to attain the necessary accuracy. A new generation of Monte Carlo tools combining accurate fixed order predictions with EW and QCD effects in both final and initial state radiation will eventually emerge. It is important that this theoretical work develops targeting applications to muon collider physics studies. Not only this offers theoretical accuracy targets as previously emphasised. The treatment of EW radiation in the very definition of the relevant observables is still to be established, in a way that does not hide relevant information on the underlying hard process by an overly inclusive treatment of radiation, that facilitates if possible the obtention of accurate predictions and that proves useful and compatible with the experimental perspectives for measurements. In addition, the observation of the novel high-energy EW phenomena and the verification of their theoretical predictions is an integral part of the physics case of the muon collider.

\subsection*{The future muon collider theory community}

It is not easy to quantify the size of the community that will address the aforementioned theoretical challenges. An estimate can be obtained by comparison with the theory community that provided tools and ensured the optimal exploitability for physics of the LHC data. 

Sustained efforts on LHC theoretical tools began in early 2000~\cite{Altarelli:2000ye}. At that time, the theory community was rather small, of around 30~theorists, and not much larger than the present-day muon collider community. The theoretical toolkit was quite small. Available resources included basic LO Monte Carlo simulations for $2 \to 2$ processes, a single NNLO inclusive prediction for Drell-Yan, and a handful of NLO predictions implemented in specialized codes leading to distributions that could be only compared with unfolded data. Within the following decade, the size of the community grew by more than a factor of 10, achieving fundamental milestones such as the automation of LO predictions (in codes such as Alpgen, MadGraph, and Sherpa) and later the automation of one-loop computations (MadLoop, Samurai and OpenLoops), the merging of LO calculations with parton showers and later the extension to NLO (MC@NLO and POWHEG), as well as the revolutionary advances in two-loop computations that enabled fully differential NNLO predictions. In addition, at the frontier between theory and phenomenology, radical innovations in jet substructure and boosted object studies opened new avenues in collider physics.

Globally, the degree of novelty of the muon collider project today is higher than the degree of novelty of the LHC project in the early 2000. The LHC was not the first proton collider, nor the first collider observing QCD jets or more generally probing strong interactions in the regime of perturbative QCD. The muon collider will instead collide muons for the first time, and first probe the high-energy regime of EW interactions as previously emphasised. This opens new opportunities for physics, to be explored by a large community of theorists and phenomenologists beyond those who will develop the muon collider tools. On the other hand, the challenges for muon collider tools development appear of comparable scale as the one posed by the LHC. We can thus envisage a community of a similar size to set up in the coming decade.

The analogy with the LHC can be also taken in order to outline the key factors that ensured the success of the LHC and will similarly ensure the success of the muon collider tools development enterprise. First, a set of clearly defined challenges. Second, the attraction of the best talents. Muon collider physics is sufficiently developed to pose clear challenges, and highly innovative, which is essential in order to attract talents. The third element of the LHC success was the anticipated influx of experimental data that spurred sustained theoretical efforts. In the case of the muon collider, this third elements stems from the increasing confidence on the collider feasibility through the acceleration of the R{\&}D program after the next European Strategy Update.

\chapter{Detector R\&D}
\label{3:det:ch}

\section{Introduction}
\label{3:det:sec:resources}

To effectively pursue the vast physics programme of the muon collider, detectors must be optimised for physics in the challenging environment created by machine-induced backgrounds. Research and development over the next decade must focus on developing the technologies, tools, and crucial expertise necessary to design and construct state-of-the-art detectors for this future machine.

At the muon collider, most detector components must simultaneously optimise position resolution, timing capability, radiation hardness, data transmission, and on-detector background rejection, all while maintaining low mass and low power consumption. Many of these technical challenges align with ongoing research in other initiatives, such as the LHC and Higgs/Top/EW-factory R\&D programmes. However, the development effort must also address challenges unique to the muon collider, including its harsh radiation environment, the need to suppress significant beam-induced backgrounds, and the requirements for high-precision calorimetry to measure significantly higher energy physics objects.

The detector development programme is now aimed at designing 10~TeV detector concepts. The early iterations of the MUSIC and MAIA detector concepts need to be refined and updated as the design of the machine progresses.

The detector R\&D programme is organised around three main areas:
\begin{itemize}
    \item Simulation studies and performance - aimed at refining the detector conceptual designs, spelling out tecnical requirements, as well as developing algorithms for the reconstruction and exploitation of the data
    \item Detector technologies - aimed at identifying, developing and prototyping the technologies that will be employed in the detectors
    \item Software and computing - aimed at building and supporting the tools required by the R\&D programme, as well as the ultimate exploitation of the data, and the management of the computing resources.
\end{itemize}

Table~\ref{tab:2:det:resource:areas} summarises the labor and M\&S (Material and Supplies) resources needed for each of the three main areas. Each area is further divided into deliverables; the simulation\&performance and technology deliverables are loosely mapping into major sub-detectors components. Each deliverable is then split in tasks, as needed, that represent the major thrusts of R\&D for each deliverable. Each area, deliverable and task is motivated and further detailed in the remaining of this chapter.

Among all tasks that have been considered, ASICs and detector magnets have been identified as critical components requiring extended development timelines and should therefore be prioritised to ensure readiness for detector construction.  

\begin{table}[h!]
\centering
\begin{tabular}{|l|c|c|c|c|c|c|c|c|c|c|}
\hline
\rowcolor{cornflowerblue!80}
\textbf{Year}&\textbf{I}&\textbf{II}&\textbf{III}&\textbf{IV}&\textbf{V}&\textbf{VI}&\textbf{VII}&\textbf{VIII}&\textbf{IX}&\textbf{X} \\
\hline
\rowcolor{cornflowerblue!30}
\multicolumn{11}{|l|}{\textbf{Simulation \& Performance}} \\
\hline
Staff & 1.8 & 3.5 & 5.3 & 7 & 7 & 7 & 7 & 8.8 & 10.5 & 10.5 \\ 
Post doc & 1.8 & 3.5 & 5.3 & 7 & 7 & 7 & 7 & 8.8 & 10.5 & 10.5 \\ 
Student & 3.5 & 7 & 10.5 & 14 & 14 & 14 & 14 & 17.5 & 21 & 21 \\ 
Material (kCHF) & 0 & 0 & 0 & 0 & 0 & 0 & 0 & 0 & 0 & 0 \\ 
\hline
\rowcolor{cornflowerblue!30}
\multicolumn{11}{|l|}{\textbf{Detector Technology}} \\
\hline
Staff & 4.9 & 9.8 & 14.8 & 19.7 & 19.7 & 19.7 & 19.7 & 24.6 & 29.5 & 29.5 \\ 
Post doc & 4.9 & 9.8 & 14.8 & 19.7 & 19.7 & 19.7 & 19.7 & 24.6 & 29.5 & 29.5 \\ 
Student & 9.8 & 19.7 & 29.5 & 39.3 & 39.3 & 39.3 & 39.3 & 49.2 & 59 & 59 \\ 
Material (kCHF) & 425 & 850 & 1275 & 1700 & 1700 & 1700 & 1700 & 2125 & 2550 & 2550 \\ 
\hline
\rowcolor{cornflowerblue!30}
\multicolumn{11}{|l|}{\textbf{Software \& Computing}} \\
\hline
Staff & 1.1 & 2.2 & 3.3 & 4.3 & 4.3 & 4.3 & 4.3 & 5.4 & 6.5 & 6.5 \\ 
Post doc & 1.1 & 2.2 & 3.3 & 4.3 & 4.3 & 4.3 & 4.3 & 5.4 & 6.5 & 6.5 \\ 
Student & 2.2 & 4.3 & 6.5 & 8.7 & 8.7 & 8.7 & 8.7 & 10.8 & 13 & 13 \\ 
Material (kCHF) & 100 & 200 & 300 & 400 & 400 & 400 & 400 & 500 & 600 & 600 \\ 
\hline
\hline
\rowcolor{cornflowerblue!30}
\multicolumn{11}{|l|}{\textbf{TOTALS}} \\
\hline
Material (MCHF) & 0.5 & 1.1 & 1.6 & 2.1 & 2.1 & 2.1 & 2.1 & 2.6 & 3.1 & 3.1 \\ 
FTE & 23.4 & 46.5 & 70.0 & 93.0 & 93.0 & 93.0 & 93.0 & 116.4 & 139.5 & 139.5 \\ 
\hline
\end{tabular}
\caption{Summary of resources for detector R\&D, divided by area, over a period of ten years. Personnel is given as average Full Time Equivalent (FTE), where students count as 50\% FTE. Material \& Supplies (M\&S) are given in thousands of CHF (kCHF). The total is the sum of the three areas.}
\label{tab:2:det:resource:areas}
\end{table}

Labor and M\&S estimates were compiled by gathering input from groups currently engaged in R\&D efforts. In areas where no active R\&D is taking place, estimates were derived from the detector R\&D programme for HL-LHC detectors. It is assumed that approximately half of the required resources will be provided by other general detector funding sources, with the remaining half sourced from the targeted Muon Collider R\&D program as detailed in these tables. The labor distribution is assumed to be approximately 1:1:2 among senior staff, postdoctoral researchers, and students, with students contributing at a 0.5 FTE level, while senior staff and postdocs are engaged full-time.  

A phased ramp-up to nominal resource levels is anticipated over four years (I–IV) in 25\% annual increments for both labor and M\&S. From V to VII, resource levels will remain steady, after which the programme is expected to diverge into two detector collaborations with distinct designs, technology choices, and possibly software infrastructures. At that point, resources will increase by an additional 50\% to support the delivery of Technical Design Reports (TDRs) for the detector designs, aligning with the anticipated technical readiness of the accelerator in the late 2030s.

\section{Simulation Studies and Performance}
\label{3:det:sec:simulation}

The simulation studies and performance work area includes the following objectives:
\begin{itemize}
\item Develop conceptual detector designs that can reconstruct high-level physics objects in the presence of beam-induced backgrounds with the accuracy needed to satisfy the needs of the physics programme.
\item Define set of requirements and later technical specifications (in conjunction with the detector technology and software/computing activities).
\item Implement a set of algorithms for event reconstruction and performance evaluation to support specification, requirements and physics projections studies.
\end{itemize}

The deliverables include simulation designs for the following subsystems: tracking detectors, calorimeters (both ECAL and HCAL), muon detectors, particle identification detectors (PIDs), forward detectors, luminosity detectors, and the machine detector interface. As well, optimisation studies integrating different elements of the overall detector design in simulation must be carried out. 

Tracking efforts will focus on several key areas for the development of the detector. First, a study of full 4D track reconstruction is crucial to improve both the accuracy and efficiency of track identification. Additionally, the tracker layout must be optimized, with particular attention to the endcap regions. This includes refining track fitting techniques and establishing clear criteria for track quality selection. Further work will focus on the technical specifications of tracker sensors and more realistic digitization algorithm for tracker channels, in order to ensure optimal performance of the tracker and guide the corresponding technology development. Finally, efforts will concentrate on optimising algorithms for vertex reconstruction to ensure accurate event classification and particle identification.

Calorimeter development will also focus on a number of key issues. Similarly to the tracker, this includes refining the digitisation algorithm for calorimeter cells to ensure more accurate and realistic simulation. In addition, efforts will be made to optimise the subtraction depositions from BIB particles and energy calibration techniques. Clustering and fake rejection algorithms will also be optimized to improve particle flow reconstruction and identification. The layout of the calorimeter will be further optimised to improve reconstruction efficiency and resolution in the forward region. Finally, investigations will be conducted into alternative technologies for the HCAL to explore potential improvements in detector performance. For example, potential benefits of having a dual readout capability in the calorimeter should be investigated and quantified. 

The outermost part of the detector features a magnet iron yoke designed to contain the return flux of the magnetic field and is instrumented with muon chambers. Studies of the return yoke and implementation of a realistic map of the magnetic field will take place. We will investigate various magnetic field configurations to identify the most effective setup for muon detection. Additionally, the layout of the muon detector will be optimised, with the design choices influenced by the selected technology (or multiple technologies). A major part of the effort will also involve the development of a muon reconstruction algorithm to enhance the precision and efficiency of muon identification and reconstruction, ensuring high-quality performance in the region close to the nozzles. 

Many open questions regarding the detector design are related to the MDI, particularly concerning the tungsten nozzles. Studying modifications to the nozzles is challenging because BIB simulations must be conducted for each configuration. There is significant interest in reducing the size of the nozzle to extend the detector's coverage beyond $|\eta|=2.5$. Another area of interest is the possibility of tagging forward muons up to $|\eta|=6$ to differentiate between WW and ZZ fusion processes by instrumenting the nozzle. A momentum measurement with about 20\% resolution could also enable inclusive Higgs measurements and searches for Higgs to invisible decays. However, high occupancies and radiation doses present significant challenges for such measurements. 

Finally, dedicated detectors for luminosity measurements need to be studied. Preliminary studies using muons from large angle Bhabha scattering in the central region suggest that a luminosity measurement with $\sim$1\% uncertainty is achievable, though the limited statistics are insufficient for real-time monitoring. 

Studies of dedicated detectors for particle identification (PID) in a wide range of momentum will be pursued to understand how such detectors could help the physics program and what requirements are needed to make a significant impact on the physics program.

Efforts will also be devoted to improving the particle flow algorithm and into tailoring it to the needs of the muon collider detectors in order to improve overall performance of higher-level physics object reconstruction. In connection to this, the current heavy-flavour tagging algorithm is very simplistic if compared to those deployed by the LHC experiments. Development of a robust flavor tagging algorithm will be pursued, incorporating advanced techniques such as machine learning to improve the accuracy of charm and bottom jet identification. Similarily, development of dedicated hadronic tau reconstruction algorithms will enable detailed studies of a broad physics program. Dedicated reconstruction setups for long-lived particles that potentially decay in various parts of the detector will help defining detector requirements and general layout. At the same time a detailed look at jet substructure observables will inform the current design to ensure best sensitivity in boosted hadronic toplogies.

\section{Detector technologies}
\label{3:det:sec:technologies}

The detector technologies work area includes the following objectives:
\begin{itemize}
\item Identify key technologies that enable to achieve requirements for each subsystem in presence of beam-induced backgrounds.
\item Prototyping and test beam setups that prove the required technology.
\item Identify and propose engineering solutions for the technology, the placement and the integration of various sub-systems.
\end{itemize}

The deliverables include mature technologies for the following subsystems: tracking, calorimeters (both ECAL and HCAL), muon detectors, PID, forward detectors, luminosity measurement, as well as engineering considerations for global detector integration and optimization issues, including the MDI. 

The key areas of priority in the development are outlined below:
\begin{itemize}
    \item Sensor technologies with exploration of novel materials and their integration with state of the art electronics and mechanical support structures.
    \item ASICs and intelligent front-end processing: current technologies are moving towards CMOS sensors or 3D-integrated devices that provide local processing capability, with extensive on-board capabilities for intelligent processing with AI/ML algorithms. We should work on co-designing detectors for Muon Colliders that leverage on these emerging capabilities.
    \item Vertex and Outer Tracker - The vertex detector and outer tracker might use different technologies and strategies. We will work on the optimization of technologies for an ultra-light vertex and outer tracker detectors that deliver the required position and timing resolution within the cooling budgets.
    \item Calorimeters: natural synergies with SiW sampling calorimeter and HGCal, and strong expertise in crystal and Dual Readout calorimeters. High precision timing will be necessary in Calorimeters for Muon Collider, and the community has expertise in this and has led timing in Silicon and Crystal calorimeters.
    \item Success of this program hinges on maintaining and expanding current support to test beam and irradiation 
\end{itemize}

The identified tasks also ensure that multiple technologies will be investigated in the first phase of the R\&D. We expect choices will solidify after the first four to six years and resources will transfer to the identified technologies needed for the chosen detectors designs. Coverage across areas for items relevant for the same sub-detectors is also ensured.

For instance, in the tracker, the evaluation included overall design optimization (including digitization at the detector front-end), sensor development, ASICs, detector mechanics, power distribution, assembly considerations, and readout. Additionally, the necessary resources for conducting radiation studies and assessing their impact on the tracker were incorporated. Given the long lead time before detector construction, multiple sensor technologies (LGAD, 3D, and micro-strip) were considered, with the intention of down-selecting at a later stage.  

\subsection*{Tracking}

Vertex and tracking detectors at a muon collider face the unique challenge of BIB particles, which significantly increase the hit multiplicity. To accurately reconstruct tracks with high precision, precision timing information on the order of a few tens of ps is essential, as it allows for effective BIB rejection and ensures a manageable readout rate. Given the extreme hit multiplicity, particularly in the vertex detector, additional techniques such as stub identification or cluster shape analysis are needed, alongside 4D tracking. Overall, the tracking detector at a muon collider shares many requirements with the Phase-2 upgrades of the ATLAS and CMS detectors for the HL-LHC, but it also introduces the added complexity of incorporating high-precision timing across all tracking detectors. Additionally, the trackers for the muon collider detector face unique requirements stemming from the need to accurately reconstruct the significantly higher energy jets than at other HEP machines. 

Solutions to these challenges require an active research and development effort to fully explore and expand designs of tracking detectors beyond those of the traditional silicon-based technologies, as well as investments and investigations of emerging technologies and energy-efficient solutions including advances in quantum technologies and sensing, material science, AI/ML, and microelectronics. Especially challenging is the requirement of high-precision timing information for finely granulated detectors. The technological solutions implemented in the timing detectors for HL-LHC experiments are too power hungry to be simply scaled down to smaller pixels, and radically new directions will be required to achieve the requirements of the muon colliders. 

As the 4D-tracking is seen as the future by the community for any discovery machine, there are many examples of R\&D that a muon collider tracking detectors could profit from: Monolithic Active Pixels Sensors (MAPS)~\cite{Deptuch:2001hh} and 3D-integrated sensors~\cite{PARKER1997328}, Low Gain Avalanche Detectors (LGADs)~\cite{Moffat:2018kxw}, and advanced ASICs technologies and low-power data transmission development. Data rates at these experiments will increase by orders of magnitude relative to the HL-LHC, and the eventual solution to this incredible data challenge will involve a carefully optimized combination of many strategies including AI/ML and on-chip processing, determination of track angle and transverse momentum from single-sensor data for selection and transmission solely of interesting clusters from high momentum tracks, and communication between layers of the trackers and distributed processing across sensors and layers. The right compromise on sensor thickness that can achieve good timing resolution still allowing signal to BIB discrimination needs to be studied as well.

In a MAPS, the sensor and readout electronics are fabricated on a single silicon wafer using standard CMOS processes that are used in the semiconductor industry to mass produce processors and memory modules. MAPS detectors are a critical ingredient for future Higgs factories due to their very good position resolution ($\sim5$~$\mu$m), low-power dissipation ($<$40 mW/cm$^2$), low-material budget ($\sim 0.05$\% of $X_0$) and low manufacturing costs in large volumes. However, unlike detectors for muon colliders, the MAPS detectors for Higgs factories do not require high-precision timing at every detection layer. 

With this additional requirement for muon colliders the power consumption of tracking detectors skyrockets, and significant R\&D efforts will be needed to achieve the specifications. The requirement of precision timing in future tracking detectors poses a challenge shared by all technologies currently conceived for the muon colliders, and power consumption per unit area needs to be reduced by about two orders of magnitude compared to compared timing detectors. A multi-prong approach to address these challenges may be required to achieve the goals, where incremental improvements to several areas are simultaneously implemented. This will likely imply improvements to the silicon detectors by using alternative technologies and advances in material sciences, to achieve larger signals from MIP deposits, and that do not necessarily require active cooling for optimal operation. An example of such a development in recent years has been implementation of 4H-SiC to manufacture LGAD sensors. 

Front-end electronics for a muon collider tracker will work very differently than the current LHC experiments, having to withstand a very large flux of particles but only having to record and process bunch-crossing every $\sim 10\mu$s. This will likely see much more space take by active processing logic than memory, in contrast to the current LHC designs.
Innovative designs to drastically lower the power consumption on ASICs should be aggressively pursued. Some approaches that are being currently considered involve usage of low power technologies such as the 28 nm process (or beyond), power and area efficient circuit designs, and use of novel, beyond CMOS structures. The competing needs for significant on-detector data processing and minimal power consumption inspire the use of neuromorphic, reconfigurable AI/ML networks for local data processing. Finally, the data reduction strategy based on computation/transmission of higher level quantities requires communication between distributed sensors in many applications. For applications demanding minimal material in the sensing
volume, low power wireless inter-sensor communication coupled with distributed AI provides a solution.

While there are some common R\&D goals for the muon collider and Higgs factory communities, there needs to be a dedicated R\&D path to overcome the specific challenges a muon collider imposes on the tracking detectors. Due to the unique nature of the BIB, detector development will heavily rely on simulation efforts and a close contact should be established between the two efforts. Due to the importance of 4D tracking in particular, it seems paramount to understand the requirements of such a system as a whole early in the process to drive the specific component development. Early construction of 4D tracking demonstrator will be crucial to understand how such a system would work and should be pursued. The R\&D cycle necessarily needs to be complemented by a robust testing and characterization program with test facilities that are suitable to testing such tracking detectors under realistic conditions.

\subsection*{Calorimeter Technologies}
The past three decades of development of calorimeter technologies has resulted in maximal resolution within constraints, higher granularity and lower backgrounds for more sensitivity to new physics. In the Muon Collider Era, we expect to have sub-nanosecond tags for each individual particle region, multi-signal/multi-dimensional discrimination to further particle identification for all particles, tens to hundreds of millions of channels, tens of Mrad tolerances, tens of GBytes of per event data, and streaming readout and reconstruction with AI/ML. Above all, the great leap for the muon collider detector calorimeter is to deliver new order of magnitude advances across the board while cleaning each event of beam-induced background. 

There is a complex landscape of potential technologies from Si-W based on-detector/MAPS readout, Cryo/Noble Liquid/LAr high granularity with cold readout, optical calorimetry with dual-readout/hybrid crystals to emerging timing-centric approaches/fast glass/Crilin. The time-domain is an important and pervasive dimension and includes (AC-)LGAD/silicon, fast glass/Crilin, LYSO/fast scintillator/SiPM technologies for dedicated sub-20 ps timing layers to several 10’s to 100’s of picosecond leading-edge discrimination for many of the full-detector calorimeter technologies. The main points for the muon collider calorimetry are as follows: 

\begin{itemize}
\item Measure hit times along trajectories and design layer geometries to allow us to track from the interaction point outward and from beam collimators inward. Timing layers/walls should be arranged to efficiently catch beam backgrounds and maintain high event quality, providing multiple time measurements along trajectories leading up to the calorimeter. 

\item  At the reconstruction level, the goal is to dig deep with high quality local data, but to save frugally and intelligently when it comes to pushing data off-detector. A strong guiding paradigm is the Particle Flow (PF) algorithm in a graph theory approach taking advantage of improved spatial, temporal and energy resolutions. 

\item This level of design can be driven by AI/ML at its core, as the most optimised interplay between analysis and intrinsic detector measurements will maximise the scientific impact of such machine.

\item Simulation is absolutely central to optimising the calorimeter in concert with PFA/PID performance.  The software needs to be able to cycle through many available options and make quantitative comparisons.  

\item A physical demonstrator should be considered as mid-term goal for the detector R\&D. This can be a relative compact but structured tracking/timing/calorimeter detector slice capable of operating in the HL-LHC environment. The demonstrator should prove in realistic high background conditions that out-of-time non-pointing particle backgrounds can be efficiently suppressed while maintaining high efficiency for IP signal particles.  
\end{itemize}

\subsection*{Muon Detectors}
A possible design for a muon system could be a composite detector made of layers of different technologies to optimize timing and tracking performance. In addition, excellent spatial resolution would allow stand-alone muons to be used to begin global reconstruction of muon tracks.
Focusing on spatial resolution, values between 50~$\mu$m and 100~$\mu$m  are obtained with Drift Tubes or Cathode Strip Chambers, however, all wire-based detectors have inherent rate limitations due to slow evacuation of ions. This limitation has been overcome in Micropattern Gaseous Detectors (MPGDs), where electrodes are created using photolithographic techniques, which allows the distance between electrodes to be reduced by at least an order of magnitude, resulting in rapid evacuation of ions combined with high spatial resolution.

Switching to temporal resolution, MRPCs~\cite{CERRONZEBALLOS1996132} have gained stability and relevance by, for example, achieving a temporal resolution of about 60 ps and an efficiency of 95 percent for a detector composed of 10 250-$\mu$m  gaps arranged in a double-stack design using glass with resistivity $5\times 10^{12}$~$\Omega$cm. However, this detector technology cannot be considered for future experiments in colliders with the current gas mixture, which has a high Global Warming Potential (GWP). The use of freons will be phased out by 2030. While promising results have been obtained for standard High Pressure Laminate RPCs with alternative gases such as Hydrofluoroolefine (HFO1234ze), the performance of MRPCs still needs to be verified with these new gas mixtures. In addition, the performance of MRPCs depends on a non-negligible fraction of SF6, which has an even higher GWP than freons. R\&D in this direction is therefore crucial.

Regarding the time resolution of a classical MPGD, it is dominated by fluctuations on the position of the first ionization cluster in the drift gap, so it can be improved by using it with a faster gas mixture, but still without being able to obtain results better than a few ns. One possible approach aimed at improving the time resolution is the one followed by the PicoSec Collaboration~\cite{BORTFELDT2018317}, which obtained a time resolution of 25~ps measured with a Ne/C2H6/CF4 gas mixture. Future studies on this new technology focus on detector stability, choice of materials and geometry, radiation resistance, and gas mixture. An alternative approach, however, is the Fast Timing Micropattern (FTM) gas detector~\cite{DeOliveira:2015bda}.

\subsection*{DAQ}
The Muon Collider also poses unique challenges for DAQ systems, with its high backgrounds and a long crossing interval, and will face unprecedented data challenges from beam-induced backgrounds that require fine-precision spatial and timing resolution leading to massive stream data rates for detector readout and control. A DAQ system must be designed that will minimize overall power consumption while providing the relevant information. Real-time online processing, likely include novel ML methods and computing architectures (ASICs, FPGAs, and beyond) of all detector subsystems, including the highest rate subdetectors, will be critical to maximize the physics output of the detector, enable autonomous operations, and accelerate time-to-physics. If on-board filtering is efficient all of the relevant data can be read out in the inter-crossing interval. This would eliminate the need for a Level-1 trigger. 

\subsection*{Detector Magnet}
\label{3:det:sec:magnet}

The detector magnet work package includes design and performance studies through simulation, along with early-stage prototyping. 

Both current MUSIC and MAIA detector concepts feature a superconducting solenoid, capable of a 4-5~T field at the interaction point. Some differences between configurations of the two magnets are present, but the general R\&D directions are common. It should be noted that for the MUSIC detector both inner and outer hadronic calorimeter with respect to the position of the solenoid designs have been considered and are being further investigated, while for the MAIA detector the solenoid is smaller because both electromagnetic and hadronic calorimeters are located outside of it. 

To reach a field of 5~T at the interaction point work will be needed to improve the cable used in the solenoid coils, but available information on aluminium stabilised, niobium titanium superconducting Rutherford cable is strongly encouraging that the magnet will be feasible well in time with the foreseen schedule for the accelerator complex. 

Some interest from one of the groups involved in the CMS solenoid development and construction has been secured for the preliminary design of the MUSIC solenoid magnet.

\subsection*{Other sub-detectors}
In addition the the main systems described above, several other smaller detectors need to be designed and their technology investigated for the unique environment. 

Traditional luminosity detectors that rely on detecting forward particles produced at the interaction point as a measurement of luminosity based on silicon, or optical fibres would likely have a very large contribution from BIB; detailed simulation studies of different layouts and technologies are needed to assess if existing designs can be adapted or new technologies should be investigated. 

Another area where significant R\&D will be needed to identify the best technology is dedicated detectors for Particle Identification (PID). For low-momentum Standard Model particles (around the ~GeV scale and less) energy loss by ionization can provide useful information and can be measured in the pixel detectors, with an accuracy in the digitization process that needs to be optimized. The system is designed with an occupancy such that BIB can be efficiently separated from signals. Similarly, the same applies for heavier Beyond-Standard-Model particles that are expected to move with relatively low velocity $\beta \simeq 0.5$. However, for higher-momentum particles, other technologies are needed, as for instance Cherenkov detectors. The use in of such technology in this environment is not straightforward and requires dedicated developments; it is not obvious at this stage that such technology choice is viable at all and alternatives should be investigated.

The physics case for detecting very-forward muons (up to $\eta\sim 6$) has been well studied. Tagging those muons already provides an excellent physics case and measuring their momentum with around $\sim 20$\% accuracy would provide further valuable input. Any detector in this area needs to be embedded in the nozzle design and sustain fluences up to $\sim 10^{16}$1 MeV n$_{eq}/$cm$^2$; in addition it needs to be able to disentangle muons from the interaction point from BIB that are showering in the nozzle. Such a (tracking) device is well beyond current capabilities and requires aggressive research and new ideas to be realised in practice. A momentum measurement based on the amount of multiple scattering in the nozzle appears to be the most straightforward way to achieve the needed precision on momentum resolution but its feasibility will need to be proven once viable technologies are identified.

\subsection*{Mechanics and Integration}
Although any particle physics detector of this complexity and size requires large mechanical engineering efforts per se, a few aspects are worth noting explicitly and are unique for this environment. 
The large shielding nozzles are expected to weigh around 30~tons on each side, and require dedicated mechanical support for each of the pieces comprising this shielding structure. In addition, the limited space available beyond the current muon system design and the closest accelerator focusing magnet implies a careful design that allows integration of each component and replacements / upgrades of parts of the detector. All of this should be accomplished minimising material that enters the line-of-sight from the interaction region within the $\pm10^\circ$ acceptance of the sensitive parts of the detector.

\section{Software \& computing for detectors}
\label{3:det:sec:software}

Detector design, development and tuning all rely on software to assess the impact of any changes on the physics performance. Developing and maintaining the necessary software stack is crucial and the amount of work necessary just for maintenance should not be underestimated.

At the same time, the challenging collision environment of a muon collider offers unprecedented opportunities to further research in the realms of data science and AI/ML. These opportunities go from the optimisation of data handling, processing and compression, to the efficient use of heterogenous computing infrastructure, and to the development of novel algorithms to interpret and analyse the electrical signals from the data acquisition chain.

The software \& computing work area includes the following objectives:
\begin{itemize}
\item Build and support the tools needed to carry out simulation studies on the conceptual detector design and detailed technology simulation studies.
\item Assess and manage computing resources needed for detector R\&D studies, across the different areas.
\end{itemize}
and the high-level deliverables include:
\begin{itemize}
\item Assess and manage computing resources needed for detector R\&D studies, across the different areas.
\item Identify, acquire and maintain computing resources (processing power and storage) and tools that allow to easily share resources and data among collaborators.
\item Assess prospects of technology with an early conceptual design of a computing model for the experiment.
\end{itemize}

Collaborating with and contributing to the common Key4hep effort~\cite{Key4hep:2022xly} will be a key ingredient for managing the necessary work with the limited available personpower early in the project. There are no hard dependency constraints for the different efforts that are discussed in this section. Hence, most of them could be started right away given the availability of developers and computing resources. Nevertheless, interconnections exist and some detector R\&D questions additionally drive prioritisation decisions.

There is an ongoing migration effort to bring generally useful but separately developed components into the Key4hep ecosystem. The bulk of this work should aim to be accomplished over the next three years, and would also benefit the wider hep community. The muon collider community will reduce overhead from having to maintain separate forks, while other communities involved in Key4hep can pick up these added components with minimal effort. This migration will only be the first step toward closer collaboration with Key4hep. A dedicated effort to contribute to core Key4hep development will also be necessary, to ensure the capability to deal with muon collider-specific challenges. Closer collaboration will facilitate onboarding of new collaborators and foster the training of the next generation of experts in this area. Adoption of the edm4hep data format~\cite{Gaede:2021izq}, in lieu of the currently used LCIO data format~\cite{Gaede:2003ip}, will ease and enhance end-stage analysis by enabling better compatibility with modern tools such as RDataFrame~\cite{Piparo:2019xdy} and the scientific Python ecosystem via coffea~\cite{CMS:2020kpn}. A broader common effort will additionally facilitate the integration and use of more specialized event generators and beam background tools, for more precise and realistic studies.

The specific software development tasks that are most critical are those needed to address  crucial detector performance concerns. The first is the establishment of 4D tracking algorithms using ACTS~\cite{Ai:2021ghi} within Key4hep. The muon collider group can play a leading role here, given that it has already established tracking algorithms using ACTS. A second effort will focus on re-establishing the necessary expertise for particle flow reconstruction, most likely using Pandora~\cite{THOMSON200925}. Facilitated by Key4hep, similar efforts have been started by other communities, e.g. for the FCC. It is highly likely that dedicated algorithms have to be developed for the muon collider.

Computing resources are already moving away from being CPU-dominated. Instead, many new HPCs provide large fractions of their computing power via GPUs or other coprocessors. These changes will necessarily impact HEP data processing. The LHC experiments are already using multithreading in production, and some compute-intensive workloads have been partially ported to run on GPUs. Similar efforts will be necessary for the muon collider in order to deal with comparable detector occupancies and, more importantly, the BIB. The muon collider community should take advantage of the development work from LHC experiments as a starting point, given the limited available person power. Nevertheless, initiating development specific to the muon collider case as early as possible is necessary to discover issues and to inform requirements for the DAQ system or on-detector reconstruction. In particular, this will require adding functionality to the Key4hep stack: asynchronous, non-blocking coprocessor calls for optimal exploitation of heterogeneous computing resources~\cite{Bocci:2020olh}. GPU simulation engines are currently being developed for the HL-LHC~\cite{Amadio:2022mqi,Johnson:2024qqt} and may prove particularly useful for the BIB simulation.

Beyond traditional algorithms, many tasks, from simulation to on-detector readout to reconstruction, may be ideally facilitated by machine learning approaches. Such techniques are growing in usage at LHC experiments, primarily for object classification~\cite{Qu:2019gqs}, but generative ML for simulation is also starting to be deployed and will be used more heavily in the HL-LHC era~\cite{ATLAS:2021pzo,Krause:2024avx}. ML-based on-detector readout and filtering~\cite{Dickinson:2023yes} may be one of the only viable BIB rejection schemes. These ML algorithms will need to be deployed on GPUs or other coprocessors to maximize the inference speed; as these algorithms are naturally suited to parallel processing, they represent an ideal way to take advantage of new computing resources. Inference can be accelerated even further by processing data in large batches, which is a different paradigm than the event-based processing used in key4hep and other collider experiment software frameworks. Batching in production workflows can be achieved with minimal disruption via the inference as a service approach, which is already being explored and even adopted by several current experiments including CMS, ATLAS, DUNE, IceCube, and LIGO~\cite{Duarte:2019fta,Krupa:2020bwg,Wang:2020fjr,Rankin:2020usv,Gunny:2021gne,Cai:2023ldc,Savard:2023wwi,CMS:2024twn}.

\chapter{Magnets R\&D}
\label{3:rd:sec:mag}

\section{Introduction}\label{Sec:MagnetR&D_Intro}
As we have outlined in the evaluation of the magnet study performed so far in Chapter~\ref{1:tech:sec:mag}, the Muon Collider poses outstanding challenges to magnet technology. These have been summarized in development targets of Table~\ref{tab:MagnetR&D_tab1}. We are aware that this is only a simplified view, it lacks many detail of the actual machine configuration, and aspects such as field quality, still under discussion. It is nonetheless useful at this stage to give a compact view of the exceptional envelope of performance to be achieved, and direct the necessary development. 

\begin{table}[!h]
    \begin{centering}
    \small	
    \setlength{\tabcolsep}{3pt}
    \centerline{
    \begin{tabular}{|rc|c|c|c|cc|cccc|}
        \hline
         \multicolumn{2}{|c|}{Complex in the} & Target & 6D & Final & \multicolumn{2}{c|}{Rapid} & \multicolumn{4}{c|}{Collider ring} \\
         \multicolumn{2}{|c|}{Muon Collider} & to & cooling & cooling & \multicolumn{2}{c|}{cycling} & \multicolumn{4}{c|}{} \\
         \multicolumn{2}{|c|}{} & capture & & & \multicolumn{2}{c|}{synchrotron} & \multicolumn{4}{c|}{} \\
         \multicolumn{2}{|c|}{}& & & & \multicolumn{2}{c|}{} & 3 TeV & 3 TeV & 10 TeV & \\
        \hline
        Magnet type& & Solenoid & Solenoid & Solenoid & Solenoid & Dipole & Dipole & Dipole & Dipole & Quad. \\
        Material options& & HTS & HTS/LTS$^{(2)}$ & HTS & HTS & (NC) & Nb-Ti & Nb3Sn & HTS & HTS \\
        Aperture&(mm) & 1400 & 60...800$^{(3)}$ & 50 & 30$\times$100 & 30$\times$100 & 160 & 160 & 140 & 140 \\
        Length &(m) & 19 & 0.08...0.3$^{(3)}$ & 0.5...1$^{(4)}$ & 5 & 2 & 4...6$^{(4)}$ & 4...6$^{(4)}$ & 4...6$^{(4)}$ & 3...9$^{(4)}$ \\
        N. of magnets& & 20 & $2 \times 3030$ & 20 & 7000$^{(6)}$ & 3000$^{(6)}$ & 1250$^{(8)}$ & 1250$^{(8)}$ & 1250$^{(8)}$ & 28 \\
        Bore field or & (T) & \multirow{2}{*}{20} &  \multirow{2}{*}{2.6...17.9$^{(3)}$} &  \multirow{2}{*}{$>$40} &  \multirow{2}{*}{$\pm$1.8$^{(5)}$} &  \multirow{2}{*}{10} &  \multirow{2}{*}{5} &  \multirow{2}{*}{11} &  \multirow{2}{*}{14} &  \multirow{2}{*}{300} \\
        gradient & (T/m) & & & & & & & & & \\
        Ramp-rate& (T/s) & SS & SS & SS & 3320...810$^{(7)}$ & SS & SS & SS & SS & SS \\
        Stored energy& (MJ) & 1400 & 5...75 & 4 & 0.03 & 3.4 & 5 & 20 & 24 & 60 \\
        Heat load& (W/m) & $2^{(1)}$ & TBD & TBD & 1200 & 5 & 5 & 5 & 10 & 10 \\
        Radiation dose& (MGy) & 80 & TBD & TBD & TBD & TBD & 30 & 30 & 30 & 30 \\
        Op. temperature& (K) & 20 & 20 & 4.5 & 300 & 20 & 4.5 & 4.5 & 20 & 4.5...20 \\
        \hline
    \end{tabular}
}
    \end{centering}
\begin{minipage}{\textwidth}
\vspace{0.2cm}
\footnotesize  Notes: $^{(1)}$ Intended as linear heat load along the conductor wound in the solenoid. The total heat load in the target, decay and capture solenoid is approximately 4 kW. $^{(2)}$ Superconducting material and operating temperature to be selected as a function of the system cost. Present baseline study is oriented towards HTS at 20 K. $^{(3)}$ The range indicated covers the several solenoid magnet types that are required for the cooling cells. Extreme values are generally not to be taken simultaneously.  $^{(4)}$ Specific optics is being studied, the length range indicated is representative.  $^{(5)}$ Rapid Cycled Synchrotrons require uni-polar swing, from zero to peak field. Hybrid Cycled Synchrotrons require bi-polar swing, from negative to positive peak field. $^{(6)}$ Considering the CERN implementation (SPS+LHC tunnels).  $^{(7)}$ Required ramp-rate decreases from the first to the last synchrotron in the acceleration chain.  $^{(8)}$ Considering a collider of the final size (approximately 10 km length).
\end{minipage}
    \caption{Summary of magnet parameters for each complex in the muon collider. }
    \label{tab:MagnetR&D_tab1}
\end{table}

We see already at first sight that key research and development objectives span a very broad range. They go from achieving very high magnetic field strengths, up to and above 40~T in solenoids and 14~T in dipoles, to managing stored energies exceeding 300~MJ in a single magnet, mitigating heat loads from nuclear interaction of muon decay at levels of several W/m, and ensuring radiation resistance up to 80 MGy local dose. Overcoming these extraordinary challenges requires the innovative integration of high-temperature superconductor (HTS) technology, optimized for efficient operation at cryogenic temperatures up to 20 K, and combined in a compact design to reduce capital expenditure.

A complete overview of the state of the art in high field magnet technology, solenoids, accelerator dipoles and quadrupoles, and associated powering systems is out of the scope of this proposal, and discussed elsewhere~\cite{Bottura:2025kuj,Bottura:2025olk}. We use here the Technology Readiness Level (TRL) as a global indication of maturity for the deployment in an accelerator project, and quantify the technology gap in terms of the difference to the required TRL. We require a TRL of at least 6 to decide on construction (technology demonstrated in relevant environment). This is the level to be set as goal for the R\&D program.
 
\section{Gap Analysis} \label{Sec:MagnetR&D_Gap}
Contrasting the demands and challenges listed in Table~\ref{tab:MagnetR&D_tab1} to the state-of-the-art in magnets relevant to the Muon Collider, we see that the development faces several technical and operational challenges, creating significant gaps that must be addressed by focused R\&D. 

All-HTS solenoids with large bore and high field, such as the concept selected for the target, decay and capture channel, are close to suitable TRL. While LTS technology has reached appropriate TRL of 8, the required novel HTS concepts still need to progress up from TRL 5 to 6 by at least 1 TRL step, to a secured TRL 6.

In the case of the all-HTS UHF solenoids with NI winding technology as considered for the final cooling, the required advances are more substantial. In this case, R\&D and demonstration is needed to advance from the present values of 3 to 4 to a minimum of TRL 6, by 2 to 3 TRL steps.

The solenoids of the 6D cooling are in the intermediate range of field and bore dimension, with stored energies in the range of several MJ. In this case it is not evident to compare to state-of-the-art, as no such large size all-HTS magnet has been built. The winding technology being considered for these solenoids, partially insulated, is comparable to that of the UHF experiments, which are at TRL of 3 to 4. In addition, structural and quench protection issues are made more severe because of the large dimensions, electromagnetic forces and stored energy. We hence consider that also in this case a progression to a secured TRL of 6 will require an increase of TRL by 2 to 3 steps.

Turning now to accelerator magnets, in the case of Nb-Ti dipoles and quadrupoles they are more-or-less ready for industrial production, so no gap can be identified here. For Nb3Sn, the required performance aligns rather well with the achievements of HL-LHC, and there is no significant TRL gap from the point of view of performance that would motivate large R\&D investments. At the same time, a large-scale production will require full scale prototyping, effective industrialization and cost optimization.

With HTS accelerator magnets, on the other hand, we are only at the beginning of developments and many issues need to be addressed, resolved and demonstrated. A significant increase in TRL is required, by 3 units to reach a secured TRL of 6.

The proposed high dB/dt (3300 T/s) resistive magnets with 1.8 T peak field and 11 kA peak current for the muon rapid cycled synchrotrons share characteristics with existing accelerator technologies but push performance beyond current implementations in a number of aspects. Although the single components, magnets and power converters, have a relatively high TRL of 6 to 8, the challenge is truly the integrated system. Lacking a system test, which is part of the proposed R\&D, the technology readiness level (TRL) for the system of magnets and powering of the RCS is an estimated TRL of 4 to 5. 

\begin{figure}
\centering
\includegraphics[width=0.65\textwidth]{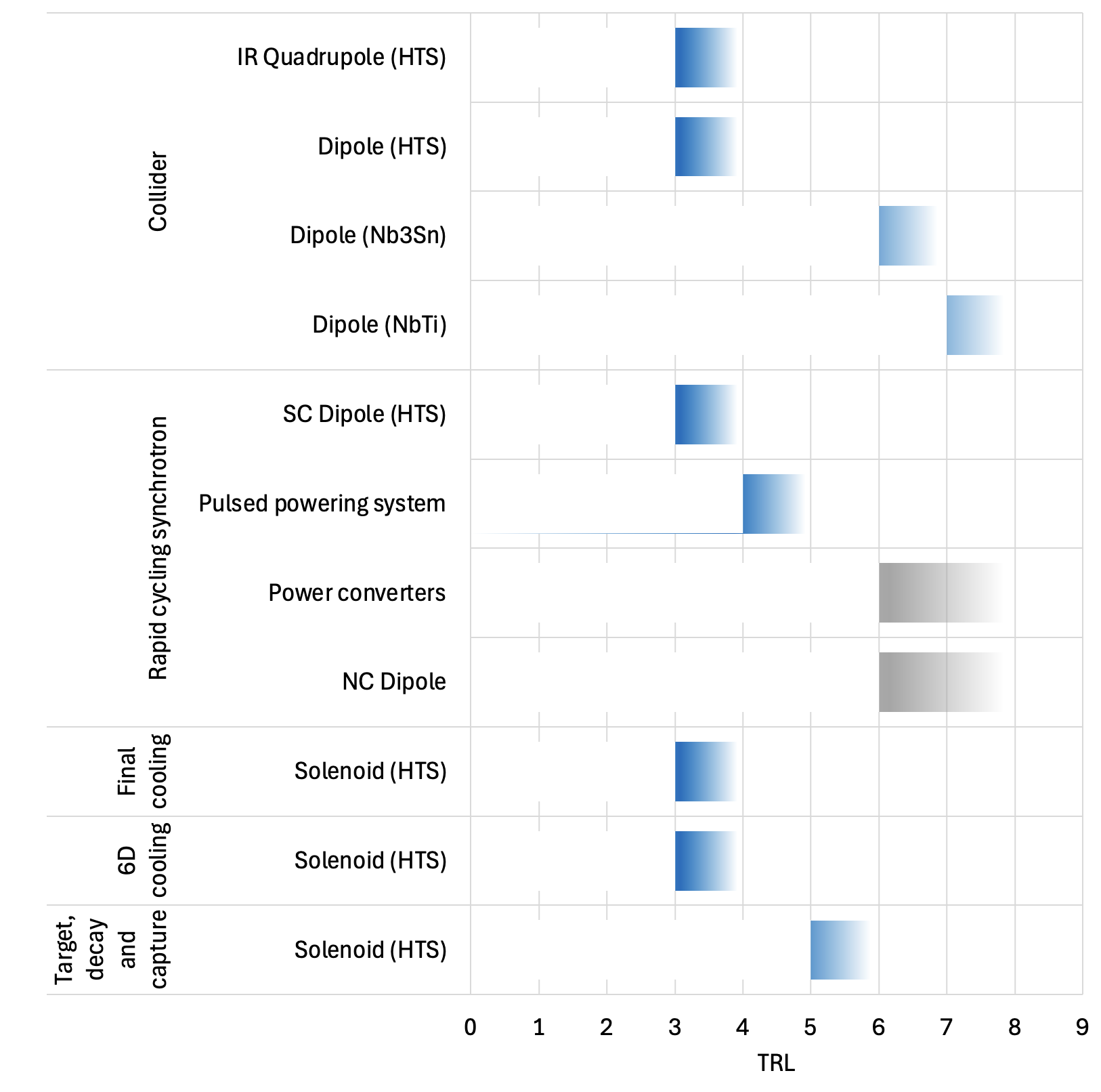}
\caption{Summary of estimated TRL for the magnets of the muon collider. Normal Conducting (NC) dipoles and power converters for the rapid cycling synchrotrons are quoted separately (grey area) and as a system. A secured TRL of 6 is needed for a decision of construction.}
\label{fig:MagnetR&D_fig1}
\end{figure}

In summary, we report in Figure~\ref{fig:MagnetR&D_fig1} a graphical representation of the present TRL for the various magnet and powering systems of a muon collider. We can use this representation to define the scale and type of R\&D required, whether still at the level of small-scale demonstrators and models, when the estimated TRL is at the level of 4 to 5, or rather towards full-scale prototypes, for values of TRL 6 and higher. The proposed R\&D activities are described in the next section.

\section{Technology Driven R\&D} 
\label{Sec:MagnetR&D_Tech}

We have built the proposal of the R\&D necessary to advance the TRL of the muon collider magnet concepts around eight technology milestones (TMs). The activities in the TM’s address the challenges identified, and are intended to fill the gap identified earlier in this proposal, from present state-of-the-art to a secured TRL 6. In practice, each TM is associated with a magnet demonstrator, except for the RCS-String which is a system test. Achieving the TM corresponds to the construction and successful test of the associated magnet demonstrator or system.  As described below, the demonstrators span a broad range of configurations, operational temperatures, and performance targets, covering all unique aspect of the collider’s infrastructure. In addition to the eight TM’s, one R\&D activity is proposed to cover material testing necessary and relevant to the specific needs of the magnets for the muon collider.

\begin{figure}
\centering
\includegraphics[width=\textwidth]{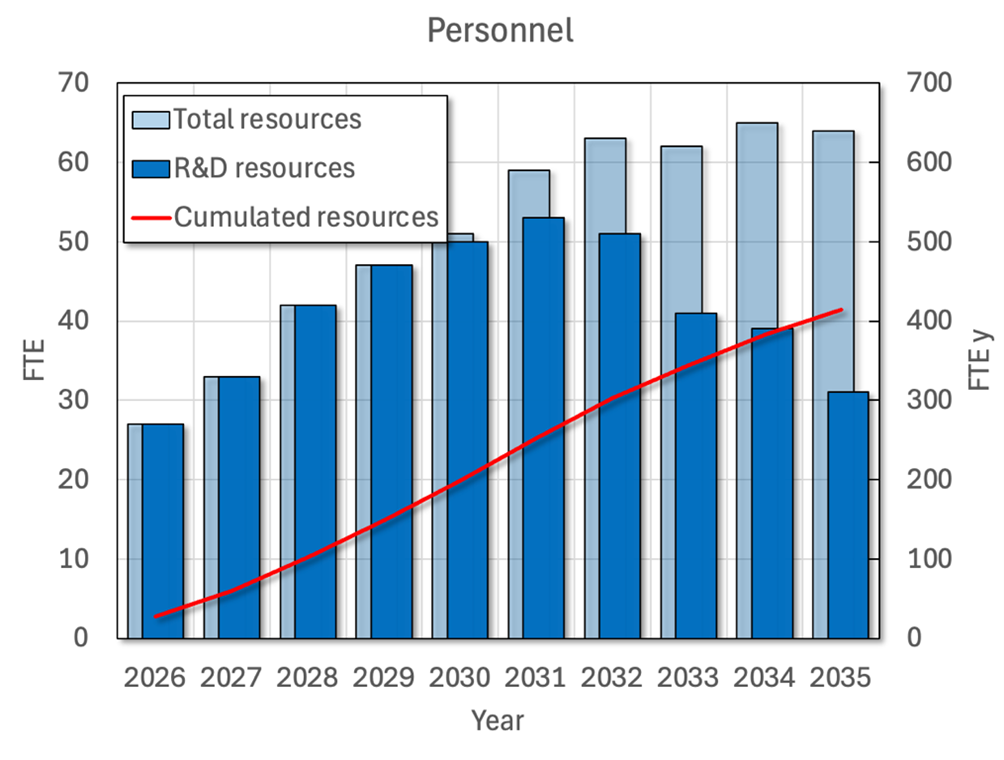}
\caption{Time profile of the personnel resources necessary to conduct the R\&D program proposed in this note, reported both as yearly values (columns, left axis), as well as cumulated over the years of activity (line, right axis). The second set of columns (shaded) reports the estimated total personnel resources that would be needed to proceed to prototyping and construction in the period after 2031.}
\label{fig:MagnetR&D_fig2}
\end{figure}

\begin{figure}
\centering
\includegraphics[width=\textwidth]{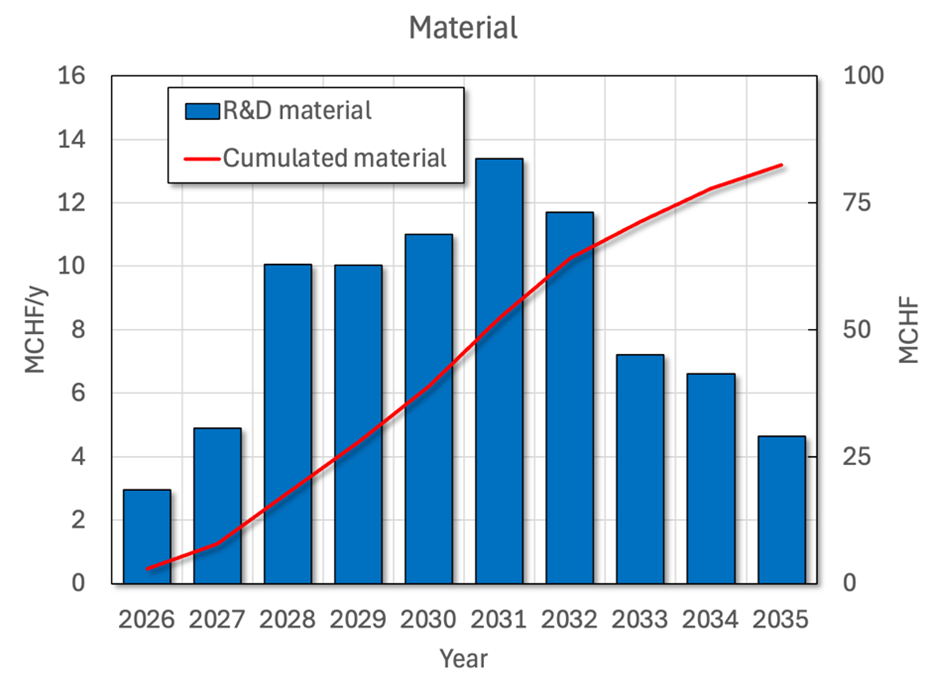}
\caption{Time profile of the material cost necessary to conduct the R\&D program proposed in this note, reported both as yearly values (columns, left axis), as well as cumulated over the years of activity (line, right axis).}
\label{fig:MagnetR&D_fig3}
\end{figure}

A summary of the personnel resources and material cost is reported in Table~\ref{tab:magnetRDsummary} and Figures~\ref{fig:MagnetR&D_fig2} and~\ref{fig:MagnetR&D_fig3}, where we report both the yearly values (in FTE and MCHF/y), as well as the cumulated values (in FTE y and MCHF). In the case of personnel (Figure~\ref{fig:MagnetR&D_fig2}), we also report the estimate of the effort that would be required to progress to the next phase of prototyping and construction, should the project proceed in this direction.

\begin{table}[h!]
\small
\centering
\begin{tabular}{|l|c|c|c|c|c|c|c|c|c|c|}
\hline
\rowcolor{cornflowerblue!80}
\textbf{Year}&\textbf{I}&\textbf{II}&\textbf{III}&\textbf{IV}&\textbf{V}&\textbf{VI}&\textbf{VII}&\textbf{VIII}&\textbf{IX}&\textbf{X} \\
\hline
\rowcolor{cornflowerblue!30}
\multicolumn{11}{|l|}{\textbf{Target solenoid demonstrator (20@20)}} \\
\hline
Staff & 0.6 & 0.9 & 0.9 & 1.5 & 3 & 4 & 3.5 & 2.1 &  &  \\ 
Post doc & 0.8 & 1.2 & 1.2 & 2 & 1.8 & 2.4 & 2.1 & 0.6 &  &  \\ 
Student & 0.6 & 0.9 & 0.9 & 1.5 & 1.2 & 1.6 & 1.4 & 0.3 &  &  \\ 
Material (kCHF) & 1000 & 2000 & 5000 & 4000 & 5000 & 7000 & 5000 & 1000 &  &  \\ 
\hline
\rowcolor{cornflowerblue!30}
\multicolumn{11}{|l|}{\textbf{Solenoid Integration Demonstrator for 6D cooling cell}} \\
\hline
Staff & 0.9 & 2.1 & 2.4 & 2.4 & 2.1 & 2.5 & 2 &  &  &  \\ 
Post doc & 1.2 & 2.8 & 3.2 & 3.2 & 2.8 & 1.5 & 1.2 &  &  &  \\ 
Student & 0.9 & 2.1 & 2.4 & 2.4 & 2.1 & 1 & 0.8 &  &  &  \\ 
Material (kCHF) & 400 & 900 & 1400 & 1700 & 1200 & 1000 & 500 &  &  &  \\ 
\hline
\rowcolor{cornflowerblue!30}
\multicolumn{11}{|l|}{\textbf{Final cooling UHF solenoid demonstrator (UHF-Demo)}} \\
\hline
Staff & 1.2 & 1.2 & 1.8 & 1.8 & 1.8 & 2.1 & 2.1 & 3.5 & 2.5 &  \\ 
Post doc & 1.6 & 1.6 & 2.4 & 2.4 & 2.4 & 2.8 & 2.8 & 2.1 & 1.5 &  \\ 
Student & 1.2 & 1.2 & 1.8 & 1.8 & 1.8 & 2.1 & 2.1 & 1.4 & 1 &  \\ 
Material (kCHF) & 300 & 300 & 500 & 500 & 500 & 750 & 750 & 1000 & 1000 &  \\ 
\hline
\rowcolor{cornflowerblue!30}
\multicolumn{11}{|l|}{\textbf{RCS magnet string and power systems (RCS-String)}} \\
\hline
Staff & 1.4 & 1.4 & 2.8 & 3.6 & 3.6 & 3 & 1 &  &  &  \\ 
Post doc & 0.4 & 0.4 & 0.8 & 0.4 & 0.4 & 0 & 0 &  &  &  \\ 
Student & 0.2 & 0.2 & 0.4 & 0 & 0 & 0 & 0 &  &  &  \\ 
Material (kCHF) & 250 & 300 & 950 & 1500 & 1500 & 1000 & 500 &  &  &  \\ 
\hline
\rowcolor{cornflowerblue!30}
\multicolumn{11}{|l|}{\textbf{Wide-aperture, steady state Nb3Sn dipole (MBHY)}} \\
\hline
Staff & 2 & 2 & 3 & 3.5 & 3.5 & 3.5 & 6.3 & 6.3 & 6.3 & 6.3 \\ 
Post doc & 1.2 & 1.2 & 1.8 & 2.1 & 2.1 & 2.1 & 1.8 & 1.8 & 1.8 & 1.8 \\ 
Student & 0.8 & 0.8 & 1.2 & 1.4 & 1.4 & 1.4 & 0.9 & 0.9 & 0.9 & 0.9 \\ 
Material (kCHF) & 300 & 500 & 750 & 845 & 750 & 750 & 1750 & 2000 & 2000 & 1500 \\ 
\hline
\rowcolor{cornflowerblue!30}
\multicolumn{11}{|l|}{\textbf{Rectangular aperture HTS dipole (MBHTS)}} \\
\hline
Staff & 1.6 & 1.6 & 2.4 & 2.4 & 2.4 & 3.2 & 4 & 3.5 & 4.9 & 2.8 \\ 
Post doc & 1.2 & 1.2 & 1.8 & 1.8 & 1.8 & 2.4 & 2.4 & 2.1 & 1.4 & 0.8 \\ 
Student & 1.2 & 1.2 & 1.8 & 1.8 & 1.8 & 2.4 & 1.6 & 1.4 & 0.7 & 0.4 \\ 
Material (kCHF) & 200 & 200 & 500 & 500 & 850 & 1500 & 1500 & 1250 & 1250 & 500 \\ 
\hline
\rowcolor{cornflowerblue!30}
\multicolumn{11}{|l|}{\textbf{Wide aperture HTS dipole (MBHTSY)}} \\
\hline
Staff & 2 & 2.4 & 2.4 & 3.2 & 4 & 4 & 4 & 4 & 4.5 & 6.3 \\ 
Post doc & 1.5 & 1.8 & 1.8 & 2.4 & 2.4 & 2.4 & 2.4 & 2.4 & 2.7 & 1.8 \\ 
Student & 1.5 & 1.8 & 1.8 & 2.4 & 1.6 & 1.6 & 1.6 & 1.6 & 1.8 & 0.9 \\ 
Material (kCHF) & 300 & 500 & 750 & 800 & 800 & 800 & 800 & 800 & 1100 & 1250 \\ 
\hline
\rowcolor{cornflowerblue!30}
\multicolumn{11}{|l|}{\textbf{Wide aperture HTS IR quadrupole (MQHTSY)}} \\
\hline
Staff & 0 & 0 & 0 & 0 & 1.5 & 2 & 2 & 2 & 3 & 4.2 \\ 
Post doc & 0 & 0 & 0 & 0 & 0.9 & 1.2 & 1.2 & 1.2 & 1.8 & 1.2 \\ 
Student & 0 & 0 & 0 & 0 & 0.6 & 0.8 & 0.8 & 0.8 & 1.2 & 0.6 \\ 
Material (kCHF) & 0 & 0 & 0 & 0 & 200 & 200 & 500 & 750 & 850 & 1000 \\ 
\hline
\rowcolor{cornflowerblue!30}
\multicolumn{11}{|l|}{\textbf{Muon Collider Magnets - Materials and methods R\&D in support of magnet demonstrators}} \\
\hline
Staff & 1.2 & 1.2 & 1.2 & 1.2 & 1.2 & 1.2 & 2.1 & 2.1 & 2.1 & 2.1 \\ 
Post doc & 0.9 & 0.9 & 0.9 & 0.9 & 0.9 & 0.9 & 0.6 & 0.6 & 0.6 & 0.6 \\ 
Student & 0.9 & 0.9 & 0.9 & 0.9 & 0.9 & 0.9 & 0.3 & 0.3 & 0.3 & 0.3 \\ 
Material (kCHF) & 200 & 200 & 200 & 200 & 200 & 400 & 400 & 400 & 400 & 400 \\ 
\hline
\hline
\rowcolor{cornflowerblue!30}
\multicolumn{11}{|l|}{\textbf{TOTALS}} \\
\hline
Material (MCHF) & 3.0 & 4.9 & 10.1 & 10.0 & 11.0 & 13.4 & 11.7 & 7.2 & 6.6 & 4.7 \\ 
FTE & 23.3 & 28.4 & 36.4 & 40.9 & 44.3 & 47.1 & 46.2 & 37.7 & 36.1 & 29.4 \\ 
\hline
\end{tabular}
\caption{Summary of resources for magnet R\&D, divided by area, over a period of ten years. Personnel is given as average Full Time Equivalent (FTE), where students count as 50\% FTE. Material \& Supplies (M\&S) are given in thousands of CHF (kCHF). The total is the sum of the nine areas.
}
\label{tab:magnetRDsummary}
\end{table}

The total personnel required for the proposed magnet R\&D program, over the ten years period, is 414 FTE y. This is rather evenly distributed, reaching 199~FTE y after five years, and a maximum just above 50 FTE in the middle of the ten years period. To be noted that towards the second half of the ten years R\&D, proficient personnel could transfer to prototyping and construction, seamlessly absorbed in the 60 to 65 FTE that would be needed in total (including residual R\&D) during this phase.

The total material costs over ten years are estimated at 82.5 MCHF (Figure~\ref{fig:MagnetR&D_fig3}), reaching 39 MCHF after 5 years, nearly half. The peak yearly expenditure, in the middle of the ten years period, is at the level of 10 to 13 MCHF/year. Note that in this case the drop in yearly expenditure is due to the fact that a transition to the following phase of prototyping and construction is not added.

\begin{figure}[!h]
\includegraphics[width=\textwidth]{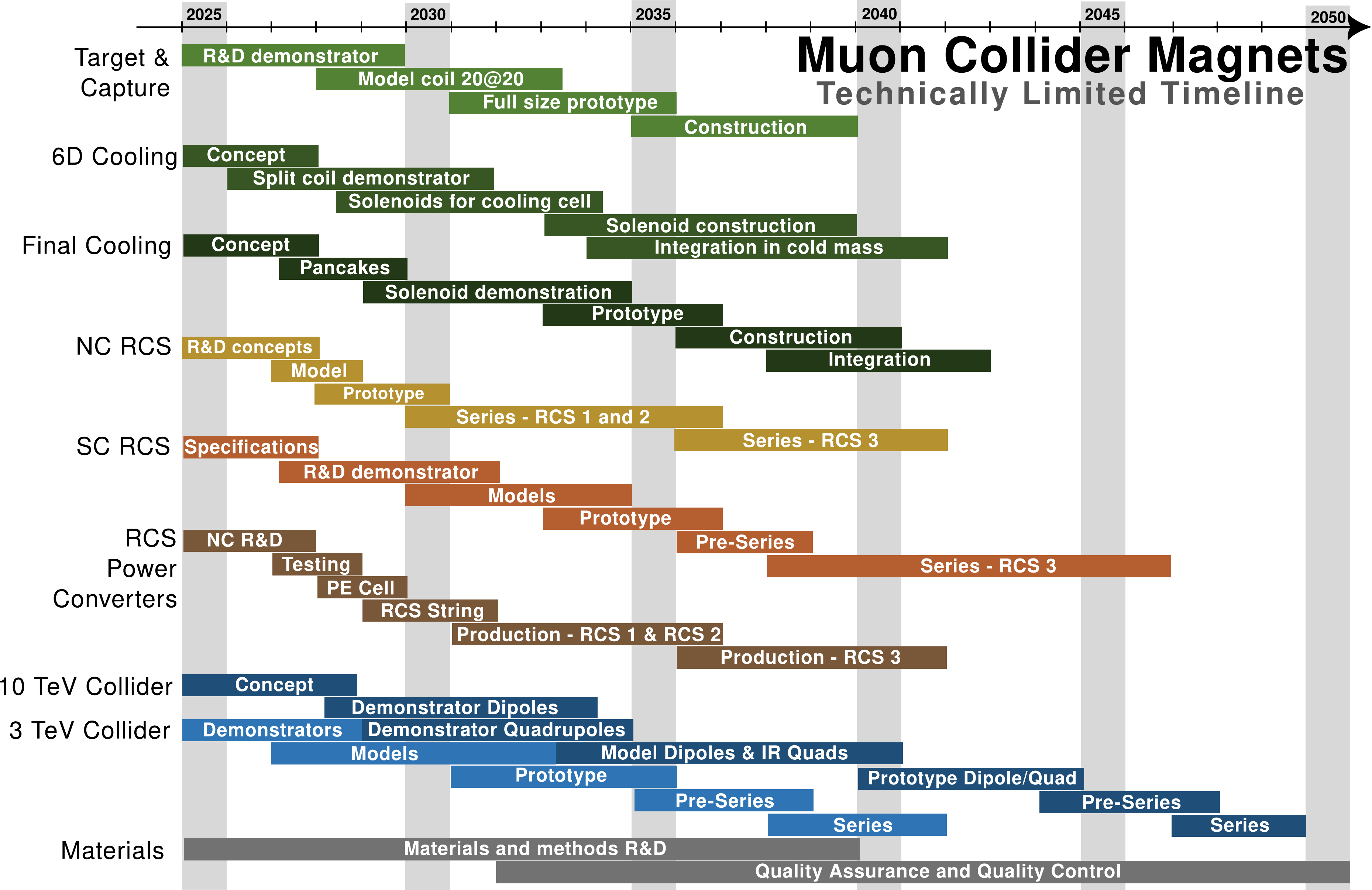}
\caption{Technically limited timeline for the Muon Cooling Magnet programme.}
\label{fig:magnet_schedule}
\end{figure}

The envisaged timelines for the eight TM is illustrated in Figure~\ref{fig:magnet_schedule}.

Below we give details on the R\&D activities that are part of the proposal.

\subsection*{20 T at 20 K target solenoid model coil (20@20)}

This technology milestones consists in developing conductor, winding and magnet technology suitable for a target solenoid. The main objective is to demonstrate stable operation and quench protection at the design point. The TM is associated with a model coil generating a bore field of 20 T, and operating at a temperature of 20 K, hence the acronym “20@20”. The estimate time to reach this milestone is 8 years, for a total cost of approximately 30 MCHF, and personnel needs of 37 FTEy.

\subsection*{Split solenoid integration demonstrator for the 6D cooling cell (SOLID)}

technology milestone consists in demonstrating operating field performance in geometrical and operating conditions relevant to the 6D muon cooling. The main objective is to demonstrate successful integration. The TM is associated with a HTS split solenoid, named SOLID, with target field on axis of 7 T, a free bore of 400 mm, a split gap of 350 mm, operating at 20 K. The estimate time to reach this TM is 7 years, for a total cost of approximately 7 MCHF, and personnel needs of 42 FTEy.

\subsection*{Final cooling UHF solenoid demonstrator (UHF-Demo)}

This technology milestone consists in demonstrating operating field performance relevant to the final muon cooling solenoid. The main objective is field level attained. The TM is associated with a prototype of the HTS final cooling solenoid, named UHF-DEMO, targeting a field of 40 T in a 50 mm bore, and total length of 150 mm. The estimate time to reach this TM is 9 years, for a total cost of approximately 5.6 MCHF, and personnel needs of 52 FTEy.

\subsection*{RCS magnet String and power systems (RCS-String)}

This technology milestone consists in demonstrating operation of a fast pulsed, energy recovery resistive magnet circuit, including the powering and energy storage infrastructure of the type designed for the rapid cycled synchrotrons. The main objectives are field tracking, field quality and energy efficiency. The TM is associated with a string made of resistive dipole magnets, generating a nominal field of 1.8 T in a 30×100 mm aperture. Four such magnets, each 5 m long, are powered in series by a fast pulsed power converter that includes energy storage in capacitor banks that store a total energy of 150 kJ. The target is to reach 4 kT/s. The estimate time to reach this TM is 7 years, for a total cost of approximately 6 MCHF, and personnel needs of 20 FTEy. 

\subsection*{Wide-aperture, steady state Nb3Sn dipole (MBHY)}

This technology milestone consists in demonstrating dipole performance at the level required for the muon collider ring, based on LTS (Nb3Sn) technology. The main objective is to demonstrate the combination of field, aperture, training memory and field quality in conditions relevant to the collider operation. TM is associated with a full-size prototype of the Nb3Sn collider dipole, named MBHY, targeting a field of 11 T in a 160 mm bore, and total length of 5 m. The operating point is in the range of 4.5 K. The estimate time to reach this TM is 11 years, marginally beyond the horizon of this proposal, for a total cost of approximately 11 MCHF, and personnel needs of 71 FTEy.

\subsection*{Rectangular aperture HTS dipole (MBHTS)}

This technology milestone consists in demonstrating dipole performance at the level required for the muon accelerator hybrid cycled synchrotron ring, based on HTS (REBCO) technology. The main objective is to demonstrate the combination of field, aperture and field quality in conditions relevant to the accelerator operation. The TM is associated with a model of the HTS accelerator dipole, named MBHTS, targeting a field of 10 T in a rectangular aperture of 100 mm width by 30 mm height, and total length of 1 m. The operating point is in the range of 20 K. The estimate time to reach this TM is 10 years, for a total cost of approximately 8.3 MCHF, and personnel needs of 60 FTEy. 

\subsection*{Wide aperture HTS dipole (MBHTSY)}

This technology milestone consists in demonstrating dipole performance at the level required for the muon collider ring, based on HTS (REBCO) technology. The main objective is to demonstrate the field performance, in combination with a wide aperture, including field quality and stability in conditions relevant to the collider operation. The TM is associated with a model of the HTS wide-aperture collider dipole, named MBHTSY, targeting a field of 14 T in an aperture of 140 mm, and total length of 1 m. The operating point is in the range of 20 K. The estimate time to reach this TM is 16 years, i.e. beyond the time span considered for the evaluation of resources. The total cost is estimated at 15.8 MCHF, and personnel needs of 126 FTEy. Over the period of interest in this proposal, i.e. ten years, the total cost is estimated at 7.9 MCHF, and personnel needs of 75 FTEy. 

\subsection*{Wide aperture HTS IR quadrupole (MQHTSY)}

This technology milestone consists in demonstrating quadrupole performance at the level required for the muon collider interaction region, based on HTS (REBCO) technology. The main objective is to demonstrate that the gradient and aperture can be achieved, possibly operating at lower temperature than the arc to increase operating margin, as collider luminosity depends critically on the performance of the IR quadrupoles. TM is associated with a model of a HTS wide-aperture interaction region quadrupole, named MQHTSY, targeting a gradient of 300 T/m in an aperture of 140 mm, and total length of 1 m. The operating point is up to 20 K, although lower operating temperature will be explored if it allows reaching the desired performance. The estimate time to reach this TM is 16 years, i.e. beyond the time span considered for the evaluation of resources. The total cost is estimated at 8.8 MCHF, and personnel needs of 60 FTEy, assuming that much of the technology will be shared with the development of the wide aperture HTS dipoles MBHTSY. Over the period of interest in this proposal, i.e. ten years, the total cost is estimated at 3.5 MCHF, and personnel needs of 27 FTEy. 

\subsection*{Materials and methods R\&D}

This line of activity comprises material testing as well as the development of design and manufacturing methods relevant to the magnets for a muon collider. The objective is to support development and testing that are typically shared among several R\&D in the proposal, and specific to the developments requested to achieve the TM’s in this proposal. The material testing and methods development provisionally included in this activity are:

\begin{itemize}
    \item High-field measurement of transport properties of REBCO conductors;
    \item Micrography, Micro-structure and mechanical properties of REBCO conductors and winding;
    \item Radiation effects in REBCO conductors;
    \item · Tailored experiments to establish design rules for HTS magnets, e.g. allowable hot-spot temperature, or allowable peak stress and strain;
Multi-physics modeling of transient electro-magnetics, mechanics and thermal fields in HTS magnets, relevant to the electro-mechanical design, operation and quench protection of NI HTS magnets.
\end{itemize}

The resources estimate for this activity over the reference period of 10 years is a total of 3 MCHF and 30 FTEy. 

\section{Synergies and Collaborations} \label{Sec:MagnetR&D_Synergies}
\begin{figure}
\centering
\includegraphics[width=\textwidth]{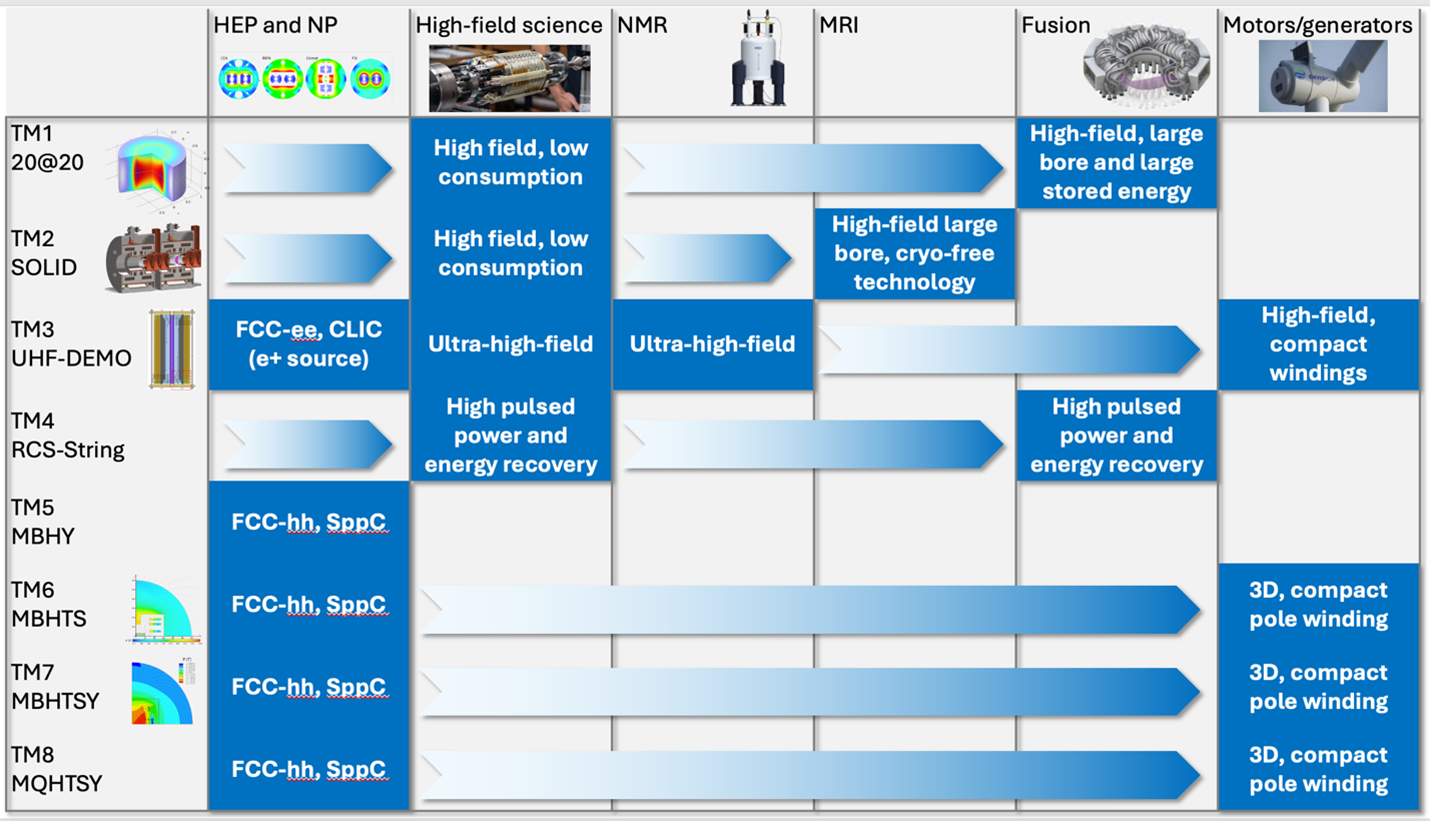}
\caption{Schematic representation of the impact of the R\&D driven by the technology milestones defined in this proposal on other programmes in HEP and other fields of application. The specific technologies of interest are indicated in the highlighted boxes.}
\label{fig:MagnetR&D_fig4}
\end{figure}

While fully relevant and specifically tailored to the muon collider, the proposed R\&D program would unfold within the scope of wider R\&D activities with connections and implications to other HEP R\&D, as well as R\&D in other fields of magnet technology and applied superconductivity. We have identified potential synergies and collaborations, as well as the opportunities for co-funding from sources other than the Muon Collider activities. We give in Figure~\ref{fig:MagnetR&D_fig4} a synoptic view of our evaluation of the expected impact. The representation in Figure~\ref{fig:MagnetR&D_fig4} is purely graphical, but suggestive, demonstrating that all proposed TM’s (rows) have a connection to other programs in HEP and other fields of application (columns).

\chapter{Accelerator R\&D}
\label{4:sec:acc}

\section{Accelerator design}
\subsection{Proton complex}
A lot of the R\&D for the proton Complex is not exclusive to the Muon Collider needs and could be developed in parallel with other needs in the accelerator community. Of course, an increased interest and support from the community would make the development more interesting and help to gather funds and, in some cases, speed up the studies or push them to the limits the Muon Collider project currently needs (in terms of parameters). Many facilities would benefit from such R\&D efforts, including SNS, CERN, JPARC, ESS, ISIS and Fermilab, to list some. In the following paragraphs we outline a list of topics that are central to the development, understanding and ultimately the construction of the proton complex for a Muon collider. The topics are separated between front-end studies, accumulation and compression, ring collimation, bunch recombination and code development and will be described in more detail in the text below.

\begin{figure*}[ht!]
    \centering
    \includegraphics[width=0.46\textwidth]{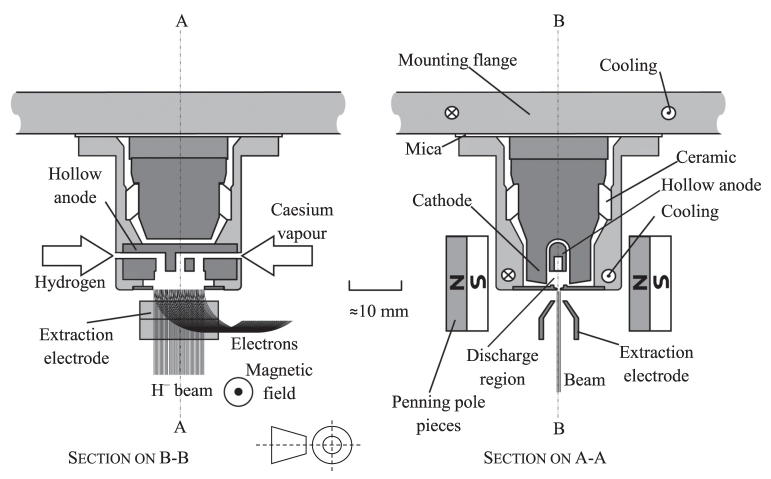}
    \includegraphics[width=0.46\textwidth]{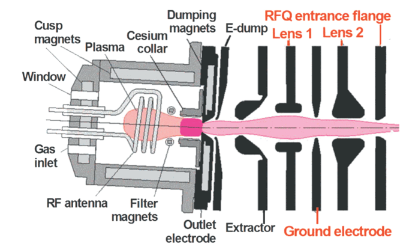}
    \caption{Schematic cross section of a Penning ion source from ISIS (left) and an RF driven source from SNS (right)~\cite{Faircloth_2018}.}
    \label{4:sec:acc:fig:protondriver:source}
\end{figure*}

High brightness, negative hydrogen ion sources are used routinely in large, accelerator-based, user facilities operating worldwide.  Negative hydrogen beams are the preferred means of filling hadronic circular accelerators like colliders and in loading storage rings as well as enabling efficient extraction from cyclotrons (see Figure~\ref{4:sec:acc:fig:protondriver:source}).  Many of these facilities are currently working on various improvement / upgrade projects, both in the near and long term, which are driving further development of their ion source and LEBT (Low Energy Beam Transport) and, in some cases, their overall front end injector system. Presently, the SNS source produces 50-60 mA H- pulses, 1 ms in length at 60 Hz (6\% duty-factor) for maintenance-free periods of ~6 months. One area of research for a Muon Collider would be to develop a MW-scale proton driver to produce a source of muons.  This would likely require an ion source capable of ~80 mA, 2.5-5 ms pulses at 5 Hz (1.25-2.5\% duty-factor), as discussed in Section~\ref{1:acc:sec:proton}. To develop such requirements, it is necessary to further test the large-aperture sources, like the one at SNS, on a test stand using these beam parameters. This would first require extending the RF pulse from 1 to 2.5-5 ms while dropping the RF. Running the source designed for higher repetition rate at 1.25-2.5\% duty-factor will also change thermal conditions which will likely alter the critical Cs distribution within the source effecting performance. Key source parameters, such as extracted beam current, emittance, and lifetime, remain to be experimentally determined, with many unknowns still to be addressed.  With still many unknowns we propose a program in order to demonstrate the feasibility of $H^-$ source capable to serve the front-end of a future Muon Collider:

\begin{itemize}
    \item Choice of source design and technology,
    \item Optimization of the current source design and simulations,
    \item Demonstration of high current (>80 mA) and long pulse extraction (>2 ms), which might involve purchase of materials to perform the experiments, and
    \item Optimization of the source with RFQ input need in mind (regarding emittance and acceptance).
\end{itemize}

RFQ design studies are complementary to source studies since they also have an impact in beam quality and maximum current in the front-end of any high power linac. This effort would cover transmission improvement studies and optimizations, going through the following topics of:
\begin{itemize}
    \item Study of the dependencies of diverse parameters in RFQ design,
    \item Design of an RFQ with high transmission (>98\%) for high current and long pulses for H- beams,
    \item Optimization of RFQ with the source output (from previous topic) and
    \item Creation of a set of specifications (electrical/mechanical/tolerances) needed for such a device.
\end{itemize}

The last item, in the front-end part, is losses driven by $H^-$ stripping, studies and experimental verification. Beam loss in a super-conducting H- linac must be kept under control to prevent heat deposition in the linac’s cavities, irradiation of the linac components, and creation of unacceptably high radiation levels in the linac tunnel. One of the important sources of beam loss, especially in the early linac sections, is the stripping of $H^-$ ions resulting from intra-beam particle collisions. Extensive operational experience with managing this kind of beam loss has been accumulated, in particular, at the SNS. Enabling accurate beam loss prediction in the MC proton driver linac may require development of appropriate analytic approaches and simulation codes and their experimental benchmark at an existing SRF linac such as the SNS.The main outcomes of this effort should be:
\begin{itemize}
    \item Simulation studies of H- stripping and mitigations in high power linacs and
    \item Measurements and benchmarks of those studies in current facilities like SNS, Linac4 and ISIS
\end{itemize}

The next block of R\&D effort is focused on accumulation and compression of high intensity bunches. Accumulating short, intense proton pulses requires charge exchange injection from an $H^-$ injector into a ring. The conventional approach is to pass the $H^-$ beam through a stripping foil. However, such foils are limited in the maximum beam power density that they can tolerate before sublimating, and provide the major source of loss in such accumulation
rings. Both issues are potentially problematic for a muon collider. Laser Assisted Charge Exchange (LACE), under development at the SNS for nearly two decades, is an alternative technology that eliminates the need for a stripping foil. First, $H^-$ atoms are converted to $H0$ by Lorentz stripping in a strong magnetic field, then neutral hydrogen atoms are excited from the ground state to upper levels by a laser, and the remaining electron, now weaker bound, is stripped in a strong magnetic field. All of these steps have been successfully demonstrated by experiments at the SNS. The full injection demonstration system is currently being designed at the SNS. Potential funding sources for its implementation are being explored. Following SNS lead, the main R\&D main focus point for this topic are:
\begin{itemize}
    \item Conceptual design of $H^-$ laser assisted charge exchange injection (LACE) for 5 GeV and 10 GeV and
    \item Demonstration of the principle at SNS accumulator ring.
\end{itemize}

\begin{figure*}[ht!]
    \centering
    \includegraphics[width=0.8\textwidth]{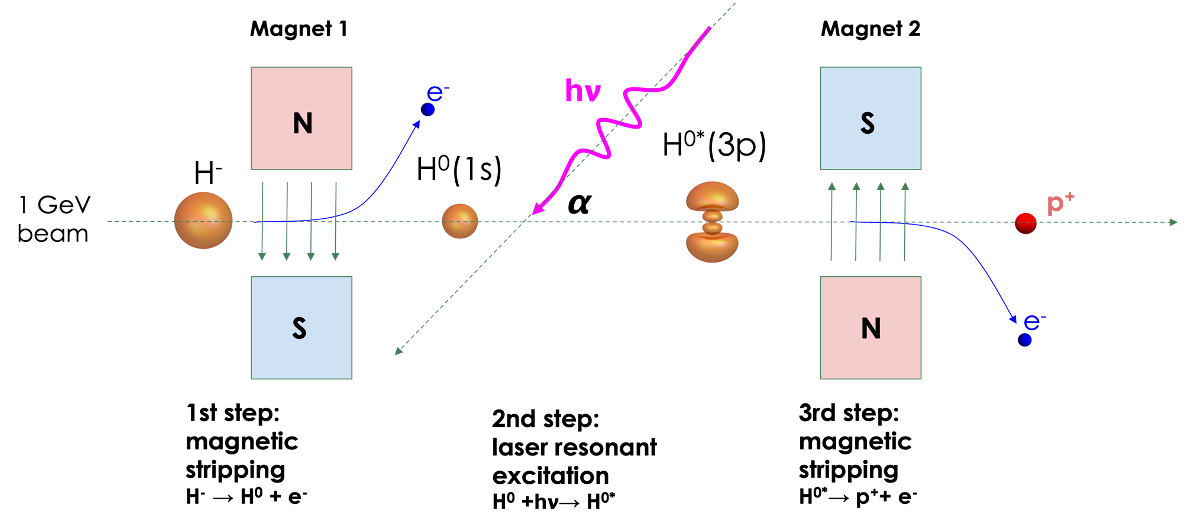}
    \caption{Steps on the laser assisted ionization process for injection at the SNS accumulator ring~\cite{Lace_article1,Lace_article2,Lace_article3}.}
    \label{4:sec:acc:fig:protondriver:lace}
\end{figure*}

There are several questions critical for the MC proton driver parameter choices including limits on the beam intensity due to space charge and bunch compressor design. One of the fundamental limits of a stored beam’s intensity is the space charge tune shift. It must stay below a certain threshold to prevent beam instability and loss.
 
The SNS accumulator ring is currently the world’s highest-power, highest-intensity proton synchrotron reaching $1\times10^{14}$ particles in an about 150 m long bunch and a space-charge tune shift of about -0.15. Some of the proton driver scenarios go beyond anything demonstrated so far in terms of the space-charge tune shift. Therefore, it is important to validate the proton driver parameter choice and bunch compression scheme experimentally.
 
It is during the bunch compression step in the bunch formation process when the space-charge tune shift reaches its maximum. One potential avenue for experimental exploration of the bunch compression limits and the corresponding beam dynamics is demonstration of bunch compression in the SNS accumulator ring. All necessary diagnostics for full bunch characterization is already available. Initial tests can be completed using the existing hardware, in particular, using the ring’s dual harmonic RF system. Barrier-bucket cavities may later be installed if needed for improved performance.
Study and verification of the feasibility of compression is also an important topics for the proton complex. Study of space charge related effects and search of potential issues during the compression of high-intensity bunches should be thoroughly investigated:
\begin{itemize}
    \item Understand in simulations and experimentally the limitations on the compression of high intensity bunches,
    \item Simulation studies of different compression schemes,
    \item Measurements and code benchmarks of bunch rotation in high space charge conditions (possible at SNS and CERN) and
    \item Improvement of the compression scheme based on recombination needs (see next topic).
\end{itemize}

The need to have a multi-bunch solution for compression, dues to high space charge effects, creates the need to recombine the bunches prior to the target. Recombination of high intensity bunches and its feasibility and limitation has the be understood in order to allow the development of the lattice and surrounding diagnostics to verify it. A study of the following topics is thus needed:
\begin{itemize}
    \item Understand limitation on the recombination of high intensity bunches,
    \item Study in simulation of possible bottlenecks faced while recombining short and space-charge limited bunches in a transfer line,
    \item Proposal, study and proof of principle (using available facilities) of bunch recombination. Evaluation of loss of quality in the process and proposal of mitigations.
    \item Improvement of the recombination scheme based on the compression studies results (previous topic) and experimental results.
\end{itemize}

Finally the collimation of beams in high power synchrotron has to be evaluated. Since collimators both at SNS, CERN and JPARC do not work as planned, it seems that a two stage collimation scheme is too naive to be applied for a ring on GeV range. This effort would work on collimation studies taking a MC proton driver as an example with the main outcomes as:
\begin{itemize}
    \item Study of different collimator setups and
    \item Calculation of some cases for CERN, SNS or JPARC to benchmark loss simulations and collimation behaviour.
\end{itemize}

The MC proton driver may operate in an extreme, previously unexplored region of bunch charge densities. Special care must be taken to model the beam dynamics under the influence of collective effects such as space charge, impedance, and electron cloud build-up. Thus, it is critical to refine and benchmark any new and existing simulation codes and their underlying models. It may require both further code development and their experimental validation.
 
A standard code that has been successfully benchmarked and applied at many places, such as SNS and JPARC, for both single particle dynamics and collective effect simulations is PyORBIT. While PyORBIT models space charge and impedance, some of its treatments make long-bunch approximations which may be invalid during bunch formation in the MC proton driver. For example, a numerical instability has been observed when computing the longitudinal space charge force during bunch compression. A 3D field solver may be necessary to resolve this issue. Electron cloud effects are not currently modelled in PyORBIT and may need to be implemented. Various models in PyORBIT and other simulation codes, such as XSuite, can then be benchmarked against relevant experimental measurements. Additionally, simulation codes capable of modeling space-charge effects must be benchmarked and cross-compared to ensure accuracy across the range of intensities required for the Proton Complex of a Muon Collider.
A campaign on the development and benchmarking of available space-charge codes in necessary and the main goals should focus on:
\begin{itemize}
    \item Comparison of dynamics of high intensity bunches in various codes (PyOrbit, XSuite, etc),
    \item Simulation studies of different accumulation/compression schemes with comparisons and
    \item Intensity and space-charge driven instability simulations, comparisons and characterization.
\end{itemize}

\subsection{Muon production and cooling systems}
The preliminary design for the muon production system will be taken forwards to develop a start-to-end simulation of the full system from the target to the end of the beam reacceleration. The cooling system design will be developed including detailed consideration of engineering constraints in areas where these constraints are likely to impinge on the performance, for example in the rectilinear cooling system. Attention will be paid to maintain discussion with the hardware development teams. Ultimately the integrated system performance will be assessed and a global optimisation will be performed to yield the best muon production system.

The integrated development of the muon production system in the light of single particle optics will enable a start-to-end simulation and combined optimisation of the cooling system.

In the first five years of the R\& plan, the conceptual design for the muon production will be completed enabling integrated system optimisation to be undertaken.
\begin{itemize}
\item Around the target area design of an extraction line for the spent proton beam will be developed. The chicane and proton absorber concept for removing beam impurities will be further developed and the level of residual radiation will be assessed.
\item The longitudinal capture scheme will be refined including assessment of collective effects such as beam loading which may compromise performance.
\item Existing concepts for separation into positively and negatively charged cooling lines will be developed accounting for realistic integration of bunching RF cavities and a large aperture solenoid chicane.
\item The rectilinear cooling system will be developed, taking into account results from the muon cooling demonstrator R\&D programme. Improved performance will be sought by a more developed matching between cooling stages. More detailed cooling cell layouts will be developed, in consultation with hardware teams.
\item The concept for merging the bunch train into a single higher intensity bunch will be reviewed. Reoptimisation for alternate beam emittances and potential novel schemes will be performed.
\item The final cooling scheme will be developed. Further studies including improved knowledge of availability of RF at low frequencies will be performed, also taking into account development of the rectilinear cooling line.
\item A scheme for reacceleration will be developed from the O(5~MeV) energy at the end of the final cooling to the 250~MeV energy that is defined for the interface to the main acceleration system. Consideration will be given for merging and handling of both charge signs.
\item A realistic model for collective effects in the absorber will be developed.
\item Alternative cooling schemes will continue to be pursued, particularly for the low energy and low emittance portion of the cooling channel (e.g. after the final cooling system).
\end{itemize}

In the second five years of the R\& plan, the integrated system will be optimised. Improved performance will be sought, including detailed calculation of collective effects and the impact of the ongoing R\&D hardware programme. Requirements for alignment will be studied and, correction schemes developed. Collective effects simulations will be validated by comparison with other codes.

\subsection{Low-energy acceleration}

No studies have been yet performed for the PA and RLA1. A design for these lower-energy accelerators should be produced next. The large acceptance requires large apertures and tight focussing, especially in the PA. Combined with moderate beam energies, this favours solenoid rather than quadrupole focusing for the entire PA linac. 

Significant progress was made in the design of the lattice for the RLA2. The design has been adjusted to a racetrack linac and now reaches the required performance figures without machine imperfections. In the next step, the impact of machine imperfections will be studied and integrated into the simulations.

The expected accelerating gradient and balancing the beam loading effect in the RF cavities plays a major role for the muon survival rate in the low energy section. 

The study of a possible rebalancing of the handover energies in the design of the PA and the RLA1 might result in an easier low-energy section design. 
The rebalancing could also ease the choice of RF frequencies for the different accelerators, as well as affect the choice of focusing magnets.

An initial 6D phase space distribution is essential for performing more realistic simulations in the low-energy acceleration chain of a muon collider. The simulations have thus far assumed a Gaussian distribution, but this assumption needs to be validated with the findings of the upstream systems. 

\subsection{High-energy acceleration}
The extremely fast acceleration of muons in RCSs extends the synchrotron concept into uncharted territory in terms of parameters, which are beyond those of present accelerators. The conventional design tools have to be carefully scrutinized and extended to make sure that they remain applicable for the muon RCSs. No existing RCS has a comparable circumference or accelerates as fast and with such a large synchrotron tune. In addition, the concept of interleaving super- and normal-conducting magnets is novel.

The R\&D activities for years I to V of the R\&D programme are described in the following paragraphs.

The longitudinal tracking and RF studies must be performed for counter-rotating muon beams in all RCSs. 
An end-to-end tracking simulation through the whole RCS chain is foreseen to ensure the consistent evolution of bunch length and longitudinal emittance, in particular at the transfers between the rings. 
The code is planned to be extended to include injection mismatch and other injection errors. With this extension, a detailed simulation with the full longitudinal distribution at the handover between RLA2 and RCS1 is planned to be performed. This could possibly highlight additional requirements for the low-energy acceleration chain or the RF systems configuration at the start of the ramp.

The present RF voltage requirements are based on a synchronous phase of about \ang{135},  which results in a comfortable bucket size. An optimization of the synchronous phase will, therefore, be performed to minimize the RF power requirements, as well as the number of cavities. The maximum power per cavity has a strong impact on the main components of the RF system, for example, the fundamental power coupler. 
As an alternative, the proposed cavity system could be maintained, using the additional margin in the synchronous phase to enable a more rapid acceleration of the muons. 
Increasing the acceleration speed would require faster ramping magnets, increasing the technical challenge in the lower energy RCS. 

The existing semi-analytical transient beam loading simulation is planned to be integrated into the longitudinal beam dynamics study to realistically reflect the change in supplied cavity voltage during the acceleration process. 
A detailed study on the effects of different RF cavities and HOM-damping schemes is foreseen. 
These studies will not only include beam dynamics aspects but will also elaborate on the RF power requirements and guide the engineering design of the RF system. 
The strong transient beam loading with the high-intensity single bunches also poses challenges to the low-level RF feedback systems. A gated 1-turn delay feedback is being considered, but the conceptual design must be created to estimate its benefit and RF power requirement.

The parameter set for the RCS chain (Table~\ref{1:acc:tab:RCS_RFpars}) has been optimized with genetic algorithms to minimize the total RF voltage installed in the entire chain and keep a total survival rate above \SI{70}{\percent}.
While the first RCS must be equipped with almost \SI{21}{\giga\volt}, the RF voltage is smaller for the following two stages, despite the fact that RCS3 is significantly larger than the first two accelerators.   The \lstinline{rcsparameters} tool, which was created for the calculation of the baseline parameters in the CERN tunnels, is used for the optimisation of the global parameters of the RCS chain,. 

The required RF voltage for a given transmission is smallest for a perfectly linear energy ramp. 
This would require that the current driving the pulsed magnets in the RCS varies linearly with time. 
While a close approximation to this is possible, that would require a very costly power source for the magnet. 
It would be less costly to drive the magnets with a current pulse that is closer to a portion of a sine wave. However, a nonlinear rise of the field must be compensated by additional RF voltage to reach the same muon survival.
In this aim, a working group was created to link cost models for the RF system and the bending magnets with their power converter. The models are still progressing. First results are very encouraging and have shown that is possible to get a quasi-linear ramp in the power supplies with an affordable cost. The cost models still need to be consolidated in details and consistency before allowing cost optimization. This effort will continue, adding also software tools to extract cost estimates.

The work on the RCS lattices started and now cover the whole RCS chain and possible limitations. A special emphasis will be on ensuring sufficient energy acceptance and a careful evaluation of possible synchro-betatron resonances. One bottleneck is the required total length of the quadrupoles because of the low available gradient with pulsed quadrupoles. The direct impact is an increase of the beam size and thus of the required apertures.  Investigation is ongoing to find solutions to reduce the required aperture. Different designs, using i.e. combined function magnets will be further investigated to optimize the acceleration efficiency and reduce the magnet aperture.
The effect of deviations in the magnetic field strength of fast ramping magnets on the beam and possible mitigation methods have to be studied. Following the simulations, a technical design of the arc cells is envisioned.   

Code development will be pushed to enable start-to-end simulations and integrate the magnetic ramp function. 
Such simulations will give a first estimate of the emittance growth during the acceleration. 
While the first focus has been on the optimization of separate beam dynamics simulations for the transverse and the longitudinal planes, full simulations of the beam dynamics in six dimensions (6D) have been performed with X-Suite.
Tracking has shown the need for fast tuners to keep the longitudinal emittance.

Furthermore, the present RF frequency for the muon acceleration of \SI{1.3}{\GHz} has been taken as an assumption, as detailed cavity designs are available for the International Linear Collider (ILC). Although the frequencies around \SI{1.3}{\GHz} are well suited in terms of achievable gradient, further studies considering bucket area, beam loading and wakefields could force a move to a lower RF frequency. Once the suitable frequency range has been fixed, the exact RF frequency remains to be determined based on arguments such as the matching in frequency with the muon pre-acceleration and cooling stages.

Beyond the refined simulation campaign for acceleration introduced above, RF manipulation schemes like bunch rotation are considered to shorten the bunch and achieve the stringed longitudinal bunch parameters at the transfer from the last RCS to the collider. 

Years VI to X will focus on tests with existing accelerators, integration studies and alternative schemes, as described in the following paragraphs.

Existing accelerators could be used as test-bed with longitudinal parameters at least approaching the parameters required for the muon RCSs. For example, synchrotron tunes as large as $0.1$ could be exercised with ions at injection energy in the CERN SPS, to benchmark the consequences of discretised synchrotron motion. Further beam tests may be possible in the existing RCS at JPARC or the \SI{8}{\GeV} Booster synchrotron at Fermilab, both accelerating proton beams within a few \num{e4} turns.

When the arc cell for the hybrid RCS is frozen, it will be of great interest to integrate a pulsed magnet alongside a superconducting magnet and RF cavity under full power to address any potential integration or crosscheck issues. This study will include the mechanical design of an hybrid cell, make the prototypes to fill one cell with one superconducting magnet, one pulsed magnet and one RF cavity including their power supply, check there is no crosstalk, and demonstrate that it possible to pulse at the same time the magnets and deliver the needed power pour stored energy replacement and phse shifting of the cavities.

An FFA accelerator is under intensive study for a proton driver of a spallation neutron source. The muon acceleration community will benefit from those developments. 
If the progress on the alternative with vFFA magnets gives a realistic table of parameters for the magnets, the demonstration of the vFFA will go through the design of the vFFa magnets, the validation of the design with small-scale prototypes, the final demonstration with a full-scale prototype within the field qulity tolerances.

Following the progress on HTS pulsed magnets with a target of \SI{500}{\tesla\per\s} \SI{3}{\tesla} fast ramping dipoles to replace the pulsed dipoles in the later RCS stages, the parameter table of the RCS will be updated and the lattice design will integrate such devices to compare the performances with the baseline.

\subsection{Collider ring}
Recent studies of the muon collider ring design described in section \ref{1:acc:sec:collider} have concentrated on the lattice design of a \SI{10}{\tera\electronvolt} centre of mass collider with a challenging $\beta^* = \SI{1.5}{\milli\metre}$. Despite significant progress, a lattice with sufficient dynamic aperture over the full required energy range has not yet been found even for a perfect machine without imperfections as alignment errors or unwanted multipolar components of the magnets. For this reason and due to lack of resources, it has not yet been possible to study several basic beam dynamics phenomena and potential limitations (with the exception of coherent instabilities caused by impedances).

On the other hand, recent studies~\cite{Vanwelde2024a} indicate that slightly relaxing $\beta^*$ (and reducing luminosity accordingly) enables designing lattices with sufficient dynamic acceptances for the perfect machine. This statement will likely hold as well after taking recent findings on magnetic field limitations of large aperture superconducting magnets into account.

These observations motivate the proposal to pursue in parallel efforts to further improve lattice designs to find a lattice with $\beta^* = \SI{1.5}{\milli\meter}$ working for the perfect machine with and to start basic studies on beam dynamics design with a lattice with relaxed performance.

A possible list of tasks related to the muon collider ring design is:
\begin{itemize}
  \item Lattice design effort to optimize a \SI{10}{\tera\electronvolt} machine with very small $\beta^*$: this activity should be continued with high priority as the outcome is needed for many other studies.
  \item Studies on machine imperfections:
  \begin{itemize}
  \item Impact of regular deformations of the machine: highest priority study as the feasibility of the proposed neutrino radiation studies and the facility siting proposals rely on the feasibility of this concept. Note that for this scheme, deformations of the machine in, amongst other parts, the challenging chromatic compensation section with large betatron functions and strong sextupoles are required. This results, e.g., in vertical dispersion generating momentum dependant coupling due to the sextupoles.
  \item Impact of alignment errors: Extrapolating from other projects requiring small $\beta^*$ and local compensation of chromatic effects, very tight tolerances have to be expected. The aim of the study is to demonstrate the feasibility of beam based measurements to assess deviations of the machine optics from the nominal one and corrections to improve. Required initial alignment tolerances will have to be agreed with teams working on alignment systems.
  \item Magnet strength errors: study of the impact of static and time varying unwanted magnetic (multipolar) fields. Beam based methods to correct in particular static unwanted magnetic fields should be investigated. A particular source of unwanted time varying magnetic fields is power converter ripple. 
\end{itemize}
  \item Beam-beam interactions occuring in the IPs: both the significant beam-beam tune shift of about $\xi \approx \, 0.08$ and the finding of non negligible increase of the luminosity in a one-pass simulation of a beam encounter indicate that beam-beam effects are relevant and need to be studies. Such studies should be carried out in a  self-consistent way and for multiple beam encounters.
  \item Scenarios for collider commissioning and operation: it will not be possible to start collider operation with the nominal optics as the small $\beta^*$ lead to large betatron functions at other locations and chromatic aberrations 
  rendering the machine sensitive to imperfections. Thus, commissioning must start with a relaxed optics. This will be followed by many iterations alternating measurements with beam to understand the machine optics and imperfections and corrective actions (slightly) changing some magnet strengths. With smaller $\beta^*$, machine stability (drifts of, e.g., element positions or magnet characteristics, vibrations ...) may become an issue. The aim of this study is to devise appropriate strategies for machine commissioning and operation and study their feasibility under realistic assumptions.
  \item Studies on beam coupling impedances and coherent instabilities are treated in another section.
\end{itemize}

\subsection{Collective effects}

\subsubsection*{Proton driver}

In the proton complex, collective effects could limit the intensity reach of the different machines and therefore impact the muon yield.

A major focus in the first five years of R\&D will be on the study of the impedance model of the accumulator and compressor rings, and the estimation of the longitudinal and transverse instability threshold.
The stability simulations should also include effects such as coherent direct space-charge and beam loading to evaluate the performance reach of the different proton complex options.

In the second part of the R\&D decade, start-to-end tracking simulations of the whole proton complex should be carried out, from the Linac to the target.
This would require interfacing different codes such as RFtrack and XSuite for the particle tracking in the different machines.

\subsubsection*{Muon cooling}

The studies of collective effects in the capture and cooling line need to be expanded to ensure that the required beam quality can be reached without issues. Some technological choices, such as the material and geometry for the absorbed as well as the RF cavities design, voltage and setup may be further constrained by these effects. The following steps will be taken in order to improve the design of the cooling line taking into account various collective effects:
\begin{itemize}
    \item Compute the wake fields generated by the beams passing through the absorber taking into account its electromagnetic properties, including dissipation due resistivity.
    Build an impedance model for the cooling line including other more standard elements.
    \item Perform beam dynamics simulations through the cooling line including space-charge, beam loading and wakefields in addition to scattering effect.
    The development of new tools for beam dynamics, or the extension of existing ones, will be necessary.
    Possibly re-design the line to maximise the overall performance of the cooling channel.
    \item Determine the behaviour of charged particles generated by ionisation during the bunch passage through the absorbers.
    The development of new tools for plasma dynamics, or the extension of existing ones, will be necessary.
    Such models inevitably involve a significant amount of high performance computing resources and would greatly profit of development on most advanced computing platforms.
    \item Develop simplified and less numerically demanding models for the interaction of the muon beam with the plasma, possibly based on fast data-driven surrogate models, and integrate them in the start-to-end simulation model of the beam dynamics along with other collective effects.
\end{itemize}

\subsubsection*{Acceleration complex}

In the RLAs, the wakefield effect of the RF cavities will be evaluated and its impact on beam dynamics simulated with tracking simulations, using RFtrack combined with Xsuite.

In the RCS, as the magnet design will evolve, the impedance model of the magnet vacuum chamber should be updated accordingly. 

The two beam effects must be studied in detail, in particular the effect of the second beam wakefield on the first beam: with the distributed RF stations required to preserve longitudinal focusing, the wakefield left by a bunch in RF cavities might not be fully decayed when the counter-rotating bunch crosses the cavity.
Developments are ongoing to simulate this effect with Xsuite tracking simulations.
Coherent beam-beam effects at the interaction points should also be included in the stability simulations, and if necessary separation schemes of the two beams should be investigated.
Once finalized, the lattice model of the RLA and RCS should be integrated in the model to investigate the effect on non-linearities on coherent beam dynamics.

Starting from R\&D year VI, the impedance model should be refined with additional elements such as the beam instrumentation, transitions between elements, or beam injection and extraction devices.
The transverse damper characteristics (bandwidth, location etc.) should be detailed in collaboration with the RF experts.

\subsubsection*{Collider}

The highest priority study in the collider ring for the first half of the R\&D period will be the investigation of the coherent beam dynamics effects resulting from the absence of RF. In particular, the effect of high-order $\alpha_p$ on longitudinal beam dynamics must be added and its impact on coherent beam stability assessed.

In the second half of the R\&D period, the impedance model should be refined with additional elements besides the magnet vacuum chamber, such as the beam instrumentation, pumping ports, transitions between magnets among others.
In particular a detail of the final focusing region, where the large Twiss $\beta$ functions will increase the impedance contribution of the elements, should be included.

\section{Machine-detector interface}

Working towards a realistic Machine-Detector Interface (MDI) design is a highly iterative process, which must consider physics performance goals, machine design requirements, integration constraints, engineering aspects, technical limitations, and other elements (e.g.\,radiation damage, heat deposition). The conceptual and technical MDI design will evolve with time and is not a one-off task because of its complexity and many dependencies. A coordinated design between detector and machine is fundamental. It is expected that maturity of the MDI design will advance once the detector and machine design progresses. This section presents a design roadmap for the next ten years, with the goal to achieve an intermediate step after five years, followed by a more detailed design and integration study towards a conceptual design report in the second five years.

A core objective of the first five-year phase is to converge on the main conceptual design features for the MDI and interaction region, including detector envelope, absorber and nozzle configuration, final focus layout, vacuum apertures, magnet apertures and field strengths. This requires background simulations, detector and physics simulations, lattice design studies, radiation damage studies, particle tracking and dynamic aperture studies, and impedance simulations; some of these studies are detailed in other sections of this report. Although a full-scale engineering design and integration study is beyond the scope of this first phase, it is nevertheless important to elaborate a first technical design and a global integration concept for the MDI, including nozzle, shielding, central beam chamber, support structures, forward detectors (including LumiCal), and accelerator equipment (vacuum system, magnets, beam instrumentation, cryogenics, etc.). Iterations of the conceptual MDI design will be essential during this design phase, in order to incorporate the findings of the first technical design and integration studies. 

As the MDI design evolves, a continuous effort to simulate the beam-induced background with Monte Carlo codes is essential. One of the key tasks for the first phase is the optimization of the shape and material composition of the nozzle using the FLUKA simulation framework. Separate studies are required for the different center-of-mass energies (3\,TeV and 10\,TeV). The optimization can rely on simple key figures (e.g.\,hit densities in the vertex detector) in order to avoid the need of a full detector simulation for each design iteration. For certain reference configurations, a full evaluation of the background in dedicated detector simulations will be essential, to provide a feedback for the nozzle design. Where applicable, alternative optimization techniques (machine learning) shall be explored. During the nozzle design process, the feasibility and benefits of instrumenting the nozzle needs to be evaluated in close collaboration with detector experts. The nozzle optimization is mainly driven by decay-induced beam losses, but the contribution of other background sources (incoherent pair production and beam halo losses) should be assessed through the same simulation framework. This requires input from other codes (e.g.\,GuineaPig) and machine studies (multi-turn tracking simulations). In this context, it is desirable to improve the description of incoherent pair production by muons in GuineaPig, or to develop a new event generator. Another important aspect of the background studies is the muon background (Bethe-Heitler muons); these muons can reach the detector even when being produced further away. A careful study of this background component is important and must include a more extended Monte Carlo model of the machine. Complementing the background studies, updated estimates of the cumulative radiation damage in the detector need to be derived for the evolving MDI layout. These estimates shall include all background contributions, including incoherent pairs and halo losses, which have been neglected so far in the radiation damage studies.

Another key task is the optimization of the interaction region layout, under consideration of the beam-induced background, optics-related constraints, shielding requirement for magnets (heat load to cold mass, radiation damage in the coils), as well as magnet technology limitations (coil aperture versus peak field). The shielding requirements for the magnets must be refined, which in turn can impact the required magnet apertures. The work prioritizes the optimization for a 10\,TeV machine, but the lessons learned can be applied to a reoptimization of MAP’s 3 TeV interaction region. 
When iterating on the interaction region design, it is important to study forward muons and assess the options for integrating forward detectors in the beam line or elsewhere in the collider tunnel. 

An important design specification for the interaction region layout is the minimum acceptable vacuum aperture in the magnets and nozzle (in terms of transverse beam sigma). A smaller aperture of nozzle and the masks embedded in the final focus magnets is beneficial for reducing the decay-induced background, but can lead to enhanced halo losses in the interaction region. It is essential to quantify the permissible level of beam halo losses per bunch crossing, which must remain below the decay-induced background. In addition, beam dynamics simulations will be essential for understanding the expected beam losses. The aperture studies must also account for possible machine imperfections. Based on these studies, the interaction region aperture (presently assumed to be 5$\sigma$, where $\sigma$ is transverse beam size) should be revisited and the need of other mitigation measures for halo losses (e.g.\,halo scraping system far from the IP) should be assessed.  

Another important task is to define the required instrumentation in the MDI region. A luminosity measurement detector is one of the key systems to integrated in the forward region. It is important to elaborate first concepts for measuring the luminosity, to evaluate suitable detector technologies (LumiCal), and to simulate the performance of such a luminosity monitor considering the beam-induced background. 

Although a full-scale engineering design and integration study of the whole MDI region can only be carried out at a later stage, it is important to address some of the key engineering and integration aspects, which have an important impact on the MDI design.
Based on the MDI model conceived in the physics background studies, a first technical design and integration concept for key systems must be elaborated. This includes a first engineering concept for the nozzle, which must address the evacuation of the deposited heat by means of a cooling system, the nozzle support structure, the nozzle segmentation and assembly, as well as the integration of the nozzle inside the detector and solenoid. In addition, a first technical design for the central beam chamber around the IP must be elaborated and the required material budget of the chamber needs to be defined; in addition, the connection to the nozzle needs to be studied. It is also important to study the integration of instrumentation inside or near the nozzle, in particular the integration of a luminosity monitor. 

Another important aspect is the integration of the triplet quadrupoles in the interaction region. Considering the proximity of the magnets to the IP ($L^*=$\SI{6}{\meter}) the required distance between nozzle and the first triplet quadrupole needs to be carefully assessed. In addition, a first integration study for the triplet magnets in the cavern and tunnel needs to be carried out, considering that the first magnets need to be tightly surrounded by heavy shielding. Another aspect is the support structure for  magnets and shielding, since the first quadrupoles might be located in the experiment cavern. 

Although the foreseen design and integration studies may not cover all the details, they will prepare the path towards a fully integrated MDI and IR design. 

The second phase, starting from R\&D year VI, shall be dedicated to an integrated engineering for most of the MDI systems, with a more detailed design of sub-systems and a more complete integration study. In particular, the technical specifications for the different MDI components, including specifications for the alignment and stabilization requirements, need to be refined. This will allow for a more detailed mechanical design of important systems like the nozzle or the central beam chamber, including correction schemes for possible misalignments. This also includes a more comprehensive analysis of the support structure by means of stability simulations (vibration modes, cross-talk, impact of ground motion etc.). As the design evolves, the the design choices need to be validated by the necessary simulations (e.g.\,impedance studies, thermo-mechanical studies). Furthermore, the design and integration of the luminosity monitors and beam instrumentation like beam position monitors must be studied in more detail. The requirements and specifications for instruments needs to be refined by assessing operational aspects such as luminosity tuning and beam monitoring. Another important aspect are the space requirements for the integration of cable shafts, cooling pipes, etc given the tight space constraints in the MDI region. Considering the complexity of the overall integration, possible options for building a (small-scale) mock-up of the central MDI region should be explored, including, for example, the nozzle and central chamber. Depending on the available resources, such a mock-up could be built after the CDR phase, i.e., after the second five-year period.

It is assumed that the technical design for the final focus triplet quadrupoles and the adjacent chicane dipoles will advance in this second five-years period. This will allow for a more detailed technical design of the radiation shielding inside magnets and magnet interconnects. It is also assumed that a first design of the cryostats and cryogenic lines for IR magnets could be available, which would enable a more detailed integration study of the entire interaction region, including the shielding configurations around magnets. First ideas about the space requirements for the handling, transport and installation of magnets in this confined region need to be elaborated. 

Radiation transport simulations, including background, heat deposition and radiation damage studies, remain an important aspect of the second phase. As the MDI and interaction region design evolves, the Monte Carlo simulation will provide an important feedback about the impact of design choices. It is expected that the geometry models for the radiation studies will incorporate more and more details as the technical design of the MDI components progresses. It is also expected that the detector design and the associated technologies will advance in this period and hence the generated background samples will be essential for assessing the expected physics performance. 

In addition to the MDI and IR design studies, it will be important to design a collimation or scraping system, which reduces the halo-induced background in the detector and also protects the detector in case of accidental beam losses. Multi-turn tracking studies will be needed to assess the efficiency of collimation techniques. In addition, it might be necessary to explore novel collimation schemes. The studies of the first period will provide the necessary input concerning the acceptable amount of halo losses in the interaction region. It will therefore establish specifications for future design studies of a halo cleaning system, which can be addressed once more resources become available.

It is expected that the listed activities will provide a MDI design and integration concept, which is sufficiently detailed for a CDR. The different engineering studies will provide enough elements to demonstrate the technical feasibility of the MDI design. In addition, the radiation transport studies and detector simulations will be crucial for demonstrating that the beam-induced background can be managed.

\section{Neutrino flux mitigation system}

To implement the beam trajectory deformation required for radiation mitigation, the collider magnets must be supported and precisely moved using jacks. These jacks must meet stringent requirements in terms of vertical stroke, angular positioning, and step accuracy to ensure proper beam alignment and stability. Given the unprecedented nature of these dynamic adjustments in an accelerator, a dedicated test program is essential to develop, validate, and optimize the proposed solutions.
This chapter outlines the specifications for the jack system, the proposed development and testing phases, and the key components necessary for a functional prototype. Additionally, it addresses the challenges associated with the relative movement of adjacent magnets, particularly concerning interconnection components such as bellows and current lead.

There are still discussions on the period length of the deformation pattern and requirements in terms of the accuracy of position and angles will be the result of thorough beam dynamics studies. For the proposed jack development, a deformation pattern of \SI{100}{m} as well as state-of-the-art positioning tolerances are assumed. This lead to the tentative list of requirements:
\begin{itemize}
    \item Magnet angular range: $\pm1$~mRad
    \item Vertical stroke: $\pm27$~mm, the beam trajectory requires $\pm25$~mm
    \item Number of lay-outs, thus jack steps per year: 200
    \item Angular position: 0.02~mRad
    \item Step tolerance: 0.004~mRad (10 m cryostat)
\end{itemize}
We propose a new jack design for collider magnets to answer the requirements. This requires a test program to develop and proof the solutions. This requires:
\begin{itemize}
    \item New jack development. Possibly done partnering with industry for acost estimate
    \item Three new jack prototypes for test mock-up.
    \item Magnet mass representation. 
    \item Control system to drive the movement of an individual jack first, later three jacks together.
    \item Components to connect three jacks together to manipulate a magnet interconnection with one electrical motor. 
    \item Mechanical and electro-mechanical components and required control system.
    \item Measurement system to verify the steps. First sensors build into the components of the magnet mover assembly. Secondly an external system to verify, this could be the laser tracker system used by CERN Survey.
\end{itemize}	

A second phase of the mechanical system development should involve test mock-up based on representative magnets and interconnection to confirm the behaviour of the system to be installed in the collider.

Moving the magnets leads to relative movement between two adjacent magnets. This leads to movement in components of the magnets’ interconnection. Although movements are small, such conditions have never been seen in an accelerator before, thus, mock-up should also incorporate possibility to test the components of the interconnection, most of all the bellows and current leads. This requires:
\begin{itemize}
    \item Cryogenic supply 
    \item Current supply
    \item Interconnection thermal shielding 
    \item Interconnection vacuum
    \item Current lead design
    \item Bellows for cryogenic lines
\end{itemize}
	
The mock-up installation first phase, the simple interconnection mock-up to test mechanical components, requires space of installation length by width with around 2 m of free working space around the installation. For further phases of the mock-up, the space requirement may be significant depending on the final magnet length. To the test the behaviour of the interconnection, a facility prepared to test cryogenic installation is required.

\section{Radiofrequency systems}

In general, the R\&D for the muon collider RF systems will have two phases of 5 years each. The first 5 years will be devoted to the development of the concept and defining main parameters of the RF systems in close contact with the beam dynamics design and the other accelerator technologies as well as taking into account cost and power considerations. During the second 5-year phase, the conceptual design of the RF systems will be developed.

\subsection{RF for muon cooling complex}

The first five years will be devoted to the conceptual design of several RF systems and a supporting experimental programme aimed at studying the operation of high-power RF structures in strong magnetic fields.

\paragraph*{Design effort}  
Together with the beam dynamics design working group, develop a concept for the following RF systems listed below together with the main challenges to be addressed during this design phase:
\begin{enumerate}
\item Front-End: Moderate gradients from a few to 20 MV/m in moderate magnetic fields ~ 3 T, many different frequencies generation and control.
\item Rectilinear cooling channels: High gradient >30MV/m in high magnetic fields >10 T at hundreds of MHz.
\item Final cooling and re-acceleration: high gradient in moderate magnetic fields ~3 T, many very low frequencies.
\end{enumerate}
Both operational and integration constraints coming for the other accelerator technologies, like magnet, cryogenics, and vacuum must be considered.  
The development of the concept must include the following points:
\begin{itemize}
\item 	Layout of the RF systems based on the beam dynamics, high gradient and integration constraints.
\item 	Final RF frequency choice will have to be done based on the beam dynamics requirements as well as on the technological and practicality considerations.
\item 	Adequate type of the accelerating structure and the RF system layout have to be developed: standing versus travelling wave with recirculation, cell-to-cell coupling, RF power coupling and waveguide network, RF tuning, potentially heavily re-entrant structure and induction LINAC for very low frequency cavity.
\item 	Decision on the RF window between individual RF cells must be taken based on the beam dynamics and technological considerations: Beam loading, emittance dilution, RF performance, thermo-mechanical stability, fabricability.
\item 	Beam loading mitigation strategy must be developed together with the beam dynamics.
\end{itemize}

\paragraph*{Experimental effort}
On the experimental side high-power RF tests in the strong magnetic field are mandatory to complete the conceptual design. In particular, an RF test stand at RF frequency in the UHF band (0.3 – 1 GHz) capable to drive at least one RF cavity to a high gradient of ~50 MV/m and equipped with a solenoid large enough to house the RF cavity and strong enough to generate high magnetic field of up to 10 – 15 T is needed to give quantitative answers on the achievable operating gradient which can be used in the muon cooling complex design. 
RF test stands at higher RF frequency in L, S, C and X-bands are very useful in studying and understanding the RF breakdown in strong magnetic field phenomena with the limitation of giving answers to inform physics models that need to be scaled to determine the achievable gradients in the UHF frequency band. Nevertheless, it provides important insights to its dependence on the material, cavity geometry, RF pulse length among others.  Thus, building and commissioning of RF test stand as well as defining adequate experimental program is of highest priority. This program should include:
\begin{itemize}
\item	Cavity wall materials and coatings study: Be, Cu, Cu-alloys, Al, HTS (REBCO), …
\item	Operating parameters: RF pulse length, temperature, …
\item   Cavity geometry and magnetic field configurations
\item	Surface treatments and coatings
\item	Testing a re-entrant cavity at UHF frequency which has a small radius to fit into the magnet bore.
\item	Exploring non-vacuum cavities, such as gas-filled or dielectric-loaded cavities.
\end{itemize}
In addition, it is essential to compliment the experimental activity with a theoretical one by re-establishing and further improving the RF breakdown models developed earlier during the MAP era. This will not only help understand the physics of the breakdown phenomena but also improve the RF cavity design process.

The second phase of the R\&D programme will refine the conceptual design of the RF systems taking into account integration constraints. 
Engineering design and prototyping of major components will be done. 
Validation of the performance of the prototype RF structures in magnetic field will be done. 

\subsection{RF for accelerator complex}
Based on the beam dynamics requirements, the optimization and design of radio frequency (RF) cavities and couplers for low-energy acceleration in a linac and RLAs and high-energy acceleration in RCSs will be done. The newly designed superconducting RF (SRF) cavities need to be integrated into cryomodules. It is foreseen that a common cryomodule concept will be developed for all cavity types and frequencies. Following that, a full-scale and full-power prototype cryomodule will be built for verifying the functionality of all components and couplers. 

Due to the different beam and machine parameters in the low- and high-energy acceleration, the impact of the optimisation will be significant. The target parameters, such as the fundamental mode frequency or requirements for suppression of higher order modes (HOMs), will be discussed with other working groups involved. For the final choice of frequency, the power requirements of the acceleration and cryomodules might play a significant role.

There are several challenges that need to be addressed. Among these challenges are: i) A very high muon bunch intensity favors larger cavity apertures and hence lower RF frequencies. A relatively small aperture of 1300 MHz cavity may not be sufficient and might require switching to lower frequency, e.g., 800 MHz (synergy with FCC-ee); ii) The high gradient operation of multi-cell 800 MHz (or similar frequency) cavities must be demonstrated; iii) Lower frequency SRF cavities ($\leq 400$~MHz) most likely will need to utilize Nb/Cu SRF technology, which requires further development; iv) Stray magnetic fields from high-field magnets may significantly degrade performance of SRF cavities. Hence, there is a need of developing efficient magnetic shielding and/or develop SRF cavities based on alternative superconductors; v) The effect of high-intensity radiation from muon decays on the performance of SRF cavities is unknown and must be studied.

During the acceleration, muons will constantly decay, resulting in many seed particles for multipacting, which might lead to breakdowns. The magnitude of this effect needs to be studied in detail and included in the shape optimisation. 

Beam dynamics studies and evaluation of the collective effects will determine the requirements to HOM suppression. For the chosen cavity geometries, HOM damping schemes will be developed to satisfy the requirements. HOM coupler and FPC (Fundamental Power Coupler) designs will follow. Depending on the powering and beam dynamics requirements, it might be necessary to adjust the number of cells per cavity to stay within the power limits of the couplers. In different parts of the RCS chain, the requirements might favour the development of accelerator-specific cavities. 

The cavity fabrication process produces imperfect cavities with pseudorandom deviations from the ideal shape. An investigation of the impact of such deviations on the cavity performance will be performed. The results of this analysis will determine fabrication tolerances. 

Special attention will be given to the investigation of cavity tuner technologies capable of providing the necessary tuning speeds and ranges with the required precision. In preparation, the impact of a non-perfect cavity detuning on the beam dynamics will be investigated. After a technology was chosen, the complete tuning system would be designed and verified with a prototype.

To reduce the impact of the high beam-induced voltage in the fundamental mode, a cavity feedback system will be implemented in the RCS chain. The properties of this system, as well as the impact on the power requirements and beam dynamics, will be studied.

Over the first 5 years, the key R\&D priorities for developing SRF cavities for muon collider are:

\begin{itemize}
\item	Perform studies of beam interaction with SRF cavities at different frequencies, including acceleration, longitudinal beam dynamics, wakefields, bunch length evolution, and energy spread control. Select RF frequencies for different accelerators.
\item	Using synergies with other programs, develop SRF cavity concept designs, select one or two most challenging to fabricate prototypes.
\item	Develop cavity treatment recipe for high gradient and low sensitivity to residual magnetic field. Demonstrate the cavity performance in vertical testing. Utilize synergies as much as possible.
\item	Continue R\&D on developing Nb/Cu for low-frequency SRF cavities.
\end{itemize}

Based on the concepts develop during the first 5 years, the conceptual designs of the RF systems for the linac, RLAs, and RCSs will be further developed in the second five-years R\&D phase. This will include the designs of cryomodules for the cavities operating at different frequencies, RF power distribution, tunnel layout, etc. A prototype cryomodule(s) will be built and tested to verify the design concepts.

While the current baseline assumes an accelerating gradient of \SI{30}{MV/m},  possible improvements in the SRF technology over the previous 5 years could increase this value significantly. Consequently, the number of required cavities could be reduced, decreasing the complexity of the integration as well as the size of the cryogenic plant. Additionally, the overall power consumption of the RF system could possibly be reduced. Thus, one iteration on the overall SRF system design might be necessary.

The integration of the RF power infrastructure into the tunnel poses an additional challenge in RCSs. Due to the distribution of the RF system over multiple stations, klystrons and auxiliary components will need to be distributed around the ring as well. The grouping of klystrons to power multiple cavities will be different in all accelerators, requiring different concepts for the combination of power electronics. The same challenge applies to the distribution of the required cryogenic lines to all RF stations.

\subsection{RF power sources}
Currently there are no RF power sources anywhere close to the power required for a muon collider at frequencies below 1~GHz, hence work is required to develop new frequency, high power sources. 
Together with the RF system design working group we will define the required parameters and specifications for the RF power sources across the whole muon collider including proton drive, muon cooling and accelerator complexes. 
Key areas of concern are the power sources for the muon cooler, which will need 352 and 704~MHz sources. While it is possible to obtain 500-1000~kW sources at the frequency this would require a very significant surface building to house all the required sources, an optimum solution would be to increase these to 24~MW and feed multiple cavities from each source. 
These specifications will then be used as a basis to develop a concept and main parameters for the RF power sources including its type and the optimum configuration for the overall RF systems. These concepts will be developed in disk model codes to evaluate performance. It is expected that either Core Stabilization Method (CSM) Klystrons or Two-stage Klystrons will be a good fit to the requirements, but new concepts will also be considered.
The most successful concepts will then be further developed towards high efficiency and lower capital and operation cost from the AC plug to RF cavity. This will include full PIC code simulations and designs for the gun, collector and magnets.
Together with the industrial partners this conceptual design will be updated considering commercial requirements ready for prototyping.

In the second 5 years we will finalise the industrial design and fabricate sub-components. At present there are no multibeam CSM klystrons or two-stage klystrons proven experimentally, so a prototype tube is required as a proof of principle. This may be based on existing commercial tubes from industrial partners, with modification to CSM or two-stage.
Once fabricated it is necessary to condition and test the RF sources under a variety of operating conditions.

\section{Target system}
\label{3:rd:sec:tar}

The target systems will be addressed in multiple work-packages, encompassing simulation studies, design and engineering of the target systems and auxiliaries, material R\&D for the different target technologies being considered, and work related with the cooling demonstrator. 
The work packages intended for the first five years of R\&D are listed below.

\paragraph*{Frontend layout \& beam-matter interaction simulations}
The Muon Collider layout in the frontend needs optimisation and to become well defined. So far, the studies allowed the optimization of the proton beam parameters and of the carbon-target design in view of maximum pion yield. They were also the basis for the shielding design (see sections \ref{1:tech:sec:tar} and \ref{1:tech:sec:rad_shield}). Important work has been done on the modelling of the chicane with the goal to create an extraction channel for the protons that pass through the target unimpeded. However, the challenges that were identified, such as high radiation load on the chicane coils, still require further study and optimisation. 
Short-term objectives include performing Monte-Carlo simulations of the entire frontend and defining the lattice from the target, through the tapering section, and to the end of the momentum selection chicane. These activities go along with other work-packages responsible for the design and calculation of the solenoid magnets (section \ref{3:rd:sec:mag}) and of the radiation shielding (section \ref{3:rd:sec:rad_shield}).  The work-package will provide the energy deposition maps for the thermo-mechanical design of the shielding and target systems, as well as the pion/muon yields and transport efficiency via the entire frontend to provide input to the cooling section design. 
Given its dependencies, the before-mentioned simulation work shall be prioritized in the first 5 years. It is however expected to be continued throughout the entire 10 years.

\paragraph*{Target system \& Beam-intercepting devices design}
The work-package will lead the technical design of the graphite target system and other BIDs. As highlighted earlier in the evaluation report, a graphite target system has been conceptually designed and engineered. While the 2 MW concept has achieved a good level of maturity, higher-power options require further study and development, if proven to be feasible. 
The priority in the first 5 years is to detail the design of the target system, including critical target aspects such as materials and manufacturing considerations. Another deliverable will be the design, integration and cooling system specification of the proton beam window, which is a critical component due to the very high yearly radiation damage. The engineering integration of the target assembly system within the cryostat, including the actively cooled proximity shielding, supporting structure and services, will also be defined.

\paragraph*{Target R\&D and testing}
No practical R\&D or testing for the Muon Collider target has been conducted so far. The uniqueness of the Muon Collider target facility, which lies in its MW-class target with nanosecond-scale pulse lengths and high beam intensity, impose a full-fledged material and design validation campaign. The objective is therefore to define and carry out the technological R\&D and beam tests required to validate the target system engineering design.
Synergies can be explored with carbon-based neutrino targets such as the T2K or LBNF targets, or other high-power targets such as the Beam Dump Facility target in terms of cooling systems technology. The main deliverable of the first 5 years will be the design and beam testing of a carbon-target for the muon collider in conditions analogous to the final facility. The use of HiRadMat at CERN was identified as the most suitable choice because of the broadly tuneable pulse and bunch schemes of the high-energy protons that can be provided to its beamline.

\paragraph*{Liquid Pb Target design and R\&D}
The work-package has the objective to assess the feasibility of a liquid lead target and propose a conceptual design of the target system. The challenges lie in the modelling of heavy liquid metals in the presence of electromagnetic fields, together with the modelling of the dynamic response of the liquid-pulse interaction.
The design and simulations of the liquid lead target concepts will be pursued, as well as the development of a circulating liquid lead target mock-up.
Beam tests at HiRadMat of a liquid lead target will be pursued to validate the shock-wave dynamics modelling in heavy liquid metals and judge about the feasibility of such technology for a 4-MW Muon Collider frontend. 

\paragraph*{Fluidized W target design and R\&D}
The objective is to address the feasibility of a fluidized W target for the Muon Collider and propose a conceptual design for its target system. Offline experience exists at RAL with circulating loops of tungsten powder and beam-tests with static powder have also been carried out in the past. However, the maturity of the technology and its applicability to the Muon Collider conditions require more research.
The deliverable will be to conclude on the feasibility of a fluidized tungsten (W) target via the detailed modelling of the system and by performing mockups and tests.	

\paragraph*{Demonstrator Facility BIDs}
Work on BIDs for the demonstrator is of high priority and has the objective to design and produce the target, absorbers and magnetic horn for the Demonstrator at CERN, as well as carry-out the beam matter interaction studies. However, a realistic spread over the study is required given the timeline of the demonstrator.
From the first five years will result the execution of the energy deposition, pion/muon yield and R2E studies for the demonstrator facility at CERN. The engineering and technical design for the BIDs and horn will also be made. The work package will also be responsible for the integration model of the demonstrator target area.

Four additional work packages are envisaged for the period starting from year VI of the R\&D programme.

\paragraph*{Target system \& Beam-intercepting devices design}
Following the initial work, a detailed design of the cooling stations for the target, proton beam window, and proximity shielding will be carried-out. Despite not exhibiting, a priori, a technological challenge which requires extensive R\&D, it may lead to significant requirements in terms of space and costs.
The definition and integration of instrumentation in the target system, the evaluation of installation, dismantling, and maintenance aspects will be addressed. The high activation of the target system components also requires the addition of waste disposal considerations in the study.
Finally, a deliverable of this work-package will be the conceptual design of the proton dump system. This can be made based on the experience of other systems at CERN, such as the HL-LHC external dumps and the SPS internal beam dump. 

\paragraph*{Target R\&D and testing}
The second five years (particularly in case not compatible with CERN’s accelerator schedule during the first 5 years) will focus on concluding the beam-tests. Moreover, a full offline mock-up of the entire graphite-target assembly will be made to test and validate the entire system. 
The characterization of the selected materials for the target under operational conditions will be pursued, including fatigue, erosion and other relevant properties.

\paragraph*{Liquid Pb Target design and R\&D}
Similarly, the conclusion of the beam-tests and offline mock-up of the liquid lead target concept will span to the 2nd half of the study. This phase will provide as deliverable the definition of the services, space requirements, and P\&ID for the liquid lead target.

\paragraph*{Demonstrator Facility BIDs}
The 2nd part of the R\&D program will conclude the engineering, technical design and production as well as the installation and commissioning of the muon production target, its cooling system, the beam dump and the magnetic horn system for the demonstrator. 

%
%
\section{Beam Instrumentation}

For all parts of the muon collider complex, from the $H^{-}$ source to the collider, the first and fundamental R\&D phase for beam instrumentation will be closely tied to the overall project R\&D. This involves defining and keeping up to date a comprehensive set of functional specifications across beam dynamics, RF, radiation protection, and other fields. 

The beam instrumentation in the RCS and the main synchrotron rings is expected to be based on more standard beam instrumentation and could be studied in a second time. In the coming years, the main R\&D activities should focus on Proton driver and Muon cooling channel beam instrumentation. 

\paragraph*{Proton Driver}

As discussed in Section~\ref{1:tech:sec:inst}, the design of beam instrumentation for the proton driver complex can leverage experience from technologies used in hadron linacs and synchrotrons up to 5-10 GeV. While existing techniques can generally fulfil future proton driver specifications with appropriate mechanical, detector, and electronics adaptations, some cases will require more substantial development. 

The key R\&D priorities for proton driver instrumentation include:

\begin{itemize}
\item Assessment of interceptive diagnostics performance and survivability (secondary emission monitors, imaging systems, flying wires) under high-brightness $H^{-}$ and proton beams

\item Development of low-density materials \cite{CNT} for hadron beam instrumentation
    \begin{itemize}
    \item Investigation of nano-material technologies for thin wire and sheet production
    \item Advanced thin coating deposition techniques, particularly for scintillating coatings
    \item Focus on diagnostics for the target area's multi-MW proton beam environment
    \end{itemize}

\item Enhancement of non-invasive diagnostic techniques
    \begin{itemize}
    \item Advanced laser stripping methods for linac $H^{-}$ beam profile measurements \cite{LEM}
    \item Refined beam gas ionization \cite{BGI}and fluorescence monitoring systems
    \end{itemize}

\item Implementation of high-bandwidth data acquisition systems for high repetition rate operation and short bunches

\item Development of modern firmware and software solutions, including machine learning and AI, to interface beam instrumentation with machine optimization and protection systems, such as hadron transmission monitoring, beam auto-steering on target, and machine interlocks for excessive beam power

\end{itemize}

In line with what mentioned in Section~\ref{1:synergies}, collaboration opportunities exist with CERN Proton Beyond Collider (PBC) projects, worldwide neutrino facilities, neutron spallation sources, and high-power fixed target facilities.

\paragraph*{Muon cooling channel}

In the ionisation cooling channels, R\&D activities should address the challenge of providing beam instrumentation for transverse and longitudinal monitoring inside and in-between cooling channels. This would include longitudinal monitors with picosecond resolution needed for timing, synchronization and beam profile measurements, possibly using Cherenkov radiation or transition radiation \cite{OTR} and streak camera \cite{StreakCamera} or electro-optical techniques \cite{DEOS}.

\section{Radiation shielding}
\label{3:rd:sec:rad_shield}

Radiation to equipment poses a significant challenge for all stages of the Muon Collider complex, from the front-end up to the collider ring. Dedicated shielding configurations must be designed for the different machines (muon production source, accelerators, collider), in particular for superconducting magnets. The main purpose of the shielding is to mitigate the risk of magnet quenches, manage the load to the cryogenic system, and avoid magnet failures due to long-term radiation damage in magnet coils and insulation. In this section, we describe the required shielding design activities for the front-end, the rapid cycling synchrotrons and the collider ring. The activities summarized here are closely linked to the target design activities in Section~\ref{3:rd:sec:tar} (for the front-end shielding), the magnet R\&D in Section~\ref{3:rd:sec:mag}, as well as the R\&D work for the vacuum and cryogenics systems (see Sections~\ref{3:rd:sec:vac} and \ref{3:rd:sec:cryo}).

The shielding studies performed so far were mostly of conceptual nature, with the goal to establish key parameters like the required shielding thickness, which in turn has an important impact on the magnet apertures. In the next five-year period, it is foreseen to optimize the shielding configurations, explore new shielding materials, and progress towards a more realistic absorber design. It is also important to refine the radiation hardness requirements for different components (superconductors, magnet insulation, etc.). Collaborative efforts with other scientific communities (e.g.\,fusion) will play an important role. The experimental results and modelling efforts by these communities will provide an important input for understanding radiation effects and improving the radiation hardness of magnets and other equipment.
  
The different shielding design studies require radiation transport simulations with Monte Carlo codes like FLUKA. The shielding design will be an iterative process, in close collaboration with optics, magnets, vacuum, and cryogenic system experts. An important objective of the next five years is to refine the technical specifications for the shielding elements. In addition, first engineering design studies, including thermal and structural simulations, and integration studies need to be carried out, in order to progress towards a technically realistic design. 

\paragraph*{Radiation shielding in front-end}

The secondary particles produced in the production target require a sophisticated radiation shielding in front-end region, both around the pion/muon production target and in the downstream chicane. It is important to find the best compromise between physics requirements (pion/muon yield) and technological aspects (shielding and target engineering, magnet design, achievable aperture and field strength).  

A first conceptual shielding design for the superconducting solenoids near the target has been developed for a proton drive beam power of \SI{2}{\mega\watt} (with a graphite rod as target). One of the key objectives for the next five-year period is to converge on the front-end layout, which requires further iterations of the target assembly and the shielding and solenoid configuration (see also Section~\ref{3:rd:sec:tar}). With the present shielding design, the displacement damage in the HTS coils reaches about \SI{1E-3}{DPA/year}, which might be too high if the defects cannot be annealed during thermal cycles of the magnets. The maximum allowed radiation damage needs to be scrutinized, based on new experimental data and modelling work for HTS, which is expected to be published by  the fusion community in the coming years. In addition, the shielding requirements for a \SI{4}{\mega\watt} source need to be developed, which has not yet been studied in detail. The shielding configuration for a \SI{4}{\mega\watt} source will depend on the chosen target technology (graphite, liquid lead, tungsten powder), since the lateral shower development and hence the radiation load on the solenoids depends on the target material and geometry. 

As the conceptual shielding design evolves, it is important to advance in parallel with further engineering design studies and integration studies since this can affect the conceptual shielding and front-end design. This includes defining in more detail the shielding segmentation and performing thermo-mechanical studies for the shielding blocks. Another important aspect is the expected radiation damage in the shielding, which can reach \SI{1}{DPA/year}. The consequences of such high damage values need to be assessed.

Another important objective for the front-end design is to develop a concept for extracting the spent proton beam, which escapes from the target and carries a non-negligible fraction of the original beam power. First attempts to design an extraction channel passing through the normal-conducting solenoids in the chicane proved to be very challenging. The remaining radiation load on the normal-conducing solenoids remains unacceptably high and alternative solutions need to be elaborated with high priority. At the same time, it will be important to study how the radiation hardness of the normal-conducting solenoids in the chicane can be improved, for example by using more radiation-resistant insulation materials.

\paragraph*{Radiation shielding in the collider}

The shielding simulations for the collider arcs performed so far were only generic studies, which assumed a string of dipoles and a very basic geometry for the shielding and magnets. In the next five-year period, it is foreseen to refine the shielding design by developing a more realistic shielding configuration and by using a real arc lattice including quadrupoles and combined-function dipoles. In addition, the shielding studies for the experimental insertions will be revisited. The objectives include the following:

\begin{itemize}
\item optimize the conceptual shielding design by means of Monte Carlo simulations (heat deposition and radiation damage), in particular the transverse cross section of the shielding inserts in dipoles and quadrupoles, the front-face shielding of magnets, and the shielding of interconnects; the contribution of beam losses to the heat deposition and radiation damage has to be studied in addition to muon decay. 
\item contribute to the iterative design studies for the radial magnet build for the different magnet types in the ring (best compromise between shielding thickness, coil aperture, shielding and magnet temperature, thermal insulation, cooling scheme). Another important aspect is the definition of the operating temperature, which requires a close collaboration between various experts (cryogenics, vacuum), see also Section~\ref{3:rd:sec:cryo}.
\item assess if the shielding requirements change for a wobbled machine configuration, which is essential for mitigating the environmental impact of neutrinos.
\item in the present concept the shielding is also the vacuum chamber, it is required to study the coating of the shielding with a conductive layer that provides an adequate impedance
\item study the stimulated gas desorption produced by the muon beam decay products impinging the chamber, and simulate its effect on the vacuum stability and the required pumping distribution to cope with this gas load.
\end{itemize}

Based on the radiation transport studies, technical specifications for the radiation shielding in collider magnets and interconnects can be derived, including tolerances and specifications for the alignment and stabilization of the shielding. It is foreseen to develop a first engineering design and carry out first integration studies for shielding inside the magnets. This includes defining the shielding segmentation and devising a cooling system for dissipating the decay-induced heat deposition (\SI{500}{\watt/\meter}). Thermo-mechanical studies will be needed in order to dimension the cooling circuit and quantify the effects of the thermal load on the shielding. Iterations with the conceptual shielding studies, for example concerning geometrical aspects or the shielding material, will be important. 

Another objective is to study the radiation levels and cumulative dose in the collider tunnel and assess radiation effects in equipment and infrastructure (e.g., cables, optical fibres, electronics). This will give a first understanding about the need of shielded alcoves and other mitigation measures such as local shielding for electronics.  

\paragraph*{Radiation shielding in the accelerator rings}

The shielding concepts for the rapid cycling synchrotrons are less advanced than for the collider. First studies showed that a radiation shielding is needed for the superconducting magnets in the accelerator rings, but not necessarily for the room-temperature magnets. However, a more detailed assessment of the particle-induced heat deposition and radiation damage is needed in order to refine the requirements for shielding inserts and the front-face shielding of the magnets. As for the collider, it would be important to perform first engineering and integration studies for RCS shielding and iterate with the conceptual shielding design studies. This will define technical specifications, including tolerances and specifications for the alignment and stabilization of equipment. In addition to the shielding for superconducting magnets, it is very likely that a radiation shielding is also needed for the RF cavities in the RCS rings; this will require similar simulation studies as for the magnets.

It is expected that conceptual design for the radiation shielding in the different machines of the muon collider complex has reached a more advanced level after the initial five-year period. The second five-year phase shall be dedicated to an integrated engineering for most of the shielding configurations, with a more detailed design of the shielding components and a more complete integration study. In particular, the technical specifications for the different shielding parts need to be refined, which will allow for a more detailed mechanical design of the shielding. This also includes a more comprehensive analysis of the required support structures. It is also assumed that the technical design for magnets (e.g.\,target solenoids, collider magnets) will advance in this second five-years period, which will allow for an integrated engineering of the different systems. Furthermore, it will be important to study the requirements for integrating beam instrumentation in the shielding.

As the design evolves, the design choices need to be validated by the necessary simulations (e.g.\,radiation studies, thermo-mechanical simulations). 

Furthermore, the prototyping needs for shielding elements need to be elaborated, in order to validate the design choices. In particular, if more resources become available small-scale prototypes of the shielding insert for collider ring magnets could be built, in order to test assembly procedures and to perform tests of the cooling circuit. At a later stage, possibly after the completion of the 10-year period, larger-scale prototypes of radiation absorbers or mock-ups of entire shielding configurations could be envisaged. 

As the design of the machines and the tunnel infrastructure progresses, more comprehensive studies of the radiation fields in the collider and RCS tunnels are needed. This will define the radiation hardness requirements for equipment to be installed in the underground areas (according to radiation levels in the tunnels and alcoves). These studies will provide an important input for the civil engineering. 

%
%
\section{Cryogenics}
\label{3:rd:sec:cryo}

The main objectives of the Cryogenics Work Package are to investigate suitable cooling strategies for the several structures requiring a cryogenic environment, with special focus on superconducting magnet structures operating at or around \qty{20}{\K}. This includes defining integration studies and experimental validation as required. Magnet structures requiring lower temperatures are extensively covered by the High Field Magnet (HFM) programme , and as such the studies  and associated tasks are not repeated here. The Cryogenics Work Package strives to provide functional requirements for local heat extraction on magnets, SRF cavities, and radiation shielding. The activities within the Cryogenics Work Package have been grouped into four main deliverables:

\paragraph*{Evaluation of relative performance and added value of cooling strategies for the magnets and absorber structures:}

As has been done for the collider ring, heat load estimation and corresponding proposals of cryogenic cooling options need to be carried out for the accelerator ring, considering the overall cryogenic electrical consumption. Concerning the Proton Driver, Front End and Cooling Channel, work will focus on heat load estimation and identification of potential showstoppers/challenging magnets.
A common thread while working on the several systems that make up the complex will be the cryogenic distribution and the exergetic analysis of the accelerator and collider rings, with the overall sustainability of the cryogenic system in mind. This exercise can be divided into cooling schemes to support magnets with an operating temperature at around \qty{4.5}{\K} (Nb$_3$Sn), and with an operating temperature at or above \qty{20}{\K} (HTS). For the latter, cooling can be achieved in principle using either helium or hydrogen as operating fluid; a separate task is allocated to investigate the suitability of hydrogen for large-scale accelerator cooling, while the practical limits of helium cooling above \qty{10}{\K} or \qty{20}{\K} need to be established; these activities are foreseen for the period 2026-2030.
All of the proposed cooling schemes will have a reduced fluid content, the impact of which needs to be studied with respect to magnet protection, stability, energy extraction in case of resistive transition, as it brings definite advantages but also new challenges. A comprehensive list of pros and cons of cooling with reduced amounts of coolant will be established to highlight the broad range of impact it can have on the accelerator's infrastructure - this work is foreseen for the second half of the ten-year period (2030-2035).
Also from 2030 onwards, an in-depth assessment of the cooling solution for the radiation absorbers in the collider ring will to be carried out to investigate the trade-off between water, CO$_2$, or other alternatives, as appropriate coolant. 

During the first 5 years, the focus will be on carrying out studies for suitable cooling schemes for the magnet cold masses, and on establishing the practical limits of helium cooling as function of heat loads and operating temperature. The second 5-year period will focus on evaluating the impact of reduced cryogen content cooling schemes on the magnet's stability, quench management, and heat extraction. Investigation of suitable cooling solutions for the tungsten absorbers will also be addressed at a later stage.

\paragraph*{Functional requirement proposals for design and integration of the heat intercepts, cold mass coils, and tungsten absorber:}

Strategies should be devised to mitigate the effects of, or even reduce the number of cold-to-warm transitions, as they will have a determinant impact on the cryogenic heat loads of the accelerator ring. 
A study of local heat extraction between the cooling fluid and the magnet coil, including optimization of used materials and interfaces, will allow us to provide the magnet designers with functional guidelines on integration of the cooling features in the coil and/or cold mass structures.
In view of the considerable heat load deposited onto the collider magnets via the absorber and cold mass supports, a thermal design optimization of the supports along with appropriate integration of heat intercepts (thermal shields) will reduce the incoming heat loads to the temperature level of the magnets, which can have a significant impact on the overall power consumption of the \qty{10}{\km} collider ring.
A preliminary investigation into the impact of movers in the cryogenic infrastructure for the collider magnets shall be started, along with a study on compact re-cooling stations in the interconnecting regions, in which the field-free regions shall be kept as short as possible.
Most activities within this task are foreseen to start between 2027 and 2028, and run for an estimated 5-year period. Two of the tasks, namely the study of the impact of movers on the cryogenic infrastructure and the study of compact re-cooling stations at the interconnecting regions, are foreseen to start around 2032, when the remaining cooling and integration studies have reached a certain maturity.

\paragraph*{Explore potential changes in high-capacity cryogenic plants/processes that can lead to noticeable savings in energy consumption:}

Centrifugal compressors to replace existing technology on high-capacity cryogenic refrigerators can lead to a potential 25\% reduction in electrical consumption. R\&D, including a pilot facility, is required to bring this technology to an appropriate maturity level in time for possible use in the Muon Collider complex. This task is divided into two parts: first, within the initial five years, a study needs to be carried out to quantify the potential energy savings of replacing existing technology on cryoplants with centrifugal compressors. Depending on the findings of such study, the second 5-year period would see the installation and commissioning of a pilot facility using such compressors at CERN, to validate the technology in a relevant environment and to gain operational experience. It is worth noting that this type of R\&D has potential benefits for any future accelerator/machine relying on cryogenics and is not exclusive to the Muon Collider.

\paragraph*{Assessment of hydrogen as a viable technical coolant for accelerator magnets in parallel with safety assessments:}

It should be evaluated whether hydrogen is an option worth investigating as a coolant for the collider ring's superconducting magnets at around 20 K, as its high enthalpy of evaporation could enable sufficient heat extraction from the cold mass in near-isothermal conditions, using relatively small mass flow rates of hydrogen in narrow, confined channels, with a very low overall hydrogen inventory. Hydrogen is an abundant resource, and one could profit from investments coming from other fields of industry to offset the current shortcomings in cost, storage, and safety assessments. Its exploration as a viable technical coolant for future colliders needs to be investigated in parallel with preliminary safety considerations, so a fair trade-off can be made.
Starting in 2027, we intend to carry out an assessment of the safety implications of using hydrogen as a coolant in confined geometries in the context of an accelerator magnet. In parallel with the later stages of this activity, starting in 2029, a conceptual study on thermal performance of forced flow of 2-phase hydrogen in cooling channels will be conducted.
If the results from both the safety assessment and the theoretical thermal performance are considered satisfactory, the study should culminate in a proof-of-principle small-scale test stand, that applies the relevant safety requirements that would be mandatory in a relevant accelerator environment; this would take place from 2030 onwards.

The resources proposed for progressing in the cryogenics-related tasks to achieve a CDR in 10 years do not reflect present commitments, but express a projection of the resources needed (not presently foreseen within TE/CRG).

%
%
\section{Vacuum}
\label{3:rd:sec:vac}

The technical design of a muon collider requires a research program to identify technical solutions and optimize the design of several areas linked to the vacuum system (windows, pumping system integration, chambers, etc.). The most relevant areas that have been identified are:
\begin{itemize}
    \item Thin beam windows for muon cooling and absorber integration
    \item Wedge shaped absorbers for liquid hydrogen
    \item Integration of vacuum equipment in the presence of strong magnetic fields in the cooling channel.
    \item Vacuum chambers for the Rapid Cycling Synchrotron (RCS), eddy currents and impedance. Conceptual pumping system.
    \item Vacuum chambers for collider, integration with tungsten shielding, impedance and dynamic pressure. Conceptual pumping system.
\end{itemize}

\paragraph*{Thin beam windows and absorbers}
As discussed in Section \ref{1:tech:sec:vac:3}, the final cooling requires thin beam windows able to deal with high brightness beams. Commercial Si$_3$N$_4$ windows have been already mechanically characterized and exposed to high brightness proton beams.

It is necessary to develop new windows, evaluate other possible materials (i.e. Be, SiC, C, etc.) and study how the thickness affects the window resistance. The setup developed for mechanical characterization of commercial silicon nitride windows (see Figure~\ref{fig:experimental_setup}) can be exploited to test other materials. Until now only commercial windows have been tested. It is necessary to develop collaborations to produce ad-hoc windows with characteristics focused on the performance at the final cooling.

Thin windows are in general difficult to manipulate and the joining techniques need to be studied. Silicon based membranes are produced on silicon frames that are usually glued on another support. This solution is not radiation hard. A new proof of concept is being carried out at CERN to test the possible transfer of silicon nitride films to a metallic substrate, using similar techniques used for the production of vacuum chambers \cite{LainAmador:2693564}. If this trial succeeds, it will be necessary to execute an R\&D program to refine the technique and characterize the properties of the new windows as mentioned earlier. Finally, it is important to characterize the expected window deflection that will involve a small radial difference of the absorber length. This errors shall be included in the beam dynamic models.

To reach the best transverse emittance at the end of the final cooling of the muon collider, hydrogen is the preferred absorber \cite{palmerMuonColliderFinal2011}. Due to the large stopping power and high density of a liquid hydrogen absorber at the low energy of the final cooling, the amount of deposited power would involve extreme pressure excursions, reaching values above 100 bars after stage 6 according to \cite{ferreirasomozaIMCC2024} and as shown in Figure \ref{fig:pressure_evolution}.

\begin{figure}[h]
    \centering
    \includegraphics[width=10cm]{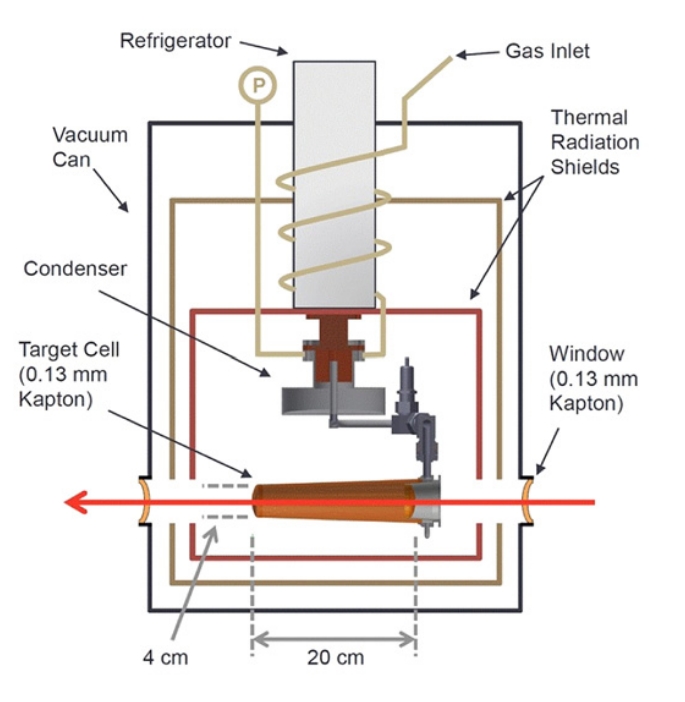}
    \hfill
    \caption{Example of a liquid hydrogen absorber for beam irradiation. Reproduced from \cite{Howell_2022}.}
    \label{fig:absorber_setup}
\end{figure}

On top of the mechanical characterization of thin membranes, it is proposed to build a cryostat to test small absorber volumes limited by different thin windows. The configuration can be similar to other experiments \cite{kendellenCryogenicTargetCompton2016, Howell_2022}, like shown in Figure \ref{fig:absorber_setup}. One cryo-cooler condenses the gas that is contained inside the absorber volume. The connection between the absorber and the cryo-cooler shall include heaters and thermocouples to control the temperature.

The volume of absorber shall be minimized to contain no more than 30 g of liquid H$_2$ inside a cryostat with a leak-tight volume of no less than 400 l. That would allow to contain the gas below 1 bar after any possible window failure and evaporation of the liquid. The system shall include a pumping system equipped with a purge to dilute the hydrogen with inert gas below the flammability limit before releasing it to the air.

The shock waves produced by the beam power deposition shall be monitored with different instrumentation like thermocouples, strain gauges, optical fibre sensors, etc. \cite{absorber_instrumentation1,absorber_instrumentation2}. The results shall be used to validate the modelling of the shock-wave generation and propagation through the absorber and its interaction with the thin windows at the extremities. The radiation hardness, signal transmission and electrical noise will be other important aspects to treat on the development of the instrumented absorber.

This setup shall be irradiated with a high brightness beam (i.e. protons) able to mimic the power deposition of the muon beam. The experiment shall be installed in a beam line able to deliver beams with size $\sigma_{RMS}<1$\,mm, to be compatible with small thin windows of maximum 10$\times$10\,mm. Each pulse shall be able to deposit from 10 up to 1000\,kJ/kg on hydrogen and achieve a similar power deposition than what is expected at the different stages of the final cooling of the Muon Collider.

The use of lithium hydride as an absorber is not excluded, and it may be required at some stages of the final cooling, either as single absorbers or as thick windows combined with liquid hydrogen to contain the pressure excursion. Prototypes of these windows shall be produced and mechanically characterized. The main difficulty with such study is the manipulation of a material that reacts violently with water and will degrade if exposed to air. That would require a facility prepared to produce, mechanize and manipulate them in inert atmosphere. Another important aspect is the brittle nature of this material, that needs to be studied under beam irradiation. To ensure leak tightness that may require to add a thin membrane that mechanically would rest on top of the LiH and will ensure the tightness of the setup.

This study should include absorbers with forced flow able to dissipate power at the rate required at the final cooling stages (approx. 50\,W) integrated in the small magnet bore. The cryogenic circuits of the different absorbers shall be optimized and a system to control the density of each cooling stage has to be developed.

This study has a very important impact on the final achievable luminosity and hence it should be prioritize during the first 5 years of the R\&D program. Also the design choices may be have an impact on the final cooling scheme.

\paragraph*{Wedge shaped absorbers for liquid hydrogen}

In Section \ref{1:acc:sec:cool:rec_cooling}, wedge shaped liquid hydrogen absorbers are proposed. It is required to simulate and build prototypes in order to confirm the viability of such an absorber. The first step of such R\&D program will require an evaluation of the maximum tolerable deflection of the limiting windows.

A set of concepts shall be firstly numerically analysed and later prototyped using different thin materials. As a first step thin metallic foils of tenths of \textmu m seem the best candidate, but the sharp angle may be incompatible with a large differential pressure. Alternative solutions to decouple the function of retaining the liquid and act as vacuum barrier may be required and shall be part of this study.

The performance of the rectilinear cooling will be impacted by the feasibility to produce wedge shaped absorbers containing liquid hydrogen. This R\&D program shall be carried out as priority during the first 5 years. Such program should validate the feasibility of the concept and the contrains for its integration in the rectilinear cooling.

\paragraph*{Integration of vacuum equipment cooling channel}
The muon cooling section is a succession of RF cavities, drift pipes and absorbers inside different solenoidal fields (see Section \ref{1:acc:sec:cool}). The mechanical apertures required at each section, position of windows, including possible Be windows inside the RF cavities, will define the volumes that need to be under vacuum. The integration of vacuum equipment and instrumentation will be very challenging. The requirements have to be defined (i.e. pressure requirements at the RF cavities) and the addition of ports for pumping in a congested area with the presence of magnetic fields has to be considered already at the conceptual phase of the cooling channel.
This work is foreseen to start from year IV of the R\&D period.

\paragraph*{Vacuum chambers for the Rapid Cycling Synchrotron}
As described in Table \ref{1:acc:tab:RCS_RFpars} the fast pulsing of the RCS would require special vacuum chambers to minimize the generation of eddy currents and have an acceptable impedance. Once the mechanical aperture is defined, it is necessary to define the chamber wall thickness that will define the minimum magnet aperture. In order to minimize the required thickness (and hence the magnet aperture), mock-up chambers have to be built to validate the numerical design. These prototypes will allow to test the use of coatings to reduce the impedance of these vacuum chambers \cite{impedance_ceramic_ch1,impedance_ceramic_ch2}. Depending on the vacuum configuration, the chamber may integrate active pumping (NEG coating) that may have an effect on the impedance budget.This work is foreseenn for the second half of the ten-year period (2030-2035).

\paragraph*{Vacuum chambers for the Collider}
The superconducting magnets at the collider will have to integrate the vacuum chamber, tungsten shielding, cooling to dissipate the power from the beam decay products, insulation vacuum, thermal shield and cold bore. To minimize the aperture it is proposed to integrate the tungsten shielding as vacuum chamber. That will require a study of the mechanical aspects of such chambers (i.e. end transitions, jonining techniques, etc.). The shielding shall also integrate the cooling system (ideally water) to intercept the power from the decay products.

To reach the impedance specifications, the tungsten chamber shall be coated with a material with low resistivity. Copper is the first candidate. The deposition of copper on tungsten surface inside a small aperture vacuum chamber has to be studied, to validate such option.

The conceptual design of the collider magnets shall include the pumping layout of the beam and insulation vacuum. This work is foreseen for the second half of the ten-year period.

%
%
\section{Radiation protection}

A key challenge of a high-energy muon collider is the neutrino flux generated by muon decays in the collider ring, which can reach the Earth's surface even at significant distances from the collider. Rare single events of these high-energy neutrinos produce secondary particle showers. The objective is to develop a comprehensive, site-independent dose assessment model to optimize collider ring designs and demonstrate that the impact of these showers is negligible. Efforts will focus on refining the dose model developed so far to provide accurate estimates of neutrino-induced radiation beyond the accelerator complex.

The current FLUKA studies will be expanded to include additional realistic geometries of relevance for the collider arcs and the interaction points. These studies will cover all relevant distances and various exit angles of the neutrino flux at the Earth's surface, including interactions with above-ground structures and other materials.
Finally, a comparison of the FLUKA simulations with other advanced Monte Carlo codes commonly used in radiation protection will be performed.

Concerning the folding of the simulation results with the collider lattice, further studies will be carried out to investigate the feasibility of the lattice, understand the impact of the mover system on beam dynamics and to refine the model around transitions between magnets and straight sections.
An important aspect which was not yet considered are the transitions between magnets and straight sections. Corrections coming from the overlaps between neutrino fluxes emerging from accelerator elements will be evaluated.

The implementation of the movers has to be further optimized and fine-tuned to ensure that beam operation is not compromised. Each individual step taken by the mover system would require interruption in the beam circulation and therefore reduce the time available for the collisions. In this context, the number of steps should be minimized given that sufficient dose mitigation is provided.

The Geoprofiler tool is set to undergo further improvements, including the development of a dedicated location optimization feature. This new tool aims to evaluate a wide range of potential configurations for positioning and orienting the collider while ensuring that the shafts of the interaction points remain within the CERN domain. The optimization process involves scanning a grid of parameters such as the accelerator's central coordinates (East, North, and Altitude) and its orientation angles in three dimensions ("slopes"). For each parameter, minimum and maximum values as well as the step size to control the granularity of the search can be defined. These parameters will generate a variety of potential configurations, which will then be tested using the Geoprofiler REST API (Application Programming Interface). For each configuration, Geoprofiler will compute the number of neutrino exit points from interaction regions and the shortest distance to neutrino exit points from the collider arcs. By applying predefined conditions to these outputs, the tool can identify the most promising configurations that minimize the neutrino-induced radiation impact. With sufficient geographical data, this algorithm can be adapted for use in evaluating potential collider locations worldwide.

For the surface maps, further investigation is needed to expand on the initial analysis of neutrino exit points along the ring. This involves refining the representation of the neutrino cone width and identifying intersecting exit points and assessing their overlaps. The in-depth analysis of the areas impacted by the neutrino flux should also consider height and depth clearances taking into account the respective neutrino flux width. 

Lastly, a comprehensive assessment of systematic uncertainties associated with the neutrino-induced dose model will be performed. This analysis will encompass uncertainties arising from neutrino interaction models, variations in interaction material properties, as well as the resolution of geographical data and the computational methods employed for ground intersection calculations. The relative contributions of these uncertainty sources will be quantified to to guide efforts aimed at improving the overall accuracy and reliability of the dose evaluation model.

\paragraph*{Planned work on conventional RP aspects}

Depending on the advancement of the muon collider and test facility designs, an RP assessment may be conducted in the future to evaluate shielding requirements and provide input relevant for the optimization of the conventional RP aspects.

Besides the unique neutrino radiation hazard, also conventional radiation protection challenges need to be addressed at an early stage of the project as they strongly determine the design of the facility. 
According to the radiation protection principles, the exposure of persons to radiation and the radiological impact on the environment must be optimised. To allow for such optimization, numerous radiation protection guidelines should be followed from the design phase onwards. Due to the high beam power of the muon collider, high prompt and residual dose rates require considerable shielding and access restrictions allowing only remote interventions in the radiologically critical areas of the complex. Also, the risk due to highly activated air and other gases as well as the impact of their release into the environment heavily influence the design. Next to that, the releases of potentially activated water and the activation and contamination of earth and groundwater surrounding the facility needs to be studied. Finally, the design should consider the minimization, processing and storage of radioactive waste.
The given aspects should be tackled to the required detail for the initial design proposal of the test facility as well as for the key areas of the muon collider complex.

The design of the test facility needs to deal with the radiation protection challenges mentioned above, particularly when aiming at a facility that is potentially upgradable to the order of 4 MW beam power. The latter would strongly drive the design choices. For example, sufficient space for adding the required shielding and other infrastructure (e.g. a morgue room) would need to be foreseen as well as civil engineering structures allowing to confine the highly radioactive zone and avoid streaming of radiation through shafts (e.g. chicanes). The depth of the facility should be chosen such to prevent ground water activation. The requirements put on a He vessel needed around the target, which is the considered a critical region, must also be investigated.

\paragraph*{Additional opportunities}

A study of particular interest, should further resources become available, is the investigation of the feasibility of reusing the LHC tunnel as the accelerator stage of the muon collider facility. Given that the majority of muon decays will occur within the collider ring, the accelerator ring has not yet been included in the neutrino-induced dose evaluation. However, the developed model provides a foundation for extending this analysis.

Finally, the tools, developed in the neutrino-induced dose evaluation process for the primary CERN-based configuration, are designed to be site-independent and can be applied to asses other potential muon collider locations.
  
\section{Software for the accelerator}

The tracking software for beam dynamics simulation, RF-Track ~\cite{bib:rftrack}, will be used to simulate and optimise the low-energy acceleration stage. The election of RF-Track for this complex, is based on the following points:
\begin{itemize}
    \item It is optimized for linear accelerators, such as the PA, RLA1 and RLA2 linacs.
    \item It can track muons of any energy and velocity and simulate their decay.
    \item It includes collective effects relevant to the proposed design such as space-charge, beam loading and beam-beam interaction.
    \item It can simulate multiple species simultaneously.
\end{itemize}
However, a complete simulation of the PA, RLA1 and RLA2, requires further work. At the present, the proposed future upgrades are:
\begin{itemize}
    \item Implementation of intrabeam scattering simulation in RF-Track: This collective effect becomes more important for low-energy and high dense beams. Therefore, its effect in the low-energy acceleration of the Muon Collider needs to be studied. 
    \item Simulation of circulating topologies: In RLA1 and RLA2, the bunches are expected to circulate several times along the linac. This effect raises concerns about the impact of long-range wakefield effects due to higher-order modes in the accelerating structures in a recirculating topology. Very few codes are able to correctly handle recirculation and long-range effects. PLACET2~\cite{placet2} was developed to simulate such recirculating topologies, and has recently been extended to simulate non-ultrarelativistic muons. However, it cannot calculate most of the collective effects required for the simulation of the low-energy subsections. For this reason, we plan to integrate the recirculation functionalities of PLACET2 into RF-Track. 
\end{itemize}
Once RF-Track is upgraded as described, it will be used for the lattice design and performance optimisation. Finally, the idea is also to improve the final cooling simulations further. In this respect, the next step is to refine the modelling of the energy straggling and introduce the dielectric constant to capture space charge effects in the absorber. 

\chapter[Muon Cooling Demonstrator Programme]{Muon Cooling Demonstrator Programme}
\label{2:demo:ch}

The Muon Cooling Demonstrator Programme will be an essential component of the muon collider R\&D programme. Muon cooling is required in order to deliver the required luminosities but it is a technology that has not been fully proven.

The muon cooling demonstrator will demonstrate
\begin{itemize}
\item Successful integration of cooling equipment.
\item Operation of the cooling equipment with beam.
\item Delivery of required beam physics performance.
\end{itemize}

\begin{figure}[htb]
    \centering
    \includegraphics[width=\textwidth]{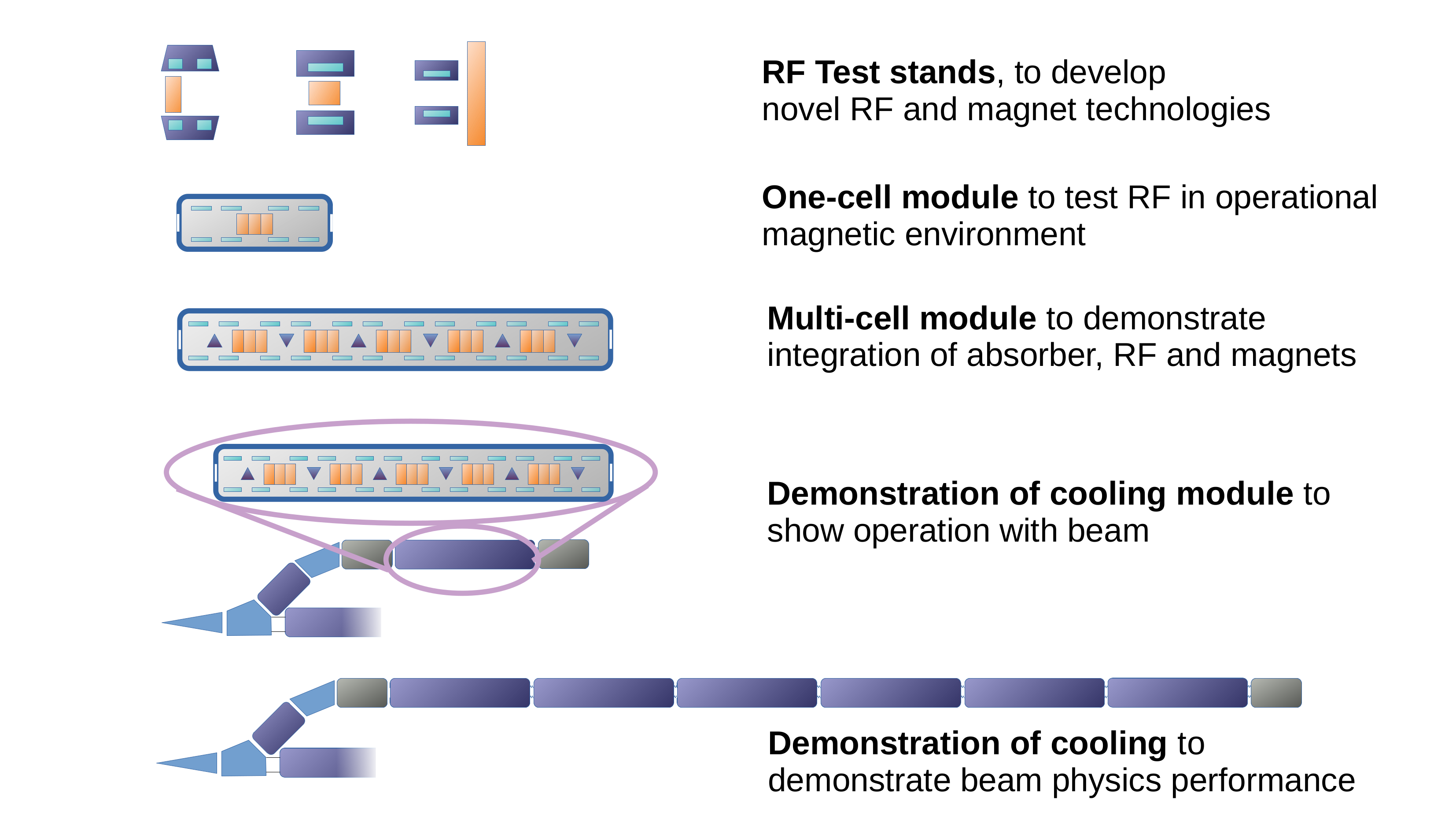}
    \caption{Schematic representation of the Muon Cooling Demonstrator Programme.}
    \label{3:acc:fig:demo_schematic}
\end{figure}

Delivery of the demonstration of muon cooling will require a programme of R\&D to understand and mitigate risks surrounding construction of the cooling lattice. The principle issues are:

\begin{enumerate}
\item The cooling cell has RF cavities and solenoids in close proximity. Solenoid fields are known to induce RF breakdown which must be understood in detail.
\item Warm-cold interfaces between adjacent RF cavities and solenoids require careful attention to thermal management.
\item Integration of ancillary equipment such as vacuum, RF power and beam instrumentation may be very challenging to implement in such a compact lattice.
\item Beam instrumentation must enable suitable commissioning of the equipment. For the beam demonstration in particular, where muon rates may be low compared to conventional beams, suitable instrumentation must still be implemented. 
\item The integrated facility must be operable in a routine manner. 
\end{enumerate}
In order to deliver this, a staged R\&D programme is envisaged, with each stage demonstrating the technology more fully:
\begin{enumerate}
\item Several RF test stands will be constructed to understand the limits to RF gradient that can be achieved in the presence of high-field solenoids.
\item A one-cell module will be implemented in order to test the operation of RF cavities in an operational magnetic environment.
\item A multi-cell module will be implemented to demonstrate integration of absorber, RF and magnets.
\item The multi-cell module will be operated with beam in order to demonstrate commissioning and operation of the cooling equipment with beam.
\item A cooling line comprising several cooling modules will be implemented to demonstrate beam physics performance.
\end{enumerate}
The programme is shown schematically in figure \ref{3:acc:fig:demo_schematic}. Each stage is described below.

\subsection*{Demonstrator Cooling Lattice}
The Demonstrator Cooling Lattice is a subsection of the cooling lattice described in section \ref{1:acc:sec:cool:rec_cooling}. Ongoing technical development of the lattice is described in section \ref{1:tech:sec:cool_cell}.

\subsection*{Relation to MICE}
The Muon Ionization Cooling Experiment (MICE) proved the ionization cooling principle based on the observation of emittance reduction in a single absorber \cite{MICE:2019jkl, MICE:2023vpa}. In MICE an ensemble of muons was passed through the absorber over a period of hours. The position and momentum of each muon traversing the apparatus was measured and the ensemble was assembled into a beam during offline analysis. The emittance of the ensemble was measured upstream and downstream of the absorber, proving the cooling principle.

As part of the Muon Cooling Demonstrator Programme, a pulsed beam of muons will be passed through a sequence of absorbers and RF cavities. Instrumentation will measure collective properties of the beam such as mean position and beam width. This will enable demonstration of satisfactory operation of the equipment, and eventually emittance reduction. The differences between the Demonstrator and MICE are summarised in table \ref{tab:mice_vs_demo}.

\begin{table}
    \centering
    \begin{tabular}{l|ll}
                      & MICE                 & Demonstrator \\ \hline
     Cooling type     & Transverse           & Transverse and longitudinal \\
     Number absorbers & Single absorber      & Many absorbers \\
     Cooling Cell     & Cooling cell section & Several cooling cells \\
     Acceleration     & No reacceleration    & Reacceleration \\
     Beam             & Single particles     & Bunched beam \\
     Instrumentation  & HEP-Style            & Accelerator-style
    \end{tabular}
    \caption{\label{tab:mice_vs_demo}The advantages of the Muon Cooling Demonstrator compared to the Muon Ionization Cooling Experiment (MICE) are listed. MICE proved the principle of ionization cooling. Demonstration of operation of a muon ionisation cooling scheme in an accelerator context is now required.}
\end{table}

\section{Demonstration of RF Operation}

Muons lose momentum in ionisation cooling when they pass through absorbers. The momentum is returned to the muons by RF cavities only in the longitudinal direction, resulting in a reduction of the muon beam emittance. Relatively high RF gradients are required in order to maintain phase stability of the muon beam, especially in the early stages of cooling where the beam has a large time and momentum spread. 

There is uncertainty in the electric field gradient that may be reliably achieved. The cavities will be immersed in strong magnetic fields. Magnetic fields are known to induce breakdown formation at lower gradients than normal cavities. This is a key technical risk for the programme that cannot be mitigated through simulation or design work. An experimental programme is required to manage this risk.

Studies have shown that the operation of cavities at high gradient requires that surface electric and magnetic fields be minimized to prevent material strain and field emission. Material strength also plays a critical factor in determining the acceptable strain induced in the material before cyclic fatigue results in degradation of the surface through the creation and motion of dislocations in the material. Doped copper alloys, and cold copper (cavities operated at cryogenic temperatures) have both been shown to achieve significantly increased high gradient performance; however, this has yet to be studied in detail in strong magnetic fields or in conditions similar to those found in a muon cooling cell. Other metals (W, Mo, Be, etc.) may also hold advantages in allowing for high gradient operation and should also be investigated. 

Experiments performed at Fermilab's Muon cooling Test Area (MTA) \cite{Bowring:2018smm, Freemire:2017htp} demonstrated the degradation in achievable gradient in a solenoid field but also demonstrated improvements to gradients by considering novel cavity materials and use of high pressure gas to insulate the cavities.  Further experiments, without magnetic field, indicate that operation at liquid Nitrogen temperature may enable higher gradients to be achieved (`cold copper' technology).

The effect was modelled by electrons, emitted from an imperfection on one side of the cavity, being focused to the other side of the cavity and causing damage. This in turn led to further electron emission, causing more damage. By insulating the RF cavities with high pressure gas or using materials that are less prone to electron emission and more resistant to damage it was thought that breakdown may be avoided. The MTA experiments indicated that gradients may be achieved, in the presence of magnetic fields, that are even higher than may be achieved in conventional copper cavities with no magnetic field.

The IMCC now requires that these results be validated with further tests, leading to a comprehensive understanding of the breakdown process. Additional tests are foreseen to examine factors such as: RF pulse length; magnetic field strength; magnetic field direction; cavity temperature; cavity geometry; cavity wall material; and RF frequency.

The results will be compared with requirements for RF breakdown rate as determined by a reliability study for the muon collider complex. This will be performed as part of the integrated facility design.

\subsection*{RF Test Stands}
In order to achieve this comprehensive experimental programme, a number of test facilities are foreseen. Several test stands are required to manage the expected throughput of cavities. Initiating the programme at multiple sites will have the added advantage of enabling study of different fabrication processes and preparing multiple sites for the eventual required RF system manufacture for the muon collider itself. Several potential host sites have expressed interest including CERN, INFN LASA, Daresbury Laboratory, and SLAC.

Initially, experiments at high frequency compared to the muon collider cooling system will be performed. High frequency cavities are smaller so that the cost of cavities, RF power and experimental infrastructure, in particular solenoids, is significantly reduced. Several sites have existing infrastructure appropriate for these frequencies. For example a compact quarter-wave resonator with a removable back plate operating at 3 GHz would enable testing of different materials with a rapid turn-around using existing solenoid and RF equipment. Such a cavity would enable research into conditioning algorithms with different materials to optimise the best route to maximise the available gradient and magnetic field that can be achieved with each material, and avoid deconditioning of brittle materials. Interchangeable backplates will be made of different materials to test how material properties impact the breakdown behaviour. 

Systematic studies can be performed including conditioning the backplates, assessing the best conditioning routine for operation in a magnetic field, studies of conditioned states, effect of pulse length, breakdown rate scaling with field and deconditioning. After testing, backplates will be put under microscopes to study the nature of the surface damage. Breakdown tests typically take 10-100 million pulses requiring significant analysis. Experimental conditions and waveforms will be recorded and comparisons made with the earlier unmagnetised experiments. The relationship between the breakdown rate and the experimental conditions will be explored with unsupervised machine learning and other modern computational approaches. 

Purchase or construction of a new solenoid having high bore, in conjunction with existing RF sources, would enable measurement of conventional cavities. This would enable validation and extension of the studies considered above. Different cavity geometries and materials could be studied.

In parallel multi-physics simulations are foreseen that capture electromagnetics, thermomechanics and beam dynamics to properly design and investigate the design and performance of these cavities to allow for optimization. Simulations of breakdown dynamics, including beam trajectories from field emission sites, multipactor effects, and plasma dynamics will be performed to help refine theoretical models and to compare with the measurements. 

As the experimental programme develops, lower frequency facilities will be developed. Lower frequency power supplies and large bore solenoids will be procured or manufactured. Due to the low frequency of operation the rf cavities and magnets may be relatively large and expensive. On the other hand the implementation of magnetic fields around the cavities provides an opportunity to study solenoid technologies which may be suitable for cooling. Several Tesla solenoid fields could be implemented with off-the-shelf magnets while higher fields will require custom magnets to be manufactured.

High-gradient operation will also require significant RF power delivered to the cavities. As seen in Table~\ref{rf:tab:cool_power} the power level will reach tens of GW peak. The implementation of RF technology accounting for the cavity losses by optimizing material selection, cavity design, and coupler design will have significant cost savings and reduction in overall operational complexity.

The resource requirement is dependent on the nature of tests undertaken and existing infrastructure which varies by site. A typical resource requirement estimate is shown in table \ref{tab:ch12:rf_test_resources}.

\begin{table}[h!]
\centering
\begin{tabular}{|l|c|c|c|c|c|c|c|c|c|c|}
\hline
\rowcolor{cornflowerblue!80}
\textbf{Year}&\textbf{I}&\textbf{II}&\textbf{III}&\textbf{IV}&\textbf{V}&\textbf{VI}&\textbf{VII}&\textbf{VIII}&\textbf{IX}&\textbf{X} \\
\hline
\rowcolor{cornflowerblue!30}
\multicolumn{11}{|l|}{\textbf{RF Test Stand}} \\
\hline
Staff & 0.1 & 0.25 & 0.25 & 0.5 & 0.75 & 0.75 & 2 & 2 & 3 & 3 \\ 
Post doc & 0 & 0.5 & 0.5 & 1 & 1 & 1 & 1 & 1 & 1 & 1 \\ 
Student & 0 & 0 & 0.5 & 1 & 1 & 1 & 1 & 1 & 1 & 1 \\ 
Material (kCHF) & 0.25 & 0.05 & 0.15 & 0.75 & 0.1 & 0.5 & 0.1 & 0.1 & 0.1 & 0.1 \\ 
\hline
\end{tabular}
\caption{Typical resource requirement for a single RF Test Stand. Existing equipment available for S-band was assumed. Purchase of a low bore magnet having modest field is made in Year I and an upgrade to L band capability in Year IV.}
\label{tab:ch12:rf_test_resources}
\end{table}

\section{One-cell module}

Within the European grant MuCol, a complete 3D model of a cooling cell will be designed (see section \ref{1:tech:sec:cool_cell}). This process will involve iterations with  beam dynamic simulations to determine an optimum between the efficiency of the process and the stresses and the forces exerted by the magnets, and the  beam loading.

On this basis, the IMCC is planning  to implement in a first phase  a one-cell module. This will enable testing of the magnetic lattice performance and the RF system performance in the simplest configuration near to operative conditions.

Preparation of the one-cell module will also necessitate developing supporting infrastructure. A klystron operating at 704 MHz will be acquired that can provide suitable peak power to reach the required gradients along with services for operation of RF and magnets.

\section{Multi-cell module}
The development of a longer test module to demonstrate integration and assembly of the superconducting solenoids, RF and other equipment is an essential stage in the delivery of the Muon Cooling Demonstrator. 

In muon cooling emittance is reduced by passing muons through an energy absorbing material. Momentum loss in the absorber results in a reduction of the momentum spread of the beam, resulting in cooling. Multiple Coulomb scattering causes an increase in momentum spread, which is in tension with the cooling effect and for a poorly conditioned beam can cause heating. The emittance reduction will be strongest for a beam with a large transverse momentum spread. This can be achieved by using a very tightly focused beam. 

Tight focusing is challenging to achieve while maintaining sufficient dynamic acceptance. The IMCC has designed a very compact solenoid lattice that provides tight focusing while maintaining a large dynamic acceptance.

Ionisation cooling does not naturally reduce longitudinal emittance. In order to reduce longitudinal emittance, the transverse and longitudinal coordinates must be coupled. This can be achieved by integrating weak dipoles with the solenoid lattice. The dipoles create and sustain a position-momentum correlation in the beam. A wedge-shaped absorber matching the position-momentum correlation removes more momentum from the higher momentum part of the beam resulting in a reduced momentum spread, at the expense of a wider beam. In this way longitudinal emittance is moved to transverse phase space where the ionisation cooling process can be effective.

Integration of this equipment into a single cooling lattice is very challenging. While similarly compact lattices such as Induction Linacs and Drift Tube Linacs have been implemented, the ionisation cooling scheme has a number of unique features that must be demonstrated. The compact and high-field solenoids produce fields that are highly overlapping. Adjacent coils have opposite polarity so that large forces act between adjacent coils that must be appropriately managed. Fringe fields are overlapping so that a large peak field is required at the coil to deliver suitable field on-axis. Adequate quench handling must be implemented so that magnets are suitably protected in the event of the quench of a nearby solenoid. 

Despite the closely-packed superconducting magnets, warm cavities must be operated with robust RF gradients as discussed above. RF power must be brought to the cavities as well as other ancillary systems such as frequency tuners and cooling. Superconducting coils must be thermally insulated from neighbouring warm equipment.

Construction of a test module will demonstrate satisfactory handling of thermal and mechanical integration issues. It will enable in-situ testing of RF cavities in the field configuration in which they will finally be operated.

\section{Muon Cooling Beam Demonstration}

Implementation of muon cooling requires that the challenges described above are met while enabling routine operation in an accelerator environment with beam. Successful operation of the equipment with beam will provide the confirmation of the ionisation cooling technology.

In order to bring beam through the equipment, a number of challenges must be addressed. The optical performance of the equipment must enable control of a pass band at the operational momentum of the cooling cell, ensuring adequate focusing and dynamical acceptance. The B-type lattices, that the Demonstrator is modelled on, operate between $\pi$ and $2 \pi$ phase advance. This second stability region is chosen because it enables tighter focusing at a cost of reduced dynamical acceptance. The beam will initially fill the dynamical acceptance of the solenoid lattice, so any reduction in dynamical acceptance will result in degraded performance.

Dynamical acceptance may be compromised by field imperfections and misalignments that must be carefully controlled even in the presence of large magnetic forces. The dipoles that provide transverse-longitudinal coupling may also introduce behaviour that can compromise the dynamic aperture. Preliminary studies indicate tolerances O(0.1~mm) without a correction scheme, which will be challenging to achieve. Steering corrections are envisaged using the coupling dipoles but no detailed study has been performed. Focusing corrections are envisaged by trimming the main solenoids. Design and implementation of this correction scheme must be considered.

Reacceleration of the beam requires appropriate operation of the RF cavities. The RF bucket size must be correctly matched to the momentum acceptance of the magnetic lattice, accounting appropriately for the energy absorber geometry. Random processes in the energy absorption process result in an RF bucket that does not have a clean edge. Transverse dynamical aperture is momentum-dependent, particularly in the vicinity of the stop-bands. This means that careful matching of the RF bucket to the transverse dynamical behaviour is required.

These corrections and matchings must all be performed in the context of an operational facility, based on measurement of the beam behaviour. Instrumentation based on electromagnetic interference caused by the beam, which is conventionally used in accelerators, may be challenging owing to the relatively low currents available for muon beams even in a bespoke facility. The currently envisaged instrumentation schemes take advantage of beam-intersecting devices, to which muons are relatively tolerant, but further details must be developed. The beam spot size is about 10-20~mm RMS while the length of each bunch in the train is roughly 100~ps. Closed orbit deviations induced by the dipoles are around $\pm$10~mm. Position and spot size measurements must be made at 1~mm accuracy or better while timing measurements must be made at the level of a few ps accuracy in order to steer the beam and phase the cavities correctly. The timing measurement in particular will be very challenging. For commissioning and routine operation of a cooling channel, this instrumentation will be required integrated into the cooling cells.

Additionally, dedicated instrumentation sections upstream and downstream of the cooling system are planned to measure emittance and match the beam into the cooling system. This will enable validation of the Demonstrator performance. The longitudinal and transverse phase space of the beam must be adequately controlled in these regions while a precise measurement of the beam properties is made. Although absorbers will not be necessary, debunching must be prevented so that the beam is transported successfully into the cooling system. Debunching can only prevented using RF cavities. Thus many of the geometrical constraints in the full cooling cells will also be present.

The operation of several cooling cells together is required to demonstrate appropriate handling of transverse and longitudinal optics. Because the lattice has a large longitudinal and transverse phase advance it is expected that any mismatch of the beam initial conditions or effects of poor lattice conditioning will be exhibited by filamentation after a few cells. For this reason a 5-cell module has been envisaged as an excellent test for commissioning, instrumentation and operation of the combined system. However such a module will not provide a measurable emittance reduction; several such modules will be required to demonstrate satisfactory emittance reduction, which is envisaged as the next stage of the programme.

Concepts for implementation at CERN or Fermilab are under development.

\section{Beam Instrumentation for the Muon Collider Demonstrator}
\subsection*{Proposed instrumentation inside the cooling cells}
The cooling cell environment presents significant challenges for instrumentation due to high magnetic fields, vacuum conditions, and complex mechanical integration within the limited available space. Additionally, a high background of low-energy electrons is expected, potentially contributing to noise. The RF cavities will operate at room temperature, while the absorbers will be maintained at cryogenic temperatures.

The required observables inside the cooling cells include position, transverse profile, and longitudinal bunch time structure. For position and transverse profile measurements, a silicon pixel detector appears to be the most suitable technology. However, minimizing the material budget is crucial to reduce beam perturbation. This project could benefit from the next generation of ultra-thin, flexible silicon pixel detectors currently being developed within the ECFA DRD framework \cite{silicon_MAPS_detector}. These detectors could potentially be mounted on the external surface of the absorbers to optimize space usage.

For longitudinal bunch time structure measurements, an active material capable of emitting Optical Transition Radiation (OTR) or Cherenkov light, read out using a streak camera\cite{StreakCamera}, could provide the required 1 ps resolution. However, integrating such a system within the cooling cell requires careful mechanical design and evaluation.

The absorbers will be filled with Lithium Hydride (LiH), which, according to preliminary calculations, will emit Cherenkov light when traversed by 200 MeV muons (the beta > 1/n condition is met). If LiH has sufficient transparency at certain emitted wavelengths, this signal could potentially be used to measure the bunch time structure or beam intensity without adding to the beamline’s material budget. However, this idea relies on assumptions about the transparency of LiH, which should be carefully studied.

\subsection*{Proposed instrumentation in between cooling cells}
Beam position, profile, bunch time structure, and intensity must also be measured between cooling cells. The primary challenge in this location is the limited space and complex mechanical integration.

For position, profile, and bunch time structure measurements, the same instrumentation as inside the cooling cells is proposed: ultra-thin, flexible silicon pixel detectors and OTR/Cherenkov materials read out with a streak camera.

For beam intensity measurements, a Fast Beam Current Transformer (FBCT) is proposed \cite{FBCT}. While FBCTs are typically designed for higher beam intensities, the CERN Beam Instrumentation group is currently conducting an R\&D program to develop an FBCT that could operate within the intensity range of the Muon Collider Demonstrator.

\subsection*{Proposed instrumentation for the upstream and downstream stations}
The environmental constraints are less stringent at the upstream and downstream stations, making it possible to measure the beam profile and position using scintillating fibers~\cite{XBPF}. These detectors provide the required performance at a lower cost than silicon pixel detectors and have already demonstrated their effectiveness in the MICE experiment.

For the remaining instrumentation, a Fast Beam Current Transformer (FBCT) is proposed for beam intensity measurement, while an OTR/Cherenkov detector read out with a streak camera is suggested for bunch time structure measurements.

\section{Demonstrator implementation at CERN}

\subsection*{The three options}

Three different sites have been considered for the implementation at CERN of a muon ionisation cooling demonstrator. They would all receive beam extracted from the PS as the optimal energy for muon production is in the range $5 - 20$ GeV. At present we are considering, for various reasons, extracting at 14 GeV. 

The proton pulse must be kept short so that the instantaneous muon current is as high as possible. Proton pulses may be extracted from the PS around 10 ns in length, which would yield 7 muon bunches in the train.
A machine development session will be proposed for the 2025 run at the PS to measure the real characteristics of the extracted beam. 

Protons will impinge on a target where they produce pions. Spent protons travel onto a beam dump. Pions emerging from the target are diverted by a dipole into a selection region. The pions are planned to be transported towards the cooling section and decay during this transfer line. Concepts for removing remnant pions and electrons are under investigation, for example by using a momentum selection chicane. The muons arising typically have rather large emittance. The IMCC has determined that a relatively late-stage cooling system will provide the best test of muon cooling, so that lower beam emittances must be delivered. In the absence of a full muon collider system, these emittances can only be provided by collimating the beam. Following collimation, a phase rotation system will create the short microbunches required for the cooling Demonstrator. An efficiency of around $10^{-5}$ muons per proton on target is expected for this system.

The possible sites are:  
\begin{itemize}
    \item A new site towards the BA1 access site to SPS, with beam coming from TT10
    \item A site reusing the old PS neutrino beam tunnel TT7
    \item A site re-using the CTF3 (ex LIL, Linear Injector of LEP) building. 
\end{itemize}

\subsection*{BA1-TT10 site}
The BA1 TT10 site has been considered for two reasons:
with the excavation of a new cavern for the target and the cooling line, the full beam power of the PS could be accepted (~80 kW)
such a cavern could be built in a place compatible with future injection from an eventual Superconducting Proton Linac (SPL), therefore reducing the cost of a future eventual muon collider. 

We did not study in details such solution since the maximum level of power available from the PS would not allow to study in more detail the effects of high muon intensities (e.g. collective effects), since the higher power can only be achieved by a higher repetition rate, not by a higher intensity per bunch. This site would therefore not bring a significant added value for the increase of cost. The only advantage would be the re-use of the infrastructure in case a muon collider would be approved for construction at CERN in the future.  

\begin{figure}
    \centering
    \includegraphics[width=1\linewidth]{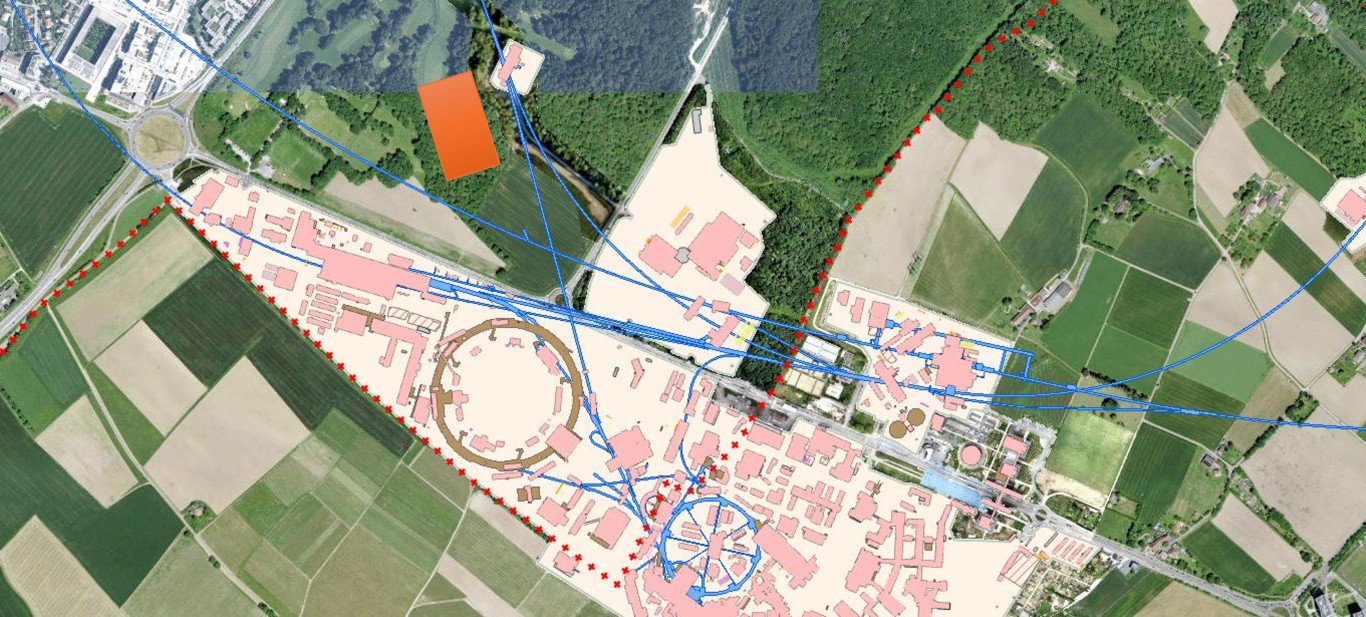}
    \caption{In orange, an approximate position for the BA1-TT10 demonstrator cavern}
    \label{fig:BA1-TT10}
\end{figure}
\subsection*{TT7 site}
TT7 is an old tunnel that was hosting the neutrino beam line and it is now used as storage. It has the advantage of a reduced cost for civil engineering compared to the BA1-TT10 site, although it will be necessary to enlarge it to place the momentum selection chicane, and to provide more space for the integration of the entire cooling line. The proton beam will be extracted from the PS with the existing extraction towards the AD and the SPS, and a simple dipole will be sufficient to deviate a bunch towards the transfer line to TT7. A study has shown that in principle existing magnets (or magnets identical to those already installed in the PS transfer lines) can be used to steer the beam at 14 GeV towards TT7. Since the initial part of TT7 is pointing upwards, it is necessary to transport the proton beam until it reaches the old target room that is horizontal. In order to have enough length for most of the pions to decay it is necessary to enlarge the target area. 

\begin{figure}[h]
    \centering
    \includegraphics[width=1\linewidth]{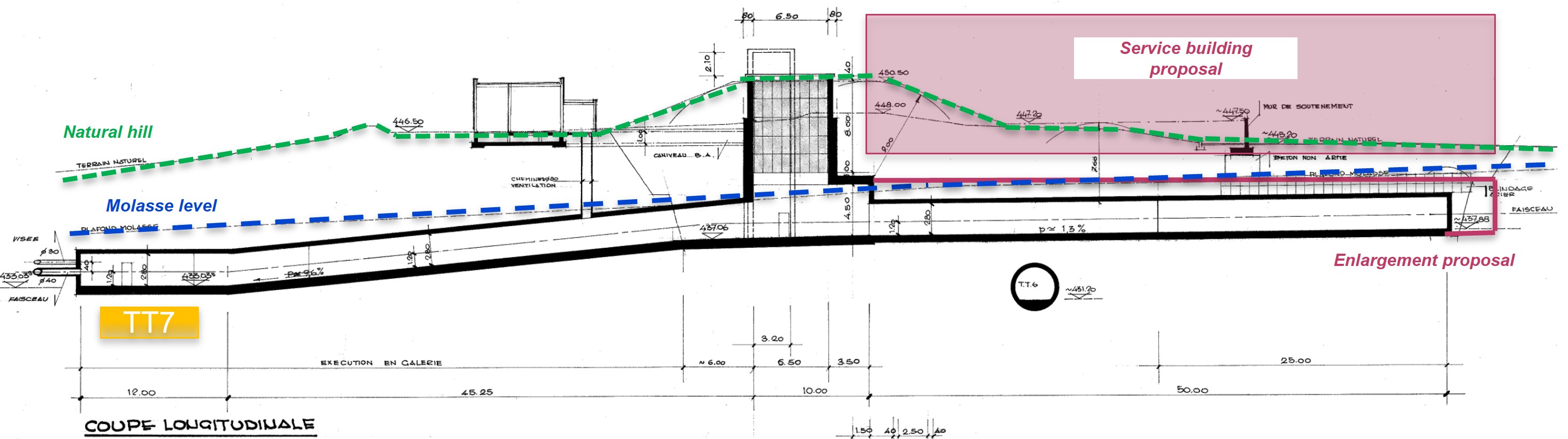}
    \caption{The TT7 tunnel seen from the side. The target and the chicane will be located at the centre under the shaft in order to avoid any radiation emission towards the surface. Beam is coming from left}
    \label{fig:TT7 - side view}
\end{figure}

Plans are being made to estimate the costs for an enlargement along the entire length of the downstream tunnel. About 50m are available to host the cooling line, that will allow installation of about 30 cells, assembled in 5-cell modules.   

\begin{figure}[h!]
    \centering
    \subfloat[\centering Civil Engineering Option 1]{{\includegraphics[width=0.4\textwidth]{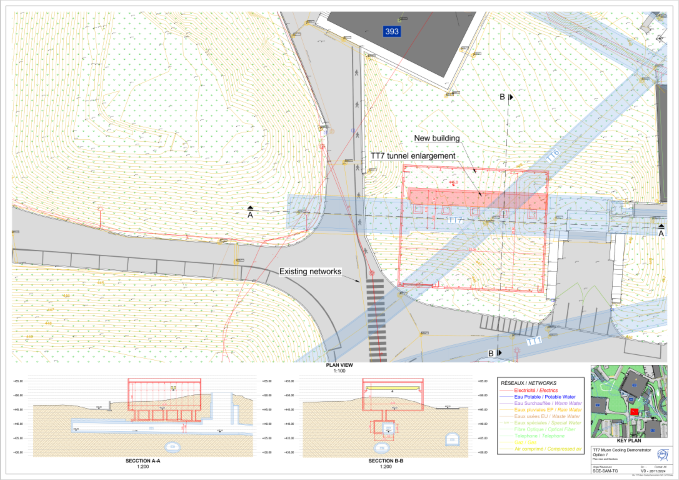} }}%
    \subfloat[\centering Civil Engineering Option 2]{{\includegraphics[width=0.4\textwidth]{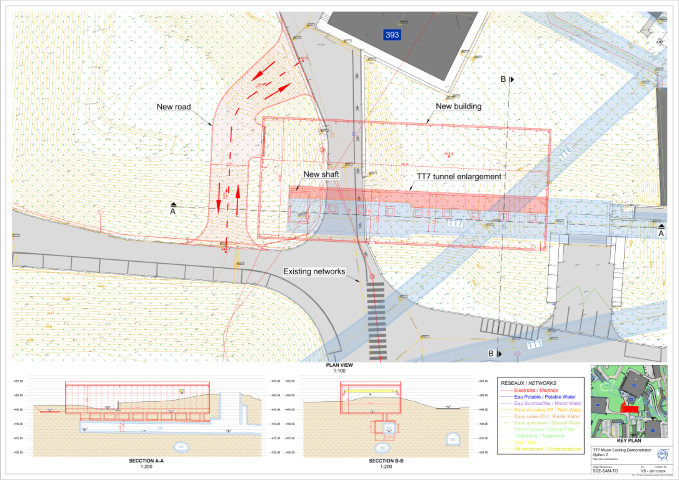} }}%
    \caption{Civil Engineering Options for Muon Cooling Demonstrator}%
    \label{fig:CivilEngOptions}%
\end{figure}

The excavation works consider a vertical increase in size by approximately 1 m and transversal increase in size of 3 m. The design of the TT7 enlargement needs to be adapted to cope with access, transport and safety requirements. Moreover, operation of the cooling demonstrator makes it necessary to install required services in a dedicated infrastructure on the surface. 
There are currently two different options proposed for the civil engineering interventions to TT7 to faciliate the Muon Cooling Demonstrator:
\begin{itemize}
  \item Option 1: Enlargement of 3m width and 1m height over 25m of TT7 starting from TT7 shaft and ending before the road that leads to AD .
  \item Option 2: Enlargement of 3m width and 1m height over approximately 50m of TT7 starting from TT7 shaft and ending at the end of the current TT7 tunnel, with an access shaft beyond that.
\end{itemize}
The two options can be seen in Figure~\ref{fig:CivilEngOptions}.

Option 2 would require a road diversion and as a result is a more complicated solution. Both solutions would have a surface building over the extent of the enlargement zone. The proposed size of this surface building would be 20m wide and 25m or 50m long (for Option 1 and 2 respectively).

A geotechnical engineering  study is currently underway to analyse these two options and assess the practicality from a civil engineering perspective. 

\begin{figure}[h]
    \centering
    \includegraphics[width=0.5\linewidth]{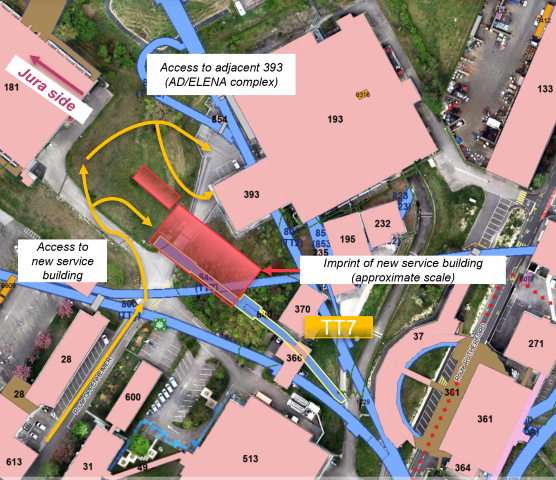}
    \caption{Top view of the the TT7 service building and different possible options to access the AD and the TT7  building}
    \label{fig:TT7 - top view}
\end{figure}

A full integration pre-study has been performed to identify possible showstoppers or critical aspects~\cite{tt7_report_24}. 
Several points shall have to be addressed if the proposal is accepted during the CDR/TDR phases, but no major showstopper was  identified. In the following a few pictures of the 3D model of the facility show a possible implementation in TT7. 

\begin{figure}[h]
    \centering
    \includegraphics[width=1\linewidth]{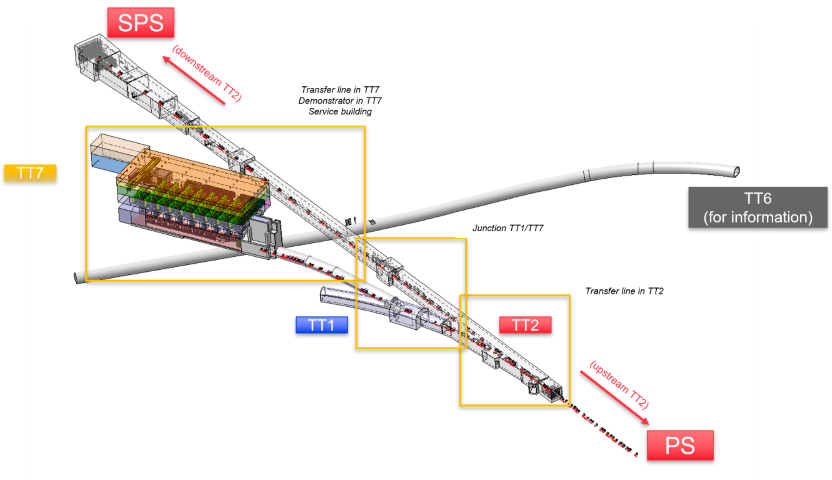}
    \caption{3D model of a possible implementation in TT7}
    \label{fig:TT7 - 3D integration model}
\end{figure}

\begin{figure}[h]
    \centering
    \includegraphics[width=0.75\linewidth]{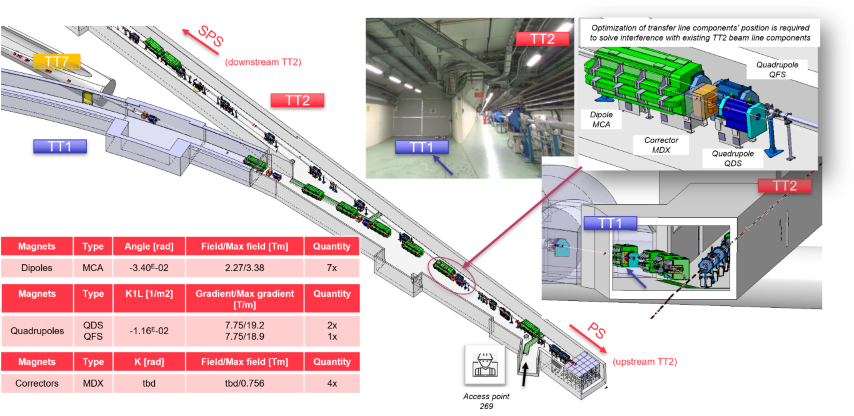}
    \caption{some aspects of the connection of the TT7 transfer line to the TT2 line coming from the PS. }
    \label{fig:TT7 3D model transfer line}
\end{figure}

\begin{figure}[h]
    \centering
    \includegraphics[width=1\linewidth]{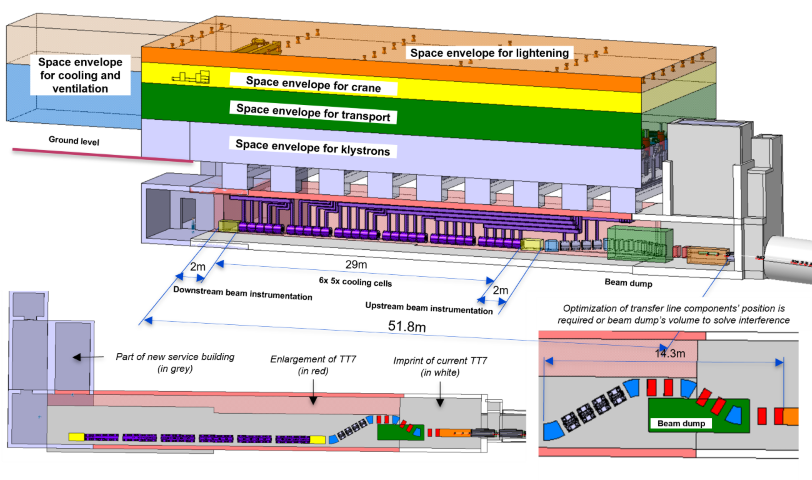}
    \caption{3D model of the TT7 cooling section and of the associated service buildings}
    \label{fig:TT7 - 3D model cooling line and service building}
\end{figure}

\subsection*{CTF3 site}
CTF3 is a building originally made to host the Linear Injector of LEP (LIL), then adapted to host the various versions of the CLIC Test Facilities (three in total, plus the present CLEAR facility). The advantage of this building is that it would require minimal interventions to make it compatible with a muon cooling linac, since it already has a klystron gallery on top, and one can build a customised target area at the centre of the old combiner ring, for a very limited budget with respect to the two previous options. The disadvantage is that there is no extraction pointing towards this tunnel, however it should be possible to add one during a long shutdown of the CERN accelerator complex. At present we believe such intervention could be carried out during LS4 (presently foreseen in 2034). 
CTF3 is also the best candidate at the moment to  test the first cooling cell, and later the 5-cell module. This activity will take place until LS4. Civil engineering for the target area can be carried out outside the shutdown period as the area is not included in the access chain to CERN accelerators, and there should not be any interference with the CLEAR facility, that is a standalone facility.

The tunnel connecting CTF3 to the PS has been closed by a concrete wall, which will be very simple to remove, partially or entirely. CTF3 provides about 100 m for the cooling line allowing more that 30 cells to be installed, or to install some test beam line to use the muon beam produced. It is to be noted that since we will select a muon beam of about 200 MeV, there will be no problem to stop any radiation with a few meters of concrete, therefore we do not expect any radioprotection issue. The beam characteristics would be the same as in TT7. A full integration and radioprotection study will be performed during 2025. the CTF3 option seems very attractive since it will most probably reduce at the minimum the cost of civil engineering to adapt the existing building. 

\begin{figure}[h]
    \centering
    \includegraphics[width=0.4\linewidth]{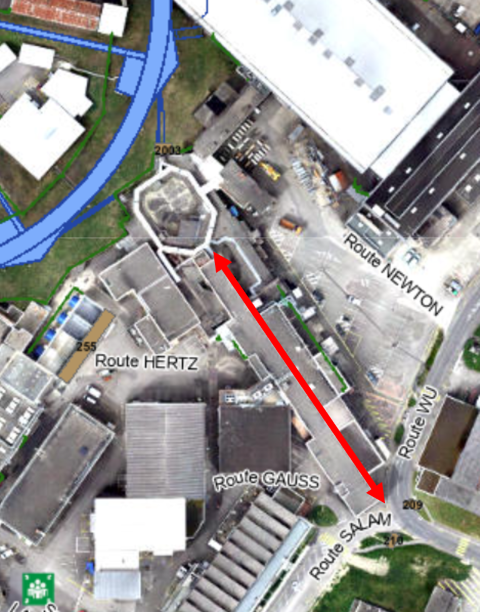}
    \caption{The CTF3 Building}
    \label{fig:CTF3 Building}
\end{figure}

\subsection*{Cost and timeline}

\begin{figure}[h]
    \centering
    \includegraphics[width=\linewidth]{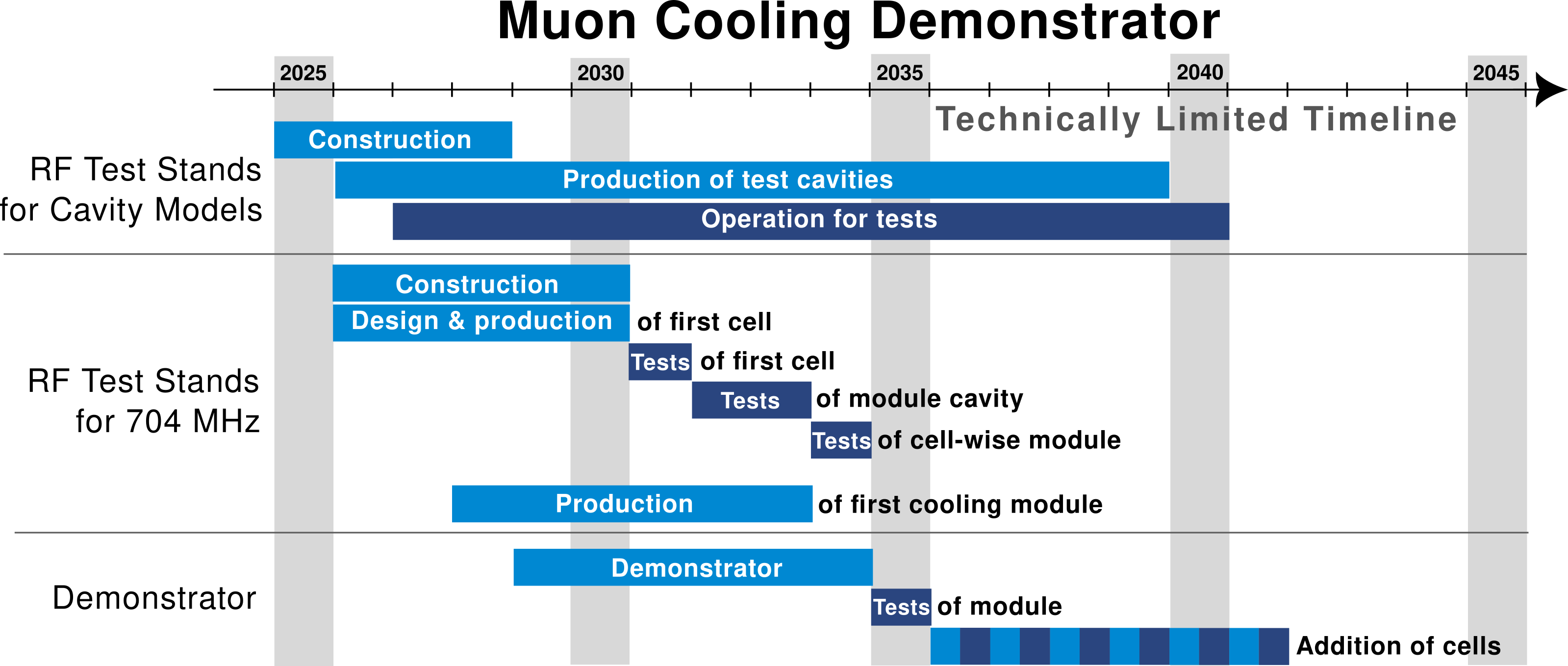}
    \caption{Estimated technically limited timeline for implementation of the Demonstrator R\&D programme in TT7.}
    \label{fig:ch12:demo_timeline}
\end{figure}

The estimated material and effort required to implement the TT7 option for the Demonstrator is listed in table \ref{tab:ch12:demo_resources}. The associated schedule is shown in figure \ref{fig:ch12:demo_timeline}.

\begin{table}[h!]
\centering
\begin{tabular}{|l|c|c|c|c|c|c|c|c|c|c|}
\hline
\rowcolor{cornflowerblue!80}
\textbf{Year}&\textbf{I}&\textbf{II}&\textbf{III}&\textbf{IV}&\textbf{V}&\textbf{VI}&\textbf{VII}&\textbf{VIII}&\textbf{IX}&\textbf{X} \\
\hline
\rowcolor{cornflowerblue!30}
\multicolumn{11}{|l|}{\textbf{One-Cell Module}} \\
\hline
Staff & 4 & 4.50 & 7 & 7 & 5.5 & 4.5 & 0 & 0 & 0 & 0 \\ 
Post doc & 5 & 6 & 5 & 5 & 5 & 4 & 0 & 0 & 0 & 0 \\ 
Student & 1 & 1 & 1 & 0 & 0 & 0 & 0 & 0 & 0 & 0 \\ 
Material (kCHF) & 600 & 2250 & 3900 & 4800 & 4350 & 700 & 0 & 0 & 0 & 0 \\ 
\hline
\rowcolor{cornflowerblue!30}
\multicolumn{11}{|l|}{\textbf{Multi-Cell Module}} \\
\hline
Staff & 0 & 0 & 0 & 5.5 & 5.5 & 6 & 7 & 7 & 7.5 & 7 \\ 
Post doc & 0 & 0 & 0 & 3 & 4 & 4 & 5 & 6 & 6 & 6 \\ 
Student & 0 & 0 & 0 & 0 & 0 & 0 & 0 & 0 & 0 & 0 \\ 
Material (kCHF) & 0 & 0 & 0 & 300 & 1500 & 7200 & 11200 & 12700 & 9900 & 6300 \\ 
\hline
\rowcolor{cornflowerblue!30}
\multicolumn{11}{|l|}{\textbf{TT7}} \\
\hline
Staff & 0 & 0 & 0 & 4.7 & 4.7 & 7 & 8.5 & 8.7 & 7.2 & 6.5 \\ 
Post doc & 0 & 0 & 0 & 4 & 5 & 5 & 5 & 6 & 6 & 6 \\ 
Student & 0 & 0 & 0 & 0 & 0 & 0 & 0 & 0 & 0 & 0 \\ 
Material (kCHF) & 0 & 0 & 0 & 300 & 2000 & 7200 & 14700 & 19700 & 21900 & 6300 \\ 
\hline
\hline
\rowcolor{cornflowerblue!30}
\multicolumn{11}{|l|}{\textbf{TOTALS}} \\
\hline
Material (MCHF) & 0.6 & 2.2 & 3.9 & 5.4 & 7.8 & 15.1 & 25.9 & 32.4 & 31.8 & 12.6 \\ 
FTE & 9.5 & 11.0 & 12.5 & 29.2 & 29.7 & 30.5 & 25.5 & 27.7 & 26.7 & 25.5 \\ 
\hline
\end{tabular}
\caption{Estimated resource requirement for the implementation of the Muon Cooling Demonstrator in TT7. The resource requirement does not include the resource for individual RF Test Stands.}
\label{tab:ch12:demo_resources}
\end{table}

 \FloatBarrier

\section{Demonstrator implementation at Fermilab}

\subsection*{Fermilab Accelerator Complex Evolution} 

Broadly speaking, the Fermilab proton complex is optimized towards the delivery of intense high-energy protons to produce a variety of secondary and tertiary beams for fundamental physics experiments. After the Proton Improvement Plan II (PIP-II) upgrade ($\sim$2030)~\cite{PIP2}, Fermilab will feature intense proton beams available for experiments at 0.8 GeV (after the PIP-II linac), 8~GeV (after the Fermilab Booster), and up to 120~GeV (after the Main Injector). At 120~GeV, the protons are delivered to the Long-Baseline Neutrino Facility (LBNF) beamline to produce neutrinos for the Deep Underground Neutrino Experiment (DUNE) program~\cite{DUNE:2020ypp}. At 8~GeV, the protons are delivered to the Muon-to-Electron Conversion Experiment (mu2e)~\cite{mu2e} program after passing through the Recycler Ring and Delivery Ring.

The P5 report~\cite{P5} affirmed support for the PIP-II and DUNE/LBNF projects in Recommendation 1b, and in Recommendation 2 also supports a newly proposed Fermilab upgrade known as Accelerator Complex Evolution Main Injector Ramp \& Targetry (ACE-MIRT). The goal of ACE-MIRT is to achieve 2+~MW at 120~GeV for the DUNE/LBNF program: firstly, by reducing the Main Injector cycle time from 1.167~s to $\sim$0.65~s, allocating a greater portion of Booster cycles to the long-baseline neutrino program and fewer to the 8~GeV program and; secondly, by improving machine reliability and modernizing accelerator infrastructure.  Next, by accelerating the pace of development for high-power low-Z neutrino targets. Lastly, with a new irradiated materials R\&D program to anticipate the impact of the higher beam power on other targetry materials (used in horns, windows, baffles, girders, cooling manifolds etc.).

Table~\ref{tab:FNALparam} displays the present, PIP-II era, and ACE-MIRT era capabilities of the Fermilab proton complex.

\begin{table}[htp]
\centering
\caption{Parameters for Fermilab proton complex. $^{\ast}$8-GeV beam power given for what is available simultaneous with 120-GeV program.}
\label{tab:FNALparam}
\begin{tabular}{| l || l |  l |  l |}
\hline
{\bf Linac} & Achieved & PIP-II & ACE-MIRT \\
\hline
Current & 20-25~mA & 2~mA & 2~mA \\
Energy & 0.4~GeV & 0.8~GeV & 0.8~GeV \\
\hline
{\bf Booster} & Present & PIP-II & ACE-MIRT \\
\hline
Intensity & 4.8e12 & 6.5e12 & 6.5e12 \\
Energy & 8~GeV & 8~GeV & 8~GeV \\
Rep. Rate & 15~Hz & 20~Hz & 20~Hz \\
8-Gev Power$^{\ast}$ & 25~kW & 80~kW & 12-24~kW \\
\hline
{\bf Main Injector } & Present & PIP-II & ACE-MIRT \\
\hline
Intensity & 58e12 & 78e12 & 78e12 \\
Cycle Time & 1.133~s & <1.2~s & $\sim$0.65~s \\
120-GeV Power & 0.96~MW & $\sim$1.2~MW & 1.9-2.3~MW \\ 
\hline
\end{tabular}
\end{table}

Simultaneous with the PIP-II and ACE-MIRT era experimental program, the Fermilab proton complex will have the capacity to accommodate additional users. The P5 report recommends (R4g/AR12) Fermilab develop a strategic 20-year plan for the Fermilab accelerator complex consistent with a long-term vision including neutrinos, flavor, and a 10 TeV pCM collider. P5 further recommends (R6) collider-specific R\&D including large-scale demonstrator project.

The ample opportunities for siting a muon cooling technology demonstrator at Fermilab are described in the next section. Fermilab longterm strategic planning will also include specific scenarios for muon collider at the Fermilab site, and some of the recent thinking on that subject was presented for IPAC~2024~\cite{Eldred_IPAC24, Stratakis_IPAC24}.

\subsection*{Fermilab Demonstrator Sites options}

\begin{figure}[htb]
  \begin{center}
    \includegraphics[width=0.8\textwidth]{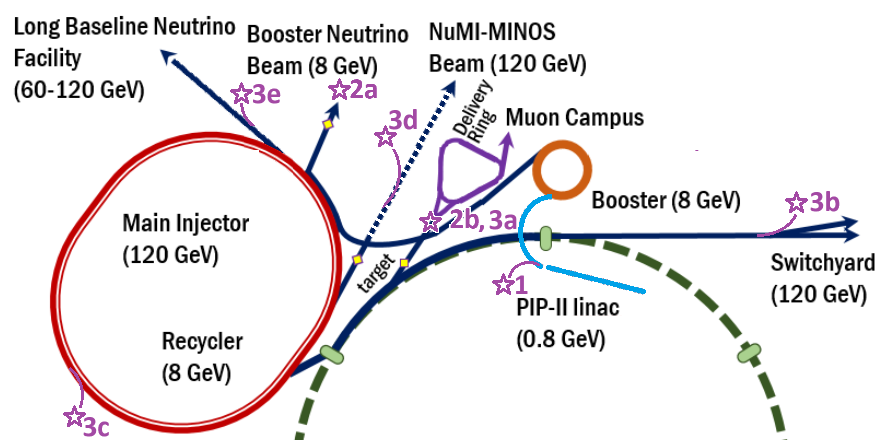}
  \end{center}
\caption{Diagram of the PIP-II/ACE era Fermilab proton complex. Magenta stars indicate candidate locations, which are described in greater detail below.}
\label{sec3c:sites}
\end{figure}

A muon ionization cooling technology demonstrator hosted at the Fermilab proton complex would benefit from the extensive infrastructure and in-house expertise in muon production, superconducting technology, and particle detectors. Moreover, there are several candidate sites at the Fermilab proton complex for the muon ionization cooling technology demonstrator, with different parameters. Figure~\ref{sec3c:sites} shows locations of candidate sites.

A brief description of the candidate demonstrator sites follows:

\begin{enumerate}
    \item 0.8-2.0 GeV H- particles from the PIP-II linac, either as a greenfield site in the Tevatron field (1) or following the proposed mu2e-II line to muon campus (not shown). The demonstrator could possibly benefit from co-location with other proposed facilities at the same site. The PIP-II beam will be available at 0.8~GeV, but the impact of proposed energy extensions of up to 2~GeV will be assessed.
    \item 8 GeV protons from the Booster, either at the present day short-baseline neutrino (2a) or muon campus site (2b). 
    \item 8-120 GeV protons from the Main Injector, using a modification P1 Muon Campus line (3a), P1 Meson line (3b), MI Abort beamline (3c), NuMI beamline (3d), or LBNF beamline (3e). The site would need to be consistent with continued ACE-MIRT era operations of LBNF.
\end{enumerate}

Several hundred kW of the 0.8~GeV PIP-II Linac beam power could be provided with no adverse impact on any other program. The linac beam would arrive in $\sim$1~ns bunches of 1.4e8 H$^{-}$ particles, with any bunch-by-bunch chopping of the 162.5~MHz beam (not to exceed 2~mA). However, the energy of the PIP-II beam may be too low to efficiently produce $\mu^{-}$ and $\mu^{+}$.

Siting the demonstrator to use 8~GeV protons, on the other hand, has the advantage of being a beam energy appropriate for a muon collider proton driver scenario. This demonstrator would therefore also serve as a (low power) test of muon production and capture for the muon collider program. The possibility of re-using existing infrastructure (tunnel, power, cryo) at the short-baseline neutrino or muon campus site will be investigated. The 8~GeV beam arrives in 81 1.2~ns bunches of 8e10 protons, separated at 19~ns intervals (for a 1.5 $\mu$s pulse).

Fermilab also offers for the ability to stack and accelerate the beam further, using the Recycler Ring and Main Injector. The Main Injector beam would be available at energies from 8-120 GeV, with up to 486 2~ns bunches of 1.6e11 protons, separated at 19~ns intervals (for a 9.6 $\mu$s pulse). The best beamline and site to deliver the Main Injector beam is still under investigation, although the MI Abort may be most promising - the beamline was historically identified as the US-based siting option for the proposed NuSTORM program~\cite{nuSTORM:2022div}.

\chapter{R\&D Programme Synergies}
\label{sec:Section08}

The muon collider R\&D will develop a number of important technologies that have application both within particle physics, to other related facilities, and beyond particle physics both in the science community and in the larger society.

\section{Technologies}
\subsection*{Magnets}

\begin{figure}[!h]
\includegraphics[width=\textwidth]{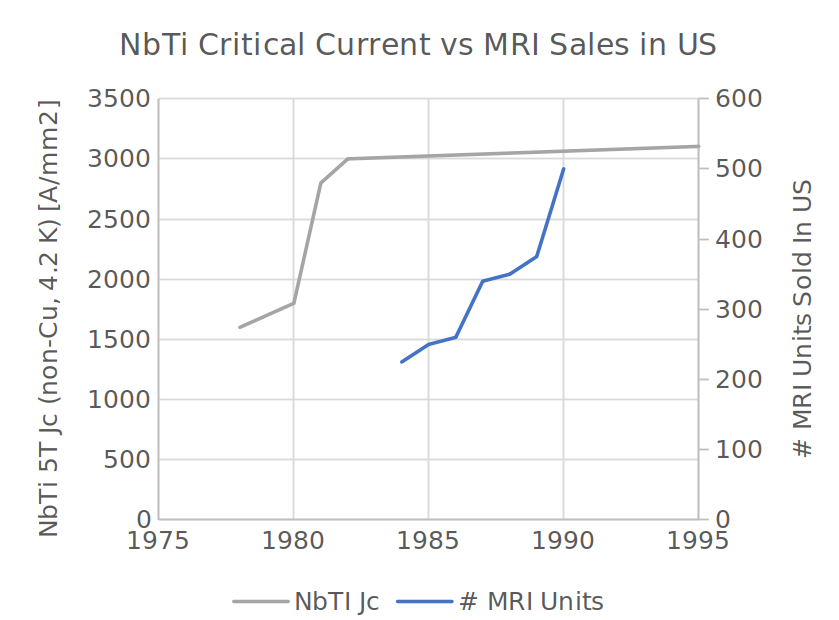}
\caption{Timeline of the performance of industrially produced Nb-Ti, measured as the superconductor critical current density, contrasted to the timeline of MRI units sold in the US. \label{fig:magnet_timeline}}
\end{figure}

Reaching  mature magnet technology required to produce, cool, accelerate and collide muon beams, is a grand challenge for accelerator magnet science. Achieving such advances requires development of new concepts for superconducting and resistive magnets, as well as suitable power converters that manage efficiently the large stored energies and powers required to drive the magnetic circuits. 

This is not new: indeed, the need for fundamental advances in accelerator magnet science has been a characteristic of the development of accelerator magnets since inception. We can recall as good examples the development of Nb-Ti, initiated on large scale by the Tevatron, through HERA, RHIC, the SSC R\&D phase and finally the LHC. A remarkable result of this endeavor is the development of industrial capability for the high performance and high quality Nb-Ti that feeds the MRI market. A demonstration of this achievement is reported in Fig.~\ref{fig:magnet_timeline}, relating the improvement in Nb-Ti critical current density Jc to the growth of the MRI production.

A similar development was necessary for Nb3Sn, to evolve the wire performance beyond the state-of-the-art achieved in the 1980’s and used for the construction of the thermonuclear fusion project ITER. This development was initiated by the DOE US Conductor Development Program in the late 1990’s and beginning of 2000’s, and is now in demonstration phase in the HL-LHC magnets. It is expected that this development may bear fruits of improved performance and production control for future applications of Nb3Sn, e.g.~in future fusion reactors, though a killer application comparable to MRI has not emerged yet.

In the case of the Muon Collider, we plan to profit from the recent advances in HTS magnets and stimulate further developments. This field of magnet science is new, very dynamic, and, most important, the challenges are synergic to several other fields and applications.

Development of the muon collider will deliver important advances in key magnet technologies:
\begin{itemize}
\item Ultra-high field solenoids - Advancing the HTS technology for high- and ultra-high-field solenoids will enhance the capability to study materials under extreme conditions. This is relevant to the creation and chacterization of new quantum states, measurement of electronic properties of components such as those used in quantum computers, the discovery of new materials, including novel superconductors, or the development of nanostructures as required to increase processing power and storage in next generation electronics, or more efficient batteries. 

Higher field will also augment the diagnostic power of NMR in structural and functional inorganic and organic chemistry, and biology. Relevant advances enabled by higher NMR fields include understanding biological processes, mapping protein structures of diseases, necessary to develop treatments, or monitoring pollutants in soil and water, including their origin.

The development of solenoids at the upper end of field and aperture at lower electrical consumption than presently possible will impact directly analytical research infrastructures for materials and life sciences, such as the European Magnetic Field Laboratory (EMFL) or the US National High Magnetic Field Laboratory (NHMFL). 

High magnetic fields made possible by reaching the goals of this R\&D will also improve analysis power and tune-ability of LS's and FEL's, and extend the reach of neutron spectroscopy instruments for the analysis of the structure of matter.

\item High field, low consumption, compact and large bore solenoids - HTS solenoid technology of the type envisaged in the 6D cooling cells will play a major role in the next step in research MRI (Magnetic Resonance Imaging), increasing resolving power, and enabling higher resolution images. HTS magnets with simplified cryogenics, possibly cryo-free, will reduce healthcare expenses and facilitate market penetration of clinical MRI, which still has not reached full exploitation for world regions where high-technology support is not readily available. 

\item High field, large bore solenoids - Widespread introduction of HTS materials in high field magnets with large bore and stored energy can lead to simpler and more efficient fusion plants. Higher field can be used to massively increase the fusion power, for a given plant size. At the same time, simplified cooling, operating at 20 K, reduces the thermal shield and space requirements in a fusion reactor, making the overall plant smaller, and easier to operate. Indeed, HTS magnet technology may be crucial to resolving the long-standing question on whether thermonuclear fusion will ever be a reliable source of energy for humanity.

\item High field, low consumption, compact dipoles and quadrupoles - Compact, medium field, low consumption HTS accelerator magnets will benefit accelerators and gantries for particle therapy, decrease their size and electrical consumption, thus facilitating the diffusion of advanced cancer treatment centers. 

HTS magnet technology with compact 3D polar windings producing high fields will benefit generators and motors, increasing their power density, reducing their weight, and making them more efficient. Even if higher energy efficiency is only a marginal benefit, the increased power density and reduced mass allows for more powerful machines which are either not possible, or very complex, from the engineering point of view. One such example is the generators used in windmills, greatly benefitting from a reduced nacelle mass. A lighter nacelle allows building a higher mast and larger rotor blades, thus also increasing the generator power.

\item Fast ramping resistive dipole systems
\end{itemize}

\subsection*{SRF technology} 
The muon collider proton driver and accelerator chain requires several 1000s of SRF cavities in the range from 300\,MHz to 1300\,MHz. Starting from LEP2, the SRF cavity and associated cryomodule technology has been developed for accelerators over the last few decades for other circular : LHC, RHIC, etc and linear : TESLA, ILC, XFEL, LCLS2, SHINE, etc accelerators for HEP and other application. Muon collider fully rely on this developments and has a lot of synergy with the ongoing SRF R\&D. Furthermore, due to very short life time of the muons, as high as possible accelerating gradient is mandatory to accelerate muons up to 10\,TeV collision energy. This motivation drives the SRF R\&D towards even high gradient not only at 1.3\,GHz but also at lower RF frequencies down to 350\,MHz. In addition to the standard bulk Nb SRF technology which is well developed, new directions potentially open the way to higher gradients including Nb3Sn, thin films, and even HTS coatings. Driven by the large scale HEP accelerator facilities the SRF R\&D has already found and will find even more other applications for light sources, nuclear physics, medical and industrial accelerators. 

\subsection*{Other technologies}
The R\&D programme to support RF cavities will yield important developments for next-generation RF sources. In particular, significant R\&D will be undertaken on normal conducting RF cavities in the frequency range 300-3000\,MHz with studies focusing on achieving high gradients even in the presence of magnetic fields.

The ability of targets to withstand high beam powers is a fundamental limitation in the performance of many proton accelerators built to create secondary particles. Fixed targets cannot support beam powers about around 2\,MW. Fluid targets can in principle withstand much higher beam powers. The first generation of fluid targets, developed with liquid metals, have been limited by heating effects in the liquid metal. Development of next generation fluid targets would alleviate this fundamental limitation in proton beam current.

The high-field magnet, RF and targetry activities together can have application in the creation of bright muon beams. This technology is, in itself, an important technique that has application in a number of different areas.

Developing solutions for cryogenic operation at temperatures above liquid is one of the most crucial and impactful. Helium cooling at 20 K, if suitably designed, can improve energy efficiency by a factor four with respect to operation at 4.2 K, and reduces the complexity and costs of cryogenics systems. This would enable broader scientific, societal and industrial applications.

\section{Applications}

\subsection*{Accelerator magnets}
A first clear synergy is the need for HTS accelerator magnet development for a future collider. The requirements for the 10\,TeV Muon Collider ring are presently beyond the state of the art and will likely need an all-HTS solution to reach peak field values of 16\,T, while operating in the temperature range of 10\,K to 20\,K with a minimal amount of cryogen. These requirements resemble closely those established recently for the hadron stage of the Future Circular Collider, FCC-hh. In fact, the development of high performance and high efficiency HTS accelerator dipoles and quadrupoles will likely profit any future collider at the energy frontier.

Lower field particle accelerators may profit from increased stability and more efficient cooling that can be achieved using HTS. The R\&D on rapid acceleration techniques will yield more efficient rapid cycling magnets that can benefit Rapid Cycling Synchrotrons.

\subsection*{Fusion}
We have identified several areas where the developments of magnet science and power converters have synergies with the needs of thermonuclear fusion. A good example is the large bore and high field HTS solenoid around the proton target in the muon production and capture channel, resembling similar magnets that produce the field swing in tokamaks (central solenoid), and the pinch solenoids in magnetic mirrors. Another example is the technology to be developed for the high field solenoids in the 6D and final cooling, based on non-insulated HTS windings to ease quench protection. Similar technology has been used in large toroidal field model coils, demonstrating successfully high field, but at the same time indicating that quench management still needs to be improved. A further example is the RF source used for heating of the fusion plasma, which is usually situated in a region having high magnetic field and suffers from the possibility of RF breakdown.

In general, future magnetically confined fusion power plants will need to reduce system complexity to become economically viable, and may require high field to reduce the size of the reactor. This field will hence profit from advances in the HTS magnet technology developed for the Muon Collider, offering simpler cryogenics at improved performance. A last point of contact among the development and studies for the Muon Collider and thermonuclear fusion is the use of power converters delivering high energy in pulses of short duration, high repetition rate and large number of pulses. This is one of the main challenges for the rapid cycling synchrotrons of the muon acceleration stage, whose pulsed power requirements resemble those required for magneto-inertial fusion.

\subsection*{Material and life science in high magnetic field}
A magnetic field induces new states of matter, and can be used to precisely study them. The ability to measure the properties of these new states, and the measurement resolution tend to grow with field, in most cases with power dependencies. The result is a constant call for higher fields in user facilities. The present state of the art of all-superconducting solenoids for material and life science is the 32\,T LTS/HTS hybrid at the National High Magnetic Field Laboratory of Florida State University (NHMFL). Higher fields can be reached presently only resorting to resistive magnet. The present highest steady state field facility is at the High Magnetic Field Laboratory of the Chinese Academy of Science (CHMFL), with a field of 45.22\,T achieved in 2022. Though functional, facilities of this class are very power-hungry, with continuous electric consumption in the range of tens of MW. The next step in high field user facility has been projected at 40\,T, and is the subject of R\&D and demonstration in laboratories in the US (NHMFL), Europe (Super-EMFL). The work on the final cooling solenoid, a 40\,T-class, 50\,mm bore all-HTS magnet, is clearly highly relevant for the companion developments in worldwide high-field magnet laboratories, promising compact, high-performance, and high-efficiency solutions to future UHF user facilities.

\subsection*{Nuclear magnetic resonance (NMR) and magnetic resonance imaging (MRI)}
The development of high-field and ultra-high-field HTS solenoids for the 6D and final cooling quoted earlier can also support development of NMR instruments. The present state-of-the-art for industrial NMR machines is the 28.2\,T, 1.2\,GHz system produced by Bruker. This magnet is an hybrid made with an LTS outsert and a HTS insert. Introducing HTS technology has not only boosted field performance, but also significantly reduced the size and consumption of the system. The next step in NMR machines is projected in the range of 30\,to 35\,T, and may profit directly from the technology developed for the HTS final cooling solenoid, exploring a range of field which is comparable, and even higher. This development may also have implications for MRI magnets, though this is presently not as direct as for NMR. The magnet program for a Muon Collider, in particular for the 6D cooling section, will need to demonstrate engineering and reliable operation of HTS solenoids in the medium to high field range (3\,T to 10\,T), and temperatures well in excess of liquid helium (10\,K to 20\,K). Success of this development will raise a definite interest in helium-free technology for MRI systems operating in the present clinical range (i.e.~up to 7\,T) but with simplified cryogenic systems, which will increase penetration of this fundamental tool of modern medicine in remote and less developed areas.

\subsection*{Low-consumption magnets for nuclear and particle physics, and medical applications}
As is the case of MRI, mentioned earlier, a successful development of HTS accelerator magnet technology for helium-free (or minimal cryogen) operation will offer energy efficient and sustainable options for the beam line magnets used in nuclear and particle physics experiments, as well as for beam production and delivery in radiation therapy. These beam line magnets often tend to require large apertures, implying large power consumption if powered with resistive coils, and may be subjected to significant heat loads from radiation, hence not within the optimal environmental conditions for LTS. Though field range and geometry may not compare directly, energy efficiency and cryogen-free (or minimal cryogen) benefits are a common denominator among the developments for the Muon Collider and beam line magnets for physics and medicine.

\subsection*{Magnets for neutron spectroscopy}
Magnetic configurations with free access to the magnet bore are useful for neutron spectroscopy, for material and life sciences. Present state-of-the-art of commercial systems is between 15\,T and 17\,T, but there is a clear call for higher fields to  uncover new phenomena. The targets set for next generation instruments range from 30\,T to 40\,T. The development of large bore, large energy and split HTS solenoids is a technology step towards the next generation of all-HTS neutron spectroscopy instruments, which could eventually integrate the improved field performance offered by the development of UHF 40\,T solenoids for the final cooling. 

\subsection*{Detector magnets for physics search}
Besides their use in colliders, research in particle physics also calls for advances in magnet technology. Such examples are the very high fields that would be needed to improve on the exclusion bounds for the mass of axions as dark mass candidates, or the need for very thin, lightweight, but large dimension solenoids and toroids to be used as detectors in space, operating in helium-free conditions. Such magnets will benefit from HTS technology, opening up options for science that cannot be reached with LTS magnets. The development in the scope of the magnet program of the Muon Collider will contribute to making this technology available.

\subsection*{Superconducting motors and generators}
The poles of superconducting motors and generators, such as those considered for marine transport and aviation, or wind turbines, are compact and non-planar to limit the dimensions and increase the power density. The challenges coming from compactness (high current density, high stress, and high energy density) and 3D shape (coil winding and joints) are similar to those of accelerator dipole and quadrupole magnets. The technical solutions developed for one field of application could directly applied to the other fields.

\subsection*{Targets for spallation neutron sources}
Spallation neutron sources have the conflicting requirement of a small beam spot size, to improve the neutron beam quality, and a high power deposition, to improve the neutron beam flux. This is typically achieved using targets comprising materials with high atomic number, but this causes increased beam damage. Flowing liquid metal targets have been used but boiling of the liquid caused cavitation and limited target performance. The fluidised tungsten target that is being studied as part of the muon collider R\&D can alleviate this fundamental limitation; a flowing target can be achieved but one that is resistant to cavitation.

\subsection*{Microwave RF sources}
Microwave sources and amplifiers use strong magnetic fields to confine the electron beam, with application in communications, RADAR, fusion heating and particle acceleration. In addition some magnetically confined fusion schemes require waveguides to be brought through a region with high magnetic field where breakdown can occur.

\begin{figure}
\includegraphics[width=\textwidth]{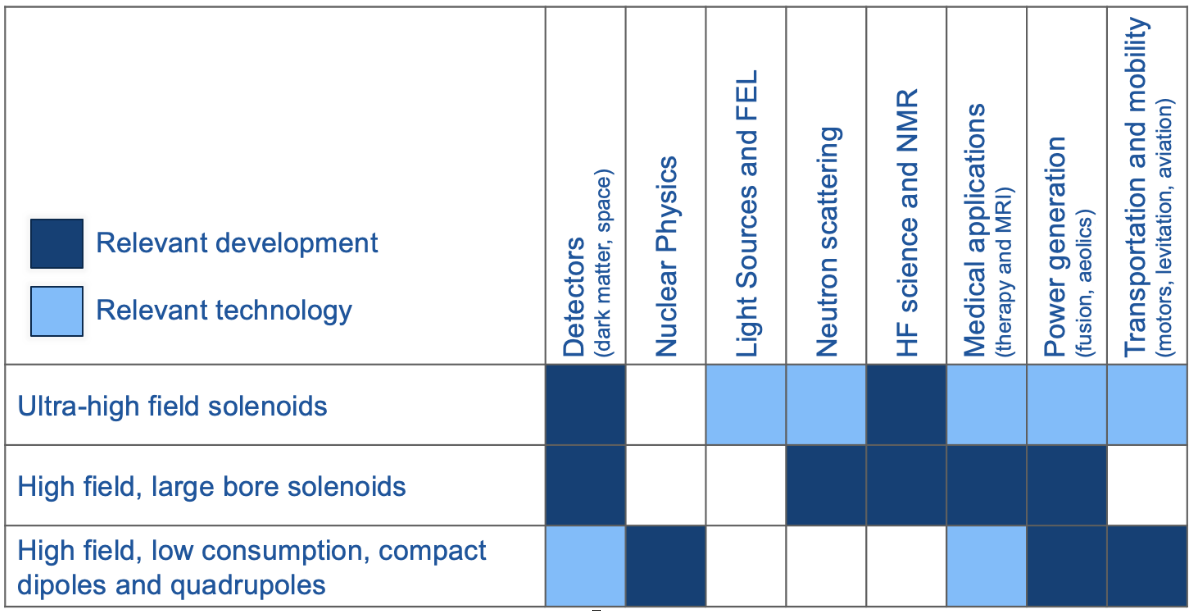}
\caption{Matrix of relevance of the developments required for a Muon Collider (rows) to other fields of science and societal applications (columns). The color shading indicates whether the developments can have direct application to other fields (dark shading) or whether the technology development is relevant, but application to other fields may require further development (light shading). 
\label{fig:magnet_matrix}}
\end{figure}

\section{Facilities}
\label{3:sec:synergy:facilities}
We have identified a number of facilities that may take advantage of, and benefit, the muon collider R\&D. 

\subsection*{Colliders}
There are several important R\&D topics that the IMCC is studying that have direct impact on other collider projects. High-Field Magnet and SRF technologies will be required by any future collider. Additionally, the muon collider detector R\&D may be a stepping stone towards achievement of the challenging parameters required by hadron colliders at next-generation energies and luminosities.

IMCC is actively working with the High Field Magnets programme, including leading an EU grant submission. Detector R\&D collaboration is proceeding through the ECFA-led DRD process.

\subsection*{Charged lepton flavour violation experiments}
Charged lepton flavour violation experiments such as mu2e and COMET require high-current and high-purity semi-relativistic muon beams. In these experiments pion production and capture is performed using targets immersed in high field solenoids.  Pions decay and the beam is cleaned in systems of long solenoids and dipoles. These pion production systems share many common challenges with the muon collider pion production system. In particular, the target and capture solenoid in mu2e and COMET has strong similarities with the muon collider target, albeit with some less demanding parameters. mu2e phase 2 and the proposed Advanced Muon Facility both represent opportunities to develop novel concepts in muon beam manipulation.

IMCC had in-depth discussions regarding possible collaboration on the target solenoid with the mu2e team. mu2e expressed significant interest in the HTS concepts that we have developed for the muon collider target. Similarly, active discussions are ongoing also with the COMET team to further strengthen this area.

\subsection*{g--2}
The proposed beam that is under study at J-PARC represents an interesting alternative route to low emittance muon beams. In this technique, positively-charged muons are stopped in a target and form muonium. A laser is subsequently directed onto the target and re-ionises the muons, resulting in a very small emittance in the outgoing positively-charged muon beam.

IMCC is seeking further discussions with colleagues from J-PARC and seek to strengthen this area.

\subsection*{nuSTORM and precision neutrino experiments}
Stored muon beams may be used as a well-characterised neutrino source. nuSTORM has been proposed as a neutrino production facility. Pions are passed into a muon storage ring. Muons arising from backwards decaying pions are captured and circulated, while remnant pions are extracted and pass onto a beam dump. The momentum difference between pion and muon beams enables injection and extraction without the use of pulsed magnets. The energy of the stored muon beam can be changed, yielding precisely characterised neutrino beams at different energies. nuSTORM will be used to study nuclear physics and neutrino scattering cross sections. This will provide a unique insight into nuclear physics. It may significantly reduce systematic uncertainties in the next generation of neutrino oscillation experiments. Additionally such a neutrino source will enable improved capability to reject exotic neutrino physics models.

nuSTORM provides an important opportunity to deliver a physics result using a high-intensity stored muon beam. While the energy and intensity is lower than the muon collider, it would be the highest stored muon beam power and so could be an important stepping stone on the way to a muon collider, were it to be approved.

An active collaboration is ongoing with the nuSTORM team. We have considered the potential to share a target, with O(200\,MeV) pions used for a demonstrator source and O(GeV) pions used for a neutrino source. Consideration is ongoing for the design of the necessary switchyard.

\subsection*{muSR and low-energy muon beams}
Highly polarised muon beams may be produced using pions stopped in a fixed target. These low energy muon beams may benefit from similar cooling techniques to the ionisation cooling system. The cooling systems are challenging because these muons lose energy much more quickly than the semi-relativistic muons used in the muon collider, so very thin absorbers are required and low-frequency or electrostatic acceleration is preferred. Issues such as breakdown at high voltage and solenoid optics are shared.

A collaboration with experts from the ISIS muon beams instruments has been established. ISIS provides one of the brightest polarised muon beams in the world. The ISIS team are interested to collaborate on advanced instrumentation concepts, for example using vertex detector-like technology as well as muon beam cooling techniques. We are collaborating on the design of a low energy muon cooling system.

\subsection*{High power proton sources}
The high-power proton source under consideration for use as the muon collider muon source shares many characteristics with proton sources used for neutron spallation and neutrino production. In particular, technical challenges such as heating in the charge exchange foil and high power targetry are shared between these facilities.

The proton driver task is led by ESS and ISIS are active members of the team. Foil heating is an area of active concern and one which we are investigating. We are also considering laser stripping techniques. More generally, the experts from ESS and ISIS are looking at issues such as space charge. The RAL target technology group, who led the construction of the T2K graphite target and are responsible for delivery of the DUNE target, are members of the collaboration.

\section*{Summary}
In summary, there are strong motivators for present and future research and development of the underlying technologies that will be required to deliver the Muon Collider, which has direct (i.e.~similar specifications) and indirect (i.e.~similar technology) relevance to other scientific and societal applications. Figure~\ref{fig:magnet_matrix} renders graphically our evaluation of the relevance of the technology developments for a Muon Collider to other fields of science and societal applications.
\part{Implementation} \label{4:impl}
\chapter{Muon Collider implementation}
\label{2:implement:ch}

\section{Overview}
The IMCC studies are based on a green-field assumption in order to understand basic physics potential, limitations and optimisation. Based on the assumption of a global R\&D effort, the possible schedule for delivery of the muon collider has been estimated. Studies have examined the potential for implementation at either CERN or Fermilab sites. The CERN siting study has successfully investigated the potential to site the muon production facility and reuse the existing tunnels for the muon acceleration complex in order to minimise the need for new tunnel construction. Siting studies at Fermilab have examined the potential to host the facility within the existing lab footprint and reuse the existing proton facility for muon production. The timeline and siting studies are described below.

\section{Timeline}
\begin{figure}[h!]
\centering
\includegraphics[width=\textwidth]{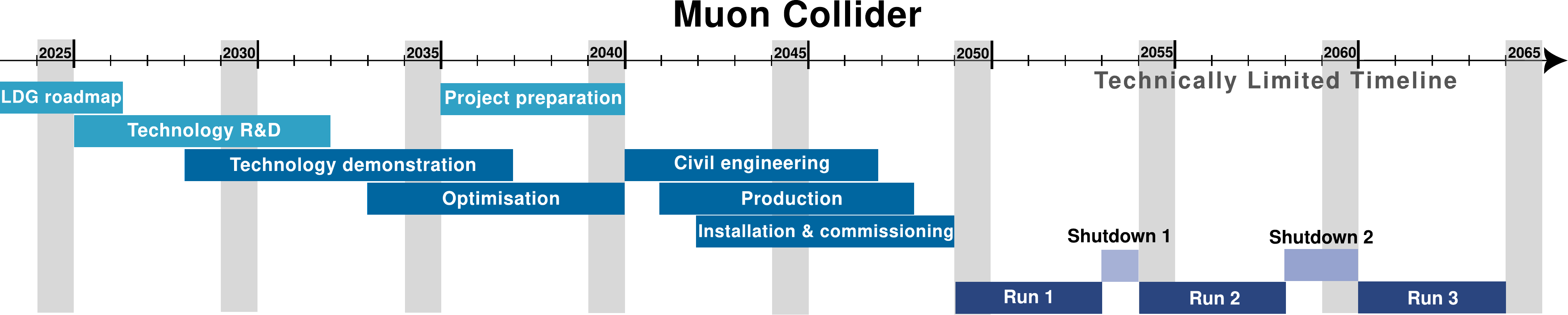}
\caption{\label{fig_schedule_facility} Technically limited schedule for the Muon Collider facility.}
\end{figure}

The IMCC has prepared a timeline that provides an estimate of when the full facility may deliver collisions. The timeline has been prepared based on estimates for:
\begin{itemize}
\item Availability of magnets: the Muon Collider will leverage the strongest \emph{available} magnets in the collider ring. Stronger dipole magnets yield improved luminosity but are not critical to achieve the target beam energy. While beam energy is considered a key requirement for the final facility and cannot be achieved without the associated appropriate magnets, the luminosity can be mitigated by other measures. 
The IMCC expects that 10~T dipoles, for example constructed using NbTi, will be available for use in the collider ring by the 2040s. Higher field dipoles, potentially using HTS, are expected to be available later. Additional R\&D is necessary to deliver solenoid magnets suitable for the muon cooling system.
\item Improved detector design: delivery of the detector will require R\&D to deliver exquisite position and time resolution while enabling sufficient data readout capabilities and background rejection despite the low-mass and low-power consumption required. These detectors will be required to operate in a harsh radiation environment, with significant beam-induced backgrounds present. Uniquely for the muon collider, high-precision calorimetry will be required to measure significantly higher energy physics objects than existing detectors.
\item Demonstration of cell integration for the muon cooling system: in order to be confident that the rectilinear cooling system may be constructed, the IMCC plans to execute a Muon Cooling Demonstration programme. The rectilinear cooling system is viewed as holding the most significant technical risk for the facility, which will be mitigated by this programme. The IMCC believes that construction of a muon cooling module will provide sufficient risk mitigation that the collider may be approved in the mid 2030s. Further beam tests will inform the construction of the facility.
\end{itemize}
The timeline shown in figure \ref{fig_schedule_facility} is technically limited. Practically the full support of a major laboratory such as CERN or Fermilab is required, for example supporting pipelining of R\&D items in order to deliver an expedited schedule and freeing up resources should new technical issues emerge that require additional effort to solve.

The existing \textbf{LDG roadmap} effort would conclude as planned in 2027. So far the community has provided resource to the IMCC roughly corresponding to the \emph{Minimal} proposal from that roadmap~\cite{https://doi.org/10.23731/cyrm-2022-001}. An initial \textbf{Technology R\&D} phase would enable the IMCC to complete the design of all components as well as begin the experimental programme to prove the performance of key systems. During the \textbf{Technology Demonstration} phase the construction of important test pieces, especially for the Muon Cooling Demonstration programme, will be performed. The results from these tests will be fed back to the accelerator design enabling the \textbf{Optimisation} phase where the overall cost and performance will be optimised in light of improved knowledge of achievable hardware parameters and delivery of a Conceptual Design Report. At this point a good estimate of performance, cost, power, carbon footprint and other sustainability metrics will be available enabling the community to make a decision to proceed with the full facility.
Further detailed timelines regarding the cooling demonstrator and the magnet production can be found in figure \ref{fig:magnet_schedule} and \ref{fig:ch12:demo_timeline}.

IMCC estimates that following the project approval it will take slightly less than a decade to construct the facility. Were the project suitably resourced, first physics could commence before 2050. The initial run would include shakedown of the facility and a ramp to the target performance. After a one year shut down, Run 2 would commence operating at full luminosity. A two year shutdown would enable installation of equipment for a luminosity upgrade, or connection to a higher energy implementation before commencement of Run 3.

\section{Implementation at CERN}
\label{2:implement:sec:cern}
\subsection*{Civil Engineering}

Civil Engineering (CE) represents a significant proportion of the implementation budget for tunnelling projects such as the Muon Collider at CERN. As a result, CE studies are of critical importance to ensure a viable and cost-efficient conceptual design from the beginning.

A key driver at CERN is to integrate the Muon Collider complex into the existing infrastructure available at CERN, thus reducing the environmental impact, costs and timescale of construction.

\subsubsection*{Layout and Placement Study}
CE design and planning play a crucial role in assessing the feasibility of the Muon Collider. Past construction at CERN indicates that the Geneva Basin's sedimentary rock, called Molasse, offers excellent tunneling conditions, while Limestone regions present more challenges. During the LEP tunnel excavation, significant issues arose due to water ingress from Limestone formations. Therefore, it is essential to position underground structures primarily in Molasse and to avoid limestone. Additionally, the tunnel orientation should aim to minimize shaft depth and reduce overburden pressure on the structures.
\newline
\begin{figure}[h!]
\centering
\includegraphics[width=\textwidth]{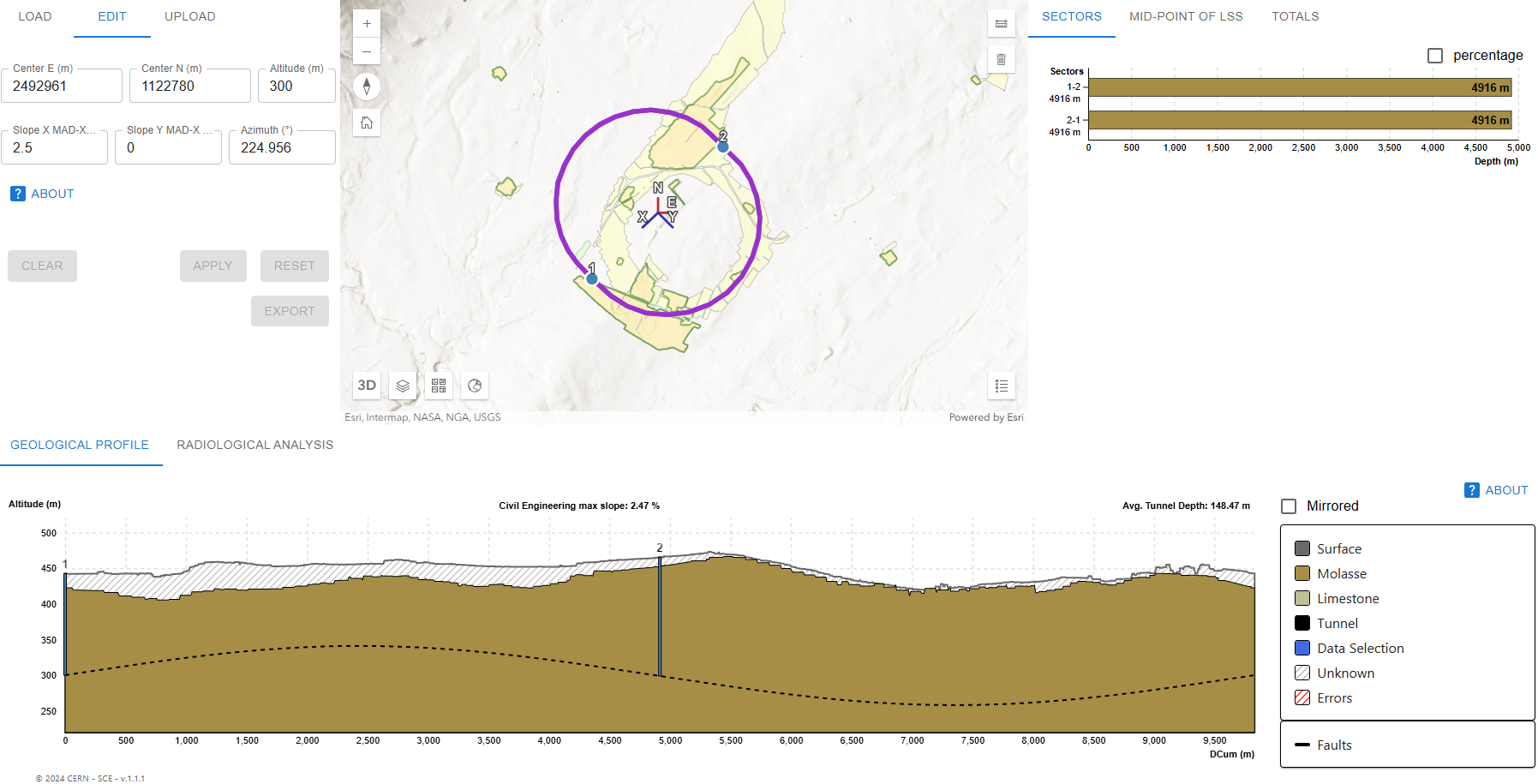}
\caption{\label{fig_Muon_Collider_Geoprofiler_Geology} Geology analysis results from Geoprofiler for a given position.}
\end{figure}

To plan the tunnel alignment and analyse the geological constraints, an application called ‘Geoprofiler’ has been developed as seen in Figure \ref{fig_Muon_Collider_Geoprofiler_Geology}. This application provides geographical visualisation of the accelerator’s footprint alongside detailed geological profiles. Within the application, users can modify the location of the tunnel alignment as well as its depth and tilt. To ensure a productive user experience, the resulting geology of the accelerator placement is displayed in real-time.

A key requirement of the application is to be able to freely modify the footprint of the accelerator and determine the most favourable position in terms of risk and cost. Additionally, radiation analysis requires high precision to position the accelerator and was added to the tool as well. Many key advancements were made during the development of the Geoprofiler tool:
\begin{itemize}
    \item Integration of the alignment data. Although currently there is no official alignment, it is likely that a MAD-X file format will be provided. Therefore, it has been made feasible to upload a MAD-X file to ‘Geoprofiler’.
    \item Several geodetic challenges were encountered. The available data in the MAD-X file first needs to be converted from a Cartesian coordinate system to a geographic one. Once the accelerator is in a geographic system, the coordinates must remain accurate even when the alignment is moved or rotated.
    \item The geological data used for the tool was collected from UNIGE (University of Geneva). There are two main components in the application to assess the geology:
    \begin{itemize}
        \item Firstly, the profile of the alignment, which displays the geological data. This includes the depth of each geological layer, the depth of the accelerator and the surface height.
        \item Secondly, the geology surrounding the shafts is displayed, from which the depths of each geological layer can be determined.
    \end{itemize}
    \item For radiological analysis, a feature to compute the intersection of radiation lines with earth surface was added.
\end{itemize}

\subsubsection*{Collider Complex}
The Collider Complex is displayed in Figure~\ref{fig_Muon_Collider_Complex}, presented below. Since the Interim Report~\cite{InternationalMuonCollider:2024jyv}, an injector complex has been designed and implemented, initiating at LINAC~4 and ultimately, injecting into a new 10\,km Collider Ring from the LHC. The LINAC 4, SPL and ARCR (Accumulator Ring Compressor Ring) are aiming to be equivelent to the Proton Driver as described in Section~\ref{1:acc:sec:proton}. The entirety of the complex's surface works would be constructed on CERN owned land across both the Meyrin and Pr\'{e}vessin sites, minimising territorial and environmental impacts.

\begin{figure}[h!]
\centering
\includegraphics[width=\textwidth]{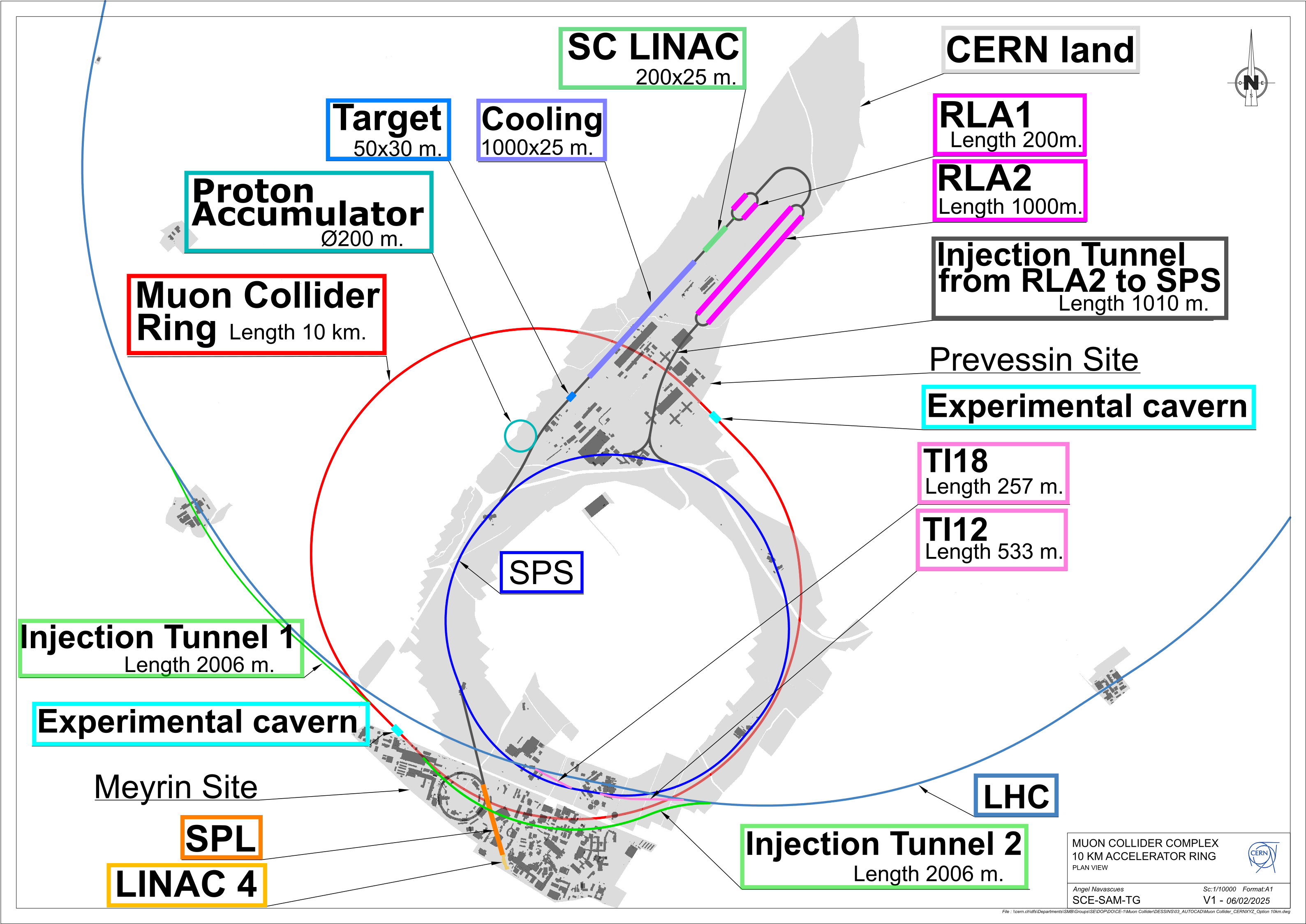}
\caption{\label{fig_Muon_Collider_Complex} Muon Collider Complex.}
\end{figure}

\paragraph*{Meyrin Site}
Initiating at the existing LINAC 4 on the Meyrin site, the complex integrates the proposed CERN Superconducting Proton Linac (SPL) design. The SPL is a linear proton machine design which would serve as part of the injector complex as a high-power proton driver for the application of muon generation. 

The transfer between the CERN Meyrin and Pr\'{e}vessin sites from the SPL travels via the SPS tunnel, this requires two junction caverns at the insertion and extraction points. The 3m diameter transfer tunnels are to be constructed using TBM (Tunnel Boring Machine).

\paragraph*{Pr\'{e}vessin Site}
The surface works have been based on the FCC-hh injection complex presented in the Mid Term report in 2024. Within the Pr\'{e}vessin site, the complex would be comprised of a series of interconnected cut and cover tunnels and surface structures, avoiding all existing CERN infrastructure, before being injected beneath the Pr\'{e}vessin site back into the SPS from RLA 2 (Recirculating Linear Accelerator 2). 

\begin{figure}[h!]
\centering
\includegraphics[width=\textwidth]{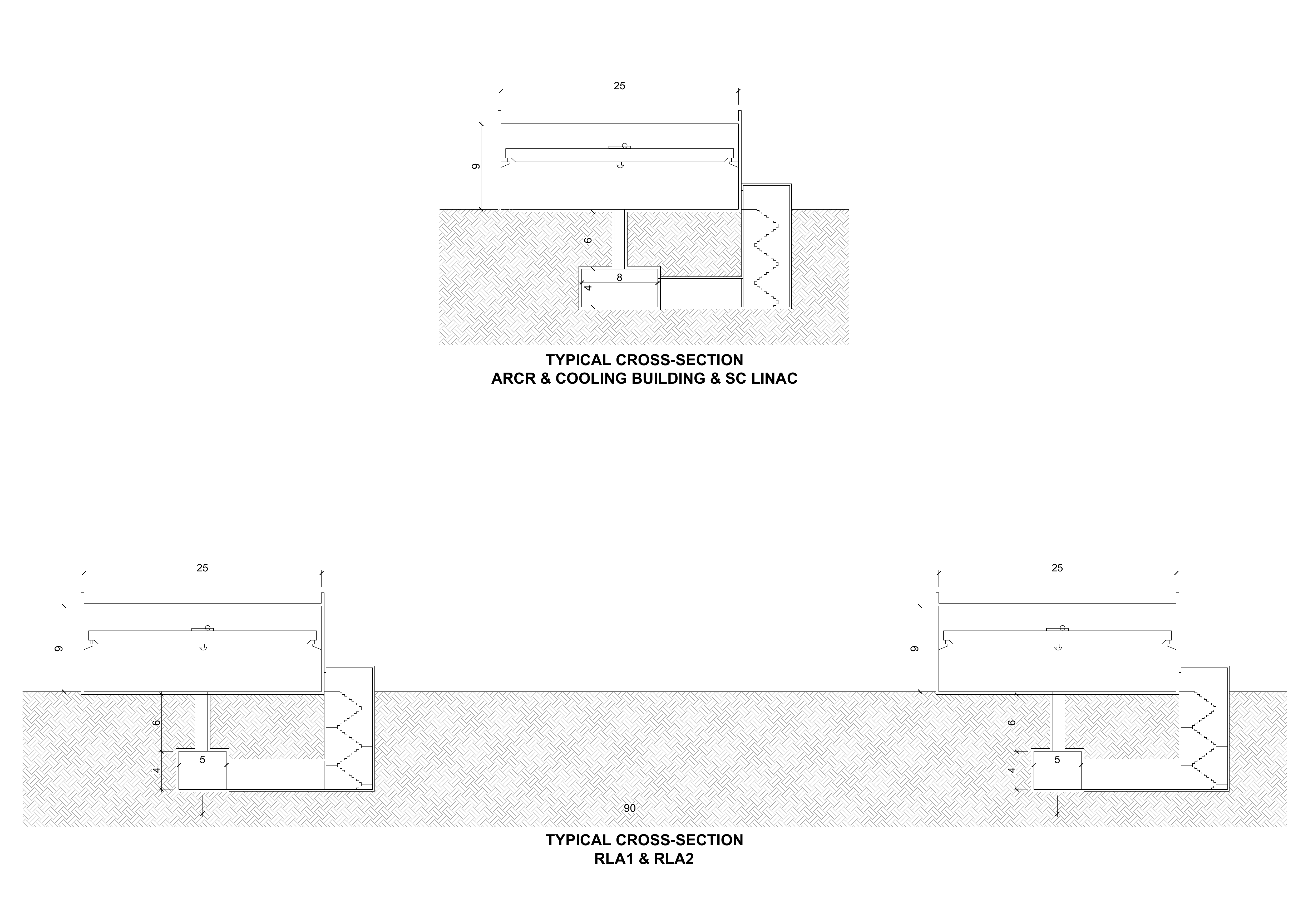}
\caption{\label{fig_Muon_Collider_Injection_Complex_Cross_Sections} Muon Collider Surface Injection Complex Cross Sections.}
\end{figure}

The Pr\'{e}vessin site would house the ARCR, Target, Cooling, SC Linac, RLA 1 and RLA 2, each connected in series via transfer tunnels. The cut and cover tunnel cross sections vary from 5\,m x 4\,m for a single beam line, to 8\,m x 4\,m for dual beam lines in parallel. Figure \ref{fig_Muon_Collider_Injection_Complex_Cross_Sections} displays the cross sections associated with each component. 

A uniform surface structure has been designed to house all components required in a surface structure above the tunnels. The surface structures will be optimised in future based on individual component requirements. The target is yet to be designed but would consist of a cavern alongside the necessary accompanying surface and subsurface structures.

\subsubsection*{Main Tunnel Sections}
\begin{figure}[h!]
\centering
\includegraphics[trim={5.5cm 0 7cm 0},clip,width=0.49\textwidth]{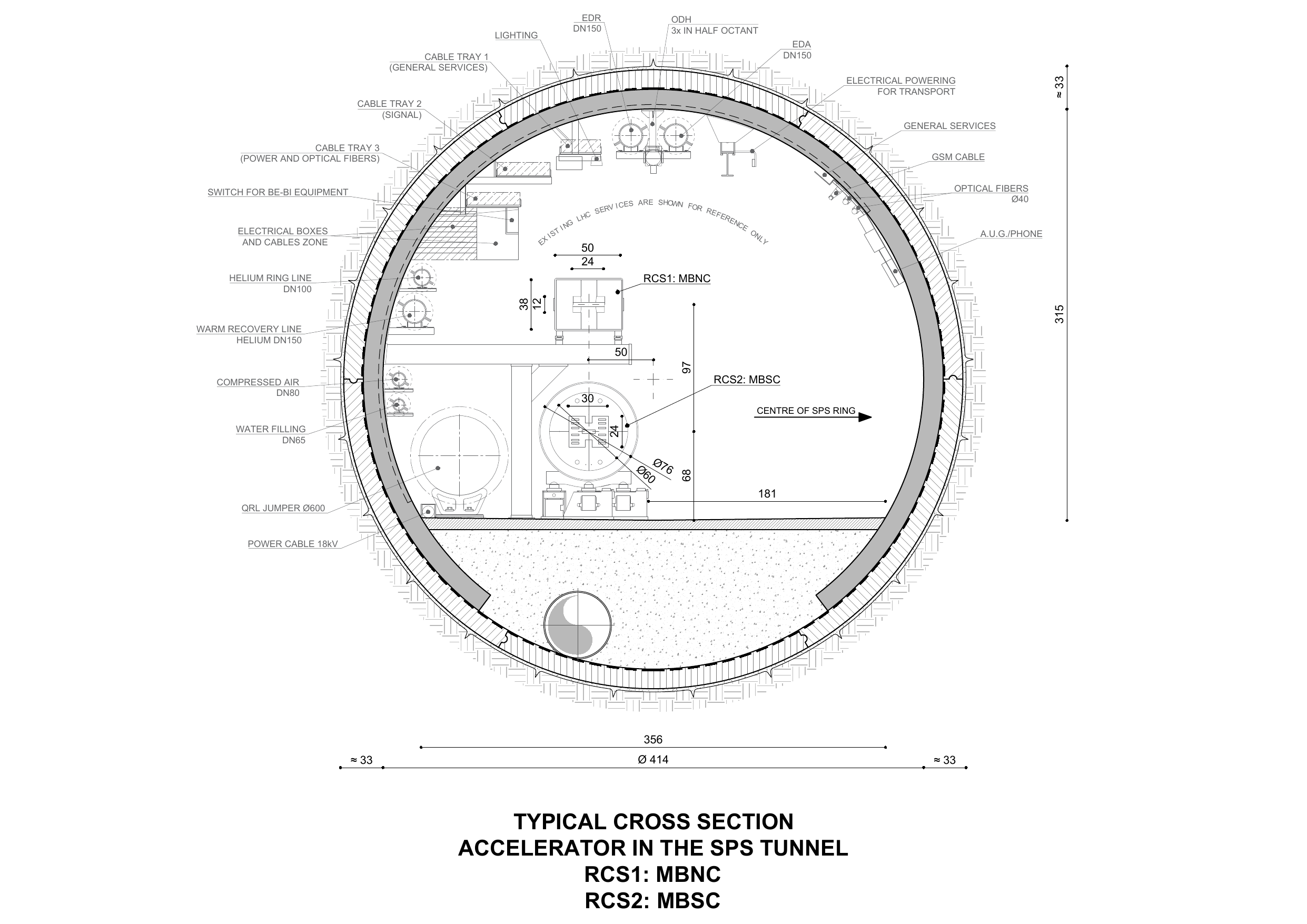}\includegraphics[trim={5.5cm 0 7cm 0},clip,width=0.49 \textwidth]{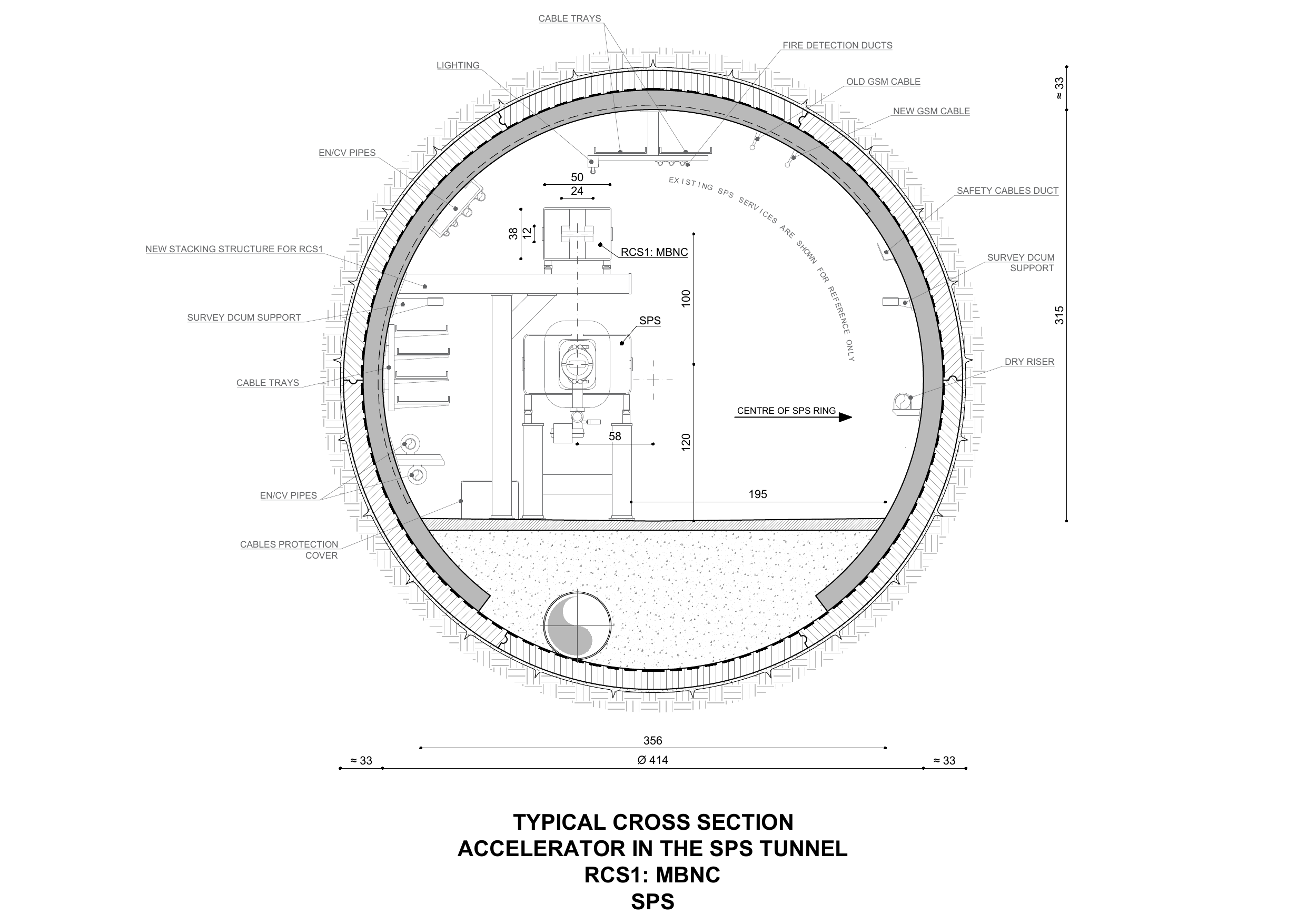}
\caption{\label{fig_Typical_SPS_Cross_Section} Left: Typical SPS Cross Section with RCS1 and RCS2 Magnets. Right: Typical SPS Cross Section with RCS1 and Existing SPS Machine.}
\end{figure}

\paragraph*{SPS}
There are two potential configurations for the SPS tunnel depending on the decision to maintain the existing SPS machine. Figure~\ref{fig_Typical_SPS_Cross_Section} (left) presents a typical SPS tunnel cross section housing RCS1 and RCS2 magnets if the SPS machine was to be removed. Figure~\ref{fig_Typical_SPS_Cross_Section} (right) then presents a typical SPS cross section if the SPS machine was to be maintained, thus the only addition to the tunnel being the RCS1 magnets.

\FloatBarrier
\paragraph*{LHC}
The SPS connects to the LHC with use of the existing TI12 and TI18 transfer tunnels, before being injected from the LHC into the Collider ring. Both transfer tunnels have 3.5m diameters, these are greater than the 3m diameter transfer tunnels proposed for the rest of the complex meaning they are well suited to house the required infrastructure. 

\begin{figure}[h!]
\centering
\includegraphics[trim={5.5cm 0 7cm 0},clip,width=0.49\textwidth]{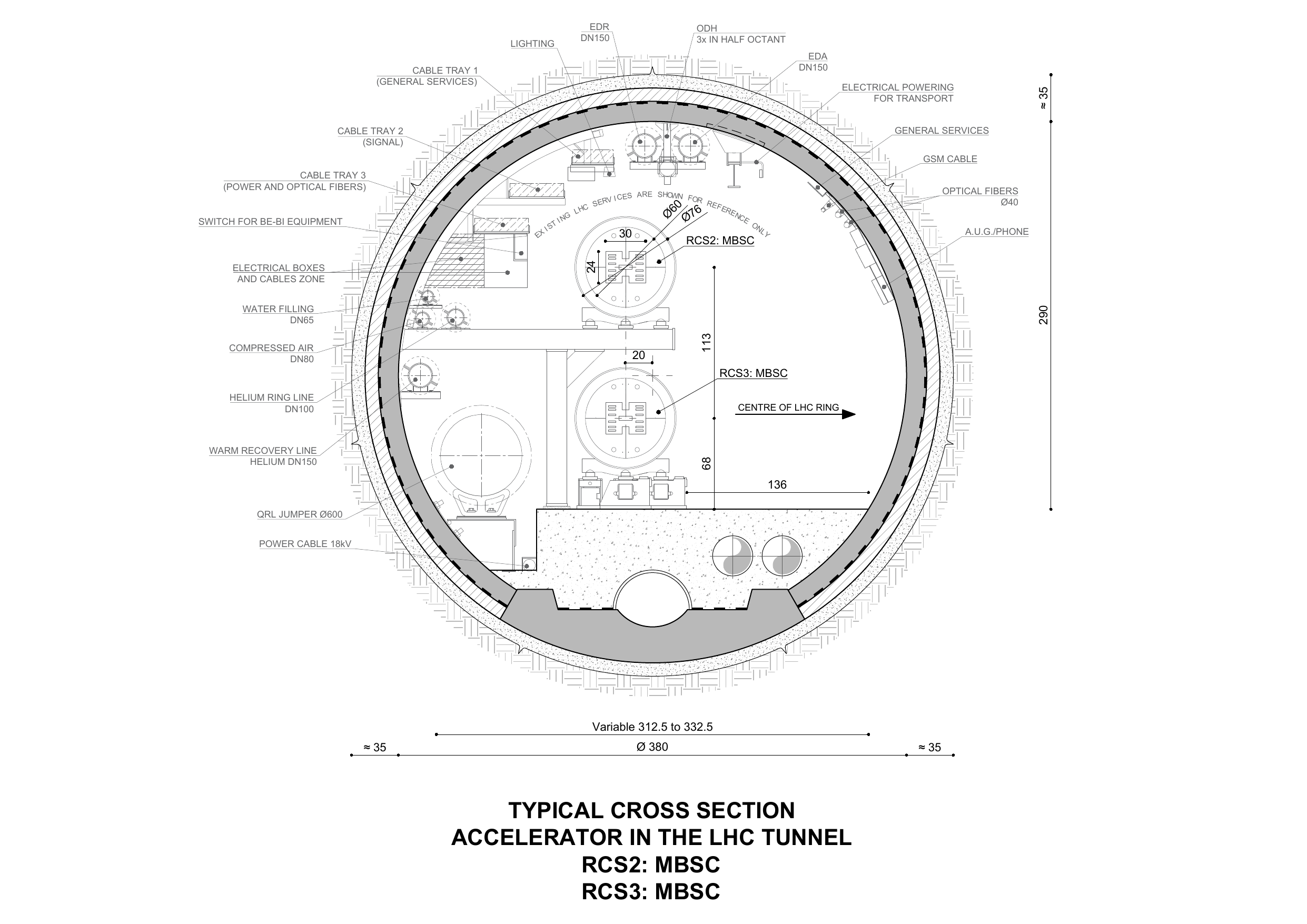}\includegraphics[trim={5.5cm 0 7cm 0},clip,width=0.49\textwidth]{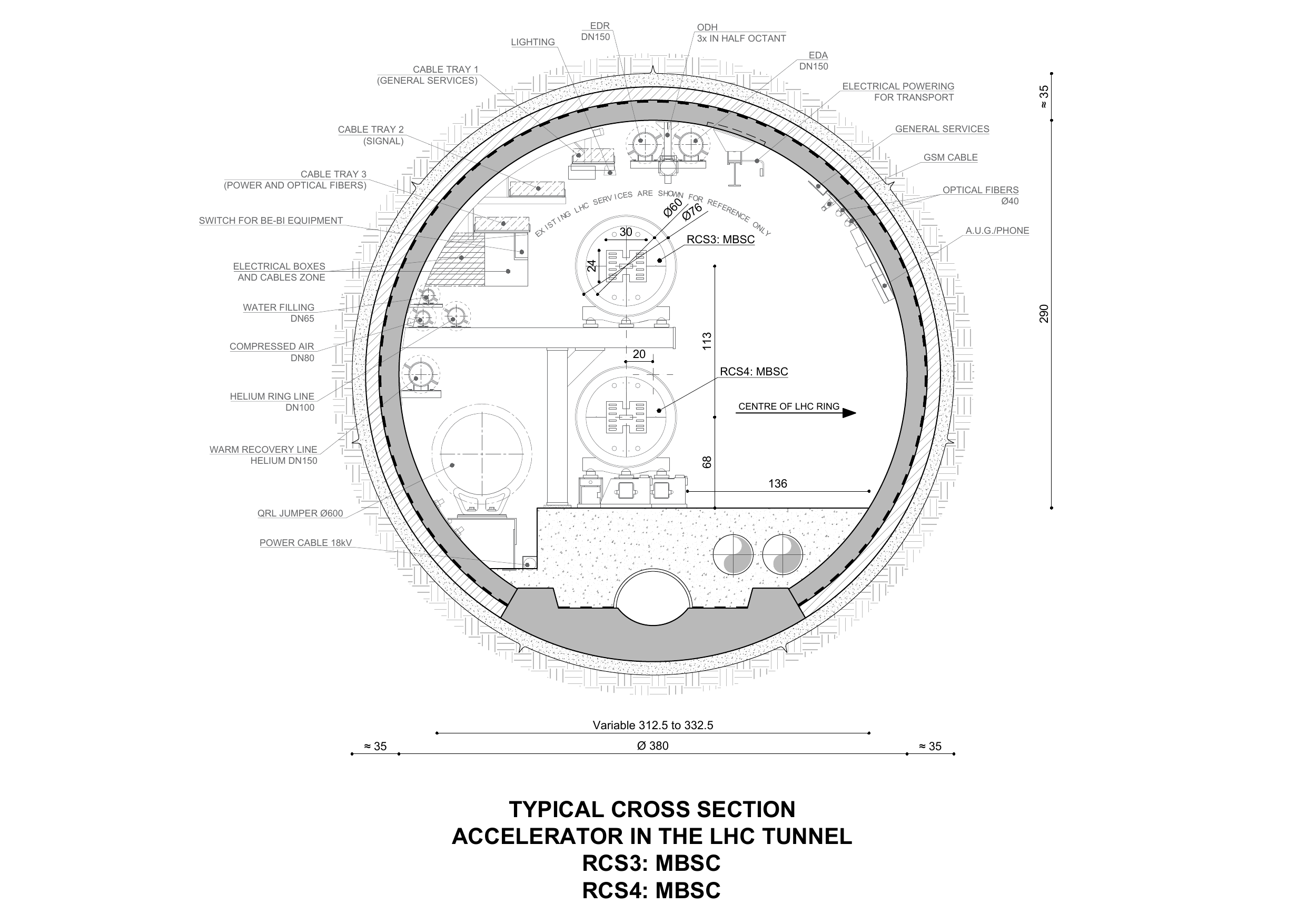}
\caption{\label{fig_Typical_LHC_Cross_Section} Left: Typical LHC Cross Section with RCS2 and RCS3 Magnets. Right: Typical LHC Cross Section with RCS3 and RCS4 Magnets.}
\end{figure}

Depending on the decision to maintain the SPS machine, two cross sections have been designed for the LHC, and are shown in Figure~\ref{fig_Typical_LHC_Cross_Section}. If the SPS machine is maintained the LHC would contain RCS2 and RCS3 magnets, if not it would contain RCS3 and RCS4 magnets. Both cross sections are presented in the following two figures.

\FloatBarrier
\paragraph*{Collider Ring}

The proposed design for the collider ring has the shape of an elongated circle, with straight sections into the 2 experimental caverns. The injection from the LHC to the collider ring begins with extraction from the straight sections of the LHC, then injecting into the Collider Ring at equidistant points either side of one of the experimental caverns. The injection must be into the outside of the collider ring to ensure a continuous transport passageway is maintained throughout the inner perimeter of the ring. The injection tunnel lengths must both be equal and can have a radius no less than that of the collider ring.

The Collider Ring typical cross section is displayed in Figure~\ref{fig_Muon_Collider_Ring_Cross_Section} below. Based off the LHC cross section, an internal diameter of 4\,m is proposed for the Collider Ring, it will be constructed using TBM and will have a pre-cast segmental lining. 

\begin{figure}[h!]
\centering
\includegraphics[width=\textwidth]{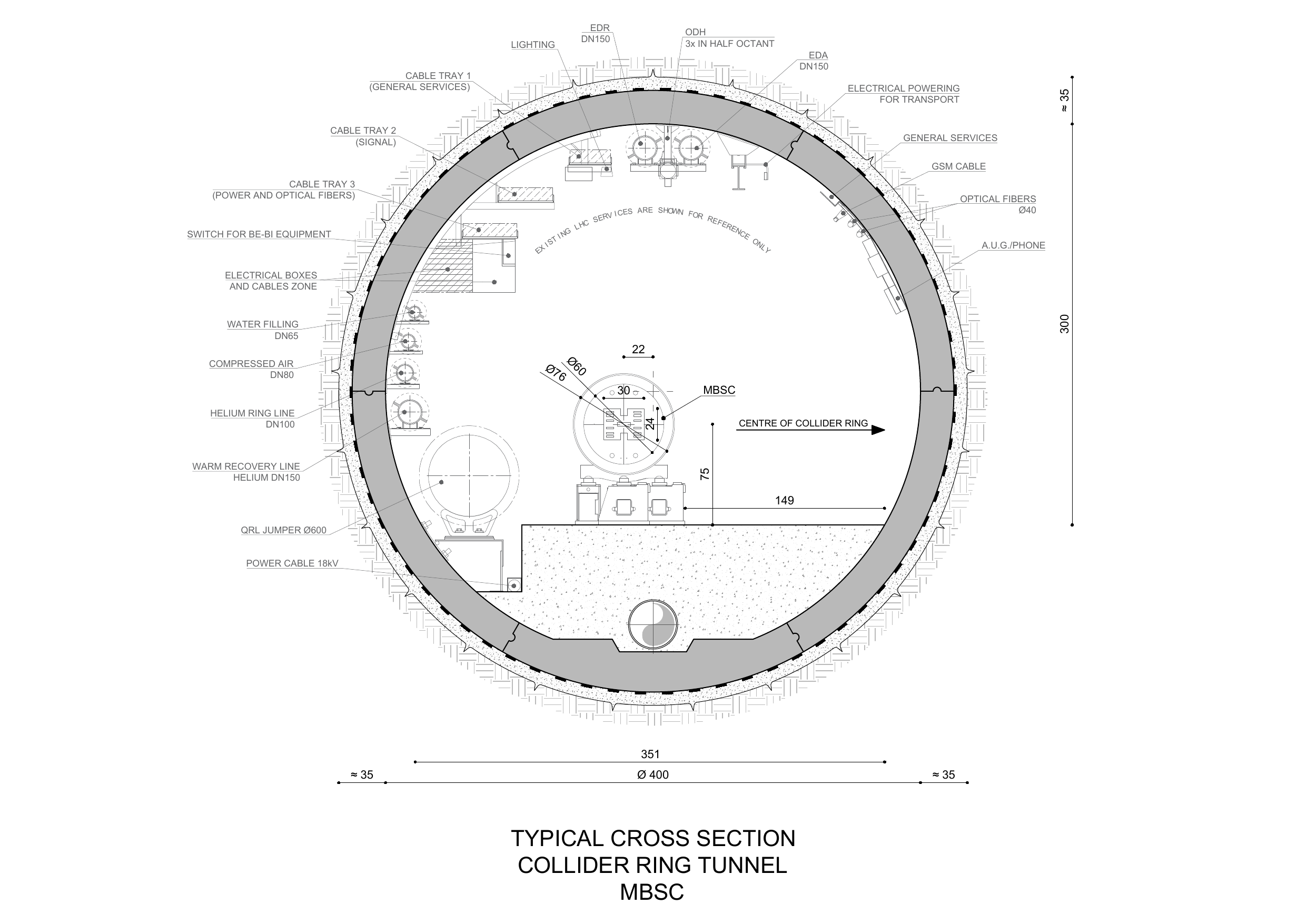}
\caption{\label{fig_Muon_Collider_Ring_Cross_Section} Cross-section of the collider ring.}
\end{figure}

Within the 10\,km collider ring there are two interaction regions each housing a single detector. Each interaction region contains the key components of an experimental cavern, a service cavern, an experimental shaft, a service shaft, and an experimental surface site based off the FCC-hh designs. The shafts have diameters of 18\,m at depths of 142\,m and 166\,m for each interaction region. The cavern floor parameters are 66\,m x 35\,m for the experimental cavern and 100\,m x 25\,m for the service cavern.

\FloatBarrier
\paragraph*{Planned Work}
The Muon Collider complex will continue to undergo further iterations on each component based on design changes and requirements. The surface complex will undergo further design including the addition of service buildings such as Cooling and Ventilation, Electricity and Water supply. 

With the main individual components determined, an initial in-house cost estimate based on primarily the FCC-hh study can be developed.

\subsection*{Infrastructure}

The implementations shown in Figure~\ref{fig_Muon_Collider_Complex} maximise the re-use of the existing infrastructure of CERN simplifying the environmental impact studies, the level of acceptance from the local population and certainly reduce cost and CO$_2$ footprint with respect to a green field implementation. 
Detailed studies for the necessity of upgrades (for instance in term of cooling power, ventilation etc...) have not been performed, but given that the accelerator is due to start at the horizon 2050, the complete renewal of all the technical systems will probably have to be foreseen anyway. Most of the civil engineering infrastructure (trenches, technical galleries etc...) is expected, on the contrary, to be almost entirely reusable.  

\subsection*{Accelerator}
\label{2:site:cern:sec}
The CERN-based muon collider would profit from the placement of the RCS chain into the SPS and LHC tunnels. Compared to a greenfield-site-based option, a significant amount of the budget could potentially be saved by not having to build the costly and long RCS tunnels. The integration of these accelerators into existing tunnels poses a number of muon collider specific challenges. 

The RF system will likely need to be spread into multiple sections to reduce the synchrotron tune to a tolerable level (more details in Section~\ref{1:acc:sec:rcs_chain}). The exact requirements depend on the momentum compaction factor and, resulting, the optics design. Due to the tunnels already existing with a set number of straight sections, in which the RF cells could be placed, the optics design might need to be adjusted to allow for the number of RF stations to be lower than the number of long straight sections in the tunnels. An alternative approach would be to distribute the RF system in the arcs of the ring, giving more flexibility in terms of the required straight section length. This approach would, however, make a distribution of the RF power and cryogenics infrastructure around the whole ring necessary.

The design of the accelerator optics comes with the additional challenge, that the radial layout needs to closely follow the structure of the tunnels. A deviation in the range of centimetres would certainly be possible, but enough clearance needs to be kept for the installation and maintenance of the infrastructure. For this purpose, a tool is planned to be developed which compares an optics design from a MAD-X file with the current lattices in the SPS and LHC. 

Additionally to the acceleration requirements, the neutrino flux from a muon-collider facility in the LHC and SPS tunnels needs to be studied. Placing the RF system into the long straight sections might cause a large flux in a concentrated area. Due to the limited depth of the SPS and LHC tunnels, the resulting dose rate at the surface may prevent such an option from being implemented. In this case, a system of movers, similar to what is envisioned for the collider ring, could be implemented. Detailed studies of the radiation doses for all energy ranges are therefore required. 

A preliminary study of acceleration parameters was performed using the same assumptions as for the greenfield study baseline. The injection energy was given by the extraction energy of RLA2 with \SI{63}{\GeV} and a bunch population of \num{2.7e12}. 
The survival rate was assumed to be \SI{70}{\%} across the high energy acceleration complex while keeping the dipole packing fraction below $0.66$ for the SPS tunnel and below $0.7$ for the LHC tunnel to retain a sufficient margin to the real machine. The baseline magnet strengths of \SI{1.8}{T} for the pulsed normal conducting magnets and \SI{10}{T} for the superconducting dipoles were kept. Based on these assumptions, a first estimation of the accelerator chain performance is given in Table~\ref{2:impl:cern:RCS:tab:RCS_key}, which was first presented in~\cite{PreliminaryParameter_MuCol5}. The given parameters were optimized to achieve the maximum ejection energy out of RCS3 while adhering to the given boundaries. The extraction energy of the third RCS can reach \SI{3.8}{\TeV} with the given configuration. 

An additional preliminary study was performed using fast-ramping high-temperature superconducting (HTS) dipole magnets in the last RCS. Within this configuration, an extraction energy of \SI{5}{\TeV} can be reached, matching the collision energy in the greenfield study. In this case, a slightly lower survival rate of \SI{68}{\%} can be reached. The RCS parameters for this study can be found in Table \ref{2:impl:cern:RCS:tab:RCS_key_HTS}. This option requires further investigation on the required dipole apertures and beam dynamics implications in the last hybrid RCS. 

The configuration presented in Table~\ref{2:impl:cern:RCS:tab:RCS_key} allows for a staged construction approach. The first two RCSs feature only normal conducting dipoles, thus requiring no fundamental research for the magnet technology. 
Using this configuration, an unprecedented lepton centre-of-mass collision energy of roughly \SI{3}{\TeV} could be reached. With additional time ready for the research on HTS pulsed magnets and the hybrid RCS concepts, the facility could later be upgraded to host a second RCS in the LHC tunnel, increasing the collision energy to \SI{10}{\TeV}. 

\begin{table}[!h]
    \centering
    \begin{tabular}{c|c|ccc} 
         Parameter&  Unit&  RCS 1 SPS&  RCS 2 LHC&  RCS 3 LHC\\\hline
         Hybrid RCS&  -&  No&  No&  Yes\\
         Circumference& m& 6912& 26659& 26659\\
         Injection energy& GeV& 63& 350& 1600\\
         Extraction energy& GeV& 350& 1600& 3800\\
         Energy ratio&  - &  5.6&  4.6&  2.4\\
         Assumed survival rate& - &  0.88&  0.86&  0.92\\
         Total survival rate& - & 0.88& 0.76&0.70\\
         Acceleration time&  ms&  0.45&  2.60&  4.42\\
         Revolution period&  \textmu s&  23.0&  88.9&  88.9\\
         Number of turns& - & 19& 29& 50\\
         Required energy gain per turn& GeV& 15.1& 43.1& 44.4\\
         Average acel. gradient& MV/m& 2.15& 1.62& 1.68\\
         Inj bunch population& $10^{12}$& 2.70& 2.38& 2.04\\
         Ext bunch population& $10^{12}$& 2.38& 2.04& 1.88\\
         Beam current per bunch& mA& 18.75& 4.29&3.68\\
         Beam power& MW& 803& 523&462\\
         Bunch length& ps& 33.2& 19.3&11.1\\ \hline
         Straight section length&  m&  2809&  8000&  8000\\
         Length with NC magnets& m& 4103& 18650& 12940\\
         Length with SC magnets& m& -& -& 5680\\
         Packing Fraction& \%& 59& 70& 70\\
         Max NC dipole field& T& 1.8& 1.8& 1.8\\
         Max SC dipole field& T& -& -& 10\\
         NC dipole ramp rate& T/s& 3320& 1400& 810\\
         Main RF frequency& GHz& 1.3& 1.3& 1.3\\
         Harmonic number& - &  29900&  115345&  115345\\
         Momentum compaction factor& $10^{-4}$& 9& 5& 3\\
    \end{tabular}
    \caption{Key acceleration parameters for the CERN-site based RCS acceleration chain from~\cite{PreliminaryParameter_MuCol5}. The parameters assume one bunch per species in counter-rotation. }
    \label{2:impl:cern:RCS:tab:RCS_key}
\end{table}

\paragraph*{RF system for the RCS layout at CERN}
\label{2:impls:cern:RCS:sec:RCS_RF_CERN}
The design of the RF system for the CERN-based RCS is based on the same assumptions as the RF system for the greenfield study. The assumptions are presented in Section~\ref{1:tech:sec:rf:highacc}.

\begin{table}[!h]
    \centering
    \begin{tabular}{lc|cccc}
         \multirow{2}{*}{Parameter} & \multirow{2}{*}{Unit} & RCS 1&  RCS 2&  RCS 3 & \multirow{2}{*}{All}\\ 
         & & SPS & LHC & LHC & \\ \hline
         Synchronous phase &  \textdegree&  135&  135&  135&  -\\
         Combined beam current ($\mu^+$, $\mu^-$) &  mA&  37.5&  8.58&  7.35&  -\\
         Total RF voltage &  GV&  21.4&  61&  62.8&  145.2\\
         Total number of cavities &  - &  686&  1958&  2017&  4661\\
         Total number of cryomodules &  -  &  77&  218&  225&  520\\
         Total RF section length &  m&  974&  2760&  2850&  6584\\ \hline
         Combined peak beam power ($\mu^+$, $\mu^-$) &  MW&  803&  523&  462&  -\\
         External Q-factor &  $10^{6}$&  0.79&  3.49&  4.07&  -\\
         Cavity detuning for beam loading comp. & kHz& -1.16& -0.26& -0.23& -\\
         Beam acceleration time & ms& 0.45& 2.6& 4.42&-\\
         Cavity filling time & ms& 0.194& 0.854& 0.993&-\\
         RF pulse length & ms& 0.644& 3.454& 5.413&-\\
         RF duty factor & \%& 0.32& 1.73& 2.71& - \\
         Peak cavity power & kW& 987& 228& 195& - \\ \hline 
         Total peak RF power & MW & 905& 569& 529& -\\
         Total number of klystrons & - & 99& 60& 54& 213\\
         Cavities per klystron & - & 7& 33& 38& -\\
         Average RF power & MW& 2.91& 10.3& 14.3&27.51\\
         Average wall plug power for RF System & MW& 4.48& 15.8& 22&42.28\\
         \midrule
         Required power for static cooling per cavity &   W  & 750   & 750 & 750  &  - \\
         Required power for dynamic cooling per cavity & kW & 0.8 & 2.0 & 3.2 & - \\
         Installed RF cryogenic plant capacity & MW & 1.5 & 7.7 & 11.6 & 20.8 \\
         Required total power for static cooling & MW & 0.5 & 1.3 & 1.3 & 3.1 \\
    \end{tabular}
    \caption[RF Parameters for the CERN-based RCS acceleration chain.]{RF Parameters for the CERN-based RCS acceleration chain. For the synchronous phase, \SI{90}{\degree} is defined as being on-crest. The power values for the cryogenic losses and plant capacity are the required room temperature equivalent input powers. The installed capacity includes a factor of 1.5 over the operational dynamic consumption. The design is based on the acceleration parameters without pulsed HTS magnets. }
    \label{3:impl:cern:RCS:tab:RCS_rf}
\end{table}

\begin{table}[!h]
    \centering
    \begin{tabular}{c|c|ccc} 
         Parameter&  Unit&  RCS 1 SPS&  RCS 2 LHC&  RCS 3 LHC\\\hline
         Hybrid RCS&  -&  No&  No&  Yes\\
         Circumference& m& 6912& 26659& 26659\\
         Injection energy& GeV& 63& 370& 1600\\
         Extraction energy& GeV& 370& 1600& 5000\\
         Energy ratio&  - &  5.8&  4.3&  3.1\\
         Assumed survival rate& - &  0.90&  0.92&  0.82\\
         Total survival rate& - & 0.90& 0.82&0.68\\
         Acceleration time&  ms&  0.38&  1.37&  12.41\\
         Revolution period&  \textmu s&  23.0&  88.9&  88.9\\
         Number of turns& - & 17& 16& 140\\
         Required energy gain per turn& GeV& 18& 76& 24\\
         Average accel. gradient& MV/m& 2.7& 2.9& 0.9\\
         Inj bunch population& $10^{12}$& 2.70& 2.43& 2.23\\
         Ext bunch population& $10^{12}$& 2.43& 2.23& 1.82\\
         Length with pulsed magnets& m& 4103& 18650& 12940\\
         Length with fixed field magnets& m& -& -& 5680\\
         Packing Fraction& \% & 62& 70& 70\\
         Max pulsed dipole field& T& 1.8& 1.8& 3.0\\
         Max SC dipole field& T& -& -& 10\\
         Pulsed dipole ramp rate& T/s& 3900& 1000& 480\\
         Main RF frequency& GHz& 1.3& 1.3& 1.3\\
         Harmonic number& - &  29900&  115345&  115345\\
    \end{tabular}
    \caption{Key acceleration parameters for the CERN-site based RCS acceleration chain using fast-ramping HTS dipoles in the last RCS. }
    \label{2:impl:cern:RCS:tab:RCS_key_HTS}
\end{table}

\section{Cost drivers and cost scale}
\label{4:sustainability:sec:cost}

The approach to the cost estimate of the Muon Collider project and the analysis of the most significant sources of cost uncertainty must necessarily start from the assessment of the technical maturity of the project itself. At the time when the estimate is performed, the Muon Collider design is still in a pre-CDR stage, where the architecture has been developed to a certain level of detail and the main beam parameters are known with some accuracy, but still several options exist that may steer the project into different directions, with considerable impact on the final cost.

\subsection*{Scope and method}

For the above mentioned reasons, the estimate is presented in the form of a cost range, which is justified partly by the level of maturity of the chosen technologies and partly due to the extrapolation of the cost information from other projects and studies. An initial choice was made to consider the installation on the CERN site as a possible first configuration for the realisation of the Muon Collider, by also assuming that it would make the best use of the existing tunnels and technical infrastructure. That would certainly keep the environmental impact related to the civil engineering realisations on the low side, in spite of some initial limitations to the performance reach of the facility that may descend from this choice. The successive analysis looks into the cost variance introduced by the adoption of beyond state-of-the-art technologies that will push the overall performance and by  configuration options that will allow the complex to deploy its full potential.

A project breakdown structure (PBS) was developed and for each area of the facility and few systems could be identified as the major contributors to the overall cost. The PBS considers four levels of complexity providing increasing detail starting from the Machine Sector down to Facility, System and Component. Workpackage leaders in charge of different technological areas were asked to provide their estimates of cost for equipment considered as the cost drivers, together with the level of maturity for the specific technology. The main cost drivers are represented by the civil engineering, the general technical infrastructure, the cryogenic system, the magnet system with power converters and by the RF system with the RF power generation. The cost of these main items is mostly estimated with a bottom-up approach by contacting recognised experts in the field and in some cases by soliciting cost estimates from industrial partners. Other sources of cost are extrapolated from previous projects.

A staging approach meets the requirement of progressively implementing technologies that have different development times, like in the case of Nb$_3$Sn and HTS magnets, as soon as they are expected to become industrially available. Nonetheless the analysis presents the cost for each stage as a complete facility, rather than evaluating the incremental cost from one stage to the next, due to the too big error margin that affects any cost estimate at the moment. This kind of estimate will become possible after that the necessary R\&D will have been produced and more accurate estimates will become possible.

Having considered the level of maturity of the project, the analysis aims at obtaining a rough order of magnitude (ROM) estimate for the different stages that have been considered, with a level of precision that combines Class IV and Class V estimates for different areas of the project. For the same reason cost uncertainties take into account technical uncertainties only and do not consider purchase uncertainties.

\subsection*{Preliminary cost estimate for CERN energy stages}

A preliminary layout has been proposed for the installation of the Muon Collider complex at CERN, mostly in the already owned area, by also making use of existing accelerators and accelerator tunnels. The study for a Neutrino Factory at CERN \cite{Gruber:2002tn} that was released in 2004 has provided useful indications about a possible layout. By adopting the proposed scheme, the SPS and LHC tunnels would be used for the acceleration stages of the muons and host the Rapid Cycling Synchrotrons (RCS). The layout presented in Figure~\ref{fig_Muon_Collider_Injection_Complex_Cross_Sections} illustrates the baseline configuration for the cost estimate.

\begin{figure}[h]
\begin{center}
\includegraphics[width=15 cm]{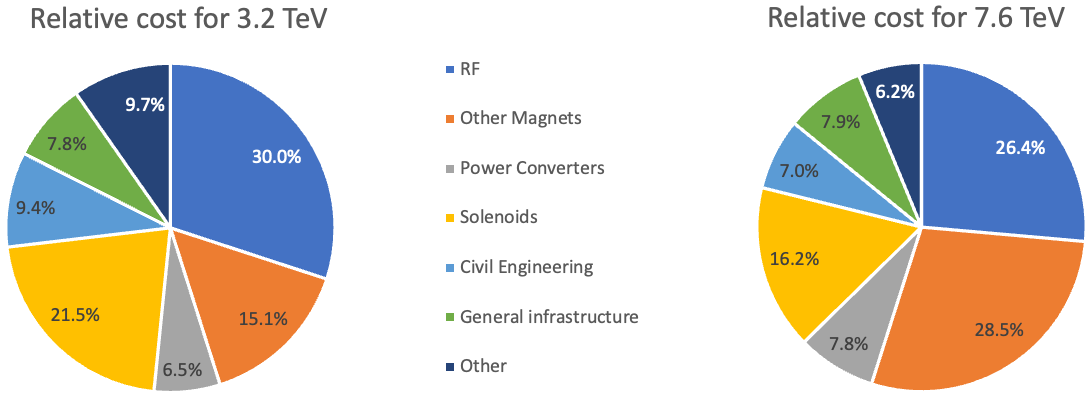}
\caption{The percentage cost distribution per technology in the Muon Collider facility at two possible CERN stages}
\label{fig:MuCol_Costdist}
\end{center}
\end{figure}

Figure~\ref{fig:MuCol_Costdist} shows the relative cost distribution associated to the main cost drivers for the envisaged scenarios of installation at CERN. For both scenarios it was assumed to reuse the existing infrastructure at CERN as much as possible, for example the SPS and LHC tunnels to host the RCS acceleration stages. Depending on the considered level of technology for the SC magnets, the energy reach and the performance of the facility may increase. The category "Other" includes the proton driver and the target complex. The level of accuracy of the present estimate justifies the fact that cost sources like computing, controls, beam instrumentation have not been considered so far. These items usually account for 5 to 7$\%$ of the accelerator cost. They could be introduced at a second stage of the costing exercise, when the estimate accuracy will reach the Class III - IV level.

The advantage of a CERN installation clearly shows up in the relatively modest cost fraction associated to the Civil Engineering contribution, together with the General Infrastructure. The estimate of cost of Civil Engineering was produced after a rather detailed study and a partial integration effort and can be considered as a reliable Class IV estimate. The highest weight for the low energy stage is represented by the cost of Radiofrequency in general, with an important contribution coming from the RF for the cooling channel, which is also a big contributor to the uncertainty of the estimate, since  its technology is still associated to a TRL of 2 and the related estimate accuracy cannot go beyond Class V.

\begin{figure}[h]
\begin{center}
\includegraphics[width=15 cm]{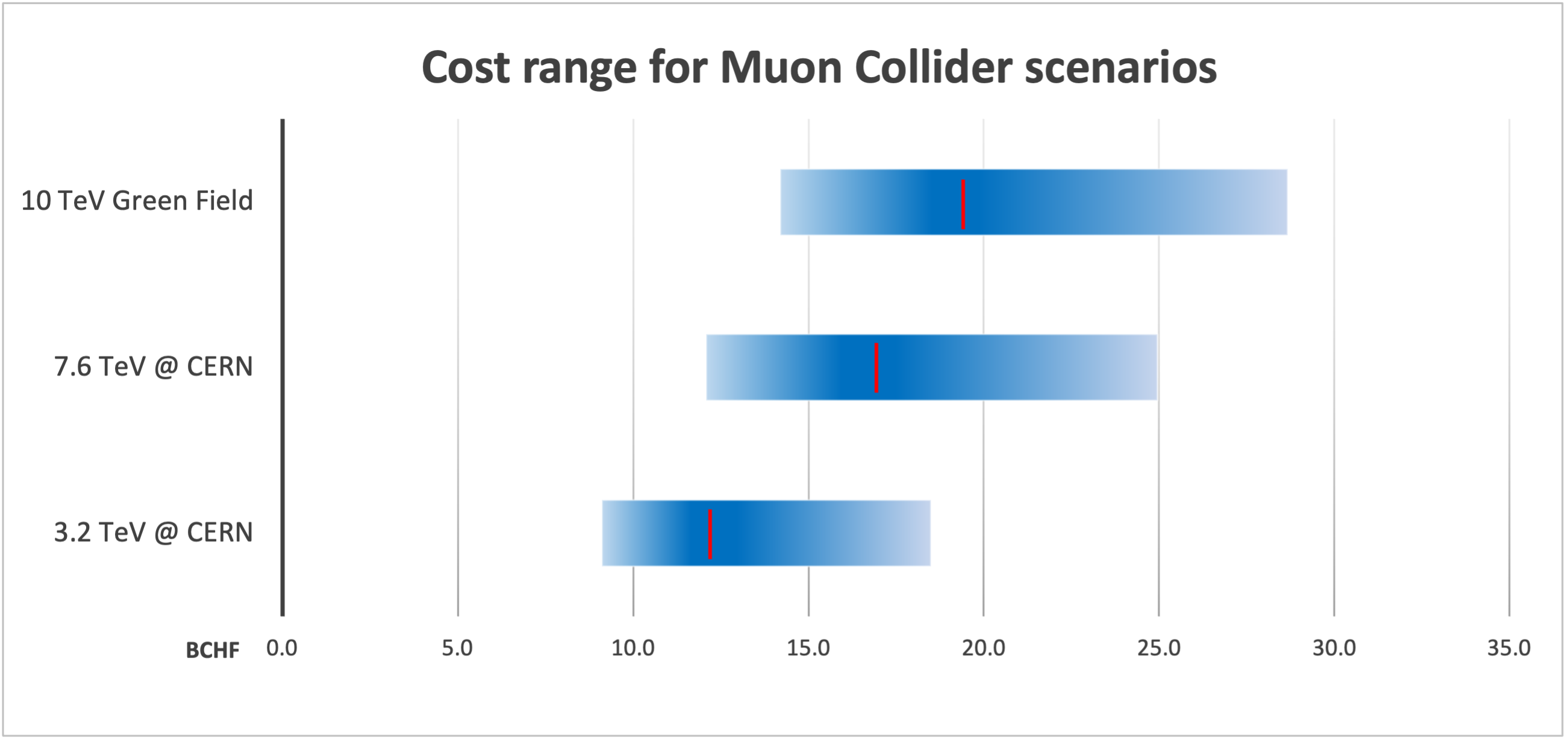}
\caption{The cost range is presented for the scenario of energy staging of the Muon Collider facility as compared to the Green Field realization. The cost for each stage represents the cost for a complete facility. The darkest part of the colour band identifies the most probable values, while the lighter sides of the band materialize the extension of the calculated uncertainty. Red markers show the position of the most probable value.}
\label{fig:MuCol_Costrange}
\end{center}
\end{figure}

In Figure~\ref{fig:MuCol_Costrange} the cost range for two scenarios (CERN installation and Green Field) is presented on a billion CHF scale. For the CERN scenario we considered two energy stages compatible respectively with the employ of Nb-Ti and Nb$_3$Sn magnets in the RCS and Collider rings. The energy limitation at 7.6\,TeV in this scenario comes from the hybrid RCS3 in the LHC ring, which could be overcome by the employ of HTS dipoles. The darkest colour band identifies the center of the cost distribution with lighter bands extending on both sides with lower probability. Red markers show the position of the most probable value.

The cost analysis shows how the cost and cost uncertainty for the different cost drivers are impacted by the level of maturity of a specific technology. Cost uncertainties for all items that were considered for this estimate belong to Class~IV and Class~V in the standard classification of project maturity levels. For Class~IV uncertainties range between -22.5\% and +35\%, while for Class~V the considered range is -35\% to +65\%.\\

The expected cost range for the realization of a 3.2\,TeV collider at CERN exhibits a reduced uncertainty, by making use of Nb-Ti arc magnets, if compared to other energy stages, where more advanced technologies for magnets are required in the acceleration stage and in the collider ring. On the other hand significant progress in the field of HTS magnets, which could eventually approach a magnet bore field of 15\,T with apertures in excess of 100\,mm, would enable the CERN installation to reach the ultimate collision energy of 10\,TeV.\\
It is worth noting that this very preliminary analysis, based on a still incomplete set of data, comes very close to what was presented by the Collider Implementation Task Force in 2021 as a contribution to the Snowmass exercise and later updated in 2023 \cite{Roser_2023}.

\subsection*{Opportunities for cost reduction}

Several areas may benefit from a substantial cost reduction in the future, depending on general progresses in the development of the HTS technology and on the opportunity to perform additional studies:
\begin{itemize}
\item The transition energy from the last recirculating linac stage to the first RCS was not optimized with respect to cost, but it was maintained at 63\,GeV for reasons related to the possibility of introducing a beam extraction to perform physics experiments at this energy. A more detailed study of the physics potential and a comparison to a cost optimized configuration may lead to a cost reduction in this area.
\item The study and introduction of other acceleration techniques, like the use of an FFAG accelerator in the accelerator chain as it was proposed in different studies for the neutrino factory, may bring to a different optimization of transition energies with a potential impact on the cost of the radiofrequency systems.
\item The cooling channel is certainly an area where considerable space exists for a cost optimization, since the existing configuration was exclusively driven by beam dynamics considerations, so far. Nonetheless the important level of uncertainty about the technical design and operation of this difficult area in the facility suggests to delay any optimization to a phase following the crucial R\&D activity.
\item The option exists to use a combination of He cryoplants for the heat loads at 4.5\,K and 20\,K, and He-Ne turbo-Brayton machines for the heat loads at 80\,K, in the 10\,TeV complex. For this option, it is not yet clear if it may represent an advantage from the perspective of cost, as investigation with industrial partners is presently ongoing and such machines are not currently available at CERN.
\item Finally, part of the muon transmission efficiency across the complex could be sacrified in exchange of a reduced cost, however this analysis would require additional time and may become an option only after that all other optimizations had been performed.
\end{itemize}

\section{Power drivers and power scale}

The inventory of power consumption for the different areas of the facility reflects the overall progress with the design study and results must be considered as still preliminary for most areas of the facility. A first estimate for power consumption could be nonetheless produced and it is summarized in Table~\ref{tab:Power} for the two energy stages of the CERN scenario and for the Green Field. The table considers the power values estimated for the nominal operation.

\begin{table}[h]
\caption{Power for the Muon Collider energy stages in the CERN scenario and Green Field. The higher collider power consumption for the 3.2~TeV CERN implementation is due to the use of low-temperature superconductor technology.}
\label{tab:Power}
\centering
\begin{tabular}{l c c c c}
 & & \textbf{CERN} & \textbf{CERN}  & \textbf{Green Field}\\
 & \textbf{Unit} & \textbf{3.2\,TeV} & \textbf{7.6\,TeV}  & \textbf{10\,TeV}\\ \hline \hline
Proton Driver & MW & 16.70 & 16.70  & 16.70 \\
6D Cooling & MW & 11.76 & 11.76 & 11.76  \\ 
RLAs & MW & 10.77 & 10.77  & 10.77 \\ \hline
RCSs & MW & 44.19 & 108.93 & 124.68  \\
Collider & MW & 10.00 & 4.10 & 4.10  \\
General Cooling and Ventilation & MW & 20.00 & 20.00  & 20.00 \\ \hline \hline
Total Power consumption & MW & 113.42 & 172.26 & 188.01  \\
\hline
\end{tabular}
\end{table}

At the moment, the power analysis includes contributions from RF power sources, RCS normal conducting magnets, cryogenics for magnets and for RF, general cooling requirements associated to buildings and cryogenics. The cooling of normal conducting RF systems is not considered here; the cooling of RF cavities in the muon cooling channel represents an unknown for the time being, just to provide an example.
With the present analysis, the RF systems represent the dominating source of power consumption, followed by the normal conducting fast ramping magnets in the RCSs. However, this preliminary inventory will certainly see integrations and adjustments along with the progress of the study.

\section{Sustainability}

The concept of muon colliders inherently involves less resources (energy, land, materials) than other collider concepts. 
All the energy of the particles is utilised in collisions, therefore it requires less energy to probe physics at the same parton centre of mass as a hadron collider. In addition, the relatively large muon mass strongly suppresses energy losses due to synchrotron radiation. 

A green field muon collider at 10~TeV requires about half the length of the tunnels (including the injectors) of a 100 TeV $pp$ collider. All the acceleration chain is entirely or partially pulsed. The electrical consumption is therefore reduced, and the collider ring, which is the only CW machine, is only about 10~km long. 

Furthermore, the IMCC based its design on technologies that should provide a reduced energy consumption with respect to the present state of the art, namely:
\begin{itemize}
    \item \textbf{Magnets based on High Temperature Superconductors (HTS)}. HTS have the potential to greatly lower the energy consumption for the same level of magnetic field (not always achievable with conventional LTS materials). It is fair to say, however, that this technology is not yet demonstrated to work reliably in an accelerator environment, therefore the need for a muon cooling Demonstrator. 
    \item \textbf{High Efficiency Klystrons}. The MC cooling channel requires low duty factor, high power (> 20 MW) RF power sources. No klystron with similar characteristics exist on the market, therefore the IMCC will, if budget is made available, develop a new klystron with a minimum of 80\% efficiency (against a typical 60\% to 65\% efficiency of presently commercially available klystrons. 
    \item \textbf{Superconducting RF cavities}. Wherever possible (for non-pulsed RF) the IMCC has foreseen superconducting RF cavities. This choice allows to reach high electric gradient for a much reduced electrical consumption with respect to room temperature RF structures. 
    \item \textbf{Efficient Cryogenics systems}. By choosing from the start to use high temperature superconductors, the IMCC will invest, as resources will become available into the development of efficient cryogenic systems at temperatures around 20K, and will therefore help pushing the boundaries for such technologies.  
    \item \textbf{Energy recovery linacs}. The IMCC presently considers in its baseline design the use of Energy Recovery Linacs, in the early stage of muon acceleration. While giving advantages in terms of fast acceleration, they also provide energy to the beam in an efficient way.
\end{itemize}

At present, the ideal muon collider is being designed considering a green field facility, not yet based on a precise geographical situation. For this reason, no considerations are being made for energy sourcing, energy recovery, water sourcing, wastewater treatment, etc. 
Such considerations will be part of the specific implementations in the various labs, in coherence with national legal frameworks and will profit from local efforts already ongoing. 
For instance, an eventual implementation at CERN will build on the experience of the HL-LHC and the FCC-ee feasibility study. 

\subsection*{Sustainability Considerations}

Studies have shown that the supply of commercial electricity continues to decarbonize on a global scale. This decarbonization is expected to accelerate with the reduction in cost of renewable technologies, in particular solar and battery storage, which has been remarkable in recent years. When considering the sustainability of future colliders, it is therefore important to consider the impact of both operations and construction on sustainability. Sustainability assessments of future colliders have shown that construction, rather than operations, may well dominate the climate impact of a new particle physics facility~\cite{arup,Bloom:2022gux,abramowicz2025linearcollidervisionfuture,breidenbach2023sustainability}. This creates a tremendous opportunity for the Muon Collider to provide a sustainable pathway to the 10 TeV parton center-of-momentum (pCM) scale. This is possible for a green field site, but is rendered even more appealing by reutilizing significant infrastructure from existing facilities, for example, CERN or Fermilab.

A life-cycle assessment of the ILC and CLIC accelerator facilities performed by ARUP~\cite{arup} to evaluate their holistic global warming potential (GWP), has so far provided the most detailed environmental impact analysis of construction. The components of construction are divided into classes: raw material supply, material transport, material manufacture, material transport to work site, and construction process. These are labelled A1 through A5, where A1-A3 are grouped as materials emissions and A4-A5 are grouped as transport and construction process emissions. The approximate construction GWPs for the main tunnels are 6.38 kton CO$_2$e/km for CLIC (5.6m diameter) and 7.34 kton CO$_2$e /km for ILC (9.5m diameter); the Muon Collider tunnel design is similar to that of CLIC, so 6.38 kton CO$_2$e/km is used here for the calculation of emissions.

While a comprehensive civil engineering report is unavailable for the Muon Collider, we estimate the concrete required for access shafts, alcoves, caverns, klystron galleries, etc. to contribute an additional 30\% of emissions, similar to what is anticipated for other colliders. The analysis indicates that the A4-A5 components constitute 20\% for CLIC and 15\% for ILC. In the absence of equivalent life cycle assessment analysis for the Muon Collider, we account for the A4-A5 contributions as an additional 25\%. For a greenfield site, a 3 TeV or 10 TeV Muon Collider would require approximately 30 km and 70 km of tunnels, respectively. However, a site reusing existing infrastructure such as CERN for 7.6~TeV center of mass would only need 15 km of tunnels primarily limited to the muon source, cooling channel and the collider ring. For these three cases, the emissions in kton CO$_2$e is on the order of 700, 300 and 150 for decreasing tunnel length, including A1-A5 contributions as described above. 

The power consumption for the Muon Collider has been estimated to be less than 200~MW for operation at 10~TeV pCM~\cite{roser2023feasibility}. This power estimate has been refined by assessing the power budget for the proton driver, muon cooling, recirculating linacs, RCSs, the collider ring and general infrastructure to be on the order of 115~MW, 190~MW and 175~MW for the 3~TeV green field, 10~TeV green field and 7.6~TeV CERN site (see Table~\ref{tab:Power}). Achieving these targets will require careful design and accounting of power budgets as the design of the Muon Collider progresses. For example, estimates of the radiation losses due to muon decays for the RCSs are of the order of 0.2~W/m/turn or 8~W/m at 50~Hz operation. While this is a tolerable amount of power loss for the beam, it will be critical to absorb this power outside of the coldest stages of the cryogenic and superconducting systems in those rings with the use of absorbers and thermal shields. 

From this peak power consumption, the average estimated power consumption over a year of operation, including standby and downtime, is approximately 0.75, 1.25 and 1.1~TWh for the 3~TeV green field, 10~TeV green field and CERN site. Assuming 16 g/kWh CO$_2$e this corresponds to approximately 70, 95 and 75 kton CO$_2$e, respectively, for a 5-year physics program. In all cases, this is significantly less than the impact of construction, emphasizing again the importance of minimizing impact due to construction when considering the sustainability of future colliders and showcasing the Muon Colliders potential for providing a sustainable pathway to the 10~TeV pCM scale.
\section{Implementation at Fermilab}
\label{2:implement:sec:fermilab}

Most elements of a muon collider design are not specific to a particular site. Two areas do have elements specific to the Fermilab site: the proton driver, since there is existing infrastructure for proton beams at Fermilab, and acceleration, since the site boundary at Fermilab is a potential limitation.
\subsection*{Proton Driver}
Fermilab currently achieves high proton beam power by accelerating protons in the main injector from 8~GeV to 120~GeV. The final energy of 120~GeV is too high of an energy for muon production for a muon collider, so the proton driver focuses on an upgrade path for the existing 8~GeV booster. The current upgrade plans (the proton intensity upgrade) focus on providing higher power at 120~GeV to DUNE, but the upgrade plans could take into consideration compatibility with a future muon collider. The proton intensity upgrade plans involve a linac energy upgrade to 2~GeV~\cite{osti_1988511}. To get to 8~GeV, the choices are between a linac to 8~GeV or an RCS. For muon collider intensities, the RCS would require intensities that are too high for 2~GeV injection~\cite{neuffer:ipac2024-tupc41}. Thus, the use of the RCS would require higher energy injection linac or further acceleration, to energies in the range of 12--24~GeV, possibly in the accumulator ring. 

Thus, a linac accelerating to 8~GeV appears favourable. This would require that the linac for the proton intensity upgrade could eventually be upgraded to be capable of higher currents, as high as 25~mA. Ideally, the accumulator ring would have bunch lengths in the 20--40~ns range, so that the compressor ring can shorten them to 1--3~ns. Directly accumulating into such short intervals in an accumulator ring with a circumference of 300--500~m as envisioned would require instantaneous beam currents as high as 300~mA, so either the accumulator ring must do some initial compression or multiple linac front ends would be needed. Scenarios assume that 4 bunches are accumulated and compressed, then extracted into separate beam lines and delivered simultaneously to the target. Charge stripping injection into the accumulator ring should be done with laser stripping to avoid foil degradation and scattering losses~\cite{neuffer:ipac2024-tupc41}.

It is important for potential future siting of a muon collider at Fermilab that muon collider requirements are considered in their proton intensity upgrade plans. Thus, muon collider proton driver studies need to be performed to provide concrete input to that process:
\begin{itemize}
    \item Since 8~GeV is a natural energy for the Fermilab site, design and simulation of 8~GeV accumulator and compressor rings should be completed to ensure that they are capable of delivering a beam with the required parameters for a muon collider, what beam parameters could be expected, and what the consequences would be for the injected beam.
    \item Design of a target station with simultaneous delivery of multiple beams needs to be undertaken to determine whether such a scenario is viable.
    \item In conjunction with the proton intensity upgrade, a detailed scenario should be designed and simulated for delivering a beam compatible with muon collider requirements.
\end{itemize}
If 8~GeV energy proves to be problematic, these studies should adjust their design parameters to a higher energy as needed.
\subsection*{Acceleration}
To keep acceleration on the Fermilab site, the circumference of a circular ring is effectively limited to 15.5--16.5~km. When using a hybrid RCS, while it may be possible to accelerate to 5~TeV in that circumference, the acceleration achieved in the stage that reaches that energy is very modest, requiring injection at 4.1~TeV or higher~\cite{capobianco-hogan:ipac2024-mopr01}. As more details of the design are taken into consideration, generally they require more space without bending, thereby reducing the energy range of that last stage or the maximum energy reached. Furthermore, to maximize the energy reach of the machine, it is beneficial to have fewer superperiods~\cite{capobianco-hogan:ipac2024-mopr01}, generally fewer than suggested in the design studies reported earlier in this document. The, an RCS design R\&D program should continue that
\begin{itemize}
    \item Produces high energy acceleration chain for the Fermilab site including as much detail as possible to get an understanding of the energy reach achievable on the Fermilab site.
    \item Performs simulations of these designs including longitudinal dynamics and collective effects, with a focus on determining the minimum number of superperiods that can be used in the design to maximize that energy reach. 
\end{itemize}

In a hybrid RCS, even for an infinitely high field in the superconducting dipoles, the average bend field cannot go beyond value that depends only on the pulsed dipole field and the energy ratio for the accelerator~\cite{berg:ffa24}. Since pulsed dipole fields are generally limited by the saturation magnetization of iron, one cannot use a hybrid RCS to accelerate to arbitrarily high energies in a given circumference simply by reaching higher fields in superconducting magnets. An alternative that would enable reaching higher energies with larger superconducting fields would be a fixed field alternating gradient accelerator (FFA). Design parameters have been considered for an FFA that would fit on the Fermilab site and accelerate to around 5~TeV~\cite{berg:ffa22}. Injection and extraction appeared challenging for that design due to the lack of a sufficiently long straight section for a septum. To create longer straights, an elliptically shaped ring is under consideration~\cite{berg:ffa24}.
\part{Physics Preparatory Group Benchmarks} \label{5:ppg}
\chapter{Introduction}
\label{ch:PPGB}

The Physics Preparatory group (PPG) will prepare the scientific contribution for the European Strategy group, based on the input received from the community. In this context, the PPG requested specific technical input on a set of benchmark topics~\cite{PPGquestions} in the subject areas that define seven physics-oriented PPG working groups. In addition, three more working groups will cover respectively ``Accelerator technologies'', ``Detector instrumentation'' and ``Computing''.

The inputs requested for the ``Accelerator technologies''~\cite{PPGquestions} are identical to those requested by the strategy secretariat for large accelerator projects~\cite{PPGacc} and  were submitted separately in the Addendum to the main 10-pages ESPPU submission.

The IMCC input on muon collider physics, detector and computing, and accelerator consists of Chapters~\ref{1:phys:ch}, \ref{1:det:ch}, \ref{1:acc:ch}, and~\ref{1:tech:ch}, together with the R\&D plans detailed in Part~\ref{3:rd}. The scope of these inputs extends beyond the topics common to multiple collider projects and that can be used to perform quantitative or qualitative comparisons. 
They are supplemented by the present Chapter, which has the aim of providing answers to specific technical questions to enable those comparisons in the context of the update of the European Strategy for Particle Physics. 

\chapter{Electroweak, Higgs and Top physics}
\label{sec:EHT}

\section{Overview}
A muon collider is a high energy electroweak collider. It offers revolutionary opportunities for Electroweak (EW), Higgs and Top physics thanks to the high available energy and the absence of QCD background. It will first and precisely explore the new high-energy regime where EW and Higgs physics unify in the SM paradigm of a spontaneously-broken gauge theory, enabling the experimental observation of its most striking and direct manifestations. The SM unification and its manifestations extend to the fermionic particles, with the top quark playing the major observational role owing to its large Yukawa coupling. Section~\ref{1:physics:sec:overview} emphasises the exploration of new phenomena within the SM as a major and  unique physics driver of the muon collider project.

The impact on the search for departures from the SM predictions of a program of measurements at the muon collider are equally striking, thanks to a rich variety of accessible observables. The present section provides an overview of the different measurements and of their sensitivity to new physics effects. Section~\ref{sec:IM} offers a detailed description of the sensitivity projections that are presented as input for the PPG assessment of the benchmark questions on Higgs, EW and Top physics. Selected physics results are described in Section~\ref{sec:PPGQ}. See also Section~\ref{1:physics:subsec:highlights}.\\[-20pt]
\paragraph*{High-energy scattering}
\noindent\\
The first class of observables are scattering cross sections for the production of SM particle pairs with an invariant mass that is nearly as high as the collider energy of 10~TeV. They can be measured with percent-level precision with the muon collider design luminosity. These cross sections feature a high characteristic energy scale, which translates into an enhanced sensitivity to BSM dynamics with very high characteristic mass scale. A typical scenario is that BSM particles with mass $M$ induce corrections of order $E^2/M^2$ to the SM predictions for observables at scale $E$, for $E<M$. Since $E=10$~TeV, detectable effects of order percent could emerge in high-energy cross sections even if $M$ is as high as 100~TeV. In an EFT description of new physics, this estimate corresponds to a order $100$~TeV exclusion reach on the scale $\Lambda$ of any dimension-six operator that produces a quadratic energy growth in the muon collider high-energy cross sections. Correspondingly, the sensitivity projections in the two upper panels of Figure~\ref{fig:EFTFIT} range from one hundred to several hundreds of TeV.

High-energy measurements at the muon collider offer a novel tool of investigation with multiple advantages in comparison with the traditional strategy based on the very precise measurement of relatively low-energy observables at the EW-scale or below. 
First, the muon collider reach is generically superior. A new physics scenario with mass $M=100$~TeV---or $\Lambda=100$~TeV EFT scale, for unit coupling---gives order $E^2/M^2=10^{-6}$ corrections to $E\simeq100$~GeV observables. It is impossible to detect such tiny departures from the SM because one part per million experimental measurements and theoretical predictions are generically unfeasible at a collider experiment. Future $e^+e^-$ colliders can reach one part per million statistical sensitivity on some observables at Tera-$Z$, but comparable theoretical, parametric and experimental uncertainties will be challenging to achieve.

Second, the muon collider measurements are more suited to turn a possible tension with the SM predictions into a conclusive evidence for new physics and to characterise the discovery. If sticking to $E\simeq100$~GeV observables, extending the sensitivity beyond current knowledge requires order $10^{-4}$ or $10^{-5}$ accuracy of the measurements and of the corresponding theoretical predictions. Many different physical effects become relevant at this high level of accuracy. Isolating genuine BSM contributions disentangling them from the mismodelling of all these many SM effects will be a challenge for future $e^+e^-$ colliders. Mismodelling is instead not a severe issue for the muon collider, given the limited percent-level measurements accuracy. 

Furthermore, a large set of high-energy observables can be studied at the muon collider including differential distributions. This will enable to characterise the observed tension, to turn it into a robust evidence for new physics and ultimately to learn about the underlying BSM model. In the EFT language, the excellent perspectives for BSM model selection at the muon collider are illustrated by the stability---see Figure~\ref{fig:EFTFIT}---of the exclusion reach of each operator under the marginalisation over the other operators. The global and single-operator limits are close because the muon collider measurements do not leave poorly constrained directions in the operator coefficients space. The measurements can thus disentangle the different operators and single out the one that is responsible for a putative departure from the SM predictions.

In spite of the strongest maximal reach on the $\Lambda$ scale, the muon collider measurements are to large extent complementary to the ones that are possible at a low-energy $e^+e^-$ collider. In fact, not all the dimension-six EFT operators grow with energy in the 10~TeV scattering processes that the muon collider can measure. Other muon collider measurements---described later in this section---can probe also the non-growing operators advancing present-day knowledge, but with order few or 10~TeV sensitivity to the scale, and not 100~TeV. A future $e^+e^-$ collider would have a slightly better reach on those operators, enabling a more complete coverage of new physics scenarios in combination with the muon collider measurements. Other examples of complementarity emerge in Higgs physics, as described later in this section. Additionally, an $e^+e^-$ Electroweak, Higgs and Top factory could be built on a shorter timescale than the muon collider. The R\&{D} path towards the muon collider is therefore not in contrast with an $e^+e^-$ project but rather highly synergetic. Especially in the scenario where tensions with the SM emerge in $e^+e^-$ measurements, technology should be ready to build a muon collider as the best option to confirm and characterize the new discovery.\\[-20pt]
\paragraph*{Vector boson fusion}
\noindent\\
The second class of measurements does not exploit the high collision energy directly, but it exploits it indirectly through the enhancement of the probability for emitting virtual vector bosons from the incoming very energetic muons. Namely, it exploits the large cross section for Vector Boson Fusion (VBF) type of reactions. 

By VBF, the muon collider will produce 10 million Higgs bosons, which can be used for the permille-level determination of the Higgs couplings to vector bosons, bottom quarks and $\tau$ leptons, and the percent-level determination of many other couplings. The projections are competitive, but again also complementary, with the sensitivity of a future $e^+e^-$ Higgs factory. The muon collider could also resolve the degeneracy between the Higgs width and the Higgs couplings by a direct measurement of the $Z$-boson fusion Higgs production cross section and in turn the absolute determination of the Higgs coupling to the $Z$. This measurement relies on the possibility of measuring the energy of the forward muons that emerge from the emission of the effective $Z$ bosons. This is possible in principle but not yet supported by detector studies. It is thus worth mentioning the absolute determination of the Higgs couplings as a further element of complementarity with a low-energy Higgs factor, had the muon collider determination turn out to be unfeasible. 

Several hundreds of thousands Higgs boson pairs will be also produced at the muon collider, in the VBF process $VV\to{hh}$. This will enable a precise and direct determination of the Higgs self-coupling at the few percent level. This is not only interesting in a comparison with the much inferior projected sensitivity of HL-LHC and low-energy $e^+e^-$. Few to ten percent is a threshold for the size of Higgs trilinear coupling modifications that can be typically expected---compatibly with other constraints on new physics including determinations of the single Higgs couplings---in generic BSM scenarios. Crossing this threshold thus offers better discovery perspectives.

The fusion of vector bosons can also produce other pairs of SM particles, including top quarks and other fermions. Or, produce pairs of vector bosons, in which case the process is denoted as a Vector Bosons Scattering (VBS). On top of the large total rate and the limited backgrounds, a remarkable feature of these measurements is that the invariant mass distribution of the final state particles extends up to high energies above 1~TeV. This offers exceptional perspectives to verify the key SM prediction of vector boson scattering Unitarization through the virtual Higgs boson exchange, but also to boost the sensitivity to EFT operators that produce growing-with-energy contribution to the vector bosons scattering or fusion amplitudes. The full breadth of a program of VBF and VBS measurements at the muon collider has not been unveiled yet. The results presented in the following section only consider the VBF production process of quarks and leptons. In a global EFT interpretation of the muon collider measurements, such VBF di-fermion processes probe operators that are not tested in high-energy scattering or in Higgs physics. Additionally, they provide the best probe of the top Yukawa coupling. \\[-20pt]
\paragraph*{Low-energy muon colliders}
\noindent\\
The IMCC does not investigate the feasibility of low-energy muon colliders, however the proposed IMCC R\&{D} program will make key progress towards the realisation of low-energy muon colliders operating at the Higgs pole of 125~GeV and/or at the threshold for $t\bar{t}$ production of 345~GeV. In fact, there is not only a technological synergy between high- and low-energy muon colliders: the very same cooling complex of the high-energy muon collider can be used to produce the beams for the low-energy colliders.

It should be noted that the physics goals of the low-energy muon colliders pose additional feasibility challenges in comparison with the high-energy collider and that addressing these challenges is not within the scope of the IMCC. Specifically, an extremely small beam energy spread is required at the Higgs pole muon collider in order to study the Higgs particle line shape, and a precise determination of the beam energy is needed in both colliders for an accurate measurement of the Higgs and top masses. However, addressing these technical feasibility challenges by dedicated studies would enable appealingly compact ($L=300$~m, for the top threshold) and cheap projects to be developed with the same facility that will deliver the high energy muon collider. The opportunities of low-energy muon colliders should thus be considered as an additional physics driver of the muon collider R\&{D} program.

\section{Input measurements}\label{sec:IM}

\begin{table}[t]
\begin{center}
\begin{tabular}{c|c|c|c}
\hline
Measurement & Description & Reference & Repository \\\hline
IM.1 & Single Higgs $\sigma\times$BR & \cite{Forslund:2022xjq} Tables~9, 10 & 
\href{https://gitlab.com/mforslund/offshell-higgs-width-at-muc}{Link}
\\\hline
IM.2 & Invisible/Inclusive Higgs & \cite{Ruhdorfer:2024dgz} Section~III & \\\hline
IM.3 & Off-shell Higgs & \cite{Forslund:2023reu} Section~2 & \href{https://gitlab.com/mforslund/offshell-higgs-width-at-muc}{Link} \\\hline
IM.4 & Double Higgs & \cite{EWMuC} & 
\href{https://github.com/xingwang1990/VBF_MuC}{Link} \\\hline
IM.5 & High-energy scattering & \cite{EWMuC} & \href{https://github.com/aglioti/muColSudakov}{Link} \\\hline
IM.6 & VBF di-fermion & \cite{EWMuC} & \href{https://github.com/xingwang1990/VBF_MuC}{Link} \\\hline
IM.7 & Low-energy MuC & \cite{deBlas:2022aow,Franceschini:2022veh} &
\\\hline
\end{tabular}
\end{center}
 \caption{Input measurements (IM) for electroweak, Higgs and top physics.
 \label{tab:IMEW}}
\end{table}

The input measurements (IM) that are relevant to address the PPG requests on electroweak physics are listed in Table~\ref{tab:IMEW}, together with the corresponding references and online repositories. The measurements are described below and used as input for the results presented in the rest of the section.\\[-20pt]
\paragraph*{IM.1: Single-Higgs.} The perspectives for single-Higgs production cross section times branching ratio measurements have been studied in~\cite{Han:2020pif,AlAli:2021let,Forslund:2022xjq,Andreetto:2024rra}, with Ref.~\cite{Forslund:2022xjq} providing the most complete and up to date results. Tables~9 and~10 of~\cite{Forslund:2022xjq} report the expected precision 
in the determination of the Higgs signal strengths in the accessible decay channels. The results in Table~9 of~\cite{Forslund:2022xjq} do not exploit the possibility of tagging the forward muons---in the rapidity range $2.5<|\eta|<6$---that emerge in the neutral $ZZ\to{h}$ VBF Higgs production process. Forward muon tagging is instead assumed in Table~10, enabling the separate determination of the Higgs signal strengths in the charged $WW\to{h}$ and neutral VBF production modes. It should be emphasised that this only requires \emph{tagging} the forward muons, which is arguably straightforward since the high energy muons produced in VBF are penetrating particles that cross the nozzles and the other elements of the collider. Other results in Higgs physics presented below (specifically, IM.2) require instead \emph{measuring} the forward muons momentum, which is less straightforward. 

All the results of Ref.~\cite{Forslund:2022xjq} include the relevant backgrounds and a parametric modelling of the detector response. Additionally, they have been validated against the available full simulation results. No theory, parametric and experimental uncertainties are included in the projections and they are not expected to play a major role given that the required (per mille) level of accuracy should not be challenging to attain at a lepton collider.\\[-20pt]
\paragraph*{IM.2: Invisible/Inclusive Higgs.} Detecting energetic forward muons and measuring their energy with order~$10\%$ resolution enables the determination of the $ZZ\to{h}$ production cross section times the branching ratio to invisible final states, as well as the absolute measurement of the Higgs $ZZ\to{h}$ production cross section by the recoil method, which in turn provides an absolute determination of the Higgs coupling to the $Z$ boson. This was studied in~\cite{Ruhdorfer:2023uea,Li:2024joa,Forslund:2023reu,Ruhdorfer:2024dgz}, and Ref.~\cite{Ruhdorfer:2024dgz} contains the most up to date results. In the context of Higgs physics, Ref.~\cite{Ruhdorfer:2024dgz} also provides a sensitivity projection for the $CP$ properties of the $hZZ$ coupling. The results include backgrounds and take into account the effects associated with the imperfections of the muon collider beam and, more importantly, the momentum resolution of the forward muon detector.\\[-20pt]
\paragraph*{IM.3: Off-shell Higgs.} If the measurement of the forward muon energy will turn out to be impossible, an absolute determination of the Higgs couplings lifting the degeneracy between modified couplings and BSM untagged Higgs decay will still be possible by the off-shell Higgs measurements of Ref.~\cite{Forslund:2023reu}. The study targets the effects of modified Higgs couplings in vector boson scattering processes $VV\to{VV}$, which are independent of the Higgs width and hence resolve the degeneracy. Backgrounds and a parametric detector simulation are taken into account. The measurements used in the analysis, namely the cross section in each analysis bin as a function of the Higgs coupling modifiers, are available at the online repository linked in Table~\ref{tab:IMEW}. \\[-20pt]
\paragraph*{IM.4: Double Higgs.} The production of two Higgs bosons in the VBF process was studied in~\cite{Han:2023njx,Buttazzo:2020uzc,EWMuC}, and Ref.~\cite{EWMuC} contains the most complete analysis of this process. The analysis exploits energy and angular information of the Higgs bosons---by a two-dimensional binning in the Higgs transverse momentum and the di-Higgs invariant mass---in order to disentangle different new physics effects encapsulated in dimension-six EFT operators. It also exploits the tagging of the forward muons to disentangle effects in charged or neutral VBF processes. The selected operators  are those that contribute to the process directly, including the operator responsible for the modification of the Higgs trilinear coupling. The link in Table~\ref{tab:IMEW} provides observed-level event yields expectations as a function of the relevant Wilson coefficients for Warsaw-basis~\cite{Grzadkowski:2010es} operators. \\[-20pt]
\paragraph*{IM.5: High-energy scattering.} The processes of highest energy emerge from the direct annihilation of $\mu^+$ with $\mu^-$ producing a pair of SM particles with an invariant mass of about 10~TeV. The perspective for measuring such high-energy scattering cross sections and their sensitivity to new physics were investigated in~\cite{Buttazzo:2020uzc,Chen:2022msz}. A complete survey of the possible final states and their interpretation in a general EFT context has been presented in Ref.~\cite{EWMuC}. 

The results are available in an online repository linked in Table~\ref{tab:IMEW}. For each final state, the repository provides the prediction of the cross section in angular bins, as a function of the relevant EFT operators in the  Warsaw basis. These truth-level cross sections have to be contracted with an efficiency/mistag matrix that provides an high-level modelling of the detector performances for the reconstruction of the highly energetic final state particles. 

On top of neutral final states---such as for example a charged lepton pair $\ell^+\ell^-$---charged final sates---such as $\ell\nu_\ell$---are also considered. The corresponding cross sections are nearly as large as the one of the neutral process counterpart because the soft and collinear emission of a charged $W$ boson is enhanced by large double logarithms. In addition, two types of neutral final states are considered. Exclusive final states are those where the emission of massive vector bosons---and relatively hard photons or gluons---is vetoed. Semi-inclusive final states allow instead for the emission of arbitrary gauge bosons radiation, in addition to the two hard particles. Results are reported for the exclusive cross section and for the cross section ``with radiation'', which is simply the difference between the semi-inclusive and the exclusive cross section. The cross sections for charged hard final states are obviously only of the semi-inclusive type as they require the emission of at least one charged $W$ boson. 

The theoretical predictions include the resummation of the leading (double) EW logarithms. The resummation effects are large and need to be taken into account for quantitatively valid results. However, tree-level cross section predictions are also reported to facilitate cross-checks. \\[-20pt]
\paragraph*{IM.6: VBF di-fermion production}
The production of a pair of quarks or leptons via vector boson fusion offers significant opportunities to probe new physics, especially for those EFT operators that lead to an energy growing behaviour in the corresponding scattering amplitudes. The production of top quark pairs was studied in~\cite{Chen:2022yiu,Liu:2023yrb} for the determination of the top quark Yukawa coupling. Ref.~\cite{EWMuC} provides (see also~\cite{Li:2025ptq} an extensive survey of the VBF di-fermion processes, namely of the final states $\mu^-\mu^+ f\bar{f},~\nu_{\mu}\bar{\nu}_{\mu} f\bar{f},~\mu^\pm\nu_{\mu} f\bar{f}'$, with $f = q,e,\tau$, by considering the effect of all the relevant growing-with-energy operators. The non-growing operators do not contribute appreciably, taking into account previous constraints and perspective constraints from other muon collider measurements. Scattering processes with muons in the final state (i.e., $f=\mu$) are not considered because the corresponding operators are probed better in the high-energy production of vector bosons from direct muons annihilation.

The results are available in an online repository linked in Table~\ref{tab:IMEW}. The cross sections are given as functions of the Warsaw-basis Wilson coefficients in bins of di-fermion invariant mass and transverse momentum. A recoil mass cut is also applied to suppress non-VBF processes, as detailed in~\cite{EWMuC}. Like for IM.5, these truth-level cross sections are to be multiplied by the efficiency/mistag matrix. For most operators, the sensitivity is driven by terms of linear order in the EFT Wilson coefficients, and the quadratic terms are omitted from the predictions. Quadratic terms are instead retained in few particular cases where an accidental cancellation of the linear (interference) term boosts their impact of the sensitivity. 

In addition to di-fermion production, di-fermion production in association with a Higgs boson, $VV\to f\bar{f}'h$, is also considered. This is advantageous to probe certain effects, as it suppresses backgrounds initiated from the photon-photon fusion processes. Due to the limited statistics, the cross sections for $f\bar{f}'h$ final states are not binned in any kinematical variables. \\[-20pt]
\paragraph*{IM.7: Low-energy muon colliders.} The cooling complex of the high-energy muon collider can be used to produce the beams for low-energy muon colliders operating at the Higgs pole of 125~GeV and at the threshold for $t\bar{t}$ production of 345~GeV. Based on available target parameters for such low-energy muon colliders, sensitivity projections for Higgs physics and top quark mass measurements have been studied in Refs.~\cite{deBlas:2022aow} and~\cite{Franceschini:2022veh}, respectively.

The 125~GeV collider would enable measurements of the Higgs properties such as the coupling to muons with half-percent precision and the determination of the Higgs boson decay width at the few percent level. To be noted that the muon collider direct measurement from the line-shape is a truly model-independent determination of the Higgs boson width unlike the one that emerges from a combined fit with Higgs couplings modifiers. Higgs line-shape studies can also measure the Higgs boson mass. However, the expected accuracy depends on the precision in the calibration of the energy of the collider, which has not been investigated yet. The purely statistical sensitivity reaches the astonishing level of one part per million, i.e. a~$0.2$~MeV error on the mass. An energy calibration at the $10^{-4}$ level would enable a determination of the Higgs mass with an error of $10$ or $20$~MeV, which would be sufficient to eliminate the corresponding parameteric uncertainties from future EW fits from low-energy measurements.

The main goal of a top-threshold collider would be the accurate measurement of the top quark mass by a threshold scan. On top of generically improving our knowledge of the SM, a determination of the top mass with less than 50~MeV uncertainty would settle the question of the SM vacuum stability and measure the instability scale. Current estimates of the theory uncertainties indicate that a 50~MeV error could indeed be achieved by a threshold scan performed at an electron-positron or at a muon collider. The top-threshold muon collider could achieve a statistical precision of 20 or 10~MeV in one year of run, with competitive advantages in comparison with the electron-positron option such as the much smaller cost and the reduced impact of ISR photon radiation. Order $10^{-4}$ precision in the calibration of the beam energy would be required, like for the measurement of the Higgs mass at the 125~GeV collider. Unlike for the Higgs line-shape measurements at the 125~GeV collider, the top mass measurement at the top-threshold collider does not pose any challenging requirement on the beam energy spread.\\[-20pt]
\paragraph*{Important missing studies.} The recent intense development of muon collider phenomenology offers a rather complete coverage of the opportunities for electroweak, Higgs and top physics. Future work will refine and consolidate the sensitivity projections in parallel with the advances on experimental physics and detector design. On the other hand, some elements of the muon collider physics potential are still to be investigated. For instance, an assessment of the potential of vector boson scattering processes is still missing, due to the large number of different channels and the complexity of the final states. These studies might have significant impact on the muon collider EFT fit, but also unveil the full potential of the muon collider on Unitarization and EW symmetry restoration tests within the SM. We also miss studies that exploit the large total number (100 and few hundred million, respectively) of $Z$ and $W$ bosons that are produced at the muon collider. They could be used for coupling measurements or for $W$ mass determination.

\section{Selected results}\label{sec:PPGQ}

\begin{figure}[t]
\begin{minipage}{0.5\textwidth}
\renewcommand{\arraystretch}{.89}
\setlength{\arrayrulewidth}{.2mm}
\setlength{\tabcolsep}{0.6 em}
\begin{center}
\begin{tabular}{c|c|c|c}
\multicolumn{4}{c}{kappa-0 Fit [\%]} \\ 
\hline
& {\makebox[23pt]{${\textsc{hl-lhc}}$}}
& {\makebox[23pt]{${\textsc{hl-lhc}}$}}
& {\makebox[22pt]{${\textsc{hl-lhc}}$}}
\\[0pt]
& \multicolumn{1}{c|}{\ } & \multicolumn{1}{c|}{\makebox[23pt]{\small{$+10\,\textrm{{TeV}}$}}} 
& \multicolumn{1}{c}{\makebox[23pt]{\small{$+10\,\textrm{{TeV}}$}}}   
\\[-2pt]\ & \ & \multicolumn{1}{c|}{\ } & \multicolumn{1}{l}{\hspace{0.5pt}+{$e e$}} 
\\ 
\hline
$\kappa_W$ & 1.7             & 0.1 & 0.1 \\ \hline
$\kappa_Z$ & 1.5             & 0.2 & 0.1 \\ \hline
$\kappa_g$ & 2.3             & 0.5 & 0.5 \\ \hline
$\kappa_{\gamma}$ & 1.9      & 0.7 & 0.7 \\ \hline
$\kappa_{Z\gamma}$ & 10      & 5.2 & 3.9 \\ \hline 
$\kappa_c$ & -               & 1.9 & 0.9 \\ \hline
$\kappa_b$ & 3.6             & 0.4 & 0.4\\ \hline
$\kappa_{\mu}$ & 4.6         & 2.4 & 2.2 \\ \hline
$\kappa_{\tau}$ & 1.9        & 0.5 & 0.3 \\ \hline\hline
$\kappa_t^*$ & 3.3 & 3.0 & 3.0 \\ \hline
\multicolumn{4}{l}{ {\scriptsize $^*$ No input used for the MuC}}\\
\end{tabular}
\end{center}
\end{minipage}
\begin{minipage}{0.5\textwidth}
\renewcommand{\arraystretch}{.89}
\setlength{\arrayrulewidth}{.2mm}
\setlength{\tabcolsep}{0.6 em}
\begin{center}
\begin{tabular}{c|c|c|c}
\multicolumn{4}{c}{kappa-3 Fit [\%]} \\ 
\hline
& {\makebox[23pt]{${\textsc{hl-lhc}}$}}
& {\makebox[23pt]{${\textsc{hl-lhc}}$}}
& {\makebox[22pt]{${\textsc{hl-lhc}}$}}
\\[0pt]
& \multicolumn{1}{c|}{\ $(\kappa_V \leq 1)$ } & \multicolumn{1}{c|}{\makebox[23pt]{\small{$+10\,\textrm{{TeV}}$}}} 
& \multicolumn{1}{c}{\makebox[23pt]{\small{$+10\,\textrm{{TeV}}$}}}  
\\[-2pt]\ &     \ & \multicolumn{1}{c|}{\ } & \multicolumn{1}{l}{\hspace{0.5pt}+{$e e$}} 
\\ 
\hline
$\kappa_W$ &         $\kappa_W \!> 0.985$        & 0.3 & 0.2 \\ \hline
$\kappa_Z$ &         $\kappa_Z > 0.987$        & 0.2 & 0.1 \\ \hline
$\kappa_g$ &            2        & 0.6 & 0.5 \\ \hline
$\kappa_{\gamma}$ &     1.6 & 0.7 & 0.7 \\ \hline
$\kappa_{Z\gamma}$ &    10 & 5.2 & 3.9 \\ \hline 
$\kappa_c$ &             -          & 1.9 & 0.9 \\ \hline
$\kappa_b$ &             2.5        & 0.5 & 0.4\\ \hline
$\kappa_{\mu}$ &         4.4    & 2.4 & 2.3 \\ \hline
$\kappa_{\tau}$ &        1.6   & 0.6 & 0.4 \\ \hline
$\Gamma_{H}$ &           --       & 1.1 & 0.6 \\ \hline\hline
$\kappa_t^*$ & 3.2 & 3.1 & 3.0 \\ \hline
\multicolumn{4}{l}{ {\scriptsize $^*$ No input used for the MuC}}\\
\end{tabular}
\end{center}
\end{minipage}
 \caption{Left: $1\sigma$ sensitivities (in \%) to Higgs coupling modifiers at the 10~TeV muon collider. The left and right panels refer to the ``kappa-0'' and ``kappa-3'' fit setups defined in~\cite{deBlas:2019rxi}. The HL-LHC sensitivity is reported for comparison and we also show the combination with a future 240 GeV $e^+e^-$ Higgs factory (taken here as the FCC-ee, extrapolating the Higgs precision of the 240 GeV run used in~\cite{deBlas:2019rxi} to 10.8 ab$^{-1}$) for an assessment of the complementarity.}
 \label{fig:3h-ch-ppg}
\end{figure}

The input described in the previous section can be employed by the PPG for a quantitative assessment of the benchmark questions. Few selected results are reported below, and in Section~
\ref{1:physics:subsec:highlights}.\\[-20pt]
\paragraph*{Higgs properties}
\noindent\\ 
Figure~\ref{fig:3h-ch-ppg} displays the sensitivity to Higgs coupling modifiers of the 10~TeV muon collider, based exclusively on the on-shell Higgs measurements (IM.1 and IM.2) described in the previous section. 
These numbers are obtained from the precisions of single Higgs measurements assuming forward muon tagging in \cite{Forslund:2022xjq}, as well as the possibility of measuring the inclusive ZBF cross section, as explained in~\cite{Ruhdorfer:2024dgz}. 
The left and right panels report the results of the ``kappa-0'' and ``kappa-3'' coupling fit setups, defined as in Ref.~\cite{deBlas:2019rxi}. The results outline the strong advance in the knowledge of Higgs properties in comparison with the projected sensitivity at the end of the HL-LHC, from~\cite{Cepeda:2019klc,deBlas:2019rxi}. The table also outlines  the complementarity of the muon collider results with the one of a future $e^+e^-$ Higgs factory running at 240-GeV run such as LEP3, or the first stages of ILC, FCC-ee and CEPC. Similar complementarity  studies could be performed by the PPG for EW and top physics and other benchmarks where the complementarity is stronger. 

Some of the Higgs coupling precisions reported in the tables in Figure~\ref{fig:3h-ch-ppg} are, in fact, limited by the projected parametric uncertainties associated to the knowledge of the SM inputs. This is the case, in particular, of $\kappa_b$. If one neglects the parametric uncertainty associated to $m_b$, where we assume a future precision of 13 MeV, the kappa-0 muon collider precision for $\kappa_b$ would improve to 0.23$\%$. 

In the kappa-3 setup, additional contributions from BSM decays to the Higgs boson width are included. 
Despite a mild decrease in the overall precision, due to the extra degrees of freedom in the fit,
it is still possible to retain per-mile level precision for most of the $\kappa$ parameters. The excellent sensitivity of the muon collider stems from the inclusive production cross section measurement in~IM.2, which enables an indirect determination of the total Higgs width, with a precision of 1.1$\%$. 

\begin{figure}[t]
\centering
\includegraphics[width=0.48\textwidth]{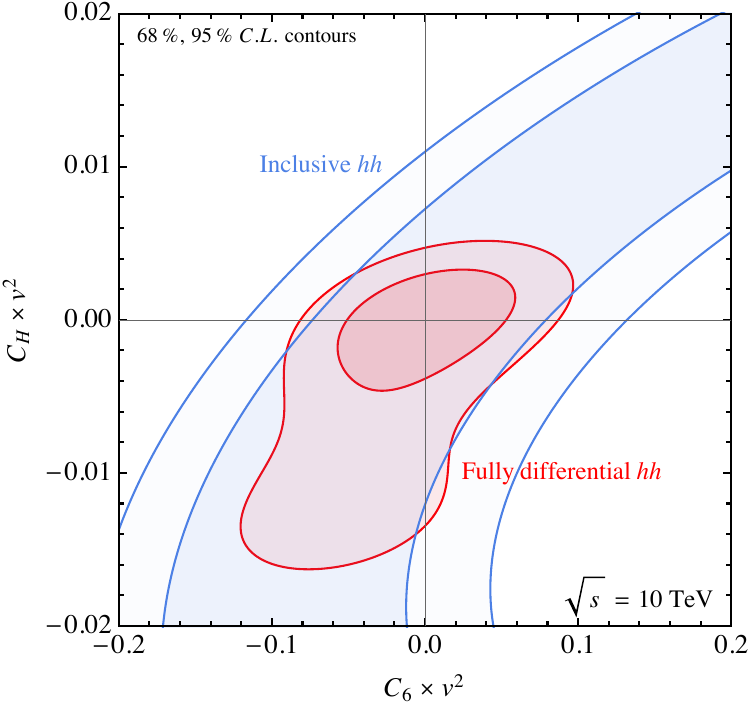}
\caption{\label{fig:constraints_hh} Constraints on the Wilson coefficients $C_6$ and $C_H$ from the inclusive $hh\to 4b$ cross-section measurement (blue), and from its differential distribution in di-Higgs invariant mass $M_{hh}$ and Higgs transverse momentum $p_{T,h}$ (red). The contours indicate 68\% and 95\% C.L. constraints (2 d.o.f.). Plot from Ref.~\cite{Buttazzo:2020uzc}.}
\end{figure}

Figure~\ref{fig:constraints_hh} illustrates the potential of the double-Higgs production process (IM.4) at the 10~TeV muon collider. The figure is taken from Ref.~\cite{Buttazzo:2020uzc} and displays the sensitivity in the plane formed by the SILH-basis Wilson coefficients (see~\cite{Buttazzo:2020uzc} for the operators definition) $C_6$ and $C_H$ normalized to the Higgs VEV $v\simeq246$~GeV. The $C_6$ operator---which is proportional to the ${\mathcal{O}}_\phi$ Warsaw operator---is completely equivalent to a modification of the Higgs trilinear coupling and the single-operator $C_6$ sensitivity corresponds to a relative accuracy $\delta\kappa_3\simeq 3.5\%$ in the determination of the Higgs trilinear coupling.

The $C_H$ operator---i.e., ${\mathcal{O}}_{\phi\Box}$ in the Warsaw basis---is equivalent to an overall rescaling of all the couplings of the Higgs boson, and the sensitivity to $C_H$ corresponds to a per mille level sensitivity to this modification of the Higgs couplings. This is comparable to the sensitivity of single-Higgs coupling measurements shown in Figure~\ref{fig:3h-ch-ppg}, in such a way that the evidence of an effect observed in single-Higgs measurements due to $C_H$ could be independently confirmed in the double-Higgs process. In the EFT fit, double-Higgs (IM.4) combined with single-Higgs (IM.1) measurements ensures an excellent global determination of ${\mathcal{O}}_\phi$, ${\mathcal{O}}_{\phi\Box}$ and of the operators such as  ${\mathcal{O}}_{\phi D}$, ${\mathcal{O}}_{\phi W}$, ${\mathcal{O}}_{\phi B}$ and ${\mathcal{O}}_{\phi WB}$ that contribute both to single- and double-Higgs production.
\\[-20pt]
\paragraph*{EFT fit}
\noindent\\   
The input measurements described in the previous section (IM.1--IM.6) can be interpreted in the EFT framework to obtain bounds on the scale of the different effective interactions contributing to those observables. We perform here a preliminary fit assuming $U(2)^3$ quark flavour symmetry and lepton family number conservation, the same assumptions indicated in PPG input request. In this fit we combine the muon collider input with current measurements and HLLHC sensitivity projections which define, following~\cite{deBlas:2022ofj}, the expected level of knowledge on EW/Higgs/Top interactions after the end of the HLLHC. 
We do a fit with a total of 56 operators which, in particular, include all those operators that induce effects that grow with the energy the processes considered at the muon collider. 
In this preliminary fit, we have ignored some operators, e.g. top-quark dipole operators (see however Ref.~\cite{Han:2024gan}), 
and neglected some non-growing effects under the assumption that they are better probed in other measurements.  

\begin{figure}[h!]
\centering
\includegraphics[width=1\textwidth]{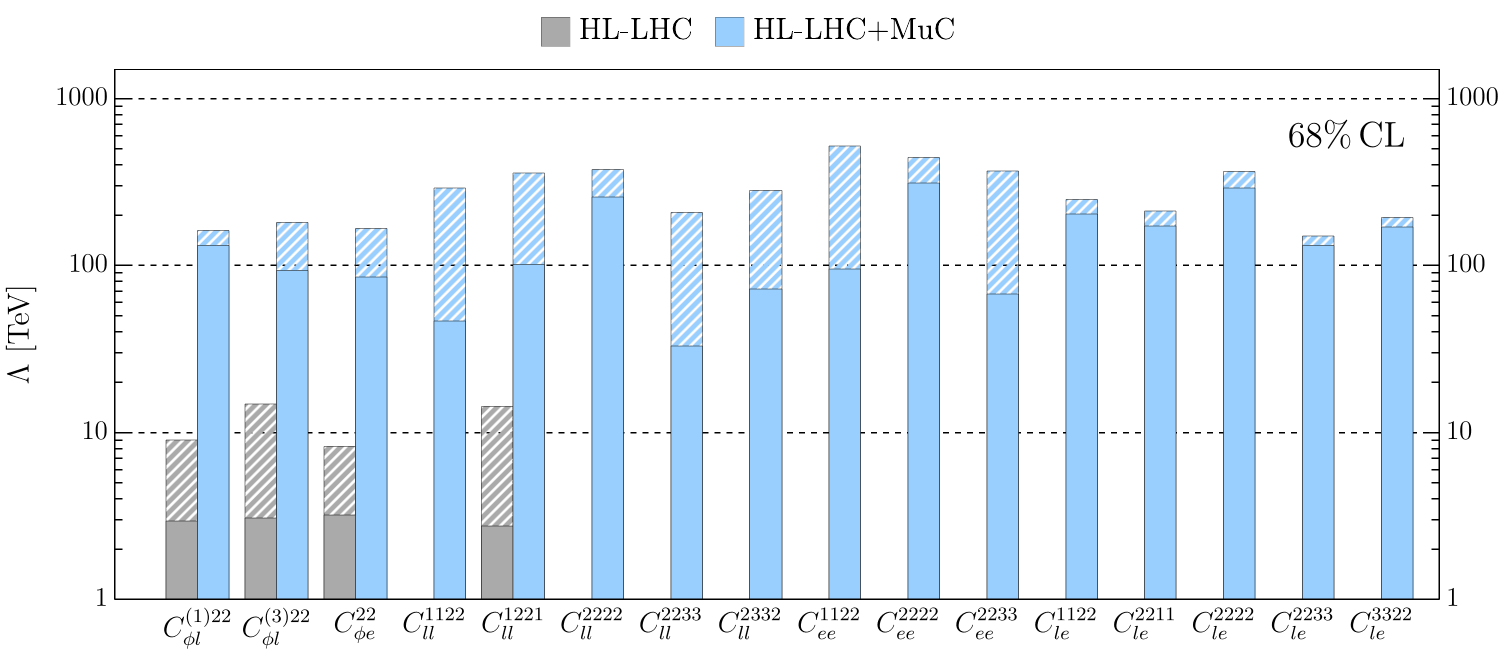}
\includegraphics[width=1\textwidth]{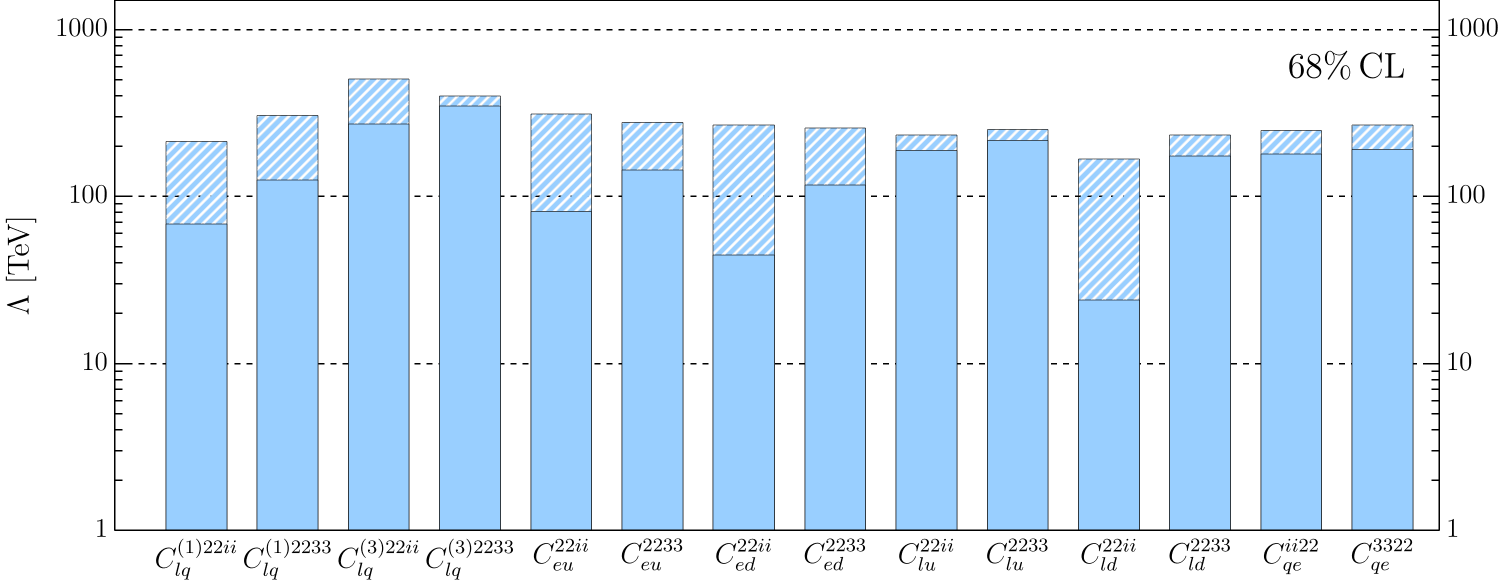}
\includegraphics[width=1\textwidth]{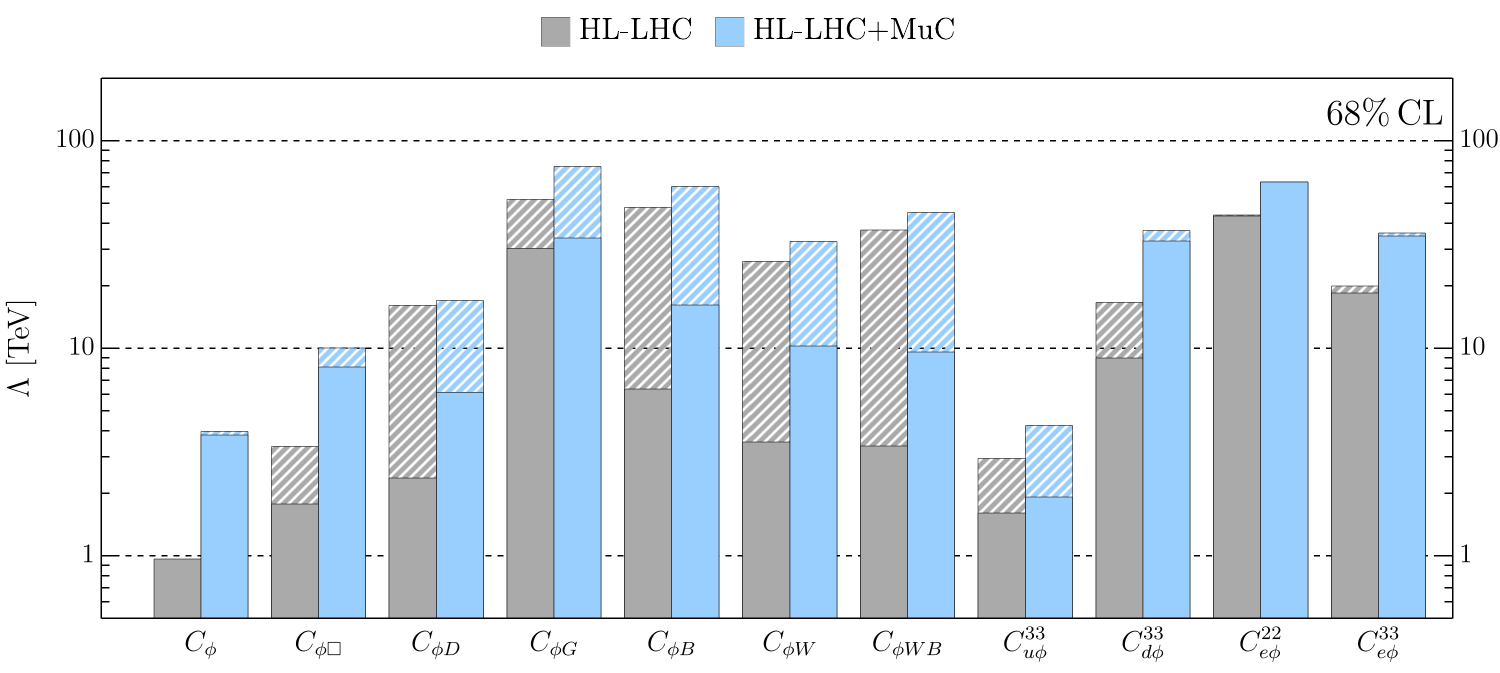}
\caption{\label{fig:EFTFIT} $68\%$~CL reach on EFT from a global fit at the 10~TeV muon collider.}
\end{figure}

The fit results are  displayed in Figure~\ref{fig:EFTFIT}. The solid bar represents the $68\%$~CL sensitivity to the operator scale, while the dashed bars correspond to the global sensitivity after marginalising over the other operators. The excellent degree of stability under marginalization stems from the rich variety of observables, nearly in one-to-one correspondence with the operators. It displays the great opportunities offered by the muon collider to probe different deformations of the SM, and to disentangle them. The fit results presented in Figure~\ref{fig:EFTFIT} have profound implications for BSM physics sensitivity, some of which are described in Section~\ref{sec:BSM}.

The first two panels of the figure report the many operators that feature a quadratic growth with energy in the direct muon annihilation processes (IM.5). The single and global sensitivities are consistently at the 100~TeV scale enabling a comprehensive probe of this very high scale new physics. The operator in the bottom panel do not grow with energy in the direct annihilation processes, while some of them grow in VBF processes. The reach is inferior, but still in most cases above the knowledge that will be available after the end of the HL-LHC. Particularly interesting is the operator ${\mathcal{O}}_\phi$, which is easily generated by heavy states mixing with the Higgs and is also one of the signatures of a composite pNGB Higgs. This is probed by Higgs coupling measurements and in double Higgs as previously discussed. The operator $C_{u\phi}^{33}$ is equivalent to a modification of the top Yukawa coupling.\\[-20pt]
\paragraph*{Electroweak and top couplings} 
\noindent\\ 
The high-energy muon collider strategy does not rely on on-shell measurements the EW fermionic couplings, 
but rather on probing a much larger variety of possible microscopic deformations from the SM that could, 
in particular, induce observables effects on the gauge boson couplings.\footnote{Actually, it could measure the $W$ and $Z$ bosons couplings by exploiting the order 100~million produced vector bosons, but this has not been studied yet.} 
For a comparison with traditional probes based on coupling measurements, it is useful to assess the implications of the muon collider tests of the underlying physics for the expected size of the coupling departures from the SM predictions.

We consider a setup where Universal new physics produces energy-growing effects in high-energy diboson final states at the muon collider. This is a very plausible theoretical setup that is encountered for instance in Composite Higgs, where the relevant interactions are provided by the SILH basis operators ${\mathcal{O}}_W=i g (H^\dagger \sigma_a \overset{\leftrightarrow}{D}\!^{\!~a}_\mu H)D^\nu W^{a\mu\nu}$ and ${\mathcal{O}}_B=i g^\prime (H^\dagger \overset{\leftrightarrow}{D}_\mu H)D^\nu B^{\mu\nu}$. In the Warsaw basis and focusing on energy growing effects entering in diboson, these two operators can be traded by two combinations of the three first operators---${\mathcal{O}}_{\phi l}^{(1)22}$, ${\mathcal{O}}_{\phi l}^{(3)22}$ and ${\mathcal{O}}_{\phi e}^{22}$---on the upper panel of Figure~\ref{fig:EFTFIT}. By restricting our global fit to the ${\mathcal{O}}_{W,B}$ operators, we obtain the following 
equivalent muon collider sensitivity to relative modifications of the EW gauge boson fermionic couplings $Zff$ and $Wff$: 
\begin{center}
\begin{tabular}{ccc | cccc }
\hline
$\delta g_L^{Z\ell\ell} $&
$\delta g_R^{Z\ell\ell} $&
$\delta g_L^{W\ell\nu} $&
$\delta g_L^{Zuu} $&
$\delta g_R^{Zuu} $&
$\delta g_L^{Zdd} $&
$\delta g_R^{Zdd} $
\\ 
\hline
 $8 \cdot 10^{-6}$ & $7 \cdot 10^{-6}$ & $3 \cdot 10^{-6}$ &
 $2 \cdot 10^{-5}$ & $ 10^{-5}$& $10^{-5}$ & $ 10^{-5}$ \\ \hline
\end{tabular}
\end{center}
The current sensitivity to the $Z$-boson couplings, from LEP, is at the order of $10^{-3}$. The statistical reach of future Tera-$Z$ factories is comparable to the muon collider sensitivity, but achieving this sensitivity would require a control of experimental and theory systematics to a similar level of the statistical uncertainties.

Similar studies could be performed for the top quark electroweak coupling, by exploring the effects of top quark compositeness~\cite{Durieux:2018ekg} that naturally emerge in the composite Higgs scenario. In the case of the top quark Yukawa coupling $y_t$ instead, the result can be directly read off from the reach on the $C_{u\phi}^{33}$ operator in Figure~\ref{fig:EFTFIT}. At the global level, the reach corresponds to a relative accuracy of around $1.5\%$ to $y_t$, which is dominated by VBF $t\bar{t}$ measurements consistently with Ref.~\cite{Liu:2023yrb}. 

A major muon collider breakthrough in top physics will be to probe lepton-top current-current four-fermion interactions at the 100~TeV scale, as well as top EW dipole operators at about 20~TeV~\cite{Han:2024gan}. To be noted that the top dipole sensitivity potential is not faithfully represented in our results because the dominant channel of $ttH$ production from direct muons annihilation is not included among the input measurements of our fit.

Additional physics opportunities associated to the top quark are illustrated in the following section on flavour physics. The high energy path to flavour physics enabled by the muon collider exploits final states with tops---including the flavour-changing $t+c$ or $t+j$ channels---for very effective probes of high-scale flavour physics, which are not only more effective but also explore complementary directions than low-energy measurements. Figure~\ref{fig:MuC_rareBdecays} illustrates this aspect in the plane formed by two flavour-violating operators. The dashed contour from the $B_s\to\mu\mu$ measurement leaves a flat direction that is lifted by the $t+c$ muon collider measurement.

\chapter{Flavour physics}
\label{sec:flavor}

\section{Overview}

\paragraph*{Flavour physics at the energy frontier}\noindent\\
A high energy muon collider offers a novel and more powerful way to test a large set of flavour violating interactions which are traditionally probed by studying rare or forbidden meson or lepton decays.
This takes 
advantage of the quartic (in case of flavour-violating terms with vanishing or negligible SM contribution) energy growth of the new physics contributions to the cross section over the SM backgrounds: $\Delta \sigma/\sigma_{\rm SM} \propto E^4 / \Lambda_{\rm FV}^4$, where $\Lambda_{\rm FV}$ is the effective scale associated to some flavour-violating operator.
By measuring $\mu^- \mu^+ \to f \bar{f}^\prime$ processes at 10 TeV of centre of mass energy, it has been shown that a muon collider is sensitive to flavour-violating contact interactions that involve two muons with scales of $\mathcal{O}(100)$~TeV. See Refs.~\cite{Azatov:2022itm,Altmannshofer:2023uci} for $bs\mu\mu$ contact terms, Ref.~\cite{Ake:2023xcz} for top quark flavour violation and \cite{AlAli:2021let,Accettura:2023ked} for Lepton Flavour Violating (LFV) interactions.

The innovative energy-driven approach to flavour violation overcomes the challenges posed by high-precision measurements and predictions, which are often the limiting factor of the traditional methods and prevent further progress due to non-perturbative SM effects or irreducible sources of experimental systematic uncertainties. It also defines a path for systematic improvement by exploiting higher and higher energy colliders when they will become available. Furthermore, it provides a formidable tool to investigate and characterise possible tensions with the SM that might emerge in low-energy flavour physics measurements.

\begin{figure}[t]
    \centering
    \includegraphics[width=0.6\textwidth]{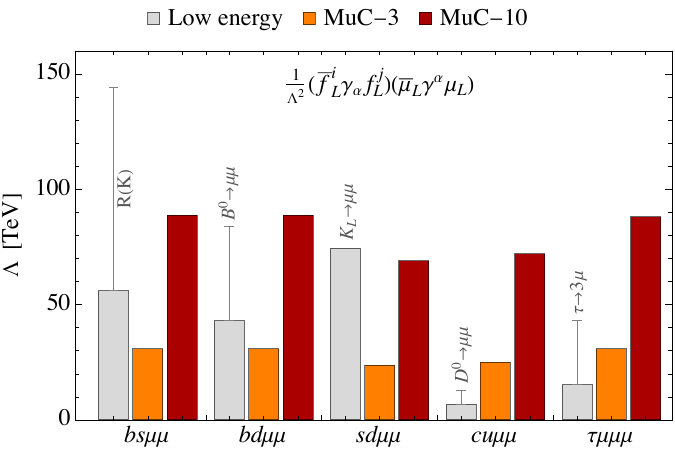}
    \caption{
    Sensitivity reach at 95\%CL to the EFT scale $\Lambda$ for flavour-violating contact interactions involving left-handed fermions, comparing low-energy rare or forbidden decays with 3 and 10 TeV muon collider with 2ab$^{-1}$ and 10ab$^{-1}$ of integrated luminosity, respectively.
    }
    \label{fig:MuC_flavour}
\end{figure}

Based on Ref.~\cite{2025:MucFlavour}, in the following we use this approach to compare the reach at a muon collider with the corresponding effects in rare or forbidden flavour-violating low-energy decays. Section~\ref{sec:ana} describes the analysis and Section~\ref{sec:res} summarizes selected results in answer to the PPG benchmark questions in the area of flavour physics.

As a first illustration of the sensitivity that is attainable at a muon collider, Figure~\ref{fig:MuC_flavour} displays the reach on the EFT scale $\Lambda$ of flavour-violating contact interactions involving left-handed fermions. The gray bars correspond to the present sensitivity from low energy observables, while the lines indicate the reach with future data---specifically, LHCb upgrade II and Belle-II with 50~ab$^{-1}$ of integrated luminosity. The orange and red bars are the expected reach of a 3 and 10~TeV muon collider with 2~ab$^{-1}$ and 10~ab$^{-1}$ of integrated luminosity, respectively. The muon collider sensitivity increases linearly with energy, a pattern that would continue with even more energetic colliders. The plot also shows that a 10~TeV muon collider would surpass the sensitivity expected from most low-energy measurements testing the same flavour-violating interactions.

\paragraph*{CKM elements determination}
\noindent\\
Beyond energy-frontier measurements, a muon collider also offers other innovative opportunities for flavour physics, specifically for the determination of the Cabibbo--Kobayashi--Maskawa (CKM) matrix elements with better precision and different methodology than the current available measurements. In fact, two completely different CKM measurements could be performed at a muon collider: one from studying the decays of the $W$ boson and one from Deep Inelastic Scattering (DIS) measurements that exploit the neutrino beam---see Section~\ref{1:phys:ch:nu}---at a dedicated fixed-target experiment. No firm projections of the attainable sensitivity are currently available. However, preliminary estimates are presented in Section~\ref{sec:CKM} based on ongoing work.

\section{Description of the analysis}
\label{sec:ana}

We consider neutral-current $\mu^- \mu^+ \to q_i \bar{q}_j$ and $\mu^- \mu^+ \to \ell_i^- \ell_j^+$ di-fermion production, where $q_i$ and $\ell_i$ are any flavour of quarks and leptons (including the $\tau$). In case of $q\bar{q}$ final states we employ charm and bottom tagging, as well as the tagging of boosted top quarks, to divide exclusive di-quark events into different tagging categories: $bb$, ${bj}$, ${cc}$, ${cj}$, ${jj}$, ${tt}$, ${tc}$, and ${tj}$, where $j$ represents a light jet (a jet not tagged as neither $b$, $c$, or $t$). These taggers are parametrised as constant efficiencies and mistag rates.
For all $q\bar{q}$ channels we consider the measurement of the total cross section in the detector acceptance region $\theta \in [10^{\degree},170^{\degree}]$. For the $\mu^- \mu^+$ final state we bin in the muon rapidity $\eta_{\mu} \in [-2,2]$---with bins of $\Delta \eta_\mu = 0.5$---in order to deal better with the forward elastic peak of the cross section. For all other $\ell_i^- \ell_j^+$ final states we separate the forward and backward regions to gain additional discriminating power on different chiral structures of new physics interactions. We assume 10~ab$^{-1}$ of integrated luminosity for the 10~TeV muon collider.

We consider exclusive cross section measurements, namely that a veto is applied to exclude the presence of massive vector bosons or relatively hard photons and gluons. In this case, the cross section for the production of the two energetic fermions can be computed at the leading order in the logarithm expansion by applying simple Sudakov double-logarithmic corrections as described in Ref.~\cite{Chen:2022msz}. With respect to the tree-level prediction, the Sudakov logarithms induce a reduction in the cross sections even up to $\mathcal{O}(50\%)$. We also reduce by a factor of 0.9 all cross sections to take into account ISR emission from muons. It should be noted that the cross section reduction in the exclusive final state corresponds to an enhancement of the vector bosons emission rate, which produces sizeable semi-inclusive~\cite{Chen:2022msz} cross sections where the extra emissions are not vetoed. Semi-inclusive di-fermion production---including charged-current di-fermion production---could be considered to further improve the sensitivity.

Within the SM, the number of expected events in all the tagging categories varies from about $10^3$ to about $10^4$. For the categories that correspond to rare SM transitions such as $bj$, the prediction is dominated by mistag. In the LFV categories with two leptons of different flavour, the main contribution to the SM cross section arises from the process $\mu^- \mu^+ \to \ell \bar\nu_\ell \bar\ell^\prime \nu_{\ell^\prime}$. After selection cuts, SM cross sections of the LVF categories are $\sigma(\tau\mu)_{\rm bkg} \approx 1.7$~fb~ \cite{AlAli:2021let}, $\sigma(\tau e)_{\rm bkg} \approx 0.08$~fb, and $\sigma(\mu e)_{\rm bkg} \approx 0.03$~fb. The statistical uncertainties on the cross section measurements in the different categories is at the percent level, and it is reasonable to assume that systematic uncertainties can be reduced at the same level not to compromise the sensitivity. 

The new physics is introduced through dimension-six Warsaw basis~\cite{Grzadkowski:2010es} EFT operators, with quarks rotation performed as in the down-quark mass basis ($Q^i = (V_{ji}^* u_L^j, \, d_L^i)^t$). We build a $\chi^2$ combining all observables and we study it as a function of the EFT operator Wilson coefficients to derive sensitivty projections.

\begin{table}[t]
\begin{center}
\begin{tabular}{|c|c|c|c|}
\hline
Coefficient & MuC-10 [TeV$^{-2}$] & $\delta(B_s \to \mu \mu)$ & $\delta(B \to K \nu \nu)$ \\\hline
$[C_{lq}^{(1)}]_{2232}$ & $[-(95.5)^{-2},~(97.0)^{-2}]$ & 3.1\% & 0.68\% \\
$[C_{lq}^{(3)}]_{2232}$ & $[-(97.0)^{-2},~(95.5)^{-2}]$ & 3.1\% & 0.68\% \\
$\left[C_{ed}\right]_{2232}$ & $[-(102)^{-2},~(102)^{-2}]$ & 2.7\% & - \\
$\left[C_{ld}\right]_{2232}$ & $[-(96.5)^{-2},~(96.5)^{-2}]$ & 3.1\% & 2.0\% \\
$\left[C_{qe}\right]_{3222}$ & $[-(100)^{-2},~(101)^{-2}]$ & 2.8\% & - \\
$\left[C_{ledq}\right]_{2223}$ & $[-(88.8)^{-2},~(88.8)^{-2}]$ & 179\% & - \\
$\left[C_{ledq}\right]_{2232}$ & $[-(88.8)^{-2},~(88.8)^{-2}]$ & 179\% & - \\\hline
\end{tabular}\\[0.5cm]
\begin{tabular}{|c|c|c|}
\hline
Coefficient & MuC-10 [TeV$^{-2}$] & $\delta(B_d \to \mu \mu)$ \\\hline
$[C_{lq}^{(1)}]_{2231}$ & $[-(95.5)^{-2},~(97.0)^{-2}]$ & 13\% \\
$[C_{lq}^{(3)}]_{2231}$ & $[-(97.0)^{-2},~(95.5)^{-2}]$ & 13\% \\
$\left[C_{ed}\right]_{2231}$ & $[-(103)^{-2},~(103)^{-2}]$ & 12\% \\
$\left[C_{ld}\right]_{2231}$ & $[-(96.5)^{-2},~(96.5)^{-2}]$ & 14\% \\
$\left[C_{qe}\right]_{3122}$ & $[-(100)^{-2},~(101)^{-2}]$ & 12\% \\
$\left[C_{ledq}\right]_{2213}$ & $[-(88.8)^{-2},~(88.8)^{-2}]$ & 1960\% \\
$\left[C_{ledq}\right]_{2231}$ & $[-(88.8)^{-2},~(88.8)^{-2}]$ & 1960\% \\\hline
\end{tabular}
\end{center}
 \caption{Sensitivity at 95\% CL of a 10 TeV MuC for $3-2$ (top) and $3-1$ (bottom) quark-flavour violating operators Wilson coefficients, from high-energy measurements. Also shown is the relative effects these values of the coefficients have on the branching ratio of the rare $B$ decays. Note that some of those operators do not contribute to $B \to K \nu\nu$.}
 \label{tab:MuC_rareBdecays}
\end{table}

\section{Results}
\label{sec:res}

Based on the analysis described in the previous section we address here the PPG benchmark requests in the area of flavour physics. See also Section~\ref{1:physics:sec:flav}.

\paragraph*{Rare FCNC decays of ${\boldsymbol{B}}$ mesons}
\noindent\\
Rare $B$ meson decays such as the $B_{s,d} \to \mu\mu$ decay and the golden-channel $B \to K^{(*)} \nu\nu$ probe the $b-s$ flavour violation in contact interactions that involve two leptons: muons in the former case or any combination of neutrinos in the latter. By $SU(2)_L$ gauge invariance, a contact interaction $bs\nu\nu$ necessarily implies also the presence of some $qq\ell\ell$ operator, with $q$ being either down or up-type quarks and $\ell$ are charged leptons.

In Table~\ref{tab:MuC_rareBdecays} we show the expected 95\%CL sensitivity on individual coefficients in the SMEFT Warsaw basis and the relative precision in the rare $B$ decays that would be necessary to probe the coefficients at the same level. Present uncertainties on the branching ratios of $B_s \to \mu\mu$, $B_d \to \mu\mu$, and $B^+ \to K^+ \nu\nu$ are at the 10\%, 64\%, and 90\% level of the corresponding SM predictions, respectively. The prospects for future measurements by the LHCb with detector upgrade-II and Belle-II with 50~ab$^{-1}$ of luminosity are approximately 4\%, 9\%, and 11\%, respectively \cite{Cerri:2018ypt,Belle-II:2018jsg}. 
In all cases, except for scalar and tensor operators, the muon collider sensitivity surpasses even the future expected one from low energy measurements. The contributions from scalar and tensor operators to the leptonic meson decays are enhanced by $1/m_\mu^2$ over the SM contribution, which grants to these observables a very strong sensitivity to these operators.

\begin{figure}[t]
    \centering
{\includegraphics[width=0.55\textwidth]{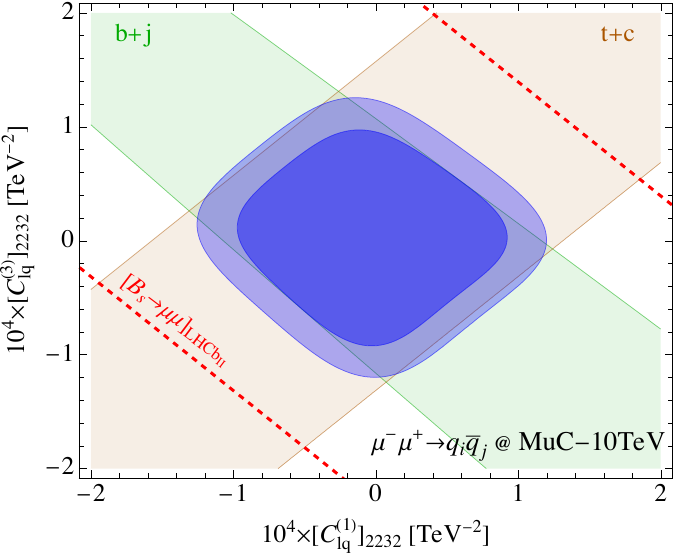}}
    \caption{68\% and 95\% CL allowed regions (dark and light blue) in the plane formed by two EFT coefficients from a global analysis of di-jet final states. Other light regions are the 68\% CL constraints from individual tagging categories. Solid and dashed red lines are the present and expected future 68\% CL sensitivity from $B_s \to \mu \mu$.
    }
    \label{fig:MuC_rareBdecays}
\end{figure}

Figure~\ref{fig:MuC_rareBdecays} shows the 68\% (light blue) and 95\% CL sensitivity region from our global analysis of all di-jet processes for a pair of flavour-violating coefficients. We also show the individual 68\% constraints from separate tagging categories, to showcase the complementarity between different channels in constraining this parameter space. The solid and dashed red lines correspond to the present and expected future (after LHCb upgrade II) 68\%CL sensitivity from $B_s \to \mu\mu$. The analogous lines from $B^+ \to K^+ \nu\nu$ are well outside the plotted region.
As can be seen in the plot, at a muon collider the $t-c$ final state, with a boosted top quark, provides a powerful probe of flavour-violation, complementary to the $b-s$ channel, which is the only one tested at a similar level in low-energy measurements. This allows a muon collider to probe more directions in the new physics parameter space.\\[-20pt]

\begin{table}[t]
\begin{center}
\begin{tabular}{|c|c|c|c|c|}
\hline
Coefficient & MuC-10 [TeV$^{-2}$] & $\text{Br}(\tau\to 3 \mu)$ \\\hline
$\left[C_{ll}\right]_{2223}$ & $ (88)^{-2}$ & $5.5 \times 10^{-12}$ \\
$\left[C_{le}\right]_{2223}$ & $ (136)^{-2}$ & $4.7 \times 10^{-13}$ \\
$\left[C_{le}\right]_{2322}$ & $ (136)^{-2}$ & $4.7 \times 10^{-13}$ \\
$\left[C_{ee}\right]_{2223}$ & $ (99)^{-2}$ & $3.4 \times 10^{-12}$ \\ \hline
\end{tabular}
\end{center}
 \caption{Sensitivity at 95\% CL of a 10 TeV MuC for SMEFT coefficients contributing to LFV in $\tau \to 3\mu$ decay. We also show the effect those coefficients would have on the $\tau \to 3 \mu$ branching ratio.}
 \label{tab:MuC_tauLFV}
\end{table}

\paragraph*{Lepton flavour violating $\boldsymbol{\tau}$ decay: ${\boldsymbol{\tau \to 3 \mu}}$}
\noindent\\
The LFV $\tau\to3\mu$ decay tests $\mu\mu\mu\tau$ contact interactions. The same interactions can also be tested at a high energy muon collider in the $\mu^-\mu^+ \to \tau\mu$ scattering process. At the muon collider, the effect of the LFV contact interaction relative to the SM scales as $E^4/\Lambda_{\rm LFV}^4$, where $E=10$~TeV. The first study of the muon collider reach was first performed in Ref.~\cite{AlAli:2021let,Accettura:2023ked}, following the work of Ref.~\cite{Murakami:2014tna} for $e^- e^+$ colliders. In Table~\ref{tab:MuC_tauLFV} we show our results~\cite{2025:MucFlavour} for the expected sensitivity of a 10~TeV MuC on LFV SMEFT coefficients, together with the required reach on the LFV $\tau \to 3 \mu$ decay branching ration required to reach the same sensitivity. The branching ratio reach at present experiments is at the level of $10^{-8}$.\\[-20pt]
\paragraph*{$\boldsymbol{\tau}$ LFU tests}
\noindent\\
The ratio of different leptonic decays of leptons allows to test Lepton Flavour Universality (LFU) in charged-currents~\cite{Stugu:1998jv,Pich:2013lsa}, with a sensitivity that is currently at the per mille level~\cite{HeavyFlavorAveragingGroupHFLAV:2024ctg}. At low energy and working at the tree level, these observables depend on a specific combination of four-lepton operators of the kind $\mathcal{O}^{V,LL/LR}_{\nu e} = (\bar\nu_i \gamma_\mu P_L \nu_j)(\bar \ell_j \gamma^\mu P_{L,R} \ell_i)$. Matching these operators to the SMEFT one obtains the following dependence of the $\tau$ LFU observables on the EFT coefficients \cite{Allwicher:2021ndi}:
\begin{equation}\begin{split}
    \left|\frac{g_\mu}{g_e}\right|^2 &\equiv \frac{\Gamma(\tau \to \mu \nu\nu)}{\Gamma(\tau \to e \nu\nu)}  \left[\frac{\Gamma_{\rm SM}(\tau \to \mu \nu\nu)}{\Gamma_{\rm SM}(\tau \to e \nu\nu)} \right]^{-1} = \\  
        & = \frac{\left| \sqrt{2}G_F + [C_{Hl}^{(3)}]_{22} + [C_{Hl}^{(3)}]_{33} - 2 [C_{ll}]_{2332} \right|^2 + \left|[C_{le}]_{2332}\right|^2}
        { \left| \sqrt{2}G_F + [C_{Hl}^{(3)}]_{11} + [C_{Hl}^{(3)}]_{33} - 2 [C_{ll}]_{1331} \right|^2 + \left|[C_{le}]_{1331}\right|^2}~, \\
    \left|\frac{g_\tau}{g_\mu}\right|^2 &\equiv \frac{\Gamma(\tau \to e \nu\nu)}{\Gamma(\mu \to e \nu\nu)} 
    \left[\frac{\Gamma_{\rm SM}(\tau \to e \nu\nu)}{\Gamma_{\rm SM}(\mu \to e \nu\nu)} \right]^{-1} = \\ 
        & = \frac{ \left| \sqrt{2}G_F + [C_{Hl}^{(3)}]_{11} + [C_{Hl}^{(3)}]_{33} - 2 [C_{ll}]_{1331} \right|^2 + \left|[C_{le}]_{1331}\right|^2}
        { \left| \sqrt{2}G_F + [C_{Hl}^{(3)}]_{11} + [C_{Hl}^{(3)}]_{22} - 2 [C_{ll}]_{1221} \right|^2 + \left|[C_{le}]_{1221}\right|^2}~, \\
    \left|\frac{g_\tau}{g_e}\right|^2 &\equiv \frac{\Gamma(\tau \to \mu \nu\nu)}{\Gamma(\mu \to e \nu\nu)} \left[\frac{\Gamma_{\rm SM}(\tau \to \mu \nu\nu)}{\Gamma_{\rm SM}(\mu \to e \nu\nu)}  \right]^{-1} = \\ 
        & =  
        \frac{\left| \sqrt{2}G_F + [C_{Hl}^{(3)}]_{22} + [C_{Hl}^{(3)}]_{33} - 2 [C_{ll}]_{2332} \right|^2 + \left|[C_{le}]_{2332}\right|^2}
        { \left| \sqrt{2}G_F + [C_{Hl}^{(3)}]_{11} + [C_{Hl}^{(3)}]_{22} - 2 [C_{ll}]_{1221} \right|^2 + \left|[C_{le}]_{1221}\right|^2}~,
\end{split}\end{equation}
where $G_F$ is the Fermi constant and $C$ are Warsaw-basis Wilson coefficients. Current measurements of these ratios enable to test modest EFT scale of about  1 - 6~TeV, depending on the particular operator considered.

\begin{table}[t]
\begin{center}
\begin{tabular}{|c|c|c|c|c|}
\hline
Coefficient & MuC-10 [TeV$^{-2}$] & $\delta(|g_\mu/g_e|^2)$ & $\delta(|g_\tau/g_\mu|^2)$  & $\delta(|g_\tau/g_e|^2)$   \\\hline
$\left[C_{ll}\right]_{1221}$ & $[-(136)^{-2},~(139)^{-2}]$ & 0 & $3.3 \times 10^{-6}$ & $3.3 \times 10^{-6}$ \\
$\left[C_{ll}\right]_{2332}$ & $[-(96)^{-2},~(100)^{-2}]$ & $6.6 \times 10^{-6}$ & 0 & $6.6 \times 10^{-6}$ \\
$\left[C_{le}\right]_{1221}$ & $[-(74)^{-2},~(74)^{-2}]$ & 0 & $3.1 \times 10^{-11}$ & $3.1 \times 10^{-11}$ \\
$\left[C_{le}\right]_{2332}$ & $[-(62)^{-2},~(62)^{-2}]$ & $6.2 \times 10^{-11}$ & 0 & $7.0 \times 10^{-11}$ \\
$[C_{Hl}^{(3)}]_{22}$ & $[-(183)^{-2},~(183)^{-2}]$ & $3.6 \times 10^{-6}$ & $6.2 \times 10^{-6}$  & $1.1 \times 10^{-6}$ \\\hline
\end{tabular}
\end{center}
 \caption{Sensitivity at 95\% CL of a 10 TeV MuC for SMEFT coefficients contributing to both LFU tests from $\tau$ decays and $\ell^+ \ell^-$ production at a MuC. We also show the corresponding relative effects these values of coefficients would have on the $\tau$ LFU tests.}
 \label{tab:MuC_tauLFU}
\end{table}

Among the operators contributing to $\tau$ LFU ratios, those that involve two muons are very effectively probed at the muon collider in the high-energy processes discussed in Section~\ref{sec:EHT}. In Table~\ref{tab:MuC_tauLFU} we show the 95\% CL sensitivity reach of a 10~TeV MuC on the corresponding Wilson coefficients, as well as the relative precision that would be necessary, in $\tau$ LFU ratios, in order to probe each of them at the same level.
\begin{figure}[t]
    \centering
{\includegraphics[width=0.55\textwidth]{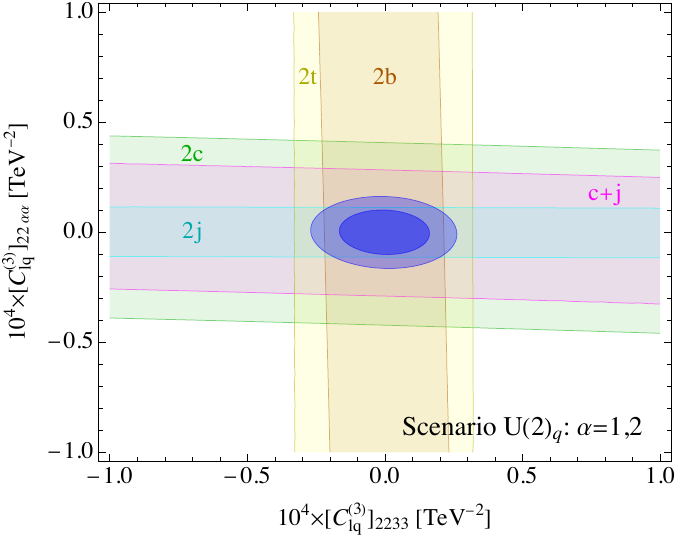}}
    \caption{
    Sensitivity region in two SMEFT coefficients relevant for the $U(2)_q$ flavour scenario from a global analysis of di-quark final states (dark and light blue are 68\% and 95\% CL regions). Other light regions are the 68\% CL constraints from individual tagging, as indicated. 
    }
    \label{fig:MuC_flavourU2}
\end{figure}
\\[-20pt]
\paragraph*{Testing flavour scenarios - $\boldsymbol{U(2)_q}$ example}
\noindent\\
Testing flavour-violating processes is not sufficient to fully uncover the flavour structure of new physics.
Equally important is understanding the structure of flavour-conserving interactions and how they depend on the different fermionic generations.
For instance, one could ask if new physics couples almost universally, as in minimal flavour violation (MFV) scenarios, or if the couplings to the third generation are larger, as in scenarios inspired by the accidental $U(2)_q \times U(2)_u \times U(2)_d$ flavour symmetries of the SM.
A high energy muon collider could probe these scenarios well above the 100~TeV scale, in interactions that involve muons.

Figure~\ref{fig:MuC_flavourU2} shows the sensitivity on two EFT coefficients that could be induced without suppression in $U(2)^3$ scenarios:
\begin{equation}   
    [C_{lq}^{(3)}]_{22\alpha\alpha} \equiv [C_{lq}^{(3)}]_{2211} = [C_{lq}^{(3)}]_{2222}~ \quad \text{and} \quad
    [C_{lq}^{(3)}]_{2233}~.
\end{equation}
Clearly, different tagging categories are sensitive to different operators, so that a non-universal scenario could be discovered. Furthermore, the multitude of different channels like $2b$, $2t$, $2c$, and $2j$ could also disentangle between the triplet, $C_{lq}^{(3)}$, and singlet, $C_{lq}^{(1)}$, operator structure, while an angular analysis could differentiate between different chiralities, as can also be seen from the global fit in the middle panel of Figure~\ref{fig:EFTFIT}.
The EFT scales corresponding to the 95\%CL sensitivity are approximately 275~TeV for the light quark generations and 215~TeV for the third generation.

\section{CKM elements determination}\label{sec:CKM}

\paragraph*{$\boldsymbol{|V_{cb}|}$ and $\boldsymbol{|V_{cs}|}$ from $\boldsymbol{W}$ decays}
\noindent\\
A 10 TeV MuC would produce approximately $\sim 10^8$ $W$ bosons, dominantly in $\gamma W$ fusion. A similar number of $W$ bosons is expected at FCC-ee or ILC, enabling a potentially very accurate determination of $|V_{cb}|$ and $|V_{cs}|$~\cite{Marzocca:2024mkc}. The muon collider has then the same statistical potential of FCC-ee and ILC, however the actual sensitivity depends on the expected bottom and charm tagging efficiencies and mistag rates (and on the possibility of strange-tagging, in the case of $|V_{cs}|$), and most importantly on the corresponding  calibration uncertainties.
Unlike at FCC-ee, tagging calibration cannot exploit a $Z$-pole run, but it might perhaps exploit the $\approx 10^9$ pairs of $b \bar{b}$ and $c \bar{c}$ produced in VBF. No dedicated study has been performed.
\begin{table}[t]
    \centering
    \begin{tabular}{|c|c|c|c|c|}
    \hline
     $\delta|V_{\rm CKM}|$  &  PDG & $\underset{\rm{(stat.~only)}}{N_{\rm incl}^{\mu,e} + R_{c,b}^{\mu,e}}$ & $N_{\rm incl}^{\mu,e}$ & $N_{\rm incl}^{\mu,e} + R_{c,b}^{\mu,e}$  \\
     \hline
     $|V_{cs}|$ & $0.62\%$ & $1.5 \times 10^{-5}$ & $0.12\%$ & $3.7 \times 10^{-4}$  \\
     $|V_{cd}|$ & $1.8\%$ & $7.6 \times 10^{-5}$ & $0.89\%$ & $0.27\%$ \\
     \hline
     $|V_{cb}|$ & $3.4\%$ & $0.10\%$ & $3.4\%$ & $0.12\%$  \\
     $|V_{ub}|$ & $5.2\%$ & $0.52\%$ & $5.2\%$ & $0.55\%$  \\
     \hline
     $|V_{ud}|$ & $3.2 \times 10^{-4}$ & $9.5 \times 10^{-6}$ & $3.2 \times 10^{-4}$ & $3.2 \times 10^{-4}$   \\
     $|V_{us}|$ & $0.36 \%$ & $2.6 \times 10^{-4}$ & $0.35 \%$ & $0.18 \%$  \\
     \hline
    \end{tabular}
    \caption{Relative precision on the absolute values of CKM elements from the PDG \cite{ParticleDataGroup:2024cfk} and after adding measurements from charged-current DIS from the MuC neutrino beam (both $\nu_\mu$ and $\bar\nu_e$).}
    \label{tab:CKM_nubeam}
\end{table}
\begin{figure}[t]
    \centering
{\includegraphics[width=0.45\textwidth]{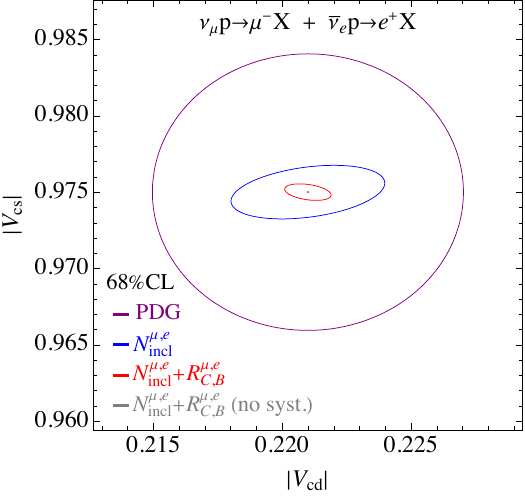}} \quad
{\includegraphics[width=0.45\textwidth]{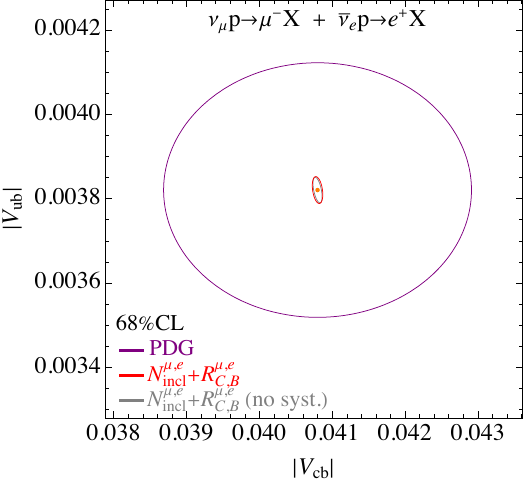}}\\[0.5cm]
{\includegraphics[width=0.45\textwidth]{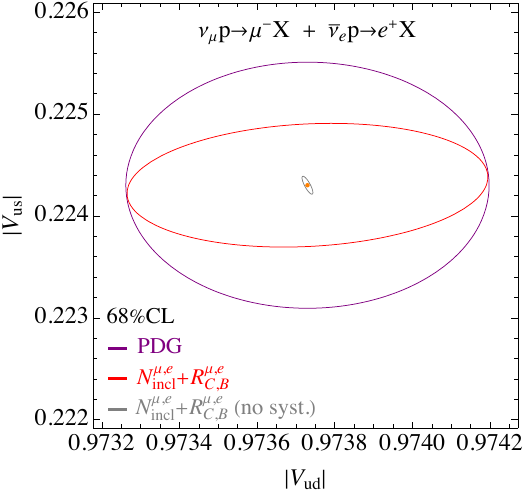}}
    \caption{
    Preliminary results for the 68\%CL sensitivity on CKM matrix elements from charged-current DIS with MuC neutrino beam.
    }
    \label{fig:CKM_nubeam}
\end{figure}
\\[-20pt]
\paragraph*{CKM measurements in neutrino DIS}
\noindent\\
Another avenue for measuring CKM elements takes advantage of the focussed and very intense beam of neutrinos produced by muons decaying in the straight sections of the collider, to perform a dedicated Deep Inelastic Scattering (DIS) experiment. This possibility was proposed long ago~\cite{Bigi:2001xb} (see also \cite{King:1997dx,Mangano:2001mj,NuTeV:2001whx,Ball:2009mk}) exploiting the neutrino beam produced by a dedicated muon ring. Here we consider instead the neutrino beam that is produced by the 5~TeV energy beams of the 10~TeV muon collider close to the interaction point, to be employed for a forward neutrino experiment as described in~\cite{InternationalMuonCollider:2024jyv} and in  Section~\ref{1:phys:ch:nu}. See also Section~\ref{sec:SI}. The preliminary results presented in this section are based on ongoing work~\cite{2025:CKMnubeam}.

On a 1~ton target mass, the neutrino beam produces around $10^9$ charged-current (CC) interactions with nucleons per year as in Figure~\ref{fig:nfl}. If the detector is placed in the direction of the $\mu^-$ beam, the neutrino beam consists of muon neutrinos and electron anti-neutrinos, with a perfectly characterised composition and energy spectrum as it emerges from the muons decay. The CC $\nu_\mu$ and the $\bar\nu_e$ interactions are easily distinguishable experimentally, as they produce a muon ($\mu^-$) and a positron, respectively. The kinematic can be fully reconstructed by the measurement of the charged lepton momentum and of the momentum of the produced hadrons. This enables the measurement of triply differential distributions in the DIS variables $x$ and $Q^2$ and in the incoming neutrino energy $E_\nu$. The yields in Figure~\ref{fig:/mucol-dis-x-q2} show that the vast majority of the events are produced in the DIS regime, with high Bjorken~$x$ and $Q^2$ as high as 100~-~1000~GeV$^2$. Reliable perturbative QCD calculations are possible in this regime, offering opportunities for a robust inference on underlying microscopic parameters such as the CKM elements. Obviously, we get access to this favourable regime thanks to the high neutrino energy.

Preliminary results for the precision on the CKM elements---not involving the top quark, which is not kinematically accessible---are reported in Table~\ref{tab:CKM_nubeam} and Figure~\ref{fig:CKM_nubeam}. The observables that are considered for the determination are---in each $x$, $Q^2$, and $E_\nu$ bin---the number of $\nu_\mu$ and $\bar\nu_e$ inclusive DIS events, $N_{\rm incl}^{\mu, e}$, as well as the ratios
\begin{equation}
    R_{c,b}^{\mu,e} = \frac{N_{{c,b}}^{\mu, e}}{N_{\rm incl}^{\mu, e}}\,,
\end{equation}
where $N_{{c,b}}^{\mu, e}$ is the number of Semi-Inclusive DIS (SI-DIS) events with $D$ and $B$ hadrons in the final state, signalling the production of a charm or a bottom quark, respectively. Current knowledge of CKM elements, taken from the PDG~\cite{ParticleDataGroup:2024cfk}, is used as prior. 

The purely statistical uncertainties, in the third column of Table~\ref{tab:CKM_nubeam}, are extraordinarily small. Parametric, theoretical and experimental systematic uncertainties will expectedly dominate the CKM element determination accuracy, with the possible exception of $|V_{cb}|$ and $|V_{ub}|$, which are probed by the relatively low statistics measurements of $b$-tagged SI-DIS processes. For a first assessment of the impact of systematics, we include PDF uncertainties using the PDF4LHC21\_40 set~\cite{PDF4LHCWorkingGroup:2022cjn}. A proton target is considered for illustration, and a constant 1\%~systematic uncertainty on the total rate prediction is included. The results shown above are obtained after marginalising over these PDF and systematic uncertainties. The results are reported on the fourth and fifth column of Table~\ref{tab:CKM_nubeam} by considering only inclusive DIS measurements and the full set of observables, respectively. The expected precision is still much superior to current knowledge for most CKM elements.

This preliminary analysis is expectedly an overestimate of the impact of PDF uncertainties because it does not include the measurements of Neutral Current (NC) neutrino DIS processes. The latter processes are insensitive to the CKM matrix element enabling an independent determination of the PDFs that is way more precise than the current PDF fits. On the other hand, the analysis currently assumes perfect charm and bottom tagging, does not employ $c$ and $b$ fragmentation functions (FF) and thus is does not account for the currently large FF uncertaintes. However, the FF can also be measured in NC neutrino SI-DIS.

While they do not provide a conclusive assessment of the expected sensitivity, the results illustrate the potential for CKM elements determination using the neutrino beam that is produced by a 10~TeV muon collider, and demonstrate that current PDF determinations are already sufficient to improve our present-day knowledge of the CKM elements. Other sources of uncertainties such as missing QCD higher orders and non-factorizable effects, and experimental calibration uncertainties, are still to be investigated.

\chapter{Strong Interactions}\label{sec:SI}

While the muon collider is a high-energy \emph{electroweak} collider, it has a strong potential also on Strong Interactions. This emerges from the direct study of 10~TeV muon collisions, from the parasitical neutrino DIS experiment with a far-forward detector as described in the previous section, and from possible low-energy colliders or muon accumulation rings that will be enabled by the muon cooling complex. 

The work on the Strong interactions potential of the muon collider project has just started and constitutes a promising arena for future progress on a number of PPG benchmarks in the Strong Interaction category.
These include:

\begin{itemize}

\item The determination of $\alpha_s(Q)$ with $Q$ in the TeV range {\it e.g.} via QCD event shapes or energy-energy correlators.
Additionally, measurement of $\alpha_s(m_Z)$ from high-precision measurements of neutrino structure functions at a far-forward detector.

\item Precision measurements of $m_W$ and $m_t$ via a threshold scan at a dedicated low-energy muon collider. In the case of $m_t$, the studies~\cite{Franceschini:2022veh} described in Section~\ref{sec:EHT} report a $\sim10$~MeV statistical precision and a 50~MeV accuracy dominated by current theory uncertainties.

\item Proton and nuclear PDF determinations: both can be accessed at a muon collider via (FASER-like~\cite{FASER:2019aik,FASER:2022hcn}) far-forward experiments, which are exposed to the high intensity and precisely-characterised neutrino beam from muon decays in the straight sections of the collider.
We provide quantitative projections for this application below.
These improved PDFs could reduce PDF-related systematic uncertainties in legacy HL-LHC measurements such as the $W$-boson mass and Higgs cross-sections.

\item QCD connections with hadronic, nuclear and astro(particle) physics: the extremely high event rates expected for a neutrino DIS far-forward experiment at a muon collider would enable a rich program of hadronic measurements such as the 3D structure of the nucleon, analogous to the Electron Ion Collider program with charged-current scattering, probing complementary partonic combinations. Neutrino DIS at a muon collider could also study hadronisation in nuclear matter, relevant in the case of nuclear targets.

\end{itemize}

\section{PDFs determination from neutrino DIS}

Consistently with Section~\ref{1:phys:ch:nu} and Ref.~\cite{InternationalMuonCollider:2024jyv}, in what follows we consider the $N_\mu=9\cdot 10^{16}$ muons that decay in the straight section of the 10~TeV collider producing the same number of muon and electron neutrinos. Nearly all these neutrinos will intercept a 20~cm radius cylindrical target placed 200~m away from the interaction region. With this choice for the geometry of the target in the transverse plane, we can safely neglect the neutrino acceptance factors and assume that all of them cross the detector. We note that the estimate $N_\mu=9\cdot 10^{16}$ is, by construction, conservative by a factor possibly as large as 10. Furthermore, we will base our results on a conservatively small target detector---described in~\cite{King:1997dx}---with a total mass of only 10~Kg. Projections for larger detectors can be obtained from rescaling, but as we show below event rates are so large even for this very compact detector that the limiting factor in the analysis will be theoretical and experimental systematic uncertainties.

\begin{figure}[h]
    \centering
\includegraphics[width=1.0\textwidth]{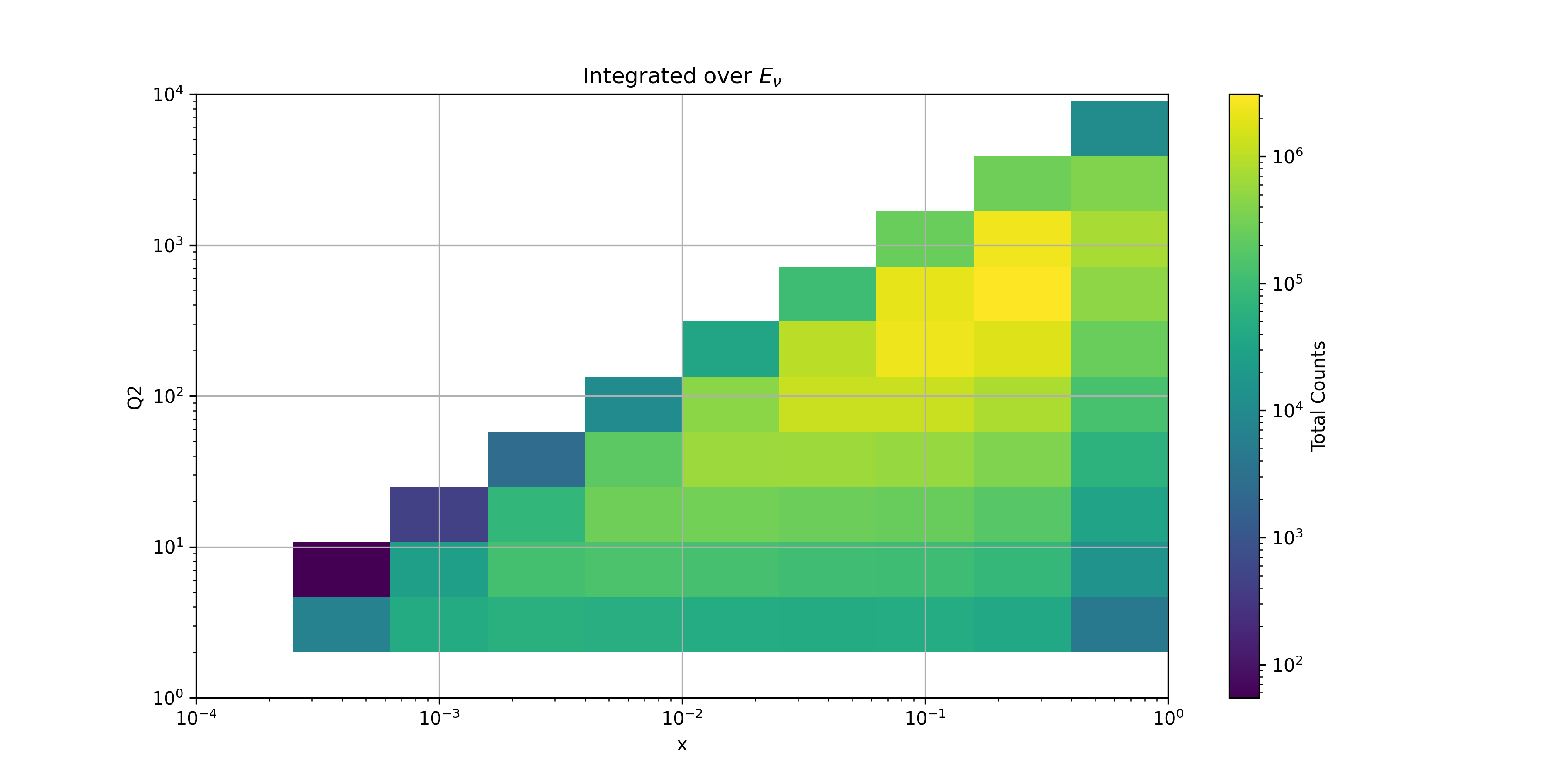}
    \caption{Predicted binned event yields for muon-neutrino DIS at a $\sqrt{s}=10$ TeV muon collider as a function of $x$ and $Q^2$. The event rates displayed assume a 10 kg neutrino target exposed to the MuC neutrino beam for one year, and are integrated over the neutrino energy range. 
    The event rate is maximal at $Q^2=500$ GeV$^2$ and $x\sim 0.3$.
    }
    \label{fig:/mucol-dis-x-q2}
\end{figure}

The energy spectrum of the neutrinos reaching this far forward detector can be computed analytically with high precision, leading to an extraordinarily well characterised neutrino beam. The neutrinos are also extraordinarily energetic, as they emerge from 5~TeV muon decays. By interacting with the nuclei, the neutrinos produce DIS events with a scale $Q^2$ that is peaked at around $Q^2=500$~GeV$^2$ and large $x\sim0.3$. This region is well inside the realm of applicability of perturbative QCD enabling precise theoretical calculations. Very large event rates are expected as shown in Figure~\ref{fig:/mucol-dis-x-q2}, covering the range $10^{-3}\le x \le 1$ and reaching up to $Q^2\sim m_Z^2$.

\begin{figure}[htbp]
    \centering
\includegraphics[width=1\textwidth]{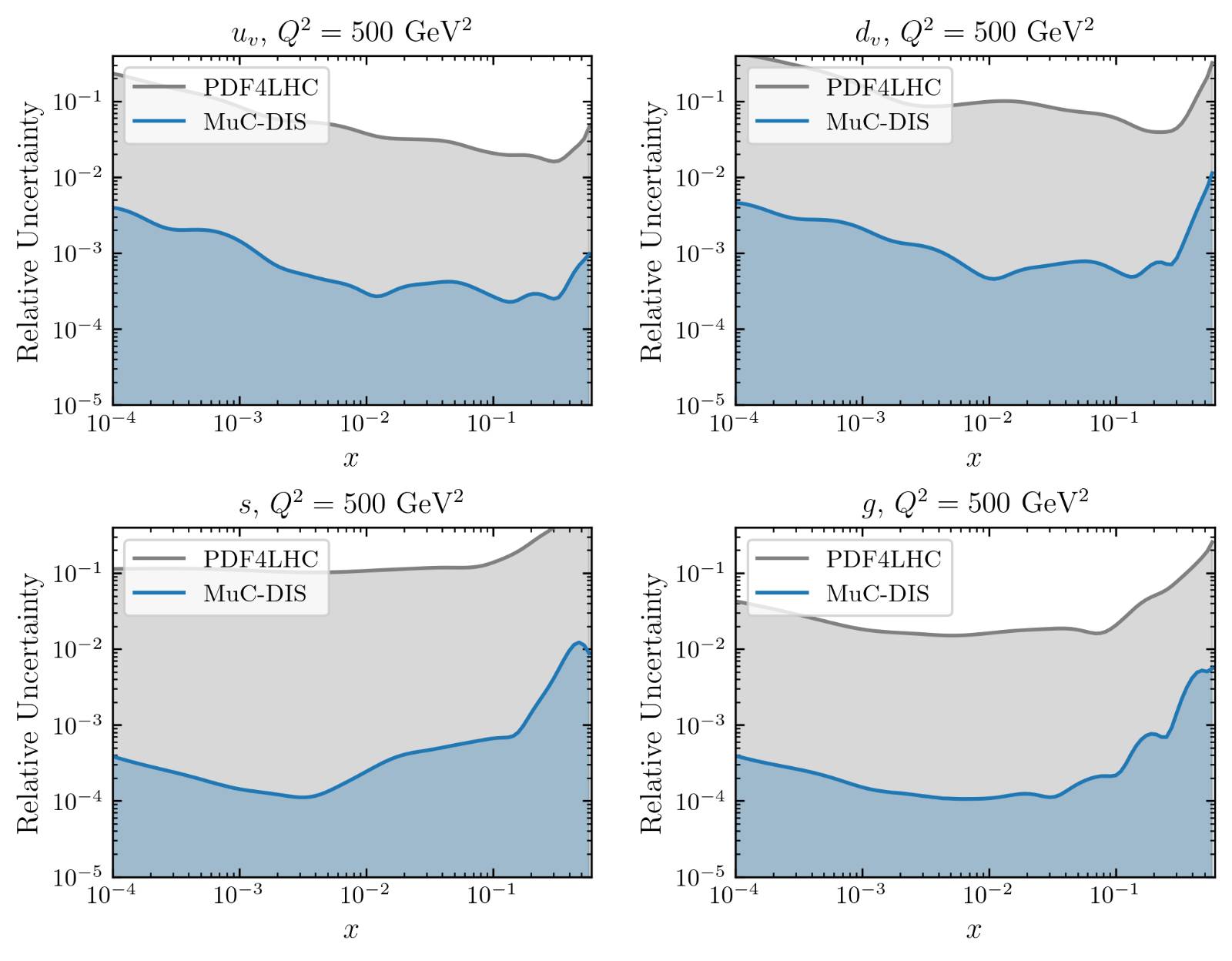}
    \caption{Reduction of PDF uncertainties thanks to neutrino DIS measurements carried out at a far-forward neutrino detector installed in the Muon Collider facility. 
    We compare the current PDF baseline, PDF4LHC21~\cite{PDF4LHCWorkingGroup:2022cjn}, with projections for MuC neutrino structure functions which consider only statistical uncertainties and which account for both
    for inclusive and charm-production data. 
    We show results for the up and down valence quark, strangeness, and gluon PDFs at $Q^2=500$ GeV$^2$ where the event rate peaks, see Figure~\ref{fig:/mucol-dis-x-q2}
    }
    \label{fig:/mucol-dis-results}
\end{figure}

The enormous statistical potential for the reduction of PDF uncertainties in comparison with current knowledge is displayed in Figure~\ref{fig:/mucol-dis-results}. The results are obtained with the procedure employed in Ref.~\cite{Cruz-Martinez:2023sdv} to quantify the (vastly inferior) potential of forward experiments exploiting the neutrinos produced at the LHC. As in~\cite{Cruz-Martinez:2023sdv}, the neutrino structure functions are simulated in NNLO QCD using {\sc\small YADISM}~\cite{Candido:2024rkr} interfaced to {\sc\small PineAPPL}~\cite{Carrazza:2020gss}.

The results of Figure~\ref{fig:/mucol-dis-results} highlight the unprecedented potential of a muon collider to operate as the ultimate experiment to constrain proton structure, surpassing any present or planned PDF-sensitive experiment. The very large event rates leads to PDF statistical uncertainties below the permille level for all flavour combinations, including the gluon, and in the complete range of $x$. Uncertainties associated with the knowledge of the neutrino fluxes are expected to have a minor impact on the results, because the composition of the neutrino beam can be very accurately simulated. 

Limiting factors that control the attainable sensitivity are probably missing higher orders and non-perturbative corrections, and/or experimental systematic uncertainties that depend strongly on the detector performances, but also on the method employed for the PDF determination. In particular, an approach based on the measurement of the neutrino structure functions---like the one of Ref.~\cite{Cruz-Martinez:2023sdv}---is strongly sensitive to the experimental resolution in the measurement of the final state particles momenta and probably not suited. New methodologies for the direct extraction of the PDF from data should be developed, as part of the future work on the design of neutrino experiments at the muon collider facility.

\chapter{BSM physics}\label{sec:BSM}

Muon colliders enable groundbreaking BSM physics exploration through distinct capabilities:

\emph{Full Energy Utilization:} As fundamental particles, muons deliver their entire centre-of-mass energy to collision processes. This can be used to produce new heavy states copiously up to the kinematical threshold, as well as to probe the high scale dynamics directly by precision high-energy measurements. A 10~TeV muon collider features a higher effective (partonic) luminosity than a 100~TeV proton collider for high-energy partonic collisions. For example, FCC-hh produces $>9$~TeV $e^+e^-$ pairs with 40~ab cross-sections versus 900~ab at a muon collider---a decisive statistical disadvantage despite FCC-hh's higher integrated luminosity. 

\emph{Multi-Scale Discovery Potential:} These facilities combine direct production reach with indirect sensitivity beyond kinematic limits. For $Y$-universal $Z'$ models, 10~TeV operation probes effective operator scales up to 500~TeV through precision $2\to2$ scattering measurements. Composite Higgs scenarios are constrained up to $m_* = 50$~TeV mass scales via di-boson final states, surpassing any other future collider projections. For HNLs, particularly through the mixing with $\nu_\mu$, muon colliders have unparalleled discovery potential in the TeV realm.

\emph{Vector Boson Synergy:} Enhanced vector boson luminosity enables dual functionality as both energy-frontier machine and precision Higgs factory. $\mathcal{O}(10^7)$ Higgs bosons enable coupling measurements competitive with dedicated $e^+e^-$ facilities, while Higgs portal scalar singlets are directly accessible through $VV \to S$ fusion processes also with mixing angles $\sin^2\gamma \sim 10^{-4}$ well below what Higgs coupling measurements can test. 

\emph{Electroweak Frontier Physics:} Unique sensitivity to high-energy SM processes provides indirect access to $\Lambda \sim 100$~TeV new physics through $(E_{\rm{cm}}/\Lambda)^2$ corrections. Precision measurements at 10~TeV achieve few-percent uncertainties, enabling a program of measurement that can at once probe BSM physics and explore new electroweak radiation phenomena that require novel exciting resummation techniques for accurate modelling.

\emph{Flavor-Specific Advantages:} Direct production of electroweak-charged states enables discovery of sleptons with $\delta_{RR}^{12} \sim 10^{-3}$ mixing parameters through $\mu^\pm e^\mp + {E}_T^{\rm miss}$ signatures. Muon-specific $g$-2 anomaly resolutions are testable via high-energy di-jet topologies already at the 3~TeV collider, with a 30~TeV machine probing electromagnetic dipole operators through $\mu\mu\to h\gamma$.

\emph{Complementary Discovery Landscape:} While FCC-hh exhibits limited superiority in direct 20~TeV $Z'$ production, muon colliders dominate indirect probes across most parameter space. They triple HL-LHC Higgs coupling precision and enable neutral scalar detection inaccessible to QCD-dominated channels. The simultaneous energy/precision combination allows full characterization of new sectors within a single facility.

This multifunctional capability establishes muon colliders as unique instruments for exploring both intensity and energy frontiers, bridging capabilities of previous collider generations through novel interaction mechanisms.

\section{New gauge forces}

\begin{figure}[t]
    \centering
\includegraphics[width=0.7\textwidth]{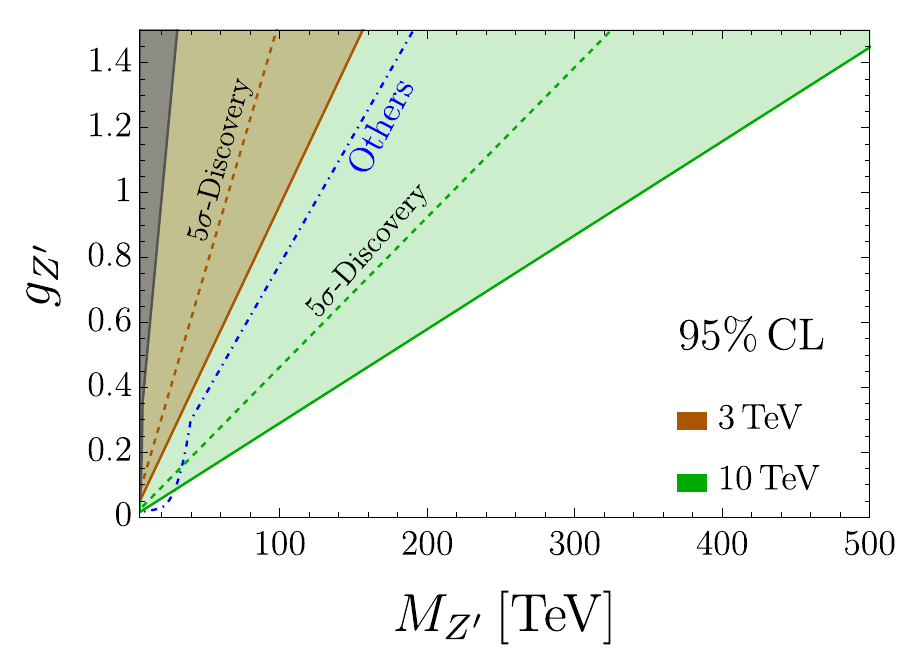}
    \caption{sensitivity projections for a $Y$-universal Z$^\prime$ model. The gray band and the blue dash-dot line represent respectively the region probed by the HL-LHC program and the sensitivity projections for all other future collider projects~\cite{Chen:2022msz}.
    }
    \label{fig:Zprime}
\end{figure}

Muon colliders can probe new gauge forces through direct resonance production, including radiative returns and electroweak PDF scanning, and as well as precision measurements. The high CM energy of muon colliders enables its superior sensitivity to new gauge forces. Here we first show a $Z'$ model projection and discuss the rescalings in other scenarios and benchmarks.  

The $Y$-universal $Z'$ model extends the Standard Model through a heavy gauge boson ($Z'$) with mass $M_{Z'}$ and coupling $g_{Z'}$ aligned with SM hypercharge \cite{EuropeanStrategyforParticlePhysicsPreparatoryGroup:2019qin,Appelquist:2002mw}. Perturbativity constraints ($Z'$ width $\leq 0.3M_{Z'}$) limit $g_{Z'} \lesssim 1.5$. Below $M_{Z'}$, the $Z'$ generates a single effective operator (analogous to $O_{2B}'$ in Warsaw basis models) that is probed up to order 100~TeV effective scale by the high-energy measurements described in Section~\ref{sec:EHT}. The operator Wilson coefficient scales as $1/\Lambda^2\sim g_{Z'}^2/M_{Z'}^2$. For unit $g_{Z'}$, $\Lambda\sim100$~TeV EFT scale sensitivity corresponds to a sensitivity to $M_{Z'}\sim100$~TeV mass. The mass reach is weaker/stronger form smaller/large $g_{Z'}$ coupling as in Figure~\ref{fig:Zprime}.

The figure shows sensitivity projections for the 3~TeV (green) and the 10~TeV (red) muon colliders. The envelope of the sensitivity attainable at any other future collider project---namely, ILC, CLIC including the 3~TeV stage and two future leptonic and hadronic colliders FCC-ee and FCC-hh---is also reported for comparison. The 3~TeV collider already matches the performances of other future colliders, and specifically of CLIC-3~TeV and FCC-hh, which have the best reach. The 10~TeV collider probes $Z'$ masses up to 500 TeV for large coupling. The vast superiority of the muon collider on this benchmark is evident in the entire parameter space, except in a narrow window of $M_{Z'} \sim 20$~TeV and small coupling where FCC-hh has better sensitivity. It should be noted that Figure~\ref{fig:Zprime} does not include the muon collider reach from the direct search of the resonantly produced $Z'$, which is relevant below 10~TeV. This was studied in Ref.~\cite{Cheung:2025uaz}.

This benchmark demonstrates muon colliders' unique capacity to explore high-scale physics through precision measurements of virtual effects, outperforming both proton colliders and lepton machines in motivated scenarios like $Z'$ extensions or composite Higgs models (see later).

Similar results hold for a large class of $Z'$ models. Integrating the $Z'$ out produces 4-fermion EFT operators that are probed in the high energy measurements as described in Section~\ref{sec:EHT}. See~\cite{Korshynska:2024suh} for a first assessment of the muon collider sensitivity to different models and of the excellent perspectives for model selection and $Z'$ couplings characterisation that stems from the excellent global sensitivity to the relevant EFT operators. See~\cite{Langacker:2008yv,Han:2013mra} for discussions on a variety of $Z'$ models and their observable impact at $e^+e^-$ colliders.

Beyond $Z'$ models, new gauge forces can be described by heavy vector triplet (HVT) models~\cite{Pappadopulo2014} (model~A), as well as other UV-complete scenarios such as L-R symmetric models~\cite{Marshak:1979fm,Mohapatra:1980qe,Dobrescu:2015qna,Brehmer:2015cia,Dobrescu:2015yba,Dobrescu:2015jvn}. Sensitivity in the ballpark of 100~TeV is generically expected in these models because integrating out the heavy resonances produces EFT operators that are accessible at the muon collider up to very high effective scale. These include 4-fermion operators as in the case of the $Y$-universal $Z'$, but also fermion-current/Higgs current and other operators that can be tested by di-boson cross section measurements. 

The presence of a charged massive vector boson triggers charged-current effective interactions that are probed at the muon collider as precisely as the neutral interactions due to the soft-collinear $W$ boson emission turning the neutral $\mu^+\mu^-$ initial state into an effective charged $\mu\nu$ initial state. This offers sensitivity to the charged heavy vector and enables discrimination between $Z'$ models and scenarios with a richer set of new gauge force mediators. All these aspects can be investigated quantitatively using the input measurements or directly the EFT fit presented in Section~\ref{sec:EHT}.

\section{Compositeness}

\begin{figure}[t]
    \centering
\includegraphics[width=0.8\textwidth]{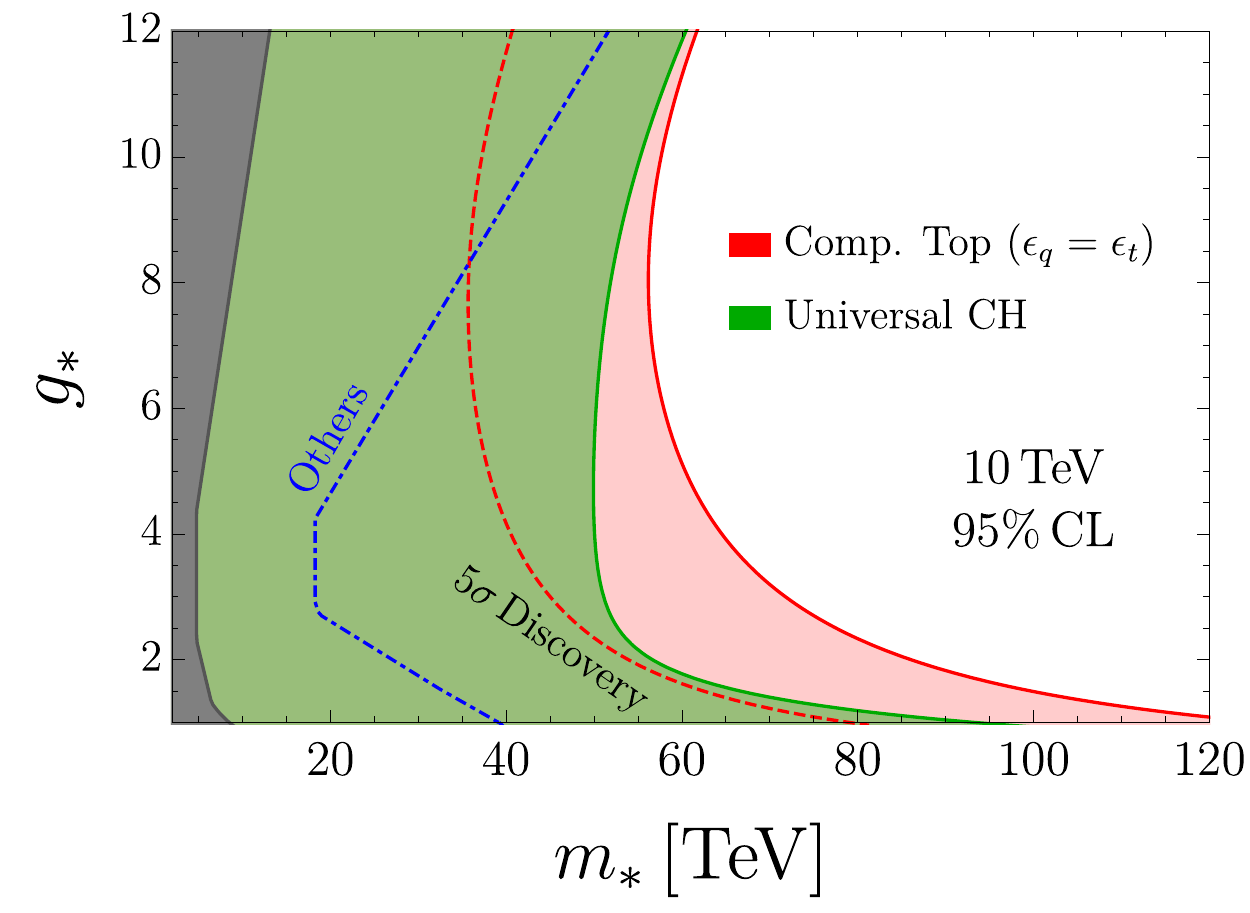}
    \caption{The $95\%$ reach on the Composite Higgs scenario from high-energy measurements in di-boson and di-fermion final states~\cite{Chen:2022msz}. The green contour display the sensitivity from ``Universal'' effects related with the composite nature of the Higgs boson and not of the top quark. The red contour includes the effects of top compositeness.
    }
    \label{fig:composite}
\end{figure}

The tension between the elementary nature of the Higgs boson and the Wilsonian interpretation of quantum field theory is a theoretical challenge that must be also investigated experimentally by probing Higgs compositeness. Figure~\ref{fig:composite} demonstrates the muon collider's unparalleled reach~\cite{Chen:2022msz,Buttazzo:2020uzc} for composite Higgs scenarios, where the Higgs emerges as a bound state of a strongly interacting sector. 

The exclusion contours span the parameter space of the coupling strength $g_*$ of the new strong sector, and its confinement scale $m_*$. The inverse of $m_*$ is the spatial extension (the radius) of the Higgs particle, analogous to the radius of the proton that is set by the inverse of the QCD confinement scale $\Lambda_{\rm{QCD}}$. A 10~TeV muon collider excludes Higgs compositeness scales up to $m_* = 50$~TeV across all plausible $g_*$ values, surpassing sensitivities projected for future electron colliders or 100~TeV proton colliders~\cite{EuropeanStrategyforParticlePhysicsPreparatoryGroup:2019qin}. 

The composite Higgs scenario considered in the figure is the one employed in Ref.~\cite{deBlas:2019rxi} matching the PPG benchmark request. The green contour represents the sensitivity to a Universal composite Higgs scenario where effects of compositeness are only visible in the interactions of the SM Higgs and of the gauge bosons. The red contour includes the effects of a large degree of compositeness for the top quark. The input measurements or the EFT fit presented in Section~\ref{sec:EHT} enable us to redo this plot using the most up-to-date muon collider sensitivity projections.

It is intriguing to notice that the muon collider attains this sensitivity breakthrough by following the same path that led to the discovery of proton and nucleon compositeness, i.e., to the discovery of a finite nucleons radius $r_N\sim 1/\Lambda_{\rm{QCD}}$. Nucleon compositeness discovery came from technological advances enabling electron-nucleon collisions with transferred energy as ``high'' as 100~MeV. This enabled us to observe for the first time the effects of the finite nucleon radius on the nucleon-photon vertex form factor because these effects scale as $r_N^2E^2\sim E^2/\Lambda_{\rm{QCD}}$ and they were too small to be observed at lower energy. By developing the muon collider technology, Higgs compositeness discovery might come from the observation of corrections to the form factor describing the interactions of the Higgs with the SM gauge bosons. In fact, the sensitivity of Figure~\ref{fig:composite} is dominated by the reach on higher-derivative corrections (such as the operators $\mathcal{O}_W$ and $\mathcal{O}_B$) to the Higgs-current interaction with the gauge field, which precisely corresponds ---in the modern EFT language---to the leading corrections to the form-factor.

A 10~TeV muon collider can directly discover Higgs compositeness, establishing a non-vanishing Higgs particle radius for a compositeness scale of around 40~TeV. If the scale is lower, below 10~TeV, it can also take the first steps towards the characterization of the new strongly-interacting sector that delivers the Higgs as a bound state---ultimately leading to a theory for the Higgs ``constituents''---by observing the resonant production of new composite resonance and charcterising their properties~\cite{Liu:2023jta}.

\section{Extensions of the minimal real scalar sector}

\begin{figure}[t]
    \centering
\includegraphics[width=0.8\textwidth]{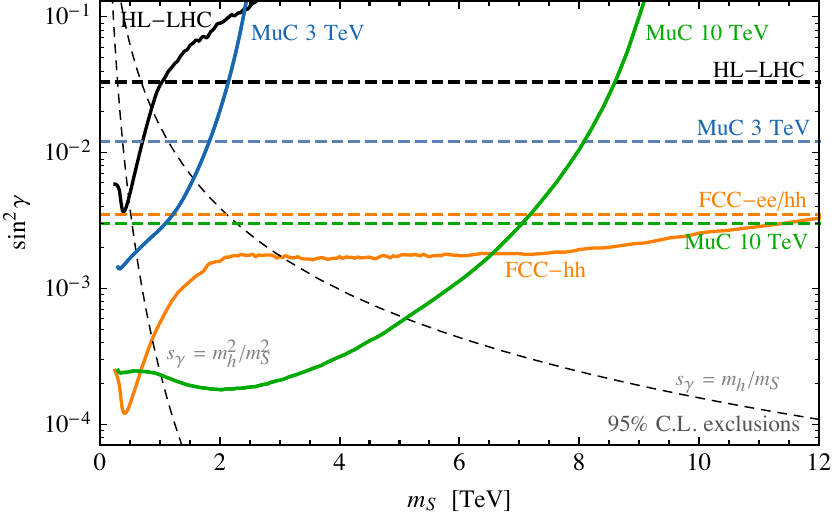}
    \caption{The 95\% C.L. sensitivity to Higgs-mixed scalar singlets at muon colliders~\cite{Buttazzo:2018qqp}, compared to HL-LHC and other future colliders~\cite{EuropeanStrategyforParticlePhysicsPreparatoryGroup:2019qin}. 
    }
    \label{fig:singlet}
\end{figure}

New physics particles may interact with the Standard Model (SM) through Higgs portal couplings rather than gauge interactions. In many beyond-SM (BSM) scenarios—including dark matter and baryogenesis models—the Higgs field mediates interactions between SM particles and new states via Vector Boson Fusion (VBF). Muon colliders excel at probing such scenarios due to their intense effective vector boson luminosity.  

Figure~\ref{fig:singlet} illustrates this capability for a benchmark model \cite{Buttazzo:2018qqp,AlAli:2021let} featuring a real scalar singlet coupled via the Higgs portal. The interaction strength is parameterized by $\sin\gamma$, quantifying mixing with the Higgs. Extensions with charged Higgs bosons (e.g., two-Higgs-doublet models) offer additional detection channels \cite{Han:2021udl,Chakrabarty:2014pja}, but even the minimal singlet scenario demonstrates unique muon collider advantages.  

The singlet-Higgs mixing reduces all Higgs couplings by $\cos\gamma \approx 1 - \frac{1}{2}\sin^2\gamma$, analogous to SMEFT operator effects. Muon colliders achieve sensitivity through:  
\begin{itemize}  
\item \textbf{Indirect probes} (dashed lines in Figure~\ref{fig:singlet})\\
Precision Higgs coupling measurements at 10 TeV MuC match $e^+e^-$ Higgs factories, while 3 TeV MuC improves HL-LHC limits by $\sim 3\times$.  
\item \textbf{Direct production:} (solid lines in Figure~\ref{fig:singlet})\\
VBF processes ($VV \to S$) exploit muon colliders' vector boson luminosity. Dominant $S \to hh \to 4b$ decays provide optimal sensitivity \cite{Buttazzo:2018qqp}.  
\end{itemize} 
The combination of indirect and direct probes at the muon collider covers more space in the plane of Figure~\ref{fig:singlet} than any other future collider project. In the plane, particular attention should be given to the regions around the two dashed grey lines that correspond to theoretically plausible parametric relations between the mixing angle and the mass. The superiority of the muon collider along these lines is remarkable.

\begin{figure*}
\centering{}
\includegraphics[width=0.6\linewidth]{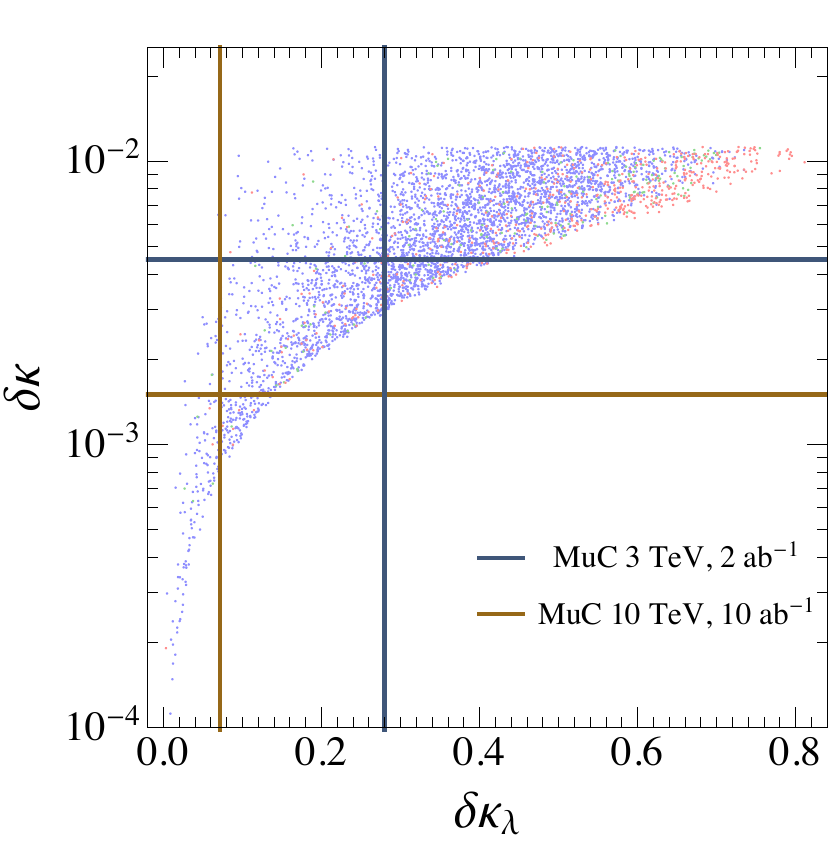}
\caption{\label{fig:EWpt-reach}Indirect reach on the SM plus real scalar singlet scenario at the muon collider.  Dots indicate points with successful first-order EWPT, while red, green, and blue dots represent signal-to-noise ratio for gravitational wave detection in the ranges $[50, +\infty)$, $[10, 50)$ and $[0, 10)$, respectively. Results adapted from~\cite{Liu:2021jyc}. See also~\cite{Accettura:2023ked}.
 }\end{figure*}

In a significant region of the parameter space, the muon collider could observe the Higgs singlet both directly and indirectly through the measurement of the Higgs couplings. This simultaneous observation will add considerable robustness to the discovery of this model. Additional signatures are possible in specific motivated models. For instance, Ref.~\cite{Liu:2021jyc} investigates models where the new singlet modifies the SM thermal Higgs potential enough to induce a strong first-order EW Phase Transition (EWPT) in the early universe. Figure~\ref{fig:EWpt-reach} shows that this scenario could also be observed through the corrections to the Higgs trilinear coupling, thanks to the excellent sensitivity of the muon collider to this parameter.

\section{Direct Searches in SUSY}
Supersymmetry provides an important and present theoretical framework for new physics searches. A simple message can be drawn here: if the underlying states are electroweak charged (such as squarks and sleptons), the muon colliders can copiously produce SUSY particles in pairs, regardless of whether R-parity is broken or not, and reach at least the kinematic limit for these states thanks to the clean lepton collider environment. Further searches on off-shell states could also enhance the reach. 

Studies using MSSM-inspired models~\cite{AlAli:2021let} show that squark production dominates over gluino production, as gluino pair production via quark annihilation with gluon radiation suffers from suppressed cross-sections when squarks are light. In contrast, stop production, particularly for $\tilde{t}_R$ and $\tilde{Q}_3$, achieves significant rates through annihilation and VBF process~\cite{Arina:2020udz} surpassing gluino rates by orders of magnitude. Slepton studies using the default MSSM framework reveal distinct patterns for $\tilde{\tau}_R$ (singlet) and $\tilde{L}_3$ (doublet) pair production via annihilation~\cite{Han:2025wdy}. Electroweakinos, such as higgsinos and winos, exhibit a complex interplay of annihilation and VBF, requiring careful treatment of neutralino-chargino mixing.

A universal energy-dependent pattern emerges across all processes: at fixed new physics mass, $s$-channel annihilation dominates at low $\sqrt{s}$, while VBF mechanisms become predominant at high energies due to enhanced vector boson luminosity. This transition mirrors Standard Model behavior but gains amplified significance in SUSY contexts given the TeV-scale mass thresholds. The squark-gluino cross-section disparity persists throughout, suggesting gluino signatures could be masked in realistic models with light squarks. These findings collectively demonstrate muon colliders' unique capacity to probe SUSY through complementary channels - leveraging both traditional annihilation processes at lower energies and VBF-dominated regimes at multi-TeV scales while addressing critical model-completeness requirements for robust BSM searches.

\section{Direct Searches in Flavour Physics}
Muon collider plays unique roles in flavor physics, both by the precision measurements across energies~\ref{sec:flavor}, and as well as direct searches on particles and states that mediate the flavor violation.  Similar to the SUSY case, if the new particles that enable flavor violation are electroweak-charged, muon colliders can efficiently pair-produce them and probe up to the kinematic threshold, except in scenarios with highly exotic decays, such as compressed spectra requiring specialized optimizations.

Charged lepton flavor violation (LFV) in the MSSM arises from off-diagonal entries in the slepton mass matrix within the soft SUSY-breaking sector. When these terms are non-diagonal in the basis where Standard Model (SM) lepton Yukawas are diagonal, physical sleptons become flavor-mixed states. This mixing induces flavor-violating interactions between sleptons, leptons, and neutralinos/charginos, manifesting in rare processes like $\mu \to e\gamma$ at the loop level. Low-energy experiments probe these effects with sensitivities extending to TeV-scale SUSY masses, contingent on the flavor structure of the model~\cite{Altmannshofer:2013lfa, Ellis:2016yje}. A high-energy muon collider complements these indirect probes by directly producing superpartners and measuring LFV processes~\cite{Homiller:2022iax}, offering unique insights into SUSY-breaking mechanisms.

In a simplified scenario with only $\tilde{e}_R$ and $\tilde{\mu}_R$ non-decoupled, the slepton mass matrix features off-diagonal terms inducing mixing. The resulting mass eigenstates decay as $\tilde{\ell} \to \ell \chi_1^0$, producing missing energy from a bino-like lightest supersymmetric particle (LSP) with mass $M_1$. For nearly degenerate sleptons, the flavor-violating parameter $\delta_{RR}^{12}$ suppresses LFV via a super-GIM mechanism, allowing lighter sleptons while evading low-energy bounds. Following~\cite{Arkani-Hamed:1996bxi}, the cross section for $\mu^+\mu^- \to \tilde{e}^\pm_{1,2} \tilde{e}^\mp_{1,2} \to \mu^\pm e^\mp \chi_1^0 \chi_1^0$ depends on mass splitting and mixing.  Current bounds from $\mu \to e\gamma$ and projected sensitivities from future experiments are surpassed by muon colliders, particularly at higher energies. For example, a 6 TeV collider with $1\text{ ab}^{-1}$ luminosity probes regions complementary to Stage-II Mu2e for $\sim$1 TeV sleptons, while higher-energy machines access parameter space beyond even PRISM/PRIME sensitivities. 

In single-light-slepton scenarios, where a selectron mixes weakly with a heavier smuon, LFV cross sections depend on $\delta_{RR}^{12}$ and $M_1$. While a 6 TeV collider partially overlaps with MEG II sensitivity, higher-energy machines (14–100 TeV) probe $\delta_{RR}^{12}$ down to $\mathcal{O}(10^{-3})$ for multi-TeV sleptons, exceeding Mu2e Stage II limits. 

Other BSM models that can induce LFV, such as heavy neutral leptons and leptoquarks, have also been studied~\cite{Mikulenko:2023ezx,Han:2025wdy}, demonstrating high sensitivity. These results underscore muon colliders’ dual role: discovering superpartners and precisely mapping their flavour structure, bridging the gap between low-energy indirect probes and high-scale SUSY breaking dynamics.

\section{Direct Searches on Heavy Neutral Lepton}
Heavy neutral leptons (HNL) can arise from various UV models. One of the primary motivations for introducing HNLs is to address the neutrino mass problem. Several BSM frameworks, such as seesaw models, propose HNLs as a natural solution to explain the smallness of neutrino masses. 

\begin{figure}[t]
    \centering
\includegraphics[width=0.8\textwidth]{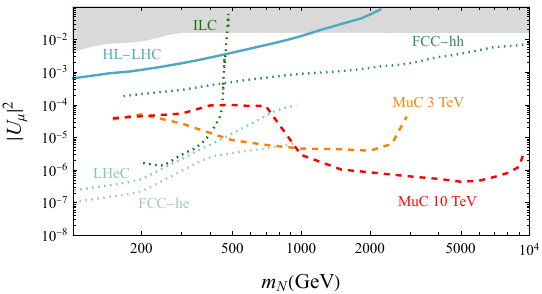}
    \caption{The 95\% C.L. sensitivity to muon-flavored HNL at muon colliders~\cite{Li:2023tbx}, compared to HL-LHC~\cite{Izaguirre:2015pga,Drewes:2019fou,Pascoli:2018heg} and and proposed future colliders (LHeC and FCC-he~\cite{Antusch:2019eiz}, FCC-hh~\cite{Pascoli:2018heg,Antusch:2016ejd},ILC~\cite{Mekala:2022cmm,Antusch:2016vyf}).
    }
    \label{fig:HNL}
\end{figure}

High energy muon collider has the unique capability to fully reconstruct TeV-scale HNL resonance and measure its mixing angle with SM sector~\cite{Li:2023tbx,Mekala:2023diu,Kwok:2023dck}. Especially, the muon-flavored HNL production at muon collider has a remarkable $t$-channel diagram. Such low-order diagram (no $s$-channel suppression) enhances the total production rate by $\mathcal{O}(10^4)$ compared to electron-flavored HNL production. Nevertheless, the VBF production and its low QCD background also make electron-flavored HNL probing sensitivity comparable with FCC-hh. As shown in Figure~\ref{fig:HNL}, either 3 TeV MuC or 10 TeV MuC has a world-leading sensitivity on probing TeV muon-flavored HNL among all current and future experiments. Furthermore, the ability to precisely reconstruct displaced vertices allows for excellent sensitivity to long-lived HNLs across a broad parameter space~\cite{He:2024dwh,Bi:2024pkk}, which can be motivated by several dark sector extensions. 

Overall, the muon collider provides an excellent environment for probing HNLs due to its high center-of-mass energy, clean experimental conditions, and unique production mechanisms. Compared to other future facilities, high energy muon collider stands out in its capability to explore both promptly decaying and long-lived HNL scenarios at the TeV scale, making it a powerful tool for unveiling the nature of heavy neutrinos and their role in neutrino mass generation.

It is also informative to note that the muon collider has special neutrino opportunities at the main detector and forward directions for new physics and dark matter~\cite{Bojorquez-Lopez:2024bsr,Adhikary:2024tvl}. 

\chapter{Dark matter and dark sectors}\label{sec:DM}

Dark matter (DM) stands as the most concrete experimental evidence for physics beyond the Standard Model, and its identity remains one of the greatest mysteries in modern science. Among the myriad of potential explanations, WIMPs (Weakly Interacting Massive Particles) are leading candidates to account for the enigmatic mass that permeates and influences the universe. Muon colliders operating at multi-TeV energies present promising opportunities to investigate TeV-scale WIMP dark matter candidates. Furthermore, dark matter may belong to a more complex dark sector characterized by rich dynamics and weak couplings to the Standard Model. These couplings, often referred to as portals, offer intriguing avenues for exploration. Muon colliders not only hold potential for uncovering WIMPs but also provide a promising arena to investigate these portals and search for particles within the dark sector, thereby enriching our understanding of the fundamental components of the universe.

\begin{figure}[h!]
    \centering
\includegraphics[width=0.8\textwidth]{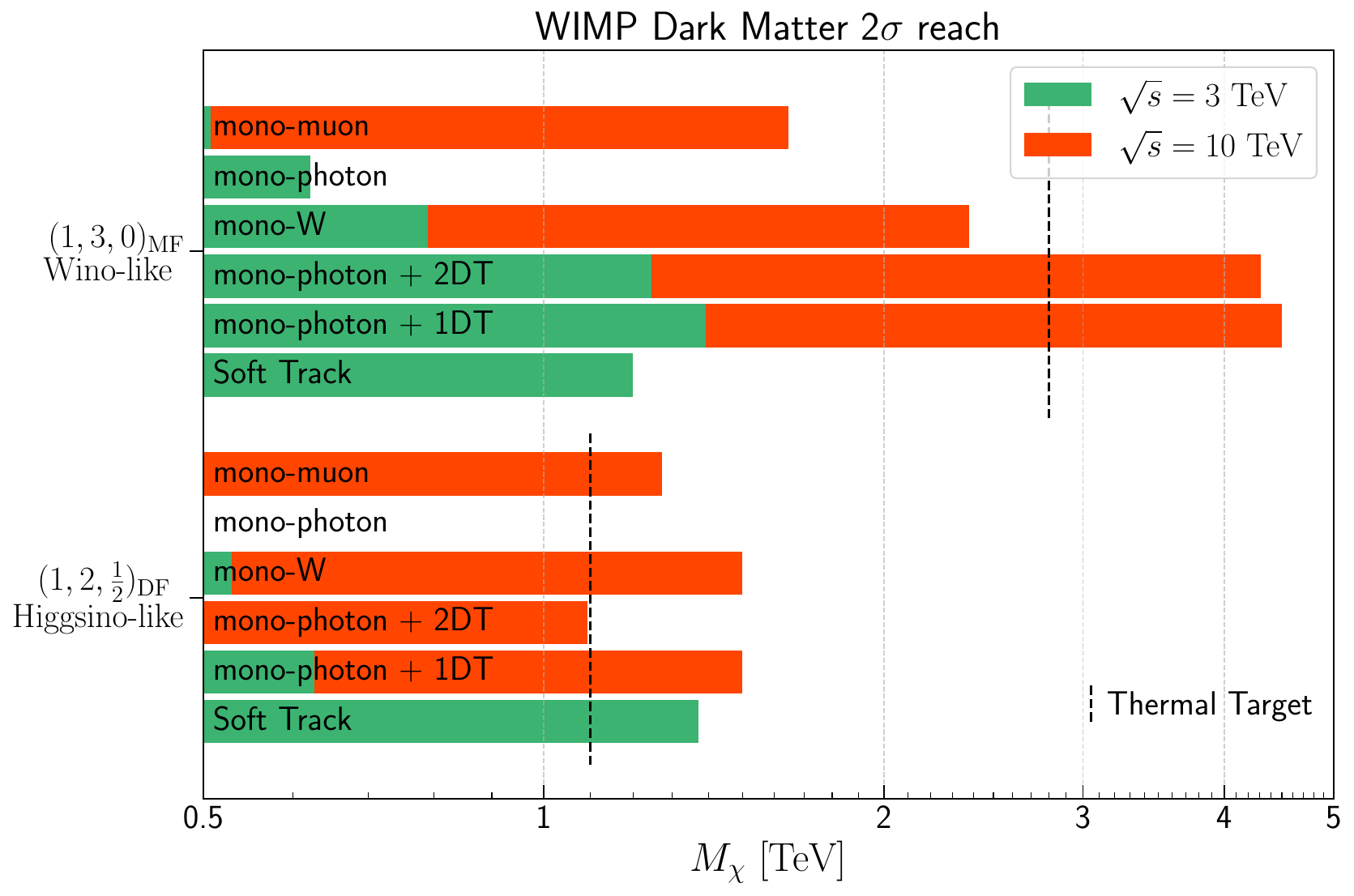}
    \caption{2$\sigma$ exclusion of fermion DM masses from different search channels. Horizontal bars for individual search channels
and muon collider energies by the different colors. The vertical bars indicate the thermal mass
targets. The reach in the mono-photon channel is taken from Ref.~\cite{Han:2020uak}, and the results of mono-W is taken from Ref.~\cite{Bottaro:2022one}. The reaches for the disappearing track are extrapolated from Ref.~\cite{Capdevilla:2021fmj}, and the sensitivity using soft tracks are studied for 3 TeV muon collider in Ref.~\cite{Capdevilla:2024bwt}.
    }
    \label{fig:MDMChannel}
\end{figure}

\section{Minimal Dark Matter}

Thermally produced WIMP is a promising candidates. Despite advances of decades of searches, some of the simplest and most compelling WIMP candidates is still far beyond the current reach. This includes the famous Higgsino (Dirac doublet) and wino (Majorana triplet) in SUSY \cite{Jungman:1995df}. This can be generalized to more multiplets under the electroweak gauge group, the so called Minimal Dark Matter scenario \cite{Cirelli:2005uq}. At the renormalizable level, their interaction with the Standard Model is fixed by the gauge symmetries. Hence, the only parameter is the mass of the dark matter, which can be fixed by the requirement that the thermal freeze-out process produce the correct amount of dark matter. This so called target mass is typically in the TeV range \cite{Bottaro:2021srh}, far beyond the reach of the LHC. 

As an important benchmark for the physics case of the muon colliders, WIMP DM reach has been the subject of a number of recent studies \cite{Han:2020uak,Capdevilla:2021fmj,Bottaro:2021snn,Han:2022ubw,Bottaro:2022one,Black:2022qlg,Fukuda:2023yui,Capdevilla:2024bwt}.  A summary of the projected reaches in shown in Figure~\ref{fig:MDMChannel}.  The general strategy falls into 3 categories. First, the signal is mono-X, where the X can be muon or electroweak gauge bosons. The production processes can be either Drell-Yan or vector boson fusion. Second, the signal has a disappearing track, resulting of the delayed decay of the charged member of the electroweak multiplet. Third, one can exploit the soft particle that is produced in the decay, producing a soft track. The latter possibility has been studied only for the 3~TeV collider~\cite{Capdevilla:2024bwt} and found more effective---see Figure~\ref{fig:MDMChannel}---than disappearing track and mono-$X$ searches. Extension of the analysis to the 10~TeV collider is currently in progress.

Even with this incomplete set of inputs, Figure~\ref{fig:MDMChannel} displays excellent performances of WIMP DM searches at the muon collider. Higgsino and Wino with thermal mass can be discovered at 10~TeV, and thermal Higgsino discovery would be possible even at a first stage of the project with 3~TeV energy. Furthermore, each thermal candidate is accessible by multiple search strategy, adding robustness to a putative discovery. Finally note that the disappearing and soft track sensitivity projections in Figure~\ref{fig:MDMChannel} are supported by detailed simulation studies~\cite{Capdevilla:2021fmj,Capdevilla:2024bwt} that fully account for the realistic experimental environment at a muon collider including the beam-induced background.

\section{Fermion portal and Leptophilic scenario}

\begin{figure}[h!]
    \centering
\includegraphics[width=0.45\textwidth]{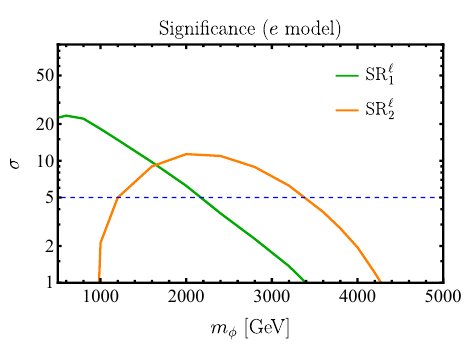}
\includegraphics[width=0.45\textwidth]{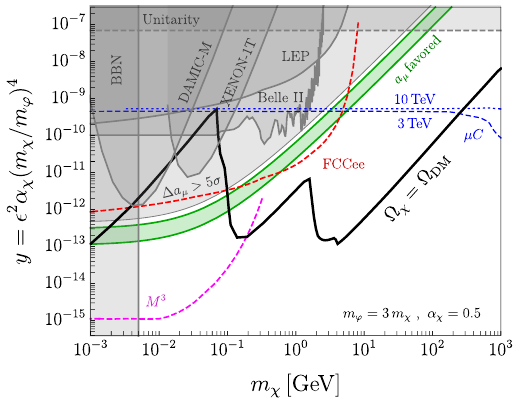}
    \caption{Left panel: Reach for models in which leptonic couplings are flavoured~\cite{Asadi:2023csb,Asadi:2024jiy}. Right panel, reach for the models where the mediator couples to lepton and dark matter separately~\cite{Cesarotti:2024rbh}. 
    }
    \label{fig:leptophilic}
\end{figure}
This is a special category in the so called DM simplified models, with a new mediator which has a significant coupling to the leptons \cite{Asadi:2023csb,Asadi:2024jiy,Cesarotti:2024rbh}. There are two cases. In the first case \cite{Asadi:2023csb,Asadi:2024jiy}, a lepton-flavor portal dark matter model in the freeze-in regime is explored. A new mediator particle is introduced in addition to the dark matter. The mediator $\phi$ has a coupling of the form
\begin{equation}
{\mathcal L} \supset \lambda_{i \alpha} \phi \bar e_i \bar \chi_\alpha,
\end{equation}
where $\bar e_i$ are SM charged leptons, and the dark matter can in principle come in different flavors labelled by $\alpha$.
Astrophysical and relic abundance constraints constrain the model’s
parameter space from all directions. The relic abundance fixes the coupling between dark matter and Standard Model leptons at each point in the parameter space. In certain regions of parameter space, the mediator decays promptly into a charged lepton and missing energy. For this region,  a simple cut-and-count analysis is applied to the mediator pair production at a 10 TeV muon collider. The selection criteria include the invariant mass of the lepton pair, their opening angle, pseudorapidity, transverse mass variable $M_{T2}$, and final-state lepton energies. Using these cuts,  two benchmark signal regions are defined, and it is demonstrated that the model remains discoverable for mediator masses up to 3.3 TeV, as shown in the left panel of Figure~\ref{fig:leptophilic}. A key aspect of the analysis is the incorporation of the non-trivial muon parton distribution function (PDF) in a muon beam, highlighting the need for
developing simulation tools that incorporate the muon PDF.

In the second case \cite{Cesarotti:2024rbh}, the mediator $\varphi$ has couplings to the dark matter and the SM as
\begin{equation}\label{eq:lag} 
{\cal L}_{\rm int} = -\frac{g_{\chi}}{2}  \varphi \chi \chi 
-  \varphi \!\sum_{\ell = e, \mu,\tau} \! g_\ell \, \bar \ell \ell ~~,~~ g_\ell = g_e \left( \frac{m_\ell}{m_e} \right).
\end{equation}
The main interest is in the parameter region of this model that yield the correct relic abundance. 
By solving the Boltzmann equation,  a curve is obtained that relates the mass of the DM candidate and the other parameters of the model. 
The parameter space of interest is above this curve, as points below it would overclose the universe. 
Two search strategies are proposed to be implemented at future lepton colliders that work to the unique advantages of the machines. 

First,  an electron-position machine (labeled FCCee on the plot) run at the $Z$-pole is considered. 
Current projections indicate that upwards of $10^{12}$ $Z$ bosons can be produced, and the leading constraint on our model would be from uncertainty bounds in the width of $Z\rightarrow \tau \bar{\tau}$, as $Z\rightarrow \tau \bar{\tau} \varphi $ for light $\varphi$ could provide a correction as $\varphi$ decays invisibly. 
Additionally, a muon collider option for a $\mu^+\mu^-$ machine operated at 3 (10) TeV with 1 (10) ab$^{-1}$ luminosity is also considered. 
The proposed analysis strategy is completely different, and instead of utilizing the precision nature of an $ee$ machine, instead the advantage comes from the high energy. 
The bond is computed using Monte Carlo simulations and a \textit{mono-photon} search; as the kinematics for the process $\mu^+\mu^-\rightarrow \gamma \varphi$ are fully determined,  a simple bump hunt is performed, modeled against the backgrounds of $\mu^+\mu^-\rightarrow \gamma + \slashed{E}$.
Ultimately, these machines have complimentary reaches despite looking for the same same model: the $ee$ machine can push to small couplings whereas the muon collider can extend to TeV and beyond mass ranges. The right panel of Figure~\ref{fig:leptophilic} displays the reaches for muon collider for this class of models.

\section{ALP Portal}
Axion-like particle (ALP) is a periodic scalar arising from a spontaneously broken symmetry. Depending on the precise symmetry structure, the symmetry under which the ALP is charged may be anomalous under the SM gauge group. Such effects can be parametrized using the following EFT operators at the dimension-5 level
\begin{equation}
	\begin{aligned}
		\mathcal{L} \supset& \frac{1}{2} (\partial_\mu a)^2 - \frac{1}{2} m_a^2 a^2 + \left(\frac{g_3}{4\pi}\right)^2 C_{GG} \frac{a}{f_a} G_{\mu\nu}^a \tilde{G}^{\mu\nu a}\\
		& + \left( \frac{g_1}{4\pi} \right)^2 C_{BB} \frac{a}{f_a} B_{\mu\nu} \tilde{B}^{\mu\nu} + \left( \frac{g_2}{4\pi} \right)^2 C_{WW} \frac{a}{f_a} W_{\mu\nu}^a \tilde{W}^{\mu\nu a} \\
		&+ \left( i c_{a\Phi} \frac{a}{f_a} Q_L^\dagger \gamma^0 \mathbf{Y_{Q}} \mathbf{\Phi} \sigma_3 Q_R + \text{h.c.} \right),
	\end{aligned}
	\label{eqn:ALPEFT}
\end{equation}
in which $\tilde{F}_{\mu\nu} = \epsilon_{\mu\nu\rho\sigma} F^{\rho\sigma}/2$, $f_a$ is the ALP's decay constant, $g_{1,2,3}$ are the SM gauge couplings, $Q_R = (u_R, d_R)$ is a tuple of right-handed quark singlets, $\mathbf{Y}_{Q} = \text{diag}(\mathbf{Y}_u, \mathbf{Y}_d)$ represents the SM Yukawa couplings, 
and $C_{BB}$, $C_{WW}$, $C_{GG}$, and $c_{a \Phi}$ are the Wilson coefficients to be probed by experiments. 

\begin{figure}
	\centering
	\includegraphics[width=\textwidth]{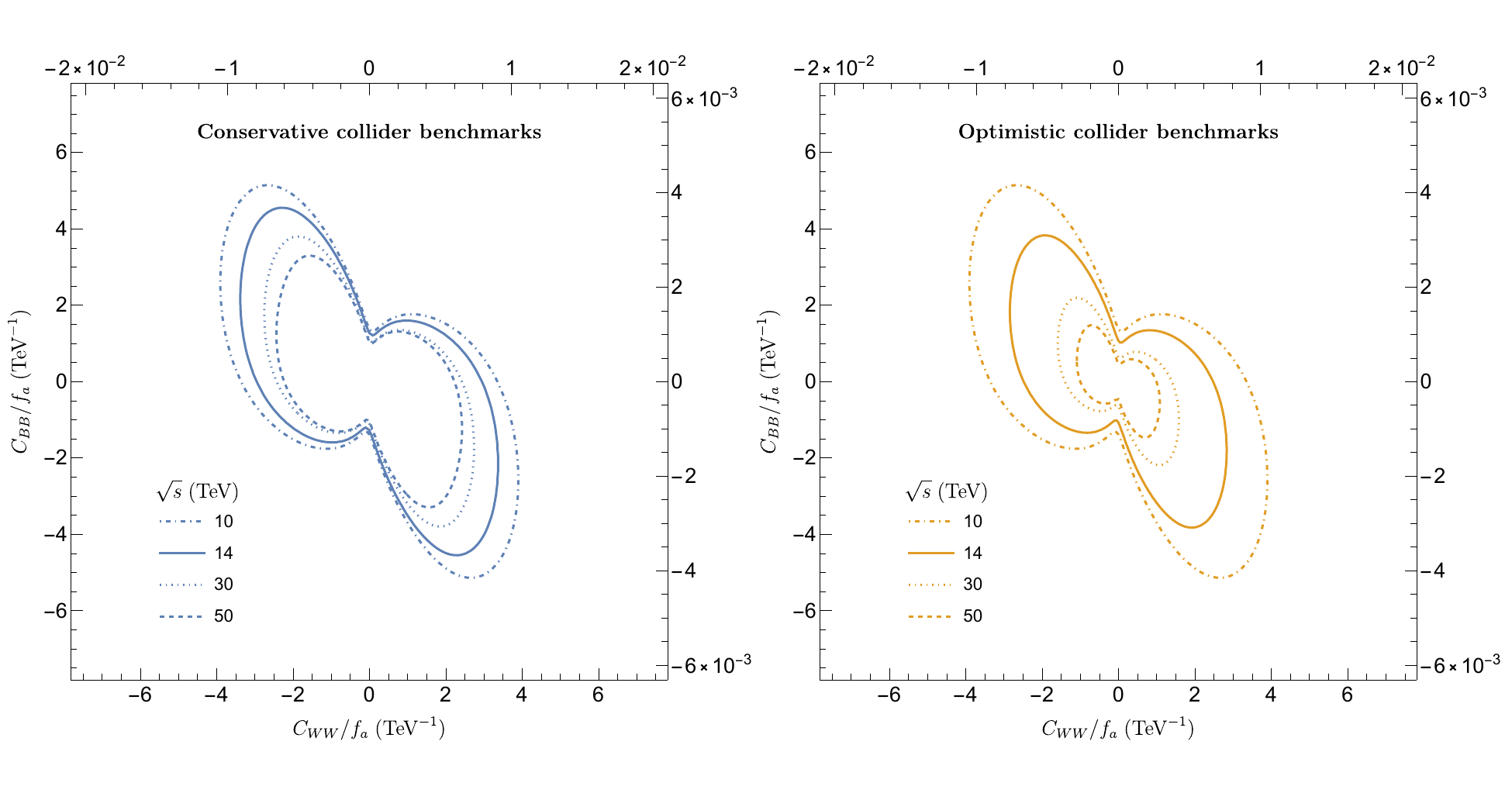}
	\caption{Projected 2$\sigma$ exclusion reach of the electroweak vector boson coupling $C_{BB}/f_a$ and $C_{WW}/f_a$ to a $1$-$\text{TeV}$ ALP of various muon collider benchmarks \cite{Bao:2022onq}: Note that the bottom and left axes indicate values of the ALP couplings defined in Eq.~(\ref{eqn:ALPEFT}), while the top and right axes indicate values of these couplings absorbing the gauge couplings and loop factors following another popular convention in Ref.~\cite{Brivio:2017ije}. The left panel assumes that the integrated luminosity $L=10\;\text{ab}^{-1}$ across all $\sqrt{s}$ whereas the right panel uses $(\sqrt{s}\;(\text{GeV}), L\;(\text{ab}^{-1})) = (10, 10)$, $(14, 20)$, $(30, 90)$, and $(50, 250)$ as benchmarks. }
	\label{fig:alp_diboson}
\end{figure}

One particularly interesting possibility is that the $aW\tilde{W}$ and $aB \tilde{B}$ couplings can be probed by a high-energy muon collider. When the ALP's mass is relatively low, the associated production $V \to Va$ is possible. This permits a clean detection on top of a relatively small SM background, such as $\mu^+ \mu^- \to \gamma\gamma\gamma$. 
With high beam energy $\sqrt{s} \sim 10\; \text{TeV}$, 
another main production channel becomes the vector boson fusion (VBF), and detections are made by the reconstruction of various di-boson channels \cite{Han:2022mzp, Bao:2022onq}. The $2\sigma$ exclusion limits for various muon collider benchmarks are shown in Figure~\ref{fig:alp_diboson}. In general, for TeV-scale ALPs, VBF production with di-photon and $\gamma Z$ final states provide the leading constraints~\cite{Bao:2022onq}, but when $C_{BB} = 0$, the $Za$ associated production also has a chance to provide comparable reach as that from VBF production~\cite{Han:2022mzp}. In either case, in the absence of the $C_{GG}$ gluon coupling, a muon collider offers a constraint on $C_{BB}$ and $C_{WW}$ that is one-order-of-magnitude stronger than HL-LHC. 

\begin{figure}
	\centering
	\includegraphics[width=\textwidth]{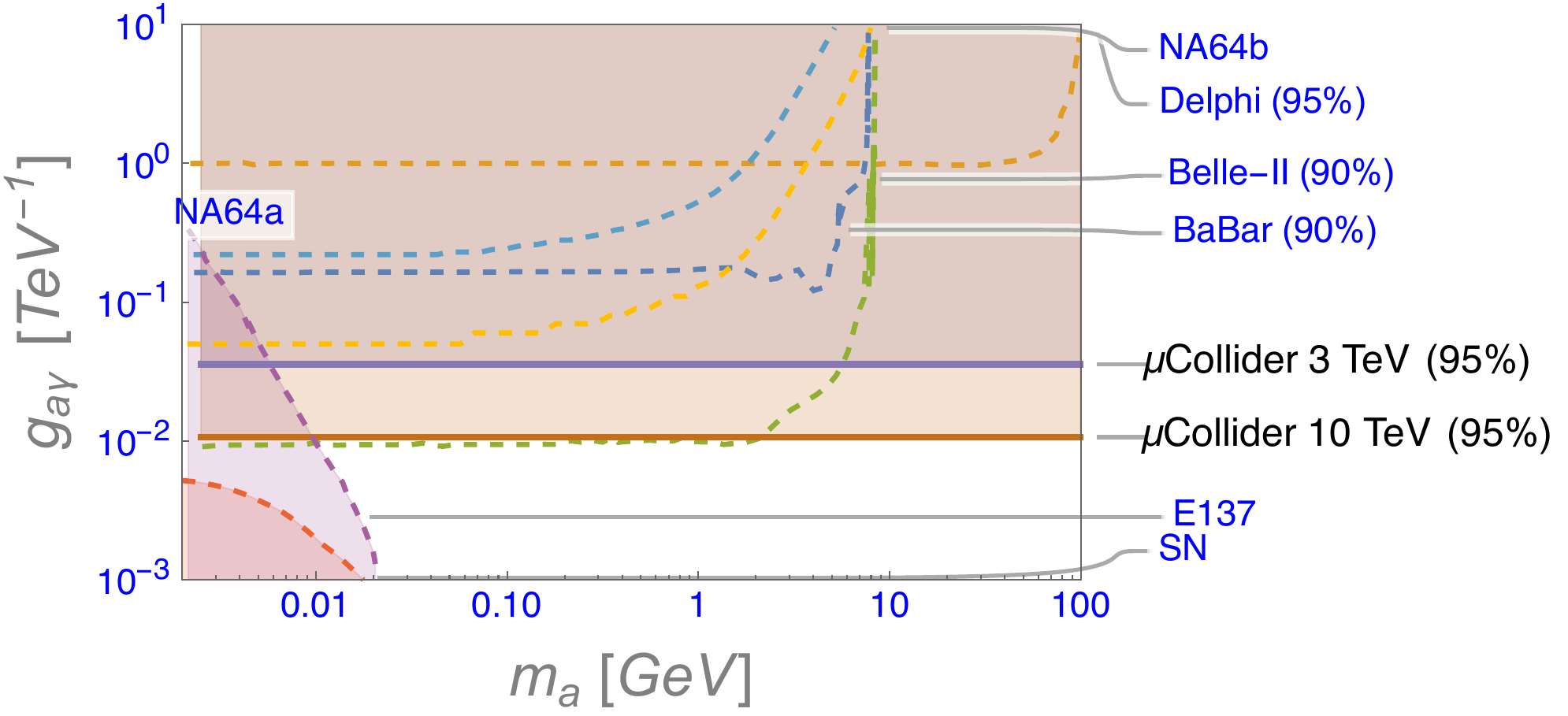}
	\caption{ALP-photon coupling $g_{a\gamma}$ probed by a muon collider as a function of light ALP mass \cite{Casarsa:2021rud}: Note that the ALP-photon coupling conventionally defined here does not include the loop-suppression factor $(e/4\pi)^2$.}
	\label{fig:alp_photon}
\end{figure}

For ALP lighter than $100\;\text{GeV}$, the associated production becomes the most relevant production channel. The signature then is mono-photon with a missing energy whose SM background is $\mu^+\mu^- \to \gamma \nu \bar{\nu}$. Although the radiative return of $Z$ gives an enhanced background for photon energy near $\sqrt{s}/2$, the muon collider can still offer a strong probe to the axion-photon coupling $g_{a\gamma} \lesssim 1.1 \times 10^{-5}\; \text{GeV}^{-1}$ at 95\% C.L. for a 10-TeV muon collider with $10\;\text{ab}^{-1}$ of data \cite{Casarsa:2021rud}. This reach is better than the current limits from BaBar and Delphi and comparable to the projected limit from Belle II as shown in Figure~\ref{fig:alp_photon}.

\begin{figure}
	\centering
	\includegraphics[width=\textwidth]{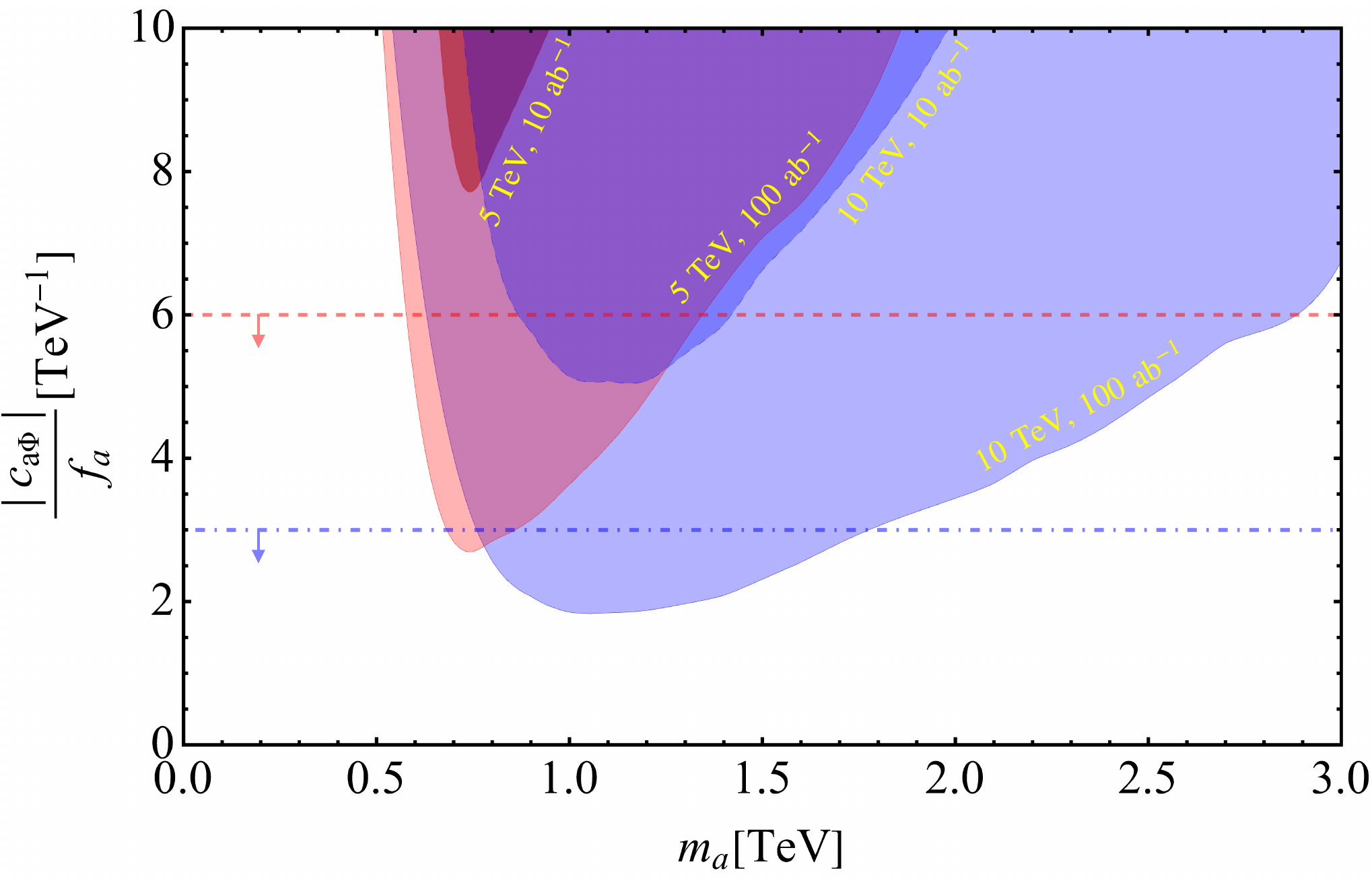}
	\caption{$5 \sigma$ detection reach of the ALP-top coupling $c_{a\Phi}$ at a muon collider as a function of light ALP mass \cite{Chigusa:2023rrz}: the dashed line shows perturbative unitarity constraints while colored contours correspond to $5\sigma$ limits for various muon collider benchmarks.}
	\label{fig:alp_top}
\end{figure}

Another promising target is the ALP-top coupling \cite{Chigusa:2023rrz, Chigusa:2025otr} for $m_a > 2m_t$. This comes from $c_{a\Phi}$ coupling in eq.~(\ref{eqn:ALPEFT}) and enjoys an enhancement from the top Yukawa. Here, both top-associated production $\mu^+ \mu^- \to t\bar{t} a$ and VBF can be relevant, and the decay channels are $a \to t\bar{t}$, $a \to VV$, and $a \to gg$. When $c_{a \Phi}$ coupling is sufficiently large compared to $C_{WW}$ and $C_{GG}$, the dominant production channel is the top-associated production. The heavy ALP resonance decays to $t\bar{t}$ predominantly. This signal is to be detected on top of SM background such as $t\bar{t} t\bar{t}$, $t\bar{t}W^+W^-$, $t\bar{t} h$, and $t\bar{t} Z$. Appropriate cuts allow us to probe this coupling for relatively heavy ALPs with a mass $m_a \sim 1$ to $3\;\text{TeV}$ as shown in Figure~\ref{fig:alp_top}. If $c_{a\Phi}$ is small relative to $C_{WW}$ or $C_{GG}$, then a richer collider phenomenology can appear as VBF production or $a \to gg$ decay becomes important. This is studied in detail in Ref.~\cite{Chigusa:2025otr}. 

\chapter{Detector instrumentation}

\section{Overview}

A summary of the technological readiness and R\&D avenues is provided in Tables~\ref{tab:ppg:det:trk-hcal}, \ref{tab:ppg:det:muon-tdaq}, and~\ref{tab:ppg:det:magnet-pid}. These tables include a summary of the key performance indicators that will ensure the detector able to extract the full physics potential from the muon-muon collisions at the highest energies. The tables are split by system and for each system a list of technologies that are expected to meet the required performance is also given, together with a brief comment on its readiness. For many systems, multiple technologies are being considered and are listed in the tables as well as in the R\&D chapters of this document.
Further details are given in the next sections for selected R\&D projects including work specific to an application at a muon collider.

\begin{table}
\centerline{
\renewcommand{\arraystretch}{1.25}
\begin{tabular}{|c|p{4cm}|p{4cm}|c|p{4cm}|c|}
\hline
 & \textbf{KPI} & \textbf{Technologies} & \textbf{TRL} & \textbf{Status} & \textbf{DRD} \\
\hline
\multirow{12}{*}{\rotatebox[origin=c]{90}{\textbf{Tracking}}} &  \multirow{7}{=}{Position precision, fast timing, radiation tolerance, data processing complexity, data transmission bandwidth, efficient power distribution, Low material budget} & LGAD & 6 & Single prototypes in relevant environments & 3 \\
& & MAPS & 6 & Single prototypes in relevant environments, first systems but in different environments & 3 \\
& & 3D sensors & 8 & Used in similar environments as systems & 3 \\
& & Planar sensors & 8 & Used in similar environments as systems & 3 \\
& & Micro-Strips & 5 & Technology tested, and similar to other validated in relevant environment & 3 \\
& & ASIC & 4 & Technology tested, and similar to other validated in relevant environment & 7 \\
& & Carbon support structures & 8 & Similar solutions already used in relevant environment & 8 \\
\hline
\multirow{8}{*}{\rotatebox[origin=c]{90}{\textbf{ECAL}}} &  \multirow{3}{=}{Segmentation, fast time-of-arrival, energy resolution, radiation tolerance} & Crilin & 6 & Large scale prototype in preparation; small scales tested in test-beams & 6 \\
& & Si-W & 8 & Extensive tests of technology, e.g. in context of CMS Phase-II upgrades & 6 \\
& & ASIC & 4 & Although very similar in concept to proven and well-tested technology & 8 \\
\hline
\multirow{4}{*}{\rotatebox[origin=c]{90}{\textbf{HCAL}}} &  \multirow{2}{=}{Segmentation, energy resolution} & MPGD & 4 & Initial prototypes and test beam results & 6 \\
& & Fe-Scintillator & 8 & Extensive tests of technology, e.g. in context of CALICE & 6 \\
\hline
\end{tabular}
\renewcommand{\arraystretch}{1}
}
\caption{Summary of the key performance indicators (KPIs), the identified technologies and their technical readiness for various systems of the muon collider detector design. Note that the key performance indicators for each system consists of a list; the list of technologies is also a list, and its order is not expected to match the order of the KPIs, but instead lists technologies that are expected to meet all relevant KPIs for that system. For each technology, a Technical Readiness Level (TRL) and a brief comment on the status and matching DRD group is also given.}
\label{tab:ppg:det:trk-hcal}
\end{table}

\begin{table}
\centerline{
\renewcommand{\arraystretch}{1.25}
\begin{tabular}{|c|p{4cm}|p{4cm}|c|p{4cm}|c|}
\hline
 & \textbf{KPI} & \textbf{Technologies} & \textbf{TRL} & \textbf{Status} & \textbf{DRD} \\
\hline
\multirow{9}{*}{\rotatebox[origin=c]{90}{\textbf{Muon system}}} &  \multirow{3}{=}{Fast time-of-arrival, radiation tolerance, fast readout} & PICOSEC & 6 & Initial prototypes and test beam results & 1 \\
& & GEM & 8 & Similar solutions already used in relevant environments & 1 \\
& & MPGD & 8 & Similar solutions already used in relevant environment & 1 \\
& & RPC & 8 & Similar solutions already used in relevant environment & 1 \\
\hline
\multirow{4}{*}{\rotatebox[origin=c]{90}{\textbf{MDI}}} &  Efficient shielding design (nozzle) & W-alloy + (C2H4, B4C, BCH2) & 3 & Simulation-based studies. Smaller-scale similar solutions deployed in other context successfully & 8 \\
\hline
\multirow{3}{*}{\rotatebox[origin=c]{90}{\textbf{Lumi}}} &  Accurate luminosity measurement & Si-W or Si & 7 & Extensive tests of technology e.g. in contex of CMS Phase-II upgrades & 6 \\
\hline
\multirow{3}{*}{\rotatebox[origin=c]{90}{\textbf{Fwd Muon}}} &  \multirow{3}{=}{Tagging capabilities, (optional) momentum resolution} & Silicon-based (TBD) & 1 & Basic idea needs demonstration and clear formulation in terms of candidate technologies & 3 \\
\hline
\multirow{8}{*}{\rotatebox[origin=c]{90}{\textbf{Electronics/TDAQ}}} &  \multirow{3}{=}{Real-time reconstruction, large banddwith data transmission, Radiation hardness} & CPU / GPU & 9 & Experience in running experiments (e.g. LHCb) & \\
& & Dedicated FPGA boards & 9 & Experience in running experiments (e.g. LHCb) & \\
& & Electrica/Optical transceivers & 9, 3 & Experience in Phase-II LHC upgrades. More novel approaches (e.g. integrated Si Photonics) need developments & \\
\hline
\end{tabular}
\renewcommand{\arraystretch}{1}
}
\caption{Summary of the key performance indicators (KPIs), the identified technologies and their technical readiness for various systems of the muon collider detector design. Note that the key performance indicators for each system consists of a list; the list of technologies is also a list, and its order is not expected to match the order of the KPIs, but instead lists technologies that are expected to meet all relevant KPIs for that system. For each technology, a Technical Readiness Level (TRL) and a brief comment on the status and matching DRD group is also given.}
\label{tab:ppg:det:muon-tdaq}
\end{table}

\begin{table}
\centerline{
\renewcommand{\arraystretch}{1.25}
\begin{tabular}{|c|p{4cm}|p{4cm}|c|p{4cm}|c|}
\hline
 & \textbf{KPI} & \textbf{Technologies} & \textbf{TRL} & \textbf{Status} & \textbf{DRD} \\
\hline
\multirow{2}{*}{\rotatebox[origin=c]{90}{\textbf{Magnets}}} &  Large field strength, homogeneous field, large aperture & Superconducting solenoid & 8 & Experience in running experiments (e.g. CMS) &  \\
\hline
\multirow{8}{*}{\rotatebox[origin=c]{90}{\textbf{PID Detectors}}} &  \multirow[t]{3}{=}{PID between 1-100 GeV} & Cherenkov & 6 & Basic analogous systems in running experiments, but needs tuning & 4 \\
& & dE/dx (Silicon-based) & 8 & Experience in running experiments, but needs tuning & 3 \\
& & ToF & 6 & Basic analogous systems in running experiments, but needs tuning & 4\\
\hline
\end{tabular}
\renewcommand{\arraystretch}{1}
}
\caption{Summary of the key performance indicators (KPIs), the identified technologies and their technical readiness for various systems of the muon collider detector design. Note that the key performance indicators for each system consists of a list; the list of technologies is also a list, and its order is not expected to match the order of the KPIs, but instead lists technologies that are expected to meet all relevant KPIs for that system. For each technology, a Technical Readiness Level (TRL) and a brief comment on the status and matching DRD group is also given.}
\label{tab:ppg:det:magnet-pid}
\end{table}

\FloatBarrier
\section{Crilin}

The Crilin calorimeter is being explored as a potential solution for the electromagnetic calorimetry requirements of a future Muon Collider. It utilizes an array of high-density crystal matrices, where each unit is read independently by two channels, each integrating a pair of Silicon Photo-Multipliers (SiPMs). This semi-homogeneous architecture provides notable advantages, including precise timing, fine granularity, longitudinal segmentation, resilience to radiation, and cost-effectiveness. In the ECAL region, simulations indicate a flux of ~300 particles/cm$^2$, mainly photons (96\%) and neutrons (4\%), with an average photon energy of 1.7 MeV. The BIB poses a significant challenge for the Muon Collider. FLUKA simulation at $\sqrt{s}=1.5$ TeV estimated a neutron fluence of 10$^{14}$ $n_{1\text{MeV}}/\text{cm}^2$ per year and a TID of 1 kGy per year in the ECAL barrel. The Crilin calorimeter~\cite{Ceravolo_2022}, a semi-homogeneous crystal design, offers a cost-effective, high-performance alternative to address BIB challenges.\\
To comply with the stringent demands of the Muon Collider, the Crilin calorimeter is designed to achieve a timing resolution below 100 ps, crucial for mitigating Beam-Induced Backgrounds (BIB) and isolating physics signals. The fine spatial resolution, defined by a cell size of $10 \times 10 \, \text{mm}^2$, enables efficient separation of background events from actual particle interactions, reducing occupancy per channel. The calorimeter comprises five detection layers, each 45 mm in length (40 mm for the crystal and 5 mm for the readout system), ensuring effective longitudinal segmentation. This segmentation is essential for distinguishing and suppressing spurious showers originating from BIB. Compared to a conventional W-Si calorimeter, the Crilin design reduces the number of readout channels and associated electronics by a factor of 10. This optimization results in a more compact and cost-efficient detection system, making it an attractive candidate for future collider experiments.
The Crilin calorimeter relies on radiation-resistant crystals, such as PbF$_2$ \cite{Cemmi_2022} and PbWO$_4$-UF \cite{crytur}, which have demonstrated stable performance under high radiation exposure. PbF$_2$ crystals retain their optical transmittance after exposure to Total Ionizing Doses (TID) up to 260 kGy, as tested at the Calliope gamma Facility (ENEA NUC-IRAD-GAM Laboratory). PbWO$_4$-UF crystals, on the other hand, withstand doses up to 1.5 MGy \cite{BTF}\footnote{Dose values measured in air.}. \\
In addition to crystal selection, SiPMs have undergone extensive radiation qualification. Hamamatsu S14160-3010PS SiPMs, with a 10 $\mu$m pixel size, exhibit only a minor increase in dark current when subjected to a neutron fluence of 10$^{14}$ n$_\text{1MeV-eq}$/cm$^2$ and a TID of 10 kGy. Given the anticipated radiation levels at the Muon Collider—approximately 1 kGy/year TID and a neutron fluence up to 10$^{14}$ n$_\text{1MeV-eq}$/cm$^2$—the use of radiation-hard materials is essential to ensure the longevity and reliability of the calorimeter system. \\

\begin{figure}[h]
    \centering
    \includegraphics[width=0.5\columnwidth]{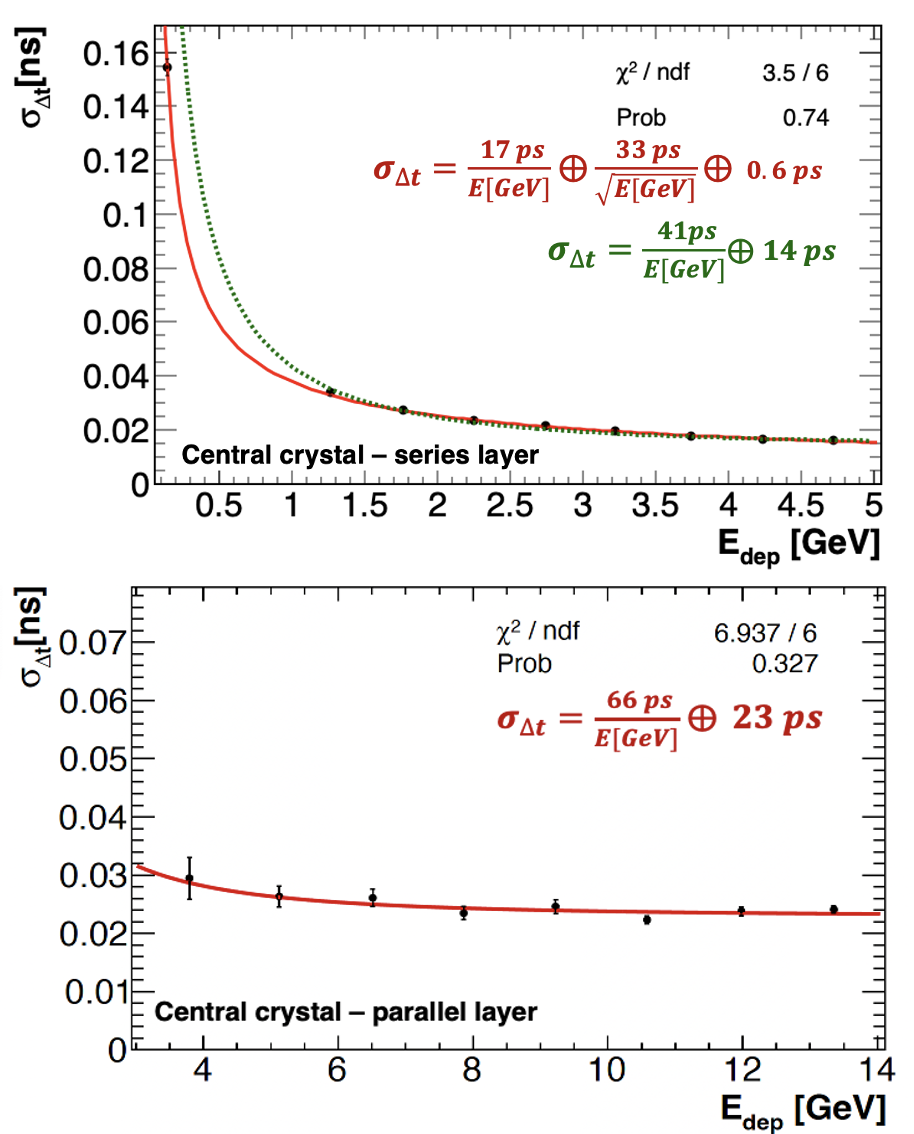}
    \caption{Time resolution as a function of the energy deposited in the crystal with the highest energy deposit. A time resolution measurement with a 450 MeV electron beam was added for the series layer (solid red line fit).}
    \label{fig:timeresosingle}
\end{figure}

Two preliminary prototypes have been developed and tested. Proto-0~\cite{Daniele_h2}, featuring two crystals and four channels, and Proto-1, integrating two layers of a $3 \times 3$ crystal matrix read out by 36 channels, demonstrated excellent time resolution (on the order of 20 ps) and good agreement with Monte Carlo simulations, validating the design's performance. For R\&D purposes, two different SiPM circuit configurations were tested: the first layer had two readout channels per crystal connected in series, while the second layer had them connected in parallel. In August 2023, Proto-1's timing performance was evaluated with a 120 GeV electron beam at CERN-SPS H2, testing different configurations by measuring time differences between the two layers and between channels within the same crystals. The central elements of the matrix, where the highest energy deposits occur, were analyzed. As shown in Figure~\ref{fig:timeresosingle}, for both the series and parallel layers, the time resolution remained below 40 ps for energy deposits greater than 1 GeV.

\begin{figure}[h]
    \centering
    \includegraphics[width=0.45\columnwidth]{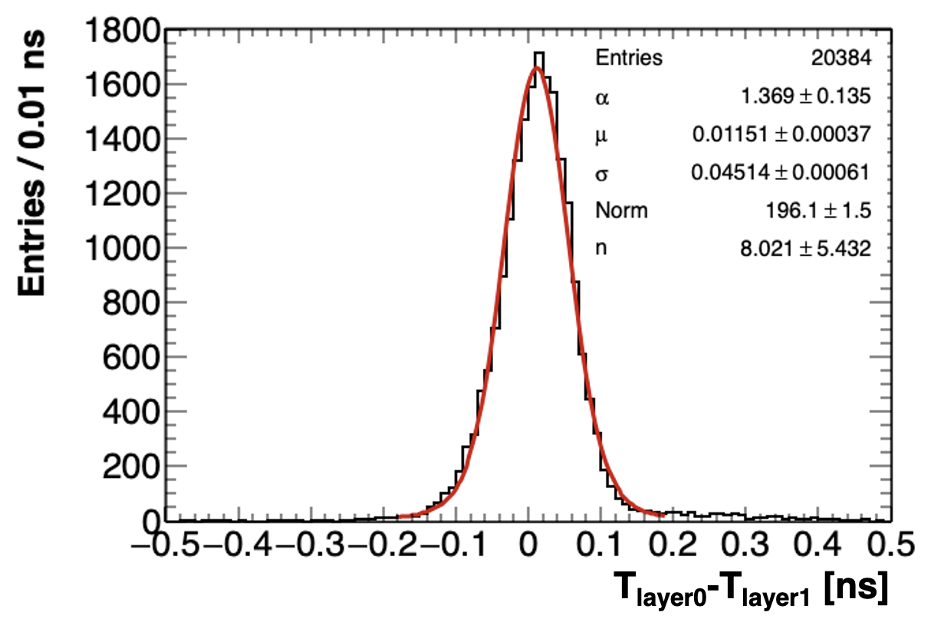}
    \caption{Time resolution for the time difference between the two most energetic crystals from different layers. A Double Sided Crystal Ball fit yields a resolution of 45 ps, mainly dominated by digitizer board-to-board jitter.}
    \label{fig:timeresolay}
\end{figure}

The time resolution achieved using the time difference between the two most energetic crystals from different layers was well within requirements. As shown in Figure~\ref{fig:timeresolay}, a Double Sided Crystal Ball fit yielded a $\sigma_{\Delta t}$ of 45 ps, primarily dominated by digitizer board synchronization jitter, measured to be $\mathcal{O}(32 \,\text{ps})$ for board-to-board cases and $\mathcal{O}(10\,\text{ps})$ for channel-to-channel cases.

\paragraph*{New Prototype}
A new prototype consisting of five layers of $7 \times 7$ crystal matrices will be fully developed, built, and tested during 2025, reaching 2 Moli\`ere Radius, 22 radiation lengths.

\begin{figure}[h!]
    \centering
    \includegraphics[width=0.3\columnwidth]{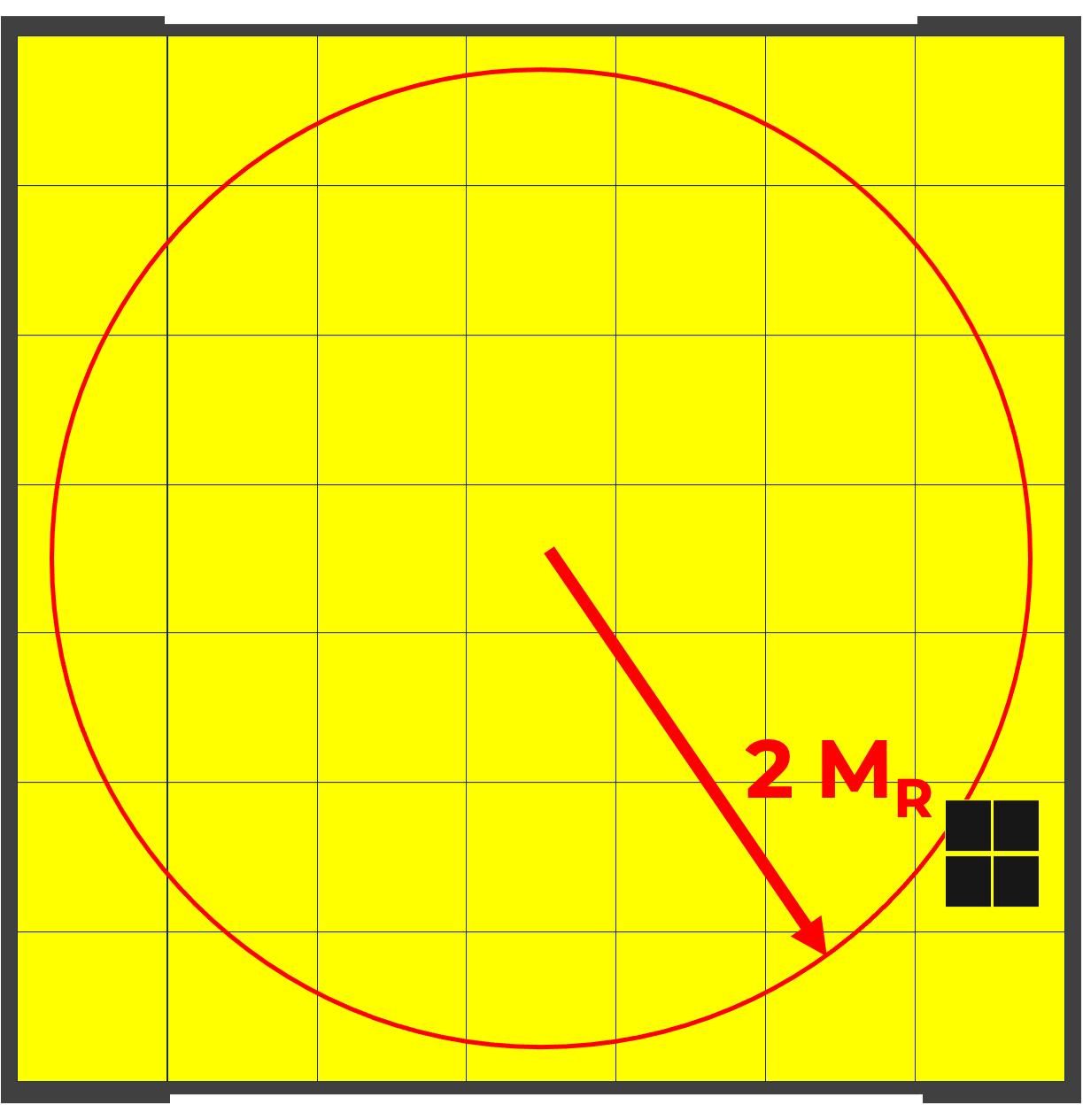}
    \caption{Sketch of the final prototype embedding five layers of $7 \times 7$ crystal matrices.}
    \label{fig:newproto_2}
\end{figure}

The mechanics and electronics will have significant changes compared to the previous prototypes. An aluminum matrix with 150 \textmu m thickness will hold the crystals, and a thicker (2.5 mm) external envelope will surround the matrix, also providing cooling through micro-channels. Moreover, a micro-coaxial Kapton strip will provide SiPM polarization and readout independently for each channel of two SiPMs in series. An overall connector will be placed at the back of the five assembled modules. A sketch of the final Crilin version is shown in Figure~\ref{fig:newproto_2}. This final version, featuring a total of 2 M$_R$ and 22 X$_0$, will provide enough coverage to finely study the energy resolution performances together with the timing information. With the beam tests planned for 2025 and 2026, we expect to obtain clear indications regarding the performance of the system. By 2027, we anticipate being ready to scale up to the full detector.

\section{Micro-Pattern gaseous detector-based hadron calorimeter}

In a Particle Flow (PF) approach, a Hadronic Calorimeter (HCAL) must guarantee an energy resolution stochastic term lower than 55\%$/\sqrt{E}$ to reach a jet energy resolution of 3\% for jets above 100~GeV, fundamental for discriminating W and Z bosons hadronic decays~\cite{THOMSON200925}. Moreover, the necessity to distinguish between neutral and charged hadrons requires a high granularity HCAL, with readout element size between $1\times 1$ and $3\times 3$~cm$^2$. Additional detector constraints arise instead from the beam induced background. The detector technology must be  radiation-hard to face the harsh environment and a good time resolution is necessary to reject BIB hits.  In a 1.5 TeV detector configuration, it was demonstrated that the BIB arrival time has a flat distribution between 8 and 16~ns, with a tail up to 20~ns, against a signal arrival time peaking at 6~ns and almost totally contained in 10~ns, allowing a rejection of the asynchronous background with an arrival time request within a 10~ns time window.~\cite{Longo:2024pk}. 

\begin{figure}[t]
    \centering
    \includegraphics[width=0.5\linewidth]{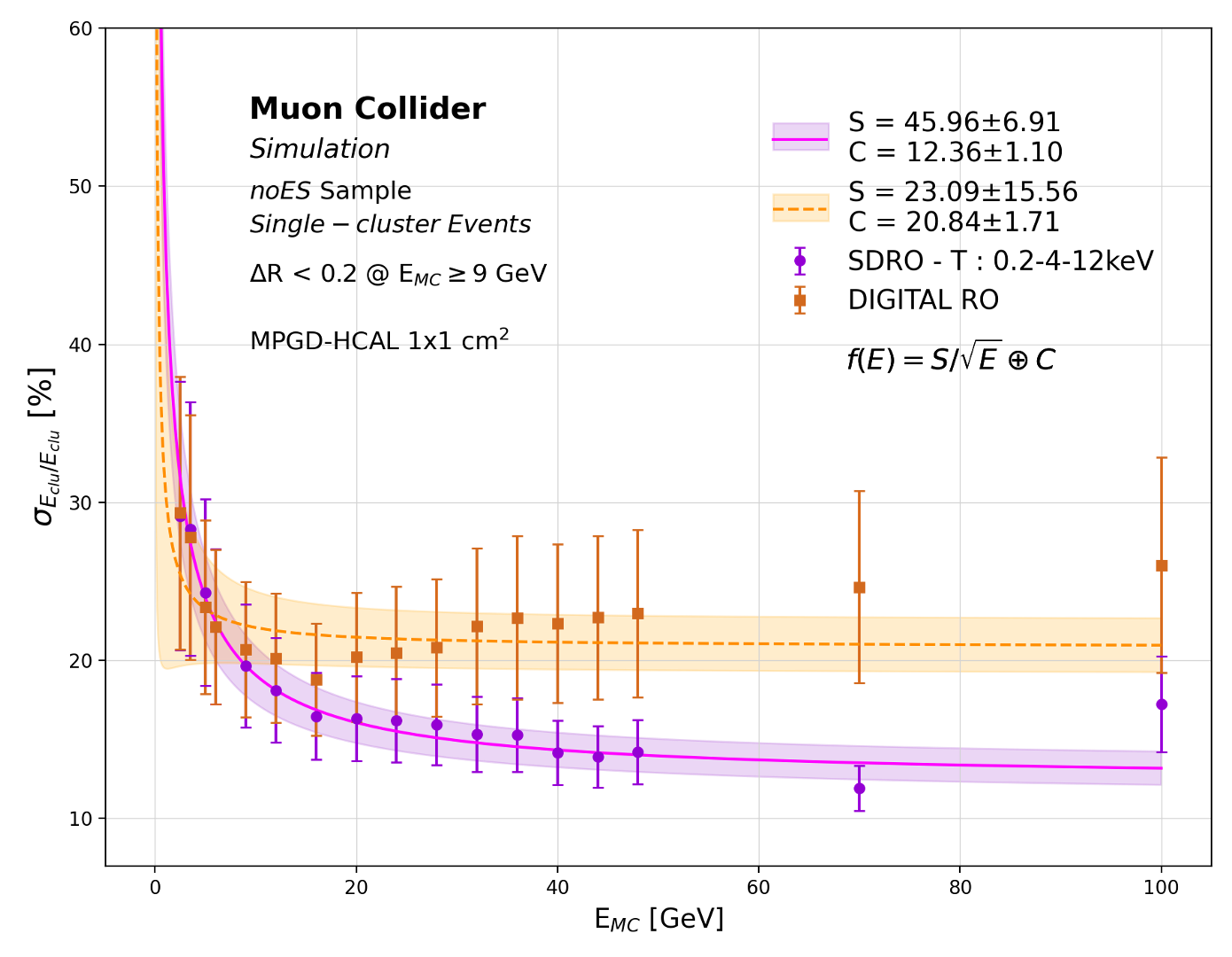}
\caption{
    Energy resolution for a digital readout (RO), where the reconstructed energy is assumed to be a function of the total number of hits in the calorimeter (in orange), and a semi-digital RO (SDRO), where instead the energy is estimated as a linear combination of the number of hits above three different thresholds (in violet): 0.2, 4 and 12~keV.  Only particle showering in the hadronic calorimeter were considered.  Both sets of data were fitted against $S/\sqrt{E}+C$~\cite{Longo:2024pk}. 
    \label{fig:MPGDHCALEnergyRes}}
\end{figure}

\begin{figure}[t]
\centering
\subfloat[4~GeV $\pi^{-}$]{\includegraphics[width=0.48\textwidth]{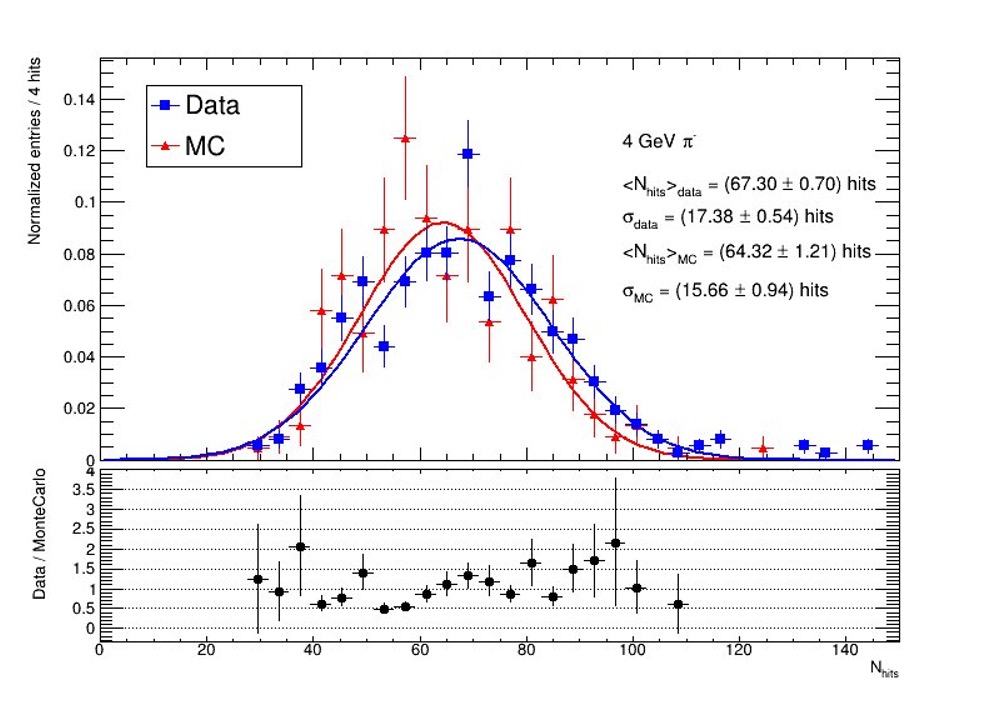}}\hfill
\subfloat[6~GeV $\pi^{-}$]{\includegraphics[width=0.48\textwidth]{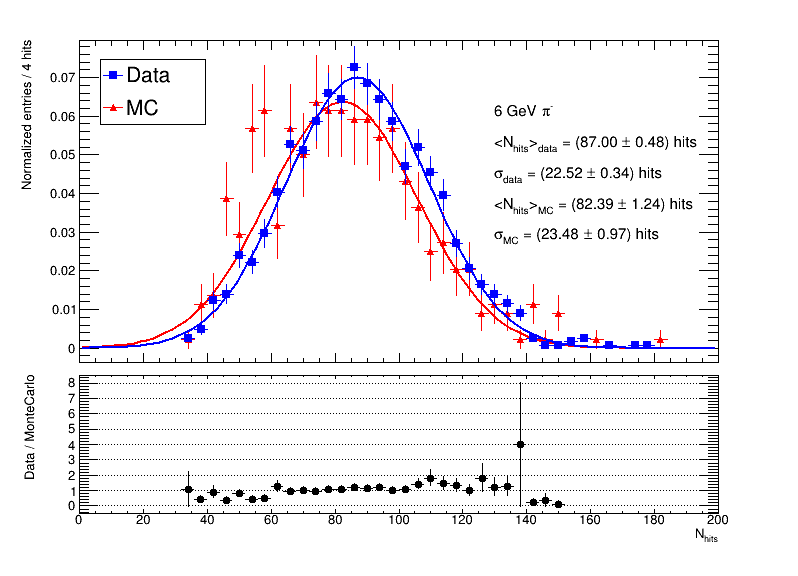}}\hfill
\subfloat[8~GeV $\pi^{-}$]{\includegraphics[width=0.48\textwidth]{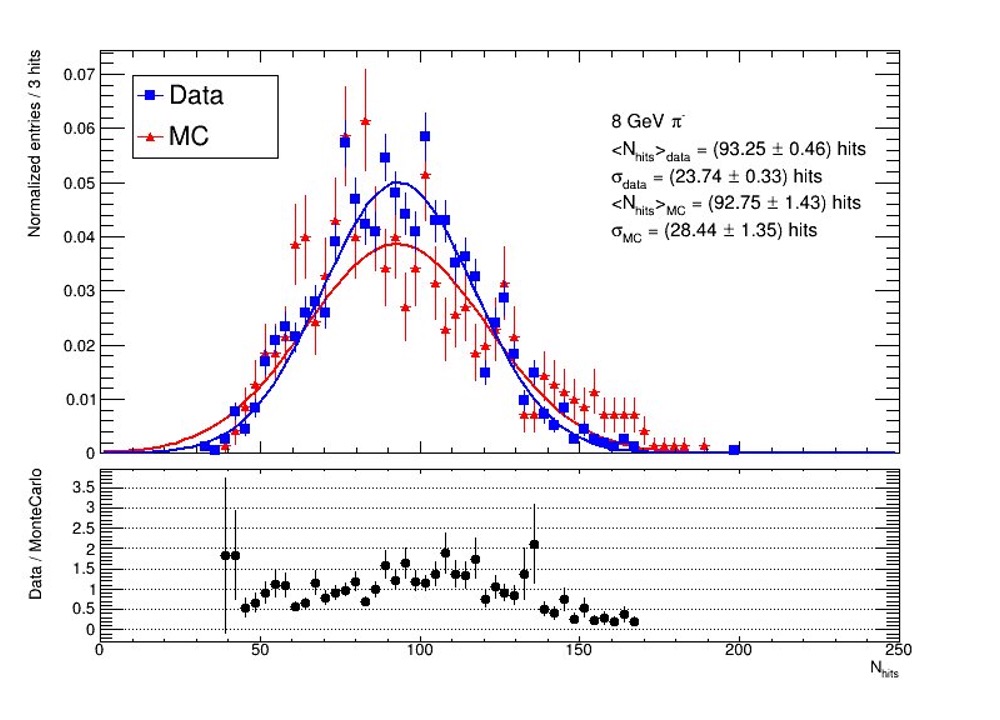}}  
     \caption{Distribution of the total number of hits in the calorimeter prototype tested at the CERN PS East Area for a 4, 6, and 8~GeV pion beam, compared with the results of a Geant4 simulation, implementing the calorimeter prototype under test~\cite{Longo:2024pk}.
     \label{fig:MPGDHCAL_ps_data_mc}}
\end{figure}

A sampling hadronic calorimeter based on resistive Micro-Pattern gaseous detectors (MPGD) can meet the HCAL requirements just described. MPGDs are able to withstand integrated charge of several  C/cm$^2$~\cite{FARINA2022}, so to not show ageing induced  by BIB,  and have a better rate capability ($O(\mbox{MHz/cm}^2)$)~\cite{micromegas_rate_capability} and spatial resolution~\cite{urwell_space_resolution},  compared to other gaseous detector technologies such as RPCs, which allows better performances in presence of BIB. Moreover,  the time resolution of $O(\mbox{ns})$ lets MPGD detectors be a suitable solution for HCAL at Muon Collider, being able to reject BIB hits from their arrival time.  Additional advantages of MPGDs lie on the possibility to easily have a highly segmented readout, with cell size of  $1\times1 $~cm$^2$, and to cover large area. Figure~\ref{fig:MPGDHCALEnergyRes} shows that a MPGD-HCAL, made of 60 layers of 20~mm of iron (absorber) and 3~mm of argon (active material) and with a readout element size of 1~cm$^2$, can achieve a stochastic energy resolution term of $46\%/\sqrt{E}$, fulfilling also the PF requirement~\cite{Longo:2024pk}. 
To pursue this MPGD-based hadronic calorimeter, a prototype made of 8 active layers has been built and tested at CERN PS Est Area with pions of few GeV and, as expected, it was observed how the total number of hits  is a function of the pions energy (Figure~\ref{fig:MPGDHCAL_ps_data_mc}). No final choice of the MPGD technology has been done and the technologies(MicroMegas~\cite{Iodice_2020}, $\mu$-RWELL~\cite{urwell} and RPWELL~\cite{Zavazieva_2023} are now under studies.

This R\&D effort is currently at the status of TRL4 and is part of the international collaborations DRD1 and DRD6. For the next years, up to 2030, it is foreseen the choice of the most outperforming MPGD technology and the electronics with the consequence realization and testing of active layers of $50\times 100$~cm$^2$ with embedded electronics, having been assumed this dimension as unit to cover the large experimental area.  At the same time, an effort on the development of the HCAL DAQ and the mechanical infrastructure (mechanical integration, powering and cooling) will start and will last up to 2035, year where a slice of the full system will be fully operative.  The software integration of such detector within the 10 TeV Muon Collider experiments framework  will continue up to 2035.

Among the MPGD technologies, $\mu$-RWELL uses fluor based gas ($CF_4$) which has large GWP; in the case where $\mu$-RWELL will be proved be the best detector technologies, studies on  a $CF_4$ substitute will be carried on in collaboration with DRD1-WG3.
\FloatBarrier

\subsection{Si-W ECAL + Scintillator based HCAL }
The design of a calorimeter for a muon collider must effectively distinguish between photons and neutrons generated by beam muon decays upstream of the interaction point (IP) and the prompt, high-energy electromagnetic and hadronic showers produced by the $\mu^{+}\mu^{-}$ collisions at the IP. The beam-induced background (BIB) primarily consists of low-energy photons ($p_{\gamma} \approx 2$\,MeV) and moderately energetic neutrons ($p_{n} \approx 500$\,MeV), which contribute significantly to the calorimeter occupancy. A substantial portion of the BIB can be suppressed using a short readout window—on the order of a few nanoseconds or less—centered around the nominal beam crossing. Achieving this requires rapid signal formation in the front-end electronics. Additionally, high lateral granularity enables advanced pattern recognition techniques to reject the non-pointing background from low-energy BIB photon and neutron showers. The calorimeter's active materials and front-end electronics must also withstand the 1-MeV-equivalent neutron fluence (somewhat lower than expected at the HL-LHC) and total ionizing dose (worse than at the HL-LHC) associated with this BIB.

Four four key parameters are identified that will enable good energy measurements at a muon collider experiment:

\begin{itemize}
    \item \textbf{High granularity} with cell size of $\sim cm^{2}$ to reduce the overlap of BIB particles in the same calorimeter cell. The overlap can produce hits with an energy similar to the signal, making harder to distinguish it from the BIB;
    \item \textbf{Good timing} to reduce the out-of-time component of the BIB. An acquisition time window of about $\Delta t = 300$~ps could be applied to remove most of the BIB, while preserving most of the signal. This means that a time resolution in the order of $\sigma_t = 100$~ps (from $\Delta t \approx 3 \sigma_t$) should be achieved. The CMS BTL based on LYSO:Ce bars with SiPM readout on each end integrated a precision $30$ ps timing layer with a high resolution crystal calorimeter~\cite{Addesa:2022qlt};
    \item \textbf{Longitudinal segmentation}: the energy profile in the longitudinal direction is different between the signal and the BIB, hence a segmentation of the calorimeter can help in distinguishing the signal showers from the fake showers produces by the BIB;
    \item \textbf{Good energy resolution} of $\frac{10\%}{\sqrt{E}}$ in the ECAL system for photons and a jet energy resolution of $\frac{35\%}{\sqrt{E}}$ are expected to be enough to obtain good physics performance. These have been demonstrated for conceptual particle flow calorimeters.
\end{itemize}

The design and technology of silicon-tungsten calorimeters are highly advanced, having been originally proposed by the CALICE collaboration and extensively studied for use in electron-positron Higgs factories. These studies have demonstrated outstanding performance, reinforcing the viability of this technology. Functional prototypes already exist, and the construction of a full-scale High-Granularity Calorimeter (HGCal)  calorimeter for the CMS experiment is currently underway.

Silicon-tungsten (SiW) calorimeters remain a highly attractive choice for muon collider detector applications due to their exceptional radiation hardness, high granularity, and precise timing resolution, which are essential for 4D event reconstruction. The SiW calorimeter is currently the default option for the MAIA detector, benefiting from the significant expertise gained through the development and construction of HGCal. The experience gained from HGCal will be instrumental in advancing the SiW technology to meet the unique challenges posed by a muon collider environment. Specifically, the technology should be able to withstand radiation doses, provide necessary energy and timing resolutions, and provide means to deliver the data from the front-end electronics to the back-end for online and offline reconstruction. 

Over the past decade, advances in precision timing and high-granularity calorimetry have brought a viable muon collider calorimeter design within reach. Prototype silicon sensors developed for the SiD detector—one of two validated designs for the e+e- colliders have achieved cell sizes as small as 13 mm$^{2}$. These prototypes have been successfully tested with cosmic rays and test beam electrons. Similar silicon sensors are being developed for the CMS detector’s High Granularity Calorimeter (HGCal) upgrade at the HL-LHC, featuring cell sizes of approximately 0.5 and 1.1 cm$^{2}$. Positron test beams have demonstrated that these sensors can achieve an electromagnetic (EM) energy resolution of 22\%/$\sqrt{E[\si{GeV}]}\,\bigoplus\ 0.6$\%. While this resolution is not as high as that of homogeneous crystal calorimeters, it is well-suited for particle flow calorimetry, optimizing both jet energy resolution and BIB rejection.

For hadronic calorimetry, compact arrays of scintillator or crystal tiles, directly bonded to printed circuit boards (PCBs) containing silicon photomultipliers (SiPMs), have emerged as a scalable and cost-effective solution. The CALICE Collaboration, which focuses on developing particle flow calorimetry for future linear $\mathrm{e^+e^-}$ colliders, demonstrated in the early 2000s that the “SiPM-on-tile” technology can achieve hadronic energy resolutions of approximately 60\%/$\sqrt{E[\si{GeV}]}$. A 2018 pion test beam validated the feasibility of constructing and operating a large-scale hadronic calorimeter prototype, which measured $72\,\mathrm{cm} \times 72\,\mathrm{cm} \times 38$ layers and contained 22,000 $3 \times 3\,\mathrm{cm}^{2}$ channels. 

Significant progress has also been made in precision timing and high-speed front-end electronics. Silicon based detectors have demonstrated a timing resolution of 30 ps for minimum-ionizing particles in test beams. A variety of scintillators with ultrafast decay times—ranging from sub-nanosecond to 10~ns—have been identified and characterized in terms of light yield, timing performance, and radiation hardness. 

Increasing the size of silicon wafers could lead to significant cost reductions, making large-scale deployment more economically viable. Investigating this option, along with potential optimizations in detector assembly techniques, will be crucial for developing an efficient and cost-effective calorimeter. The unique bunch structure of a muon collider beam compared to the LHC introduces new challenges in data readout, compression, and transmission. The timing characteristics of the muon collider necessitate the study of novel readout schemes to efficiently handle the data flow while minimizing latency and power consumption. Different compression and transmission strategies must be explored, with the goal of identifying a couple of viable solutions that can be implemented in the final detector design. 

\section{GEM and PICOSEC}

Two technologies are being developed aiming for an installation in the muon system: gas electron multipliers (GEM) and PICOSEC.

The technology used in GEM detectors is very mature (TRL8--9) and has been deployed in large scale detectors, such as CMS. It is already possible to build large modules of about 2~m $\times$ 1~m size that can operate at high-rate ($>100$~kHz/cm$^2$), have good radiation resistance and can operate with low-global warming potential gases. The main R\&D work is focused on demonstrating the operation with gas-recycling systems to further reduce the GWP. 
However, the time resolution of GEM detectors is of the order of 4-5~ns, which makes this technology impractical for instrumenting regions of the detector irradiated by large fluxes of BIB particles, such as the forward region.

The PICOSEC technology Micromegas detector~\cite{Bortfeldt:2017daz} combines a Cherenkov radiator, a photocathode and a Micromegas-based amplification stage into a high-precision timing detector. Incoming particles create Cherenkov light in the radiator, which is converted to primary photoelectrons by the photocathode. These electrons are pre-amplified in a drift region and finally amplified by the Micromegas. A time resolution of 24~ps for Minimum Ionising Particles (MIPs) has been measured in several test beam campaigns in recent years.

The current size of the PICOSEC detectors (10~cm $\times$ 10~cm) has to be scaled up to improve the cost-effectiveness when used in the detectors and to ease integration. The scaling size is driven by the radiator crystal dimensions and MM uniformity. 
The current state of the art suggests the tiles configuration as the best viable option. Studies are underway as concerning optimization of tile coupling in the mechanical frame.

Substantial PICOSEC R\&D efforts are ongoing targeting functioning with eco-gases, photocatode materials, gap thickness, detector fields, improvement of stability (with use of resistive technologies like $\mu$-RWELL, resistive (DLC) photocatodes), and multichannel readout.

The foreseen timescales for the project are reported in Table~\ref{tab:ppg:picosec}. 
\begin{table}[!h]
    \centering
    \begin{tabular}{|l|l|}
    \hline
    Milestone & Timescale \\
    \hline
       TRL3 for large size detector (2m $\times$ 1m) & $t_0$ + 3/4 years \\
       TRL5 for large size detector  & $t_0$ + 5/6 years (following R\&D assessments) \\
       Initial prototype (lab test, beam test) & $t_0$ + 7/8 years\\
       Slice of full system & $t_0$ + 10/11 years \\
    \hline
    \end{tabular}
    \caption{Expected timescales to develop a large-scale PICOSEC detector.}
\label{tab:ppg:picosec}
\end{table}

The main PICOSEC results in prototypes were obtained with a high-GWP gas mixture: Ne, C$_2$H$_6$, CF$_4$ (80/10/10) with GWP 740 driven by CF$_4$. R\&D work is targeting the use of lower GWP gases. For example, using a mixture of Ne and iC$_4$H$_10$ (90/10), with GWP 0.34, demonstrated time resolutions compatible with standard mixtures.

\chapter{Computing}

This section describes which assumptions about running and data taking were used in our resource estimates.
Given that our current reconstruction and simulation pipelines are far from optimised, we based some of our assumptions on the projections of ATLAS~\cite{atlas-hl-hlc-comp-cdr} and CMS~\cite{cms_comp_model_update_22,
cms-pub-comp-res} for HL-LHC.
Where possible, we have cross-checked this with timing studies in our current workflows. Projecting those into the future, we roughly arrive in the same order of magnitude.
We have chosen the HL-LHC as basis for computing projections because we assume that the experimental conditions at PU 200 come closest to what we expect in the presence of BIB.
Depending on the actual lattice of the accelerator we expect roughly 1 to 1.5 million hits / clusters in the tracking detectors, which is about a factor of 2-3 more compared to what the ATLAS ITk is projected to handle~\cite{ATLAS:2024rnw}.
Additionally, we have for now assumed that the higher granularity of our calorimeter system does not significantly affect the reconstruction time, or at least not be a driving factor for it.
This seems to be consistent with the fact that CMS quotes below 10~\% of reconstruction time that is spent in HGCAL reconstruction~\cite{cms_comp_model_update_22}.
Given that the final granularity is still under study, we don't think we can make a more detailed estimate and will have to return to this once it becomes possible.

In all of the projections below we have assumed that storing and processing data for calibration purposes is negligible and that it is already encompassed in the foreseen (re-)processing scheme.

We would also like to point out that almost all of the projections and estimates below are based on the assumption that there is a more or less seamless transition from learnings from HL-LHC to our software stack.
This is most likely optimistic, as simply porting solutions from HL-LHC will require non-negligible amounts of personpower, which might still be bound by HL-LHC operations until its shutdown.
Furthermore, it is not obvious that all the solutions will apply directly.

For simulation, we assumed that roughly ten times the number of events that are recorded should be generated and simulated.
Based on previous simulation studies and a target integrated luminosity of 10~$\text{ab}^{-1}$, we arrive at a number of $10^{11}$ simulated events.

\section{CPU resources}

Table~\ref{A:computing-resources:tab:running-cpu} shows the projected per event times for different processing steps as well as the total projected resource requirements to run them.
More detailed assumptions and explanations are given in the text below.
Assuming that all recorded events and simulated events are re-processed twice, we project the total amount of CPU resources to be about 20000 kHEPScore\footnotemark[1]-years.
The large majority of time in this case is spent in the (prompt) reconstruction of data events.

\begin{table}[h!]
\centering
\begin{tabular}{l r r r}
    & HEPscore\footnotemark[1] s / event & events & total CPU [kHEPscore\footnotemark[1]-years] \\ \hline
  Generation\footnotemark[2] & 640 & \multirow{3}{*}{$10^{11}$} &  $2.03 \cdot 10^3$ \\
  Simulation\footnotemark[3] & 550 & & $1.74 \cdot 10^3$ \\
  Reconstruction (sim)\footnotemark[4] & \multirow{2}{*}{1130} & & $3.58 \cdot 10^3$ \\
  Reconstruction (data)\footnotemark[4] \footnotemark[5] &  & $5\cdot 10^{12}$ & $1.79 \cdot 10^5$\\ \hline
  Re-processing (sim) & 1680 & $2 \cdot 10^{11}$ & $1.07 \cdot 10^4$ \\
  Re-processing (data) & 1130 & $2 \cdot 5 \cdot 10^{9}$ & $7.17\cdot10^2$ \\ \hline
  \textbf{Total} & & & \boldmath{$1.98\cdot{10}^{5}$}
\end{tabular}
\caption{Projected execution times per event (in HEPscore seconds) for different steps of the processing chains, the number of events they are projected to be run on and the resulting total (in kHEPscore-years) time for processing them once. The assumptions for these numbers are explained in more details in the text.}
\label{A:computing-resources:tab:running-cpu}
\end{table}

\section{Storage}

Table~\ref{A:computing-resources:tab:running-storage} shows the estimated storage needs for a five year running period.
For these projections we have assumed that we will operate on a three data tier model, where the raw data from the detector (RAW) will be stored on tape.
This will be accompanied by a RECO tier (labeled RECO / AOD in Table~\ref{A:computing-resources:tab:running-storage}) which should still have all the necessary information to re-run large parts of the reconstruction.
From this RECO tier we aim at a very lightweight analysis tier that should support the large majority of all analyses, comparable to the current DAOD\_PHYSLITE or nanoAOD formats from ATLAS and CMS.
We don't foresee any sizable contributions from intermediate data tiers.

As shown in Table~\ref{A:computing-resources:tab:running-storage} the large majority of storage needs is driven by simulation.
Clearly it will not be possible to store all of the simulated events at SIM level (corresponding to RAW) on tape, but even the reconstructed MC events (SIM RECO) are close to what HL-LHC will produce for ATLAS and CMS.
Storing only data should be manageable and the analysis tiers are projected to be almost negligible in total size.

\begin{table}[h]
    \centering
   \begin{tabular}{l r r r}
        & size [MB] / event & events & total size [PB] \\ \hline
    RAW\footnotemark[6] & 80 & \multirow{3}{*}{$5\cdot10^{9}$} & 400 \\
    RECO / AOD\footnotemark[7] \footnotemark[8] & 20 & & 100 \\
    analysis\footnotemark[9] & 0.005 &  & 0.03 \\
    SIM\footnotemark[10] & 250 & \multirow{3}{*}{$10^{11}$} & 25000 \\
    SIM RECO\footnotemark[11] & 40 & & 4000 \\
    SIM analysis\footnotemark[11] & 0.01 & & 1.0 \\ \hline
    \textbf{Total} & & & \boldmath{$29501$} \\
   \end{tabular} 
   \caption{Projected storage requirements for the different foreseen data tiers per event and total after five years of running. The assumptions and numbers of events for the different data tiers are explained in more details in the text.}
   \label{A:computing-resources:tab:running-storage}
\end{table}

\subsection*{Assumptions and other details}
\begin{itemize}
    \item[{\footnotemark[1]}] For the numbers that are based on HL-LHC projections we simply assume that HS06 translates to HEPscore values one-to-one, i.e. we take the HS06 values and quote them as HEPscore here.
    For numbers we obtained in our own studies we applied the corresponding factors to the measured CPU runtimes to arrive at HEPscore values.
    \item[{\footnotemark[2]}] The event generation time (of the hard scatter event) is effectively assumed to be the same as the one that CMS projects~\cite{cms_comp_model_update_22}. 
    We haven't got a better estimate here, but the lower complexity in QCD diagrams might be at least partly counteracted by the increased number of EW diagrams to consider.
    \item[{\footnotemark[3]}] This is an average time of projections of CMS full simulation~\cite{cms_comp_model_update_22}, and two ATLAS fast sim methods~\cite{atlas-hl-hlc-comp-cdr}.
    For the muon collider the simulation of the hard scatter event is assumed to be rather quick, but the overlay of BIB will be a main challenge. 
    We assume that we should be able to leverage many of the lessons learned in pile-up (pre)-mixing at HL-LHC.
    \item[{\footnotemark[4]}] Average of ATLAS projections for PU-200~\cite{atlas-hl-hlc-comp-cdr} and CMS projections for PU-140 and PU-200~\cite{cms_comp_model_update_22}. 
    We note that already between these estimates there is a factor of 5 between ATLAS and CMS projections.
    This is about a factor of 2 faster than our currently available reconstruction (including BIB).
    \item[{\footnotemark[5]}] For these projections we assume that we read out every event at 100~kHz, even though we only store a fraction of them afterwards.
    \item[{\footnotemark[6]}] Based on the estimates of roughly 40~MB for the tracking detector and another 40~MB for the calorimeter system used in~\cite{Accettura:2023ked}. Estimating the event size by multiplying HL-LHC projections by a factor of 10 would arrive at 50~MB~\cite{cms_comp_model_update_22,CERN-LHCC-2022-005}.
    \item[{\footnotemark[7]}] Following estimates from HL-LHC that suggest that a reduction from RAW to RECO / AOD by a factor 3 - 5 is possible.
    \item [{\footnotemark[8]}] We assume that there will be some more events stored in this data tier that will not be available as RAW on tape, but since there is no well established operational model yet, we consider them negligible in size for now.
    \item[{\footnotemark[9]}] Assuming that we can get something similar to DAOD\_PHYSLITE or nanoAOD in size.
    \item[{\footnotemark[10]}] This already assumes that R\&D is finished and that we can trim MC information quite heavily. Current simulation studies have events that are around 400~MB / event. A factor 3 between data and MC also seems to be in line with current experience and projections from ATLAS and CMS.
    \item[{\footnotemark[11]}] Here we assume that we can trim everything and are effectively left with twice the size due to the double structure in MC and data.
\end{itemize}

\part*{Acknowledgements}
This work was supported by the EU HORIZON Research and Innovation Actions under the grant agreement number 101094300.
Funded by the European Union (EU). Views and opinions expressed are however those of the author(s) only and do not necessarily reflect those of the EU or European Research Executive Agency (REA). Neither the EU nor the REA can be held responsible for them.

\printbibliography

\end{document}